\documentclass[twocolumn,tighten,times]{aastex631}
\hypersetup{linkcolor=magenta,citecolor=blue,filecolor=cyan,urlcolor=cyan}
\usepackage{rotating}    % For rotating elements
\usepackage{tablefootnote}
\usepackage{hyperref}
\usepackage{enumitem}%,showframe}
\usepackage{mathrsfs}
\usepackage{amsmath}

\usepackage{graphicx}    % For rotating elements
\usepackage{longtable}   % For tables that span multiple pages

\received{---}
\revised{----}
\accepted{----}

\begin{document}

\title{Broad-Line Region Characterization in Dozens of Active Galactic Nuclei Using Small-Aperture Telescopes}

\author[0000-0001-9704-690X]{Catalina Sobrino Figaredo}
\affiliation{Department of Physics, Faculty of Natural Sciences, University of Haifa, Haifa 3498838, Israel}
\affiliation{Haifa Research Center for Theoretical Physics and Astrophysics, University of Haifa, Haifa 3498838, Israel}

\author[0000-0002-4830-7787]{Doron Chelouche}
\affiliation{Department of Physics, Faculty of Natural Sciences, University of Haifa, Haifa 3498838, Israel}
\affiliation{Haifa Research Center for Theoretical Physics and Astrophysics, University of Haifa, Haifa 3498838, Israel}

\author[0000-0002-7284-0477]{Martin Haas}
\affiliation{Ruhr University Bochum, Faculty of Physics and Astronomy, Astronomical Institute (AIRUB), 44780 Bochum, Germany}

\author[0000-0001-8391-0883]{Michael Ramolla}
\affiliation{Carl Zeiss Industrielle Messtechnik GmbH,Carl-Zeiss-Str. 4-54,73447, Oberkochen, Germany}

\author[0000-0002-9925-534X]{Shai Kaspi}
\affiliation{School of Physics and Astronomy and Wise Observatory, Raymond and Beverly Sackler Faculty of Exact Sciences, Tel-Aviv University, Tel-Aviv 6997801, Israel}

\author[0000-0002-5854-7426]{Swayamtrupta Panda}\thanks{Gemini Science Fellow and CNPq Fellow}
\affiliation{Laborat\'orio Nacional de Astrof\'isica, MCTI, Rua dos Estados Unidos, 154, Bairro das Na\c c\~oes, Itajub\'a, MG 37501-591, Brazil}
\affiliation{International Gemini Observatory/NSF NOIRLab, Casilla 603, La Serena, Chile}

\author[0009-0008-3300-0089]{Martin W. Ochmann}
\affiliation{Ruhr University Bochum, Faculty of Physics and Astronomy, Astronomical Institute (AIRUB), 44780 Bochum, Germany}
\affiliation{Institut für Astrophysik und Geophysik, Universität Göttingen, Friedrich-Hund Platz 1, D-37077 Göttingen, Germany}

\author[0000-0003-3173-3138]{Shay Zucker}
\affiliation{Department of Geosciences, Raymond and Beverly Sackler Faculty of Exact Sciences, Tel Aviv University, Tel Aviv 6997801, Israel}

\author[0000-0002-7538-3072]{Rolf Chini}
\affiliation{Ruhr University Bochum, Faculty of Physics and Astronomy, Astronomical Institute (AIRUB), 44780 Bochum, Germany}
\affiliation{Universidad Católica del Norte, Instituto de Astronomía, Avenida Angamos 0610, Antofagasta, Chile}
\affiliation{Polish Academy of Sciences, Nicolaus Copernicus Astronomical Center, Bartycka 18, 00-716 Warszawa, Poland}

\author[0009-0007-2929-8639]{Malte A. Probst}
\affiliation{Institut für Astrophysik und Geophysik, Universität Göttingen, Friedrich-Hund Platz 1, D-37077 Göttingen, Germany}

\author[0000-0002-0417-1494]{Wolfram Kollatschny}
\affiliation{Institut für Astrophysik und Geophysik, Universität Göttingen, Friedrich-Hund Platz 1, D-37077 Göttingen, Germany}

\author{Miguel Murphy}
\affiliation{Universidad Católica del Norte, Instituto de Astronomía, Avenida Angamos 0610, Antofagasta, Chile}



\begin{abstract}

We present the results of a nearly decade-long photometric reverberation mapping (PRM) survey of the H$\alpha$ emission line in nearby ($0.01\lesssim~z~\lesssim0.05$) Seyfert-Galaxies using small ($15\,\mathrm{cm}-40\,\mathrm{cm}$) telescopes. Broad-band filters were used to trace the continuum emission, while narrow-band filters tracked the H$\alpha$-line signal.  We introduce a new PRM formalism to determine the time delay between continuum and line emission using combinations of auto- and cross-correlation functions. We obtain robust delays for {33/80} objects, allowing us to estimate the broad-line region (BLR) size. Additionally, we measure multi-epoch delays for 6 objects whose scatter per source is smaller than the scatter in the BLR size-luminosity relation. Our study enhances the existing H$\alpha$ size-luminosity relation by adding high-quality results for 31 objects, whose nuclear luminosities were estimated using the flux-variation gradient method, resulting in a scatter of 0.26\,dex within our sample. The scatter reduces to 0.17\,dex when the 6 lowest luminosity sources are discarded, which is comparable to that found for the H$\beta$ line. Single-epoch spectra enable us to estimate black hole masses using the H$\alpha$ line and derive mass accretion rates from the iron-blend feature adjacent to H$\beta$. {A similar trend, as previously reported for the H$\beta$ line, is implied} whereby highly accreting objects tend to lie below the size-luminosity relation of the general population. Our work demonstrates the effectiveness of small telescopes in conducting high-fidelity PRM campaigns of prominent emission lines in bright active galactic nuclei.

\end{abstract}


\section{Introduction}

Active galactic nuclei (AGN) are among the most luminous sources in the universe \citep{2017A&ARv..25....2P}, and mark a phase in the lifetime of galaxies, where their central supermassive black holes (SMBH) are rapidly growing by the accretion of material from their immediate environs \citep{1982MNRAS.200..115S,2022MNRAS.513.4770F}. Therefore, AGN provide a means by which the cosmic SMBH census may be quantified out to high redshifts, and SMBH-galaxy co-evolution may be traced \citep{2013ARA&A..51..511K}. This has profound implications for understanding SMBH growth in the general context of structure formation in the universe, as well as for gravitational-wave signals in the NANOGrav and LISA era \citep{NanoGRAV, 2022arXiv220306016A}.

SMBH mass estimations in AGN are based on two observables: the size of the broad line region, $r_\mathrm{BLR}$, and a measure of the velocity dispersion of the lines, $\sigma_\mathrm{BLR}$. Assuming that the latter is a proxy for the virial speed at the location of the line-emitting gas, then up to (often loosely constrained) geometrical factors, the SMBH mass may be estimated as, $M_\mathrm{SMBH}\sim \sigma_\mathrm{BLR}^2r_\mathrm{BLR}/G$, where $G$ is Newton's gravitational constant \citep{2000ApJ...540L..13P}. While $\sigma_\mathrm{BLR}$ may be estimated from the mean or the root-mean-square (RMS) spectrum of the target, $r_\mathrm{BLR}$ is too small to be spatially resolved in all but for a few cases and particular emission lines \citep[e.g.,][]{2018Natur.563..657G}, and reverberation mapping is the main means by which the BLR size is assessed \citep{2013ApJ...767..149B}.

Reverberation mapping (RM), as applies to AGN, is a technique in which light echoes are used to place constraints on the geometrical attributes of the BLR \citep[][and references therein]{2021iSci...24j2557C}. This is supported by the fact that the BLR is photoionized by the AGN continuum emission, and line emission therefore responds to continuum fluctuations. In its simplest form, RM is used to measure the size of the BLR by quantifying the time-delay (the first moment of the transfer function), $t_\mathrm{BLR}$, between line and continuum emission. This timescale is interpreted as the light crossing time across the BLR since radiative reprocessing time-scales are comparatively very short due to the high gas densities involved, and since en-route photon diffusion is negligible at the implied optical depths \citep{1993PASP..105..247P}. Therefore, $r_\mathrm{BLR}=ct_\mathrm{BLR}$, where $c$ is the speed of light. 

Meaningful time-delay measurements between line and continuum emission are available for $\sim 10^2$ sources and are mostly associated with the H$\beta$ line \citep{2004ApJ...613..682P, 2023ApJ...948L..23W}, which is easily observable from the ground, and for which the host-galaxy contamination may be mitigated using the adjacent [OIII]\,$\lambda 5007$ line \citep{1992PASP..104..700V}. These revealed an intriguing relation between the size of the BLR and the source optical luminosity, $L_\mathrm{opt}$, which is consistent with $r_\mathrm{BLR}\propto L_\mathrm{opt}^{1/2}$ \citep{2013ApJ...767..149B}. This relation forms the basis for current prescriptions for estimating SMBH masses also in sources for which RM results are unavailable, thereby leading to the current cosmic SMBH census \citep{2004ApJ...601..676V,2009ApJ...699..800V,2012AdAst2012E...7K}. 
{Additionally, radius-luminosity relationships have also been established for continuum wavebands beyond the optical, such as X-rays \citep{2010ApJ...723..409G} and infrared \citep{2011MNRAS.413L.106L}.}
It has been further proposed that the size-luminosity relation may be used as a standard ruler for cosmology \citep[][and references therein]{2011ApJ...740L..49W,2019ApJ...883..170M, 2023FrASS..1030103P}. 

There are, however, several limitations to the aforementioned use of quasars for cosmology and the SMBH census: 1) there is non-negligible scatter in the size-luminosity relation which is further exacerbated when attempting to cross-calibrate the relations for different sets of lines \citep{2016MNRAS.461..647C}, 2) different AGN sub-types may be characterized by different size-luminosity relations \citep{2016ApJ...825..126D}, 3) local size-luminosity relations for low-luminosity sources are extrapolated to much more luminous targets and at high redshifts, where they may not hold, at least not without further corrections \citep[e.g.,][who considered corrections due to the Eddington ratio]{2020ApJ...903..112D}. Interestingly, RM campaigns of individual sources, which undergo long-term luminosity variations, show an intrinsic size-luminosity relation for their BLR, whose slope may be different than the extrinsic (multi-source) relation, and shedding light on the physical origin of the BLR \citep{2016ApJ...827..118L}. Therefore, increasing the sample size and diversity of AGN with good RM data, with repeated campaigns for individual sources, is crucial for properly assessing the SMBH census, and uncovering AGN physics. 

To perform RM and obtain line-to-continuum time-delays, high-cadence spectroscopic observations are often employed whereby spectroscopic decomposition is used to separate line and continuum signals at each epoch, and multiple epochs are used to form their respective light curves \citep{2000ApJ...533..631K}. The need for high-cadence spectroscopic data means that RM is a challenging technique to implement, especially for faint targets, as telescope time is scarce. This shortcoming of RM motivated several groups to propose photometric RM (PRM) as a means to achieve high-quality RM using smaller telescopes for which telescope time is more abundant \citep{2011AaA535A73H,2012ApJ...747...62C}. Specifically, both narrowband (NB) and broadband photometric RM versions have been proposed and implemented for a handful of sources, as well as statistically for numerous sources using general-purpose survey data  \citep{2023MNRAS.522.2002P,2024ApJ...968L..16P}{}. 

The typical implementation of PRM requires monitoring in two bands\footnote{Single-band versions of PRM have been proposed, which rely on the statistical properties of quasar continuum variability \citep{2016ApJ...819..122Z}}, one of which is used as a proxy for the time-varying ionizing continuum, and the other includes in addition to the ionizing continuum also the contribution of an emission line \citep{2023A&A...675A.163C}. To constrain the line-to-continuum time-delay one may use the slew of statistical tools available for spectroscopic RM if the pure line light-curve may be extracted based on prior knowledge of the relative contribution of the emission line to the band \citep{2011AaA535A73H}. Often, however, all photometric bands contain a finite contribution of emission lines to their signal, and the contribution of the varying component of the emission line to the line-rich band is poorly constrained. In such cases, other statistical measures may be used to simultaneously constrain both the line-to-continuum time-delay and the relative contribution of the line to the band \citep[][but see \citealt{2012ApJ...747...62C} for the case of broadband data]{2013ApJ...769..124C,2016ApJ...819..122Z}. 

Here we present the results of PRM campaigns on 80 AGN targeting the H$\alpha$ line, which is significantly brighter than H$\beta$ and lies in a spectral range relatively free of iron-emission contamination. The campaigns were performed between 2010 and 2018. Each campaign typically lasts 5-6 months, and 12 AGN have been monitored in 2-3 campaigns. %a decade-long PRM campaign of 40 AGN, several of which have multiple epochs $t_\mathrm{BLR}$ measurements. 
Section~\ref{sec:sample_obs_lc} describes the sample selection, the available observations, and the light-curves extraction and their characteristics. In Section~\ref{sec:methods} we outline the methods used: the correlation formalism used for PRM time lag determination, the host-galaxy subtraction, and black hole mass estimation. Results for the AGN luminosities, H$\alpha$ BLR sizes, BH masses as well as size-luminosity relation for the sample are given in Section~\ref{sec:results}. We discuss the implications of our results for AGN physics and provide a summary in Section~\ref{sec:summary}.

\section{Data}\label{sec:sample_obs_lc}

\subsection{The Sample} \label{sec:sample_selection}

Our sample is selected from the 13th edition of the AGN catalog of \cite{2010A&A...518A..10V} using the following criteria: 1) object visibility at our observatory on Cerro Murphy\footnote{Cerro Murphy is a subsidiary summit of Cerro Armazones in the Chilean Atacama desert. Several telescopes have been installed by Universidad Cat\'olica del Norte and Ruhr-University Bochum since 1995. Since 2020 the observatory is run by the Nikolaus Kopernikus Center, Warsaw, Poland. Hosted within ESO’s Paranal Observatory, now its name is {\it Rolf Chini Cerro Murphy Observatory}.}; 2) $\mathrm{V}<16$\, mag (including the host galaxy); 3) the redshift range is such that H$\alpha$ shifts into available narrow-band filters centered at 6700\AA, 6800\AA, 6900\AA, or 6720\AA ([SII]). This yields a sample of 80 Seyfert galaxies with $0.015 < z < 0.05$ and declination $\lesssim +25^\circ$, which were monitored with varying degree of success. 

We acquired data for a total of 80 AGN. In the following analysis, we rejected objects for which the observational conditions were not favorable and the observation cadence was sub-optimal. Approximately 60 objects remained viable for reverberation analysis. After filtering out objects with noisy light curves and low variability {(those for which the fractional variability F$_{\rm var}$ is lower than 1\% for both continuum and NB light curves; see Appendix~\ref{app:lc} for details)}, the final sample consists of 48 objects, with 12 of them observed across multiple epochs. {This results} in a total of 65 RM light curves to be analyzed. Details about filtering the original dataset, as well as information on variability and cadence, are provided in Appendix~\ref{app:lc}. 
 Table~\ref{tab:sample} lists the sample with the redshift and luminosity distance values taken from \href{https://ned.ipac.caltech.edu}{NED}\footnote{The NASA/IPAC Extragalactic Database (NED) is funded by the National Aeronautics and Space Administration and operated by the California Institute of Technology.} assuming a cosmology of $H_0 = 67.8\,\mathrm{km~s^{-1}Mpc^{-1}}$, $\Omega_M = 0.308$, $\Omega_{{\Lambda}} = 0.692$ \citep{2016A&A...594A..13P}. Galaxy type and $V$ magnitudes were taken from \cite{2010A&A...518A..10V}. We note that independent reverberation-based SMBH masses have been reported in the literature for $\sim 10$ sources in our sample (see below). 

\begin{table*}
\caption{Summary of the observation sample (alphabetically ordered). Redshifts and Luminosity distances were taken from NED, and source type and $V$-mag values from \protect\citet{2006AaA...455..773V}. {The estimated fraction of the broad H$\alpha$ line captured by the NB filter H$\alpha_{\rm NB}$ is given in percentage and the available} spectroscopic data for H$\alpha$ is denoted in the last column.} 
\label{tab:sample} 
\renewcommand{\arraystretch}{0.65}
\begin{tabular*}{\textwidth}{lp{1.2cm}p{1cm}p{0.5cm}p{0.7cm}p{0.7cm}p{0.7cm}p{0.7cm}p{0.7cm}cccl}
\toprule
{Object} & {$z$} & {D$_{\rm L}$} & \multicolumn{3}{c}{RA} & \multicolumn{3}{c}{DEC} & {Type} & {Vmag} & {H$\alpha_{\rm NB}$}&{Available} \\
{} & {} & {[Mpc]} & {h} & {min} & {sec} & {$^{\circ}$} & {\arcmin} & {\arcsec} & {} & {} & {[\%]}&{spectra} \\
\hline
1H2107-097 & 0.02698 & 117 & 21 & 09 & 9.9 & -09 & 40 & 15 & S1.2 & 14.39 & 30&6dF NED\\
3C120 & 0.03301 & 149 & 04 & 33 & 11.1 & +05 & 21 & 15 & S1.5 & 15.05 & 54& FAST/HET\\
AKN120 & 0.03271 & 148 & 05 & 16 & 11.4 & -00 & 08 & 59& S1 & 14.59 & 55&6dF/HET\\
CTSG03\_04 & 0.04002 &181&  19 &38 & 04.3& -51 & 09 & 49 & S1.2&15.2 & 54&6dF NED\\
ESO141-G55 & 0.03711 & 168 & 19 & 21 & 14.3 & -58 & 40 & 13 & S1.2 & 13.64 & 53 &SALT/6dF\\
ESO323-G77 & 0.01501 & 71.3 & 13 & 06 & 26.2 & -40 & 24 & 52 & S1.2 & 13.42 & 36&BAT DR1 \\
ESO374-G25 & 0.02367 & 111 & 10 & 03 & 23.6 & -37 & 33 & 39 & S1 & 15.29 & 40&SALT\\
ESO399-IG20 & 0.0250 & 110 & 20 & 06 & 58.1 & -34 & 32 & 55 & NLS1 & 14.51 &50 &SALT\\
ESO438-G09 & 0.02401 & 113 & 11 & 10 & 48 & -28 & 30 & 4 & S1 & 14.17 & 40& SALT\\
ESO490-IG26 & 0.02485 & 114 & 06 & 40 & 11.8 & -25 & 53 & 38 & S1 & 15 & 20& 6dF NED\\
ESO511-G030 & 0.02239 & 104 & 14 & 19 & 22.3 & -26 & 38 & 41 & S1 & 14.9 & 65& 6dF NED\\
ESO549-G49 & 0.02627 & 117 & 04 & 02 & 25.8 & -18 & 02 & 52 & S1 & 14.2 &13 & 6dF NED\\
ESO578-G09 & 0.03502 & 163 & 13 & 56 & 36.7 & -19 & 31 & 44 & S1 & 15.2 & 55& FAST\\
F1041 & 0.03347 & 148 & 23 & 17 & 30.2 & -42 & 47 & 05 & S1 &15.2 & 58 & 6dF NED\\
HE0003-5023 & 0.0345 & 149 & 0 & 05 & 43.1 & -50 & 06 & 55 & S1 & 14 &-- & (1)\\
HE1136-2304 & 0.0270 & 127 & 11 & 38 & 51.2 & -23 & 21 & 35 & CL & 17.4 &34 & SALT\\
HE1143-1810 & 0.03295 & 155 & 11 & 45 & 40.4& -18 & 27 & 15& S1.5& 14.7$^*$& 70& 6dF NED  \\
HE2128-0221 & 0.05248 & 236 & 21 & 30 & 49.9&  -02& 08 & 14& S1$^{}$ & 17.4$^{*}$& 48&6dF NED \\
IC4329A & 0.01605 & 75.9 & 13 & 49 & 19.3 & -30 & 18 & 34 & S1.2 & 13.66 & 46&6dF NED\\
IRAS01089-4743 & 0.02392 & 105 & 01 & 11 & 09.7 & -47 & 27 & 37 & S1& 14.53& 37&  6dF NED \\
IRAS09595-0755 & 0.055 & 246.9 & 10 & 02 & 0.1 & -08 & 09 & 41 & S1 & 14.64 & 60& FAST \\
IRAS23226-3843 & 0.03590 & 159 & 23& 25 &24.2 & -38& 26 & 49 & S1&14.24  & 38& BAT \\
MCG+03-47-002 & 0.04000 & 180 & 18 & 27 & 14.7 & +19 & 56 & 19 & S1& 15.3 & --& 6dF NED \\
MCG-02.12.050 & 0.03600 & 164 & 04 & 38 & 14.1 & -10 & 47 & 45 & S1 & 15 & 55&FAST \\
MRK1239 & 0.01993 & 94.7 & 09 & 52 & 19.1 & +01 & 36 & 44 & NLS1 & 14.39 & 70& 6dF (2) \\
MRK1347 & 0.04995 & 234 & 13 & 22 & 55.5 & +08 & 09 & 42 & S1 & 14.59 & 66&FAST\\
MRK335 & 0.02578 & 111 & 0 & 06 & 19.5 & +20 & 12 & 11 & NLS1 & 13.85 & 50& FAST\\
MRK509 & 0.0344 & 152 & 20 & 44 & 9.7 & -10 & 43 & 24 & S1.5 & 13.2 &57&FAST/SALT\\
MRK705 & 0.02879 & 135 & 09 & 26 & 3.3 & +12 & 44 & 3 & S1.2 & 14.6 &48 & KPNO/ (3)\\
MRK841 & 0.03642 & 168 & 15 & 04 & 1.2 & +10 & 26 & 16 & S1.5 & 14.27&48 &FAST\\
NGC1019 & 0.02434 & 106 & 02 & 38 & 27.4 & +01 & 54 & 28 & S1.5 & 14.95 & 37& FAST/SALT\\
NGC4726 & 0.02543 & 120 & 12 & 51 & 32.3 & -14 & 13 & 17 & S1 & 14.2 & -- & 6dF NED/(4){}\\
NGC5940 & 0.03408 & 157 & 15 & 31 & 18.1 & +07 & 27 & 27 & S1 & 14.9 & 54&FAST\\
NGC6860 & 0.01488 & 65.3 & 20 & 08 & 47.1 & -61 & 06 & 0 & S1.5 & 13.53 & 50 &6dF NED\\
NGC7214 & 0.02385 & 103 & 22 & 09 & 07.6 & -27 & 48 & 34 & S1.2 & 14.10 & 37&6dF NED\\
NGC7469 & 0.01627 & 67.2& 23 & 03 & 15.6& +08 & 52 & 26& S1.5& 13.04 & 55&BAT/(5) \\
NGC7603 & 0.02876 & 124 & 23 & 18 & 56.6 & +00 & 14 & 38 & S1.5 & 14.01 & 35& FAST/SALT\\
NGC985 & 0.04314 & 193 & 02& 34& 37.7& -08 & 47 & 15 & S1.5& 14.28 &45 & FAST \\ 
PG1149-110 & 0.0490 & 230 & 11 & 52 & 3.5 & -11 & 22 & 23 & S1.2 & 15.46 & 50 &BAT DR2 \\
PGC50427 & 0.02346 & 109 & 14 & 08 & 6.7 & -30 & 23 & 53 & S1.5 & 15.3 &66&SALT\\
PGC64989 & 0.01937 & 83.5 & 20 & 34 & 31.4 & -30 & 37 & 29 & S1 & 13.3&53 &FAST\\
RXSJ06225-2317 & 0.03778 & 174 & 06 & 22 & 33.4 & -23 & 17 & 42 & S1 & 14.85 & 60& FAST/SALT \\
RX J1103.2-0654 & 0.02606 & 123 & 11 & 03 & 15.8 & -06 & 54 & 10 & S1 & 13.34 & 40& FAST/SALT\\
RXSJ17414+0348 & 0.0230 & 103 & 17 & 41 & 28.1 & +03 & 48 & 51 & S1 & 15.3 & 50&SALT \\
UGC12138& 0.02509 & 107 & 22 & 40 & 17.0 & +08 & 03 & 14 & S1.8 & 14.45 & 30&BAT DR2 \\
UM163 & 0.03343 & 146 & 23 & 39 & 32.3 & -02 & 27 & 45 & S1.5 & 14.86&44 &6dF NED\\ 
WPVS48 & 0.0370 & 173 & 09 & 59 & 42.6 & -31 & 12 & 59 & NLS1 & 14.78 & 60&FAST/SALT\\ 
WPVS007 & 0.02861 & 127 & 0 & 39 & 15.9 & -51 & 17 & 1.5 & NLS1 & 15.28 & 45&6dF (2) \\ \hline
\end{tabular*}
\tablecomments{(1) \cite{2015AuA...580A.113T}, (2) \cite{2018AaA...615A.167C}, (3) \cite{2022ApJ...935...72M}, (4) \cite{2019ApJ...872..134Z}, (5)  \cite{2006ApJS..164...81M}. $^{*}$ Instead of $V$-mag, $B$-mag from \protect\cite{2009MNRAS.399..683J}.}
\end{table*}

\subsection{Photometric Observations \& Light-curve Extraction}

\begin{figure}
 \centering
 \includegraphics[width=\columnwidth]{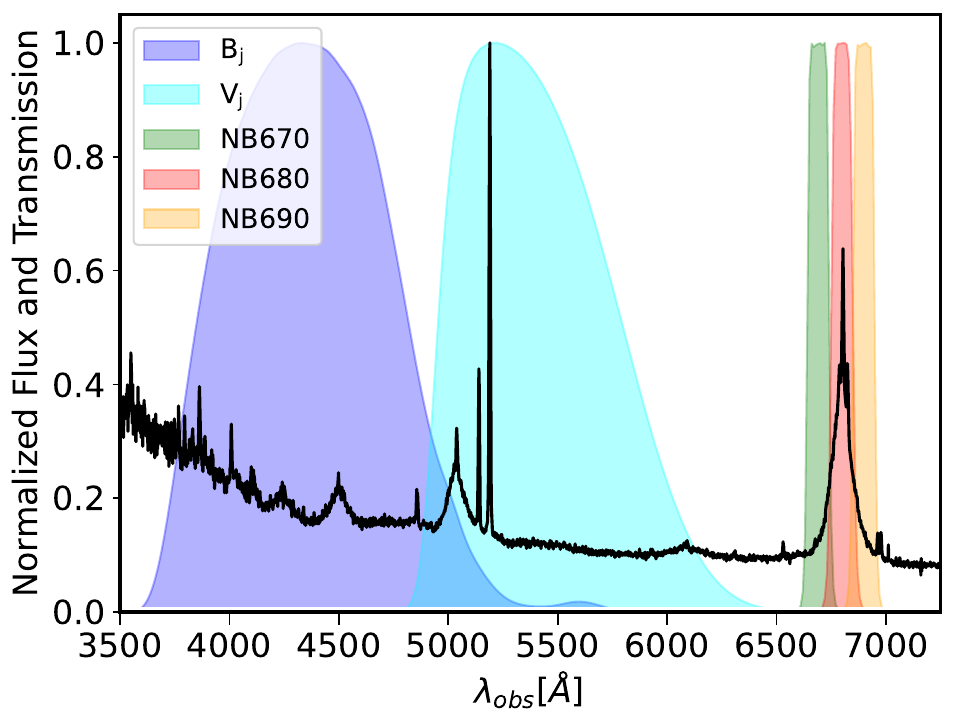}
 \caption{Normalized spectrum of Mrk\,841 at $z = 0.03642$. The normalized transmission for the different photometric filters are overplotted. The continuum signal is traced in this work by the broad bands, while the H$\alpha$ line is covered by the NB680 filter. The significantly different relative contributions of emission lines to the broad and narrow bands allows for PRM analysis. Note that partial coverage of the emission line may occur for specific targets (see also \S\,\ref{app:spec}).}
 \label{fig:mrk841_spc_filter}
\end{figure}

Photometric observations were robotically carried out at Cerro Murphy \citep{2016SPIE.9911E..2MR} using the RoBoT telescope, \citep[][a twin refractor with 15\,cm apertures, formerly called VYSOS~6]{2012AN....333..706H}, BESTII \citep[][a 25\,cm aperture telescope]{2009A&A...506..569K}, and  the BMT \citep[][a 40\,cm aperture telescope, formerly called VYSOS~16]{2013AN....334.1115R}. 

In our study the AGN continuum is tracked with the broadband Johnson filters $B$\,($\left < \lambda \right > \simeq $4330\AA) and $V$\,($\left < \lambda \right >\simeq $5500\AA). When available, we opt to use $B$-band data as the H$\beta$ line contribution is negligible. For $0.015<z<0.05$, H$\alpha$ emission may be traced with the available narrowband (NB) filters to various degrees of completeness; see Fig.~\ref{fig:mrk841_spc_filter} for the case of Mrk841 and Appendix \ref{app:spec} for all the sources in our sample. {We find that, on average, the H$\alpha$ contribution to the flux in the NB is comparable to that of the continuum.} Additionally, we carried out observations in the Johnson $R$ filter ($\left < \lambda \right >\simeq $7000\AA) and the $r_s$ Sloan filter ($\left < \lambda \right >\simeq $6200\AA).

The observations cover the years 2010-2018 with campaigns on individual sources being 5-6\,months long on average, which suffices to search for BLR delays $< 100$\,days, and hence consistent with the delays expected for our targets having an optical luminosity $L_{5100}=\lambda L_\lambda (5100\,\text{\AA}) < 10^{44.5}\,\mathrm{erg~s^{-1}}$ \citep[][but see \citealt{2024ApJS..272...26S}]{2000ApJ...533..631K,2013ApJ...767..149B}. With an average cadence of 3\,days, the time series should lead to reliable delays for all sources with $L_{5100} > 10^{42}\,\mathrm{erg~s^{-1}}$ \citep{2013ApJ...767..149B}. For 11 sources in our sample, multi-campaigns are available thus allowing to test the stability of our results and search for possible long-term variations in the BLR properties \citep{2006MNRAS.365.1180C}. 

Light curves were obtained using calibration stars close to the object. We first create a normalized light curve by comparing the flux of the object with 20-40 field stars that are stable and show no variation during the observing period. The uncertainty estimation is done by the median of the error of the calibration stars together with the error of the object. Then, for the absolute flux calibration, we used standard stars from \citet{2009AJ....137.4186L} observed on the same night, as well as non-variable stars from the Pan-STARRS1 (PS1) catalog\footnote{https://catalogs.mast.stsci.edu/} in the same image as the AGN. Cross-calibration of data from different telescopes was accomplished by determining an optimal relative rescaling factor between the datasets such that von Neumann's mean square successive difference \citep{von_neumann41} for the combined interlaced time-series is minimized; see also Appendix \ref{app:lc}. All fluxes were corrected for the foreground Galactic extinction values of \cite{2011ApJ...737..103S}.

The effective seeing for our observations was 2\arcsec - 3\arcsec, which is significantly worse than the typical atmospheric seeing at the site ($\lesssim
0.7$\arcsec)\footnote{\href{https://elt.eso.org/about/location/}{https://elt.eso.org/about/location/}}, and is caused by dome-made seeing since the telescopes are mounted in a bungalow with a flat rooftop. All light curves were obtained for an aperture diameter of 7.5\arcsec, which maximizes the signal-to-noise ratio while minimizing the influence of the host galaxy.  An exemplary light curve for Mrk\,841 is shown in Fig. \ref{fig:841_lc} while the full set of light curves is given in Appendix~\ref{app:lc}.

\begin{figure}
 \centering
 \includegraphics[width=\columnwidth]{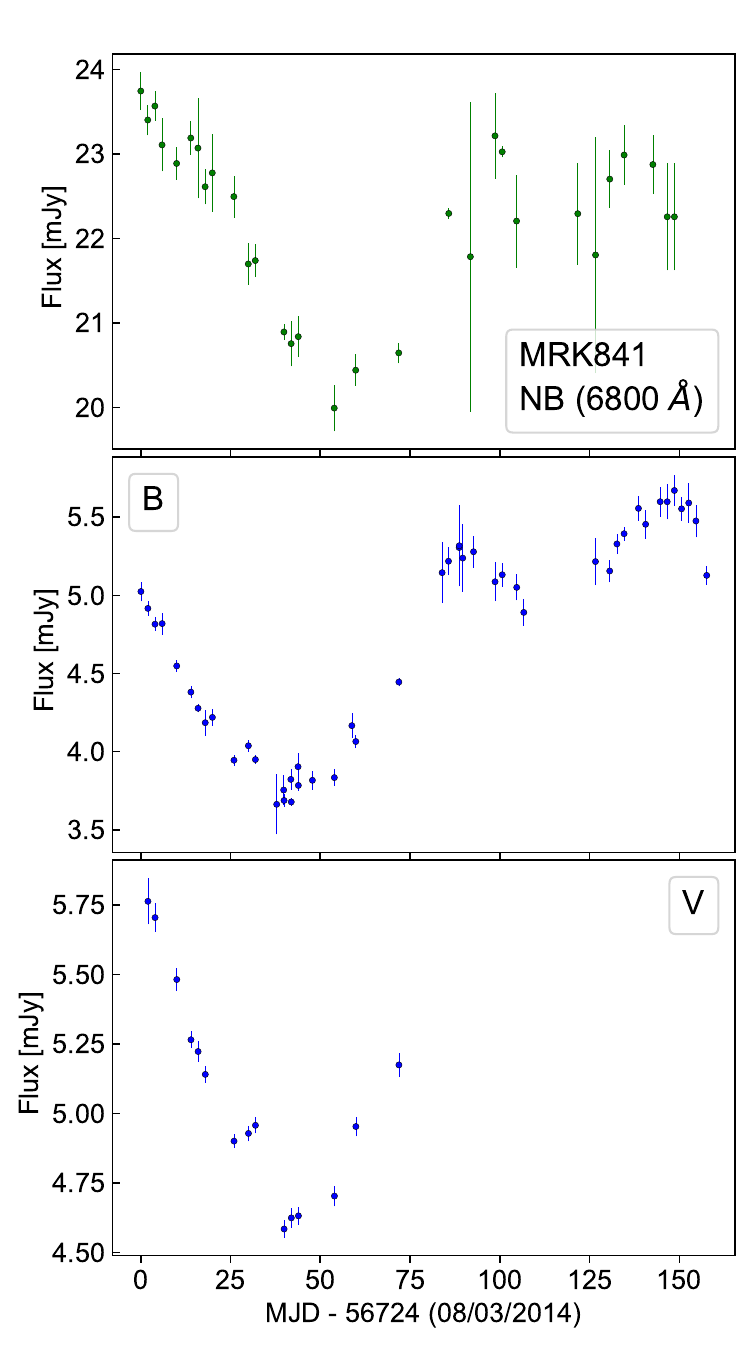}
 \caption{Photometric light curves for Mrk\,841. The first deep minimum for the NB680 band appears to be delayed by 20 days with respect to the adjacent broadband data due to the significant contribution of the broad H$\alpha$ line to the flux in the narrow band. Moreover, the continuum ($B$ or $V$) shows a strong rise between 50 and 75 days, whereas in NB680, the rise is shallower, indicating that the H$\alpha$ line may still be declining. As the contribution of the H$\alpha$ line to the band is only partial, direct cross-correlation techniques are inadequate to recover the time-delay associated with the line, and bivariate correlation analyses methods (Eq. \ref{model0}) are more suitable.}
 \label{fig:841_lc}
\end{figure}

\subsection{Spectroscopic Data}\label{sec:methods:bh}

We have spectra covering the H$\alpha$ line for all sources in our sample. Of these, we acquired 24 spectra ourselves, while the remaining were obtained from the literature. Single-epoch spectra were contemporaneously obtained using the Southern African Large Telescope \cite[SALT;][]{buckley06} between 2012 and 2013 and using the Tillinghast telescope at the Whipple Observatory between 2014 and 2015. SALT observations utilized the Robert Stobie Spectrograph \cite[RSS;][]{kobulnicky03} with the EG\,21, EG\,274. Feige\,110, G\,24-9, G\,93-48, Hiltner\, 600, LTT \,1020, and LTT\,4364 stars used for flux calibration.  Tillinghast-telescope observations utilized the FAST Spectrograph \cite[FAST;][]{1998PASP..110...79F} with BD+174708, BD+284211, Feige\,34, Feige\,110, and HD\,84937 used as standard stars for flux calibration. Additionally, a single-epoch spectra was taken for AKN\,120 in 2019 with the orange channel of the second-generation Low Resolution Spectrograph \cite[LRS2;][]{chonis16} on the upgraded Hobby-Eberly Telescope \cite[HET;][]{ramsey98,hill18}. The spectra were reduced with the automatic HET pipeline, \href{https://github.com/grzeimann/Panacea}{Panacea}.

Additionally, a couple of calibrated spectra were taken from the Swift BAT AGN Spectroscopic Survey \cite[BASS DR1,DR2; see][]{2017ApJ...850...74K,2022ApJS..261....2K}, and a few objects in our sample are included in the 6dF catalog \citep{2009MNRAS.399..683J}. Generally, these data are not flux calibrated, except for two NLS1s \citep{2018AaA...615A.167C}. Single objects with calibrated spectra were retrieved from individual publications.
Table~\ref{tab:sample} lists the H$\alpha$ spectra available for each of the objects in our sample and the complete log of observations is given in Table~\ref{tab:spectroscopy_log}.

\section{Methods}\label{sec:methods}

\subsection{Time-delay Determination}\label{sec:time_det}

\begin{figure*}
 \includegraphics[width=0.5\textwidth]{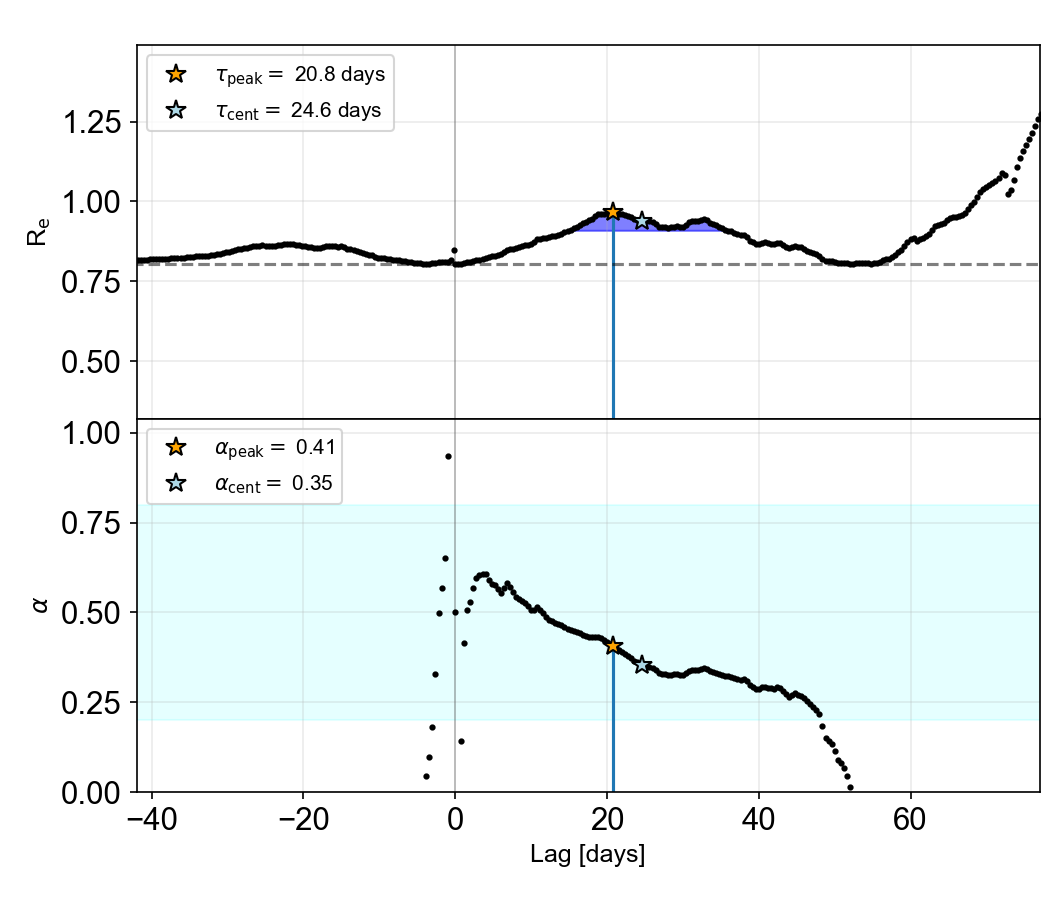}
 \includegraphics[width=0.5\textwidth]{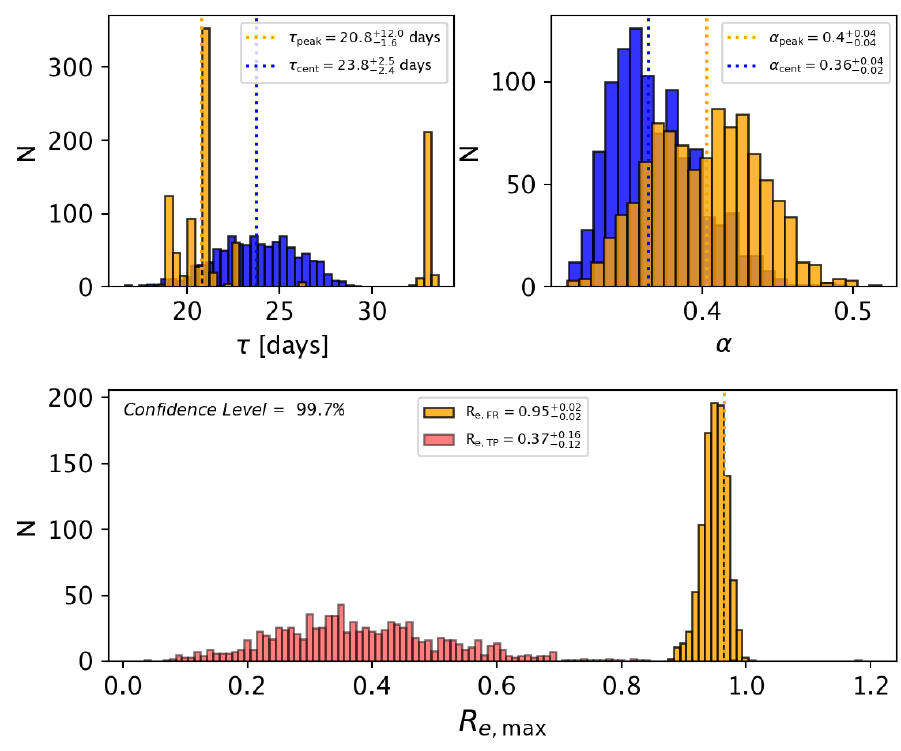}
\caption{Left: Correlation between the $B$-band and NB680 light curves for MRK841. The correlation coefficient ($R_{\rm e}$) is shown at the top and the corresponding value for $\alpha$ is at the bottom. The value for the peak delay with its respective $\alpha$ value is marked with an orange star.  The range for calculating the centroid is colored with a blue area and the corresponding lag is marked with a light-blue star.
The centroid calculation is performed within a range defined as $\geq 0.8$ times the peak $R_{\rm e}$ value, akin to the conventional method of RM. Right: the upper two panels illustrate the implementation of the flux-randomization (FR) algorithm for error estimation. The peak value is highlighted in orange, while the centroid is depicted in blue, with the final values labeled. The lower panel presents the histogram for the maximal correlation values $R_{\rm e,max}$  obtained via the FR method ($R_{\rm e,FR}$) is shown in orange with a value that approximates 0.95. In contrast, the histogram for the time-permuted version ($R_{\rm e, TP}$) is displayed in red, yielding an average value of approximately 0.4. This indicates a high level of confidence in the results (see text). }
\label{fig:corr_example}
\end{figure*}

In the common implementation of BLR RM, continuum and line signals are spectroscopically separable by their different spectral correlation scales. The time delay between the emission-line light curve and the continuum light curve may then be deduced via the interpolated cross-correlation function \citep[ICCF; see ][]{1986ApJ...305..175G,1998PASP..110..660P}. This seeks to maximize the Pearson correlation coefficient by shifting the continuum and emission-line time-series with respect to each other. Nevertheless, PRM cannot directly distinguish emission-line from continuum processes as only their combined signal is measured. By design, the NB filter in our study carries both continuum and emission-line contributions to the signal with the latter having a substantial contribution in comparison to broadband data (Fig. \ref{fig:mrk841_spc_filter}). To apply standard RM formalism to such data, a suitable proxy for the continuum light curve must be identified, and subtracted from the light curve in the band, which traces the combined continuum- and line-emission signals. This can be done if simultaneous observations are acquired in adjacent (either broad or narrow) continuum-richer bands. We refer the reader to 
\cite{2018AaA620A137R} and \cite{2020AJ....159..259S} for a successful implementation of this approach to PRM. More generally, however, uncertainties in the proper extraction of the emission-line signal from photometric data may arise when the observations in the line-rich and line-poor bands are not obtained simultaneously, or when a reliable proxy for the pure continuum signal is challenging to identify. This can happen, for example, when the BLR contribution to the signal in the continuum-rich band is non-negligible (e.g., using $B$-band data, which is sensitive to the rising small blue bump; see Fig. \ref{fig:mrk841_spc_filter}) so that over-subtraction of the continuum signal from the line-rich band may occur. An additional complication to consider relates to the fact that the contribution of the emission lines to the flux may not reflect on their contribution to the reverberating signal, which may be better traced by the RMS flux. Lastly, for significantly removed bands, interband continuum time-delays may affect the recovered delays \citep[][see their Fig. 11]{2014ApJ...785..140C}.

To overcome some of the aforementioned complications in the classical implementation of PRM, we generalize the RM formalism so that it allows for the decomposition of line and continuum signals at the light curve level using either broadband, NB, or spectroscopic data \citep[][for a similarly motivated study that uses a different implementation see \citealt{2016ApJ...819..122Z}]{2013ApJ...769..124C}. When applied to our data, the light curve traced by the NB is modeled as
\begin{equation}
F_\mathrm{NB}(t)=(1-\alpha) F_\mathrm{BB}(t) +\alpha F_\mathrm{BB}(t-\tau),
\label{model0}
\end{equation}
where $\alpha,\tau$ are model parameters, and $F_\mathrm{BB}$ is the broadband light curve for which the BLR contribution is presumably weak (note that the host-galaxy contribution to the signal is approximately constant and hence discarded). Solutions for the echo are obtained via a best-fit criterion between the model and the NB data. Specifically, $t_\mathrm{BLR}$ derives from the time-delay, $\tau$, which maximizes the Pearson correlation coefficient, $\tau_\mathrm{peak}$. The solution method adopted here is analogous to the two-dimensional correlation algorithm for the detection and characterization of spectroscopic binaries \citep[TODCOR;][]{1994ApJ...420..806Z} and is applied to PRM by transforming its wavelength dependence to time dependence. Importantly, the formalism uses combinations of CCFs and ACFs, and may thus be implemented using customary tools of trade, in which case the Pearson correlation coefficient is given by (see appendix~\ref{app:simulations} for derivation and further details): 
\begin{equation}
{\textstyle
R_{\rm e}(\tau)=\sqrt{\frac{{\rm CCF}^2(0)-2{\rm CCF}(0){\rm CCF}(\tau){\rm ACF}_\mathrm{BB}(\tau)+{\rm CCF}^2(\tau)}{1-{\rm ACF}_\mathrm{BB}^2(\tau)}}},
\label{R_E}
\end{equation} 
where $\mathrm{ACF_{BB}}$ is the autocorrelation function of the broadband-filter light curve, and the $\mathrm{CCF}$ term has the usual meaning.

Throughout this work, we implement our formalism using the interpolated CCF (ICCF) approach \citep{1986ApJ...305..175G,1998PASP..110..660P} with fixed (0.2\,days) time steps, which are below our observing cadence for all sources. Following \citet{1999PASP..111.1347W}, re-normalization and de-trending are implemented at every time step. The search window for a peak in the correlation function satisfies  $\tau_\mathrm{peak}\in [-2/3,2/3]\times t_\mathrm{duration}$, where $t_\mathrm{duration}$ is the total span of the time-series. This is justified in cases where the sampling is quasi-regular throughout the time-series, and the light curves in the different bands overlap over much of the observing period. As customary in RM studies (see also Appendix~\ref{sec:app_time} and \ref{app:simulations}), $t_\mathrm{BLR}$ is identified with either the peak of the correlation function or with its center of mass around the peak; in the latter case, the centroid calculation is carried out within a time range around the peak that is bracketed by $R_{\rm e}$ values greater than $80\%$ of the peak value (see Fig~\ref{fig:corr_example}). This results in the centroid time delay, $\tau_\mathrm{cent}$.

We note that the implementation of our PRM algorithm may yield values for the correlation coefficient, which are larger than unity (note the rising peak at $>65$\,days lag in Fig. \ref{fig:corr_example}). This results from our choice to a) use common adaptations for the ICCF, which would make the formalism easily implementable, and b) reduce the number of interpolated points to a minimum, thereby mitigating artificial structure in the correlation function. As verified by simulations (Appendix~\ref{app:simulations}), the effect is exacerbated when the search window for time delays extends to a significant fraction of the time-series (for the case of Mrk841 values exceeding unity are obtained for lags that are $\gtrsim 50\%$ of the total duration of the campaign), and the number of overlapping points between the bands is significantly reduced due to the relative time-shifts employed. Significantly non-uniform sampling of the time-series is a further factor contributing to this effect. For more information refer to Appendix~\ref{app:simulations}. To prevent correlation values from exceeding unity, a commonly applied time-stamp interpolation should be applied to all the terms used to calculate the correlation coefficient, namely $\mathrm{CCF}(0),~\mathrm{CCF}(\tau)$, and $\mathrm{ACF}(\tau)$ (see Eq. \ref{R_e}), which could result in the signal being dominated by interpolated data, thus potentially leading to spurious peaks, which we wish to avoid.

\subsubsection{Incorporating bounds on line emission in photometric data}

Useful constraints on the model concern the bounds on the value of $\alpha$ in the model (Eq. \ref{R_E}). Physically, $\alpha\in [0,1]$ hence peaks in the correlation function, which are associated, for example, with $\alpha<0$ solutions may be discarded (e.g., the rising peak at long delays for Mrk\,841; see Fig. \ref{fig:corr_example}). 
 
Refined estimates for the time-delays may be obtained by considering further information about the value of $\alpha$. For example, noting that $0.2<\alpha<0.9$ for NB data \citep[][for comparison, $\alpha<0.1$ for broadband data]{2011AaA535A73H,2012ApJ...747...62C} can be used to discard spurious peaks in $R_{\rm e}$ and obtain a more meaningful solution. Specifically, if $R$-band observations, which overlap with our NB bands, are available then we estimate the value of $\alpha$ by photometric means such that 
\begin{equation}
 \alpha_{\rm phot} = 1 - \frac{\langle F_R \rangle }{\langle F_{\rm NB} \rangle},
 \label{eq:phot_alpha}
\end{equation} 
where, $F_{\rm R}$ is the flux density in the $R$ band, $F_{\rm NB}$ the flux density in the NB and $\langle . \rangle$ denotes averaging over the time series. Broadbands other than $R$ may also be used to estimate $\alpha_\mathrm{phot}$ although the degree to which these data may serve as a proxy for the continuum emission under the H$\alpha$ line is unclear, and should be verified on a case-by-case basis, especially when large apertures are employed. The photometric estimate for $\alpha$ is likely a lower limit and $\alpha\gtrsim \alpha_\mathrm{phot}$ due to the contribution of the host galaxy within the aperture. For NB data, $\alpha<0.3$ only in cases where the emission line is partially covered by the band or if during the observation campaign, the varying part of the emission line is particularly weak.

\subsubsection{Uncertainty and confidence-level estimations}\label{sec:conf_level}

We estimate the uncertainty on $t_\mathrm{BLR}$ using the flux-randomization method \citep{1998PASP..110..660P} with 1000 repetitions, and calculating the correlation peak (or centroid) distribution, from which 15\% and 85\% percentile bounds are calculated, and the measurement uncertainties evaluated. We do not use the random subset selection approach, which results in significantly over-estimated uncertainties for $\alpha <1$ but instead introduce a confidence-level estimation procedure \citep{2013ApJ...769..124C}.

As the delayed signal in PRM of the BLR contributes only partly to the NB signal, we wish to ascertain whether the detection of the delayed component at the light curve level is robust. To this end, we repeatedly calculate $R_e$ assuming $F_\mathrm{NB}(t)=(1-\alpha) F_\mathrm{BB}(t) +\alpha \mathcal{F}_\mathrm{BB}(t-\tau)$, where $\mathcal{F}_\mathrm{BB}$ is a randomly time-permuted version of $F_\mathrm{BB}$ (c.f. Eq. \ref{model0}). The confidence level is then given by the percentage of iterations where the maximum correlation value $R_{\rm e, max}$ for the flux-randomized (FR) light curves ($R_{\rm e, FR}$) exceeds the one obtained with the time-permuted versions ($R_{\rm e, TP}$). 

A typical implementation of the above formalism for the case of  Mrk\,841 is shown in figure \ref{fig:corr_example} for which the $B$ band was used as a proxy for the continuum light curve, and the NB6800 band was used to trace the H$\alpha$ line. 
Figure~\ref{fig:corr_example} on the right shows the application of the FR algorithm for the error estimation. The upper two panels show the results for the delay ($\tau$) and the line contribution ($\alpha$), which results in  $\tau_{\rm peak} = 20.8^{+12.0}_{-1.6}$ days, and  $\tau_{\rm cent} = 23.7^{+2.4}_{-3.5}$ days, where the former (latter) is based on the peak (centroid) statistics. Identifying $t_\mathrm{BLR}$ with either of the measures leads to consistent conclusions. The lower panel of the Figure shows the results for the maximal correlation coefficient value $R_{\rm e,max}$ obtained via the FR method $R_{\rm e,FR} (\sim$0.95) and the time-permuted version $R_{\rm e,TP}$ ($\sim$0.4), which leads in this example to a high confidence level.

\begin{figure}
 \centering
 \includegraphics[width=0.95\columnwidth]{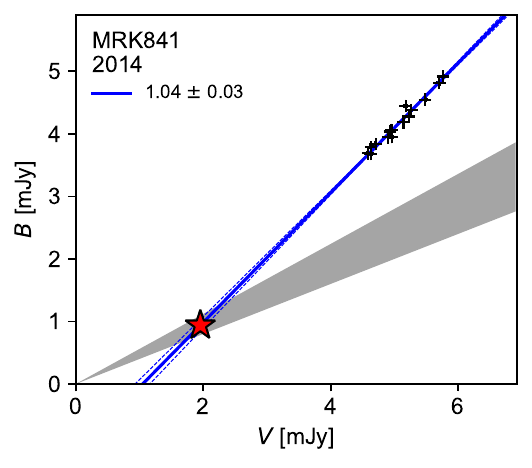}
 \caption{An example of a flux-flux diagram for Mrk\,841 exhibiting variability behavior that is consistent with a linear trend between the $B$ and the $V$ bands (plus symbols). A linear fit (blue line) intersects the galaxy-colors wedge (\citealt{2010ApJ...711..461S}; gray area) at the host-galaxy flux levels within the aperture (red star), and should be subtracted to reveal the net AGN flux (see text).}
 \label{fig:841_fvg}
\end{figure}

\subsection{Host-galaxy--light Subtraction}\label{sec:methods_host}

We apply the Flux Variation Gradient (FVG) method, proposed by \citet[][see also \citealt{1992MNRAS.257..659W,1997MNRAS.292..273W,2010ApJ...711..461S,2011AaA535A73H,2022A&A...657A.126G}]{1981AcA....31..293C} to subtract the host galaxy component from the total flux {and calculate the net AGN luminosity}. This method uses the fact that the time-varying optical emission of Seyfert galaxies shows little to no color changes over BLR light-crossing timescales, and that host colors are also fixed over the relevant timescales. In that case, we can disentangle both components by using flux-flux diagrams \citep{2015A&A...581A..93R}. Specifically, the intersection between AGN and host slopes, which are set by their colors, provides the contribution of the host component to the observed flux in each of the bands. In our sample, we mostly have observations in $BVRr_s$, and we construct flux-flux diagrams for their combinations. If available, we opt for the $BV$ diagrams to avoid $R$-band contamination by the H$\alpha$ line. We use the Ordinary Least Squares Bisector (OLSB) method to fit the flux-flux slopes \citep{1990ApJ...364..104I,1997MNRAS.292..273W,2015A&A...581A..93R}.

The contribution of the host-galaxy light within the aperture delineates wedges in the flux-flux diagram whose margins are set by the galaxy colors. These are taken from the sample of AGN host galaxies in \citet{2010ApJ...711..461S}, which results in the following slope (i.e., colors) ranges per filter combination: $BV = 0.48 \pm 0.08$, $Br_s = 0.35\pm 0.07$, $BR = 0.29 \pm 0.07$ and $VR = 0.71\pm 0.03$. The intersection of the best-fit linear trend of the flux-flux data with that expected given the host colors allows us to estimate the contribution of the latter to the bands. The host galaxy color range used in our analysis is an average across different types of galaxies (refer to Table~1 in \citet{2010ApJ...711..461S}). Since the objects have redshift from 0.01 until 0.05, there is minimal impact on the host color. The range of host slopes across different redshifts falls within the host's error margin. Therefore, we assume that the previously mentioned host colors apply to all objects. While this is a suitable approach for our purposes, it's important to note that the specific choice of host galaxy color and its subtraction can impact the final host-subtracted flux values and finally the luminosity value.

A typical FVG diagram is shown for Mrk841 (Fig. \ref{fig:841_fvg}), where the contribution of the host to the total flux within the aperture is {almost 20\% in $B$ and 40\% in $V$ showing an increase to longer wavelengths as expected}. 

\begin{figure}
 \centering
 \includegraphics[width=\columnwidth]{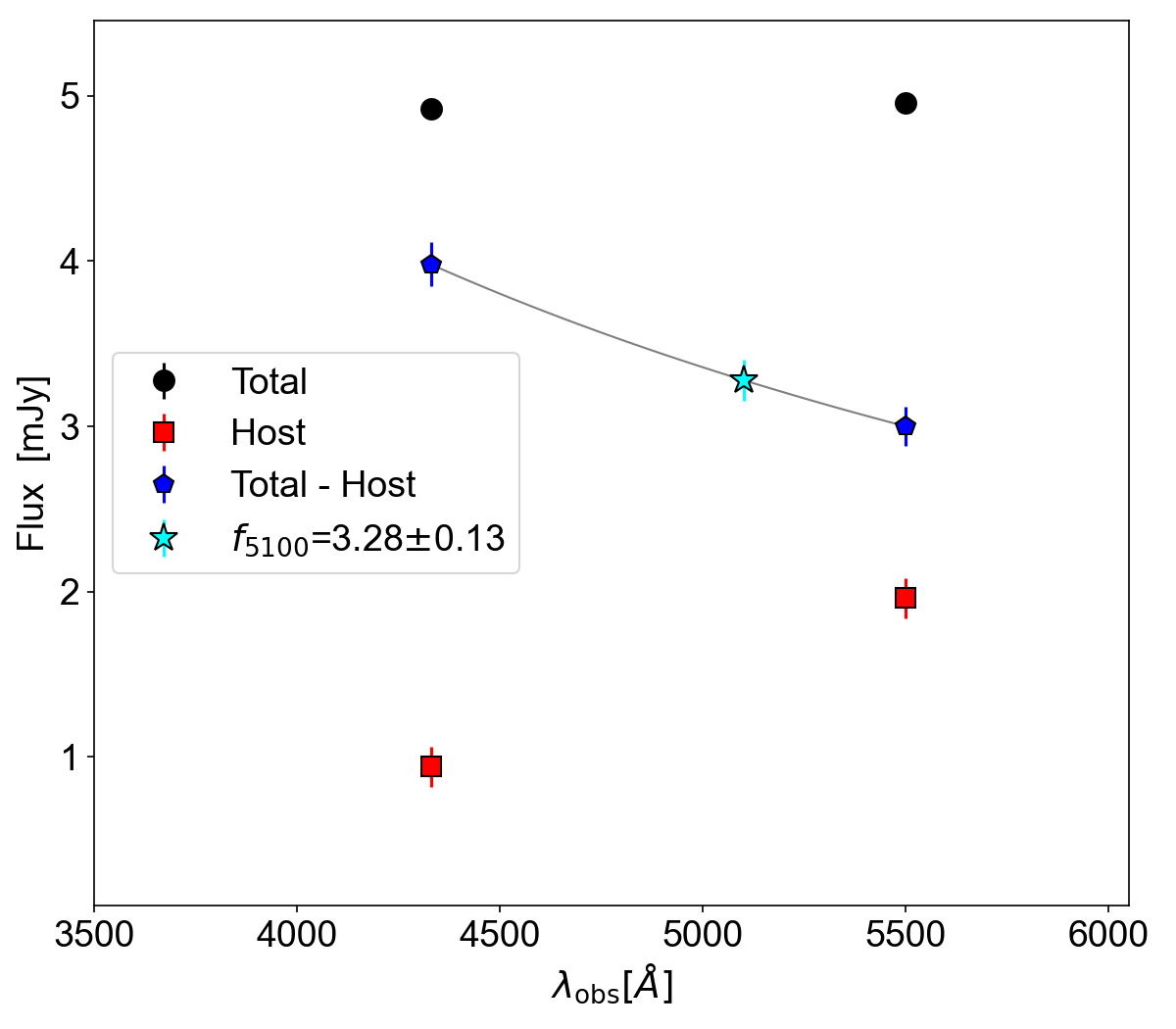}
 \caption{An example of host galaxy subtraction and flux interpolation at 5100\AA\ for Mrk841. Black circles represent the median total flux, with errors reflecting the median of all errors. Host galaxy values derived from the FVG method are shown as red squares, where error bars indicate the range of intersection between AGN and host slopes. Host subtracted fluxes trace the nuclear emission, and are depicted in blue. A powerlaw fit of the form $f_\nu\propto\lambda^{\beta}$ is illustrated by a solid line. The final host-subtracted value from this fit is denoted by a cyan star, accompanied by its corresponding flux label. }
 \label{fig:841_ad}
\end{figure}

With the FVG for 39 sources robustly constrained (see Appendix~\ref{app:fvg}), we can estimate the net AGN luminosity, $L_{5100}$. Sources for which the FVG is ill-defined and the intersection between host and AGN transects is doubtful, we report an upper limit for the luminosity, this is the case for 8 of our sources. For objects with multi-epoch data, we take the average host flux from the combined FVG (after verifying that the values are not markedly different between epochs and between different filter combinations). On average the $BV$ AGN slopes (colors) are $1.0 \pm 0.2$, in agreement with the color studies carried out in \cite{1992MNRAS.257..659W} and \citet{ 1997MNRAS.292..273W}, where the $BV$ slopes are close to unity. The $BR$ slopes show a slightly shallower average color of $0.9 \pm 0.3$ (see Section~\ref{sec:res_fvg}).

To estimate the optical luminosity at 5100\AA\, $\lambda L_{5100}$, we interpolated the host-subtracted flux using a power-law form ($F_\nu\propto\lambda^{\beta}$) with the adopted cosmology (see Table \ref{tab:bh}). Figure~\ref{fig:841_ad} demonstrates the scheme for Mrk841 resulting in a power-law slope, $\beta \simeq -1.2$, which is bluer than predicted by standard thin-disk models for which $\beta = -1/3$ \citep{1973A&A....24..337S}, and likely results from the significant contribution of the small blue bump to the $B$-band in this source (see Fig. \ref{fig:mrk841_spc_filter}).

The mean luminosity of our sample is $\mathrm{log} \left ( \left < L_{5100} \right> \right )\simeq 43.6$ with 68\% of the sources in the range $43<\mathrm{log} \left ( L_{5100} \right ) <44$.  Thus, sources in our sample bridge the luminosity gap between sources with H$\alpha$ RM measurements in the \citet{2010ApJ...716..993B} and those in the \citet{2024ApJS..272...26S} samples.

\subsection{Broad-line Characterization}

\begin{figure}
 \centering
 \includegraphics[width=\columnwidth]{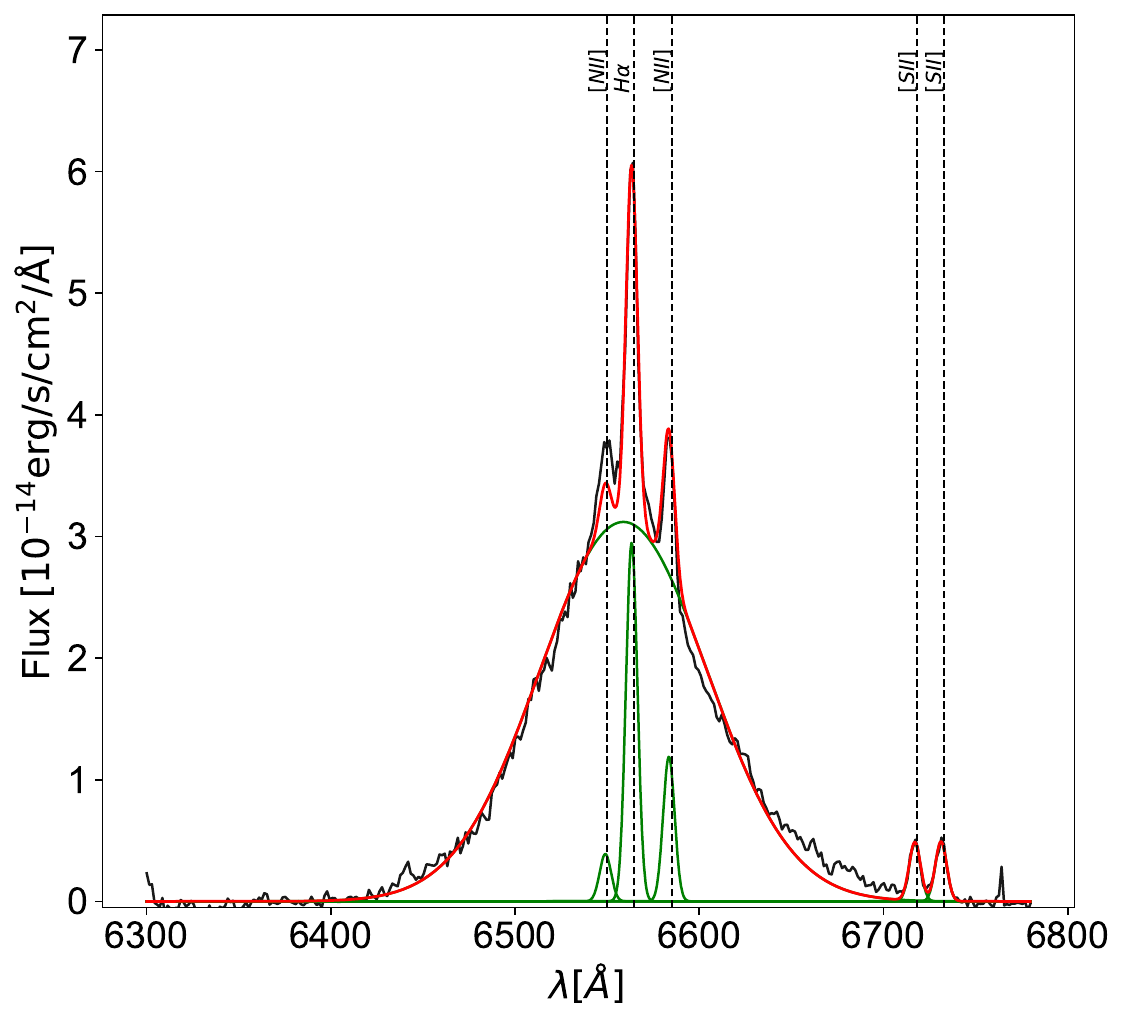}
 \caption{An example of a continuum subtracted FAST spectrum of Mrk\,841 showing the broad H$\alpha$ emission line and the narrow [SII] and [NII] lines. A best-fit model to the data is shown in red, which consists of broad- and narrow-line components (green curves). The FWHM of the broad component is used for BH mass estimatations.}
 \label{fig:841_spec}
\end{figure}

We apply the \href{https://github.com/pyspec}{\sc pyspec} package to determine the full width at half maximum (FWHM) of the broad H$\alpha$ line using available spectroscopic data (Table \ref{tab:sample}). Specifically, we subtract the continuum contribution by fitting a linear trend between flanking continuum windows. We then fit the residual emission by the sum of Gaussian curves. The combined fit often includes two narrow Gaussians for the [SII]\,$\lambda\lambda 6718,6732$, two narrow Gaussians for the [NII]\,$\lambda\lambda 6548,6584$, and narrow and broad Gaussian components for the H$\alpha$ line. Further, we set the narrow components to match those of the [SII] line (or the [NII] line when data quality is insufficient). Upon subtraction of the narrow components, the broad H$\alpha$ emission component may be fit by single or double broad Gaussian components. The FWHM of the line is estimated in two complementary ways: 1) from the fit of the broad components, or 2) from the observed broad-line profile after subtraction of the narrow lines. The quoted FWHM value is an average of both measurements and the error is the deviation between both estimations. A typical example of the fit quality is shown in figure \ref{fig:841_spec} with the full set of spectroscopic data shown in Appendix \ref{app:spec}.%, where additionaly the filter coverage of the line is overplotted.

{Table~\ref{tab:sample} includes the estimated fraction of the flux in the NB filter, which is contributed by the broad H$\alpha$ line as derived from the available spectra; c.f. the larger values deduced from photometry using Eq. \ref{eq:phot_alpha}, which include the narrow component emission (see Table \ref{tab:alpha_results}). Spectral details per source are provided in Appendix~\ref{app:spec}. We find this contribution to range from 40\% to 90\%, depending on the object's redshift, the NB chosen, the width of the line, and its equivalent width (see Appendix~\ref{app:spec}). We note, however, that the contribution of the emission line to the light curve may be different than the spectroscopically deduced value, which relies on a single-epoch measurement.}

We do not report FWHM values for two of the objects, IRAS23226-3843, and MCG+03-47-002, due to a very weak emission line and noisy spectrum. For NGC4726, in the 6dF available spectra we could not fit any line, therefore we quoted the FWHM provided in \cite{2019ApJ...872..134Z}. For the object HE0003-5023 we adopt the FWHM value reported in \cite{2015AuA...580A.113T}. In addition to the objects already classified as NLS1 in the catalog \citep{2006AaA...455..773V}, we have identified NLS1 candidates exhibiting FWHM values below 2000 km s$^{-1}$. These candidates include: IRAS01089-4743, RXSJ06225-2317, MRK705, MRK1347, HE2128-0221, and NGC7469.

In those high-quality spectra that include the H$\beta$ region, we fit an iron template to extract the optical Fe{} signal. Taking inspiration from {\sc PyQSOFit} \citep{2018ascl.soft09008G}, we employ the optical Fe\,{\sc ii} emission template, covering 3686-7484\AA, from \citet{1992ApJS...80..109B}, and fit it to the spectra employing normalization, broadening, and wavelength shift. We refer the readers to \citet{2024ApJS..272...13P, 2024ApJS..272...11P} for more details. {Values for the relative strength of the optical Fe{} signal, $R_{\rm Fe}$, are presented in Table~\ref{tab:results_rfe}, for all objects where fitting was successful. Examples of the fitting process are shown in Figure~\ref{fig:rfe}.}

\section{Results} \label{sec:results}

\subsection{BLR time Lag}

\begin{figure}
\centering
\includegraphics[width=0.99\columnwidth]{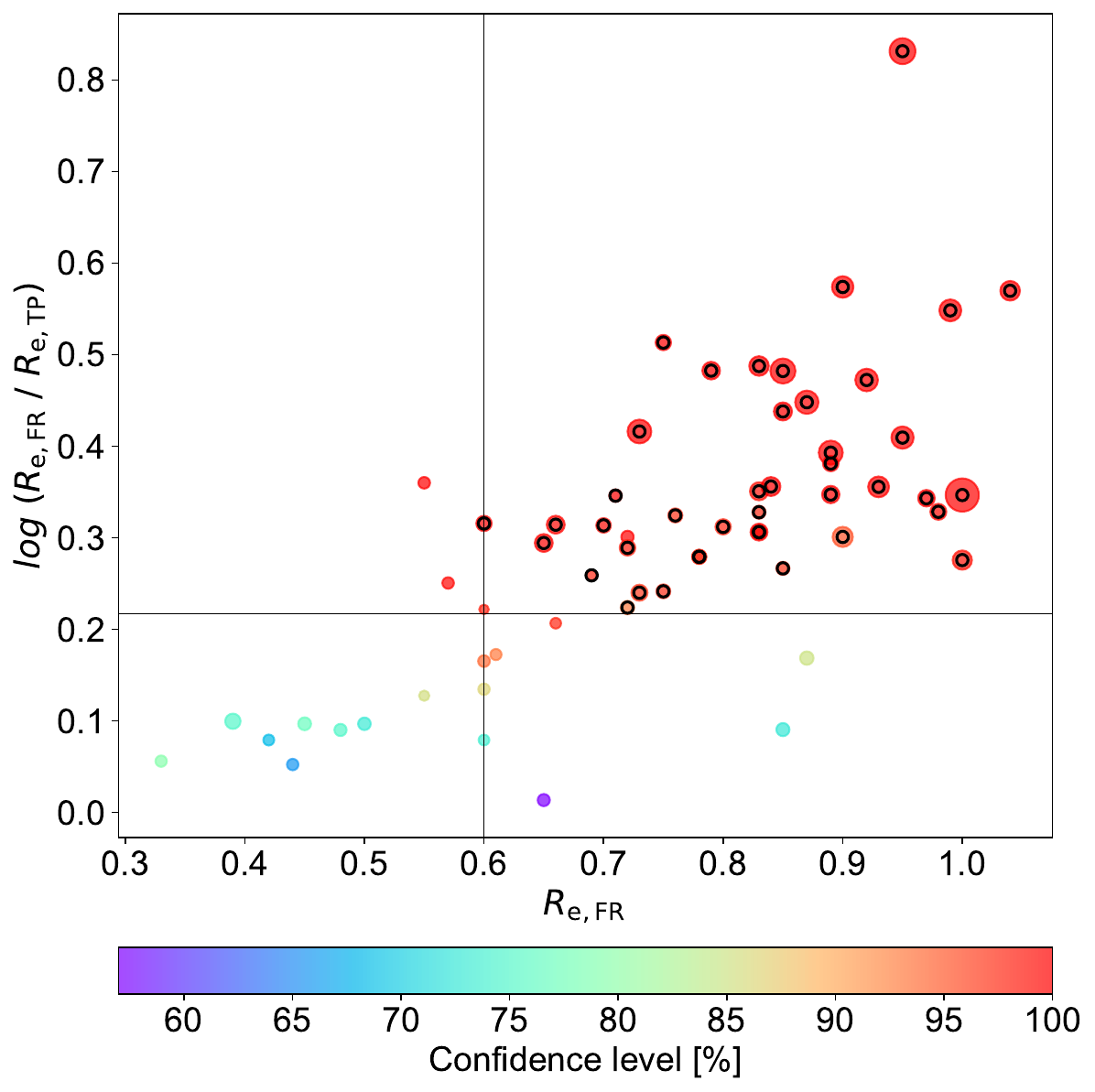}
\caption{Ratio $ R_{\rm e, FR}/R_{\rm e, TP} $ against $ R_{\rm e, FR}$ (see Section~\ref{sec:conf_level} for details), with color-coded according to the confidence level and size indicating the quality of the Continuum and NB light curve (being larger for better quality of the light curve, based on the $\eta$ value). Criteria of $ R_{\rm e, FR}\geq 0.6 $ and ratio $ R_{\rm e, FR}/R_{\rm e,TP} \geq 1.7 $ are marked with vertical and horizontal lines respectively. Good delay candidates are highlighted with a black edge.}
 \label{fig:conf_stat}
\end{figure}

Applying our lag-estimation algorithm to all targets and epochs (refer to Table \ref{tab:results}), we find that $ \tau_\mathrm{cent} \simeq \tau_\mathrm{peak}\equiv (1+z)t_\mathrm{BLR}$, where $z$ is the source redshift. Correlation functions, from which the time delays associated with light reverberation in the BLR, are shown for all sources and all epochs in Appendix~\ref{app:corr}. Approximately 80\% of the lag measurements in our final sample exhibit a confidence level exceeding 90\%.

Figure~\ref{fig:conf_stat} illustrates the quality of the time delays. Out of the 65 initial light curves, 61 were utilized for determining the time delay. 
We limit our analysis to sources that satisfy $R_{\rm e, FR} \geq 0.6$ and $R_{\rm e, FR}/R_{\rm e, TP} \gtrsim 1.6$. These criteria are indicated in Figure~\ref{fig:conf_stat} with vertical and horizontal lines, respectively. The upper-right region of the plot denotes the area of reliable delay estimations. Furthermore, time delay measurements with a confidence level above 90\%, representing accurate delay estimations, are highlighted with a black border.
The source IRAS23226-3843 meets all the criteria but is excluded from the final results due to its noisy light curves ($\eta \sim 0.9$, see Equation~\ref{eq:vn}) for both the continuum and NB band. Additionally, the broad component of the H$\alpha$ line is very weak. Consequently, this object is excluded from further analysis. Finally, we derived 40 reliable delays from the initial 61 light curves, accounting for 67\% of the analyzed data.

Of the initial 12 sources observed across multiple epochs, average delay values for 6 are presented in Table~\ref{tab:results_good} together with the results showing high confidence values. 
Lag determination for 3C120 and ESO374-G25 was conducted for only one observing epoch due to low variability and noisy light curves in some epochs. The first two epochs of MRK335 exhibit low confidence levels and a low ratio between $R_{\rm e, FR}/R_{\rm e, TP}$, likely due to sparse sampling and short duration of the light curves. Therefore, we report only one epoch delay for MRK335 from the observation year 2014. Similarly, HE1136-2304 shows a low $R_{\rm e, FR}/R_{\rm e, TP}$ ratio for its second and third observing epochs; hence, we report the optimal delay found for the 2015 season. 
For PGC64989, the first observing epoch also exhibits a low confidence level, leading us to discard this delay measurement. For the object RX J1103.2-0654, none of the epochs provided adequate data quality to report any delay. For details, we refer to the light curves and correlation functions in the Appendices~\ref{app:lc} and~\ref{app:corr}.

{For corroborative purposes, we utilized the JAVELIN code in photometric mode \citep{2016ApJ...819..122Z} to provide an independent estimate of the emission-line time-delays, and find good consistency (see Fig.~\ref{fig:delay_comparison} in Appendix~\ref{app:peak_rl}).}

\subsection{AGN-emission properties}\label{sec:res_fvg}

\begin{figure*}
 \centering
 \includegraphics[width=0.66\columnwidth]{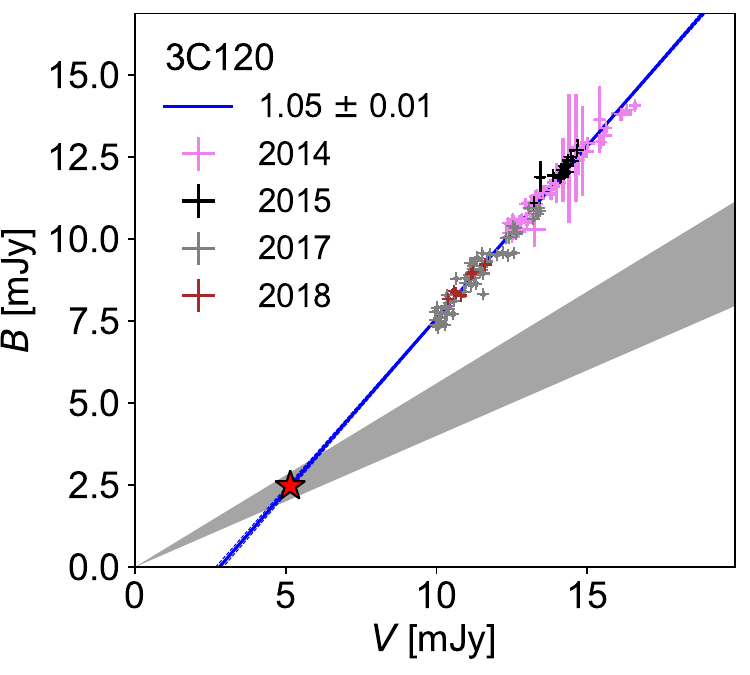}
 \includegraphics[width=0.66\columnwidth]{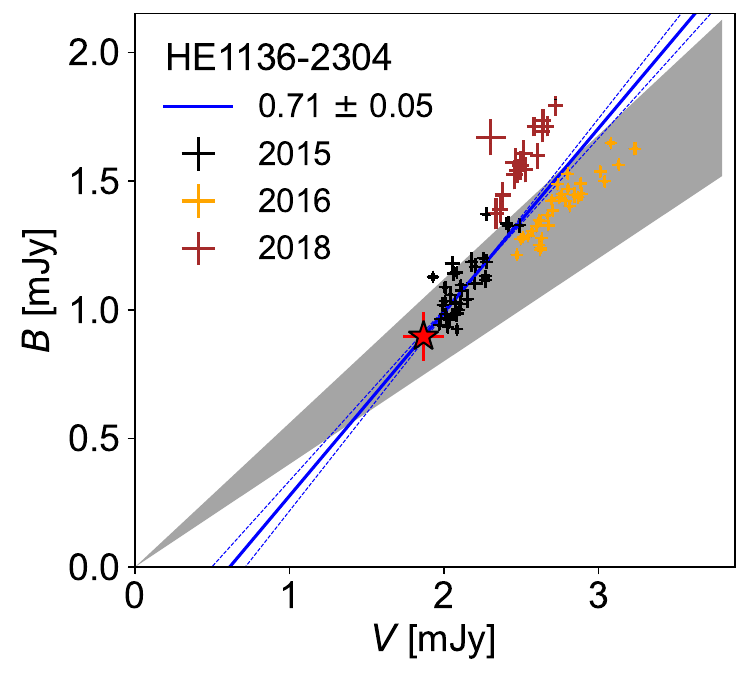}
  \includegraphics[width=0.66\columnwidth]{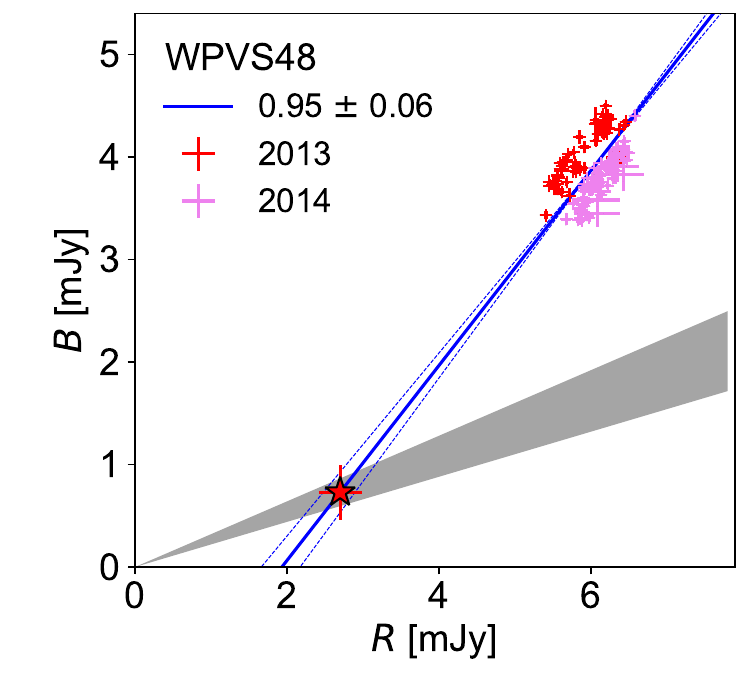}
 \caption{Multi-epoch $BV$ flux-flux diagrams for the objects 3C120 (left), HE1136-2304 (middle) and WPVS48 (right). The FVGs are contructed by the combination of different observation campaigns. Fluxes are plotted as crosses and the different colors mean different observing years. The bisector-fit for the AGN slope is plotted as blue lines and the $BV$ and $BR$ host slope from \cite{2010ApJ...711..461S} as a colored gray area. The red star marks the intersection between AGN and host slope and represents the assumed host value for the object.}
 \label{fig:fvg_3c_he}
\end{figure*}

We construct flux-flux diagrams per source for all available filter combinations (see Appendix~\ref{app:fvg}). Summing over all objects and epochs, we have 25 $BV$ AGN slopes and 52 $BR(r)$ AGN slopes with mean values of $1.02 \pm 0.22$ (for $BV$) and $0.90 \pm 0.30$ (for $BR$), which are in agreement with previous findings \citep{1992MNRAS.257..659W}. The $BR$ slopes distribution shows a larger scatter and lower values compared to the deduced $BV$ slopes, which could be explained by the contribution of (time lagging) H$\alpha$ emission to the $R$ band, and/or by nuclear reddening. 

Multiple observing campaigns were carried out for 12 targets in our sample, and allow to look for changes in AGN colors over years timescales. Among those sources, 8 have observations with the same filters, thus we can construct a multi-epoch FVG. Figure \ref{fig:fvg_3c_he} shows multi-epoch FVGs for 3C120, HE1136-2304, and WPVS48. These demonstrate three types of time dependent behaviours observed across the sample. The first (termed constant FVG) is consistent with AGN colors remaining fixed over time and across different luminosity states of the source \citep[see also][]{2015A&A...581A..93R}. The second behavior (termed varying FVG) shows a change of slope (i.e., color) of the varying nuclear component, which may be associated with variations in the properties of the accretion flow \citep{2019MNRAS.483L..17D} or with diffuse BLR emission, which may contribute to the optical band \citep{2019NatAs...3..251C}, or with time-varying nuclear reddening, or a combination thereof. Interestingly, such behavior is only seen in HE1136-2304, which is classified as a changing look AGN \citep{kollatschny18,2018A&A...618A..83Z}. The third type of time-dependent behavior is associated with FVG offsets, which do not affect the slopes that remain   consistent between individual campaigns. The most likely explanation is a change to our observing setup in-between campaigns, which is known to have occurred but was not adequately documented. This demonstrates the limitations of the FVG method in accurately subtracting the host signal.

{We find that the average host contribution to the $B$ band in our sample is $\simeq$30\%. The typical host contribution to this band exhibits a slightly decreasing trend with redshift, which results from our sample being selected for significant variability (among other criteria; see \S \ref{sec:sample_selection}). All host and AGN fluxes are listed in Table~\ref{tab:ad_all}.}%and an example for the host fraction redshift dependece in the $B$ filter is shown in Figure~\ref{fig:host_z}.}

Lastly, we note that within the confines of our sample and wavelength coverage, we do not find evidence for the "bluer when brighter" effect \citep{2014ApJ...792...54S}.

\subsection{BH mass}\label{sec:bh}

We estimate the mass of the central SMBH via
\begin{equation}
   M_\mathrm{BH} = \langle f \rangle \frac{c t_\mathrm{BLR}  {\sigma_\mathrm{BLR}^2}}{G}, 
\end{equation} 
where $t_{\rm BLR}$ is light crossing time across the H$\alpha$ emission region as deduced from the correlation analysis (\S\ref{sec:time_det}), and parametrize the mean velocity $\sigma_\mathrm{BLR}$ of the BLR with the FWHM of the broad component of the H$\alpha$ emission line as measured from single-epoch spectra. $\langle f \rangle$ is the population-mean geometrical scale factor, which is of order unity. Depending on the BLR geometry and dynamics \citep{2004ApJ...613..682P,2000ApJ...540L..13P}, $\langle f \rangle$ is often bracketed in the range $2-7$ \citep{2004ApJ...615..645O,2012ApJS..203....6P,2017ApJ...838L..10B,2013ApJ...773...90G,2011MNRAS.412.2211G}, but may substantially differ on an object by object basis \citep{2018ApJ...866...75W}. For tractability we set $\langle f \rangle = 1$, which is consistent with the approach taken by \citet{2019ApJ...883..170M} for the H$\beta$ line.

All estimated BH masses are listed in Table~\ref{tab:results}.  We collect previous BH mass estimates from the literature for comparison with our results, focusing on BH masses estimated via RM. The catalog by \cite{2015PASP..127...67B} provides an updated database of BH masses derived from spectroscopic RM, including six sources that are included in our sample. The BH masses in this database are primarily calculated using an average of all available RM lines and their $\sigma_{\rm line}$, with an adopted scale factor $f=4.3$. To align with our results, we gather data from various studies and recalculate the BH masses using their reported centroid time lag of the $H\beta$ line and its FWHM. We choose the H$\beta$ line since it is available for all the sources, whereas H$\alpha$ is only available for two of them. In our recalculation, we adopt $f=1$, following this work approach. In cases where multiple RM estimations are available, we average the BH masses obtained from the different studies. Furthermore, BH mass estimates for the objects HE1136-2304 and IC4329A were provided by two individual studies: \cite{kollatschny18} and \cite{2023ApJ...944...29B} and NCG985 and Mrk841 are within Lick AGN Monitoring Project in \cite{2022ApJ...925...52U}. For those we conducted identical recalculations using the H$\beta$ line and its FWHM. All recalculated BH masses are summarized in Table~\ref{tab:bh}. This approach ensures a standardized comparison between the BH masses derived from RM in previous studies and those obtained in our work.

\begin{figure}
 \centering
 \includegraphics[width=\columnwidth]{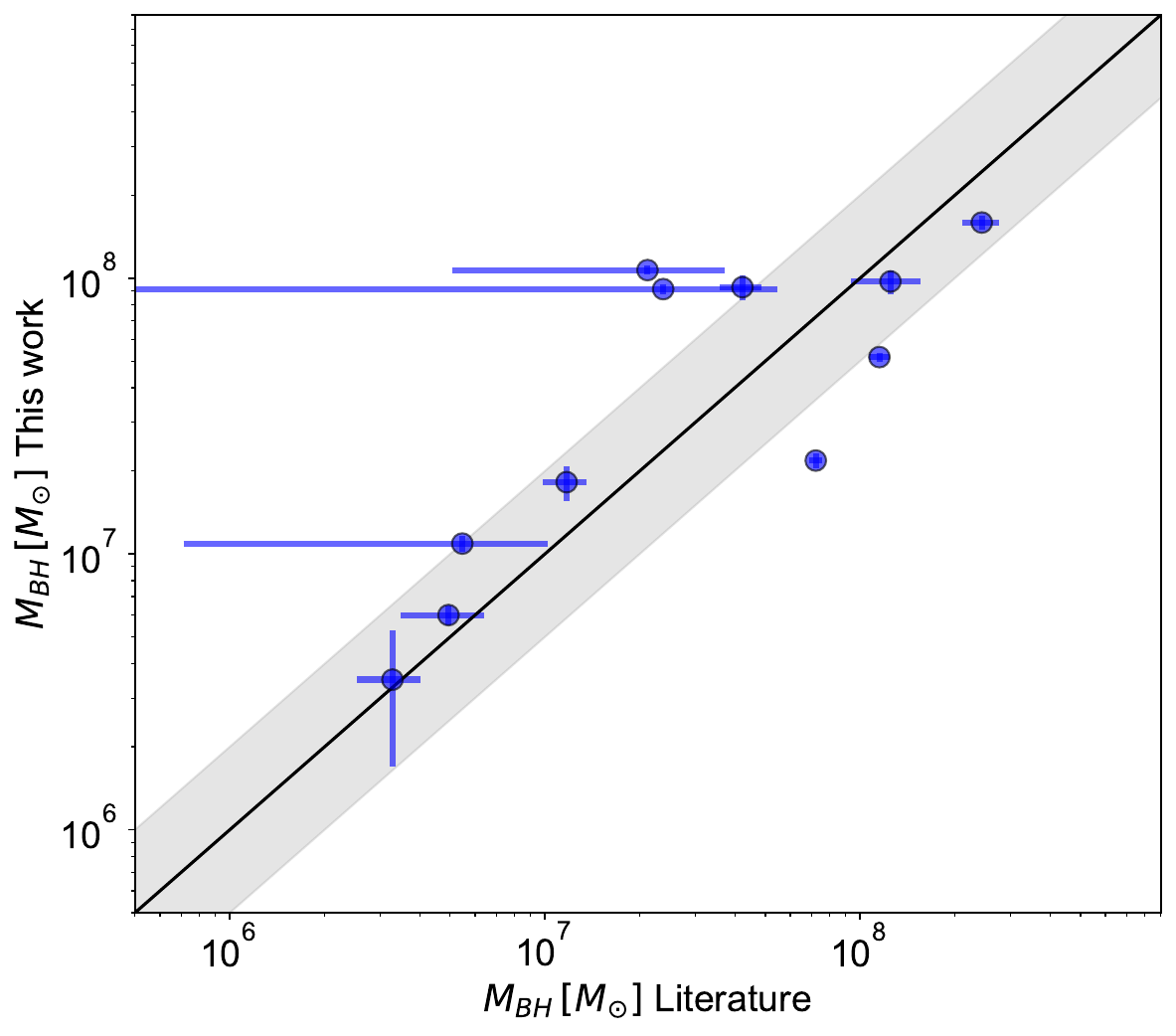}
 \caption{Comparison between BH masses estimated in this work and previously reported BH masses (after adjusting for the same $\langle f \rangle$ factor (see text)). The resulting scatter is at the 0.3\,dex level, which is typical in similar studies.}
 \label{fig:mbh}
\end{figure}

Figure~\ref{fig:mbh} shows the comparison between this work and the re-calculated BH masses. The y-axis represents $M_{\rm BH}$ from the current work and x-axis from the literature. The black line represents a one-to-one correlation between the two estimations and the grey shaded area an uncertainty of 0.3dex. It shows that BH masses obtained in this work from the H$\alpha$ line RM and from single epoch spectra are mostly in agreement with previous studies, and within the scatter typical in similar studies \citep{2004MNRAS.352.1390M}.

\begin{table}
\renewcommand{\arraystretch}{0.85}
\caption{This work and previous reported BH masses.}
\begin{tabular*}{\columnwidth}{lp{1.9cm}p{1.9cm}c}
\toprule
  &  {This work}&  \multicolumn{2}{c}{Literature}  \\ \hline
Object & $M_{\rm BH} $& $M_{\rm BH,Lit}$ & Ref. \\ 
  &  $[ 10^{7} M_{\sun} ]$ & $[ 10^{7} M_{\sun} ]$ &  \\ \hline
3C120  & $9.27^{+0.96}_{-0.96}$& $4.23\pm0.63$ & (1) (2) (3) \\
AKN120  & $15.9^{+0.82}_{-0.94}$ & $24.31\pm3.24$& (1)\\
HE1136-2304  & $2.18^{+0.14}_{-0.06}$ & $2.11\pm1.60$ & (4)\\
IC4329A & $10.69^{+0.38}_{-0.38}$ & $7.22\pm0.36$& (5) \\
MRK335 & $0.60 ^{+0.04}_{-0.05}$ & $0.49\pm0.15$  & (1) (3) (6) \\
MRK509& $5.17^{+0.2}_{-0.2}$& $11.50\pm0.88$ & (1)\\
NGC5940& $1.82^{+0.26}_{-0.22}$  & $1.17\pm0.18$ & (7)\\
NGC7469  & $0.35^{+0.13}_{-0.18}$&  $0.32\pm0.07$  & (1) (8)  \\ 
{Mrk705} & {$1.09^{+0.07}_{-0.05}$} & $0.54\pm0.47$& {(9)}\\
%Mrk841 & $9.72^{+0.92}_{-0.89}$ & $12.19\pm5.22$ & (10) \\
{Mrk841} & $9.72^{+0.92}_{-0.89}$ & {$12.47\pm3.10$} & {(9)} (10) \\
NGC985 & $9.12^{+0.32}_{-0.37}$ & $2.37\pm3.10$ & (10) \\
\toprule
\end{tabular*}
\tablecomments{Ref: (1) \cite{2004ApJ...613..682P}, (2) \cite{kollatschny14}, (3) \cite{2012ApJ...744L...4G}, (4) \cite{kollatschny18}, (5) \cite{2023ApJ...944...29B},  (6) \cite{2014ApJ...782...45D}, (7) \cite{2013ApJ...769..128B}, (8) \cite{1998ApJ...500..162C},  (9) \cite{2022ApJS..262...14B}, (10) \cite{2022ApJ...925...52U}. }
\label{tab:bh}
\end{table}

\begin{table*}
\caption{Time delays, luminosities, $M_{\rm BH}$ and accretion rate estimations. Time lags are expressed in observers frame. $L_{5100}$ notes the restframe luminosity at 5100\AA\ and the accretion rate estimators log$\dot{\mathscr{M}}$  and log$\dot{\mathscr{M}}_{R_\mathrm{Fe}}$ are obtained from Equation~\ref{mmdot} and ~\ref{mmdot_rfe} respectively. } 
\label{tab:results_good}
\renewcommand{\arraystretch}{0.85}
\begin{tabular*}{\textwidth}{lp{1.5cm}p{1.8cm}p{1.8cm}p{2.5cm}p{1.7cm}rr}
\toprule
{Object} & {$z$} & $\tau_{\rm cent, H\alpha}$ & $\tau_{\rm peak, H\alpha}$  &  $L_{5100}$ & $M_{\rm BH}$ & $\mathrm{log}\dot{\mathscr{M}}$ & $\mathrm{log}\dot{\mathscr{M}}_{R_\mathrm{Fe{}}}$ \\
&  &  [days] & [days] & [$10^{43}$erg s$^{-1}$] & [$10^{7}M_{\sun}$] & & \\ \hline
1H2107-097 & 0.02698 & $15.7^{+4.2}_{-4.2}$ & $11.2^{+1.4}_{-1.4}$ & $4.77 \pm 0.12$ & $1.63^{+0.45}_{-0.45}$ & $1.34^{+0.24}_{-0.24}$ & $-0.55\pm 0.01$\\ 
3C120 & 0.03301 & $57.1^{+5.9}_{-5.9}$ & $55.2^{+30.0}_{-4.4}$ & $14.34 \pm 0.37$ & $9.27^{+0.99}_{-0.99}$ & $-0.48^{+0.09}_{-0.09}$ & $-1.88\pm 0.22$\\ 
AKN120 & 0.0327 & $28.1^{+1.4}_{-1.6}$ & $18.6^{+23.0}_{-1.4}$ & $11.91 \pm 0.43$ & $15.98^{+0.82}_{-0.94}$ & $-1.85^{+0.05}_{-0.05}$ & $0.56\pm 0.0$\\ 
CTSG03\_04 & 0.04002 & $17.8^{+0.9}_{-0.9}$ & $14.8^{+0.4}_{-1.4}$ & $2.47 \pm 0.41$ & $3.11^{+0.16}_{-0.16}$ & $-0.93^{+0.09}_{-0.09}$ & $-2.19\pm 0.43$\\ 
ESO374-G25 & 0.02367 & $11.2^{+1.0}_{-2.0}$ &  $7.6^{+14.2}_{-2.5}$ & $<1.58^{*}$  & $4.31^{+0.39}_{-0.79}$ & -- & $1.47\pm 0.001$\\
ESO399-IG20 & 0.025 & $19.6^{+0.4}_{-0.8}$ & $24.4^{+0.8}_{-5.2}$ & $2.16 \pm 0.45$ & $1.27^{+0.03}_{-0.05}$ & $0.65^{+0.1}_{-0.11}$ & $-0.3\pm 0.01$\\ 
ESO578-G09 & 0.03502 & $19.5^{+0.6}_{-0.6}$ & $19.6^{+2.0}_{-12.8}$ & $3.39 \pm 0.17$ & $9.7^{+0.31}_{-0.31}$ & $-2.74^{+0.04}_{-0.04}$ & $-0.31\pm 0.01$\\ 
F1041 & 0.03347 & $15.7^{+0.7}_{-1.0}$ & $16.4^{+0.4}_{-0.0}$ & $1.04 \pm 0.3$ & $4.03^{+0.19}_{-0.27}$ & $-2.75^{+0.15}_{-0.15}$ & --\\ 
HE1136-2304 & 0.027 & $9.1^{+0.5}_{-0.2}$ & $10.6^{+1.0}_{-1.0}$ & $0.22 \pm 0.14$ & $2.18^{+0.12}_{-0.05}$ & $-3.85^{+0.31}_{-0.31}$ & $-1.42\pm 0.07$\\ 
HE1143-1810 & 0.03295 & $17.5^{+2.5}_{-2.4}$ & $21.2^{+0.2}_{-1.2}$ & $5.01 \pm 0.44$ & $1.53^{+0.23}_{-0.22}$ & $1.55^{+0.14}_{-0.13}$ & $-1.83\pm 0.23$\\ 
HE2128-0221 & 0.05248 & $8.3^{+0.7}_{-0.9}$ & $9.2^{+5.6}_{-5.8}$ & $2.43 \pm 0.21$ & $0.44^{+0.04}_{-0.05}$ & $2.96^{+0.09}_{-0.11}$ & $0.12\pm 0.01$\\ 
IC4329A & 0.01605 & $22.7^{+0.8}_{-0.8}$ & $13.4^{+14.0}_{-0.6}$ & 2.87* & $10.69^{+0.38}_{-0.38}$ & -- & $-1.55\pm 0.153$\\
MCG+03-47-002 & 0.04 & $16.8^{+0.4}_{-0.5}$ & $18.8^{+0.0}_{-0.2}$ & $0.66 \pm 0.52$ & -- & -- & --\\ 
MRK1347 & 0.04995 & $13.8^{+4.6}_{-1.7}$ & $21.2^{+1.0}_{-16.8}$ & $7.41 \pm 0.97$ & $0.64^{+0.22}_{-0.08}$ & $3.87^{+0.31}_{-0.13}$ & $0.76\pm 0.0$\\ 
MRK335 & 0.02578 & $12.0^{+0.9}_{-1.1}$ & $11.2^{+0.2}_{-5.2}$ & $3.99 \pm 0.22$ & $0.6^{+0.05}_{-0.06}$ & $3.09^{+0.07}_{-0.09}$ & $-1.06\pm 0.04$\\ 
MRK509 & 0.0344 & $22.9^{+0.8}_{-0.8}$ & $24.4^{+0.4}_{-1.6}$ & $17.31 \pm 1.53$ & $5.17^{+0.19}_{-0.19}$ & $0.97^{+0.05}_{-0.05}$ & $-1.59\pm 0.11$\\ 
MRK705 & 0.02879 & $15.5^{+1.0}_{-0.7}$ & $11.6^{+0.0}_{-0.8}$ & $3.21 \pm 0.48$ & $1.09^{+0.07}_{-0.05}$ & $1.56^{+0.09}_{-0.08}$ & $-0.44\pm 0.01$\\ 
MRK841 & 0.03642 & $23.8^{+2.5}_{-2.4}$ & $20.8^{+12.0}_{-1.6}$ & $6.75 \pm 0.27$ & $9.72^{+1.06}_{-1.02}$ & $-1.71^{+0.1}_{-0.09}$ & $-1.8\pm 0.18$\\ 
NGC5940 & 0.03408 & $5.9^{+0.8}_{-0.7}$ & $5.2^{+0.8}_{-0.4}$ & $2.26 \pm 0.54$ & $1.82^{+0.26}_{-0.22}$ & $0.002^{+0.17}_{-0.16}$ & $0.32\pm 0.0$\\ 
NGC7214 & 0.02385 & $6.9^{+5.2}_{-0.9}$ & $5.2^{+0.4}_{-0.4}$ & $2.1 \pm 0.41$ & $1.77^{+1.37}_{-0.24}$ & $-0.05^{+0.68}_{-0.15}$ & $3.16\pm 0.0$\\ 
NGC7469 & 0.01627 & $9.6^{+3.5}_{-4.8}$ & $16.0^{+0.0}_{-0.4}$ & $3.12 \pm 0.31$ & $0.35^{+0.13}_{-0.18}$ & $3.8^{+0.33}_{-0.44}$ & $-1.11\pm0.06$\\ 
NGC7603 & 0.02876 & $35.1^{+1.5}_{-1.3}$ & $36.8^{+4.0}_{-1.2}$ & $9.04 \pm 1.22$ & $22.34^{+0.98}_{-0.85}$ & $-2.93^{+0.08}_{-0.07}$ & $0.31\pm 0.0$\\ 
NGC985 & 0.04314 & $22.2^{+0.7}_{-0.8}$ & $24.0^{+0.4}_{-0.4}$ & $10.2 \pm 2.3$ & $9.12^{+0.3}_{-0.34}$ & $-0.96^{+0.11}_{-0.11}$ & $-2.1\pm 0.35$\\ 
PGC64989 & 0.01937 & $26.0^{+0.3}_{-0.3}$ & $26.8^{+0.2}_{-2.4}$ & $0.53 \pm 0.02$ & $5.37^{+0.06}_{-0.06}$ & $-4.33^{+0.02}_{-0.02}$ & $-1.59\pm 0.13$\\ 
RXSJ06225-2317 & 0.03778 & $19.5^{+0.2}_{-1.4}$ & $20.0^{+0.0}_{-0.4}$ & $3.23 \pm 0.46$ & $0.84^{+0.01}_{-0.06}$ & $2.1^{+0.07}_{-0.1}$ & $-1.98\pm 0.27$\\ 
UM163 & 0.03343 & $10.9^{+0.4}_{-0.5}$ & $10.0^{+0.8}_{-0.4}$ & $2.12 \pm 0.31$ & $4.97^{+0.19}_{-0.24}$ & $-2.1^{+0.08}_{-0.08}$ & $-0.81\pm 0.05$\\ 
WPVS007 & 0.02861 & $10.6^{+0.9}_{-1.0}$ & $10.4^{+0.0}_{-0.4}$ & $2.56 \pm 0.21$ & $0.49^{+0.04}_{-0.05}$ & $2.82^{+0.09}_{-0.09}$ & $4.24\pm 0.0$\\ \hline
ESO141-G55 & 0.03711 & $19.6^{+2.4}_{-2.4}$ & $19.2^{+4.4}_{-4.4}$ & $19.35 \pm 2.58$ & $9.15^{+1.17}_{-1.17}$ & $-0.002^{+0.13}_{-0.13}$ & $-0.62\pm 0.01$\\ 
ESO438-G09 & 0.02401 & $12.1^{+0.6}_{-0.6}$ & $10.2^{+6.2}_{-6.2}$ & $3.13 \pm 0.66$ & $1.23^{+0.06}_{-0.06}$ & $1.28^{+0.11}_{-0.11}$ & $5.19\pm 0.0$\\ 
ESO511-G030 & 0.02239 & $20.1^{+1.2}_{-1.2}$ & $17.9^{+0.4}_{-0.4}$ & $1.0 \pm 0.5$ & $5.15^{+0.3}_{-0.3}$ & $-3.3^{+0.25}_{-0.25}$ & $-1.11\pm 0.06$\\ 
PGC50247 & 0.02346 & $21.0^{+1.4}_{-1.4}$ & $19.9^{+1.6}_{-1.6}$ & $1.02 \pm 0.24$ & $2.27^{+0.16}_{-0.16}$ & $-1.63^{+0.13}_{-0.13}$ & $-1.57\pm 0.13$\\ 
RXSJ17414+0348 & 0.023 & $17.8^{+1.9}_{-1.9}$ & $17.8^{+5.6}_{-5.6}$ & $2.69 \pm 0.69$ & $1.69^{+0.18}_{-0.18}$ & $0.41^{+0.16}_{-0.16}$ & $0.98\pm 0.0$\\ 
WPVS48  & 0.037 & $20.3^{+4.6}_{-4.6}$ & $20.7^{+5.0}_{-5.0}$ & $5.59 \pm 2.49$ & $1.41^{+0.34}_{-0.34}$ & $1.87^{+0.3}_{-0.3}$ & $-0.75\pm 0.02$\\ \hline
\end{tabular*}
\tablecomments{* not possible to obtain host-subtracted fluxes, $L_{5100}$ reported for the 5100\AA~interpolated observed fluxes }
\end{table*}

\subsection{The size-luminosity relation for the H\texorpdfstring{$\alpha$}{alpha} emitting BLR}\label{sec:rl}

\begin{figure*}
\includegraphics[width=\textwidth]{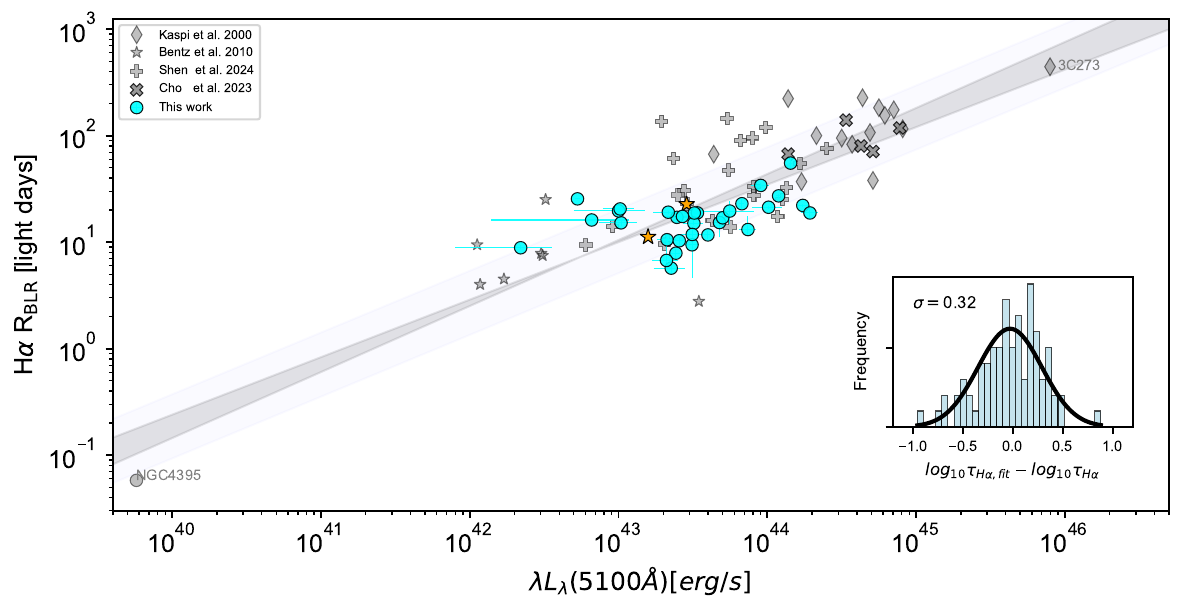}
\caption{H$\alpha$ $r_{\rm BLR}$-L diagram. Cyan circles show the results obtained in this work for sources with reliable delays and host-subtracted luminosities (orange stars denote sources for which host subtraction was not successful and the total luminosity is shown instead). Grey diamonds, stars, pluses and crosses represent previous results from \cite{2000ApJ...533..631K}, \cite{2010ApJ...716..993B}, \cite{2024ApJS..272...26S} and \cite{2023ApJ...953..142C} respectively. The result for NGC4395 from \cite{2021ApJ...921...98C} is marked with a circle and labeled, as well as 3C273 at the high luminous end. Grey shaded area represents the best linear-fit with the value $K = 1.51\pm0.05$ and $\gamma=0.57\pm0.05$ (see text). The inset shows the distribution of the difference between the RM delays and the expected value from the best-fit. The total scatter for the relation is $\sigma=0.32$dex, which is also shown as light gray area around the best fit in the main panel.}% while for only our sample is of $\sigma=0.29$.}
\label{fig:r_l_ha}
\end{figure*}

Figure~\ref{fig:r_l_ha} shows the resulting size-luminosity relation for the sources in our sample for which time delays were obtained ($r_\mathrm{BLR}=ct_\mathrm{BLR}$). Sources with host-subtracted AGN luminosity estimations are denoted by cyan circles. Those include single-epoch delays as well as the average values for multi-epoch observations.  Additional 2 sources, for which $r_\mathrm{BLR}$ values were deduced but reliable host-subtracted source luminosities are not available (IC4329A and ESO374-G25), are depicted as orange stars in Fig.\ref{fig:r_l_ha}.

Figure~\ref{fig:r_l_ha} also incorporates previously reported size measurements based on the RM of the H$\alpha$ line: 7 low-luminosity sources from \cite{2010ApJ...716..993B,2013ApJ...767..149B}; 14 high-luminosity sources (PG quasars for which the host contribution to the flux is negligible) as documented in \citet{2000ApJ...533..631K}; recent findings for 23 intermediate-luminosity sources from the SDSS-RM project \citep[][see also \citealt{2017ApJ...851...21G}]{2024ApJS..272...26S}; 5 high-luminosity AGN recently studied by \citet{2023ApJ...953..142C} and the intermediate-mass AGN NGC4395 presented in \citet[][see also \citealt{2012ApJ...756...73E}]{2021ApJ...921...98C}. All the values from the literature are listed in Table~\ref{tab:lit_ha} in the Appendix. Combined, this totals in 82 sources with varying measurement quality comprising the size-luminosity relation for H$\alpha$, which covers 4\,dex in source luminosity. 

A best-fit relation for the combined sample is shown in Figure~\ref{fig:r_l_ha} with a grey area , which is of the form:
\begin{equation}
\mathrm{log}(r_\mathrm{BLR,ld}) = K + \gamma~\mathrm{log}(L_{5100,44})
\label{eq:r_ha}
\end{equation}
where $r_\mathrm{BLR,ld}=r_\mathrm{BLR}/1$\,light-days, $L_{5100,44}=L_{5100}/10^{44}\mathrm{erg~s^{-1}}$. We employed orthogonal distance regression (ODR) fitting \citep{1990ApJ...364..104I} to determine the optimal parameters $K$ and $\gamma$, considering uncertainties in both luminosity and time lag and found $K=1.51\pm0.05$ for the normalization constant and $\gamma=0.57\pm0.05$ for the slope. 
The latest findings from \citet{2023ApJ...953..142C} report values of $K=1.59 \pm 0.05$ and $\gamma = 0.58 \pm 0.04$, nearly identical to this work. The luminosities in our sample span the mid-range and generally exhibit smaller sizes compared to SDSS values, but the overall slope is not affected.

These values are in good agreement with the best-fit values deduced by \citet[][see their Table 14]{2013ApJ...767..149B} for the H$\beta$ line with $K = 1.527 \pm 0.031$ and $\gamma = 0.533 \pm 0.035$. Restricting the fit to sources in our sample yields a flatter relation with $\gamma = 0.17 \pm 0.04$, which is dominated by the lowest luminosity targets having larger luminosity uncertainties. Further restricting to the fit to the 29 most luminous sources in our sample with $\lambda L_\lambda(5100\,\text{\AA})>1.5\times 10^{43}\,\mathrm{erg~s^{-1}}$ yields a slope $\gamma = 0.48 \pm 0.08$.

The residual scatter of the full H$\alpha$ sample around the best-fit H$\alpha$ relation is $\simeq 0.32$\,dex (see distribution in Fig. \ref{fig:r_l_ha}), which is larger than the value deduced for the H$\beta$ line ($\simeq 0.19$\,dex \citealt{2013ApJ...767..149B}). In comparison, the residual scatter in our sample of 31 sources is 0.26\,dex, which is lower than that which characterizes the SDSS-RM sample (0.34\,dex). Further restricting the analysis to the 29 brightest sources, yields a scatter of $\simeq 0.17$\,dex, which is comparable to that reported for the H$\beta$ line.

We note that the lags in our sample tend to lie below the current size-luminosity relation, by $\sim 9$\%, on average, which could be, at least partly, due to over-estimated nuclear luminosities for some of the sources (see above and Appendix~\ref{sec:comments_objects} for comments on individual objects). 
{To check whether the offset to shorter lags is potentially affected by our neglect of interband continuum time-delays across the optical band \citep{2019NatAs...3..251C,2022A&A...659A..13F,2023A&A...672A.132F}, we searched for a dependence between the H$\alpha$ lag and the ratio of the H$\alpha$ broad-line flux to the total flux in the NB filter, based on the supporting optical spectroscopy (see Table~\ref{tab:sample} and Appendix~\ref{app:spec}). Our analysis shows no such dependence regardless of the time-delay measurement scheme used (i.e., the formalism introduced here or JAVELIN). While the possibility that interband continuum time-delays affect a small fraction of our sample cannot be ruled out, it is unlikely to be noticeably affecting the bulk of the measurements (see Appendix \ref{app:peak_rl}).} Lastly, comparing the H$\alpha$ to existing H$\beta$ size measurement shows broad consistency with previous studies (see \S4.4.3).

Lastly, we observe that the H$\alpha$ line is somewhat broader than the NB filter for most of our targets (see Appendix \ref{app:spec}), and the signal in the extended high-velocity line wings, which typically contribute $\sim 10\%$ to the total line profile, is not traced by the data. As high-velocity gas emission originates from the innermost BLR \citep{2003A&A...407..461K,2018ApJ...866..133D,2024ApJ...972L...7F}, our inferred time-delays for the bulk of the emission line may therefore be biased to longer delays. A detailed account of the effect requires knowledge of the BLR geometry and kinematics, which are poorly understood, and are beyond the scope of this paper. Nevertheless, a case study of 3C\,120, for which $\sim 20$\% of the (high-velocity) line had been truncated by the bandpass, and a detailed kinematic modeling implies a negligible bias at the per-cent level \citep[][see their Fig. 3]{2014A&A...568A..36P}. An independent estimate of the potential bias can be obtained by considering the velocity resolved lags for the H$\beta$ line in NGC\,4151 \citep[][but note that these can be time-dependent as found by \citealt{2023MNRAS.520.1807C}]{2018ApJ...866..133D}, which should follow the kinematics of the H$\alpha$ line, and weighting the time-delays by the RMS line profile. In this case we find that centering on the line core within a velocity interval corresponding to the NB width results in a bias at the level of $\sim 7$\% ($\sim 20$\%) for the NB (NB\, SII) filters, which is of order the lag measurement uncertainty for our sample, and thus does not account for the residual scatter in the size-luminosity relation found here. 

\subsubsection{Testing for an intrinsic size-luminosity relation and lag stability}

\begin{figure}
 \centering
	\includegraphics[width=1.\columnwidth]{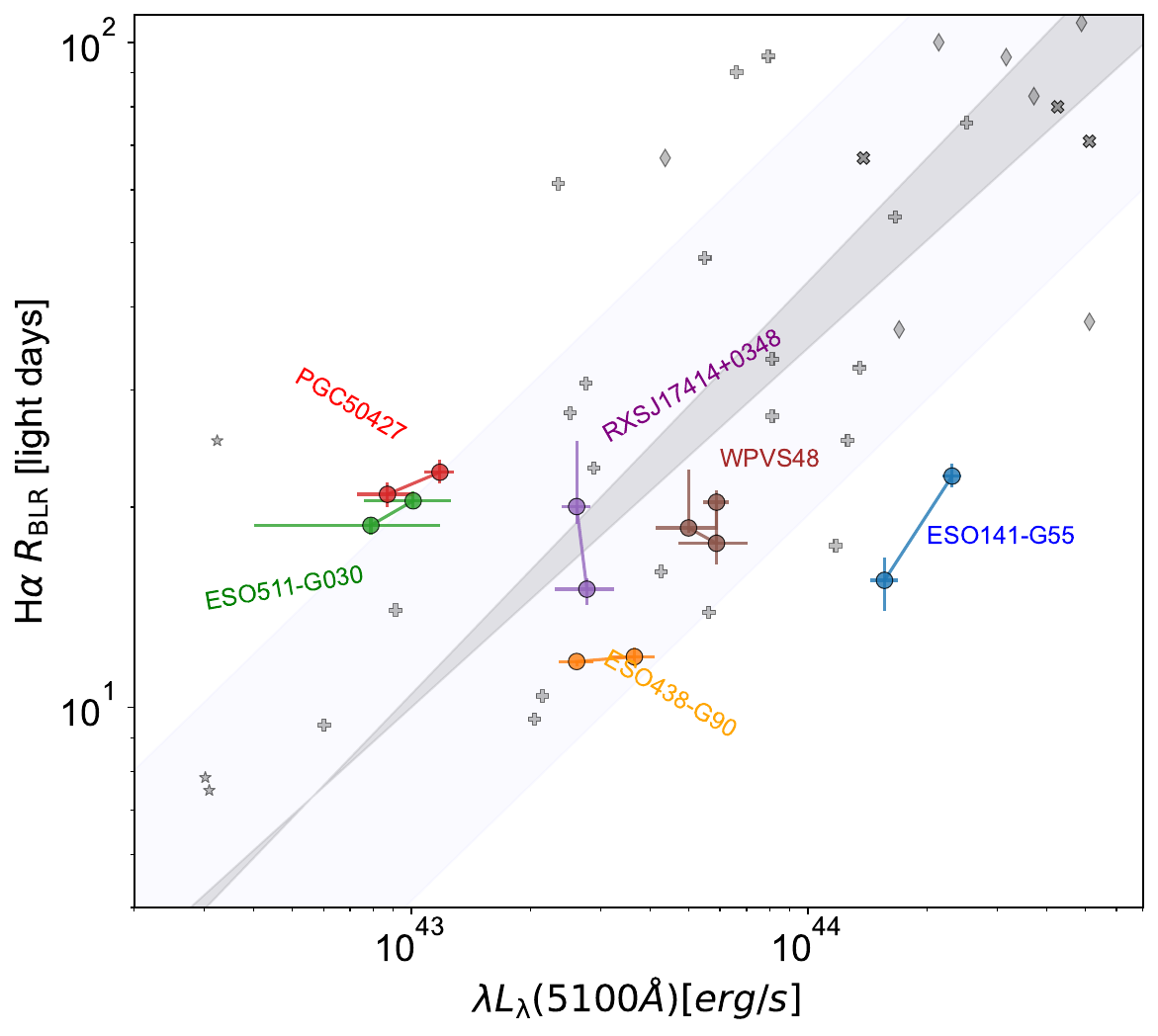}
 \caption{{The BLR size-luminosity relation for multi-epoch sources. The results for our objects are shown with different colors, connected with lines and labeled. Values for single-epoch results for H$\alpha$ from the literature are also shown (same as Figure~\ref{fig:r_l_ha}).} The tight clustering of the measurements per object indicates the stability of our PRM approach. The modest luminosity variations of the sources preclude a detailed analysis of the intrinsic size-luminosity relation.}
 \label{fig:r_l_multi}
\end{figure}

Our campaign spans a long enough timeline so that  intrinsic changes to the BLR size in response to substantial luminosity variations of the sources could be detected \citep{2006MNRAS.365.1180C}. 
Multi-epoch campaigns spanning two or more years were conducted for 12 sources in our sample, yielding H$\alpha$ delays for 6 of them (refer to the Table~\ref{tab:results_good} for averaged values and Table~\ref{tab:results} for single-epoch results). These multi-epoch observation campaigns provide independent size-luminosity measurements, as illustrated in Figure~\ref{fig:r_l_multi}.

The lag-measurement uncertainties and the modest luminosity variations of the sources do not allow to confidently quantify an intrinsic size-luminosity relation. Time-delay differences between epochs of individual sources are consistent with a scatter of $\sim0.10$ on average, which is significantly smaller than the scatter deduced for the entire sample.

\subsubsection{The size-luminosity relation and the accretion rate}\label{sec:rblr}

Previous works based on the RM results for the H$\beta$ line found that high accreting sources tend to lie below the population-average $r_{\rm BLR}-L$ relation \citep{2016ApJ...825..126D,2018ApJ...856....6D,2019ApJ...883..170M, 2023FrASS..1030103P}. Using the single epoch spectra, we estimate the value for the BH mass (see Section~\ref{sec:bh}) and calculate the {dimensionless} accretion rate of our sample as described in \cite{2016ApJ...825..126D}:
\begin{equation}
 \dot{\mathscr{M}} = 20.1\left[\frac{L_{5100,44}}{\mathrm{cos}(\theta)}\right]^{3/2}M_{\mathrm{BH},7}^{-2}
 \label{mmdot}
\end{equation} 
where $\theta$ is the AGN inclination, and $M_{\mathrm{BH},7}=M_\mathrm{BH}/10^7\,M_\odot$. We set $\mathrm{cos}(\theta)=0.75$ as in \cite{2016ApJ...825..126D} taking into account the mean AGN disk inclination. This is justified for our sources over the optical range for which the optical emission, after host-subtraction, is thought to be dominated by the self-similar part of the disk. $ \dot{\mathscr{M}}$-values are given in Table \ref{tab:results_good}, implying that, on (geometrical) average, the sources in our full sample (Table~\ref{tab:results}) emit with $ \dot{\mathscr{M}}\sim 0.6$, but with $\sim 20$\% of the sources exceeding unity and reaching up to $\dot{\mathscr{M}}\sim 40$, and hence deep within the super-Eddington accretion limit.

\begin{figure}
\includegraphics[width=1.\columnwidth]{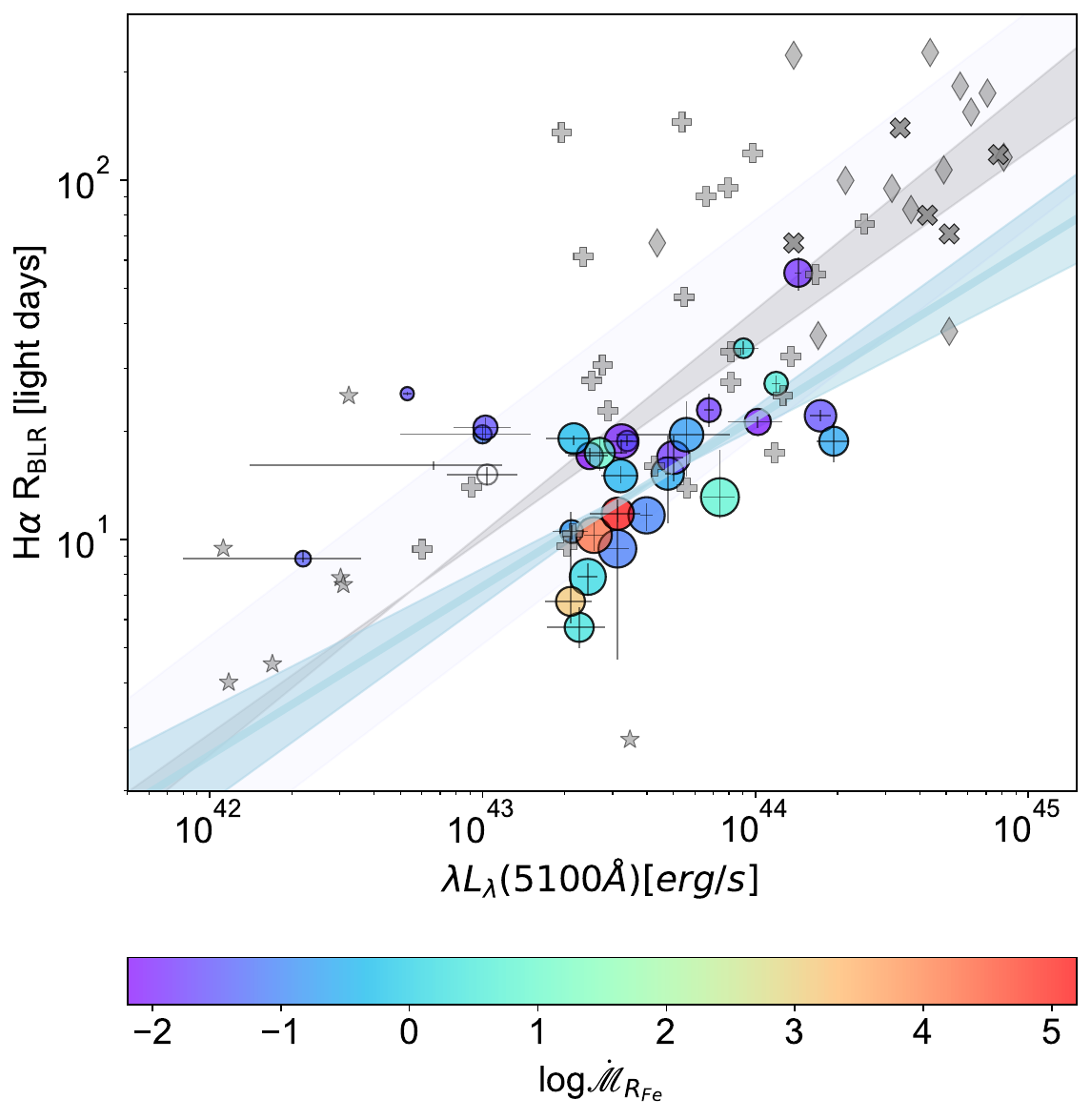}
\caption{  The BLR size-luminosity relation for the H$\alpha$ line color coded by the dimensionless accretion rate, $\dot{\mathscr{M}}$ (see text), as estimated by $R_\mathrm{Fe}$ (see Equation \ref{mmdot_rfe} and note that the method, while maintaining monotonicty, likely overestimates the highest accretion rates by a large factor). There is a tendency for highly accreting sources to lie below the population mean, which is in agreement with studies based on the H$\beta$ line. The size of the symbols corresponds to the value of $\dot{\mathscr{M}}$ (in log units) as deduced by Eq. \ref{mmdot} (see text for an inherent bias with this method). The blue shaded area represents the best fit for our sample with luminosities $>1.5\times10^{43}$erg s$^{-1}$.}
\label{fig:r_l_acc}
\end{figure}

While it is tempting to locate our highly accreting sources on the $r_\mathrm{BLR}-L$ relation, we note a natural tendency for sources in our sample with shorter time-delays (hence smaller black hole masses) to exhibit a higher accretion rate for a given source luminosity. We, therefore, opt to use a second estimator, which is independent of our BLR size measurements and is based on the ratio of the iron-blend flux to the broad H$\beta$-line flux, $R_\mathrm{Fe}$ \citep{2001ApJ...558..553M, 2019ApJ...882...79P}. Specifically, we consider the scaling of \citet{2016ApJ...818L..14D} wherein
\begin{equation}
    \dot{\mathscr{M}} \simeq \mathrm{exp} \left [7.7(R_\mathrm{Fe}-0.66) \right ].
\label{mmdot_rfe}
\end{equation}
The above linear relation likely breaks at the highest $R_\mathrm{Fe}$ values, and results in severely over-estimated mass-accretion rates, and yet it preserves monotonicity, and is not biased by our time-lag measurements. The resulting $r_\mathrm{BLR}-L$ shows the tendency for objects with high $\dot{\mathscr{M}}$ values to lie below the general population (Fig. \ref{fig:r_l_acc}), a result which is consistent with previous studies using the H$\beta$ line \citep{2018ApJ...856....6D}, but is corroborated here for the first time using a different emission line, and applying a different RM technique for an independent sample of sources. Using a permutation scheme which randomly reassigns $\dot{\mathscr{M}}$ values to sources we find that the formal confidence level of the result is $>98$\%. We further note that the result holds also when considering $\dot{\mathscr{M}}$ estimates derived by Eq. \ref{mmdot} further implying that our lag measurements are meaningful.

\subsubsection{H\texorpdfstring{$\alpha$}{alpha} to H\texorpdfstring{$\beta$}{beta} lag ratio}
\begin{table}
\centering
\caption{H$\alpha$ (this work) and H$\beta$ (literature) restframe delays.}
\begin{tabular*}{\columnwidth}{llp{1.7cm}p{1.7cm}c} 
\toprule
Object & $z$ & $\tau_{H\alpha}$ & $\tau_{H\beta}$ & Ref. \\ 
 & & [days] & [days] & \\ \hline 
3C120 & 0.03301 & $55.3^{+5.9}_{-5.9}$ & 27.9$^{+7.1}_{-5.9}$ & (1) \\ 
AKN120 & 0.0327 & $27.2^{+1.4}_{-1.4}$ & 37.1$^{+4.8}_{-5.4}$ & (2) \\ 
HE1136-2304 & 0.027 & $8.9^{+0.5}_{-0.5}$ & 7.5$^{+4.6}_{-5.7}$ & (3) \\ 
IC4329A & 0.01601 & $22.3^{+0.8}_{-0.8}$ & 16.3$^{+2.6}_{-2.3}$ & (4) \\ 
Mrk335 & 0.02578 & $11.7^{+1.0}_{-1.0}$ & 8.7$^{+1.6}_{-1.9}$ & (5) \\ 
Mrk509 & 0.0344 & $22.1^{+0.8}_{-0.8}$ & 79.6$^{+6.1}_{-5.4}$ & (2) \\ 
{Mrk705} & 0.02879 & $15.1^{+1.0}_{-0.7}$& {$4.6^{+4.8}_{-3.4}$} & {(6)}\\ 
Mrk841 & 0.03642 & $23.0^{+2.5}_{-2.5}$ & 11.2$^{+4.8}_{-1.9}$ & {(7)} \\
NGC5940 & 0.03408 & $5.7^{+0.8}_{-0.8}$ & 5.70$^{+0.90}_{-0.82}$ & {(8)} \\ 
NGC7469 & 0.01627 & $9.4^{+3.5}_{-4.5}$ & 10.8$^{+3.4}_{-1.3}$ & {(9)}\\ 
NGC985 & 0.04314 & $21.3^{+0.7}_{-0.7}$ & 7.4$^{+9.7}_{-9.4}$ & {(7)} \\ \hline
\end{tabular*}
\label{tab:ha_hbeta}
\tablecomments{Ref: (1) \cite{kollatschny14}, (2) \cite{2004ApJ...613..682P}, (3) \cite{kollatschny18}, (4) \cite{2023ApJ...944...29B}, (5) \cite{2016ApJ...825..126D}, (6) \cite{2022ApJS..262...14B}, (7) \cite{2022ApJ...925...52U}, (8) \cite{2013ApJ...769..128B}, (9) \cite{2014ApJ...795..149P}.}
\end{table}
Some of the sources in our sample have published H$\beta$ time-delays from previous campaigns. A comparison between non-contemporaneous  H$\alpha$ and H$\beta$ time-delays measurements is not meaningful on an object-by-object basis, but can be instructive across the sample as a whole. The objects from our study, with corresponding values and references is listed in Table~\ref{tab:ha_hbeta}.
For our sample, we determine the H$\alpha$ to H$\beta$ ratios for MRK335, 3C120, AKN120, NGC5940, MRK509, NGC7469, HE1136-2304, IC4329A, NGC985, MRK841 {and MRK705} as 1.34, 1.98, 0.73, 1.00, 0.27, 0.87, 1.18, 1.37, 2.87, 2.05, and {3.28} respectively. There are a few luminous ($L_{5100} > 10^{44}$ erg s$^{-1}$ ) sources with an unusually small $R_{\rm BLR}$ compared to the predictions of the $r-L$ relation. This may result from the limited duration of our monitoring campaigns, which are generally shorter than 150 days. For example, with a previously reported H$\beta$ lag of $\simeq 80$\,days \citep{2004ApJ...613..682P}, MRK509 exhibits an unusually low H$\alpha$ to H$\beta$ lag ratio. If the average H$\alpha$ to H$\beta$ ratio of about 1.3 holds also for this source, then the predicted H$\alpha$ lag would be $\simeq 100$\,days, which would not be robustly detected by our 2014 campaign (see figure 27 in Appendix~\ref{app:corr}).
When considering only the objects included in our study, the median is {$1.34 \pm0.87$}, and the average is {$1.54\pm0.87$}. Despite variations in observational epochs among objects, H$\alpha$ ratios are generally found to be slightly larger than those of H$\beta$. For comparison, the median ratio of H$\alpha$ to H$\beta$ across all sources in the samples from \cite{2000ApJ...533..631K}, \cite{2010ApJ...716..993B} and \cite{2024ApJS..272...26S} is 1.39$\pm$0.85, with an average of 1.57$\pm$0.85. The agreement between our results and previous studies suggest that the biases associated with the partial coverage of the emission line by the NB (see \S\ref{sec:rl}) are subordinate given the additional sources of uncertainty in this work.

\section{Summary and conclusions}\label{sec:summary}

We present the results of a first-of-its-kind multi-epoch PRM survey of a sample of 80 AGN, which was carried out over the course of about a decade (2010-2018) using small aperture ($\gtrsim 0.1$\,m-class) telescopes at Cerro Murphy in the Atacama Desert. The photometric data were augmented with single-epoch spectroscopy, and allow us to characterize the properties of the H$\alpha$ emitting BLR in those sources. Below is a summary of the main results of the paper:

\begin{itemize}[leftmargin=*]
\item Using small aperture telescopes, PRM can provide highly competitive results compared to spectroscopic RM (SRM), which typically uses larger ($\gtrsim 1$\,m) aperture telescopes. This is demonstrated in this work for the prominent H$\alpha$ line in nearby ($0.01<z<0.05$) sources whose line emission coincides with a set of off-the-shelf narrowband filters. 
\item We introduce a new PRM correlation scheme, which is drawn from spectroscopic binary identification, and implement it using a combination of (interpolated) auto- and cross-correlation functions, which are commonly used in RM studies. The method quantifies the properties of the delayed signal being the time-delay and the relative contribution of the delayed component to the narrowband data, but is generally applicable to PRM and SRM data. Applying it to our sample, we are able to successfully measure time-delays between line and continuum emission (traced by broadband photometry) in $\simeq 60$\% of our sources.
\item Augmenting our photometric data with single-epoch high-quality optical spectra, and using detailed multi-component fits to the H$\alpha$ emission line, we estimate the BH masses for our sample, which are consistent with previously reported estimates for 10 of our targets to within 0.3\,dex. Fitting for the H$\beta$ line and the iron-blend emission in its vicinity, we are able estimate the mass accretion rate for most objects in our sample in a way which is independent of our lag measurements.
\item To estimate the nuclear luminosity of our sources, we apply the FVG method for host-galaxy subtraction based on our set of photometric bands. Successful host subtraction was achieved for 84\% of our sample. For sources with multi-epoch data, we generally find consistent results, but note a few exceptions in our data, which can be traced to observational effects (filter changes) or to physical effects (changing-look sources).
\item We add a total of 31 objects to the $r_{\rm BLR}-L$ relation for H$\alpha$. The scatter for the total sample is $\simeq 0.32$\,dex and for our sample is $\simeq 0.26$\,dex. For the luminous ($L_{5100}>10^{43}\,\mathrm{erg~s^{-1}}$) sources in our sample, the scatter around the best-fit relation is $\simeq 0.17$\,dex, and hence comparable to that reported for the H$\beta$.
\item Taking our measured H$\alpha$ time-delays and comparing to H$\beta$ lags from the literature, we find an average $\tau_{\rm H\alpha}/\tau_{\rm H\beta}$ of $1.36\pm0.71$, which agrees for the total compilation with previous studies with $1.57\pm 0.85$. This consistency shows that the time-delays measured here are not significantly under- or overestimated.
\item We report multi-epoch time-delays for 6 sources in our sample, which are found to be stable with an overall scatter of $\simeq 0.04$\,dex, which is much smaller than that which characterizes the size-luminosity relation ($\gtrsim 0.2$\,dex depending on the sample used) thereby suggesting another origin for the latter perhaps associated with physical differences between the sources (see below) or to residual uncertainties in luminosity determination). Given the modest luminosity variations of individual sources during our campaign, we cannot robustly constrain an intrinsic size-luminosity relation. 
\item We corroborate, for the first time using an emission line other than H$\beta$, the trend whereby highly accreting sources tend to lie below the general-population BLR size-luminosity relation. Specifically, the trend is observed for two independent methods for assessing $\dot{\mathscr{M}}$, and their consistency further supports our lag measurements. 

\end{itemize}

\section*{Acknowledgements}

The observations greatly benefitted from the help of Christian Westhues, Moritz Hackstein, Francisco Pozo Nu\~nez, Zohreh Ghaffari, Bart\l omiej Zgirski, Marek G\'orski, Piotr Wielg\'orski, Paulina Karczmarek, Weronika Narloch, Hector Labra, Gerardo Pino, Roberto Mu\~noz, Francisco Arraya. We thank Victor Oknyanskyi, Saar Katalan, Carina Fian, Sina Chen, Eliran Daniel for fruitful discussions. This research has made use of the VizieR catalogue access tool, CDS, Strasbourg, France (DOI : 10.26093/cds/vizier). The original description of the VizieR service was published in 2000, A\&AS 143, 23. Some of the observations reported in this paper were obtained with the Southern African Large Telescope (SALT). This work was partly supported by grants from the German Research Foundation (HA 3555/14-1, HA 3555/14-2, CH 71/34-3, KO 857/35-1) and by grants from the Israeli Science Foundation (2398/19, 1650/23). Computations made use of high-performance computing facilities at the University of Haifa, which are partly supported by a grant from the Israeli Science Foundation (grant 2155/15). SP acknowledges the financial support of the Conselho Nacional de Desenvolvimento Científico e Tecnológico (CNPq) Fellowships 300936/2023-0 and 301628/2024-6. SP is supported by the international Gemini Observatory, a program of NSF NOIRLab, which is managed by the Association of Universities for Research in Astronomy (AURA) under a cooperative agreement with the U.S. National Science Foundation, on behalf of the Gemini partnership of Argentina, Brazil, Canada, Chile, the Republic of Korea, and the United States of America. MWO gratefully acknowledges the support of the German Aerospace Center (DLR) within the framework of the ``Verbundforschung Astronomie und Astrophysik'' through grant 50OR2305 with funds from the German Federal Ministry for Economic Affairs and Climate Action (BMWK).
%%%%%%%%%%%%%%%%%%%%%%%%%%%%%%%%%%%%%%%%%%%%%%%%%%
\section*{Data Availability}

We provide all light curves as a FITS file \href{https://zenodo.org/records/14096484}{here}.

\appendix
\counterwithin{figure}{section}
\counterwithin{table}{section}

\section{Light curves}\label{app:lc}

\begin{figure}
 \includegraphics[width=0.48\columnwidth]{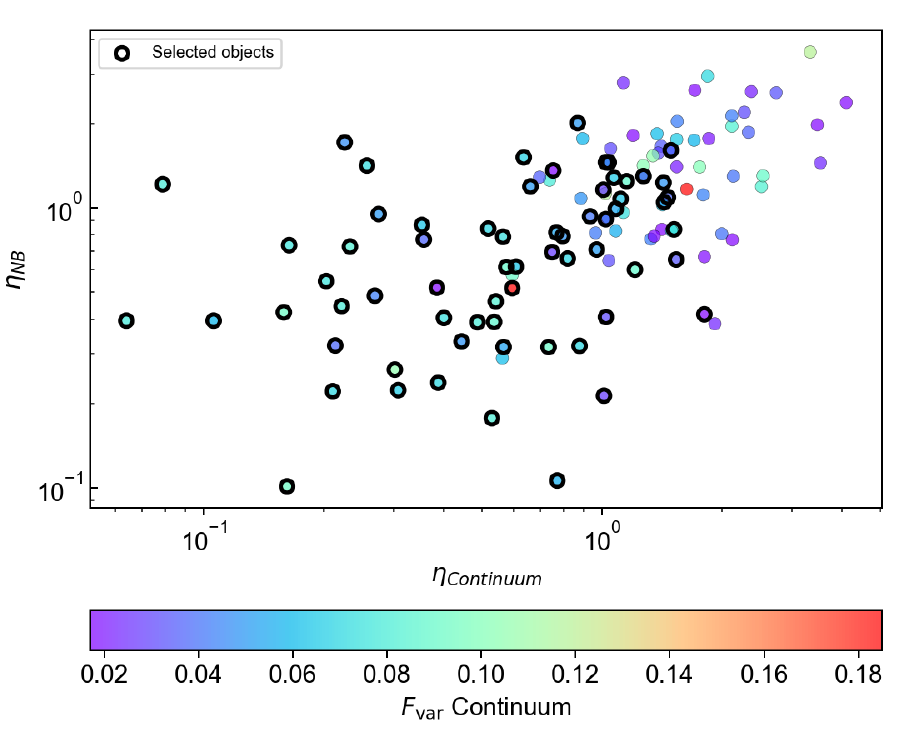}
 \includegraphics[width=0.48\columnwidth]{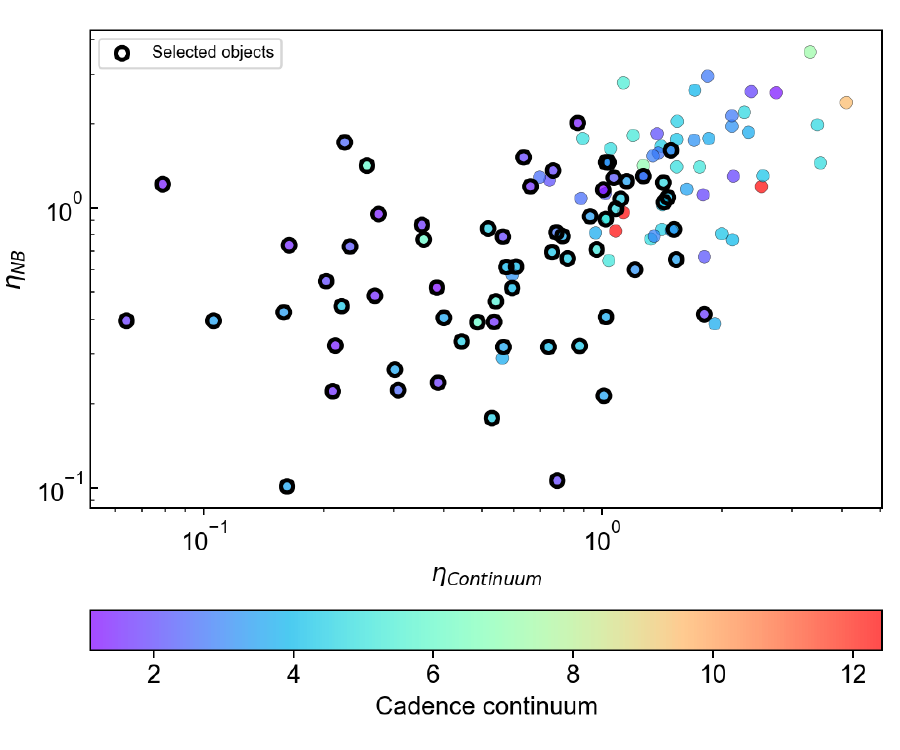}
 \caption{Observation and selection sample: $\eta$ values for NB light curves versus the continuum band light curves. Values are color-coded based on $F_{\rm var}$ (left) and cadence (right) of the continuum. Objects marked with a black edge mark the final selected light curves for the RM analysis. Same process was done for $F_{\rm var}$ and cadence of the NB band (not shown).}
 \label{fig:vnrm_std}
\end{figure}

{This Appendix presents the absolute flux calibrated optical light curves for the 48 Seyfert galaxies used in this study. Optical filters are $BVR(r)$ and the corresponding narrow band (NB) centered at 670nm, 680nm, 690nm or SII($\lambda$672\,nm) depending on the redshift. Time is expressed in modified Julian day (MJD) and fluxes in mJy.} For a discretely measured light curve at time stamps $t_i$ ($i\in [1:n]$) with fluxes $f_i$ and measurement uncertainties, $\delta f_i$, we define 
the excess variance $F_{\rm var}$ \citep{2003MNRAS.345.1271V}, and the regularity measure, $\eta$: 

\begin{equation}
 F_{\rm var} = \frac{\left (\sigma^2 - \Delta^2\right)^{1/2}}{\bar{f}} \text{ ,}\hspace{1cm} \sigma^2 = \frac{1}{n-1}\sum_{i=1}^{n}\left(f_i - \bar{f}\right)^2 \text{ ,}\hspace{1cm} \Delta^2 =  \frac{1}{n}\sum_{i=1}^{n} \delta f_i^2
 \label{eq:fvar}
\end{equation}

\begin{equation}
 \eta = \frac{\delta^2}{\sigma^2} \text{,}\hspace{1cm} \delta^2 = \frac{1}{n-1}\sum_{i=1}^{n-1}\left(f_{i+1} - f_{i}\right)^2,
 \label{eq:vn}
\end{equation}
where $\bar{f}$ is the mean of the light curve.
The regularity measure $\eta$ (Equation~\ref{eq:vn}) represents the ratio between the mean square successive difference (or the von Neumann estimator $\delta^2$) and the variance. The $\eta$ value serve as indicator for the regularity of the light curve, which is sensitive to the measurement-noise level and to the sampling of the process by the time-series. Smaller $\eta$ value imply better quality data, indicating less random fluctuations due to noise or sparse sampling of a rapidly varying signal. In both equations~\ref{eq:fvar} and~\ref{eq:vn}, $f$ represents the normalized flux of the detrended light curve. 
{In addition, we define the cadence of the light curve as the total duration of the observation campaign divided by the number of observations, noting that when large observation gaps exist, the cadence might be overestimated compared to the median of the time step of the light curves.} We list the above light-curve properties for the continuum ($B$ or $V$) as well as for the NB of each object in Table~\ref{tab:lc_properties}.

{The preliminary selection criteria for identifying objects and light curves suitable for Reverberation Mapping (RM) are as follows: 1) Record the values of $F_{\rm var}$, $\eta$, and cadence for both continuum and Narrow Band (NB) light curves. 2) Prioritize light curves where $F_{\rm var}$ exceeds {1\%} for both continuum and NB. 3) Select light curves with $\eta$ values below unity, with a few exceptions made upon eye inspection of the light curves, and ensure that the cadence is within 10 days. This choice is motivated by the luminosity and redshift of the Seyfert galaxies, where time delays are predicted to be of a similar order.  
The outlined selection procedure is depicted in Figure~\ref{fig:vnrm_std}, which illustrates $\eta$ values for NB versus those for the continuum band. The figures are color-coded based on $F_{\rm var}$ (on the left) and cadence (on the right). In total, observations were conducted for 80 objects, comprising a total of 120 light curves, accounting for observations across multiple epochs. The ultimately chosen light curves are delineated with a black border, constituting the final sample presented in Table~\ref{tab:sample}. This final sample comprises 48 objects, with 12 of them observed across multiple epochs, resulting in a total of 65 light curves. } All the light curves are given in a electronic version.

\begin{deluxetable}{llcc|llcc}
\renewcommand{\arraystretch}{0.70}
\tablecaption{Light curves properties for each source ($F_{var}$ in per-cent, $\eta$, and cadence in days) for the continuum ($B$ or $V$) as well as for the NB filter. }\label{tab:lc_properties}
 \tablehead{Object & Year & $B$ or $V$ Continuum  & NB Filter & Object & Year & $B$ or $V$ Continuum & NB Filter \\ &  &  $F_{\rm var}$/$\eta$/cadence& $F_{\rm var}$/$\eta$/cadence &  &  & $F_{\rm var}$/$\eta$/cadence & $F_{\rm var}$/$\eta$/cadence}
 \startdata
1H2107-097 	& 	2012	& 	8.0	/	0.222	/	4.3	& 	3.3	/	0.445	/	4.9	&	MRK335	 &	2010	 &	4.7	/	0.226	/	2.5	 &	1.7	/	1.712	/	4.4	 \\
3C120 	 &	2014	 &	7.8	/	0.064	/	1.8	 &	1.0	/	0.395	/	2.4	&	MRK335	 &	2011	 &	7.2	/	0.257	/	5.9	 &	0.7	/	1.416	/	4.1	 \\
AKN120	 &	2018	 &	4.5	/	0.188	/	1.4	 &	1.5	/	0.322	/	2.3	&	MRK335	 &	2014	 &	1.3	/	0.385	/	1.2	 &	1.0	/	0.518	/	1.9	 \\
CTSG03\_04	 &	2013	 &	10.5	/	0.302	/	3.6	 &	3.0	/	0.264	/	3.4	&	MRK509	 &	2014	 &	5.0	/	0.275	/	1.4	 &	1.3	/	0.949	/	2.2	 \\
ESO141-G55	 &	2013	 &	5.9	/	0.106	/	3.5	 &	1.9	/	0.395	/	3.7	&	MRK705	 &	2013	 &	8.1	/	0.529	/	4.5	 &	3.8	/	0.177	/	6.1	 \\
ESO141-G55	 &	2015	 &	7.6	/	0.203	/	2.0	 &	2.2	/	0.547	/	2.1	&	MRK841	 &	2014	 &	9.2	/	0.159	/	3.4	 &	3.8	/	0.423	/	4.9	 \\
ESO323-G77	 &	2015	 &	1.7	/	0.754	/	1.9	 &	1.9	/	1.36	/	1.9	&	NGC1019	 &	2011	 &	5.4	/	1.083	/	4.7	 &	2.2	/	0.991	/	4.0	 \\
ESO374-G25	 &	2011	 &	6.3	/	0.608	/	4.0	 &	6.5	/	0.616	/	3.9	&	NGC4726	 &	2013	 &	3.1	/	1.021	/	4.8	 &	2.6	/	0.91	/	3.9	\\
ESO399-IG20	 &	2011	 &	8.0	/	0.487	/	6.0	 &	4.8	/	0.39	/	5.5	&	NGC5940	 &	2014	 &	7.7	/	1.516	/	4.2	 &	3.9	/	0.837	/	5.2	 \\
ESO438-G09	 &	2011	 &	5.1	/	0.969	/	5.1	 &	4.2	/	0.709	/	4.8	&	NGC6860	 &	2015	 &	3.6	/	1.27	/	2.5	 &	6.2	/	1.02	/	2.8	\\
ESO438-G09	 &	2015	 &	6.4	/	0.353	/	1.7	 &	4.6	/	0.868	/	2.0	&	NGC7214	 &	2011	 &	3.8	/	1.463	/	4.0	 &	3.3	/	1.089	/	3.8	 \\
ESO490-IG20	 &	2011	 &	2.7	/	1.431	/	4.0	 &	2.5	/	1.047	/	4.0	&	NGC7469	 &	2012	 &	2.8	/	0.749	/	4.2	 &	0.9	/	0.694	/	3.9	 \\
ESO511-G030	 &	2013	 &	6.9	/	0.518	/	4.5	 &	3.0	/	0.842	/	6.7	&	NGC7603	 &	2014	 &	7.0	/	0.211	/	1.6	 &	5.4	/	0.221	/	3.7	 \\
ESO511-G030	 &	2014	 &	9.1	/	0.734	/	4.1	 &	6.0	/	0.318	/	4.7	&	NGC985	 &	2014	 &	8.9	/	0.575	/	3.9	 &	1.3	/	0.614	/	4.0	 \\
ESO549-G49	 &	2012	 &	3.7	/	1.488	/	4.6	 &	3.7	/	1.605	/	4.9	&	PG1149-110	 &	2013	 &	6.4	/	1.113	/	4.4	 &	2.4	/	1.075	/	3.7	 \\
ESO578-G09	 &	2014	 &	5.3	/	0.566	/	3.6	 &	3.6	/	0.318	/	4.1	&	PGC50427	 &	2011	 &	8.3	/	0.541	/	5.5	 &	3.1	/	0.463	/	4.8	 \\
F1041 	 &	2013	 &	6.9	/	0.878	/	4.2	 &	4.2	/	0.321	/	5.0	&	PGC50427	 &	2014	 &	6.9	/	0.82	/	4.4	 &	2.3	/	0.658	/	4.3	 \\
HE0003-5023	 &	2014	 &	4.3	/	0.269	/	1.5	 &	2.5	/	0.485	/	1.8	&	PGC64989	 &	2013	 &	2.8	/	1.011	/	3.5	 &	3.5	/	0.213	/	3.3	 \\
HE1136-2304	 &	2015	 &	7.4	/	0.564	/	1.9	 &	4.2	/	0.788	/	1.9	&	PGC64989	 &	2014	 &	4.8	/	0.496	/	1.8	 &	6.3	/	0.106	/	2.2	 \\
HE1136-2304	 &	2016	 &	6.8	/	0.636	/	1.9	 &	2.9	/	1.514	/	2.1	&	RXSJ06225-2317	 &	2013	 &	5.4	/	0.768	/	2.7	 &	3.6	/	0.816	/	3.5	 \\
HE1136-2304	 &	2018	 &	4.9	/	0.796	/	3.7	 &	0.9	/	0.792	/	3.1	&	RXSJ1103.2-0654	 &	2011	 &	4.9	/	1.426	/	4.7	 &	6.6	/	1.229	/	4.8	 \\
HE1143-1810	 &	2016	 &	8.2	/	0.233	/	2.2	 &	0.8	/	0.727	/	2.4	&	RXSJ1103.2-0654	 &	2014	 &	8.8	/	1.209	/	3.2	 &	2.4	/	0.601	/	3.8	\\
HE2128-0221	 &	2016	 &	5.0	/	0.661	/	1.6	 &	3.7	/	1.192	/	1.6	&	RXSJ17414+0348	 &	2012	 &	9.2	/	0.162	/	3.6	 &	7.3	/	0.101	/	4.3	 \\
IC4329A	 &	2015	 &	8.7	/	0.536	/	1.6	 &	3.2	/	0.391	/	1.9	&	RXSJ17414+0348	 &	2014	 &	4.8	/	0.444	/	4.5	 &	3.9	/	0.333	/	4.1	 \\ 
IRAS01089-4743	 &	2013	 &	6.7	/	0.64	/	3.4	 &	1.4	/	1.09	/	3.7	&	UGC12138	 &	2012	 &	2.2	/	1.039	/	2.6	 &	2.5	/	1.452	/	3.3	 \\
IRAS09595-0755	 &	2013	 &	11.5	/	0.67	/	3.5	 &	3.8	/	0.63	/	3.5	&	UM163 	 &	2013	 &	7.4	/	0.401	/	3.6	 &	3.1	/	0.404	/	3.3	 \\
IRAS23226-3843	 &	2013	 &	4.1	/	0.933	/	3.4	 &	1.8	/	0.932	/	3.8	&	WPVS007	 &	2012	 &	7.6	/	0.164	/	1.5	 &	3.7	/	0.735	/	1.9	 \\ 
MCG+03-47-002	 &	2013	 &	19.0	/	0.595	/	4.0	 &	5.3	/	0.516	/	4.4	&	WPVS48	 &	2013	 &	6.6	/	0.308	/	2.5	 &	2.4	/	0.223	/	3.8	 \\
MCG-02.12.050 	 &	2014	 &	3.4	/	1.534	/	3.5	 &	2.4	/	0.653	/	2.6	&	WPVS48	 &	2014	 &	7.0	/	0.388	/	1.8	 &	1.7	/	0.237	/	2.0	 \\
MRK1239	 &	2015	 &	1.0	/	1.806	/	1.8	 &	1.9	/	0.416	/	2.0	&	WPVS48	 &	2018	 &	3.7	/	0.357	/	6.0	 &	1.8	/	0.77	/	2.1	\\
MRK1347	 &	2014	 &	3.1	/	1.023	/	3.7	 &	1.3	/	0.407	/	4.6	&		\\														
 \enddata
\end{deluxetable}

\includegraphics[width=0.33\columnwidth]{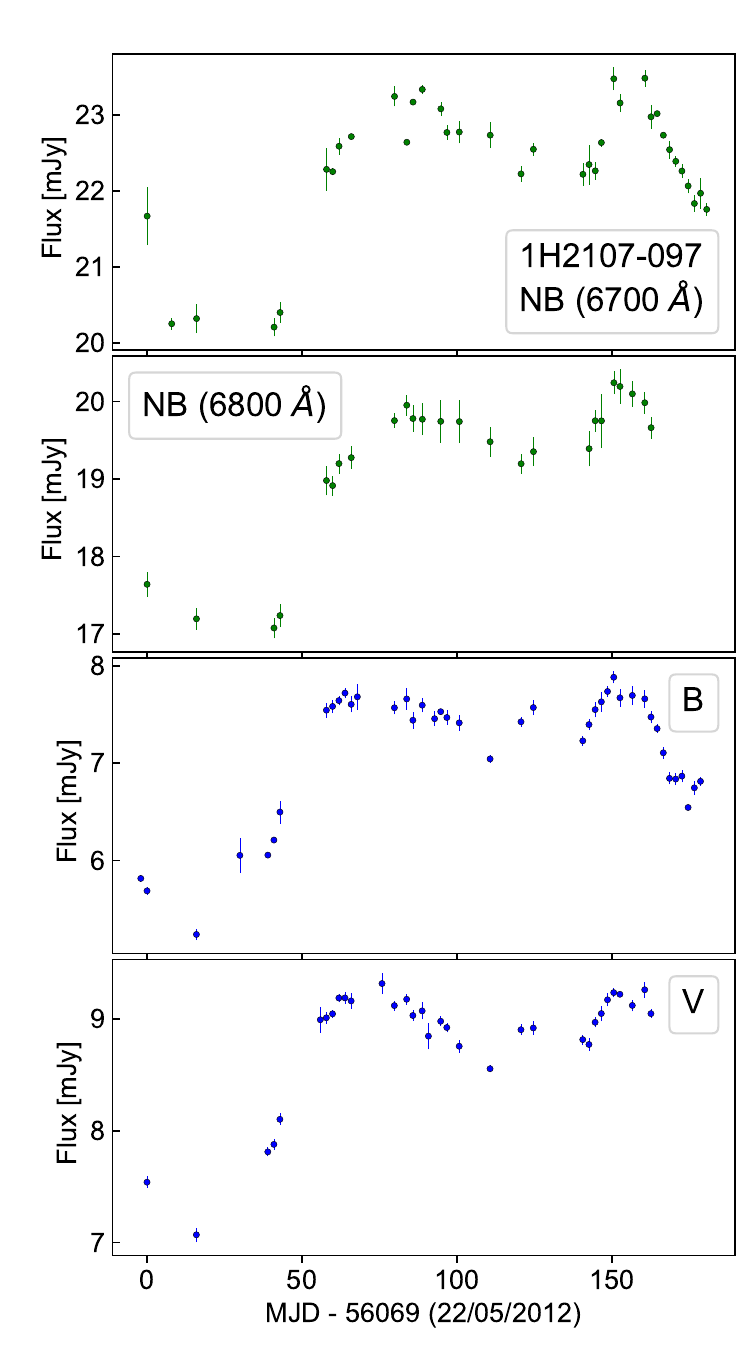}
\includegraphics[width=0.33\columnwidth]{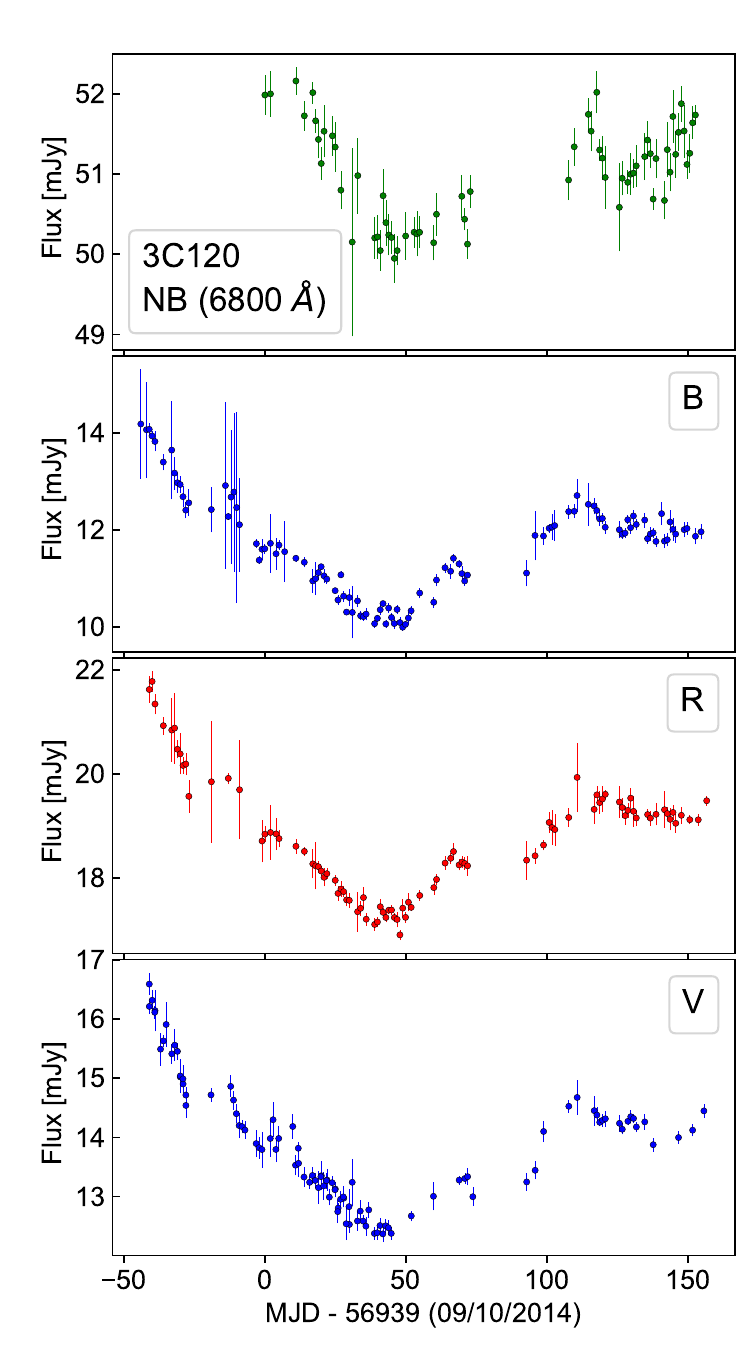}
\includegraphics[width=0.33\columnwidth]{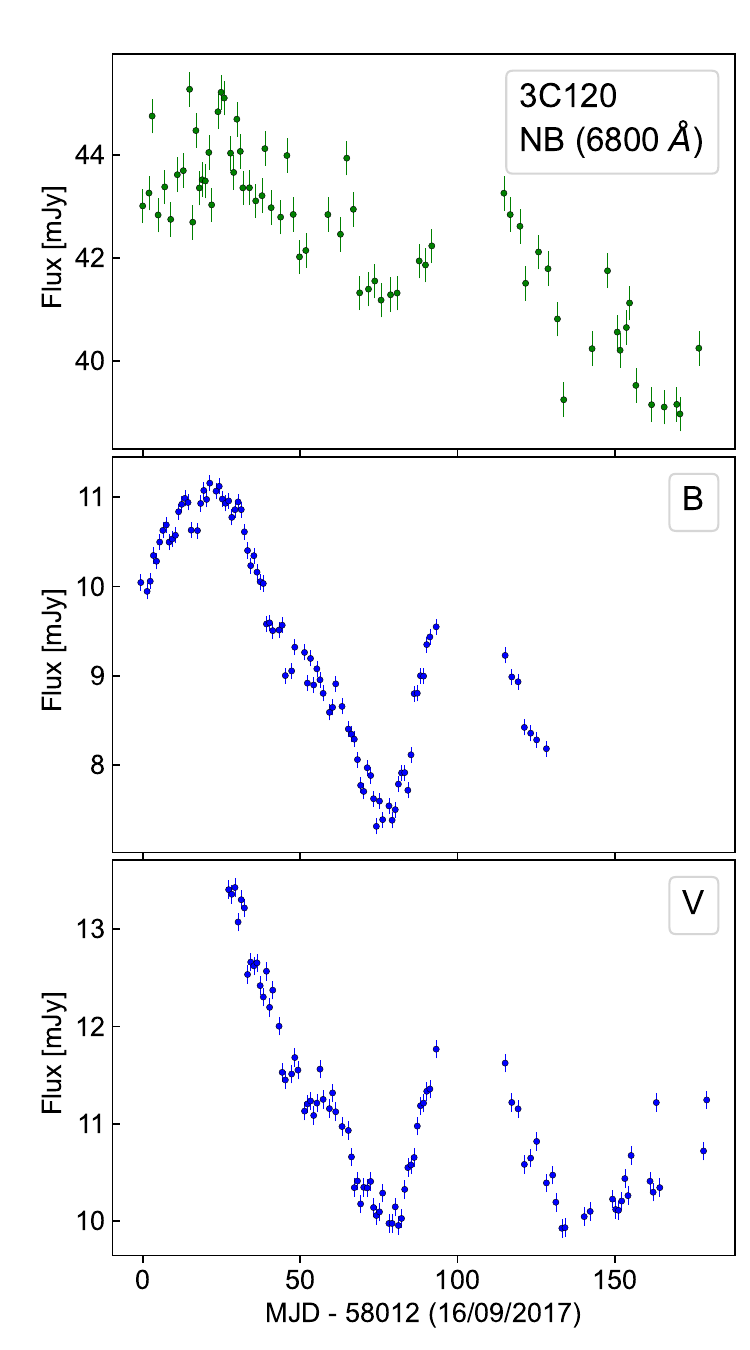}

\includegraphics[width=0.33\columnwidth]{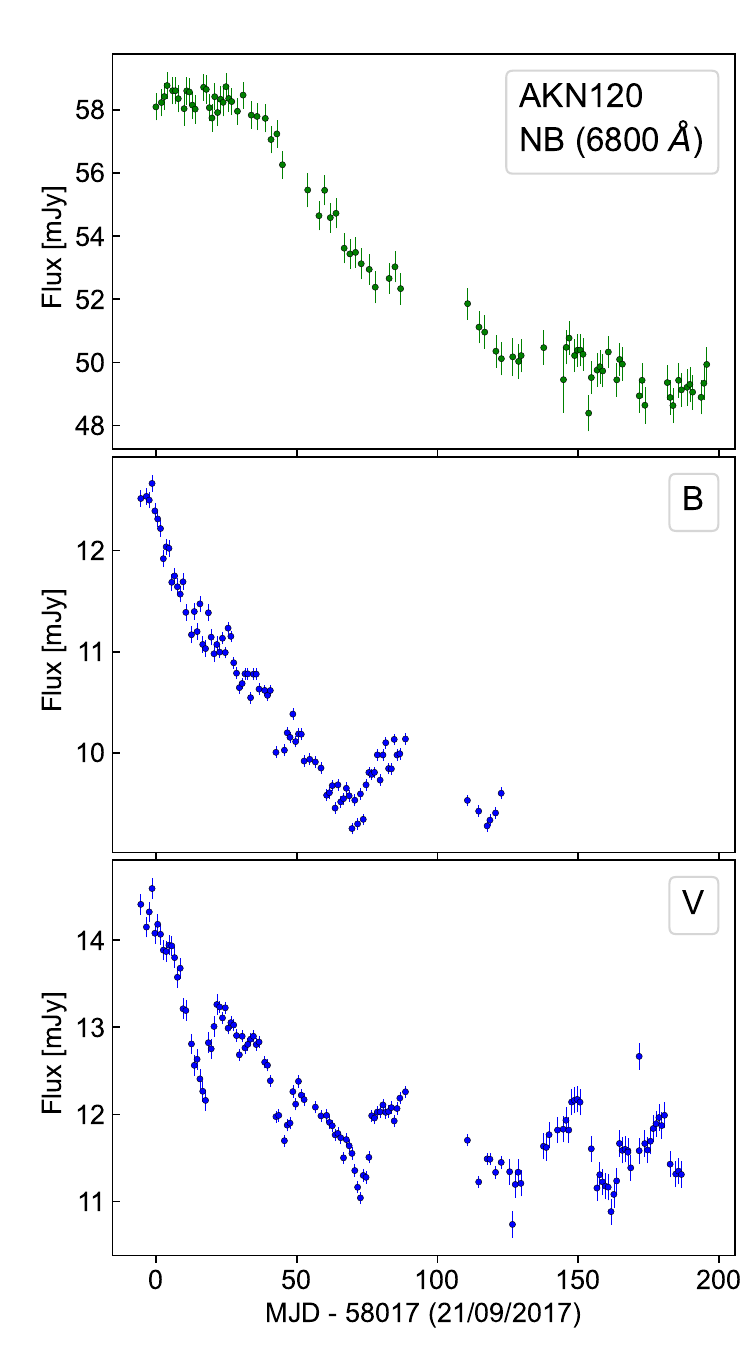}
\includegraphics[width=0.33\columnwidth]{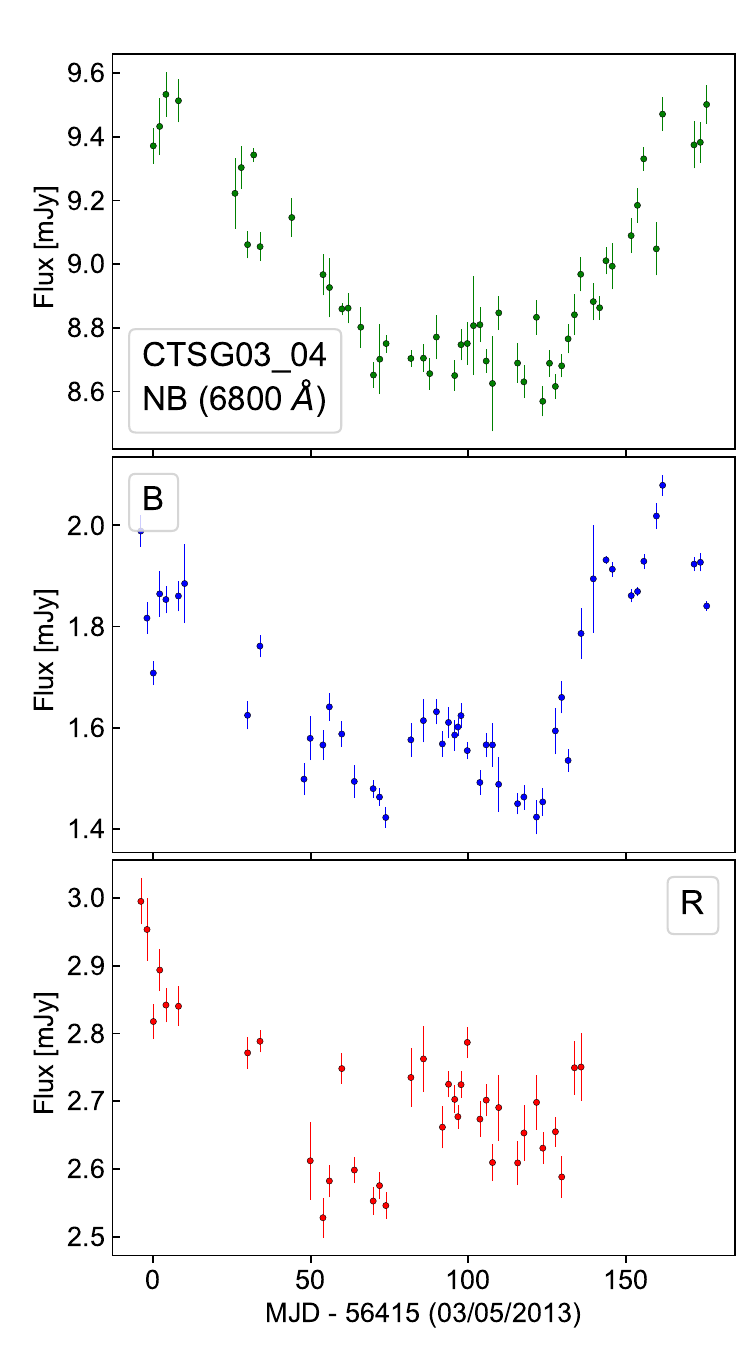}
\includegraphics[width=0.33\columnwidth]{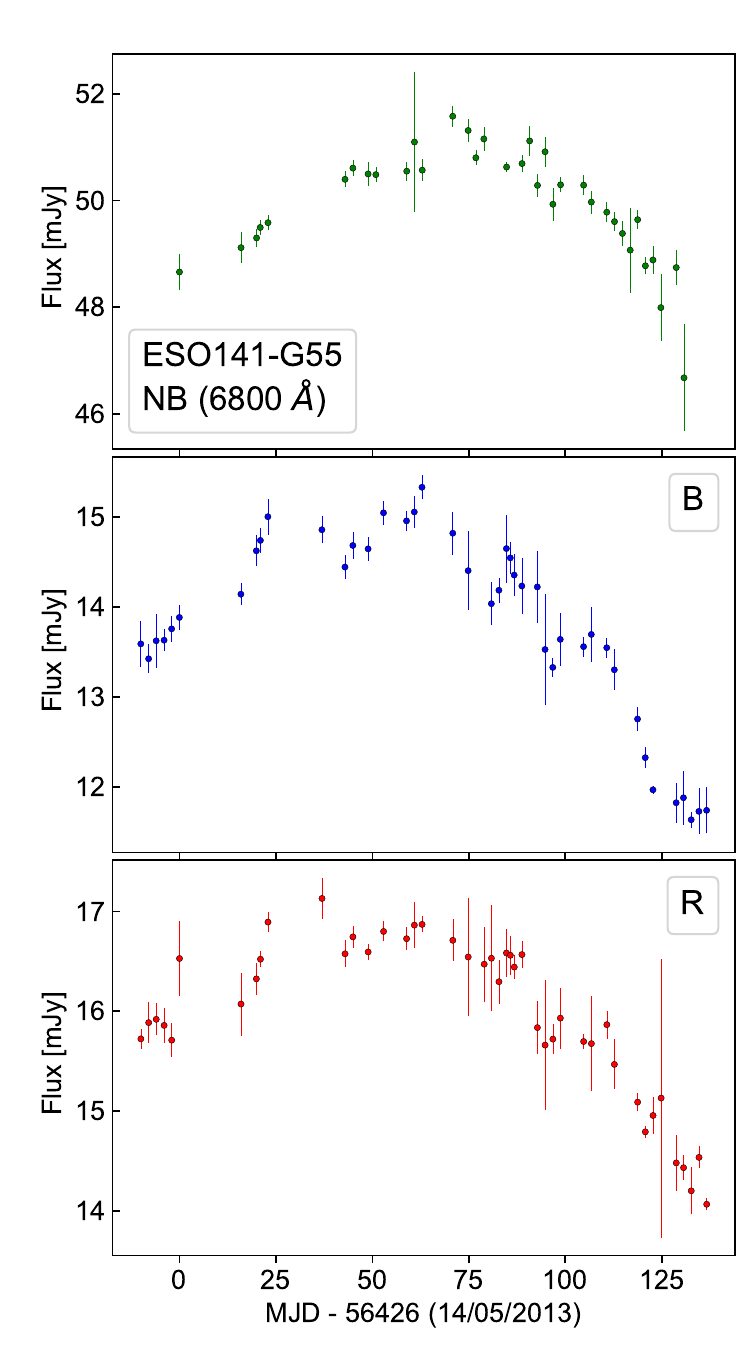}

\includegraphics[width=0.33\columnwidth]{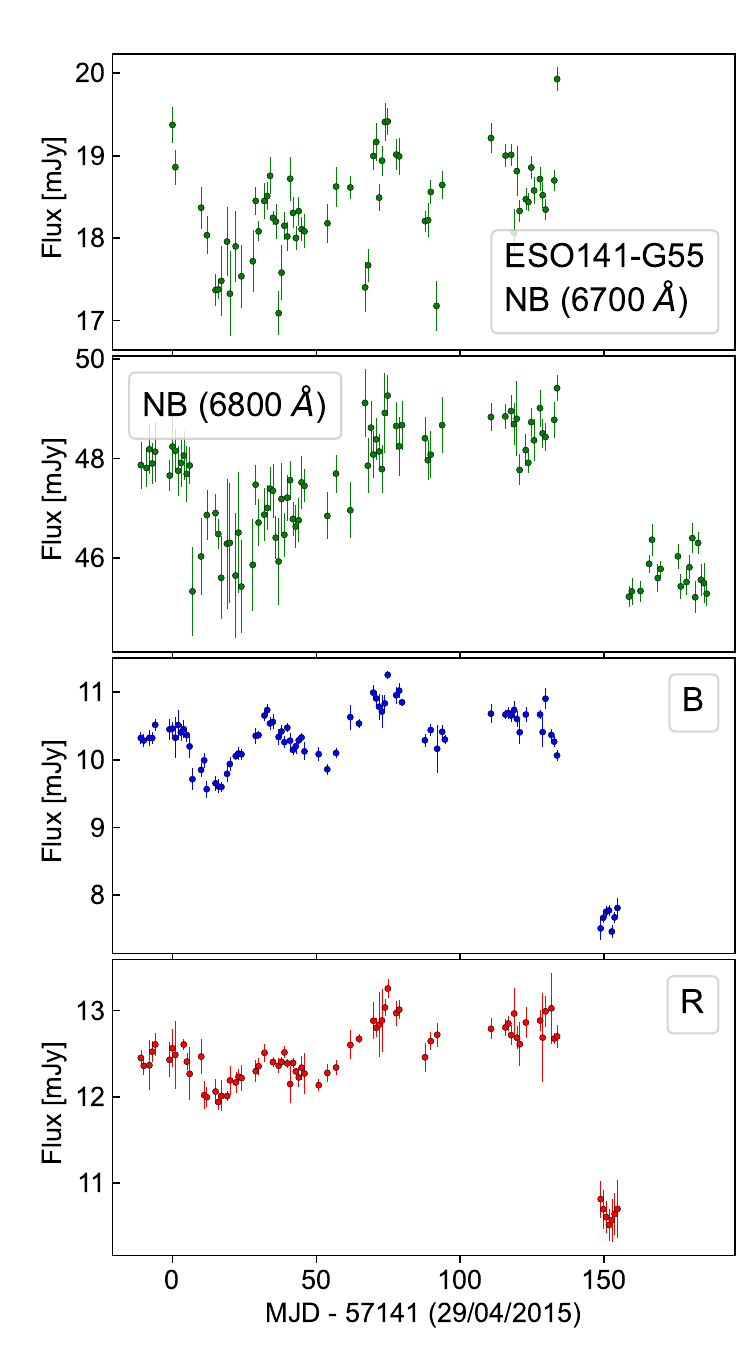}
\includegraphics[width=0.33\columnwidth]{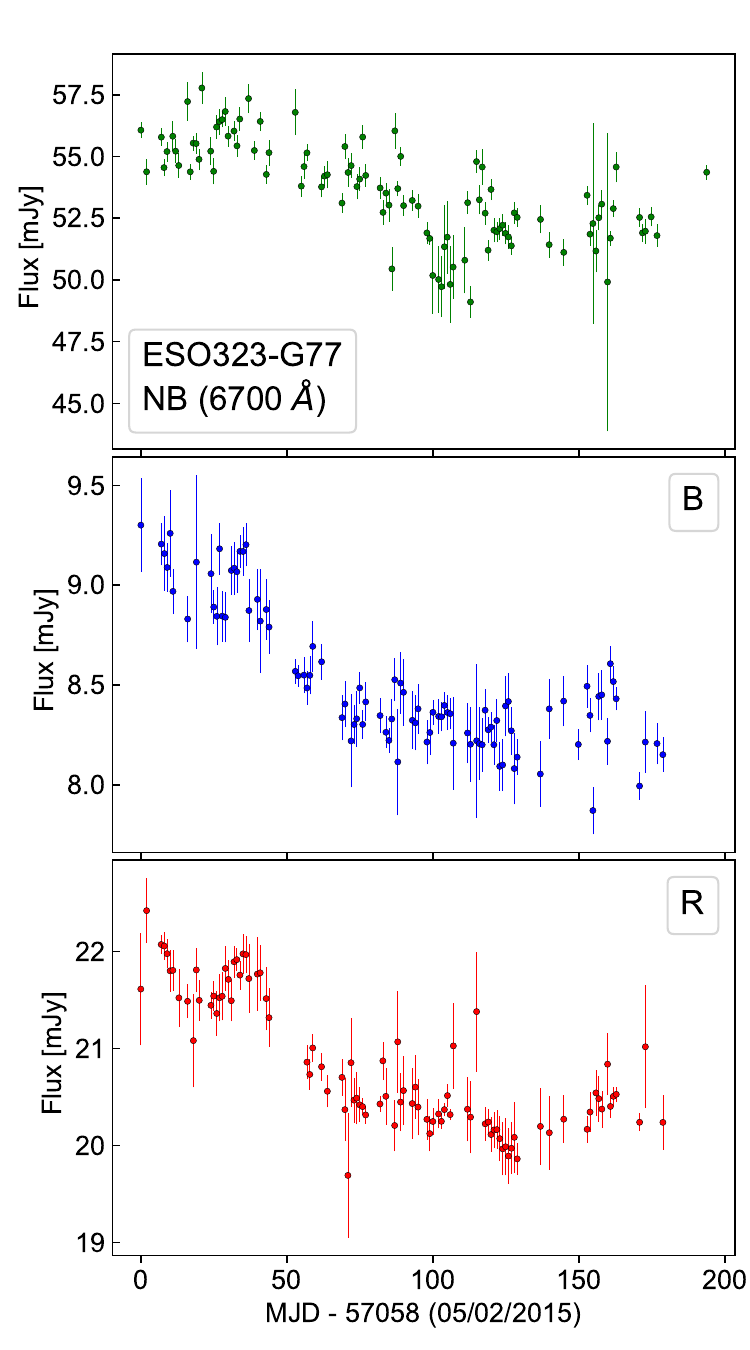}
\includegraphics[width=0.33\columnwidth]{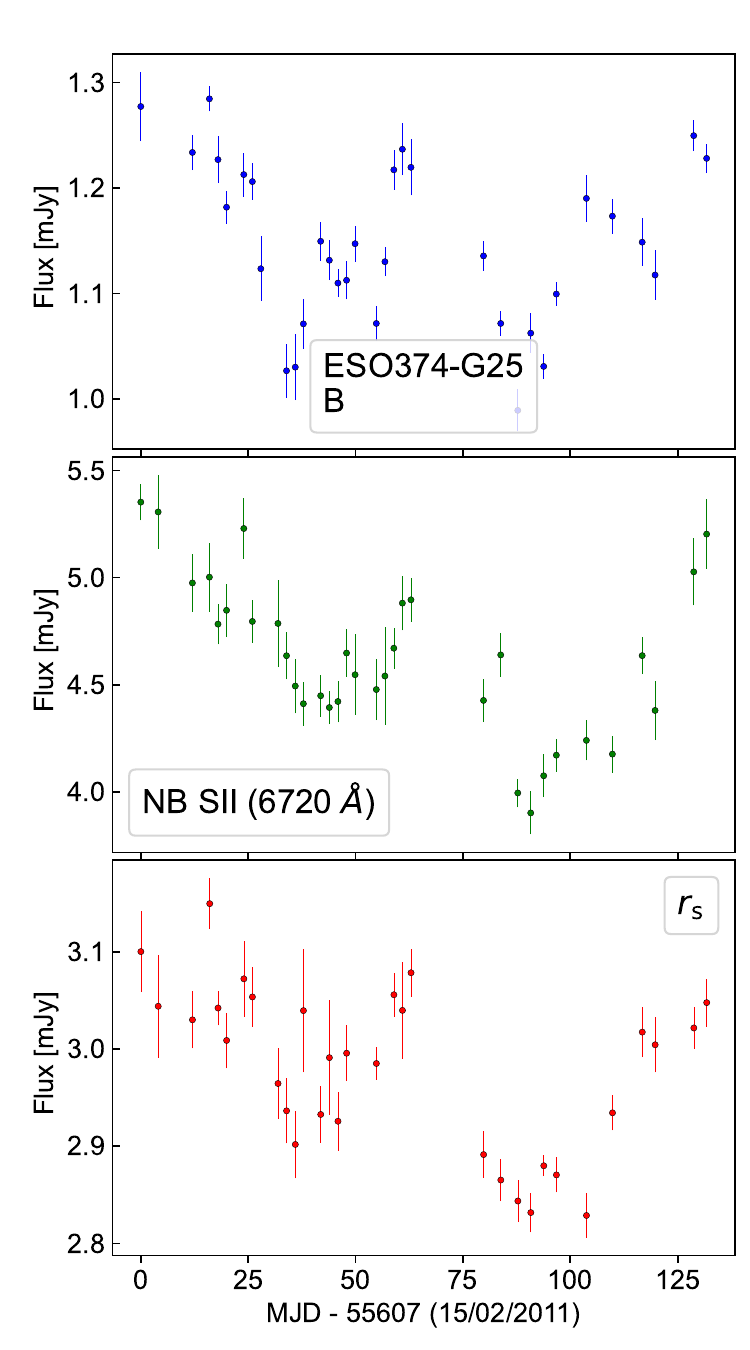}

\includegraphics[width=0.33\columnwidth]{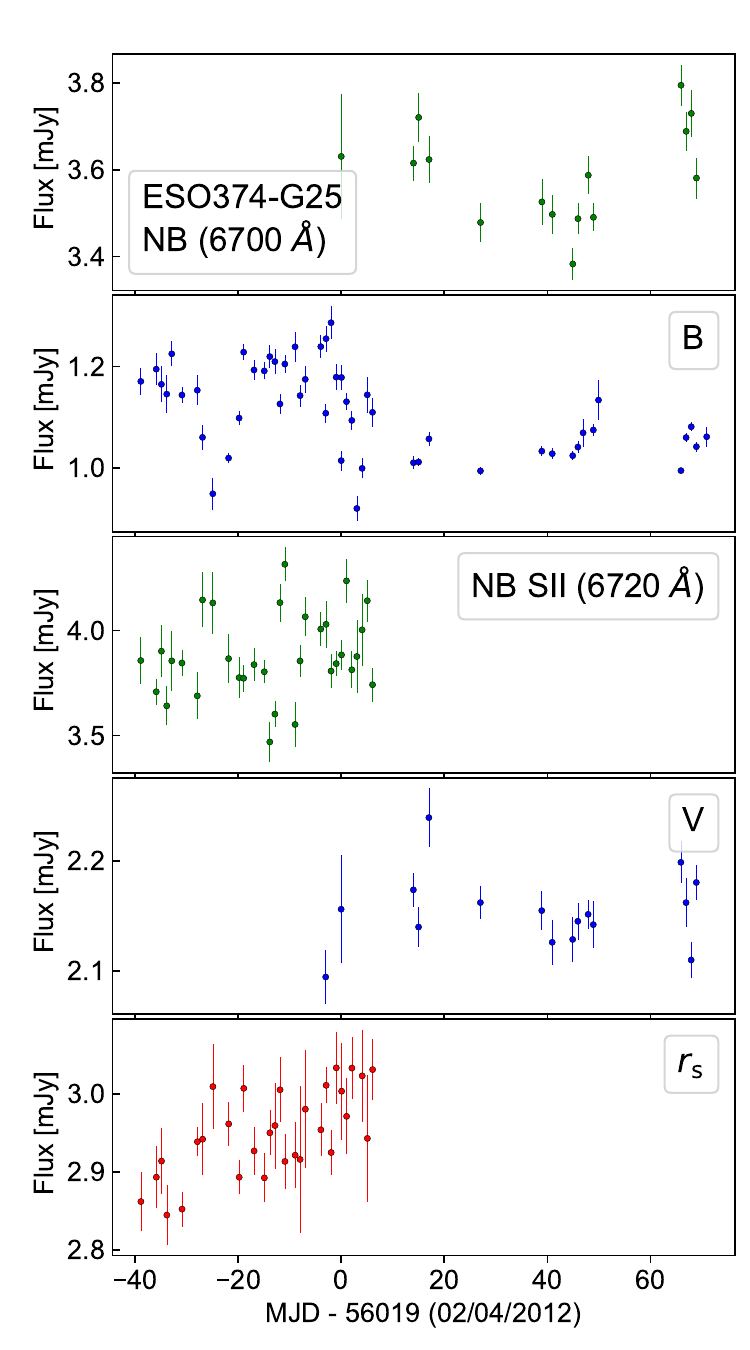}
\includegraphics[width=0.33\columnwidth]{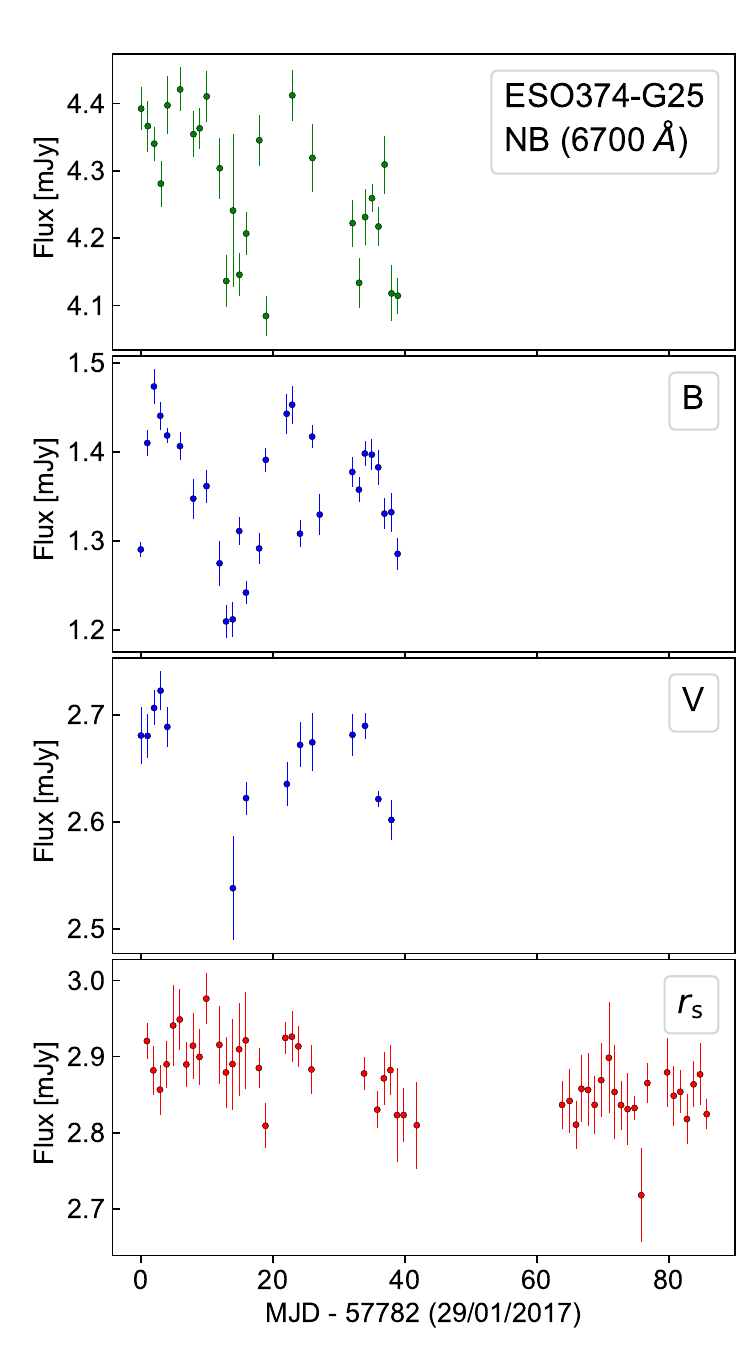}
\includegraphics[width=0.33\columnwidth]{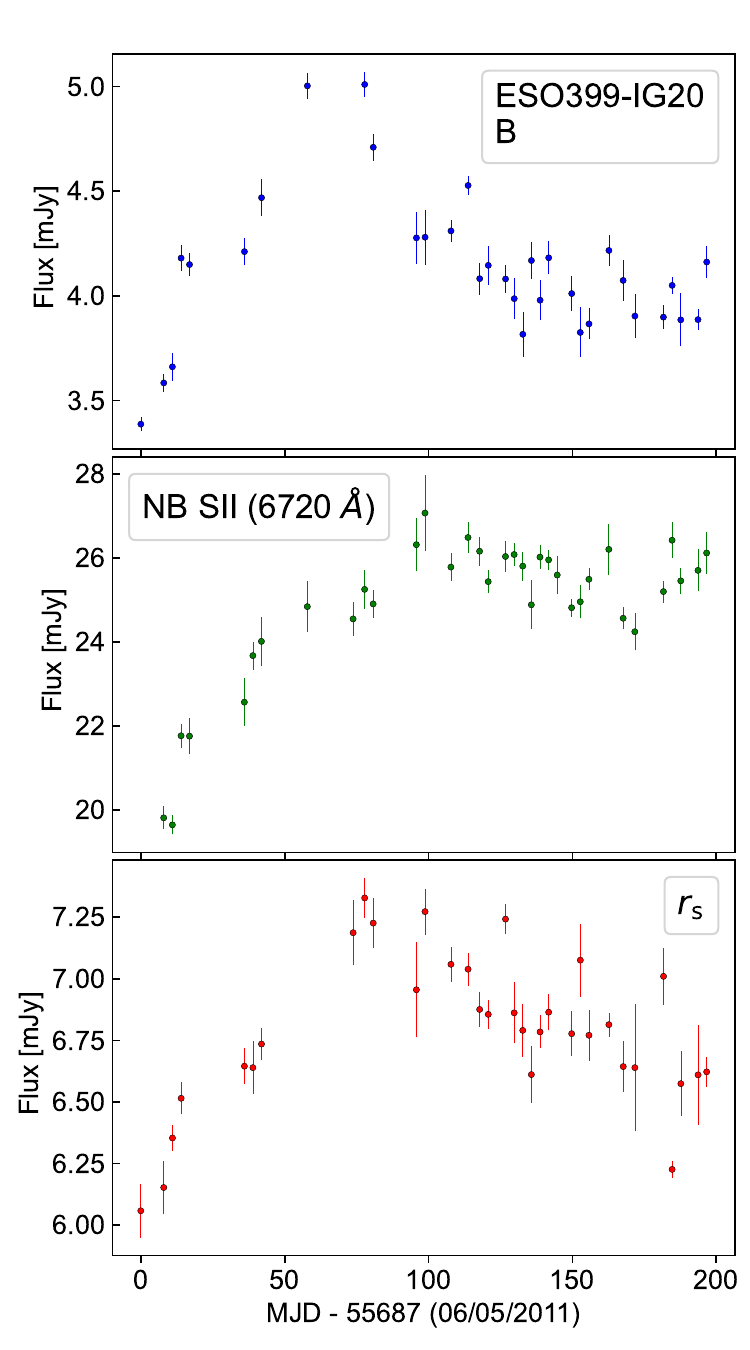}

\includegraphics[width=0.33\columnwidth]{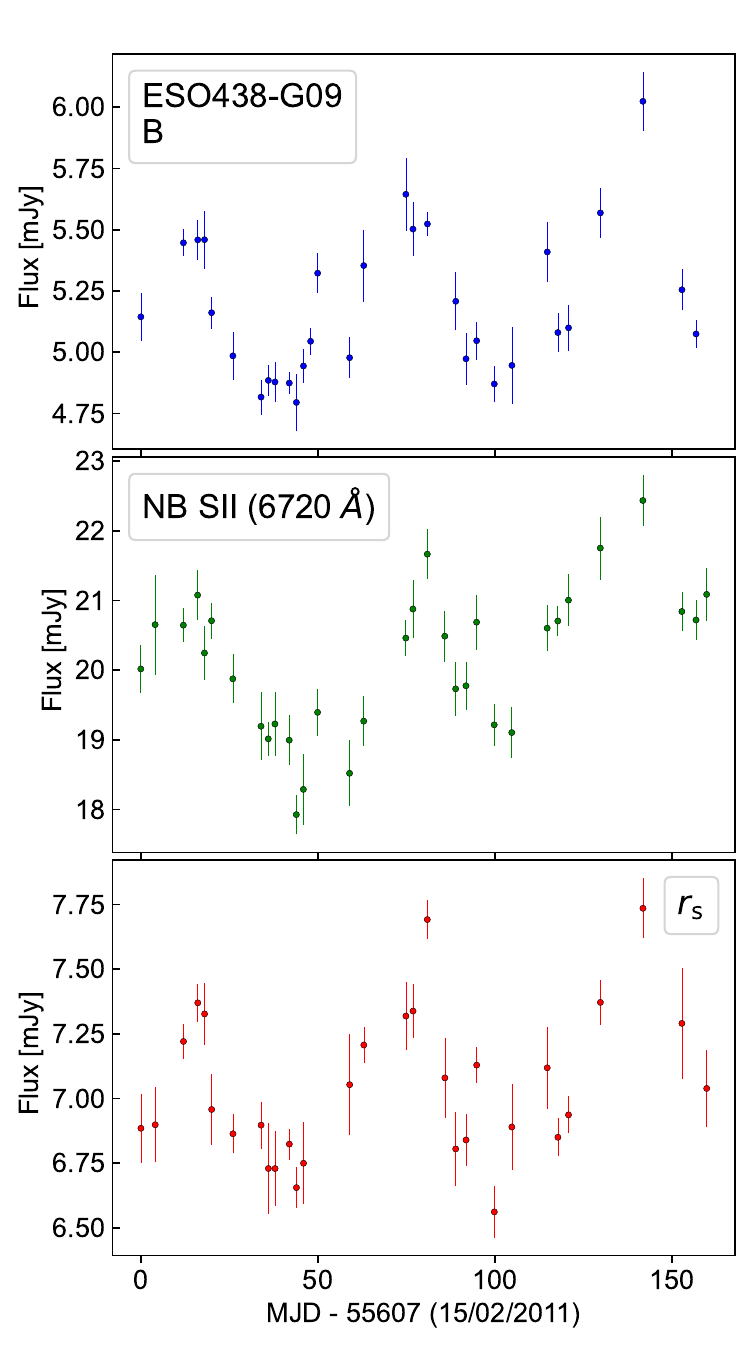}
\includegraphics[width=0.33\columnwidth]{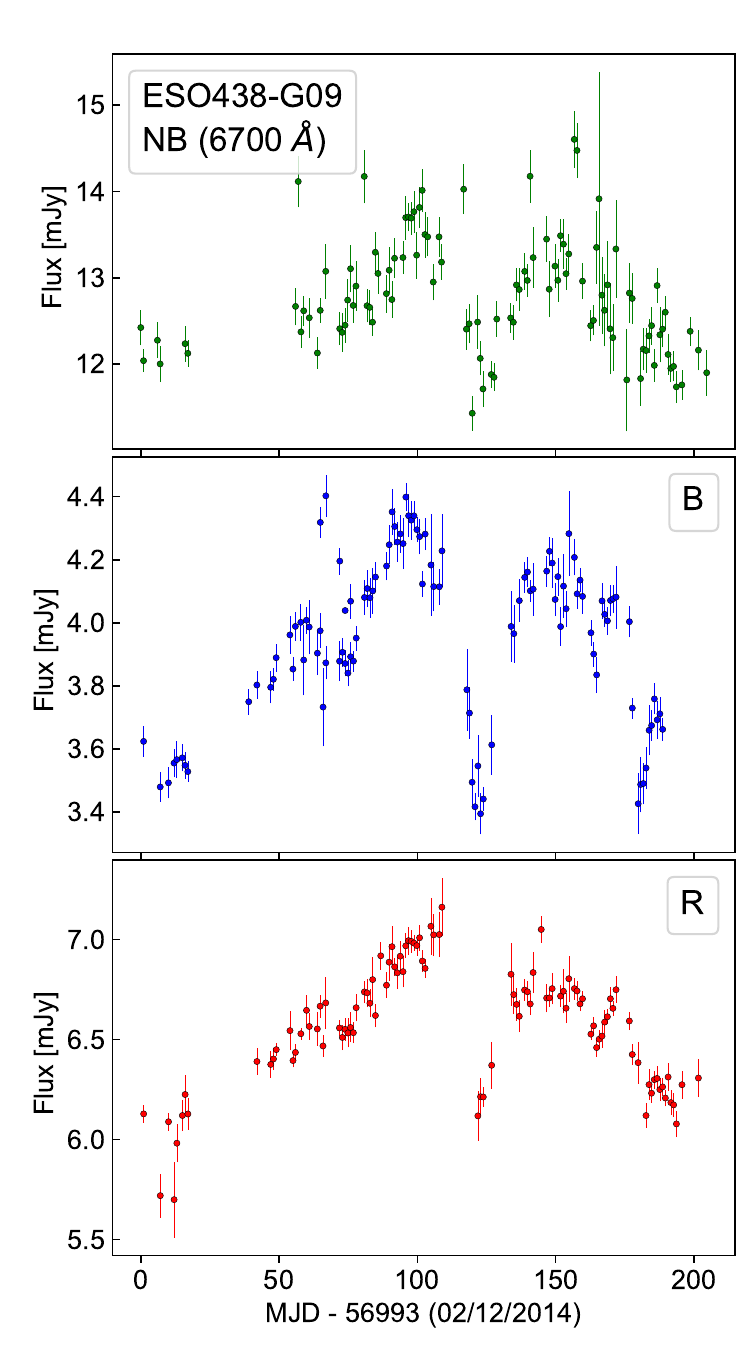}
\includegraphics[width=0.33\columnwidth]{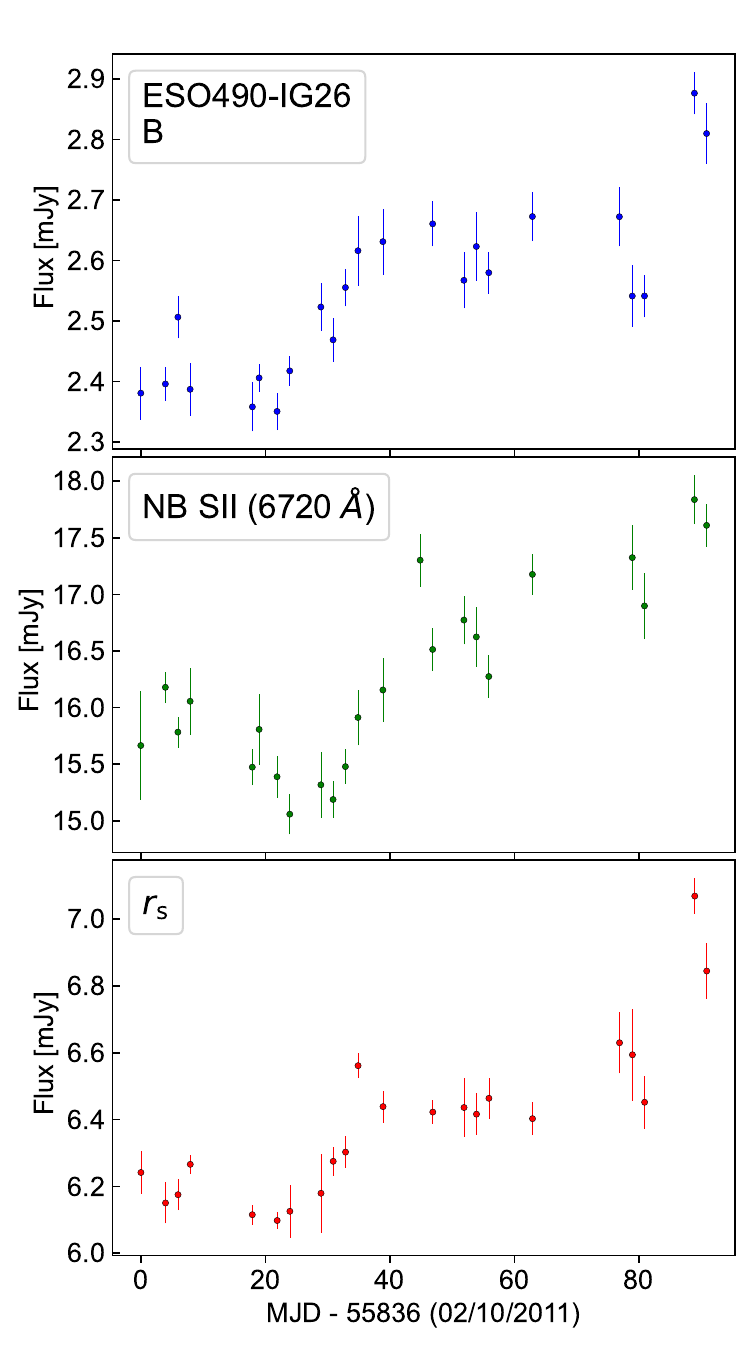}

\includegraphics[width=0.33\columnwidth]{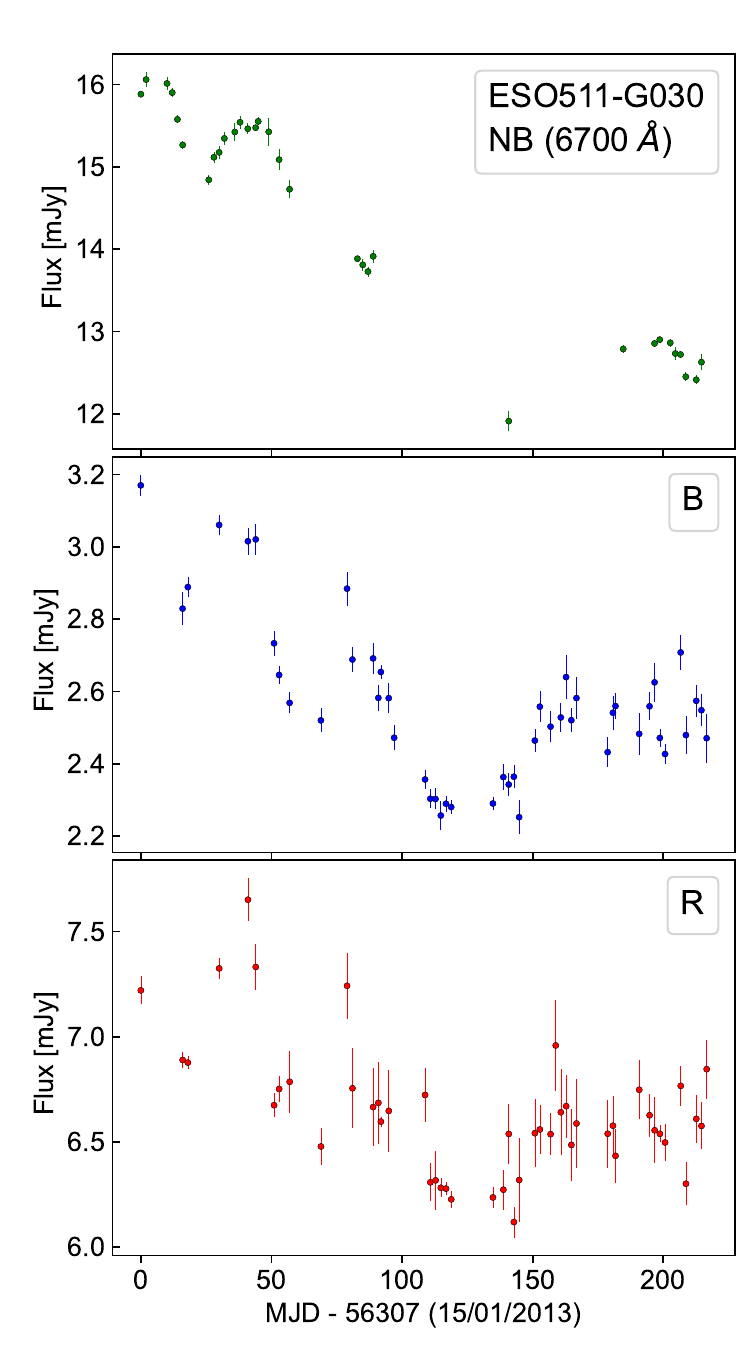}
\includegraphics[width=0.33\columnwidth]{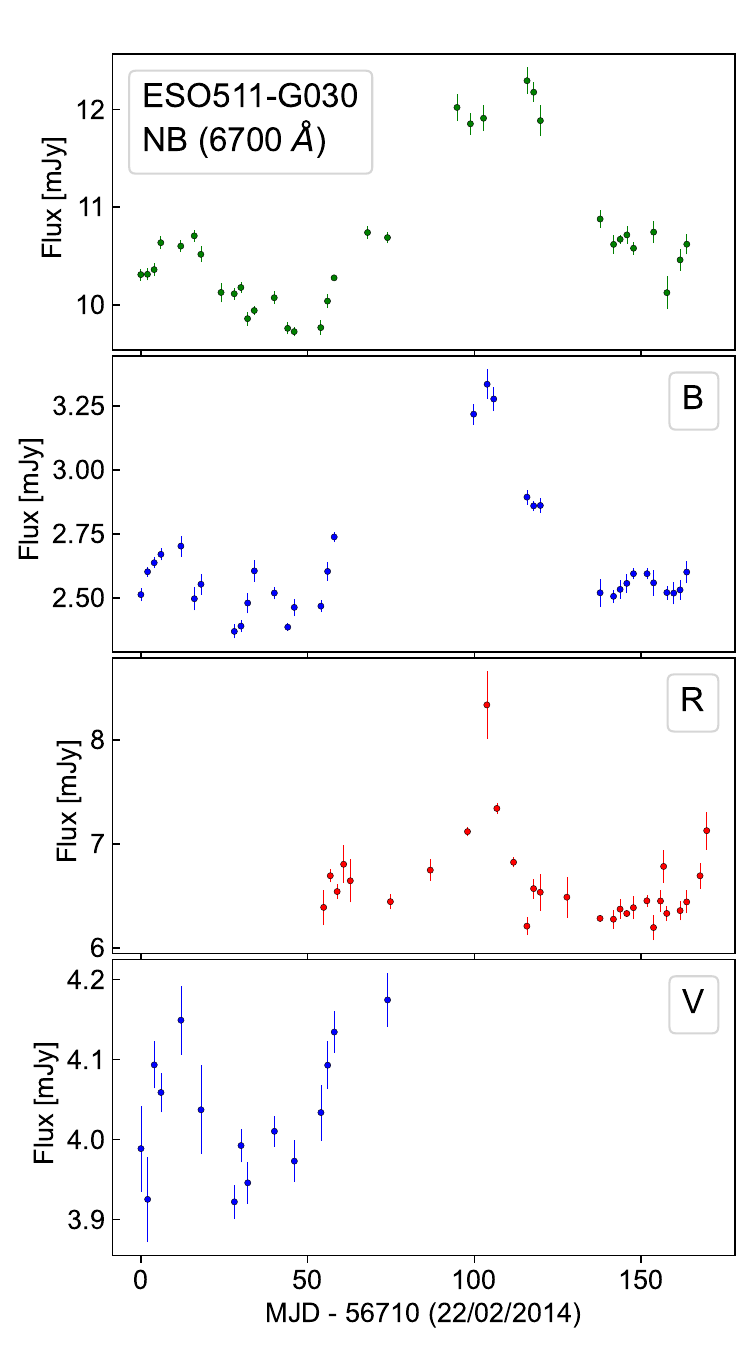}
\includegraphics[width=0.33\columnwidth]{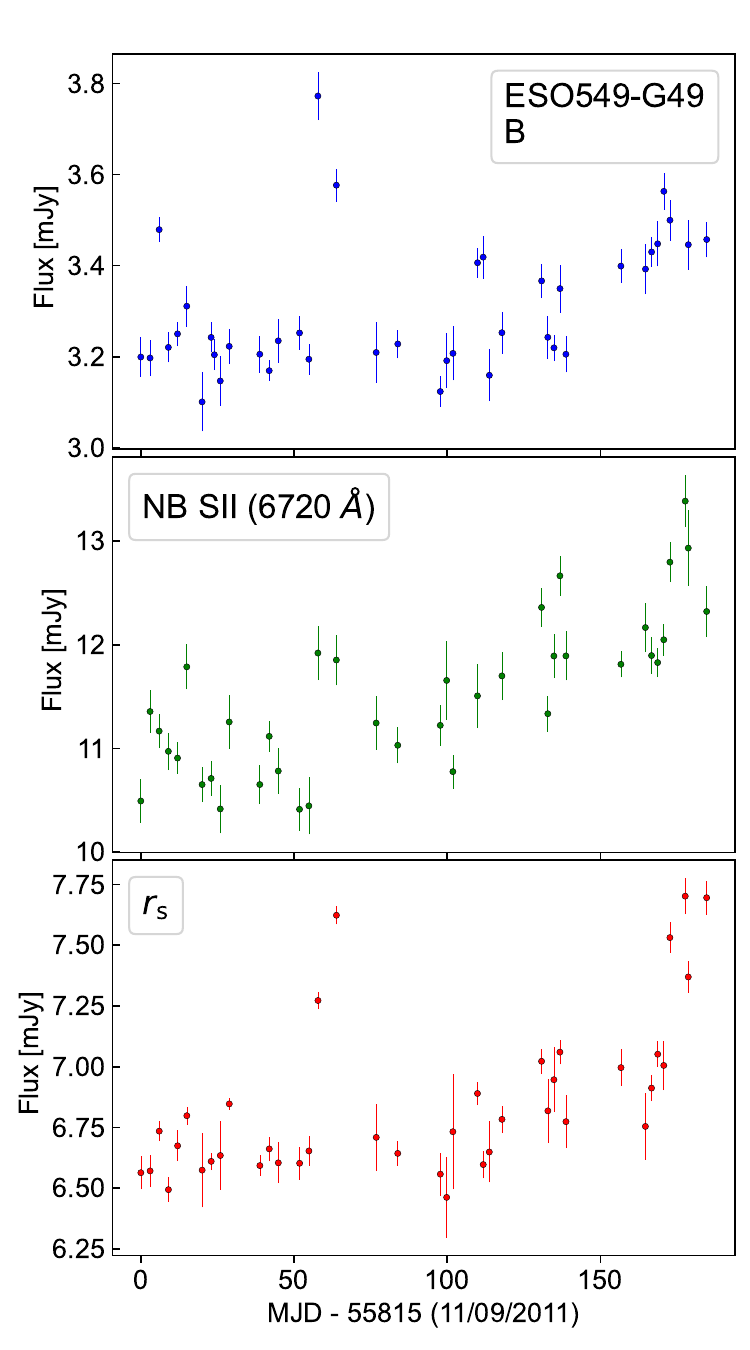}

\includegraphics[width=0.33\columnwidth]{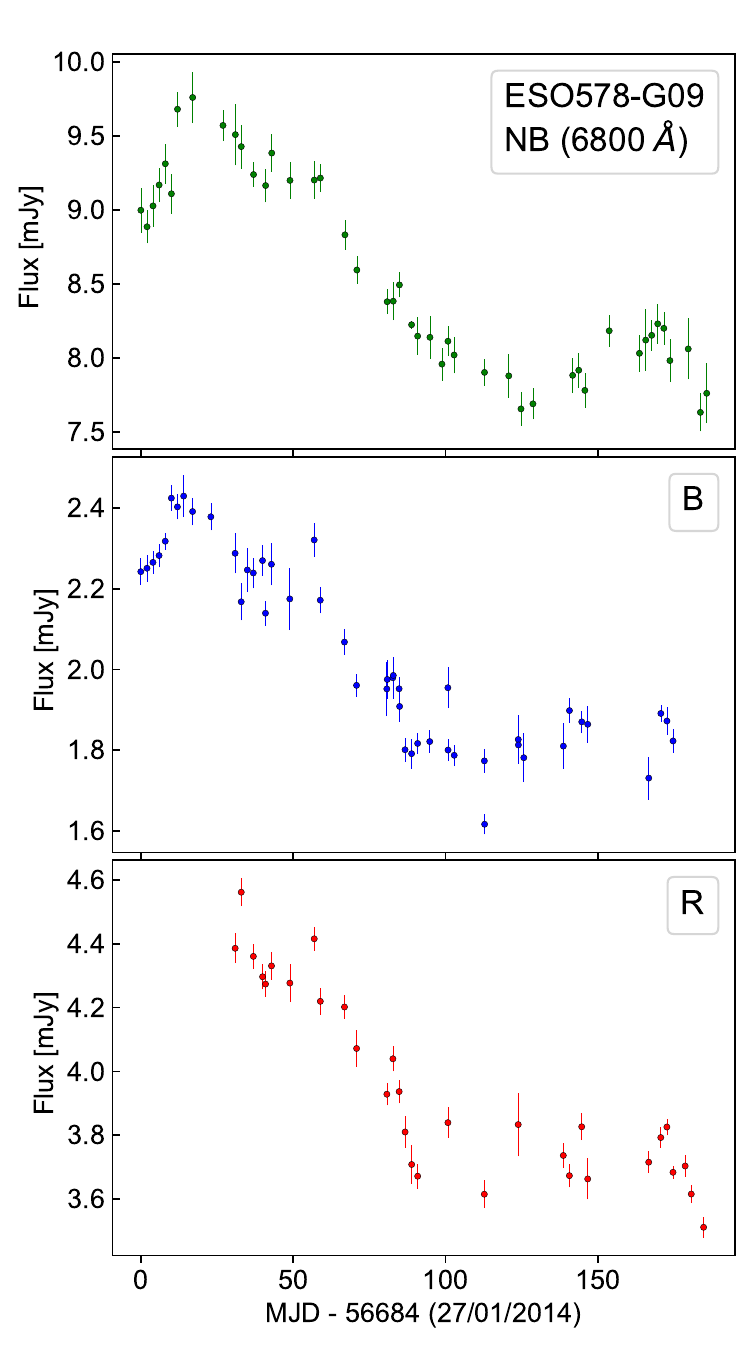}
\includegraphics[width=0.33\columnwidth]{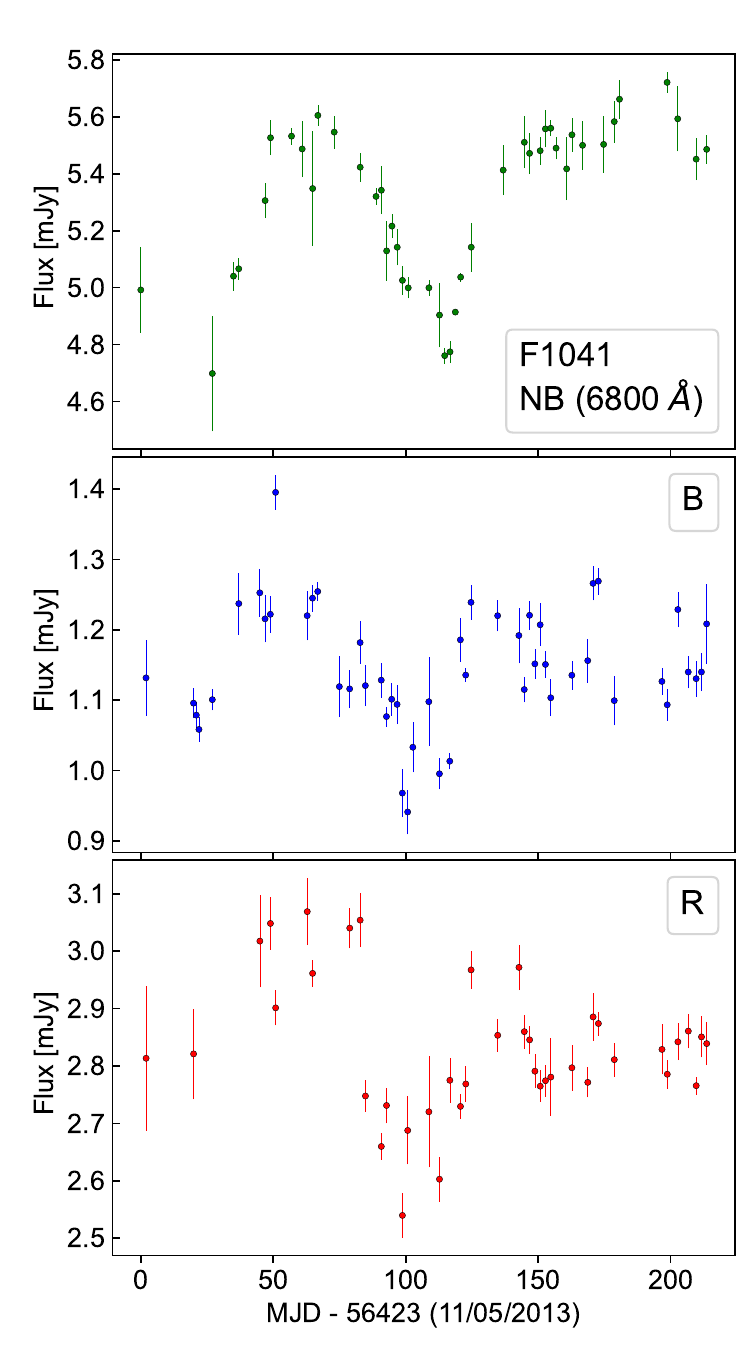}
\includegraphics[width=0.33\columnwidth]{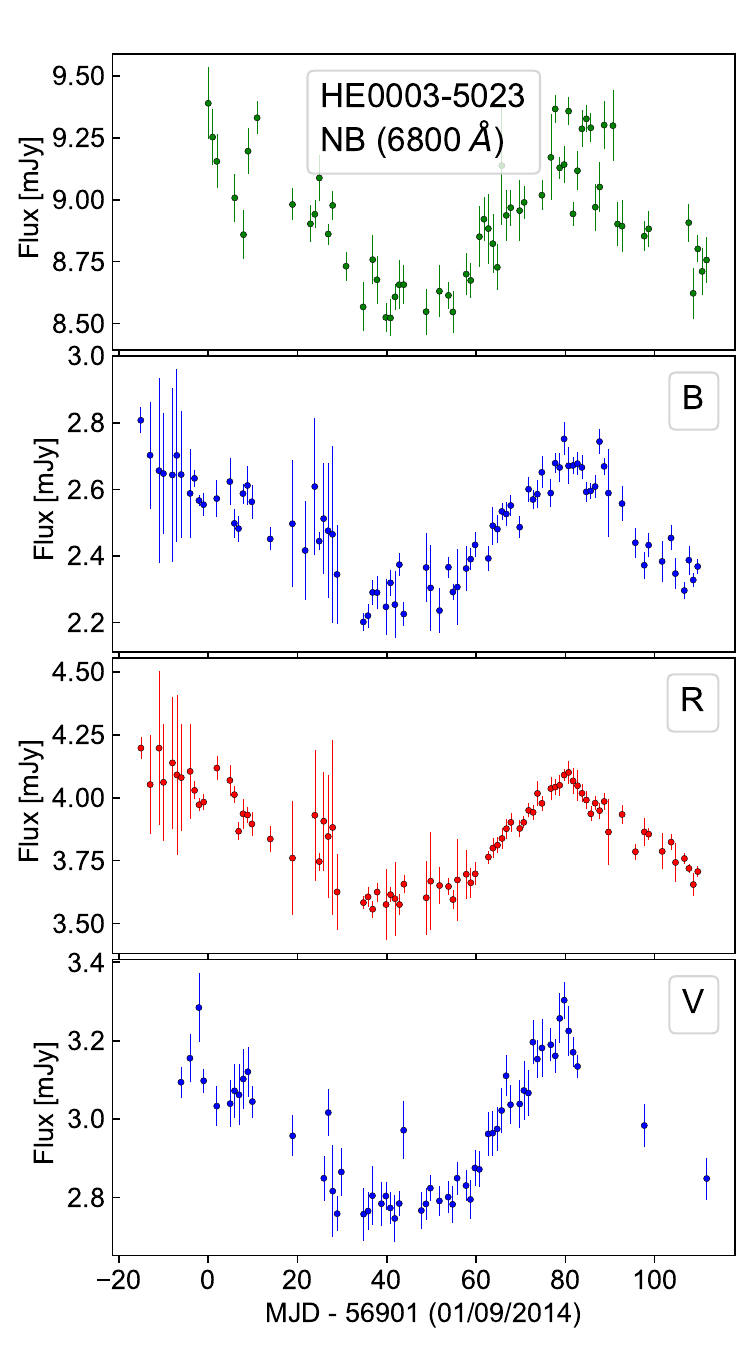}

\includegraphics[width=0.33\columnwidth]{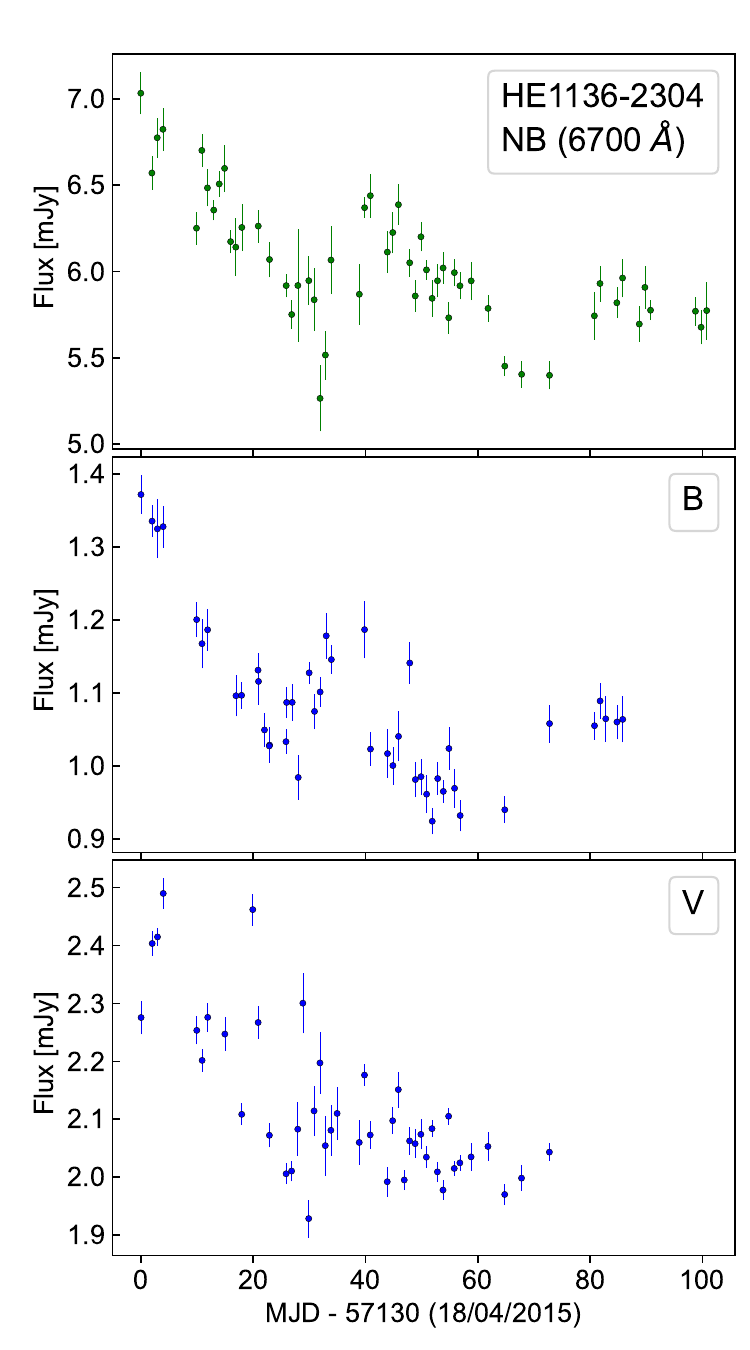}
\includegraphics[width=0.33\columnwidth]{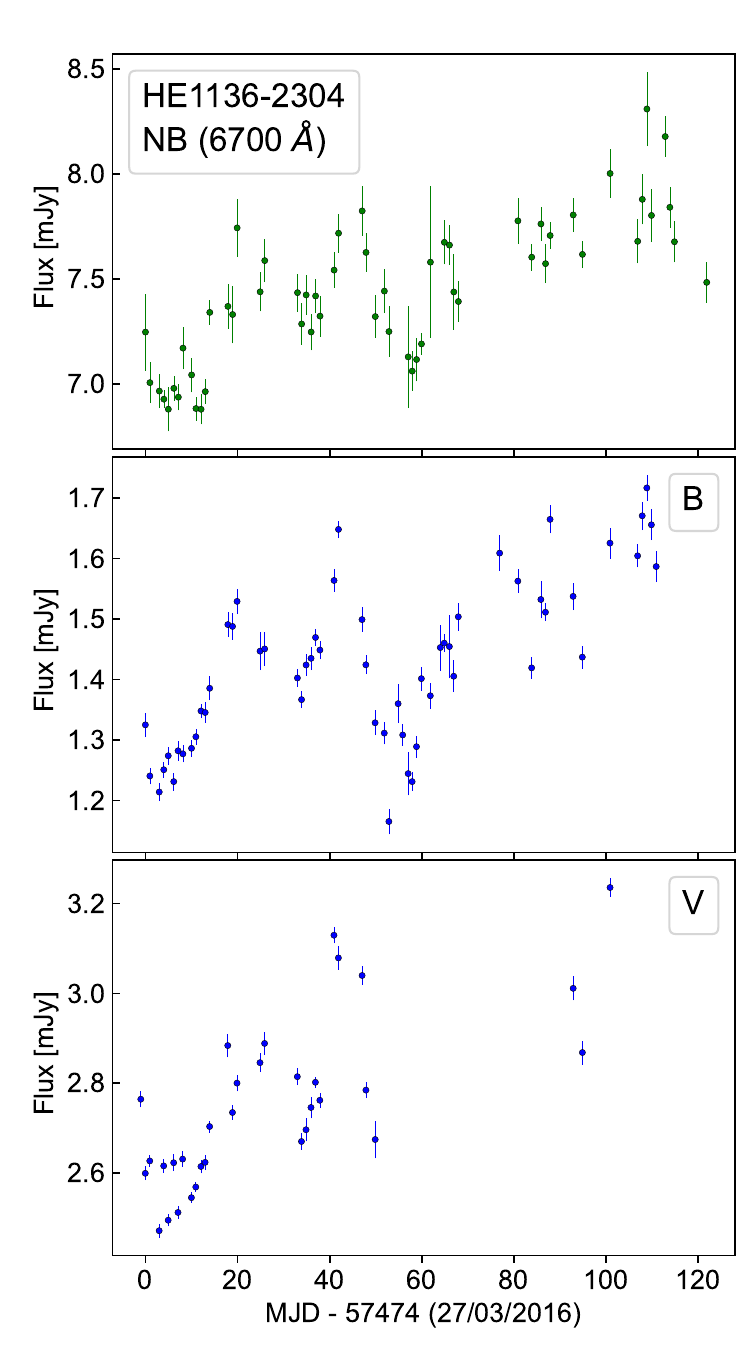}
\includegraphics[width=0.33\columnwidth]{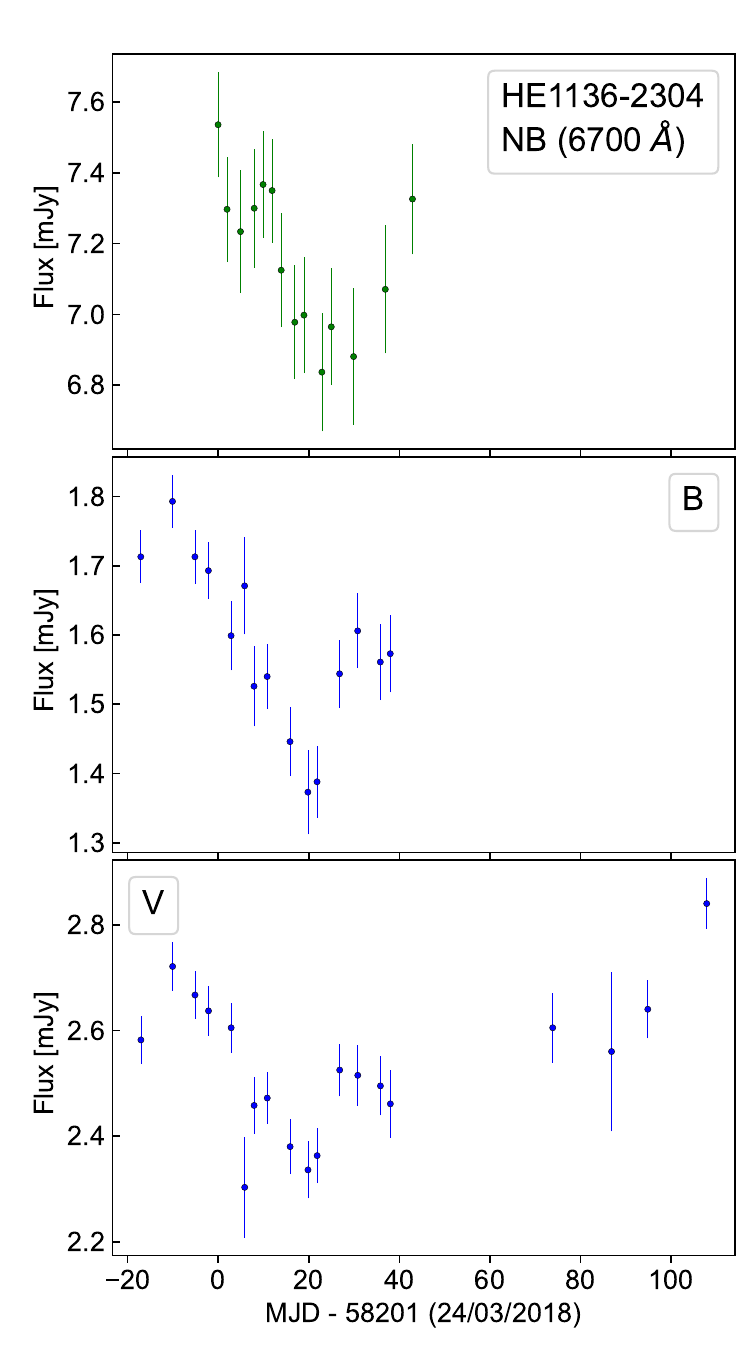}

\includegraphics[width=0.33\columnwidth]{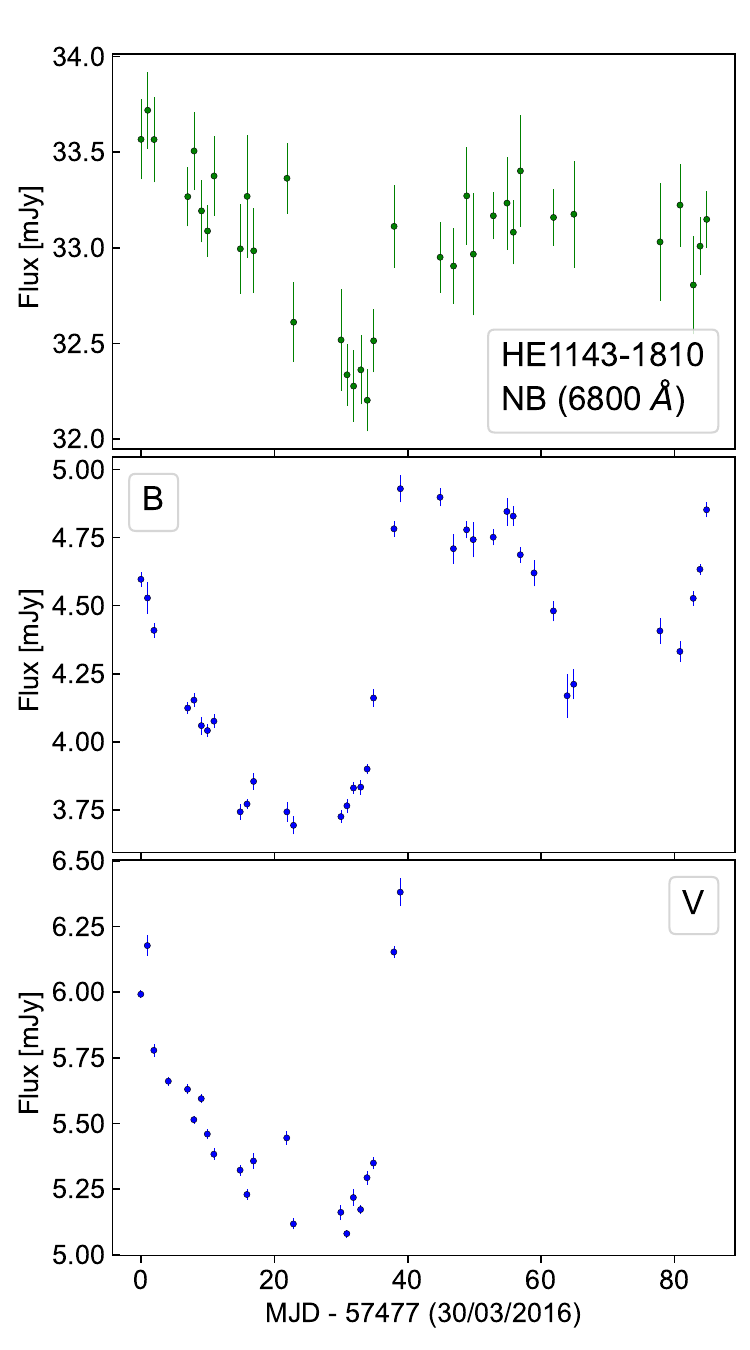}
\includegraphics[width=0.33\columnwidth]{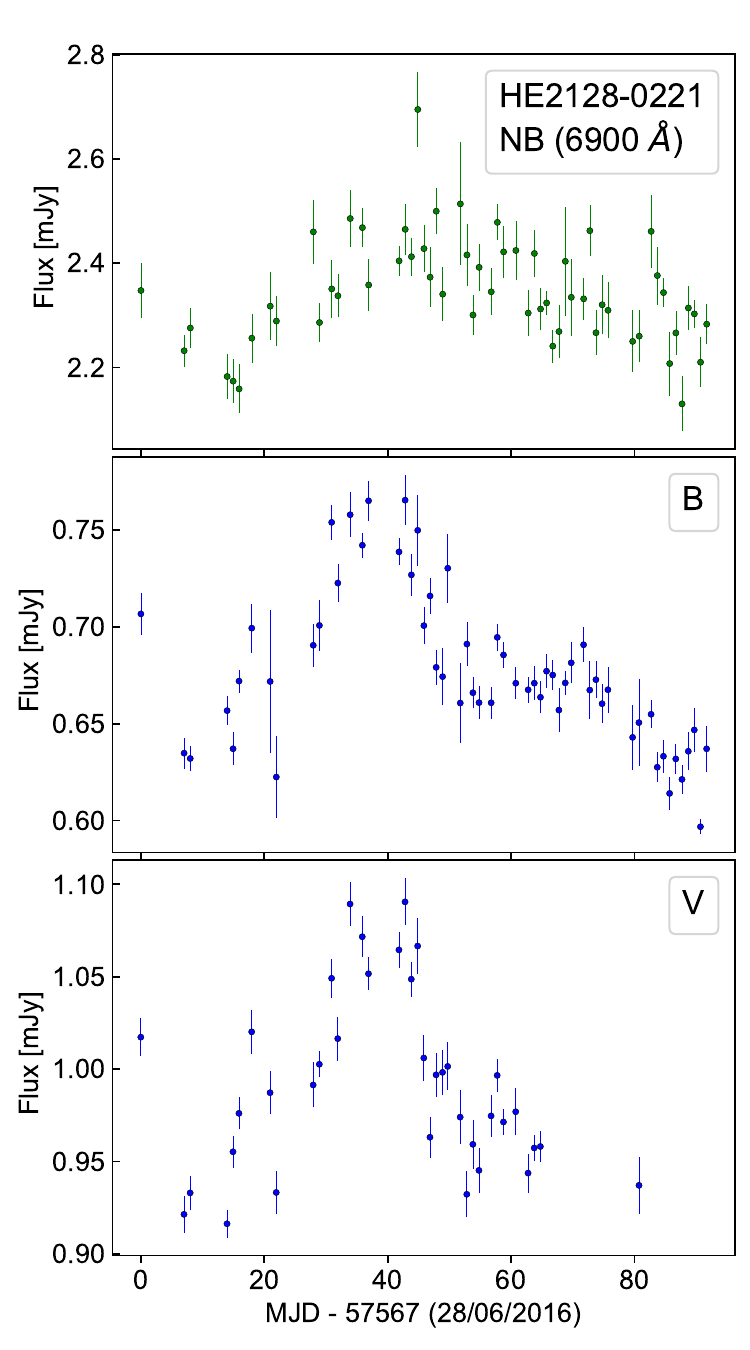}
\includegraphics[width=0.33\columnwidth]{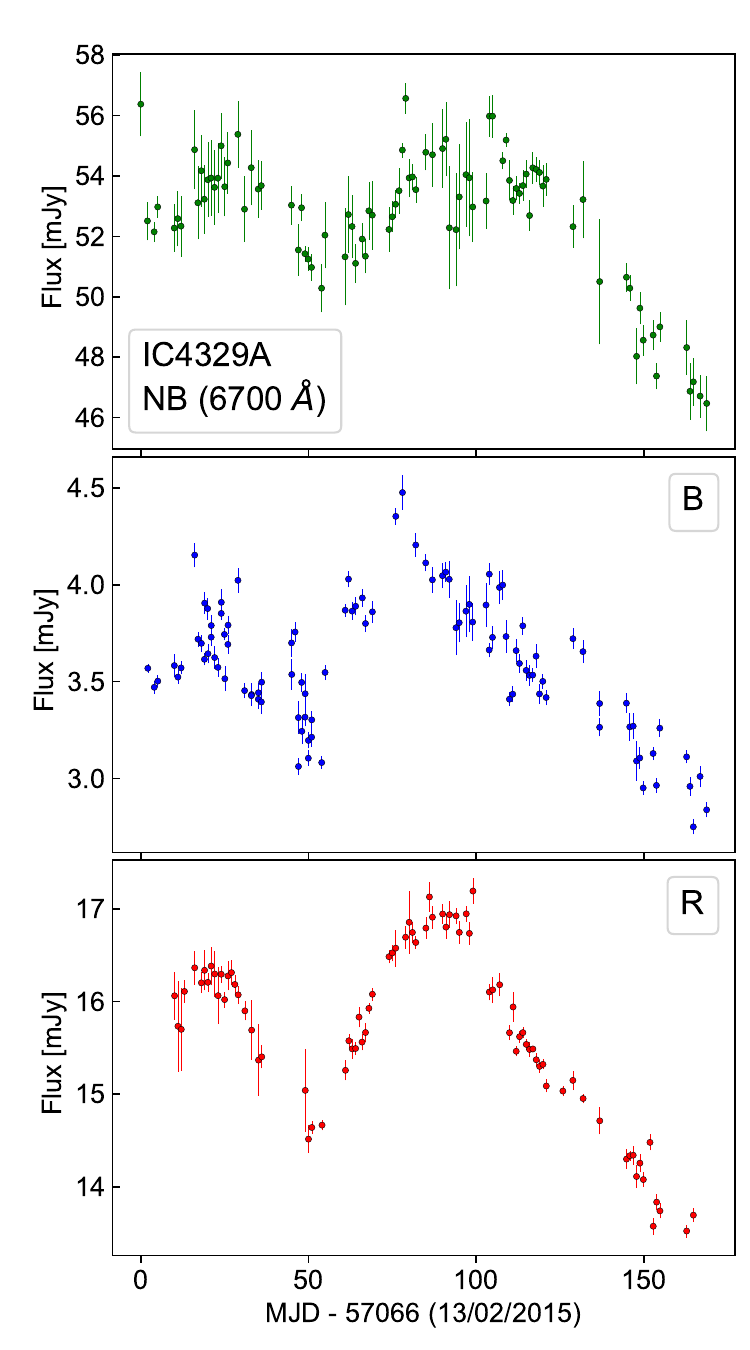}

\includegraphics[width=0.33\columnwidth]{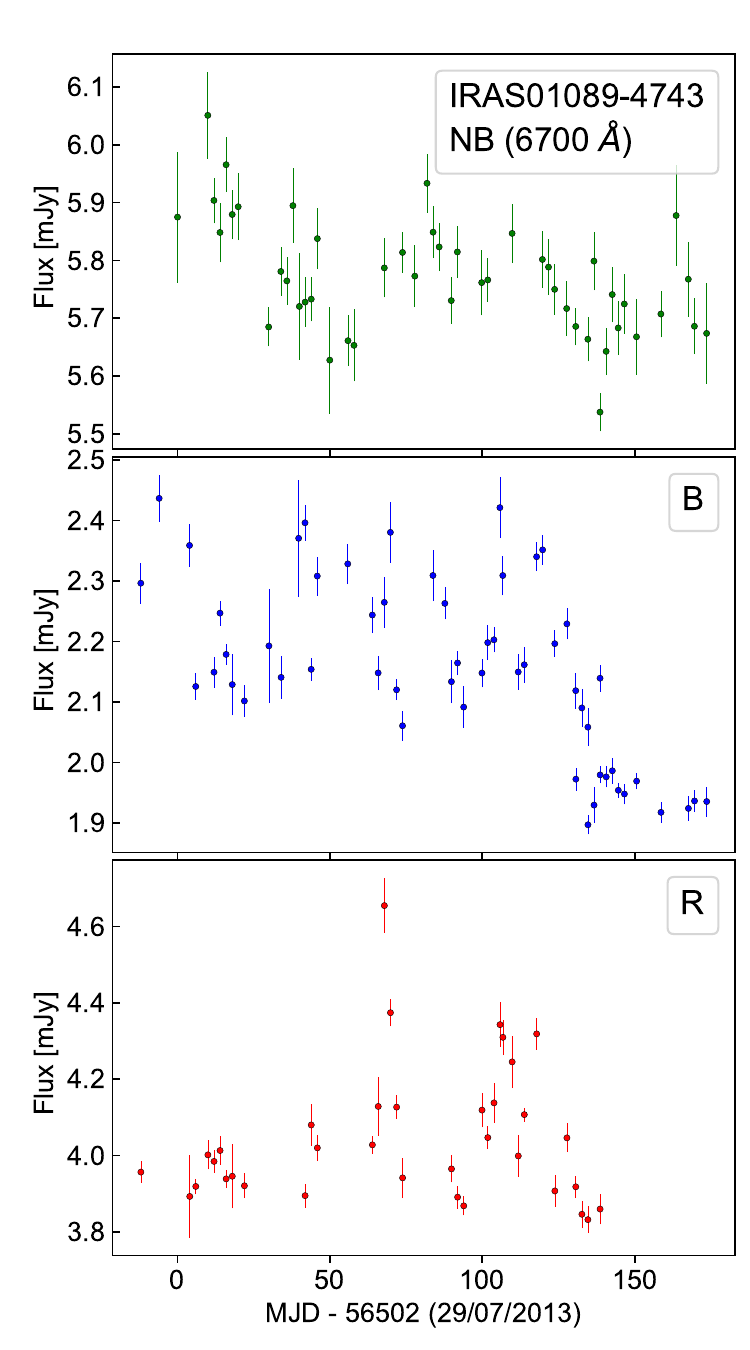}
\includegraphics[width=0.33\columnwidth]{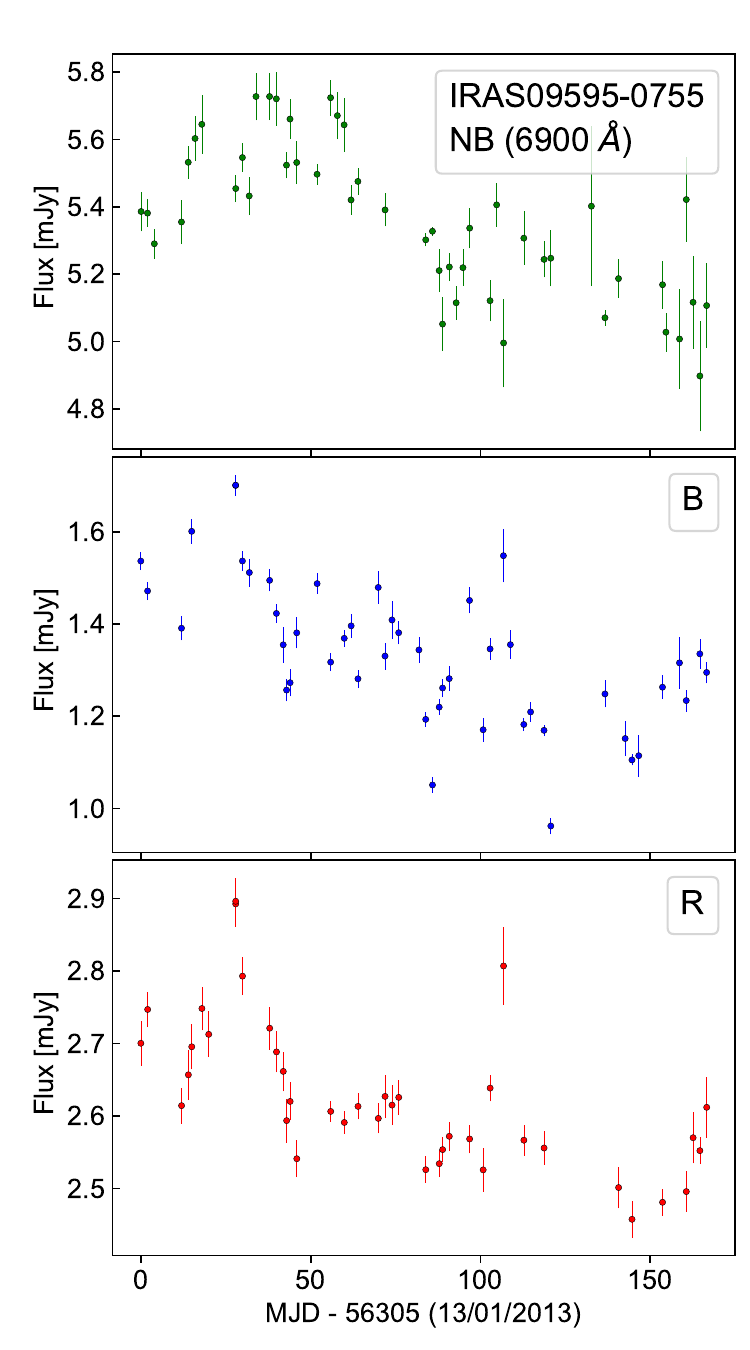}
\includegraphics[width=0.33\columnwidth]{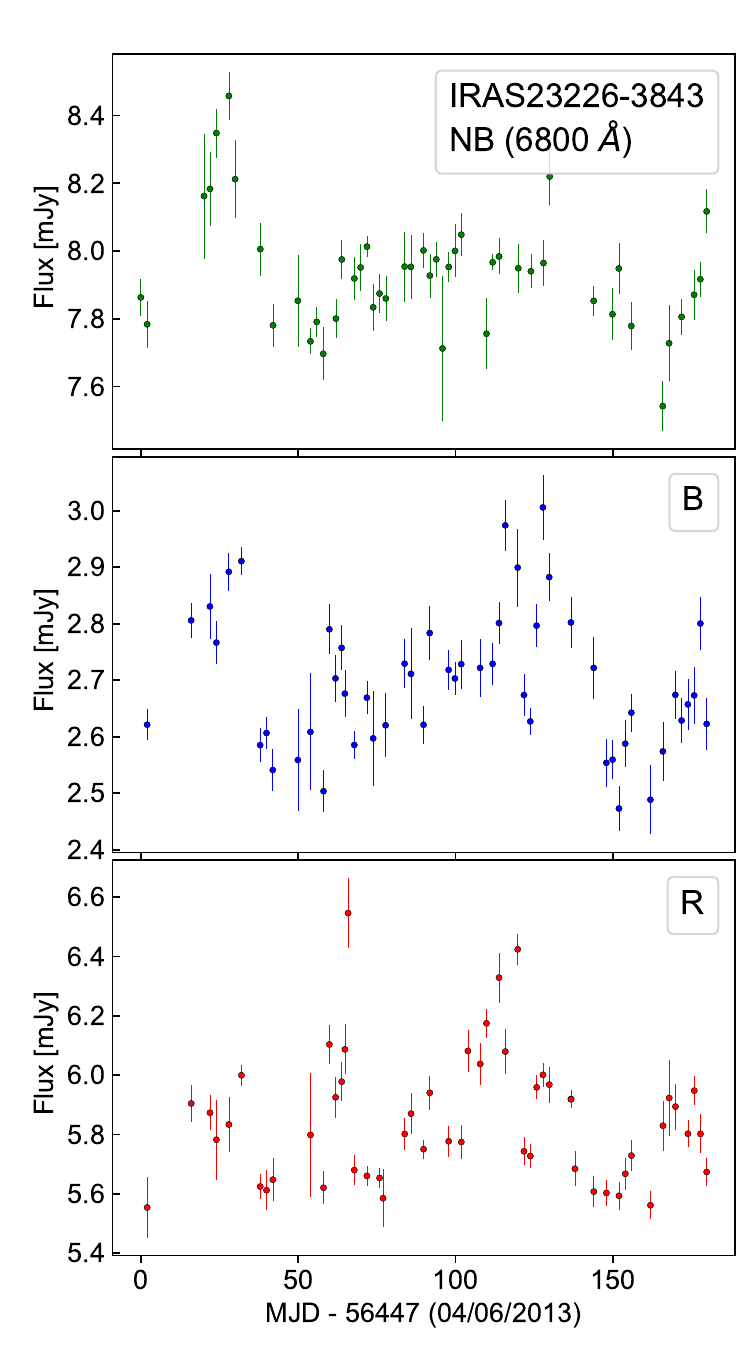}

\includegraphics[width=0.33\columnwidth]{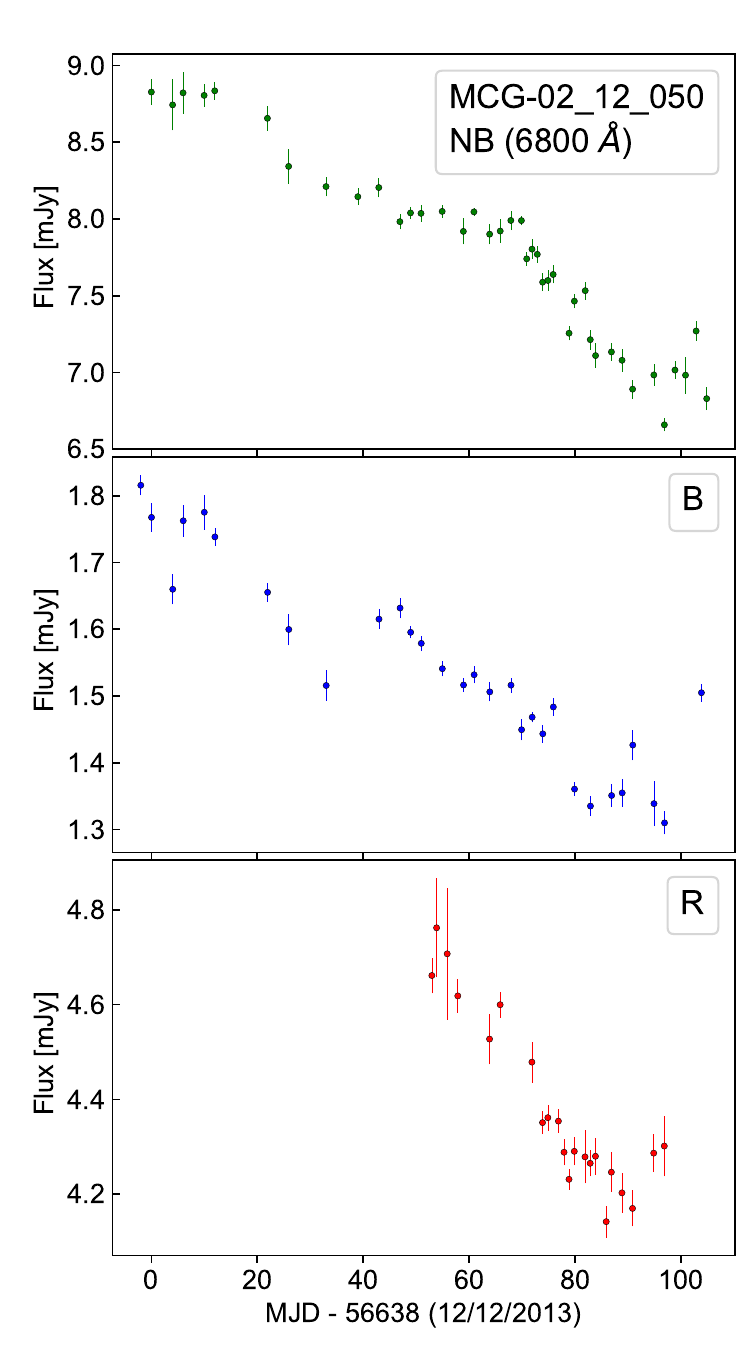}
\includegraphics[width=0.33\columnwidth]{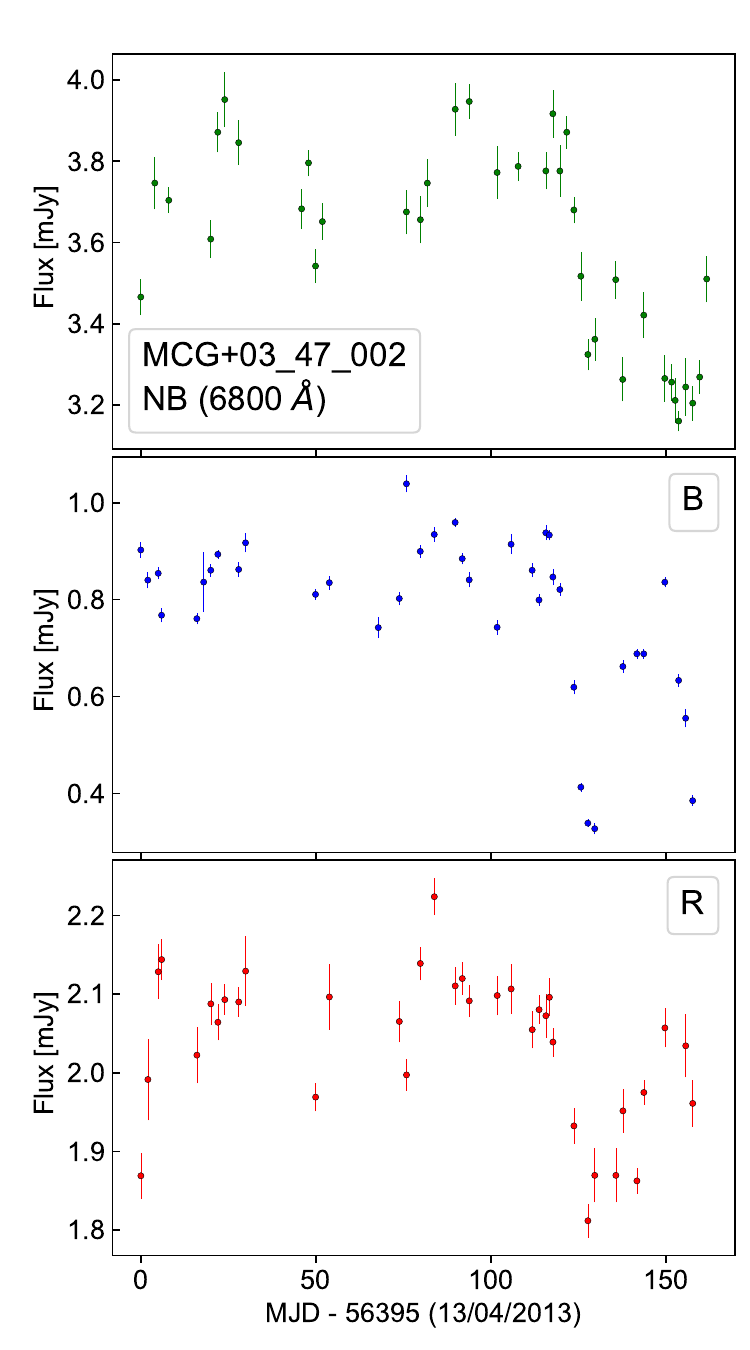}
\includegraphics[width=0.33\columnwidth]{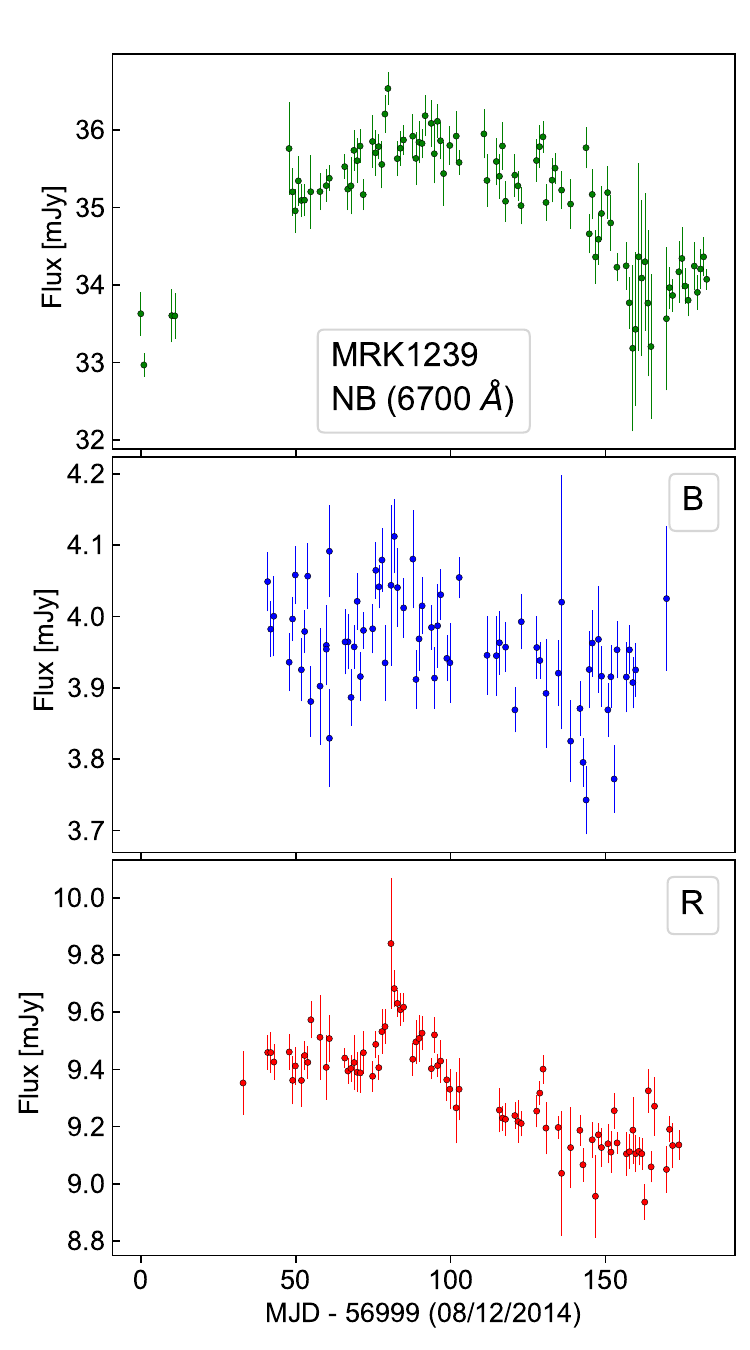}

\includegraphics[width=0.33\columnwidth]{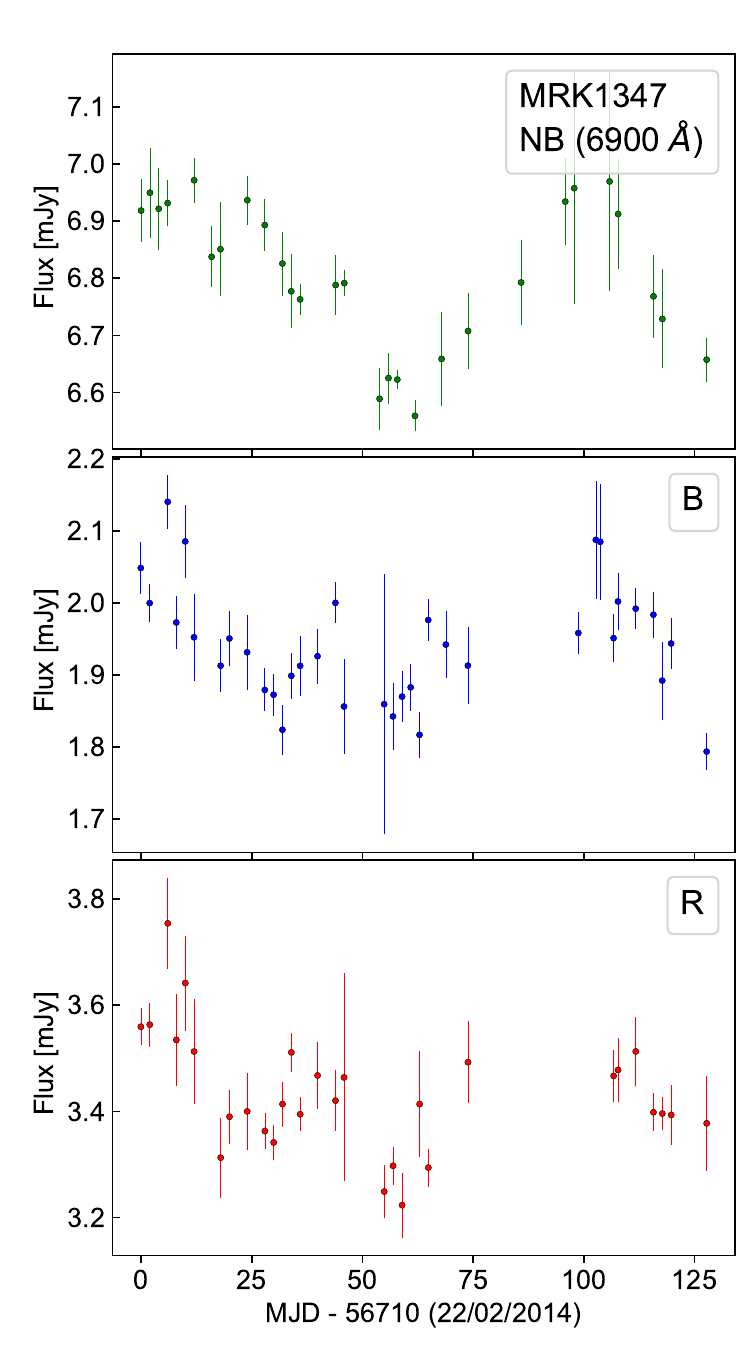}
\includegraphics[width=0.33\columnwidth]{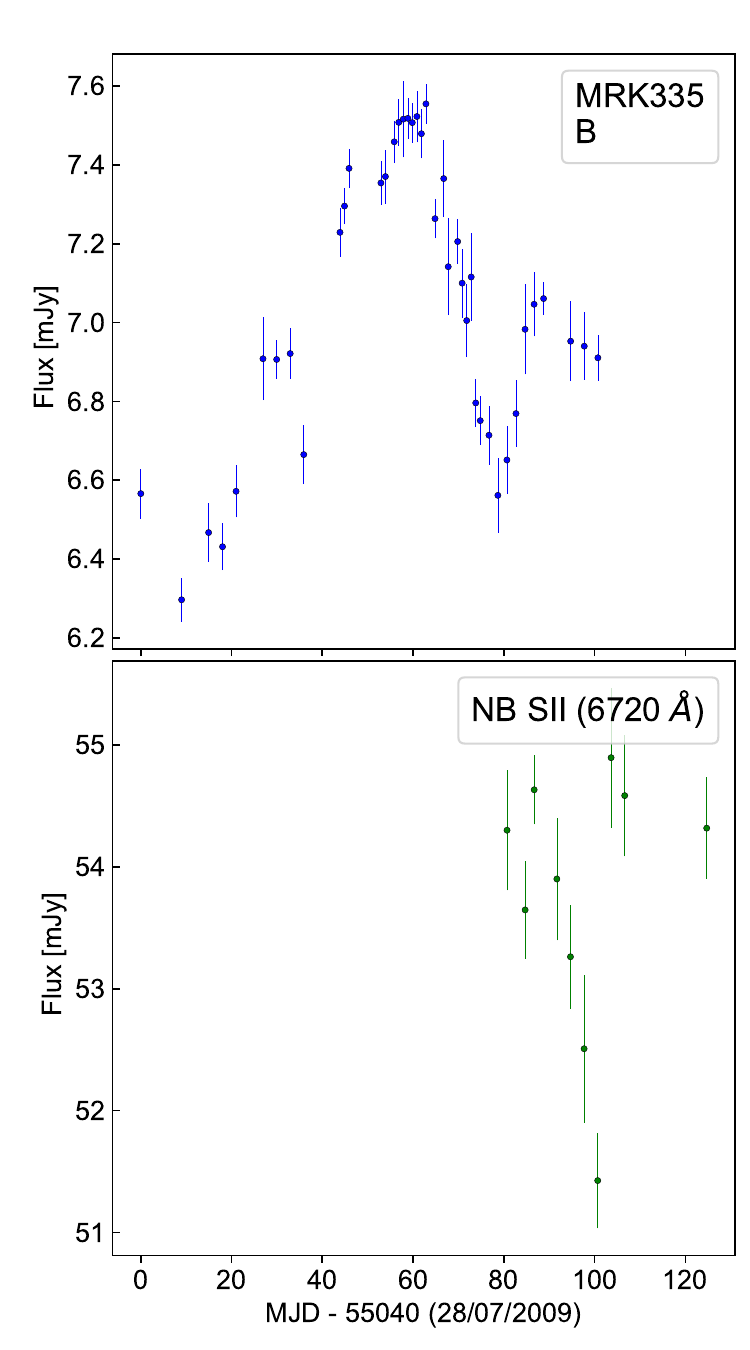}
\includegraphics[width=0.33\columnwidth]{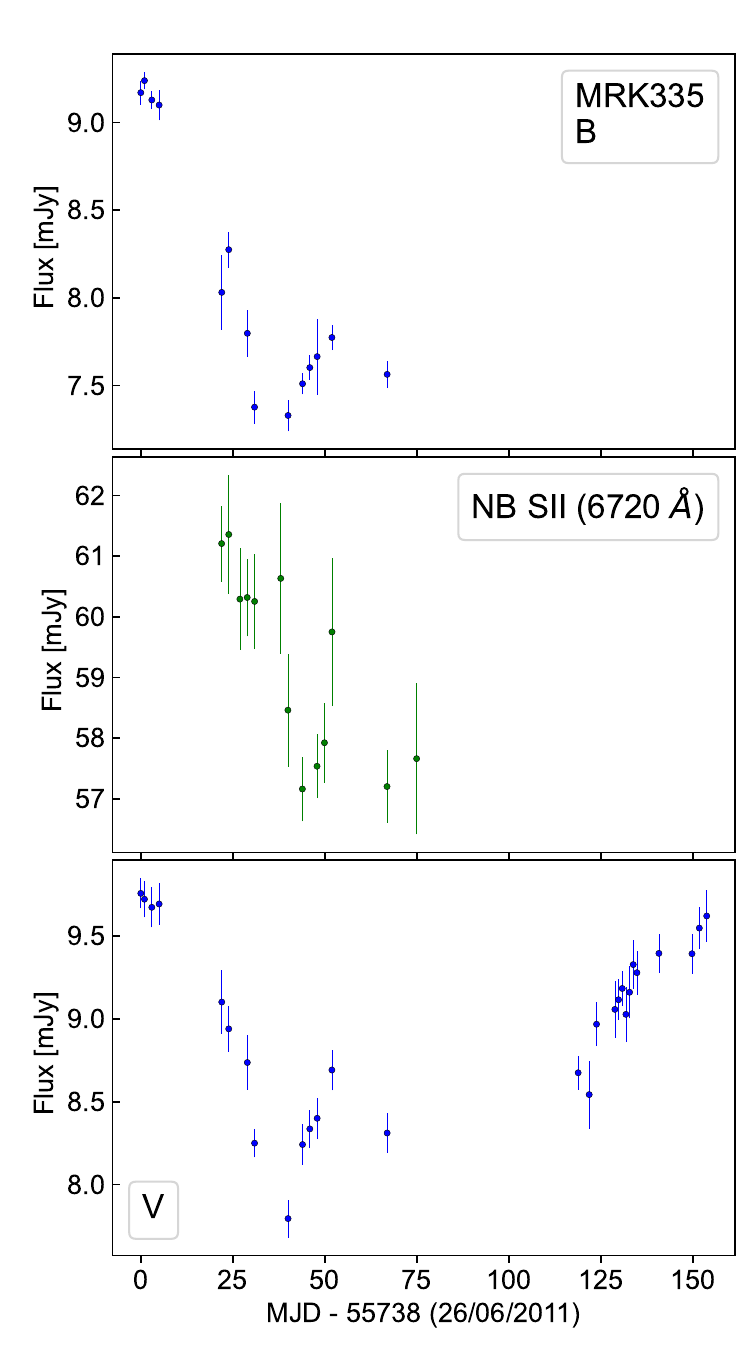}

\includegraphics[width=0.33\columnwidth]{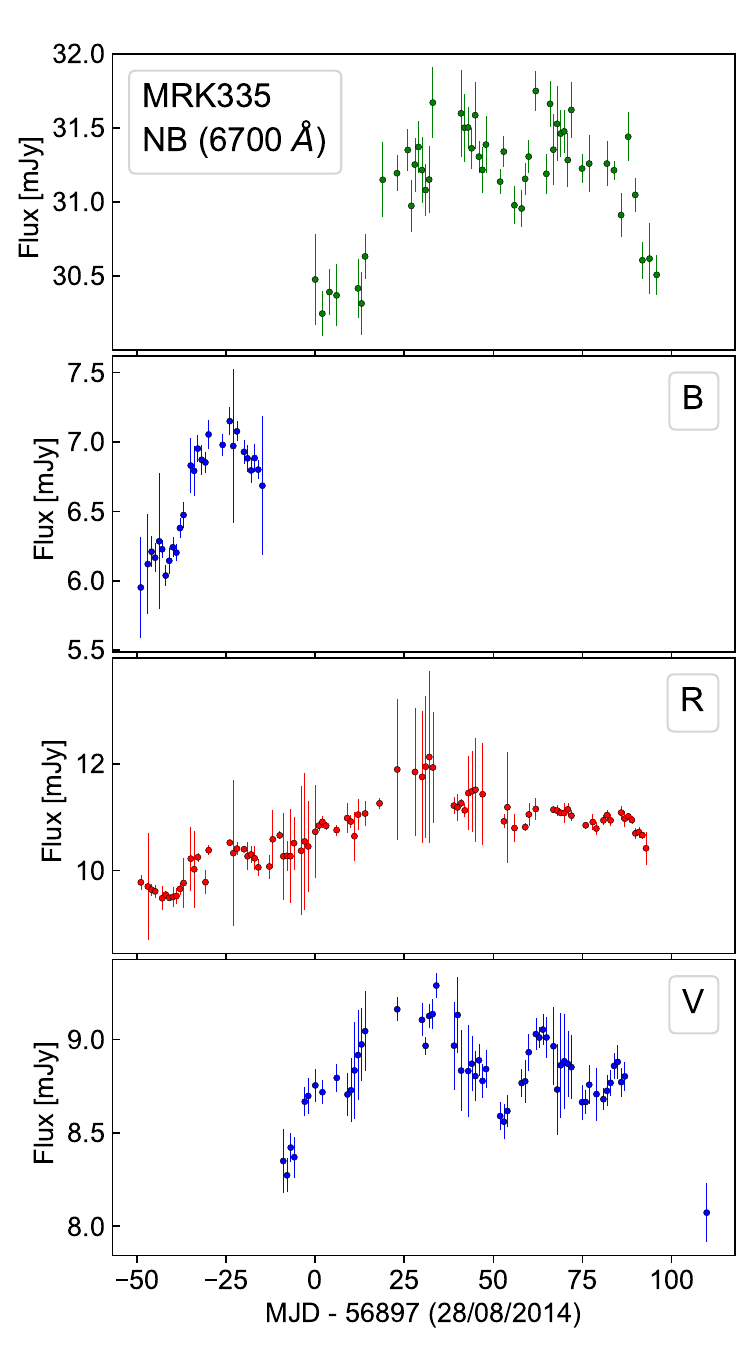}
\includegraphics[width=0.33\columnwidth]{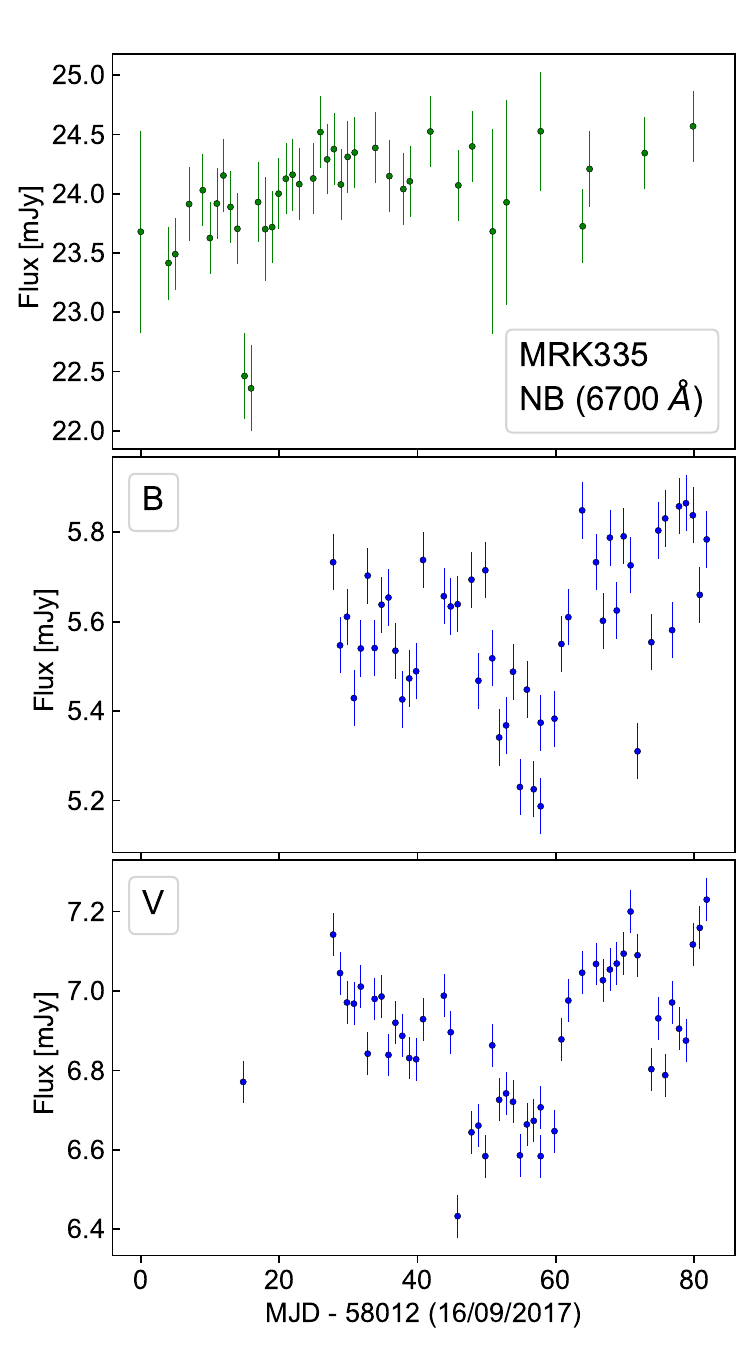}
\includegraphics[width=0.33\columnwidth]{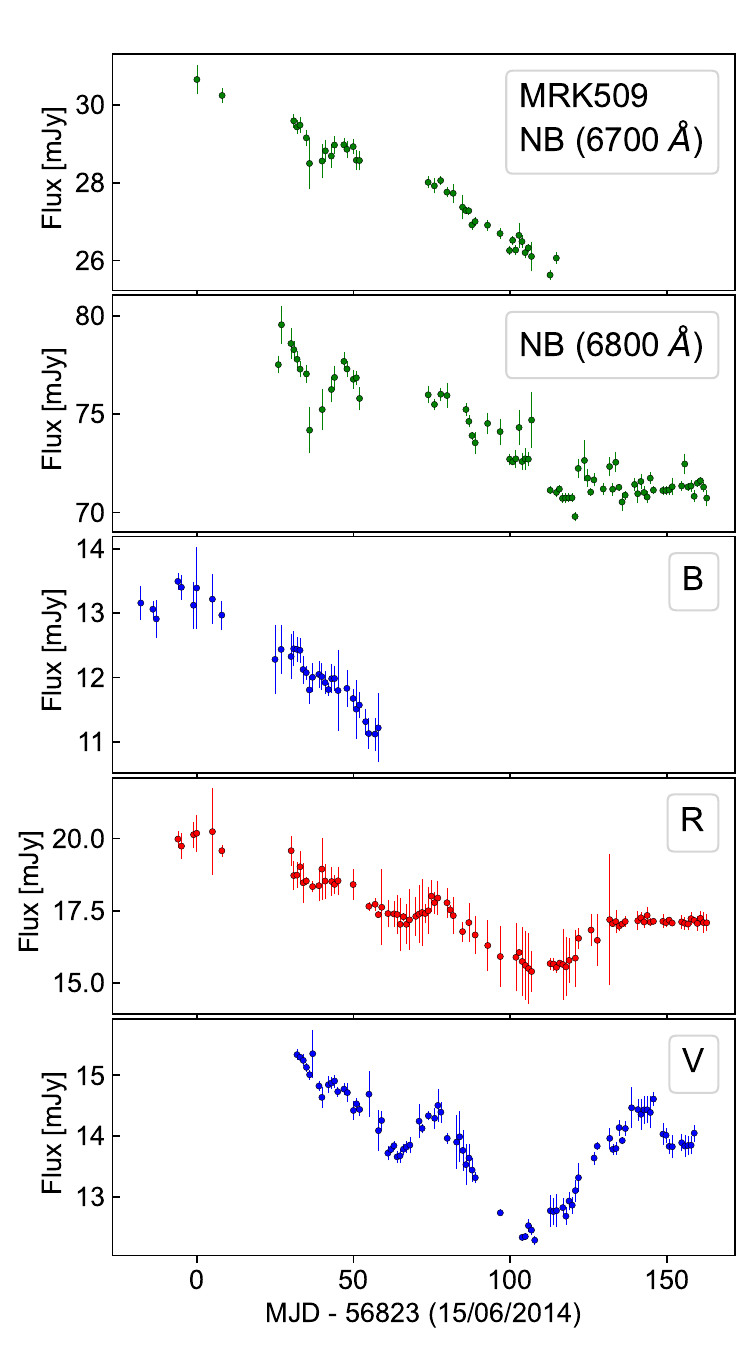}

\includegraphics[width=0.33\columnwidth]{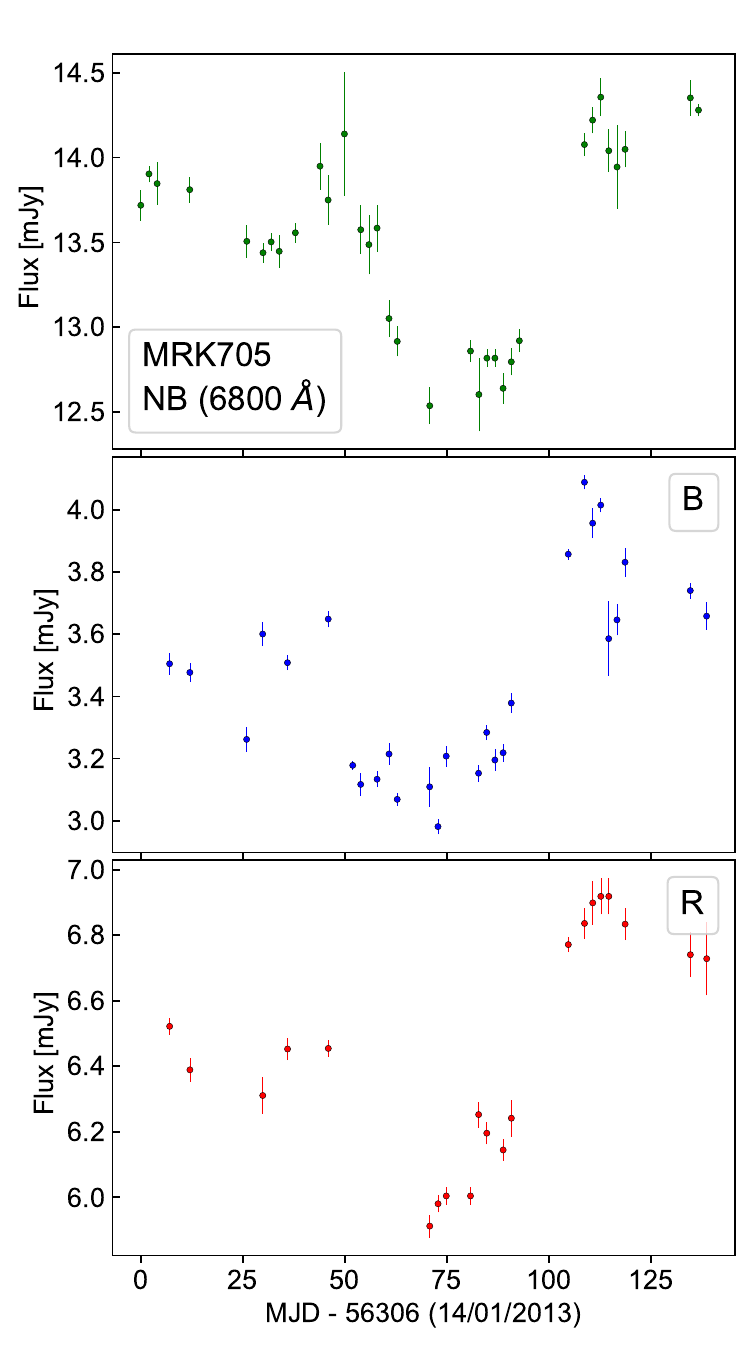}
\includegraphics[width=0.33\columnwidth]{fig_pdf/MRK841_2014_LCS.pdf}
\includegraphics[width=0.33\columnwidth]{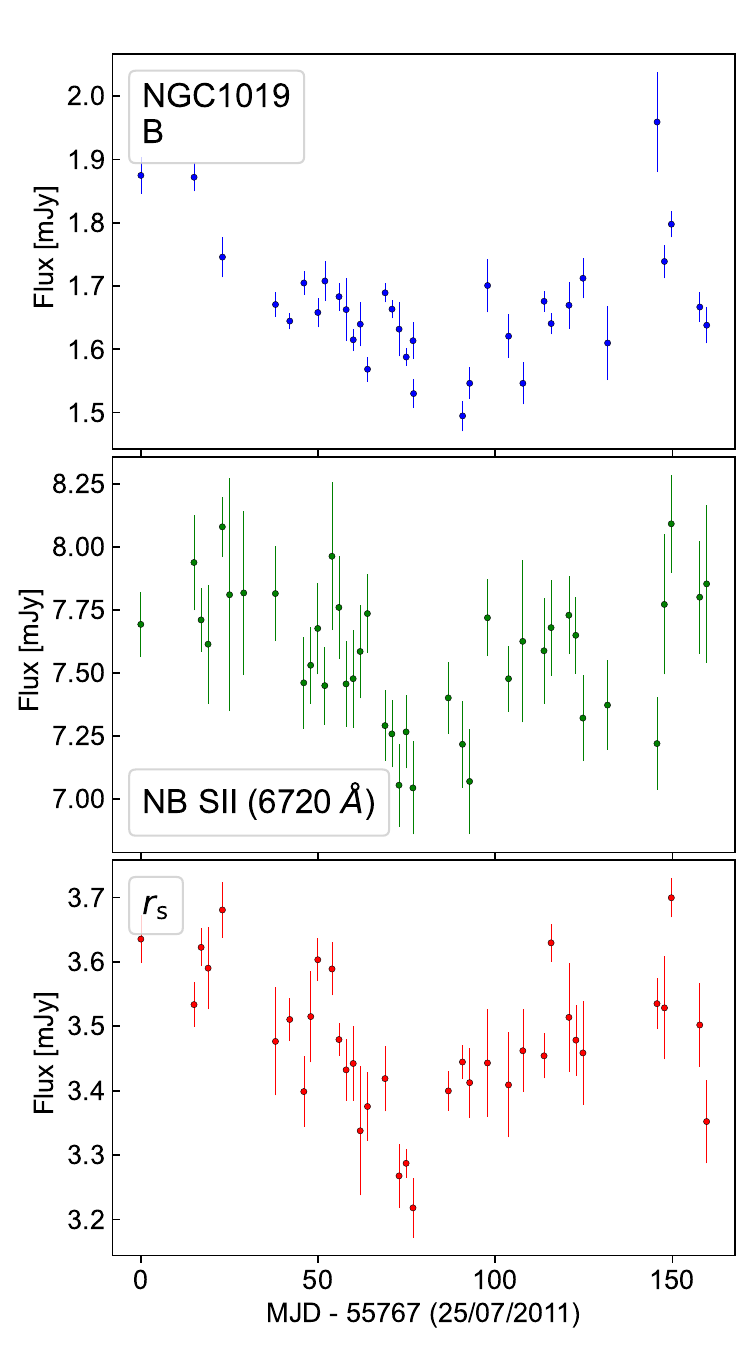}

\includegraphics[width=0.33\columnwidth]{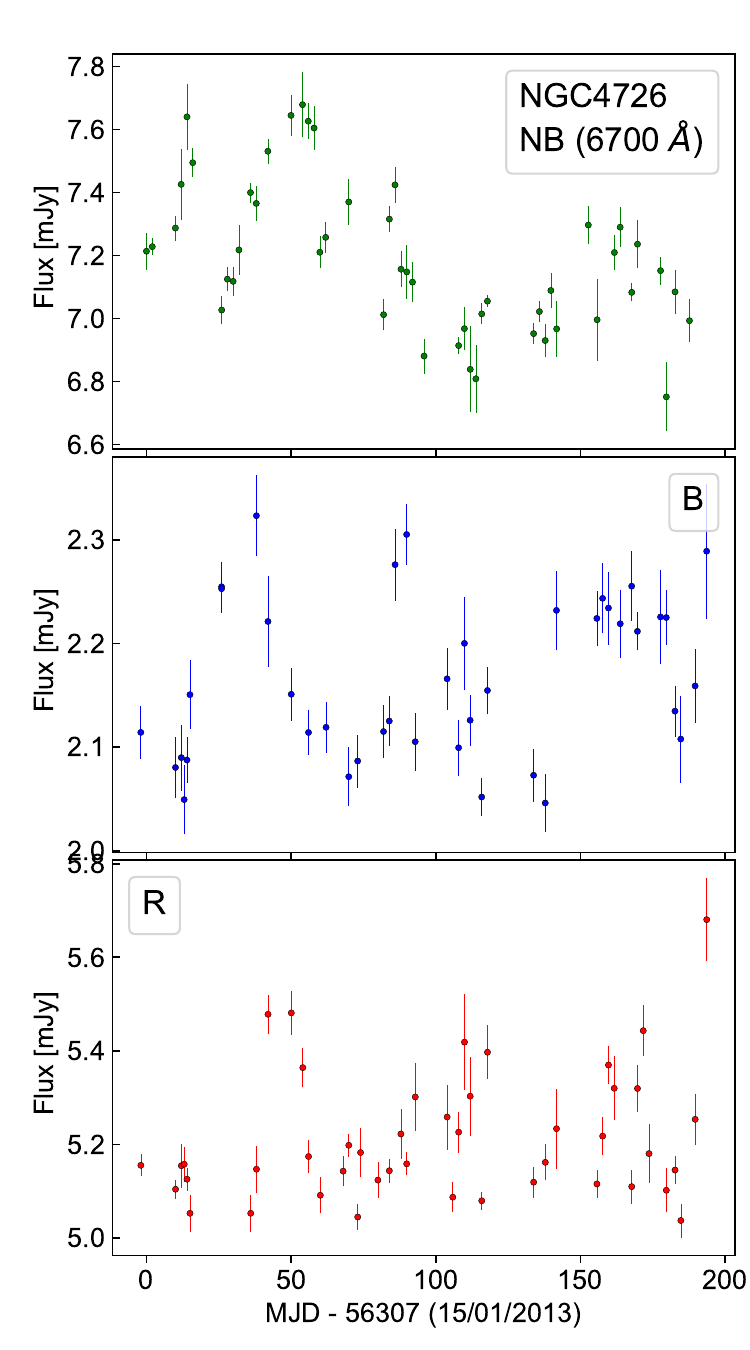}
\includegraphics[width=0.33\columnwidth]{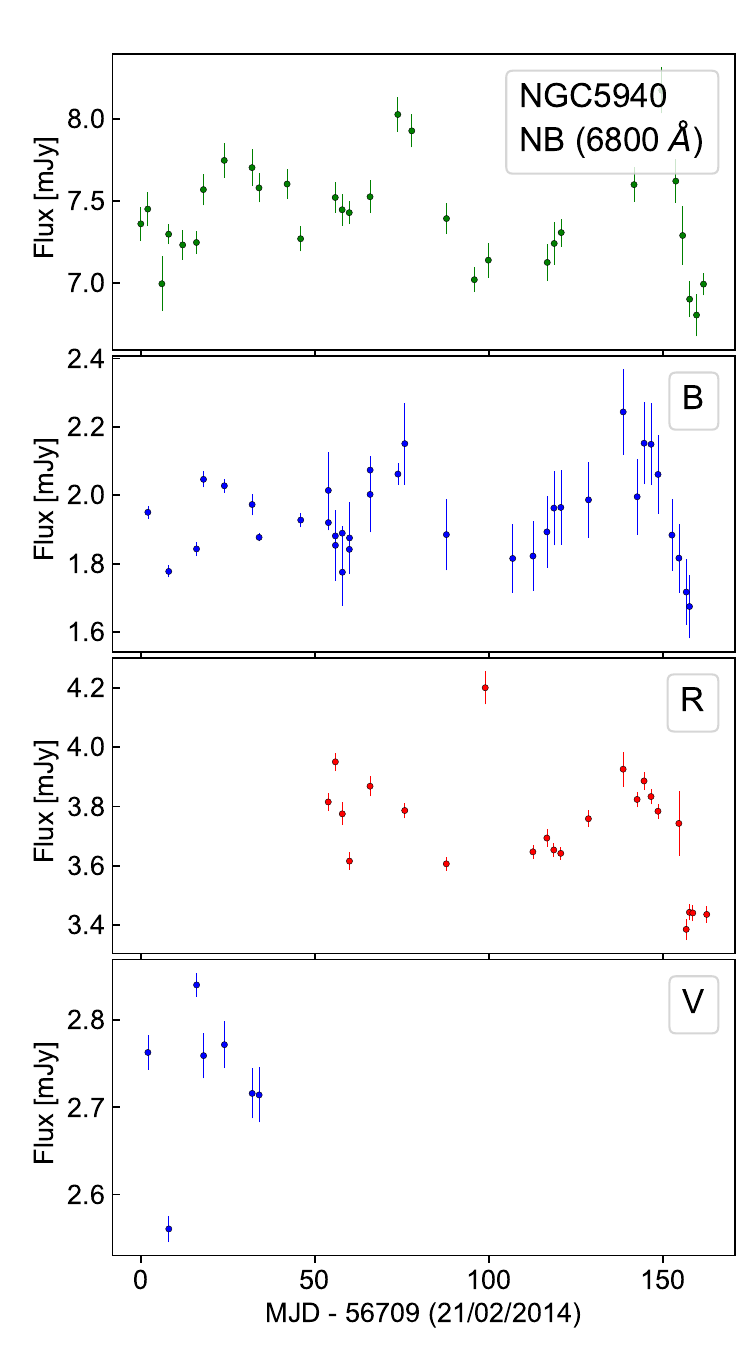}
\includegraphics[width=0.33\columnwidth]{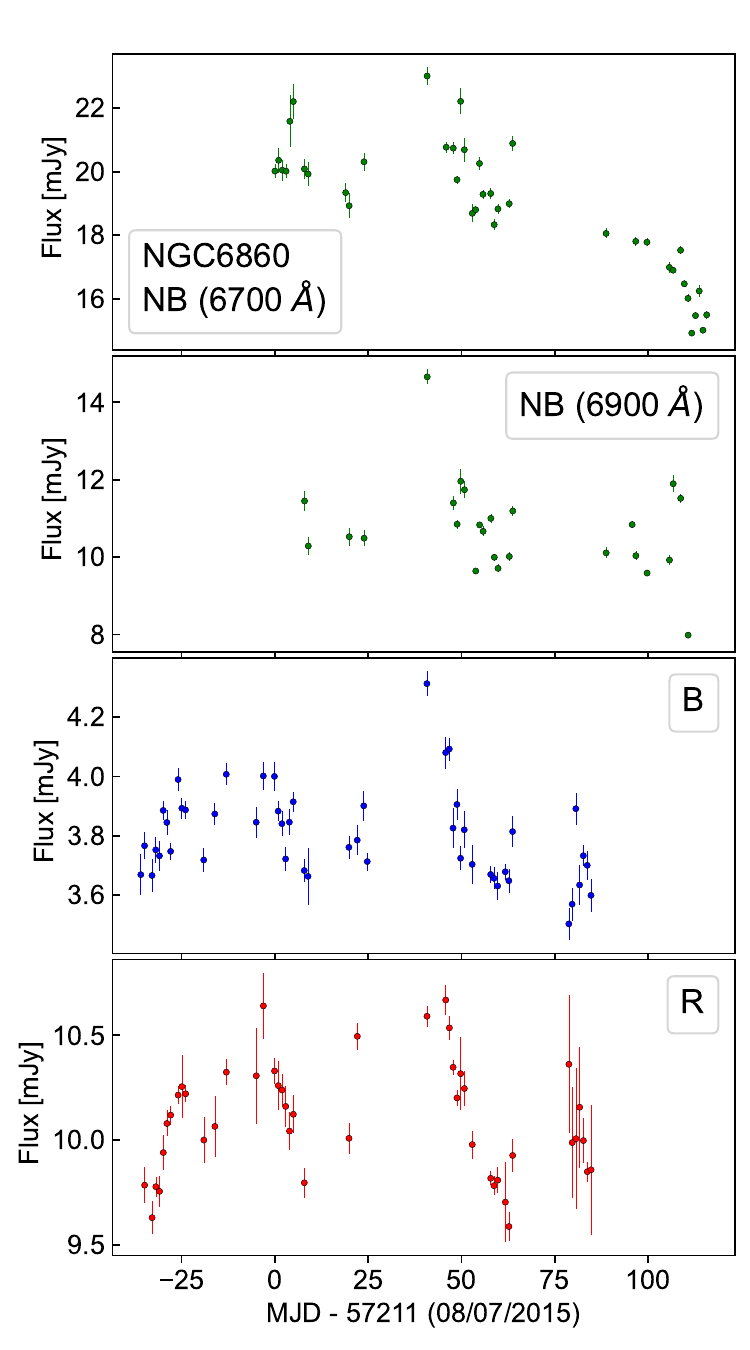}

\includegraphics[width=0.33\columnwidth]{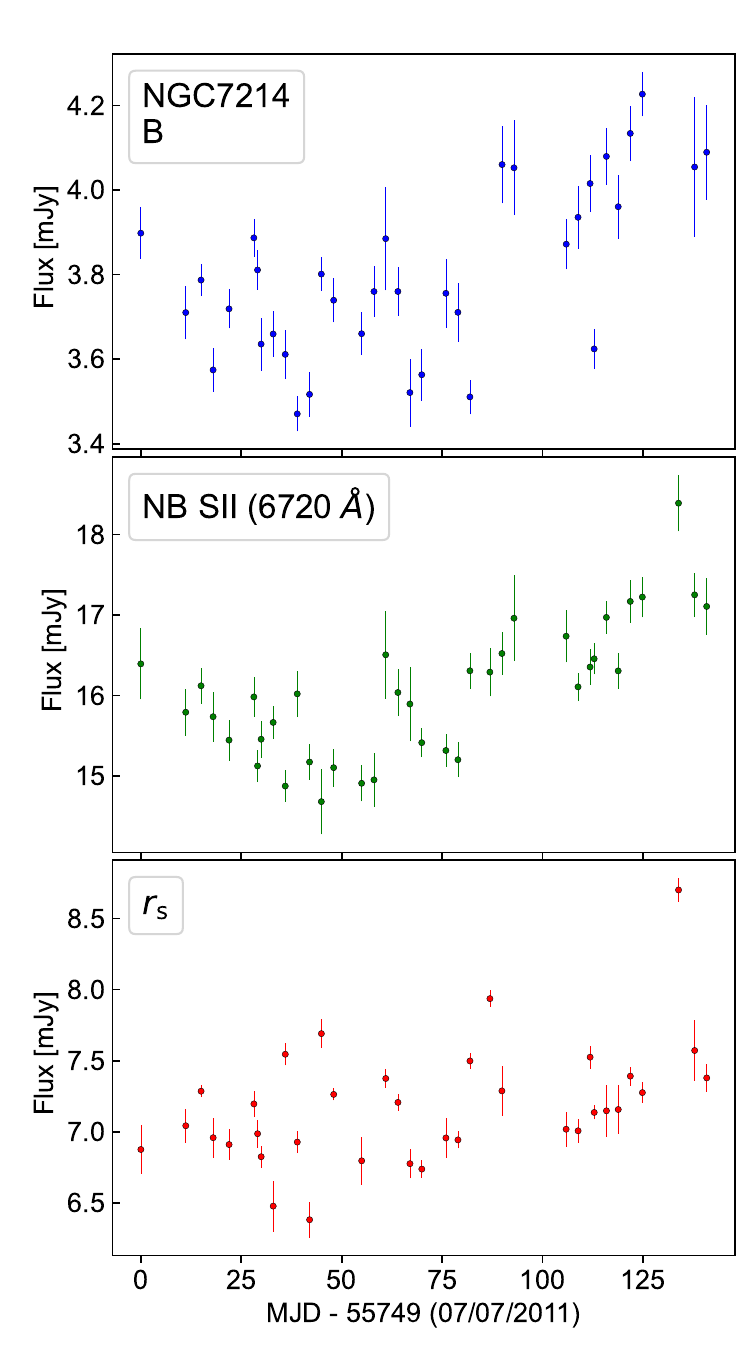}
\includegraphics[width=0.33\columnwidth]{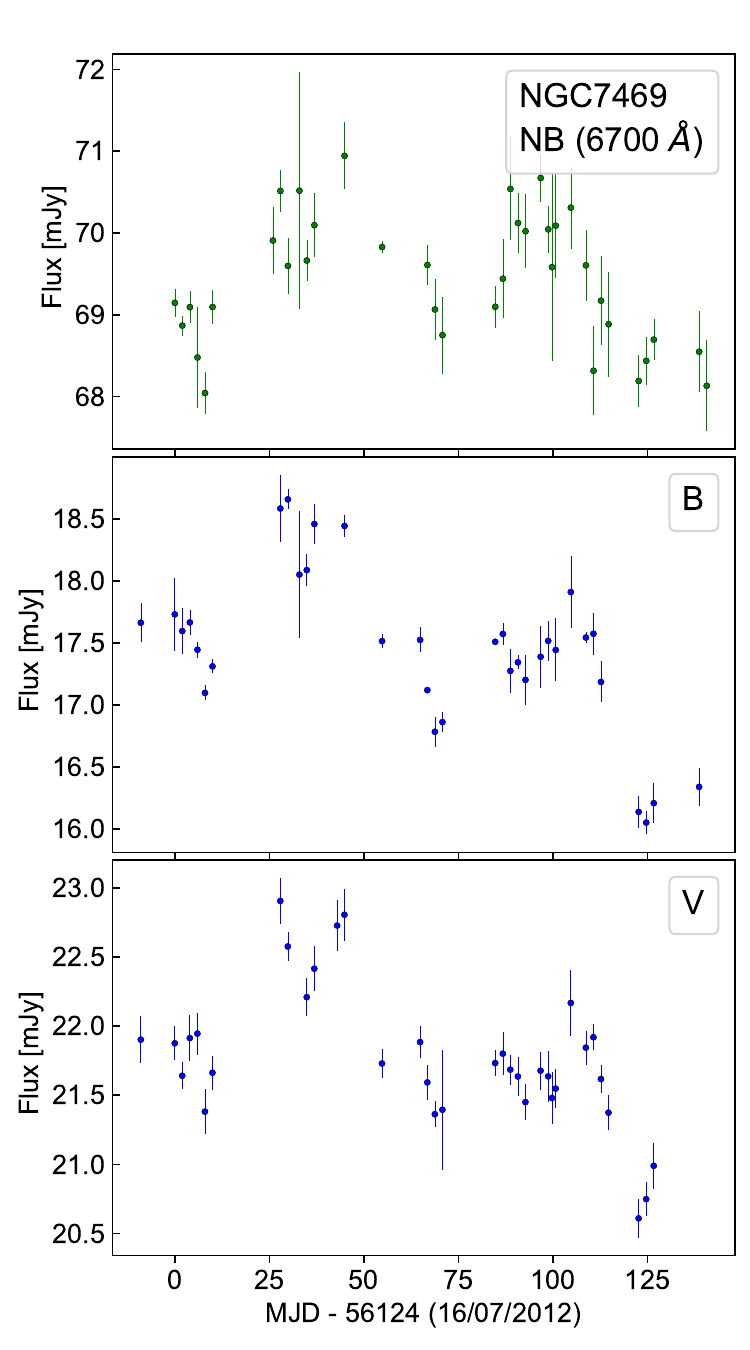}
\includegraphics[width=0.33\columnwidth]{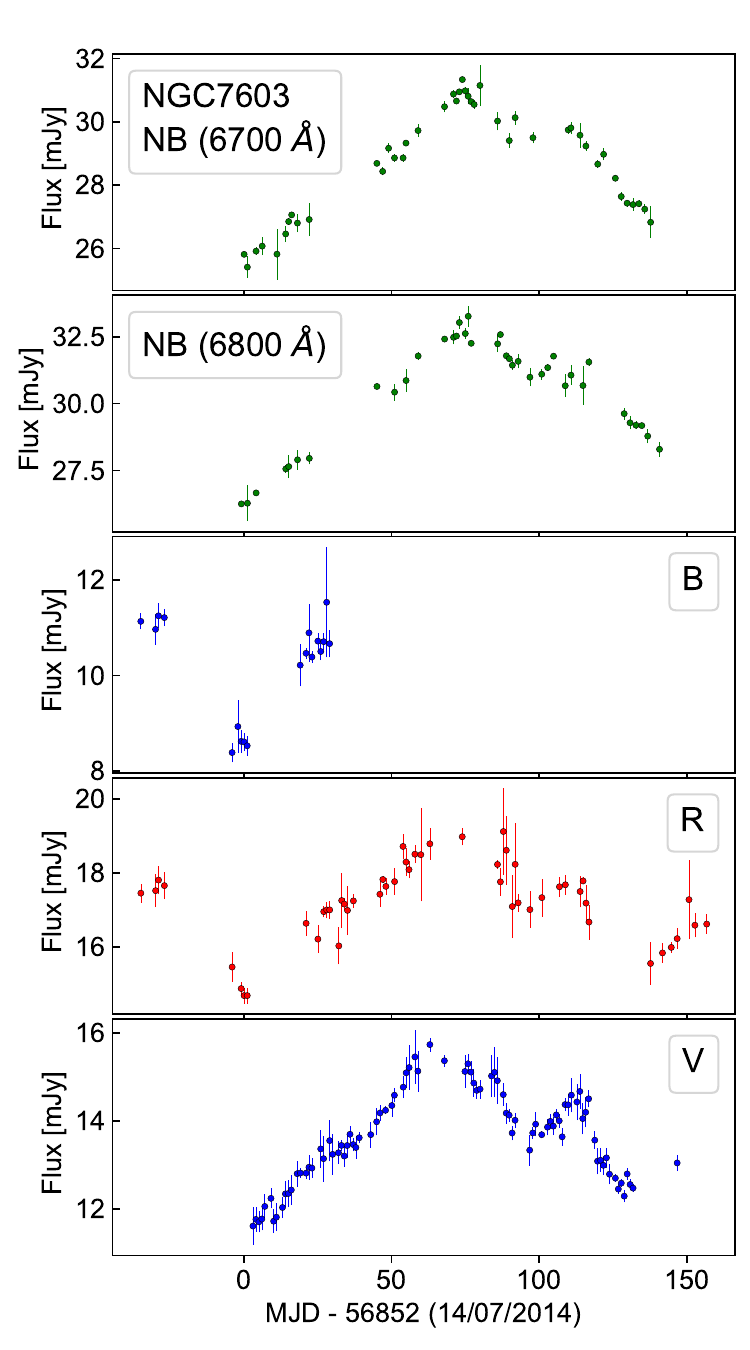}

\includegraphics[width=0.33\columnwidth]{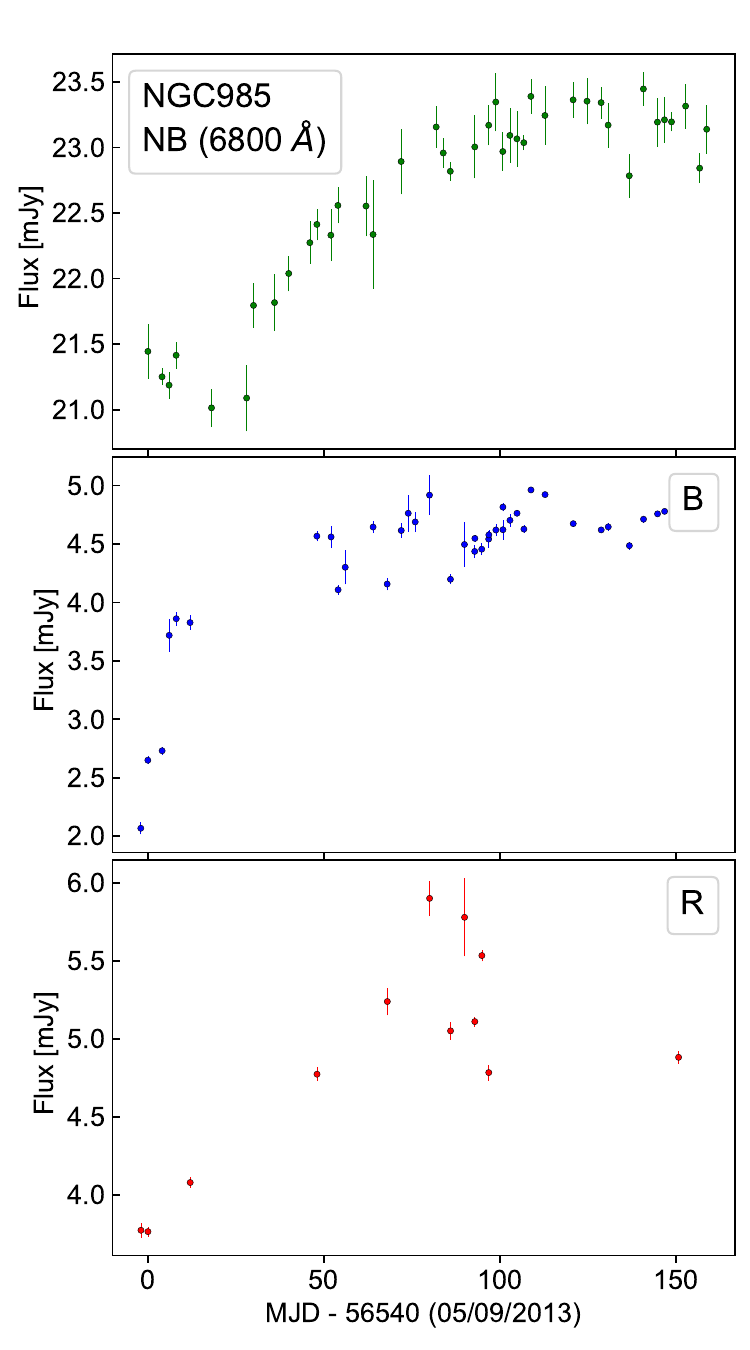}
\includegraphics[width=0.33\columnwidth]{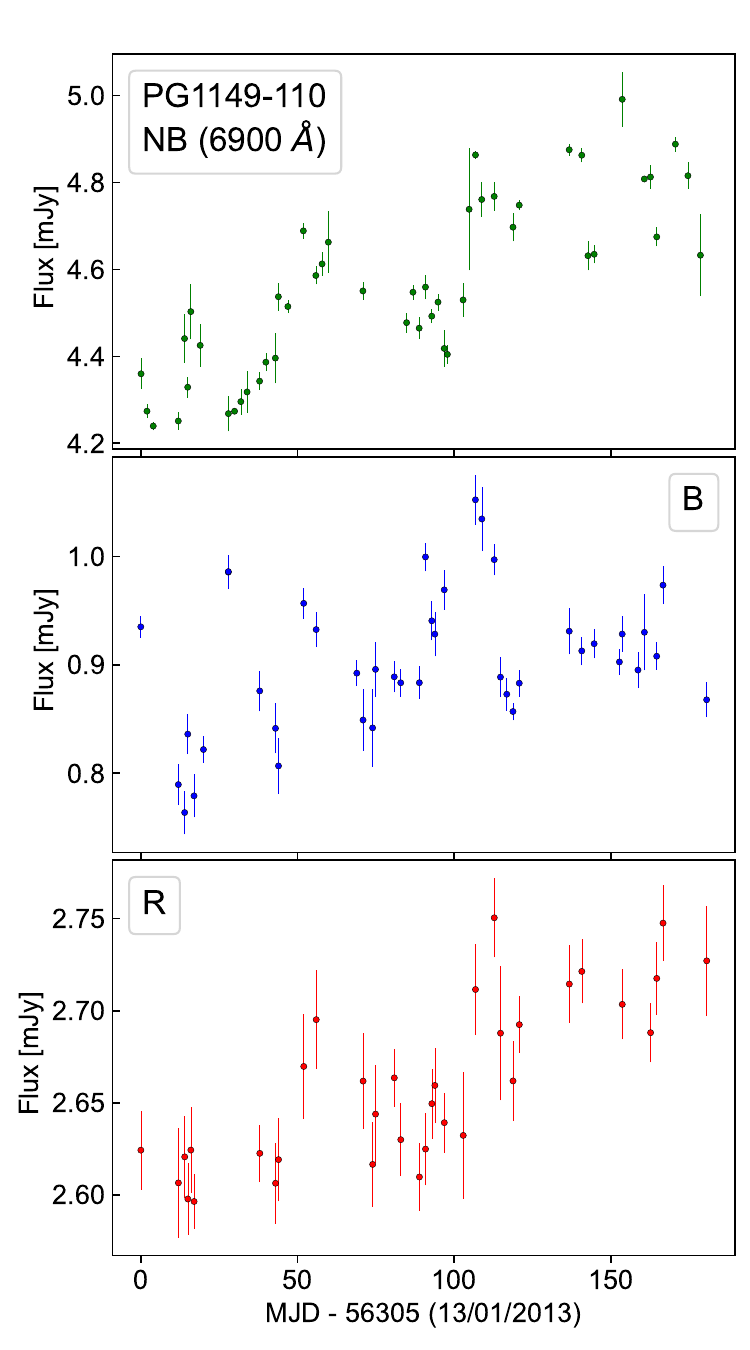}
\includegraphics[width=0.33\columnwidth]{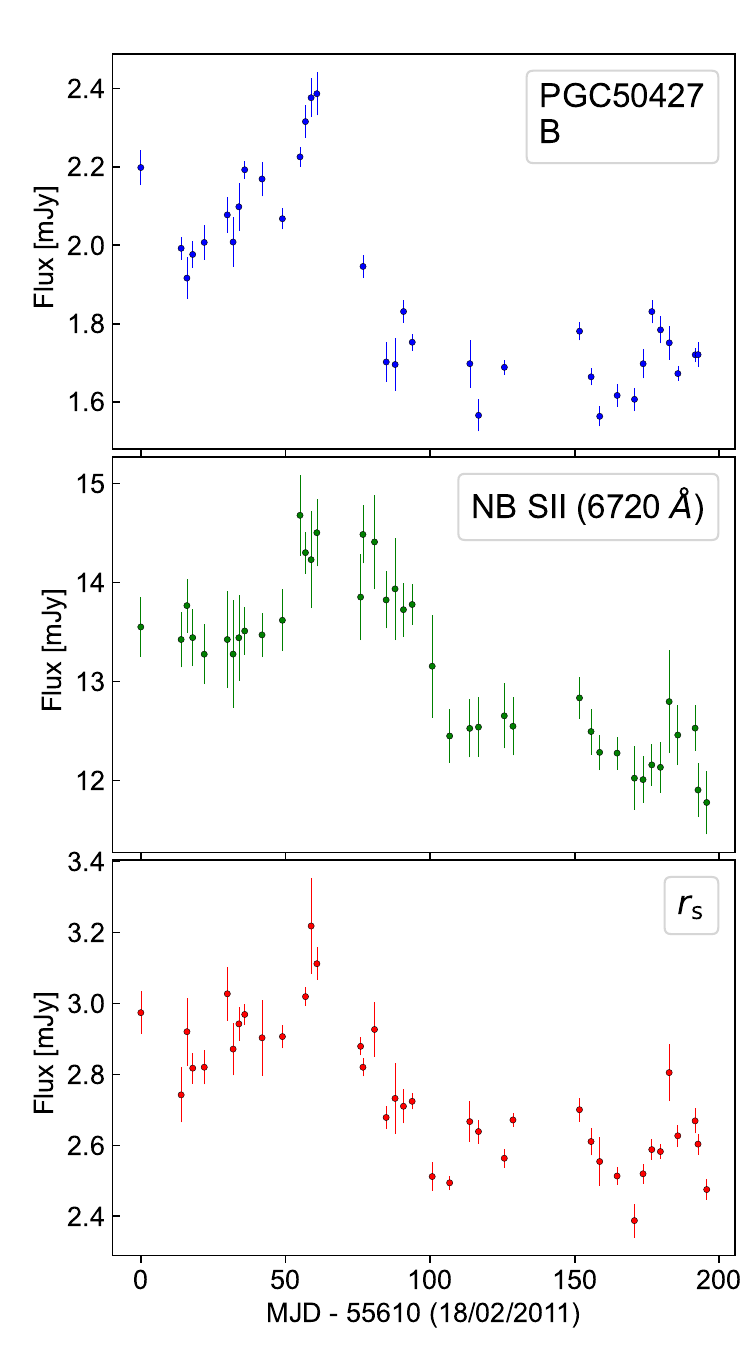}

\includegraphics[width=0.33\columnwidth]{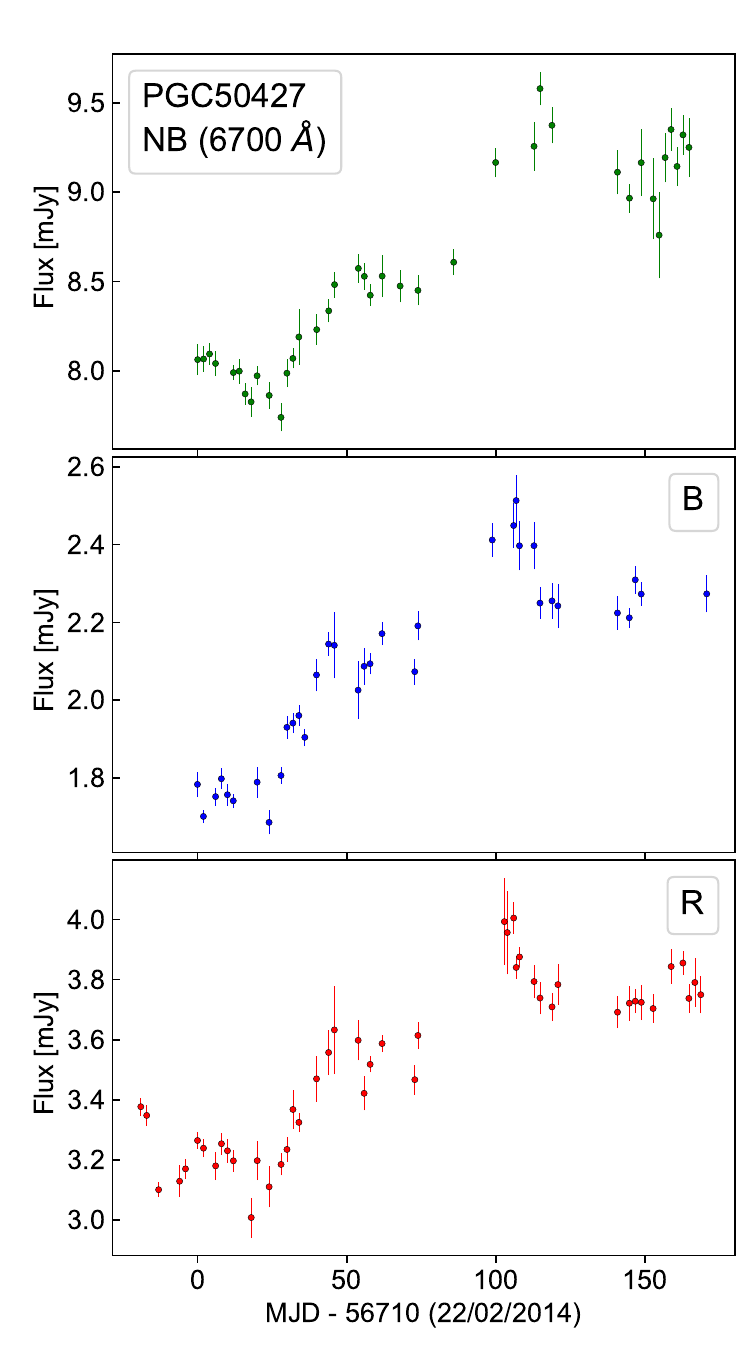}
\includegraphics[width=0.33\columnwidth]{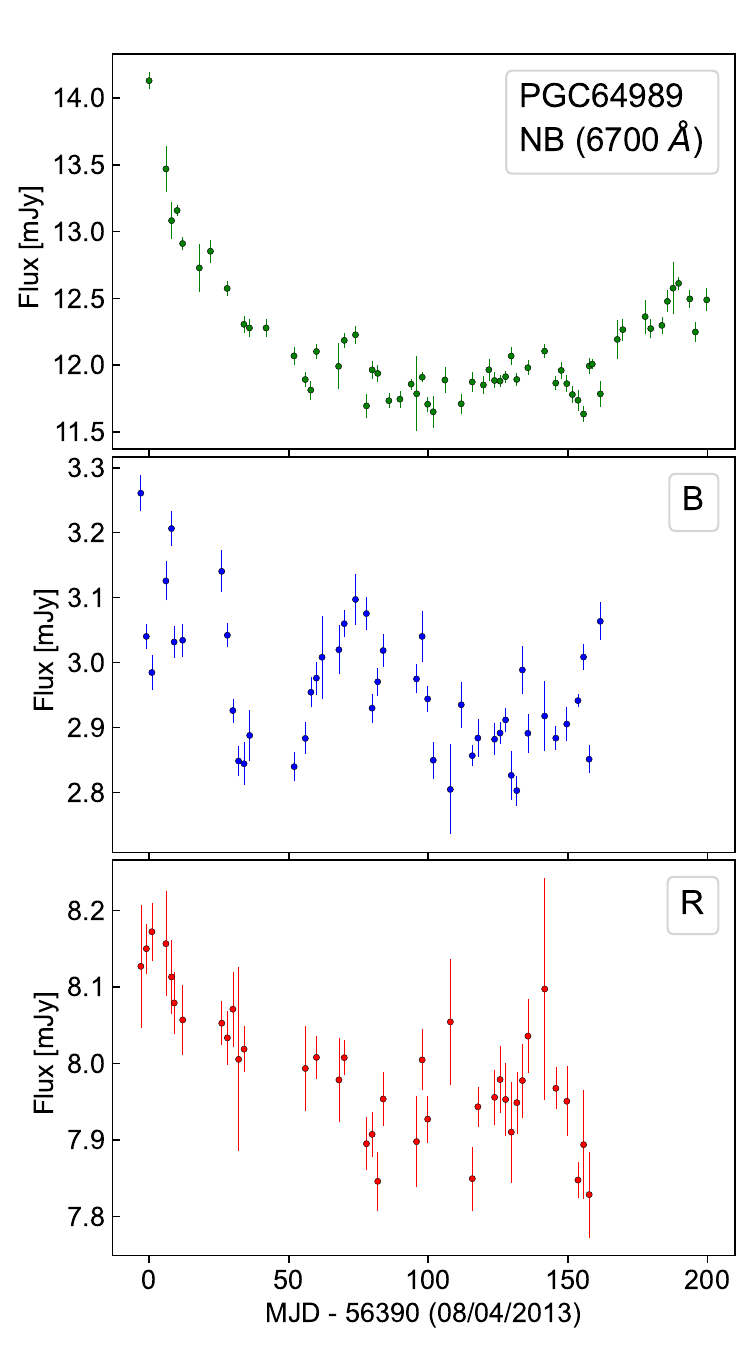}
\includegraphics[width=0.33\columnwidth]{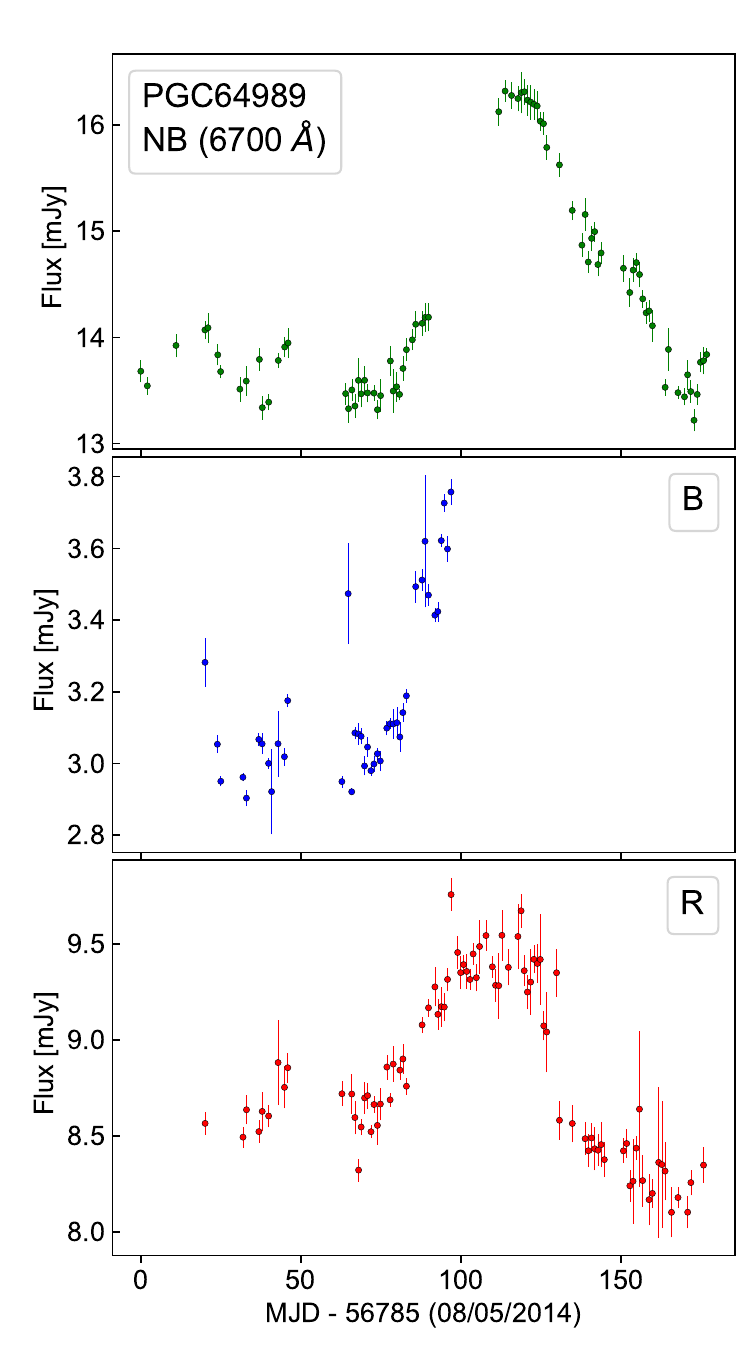}

\includegraphics[width=0.33\columnwidth]{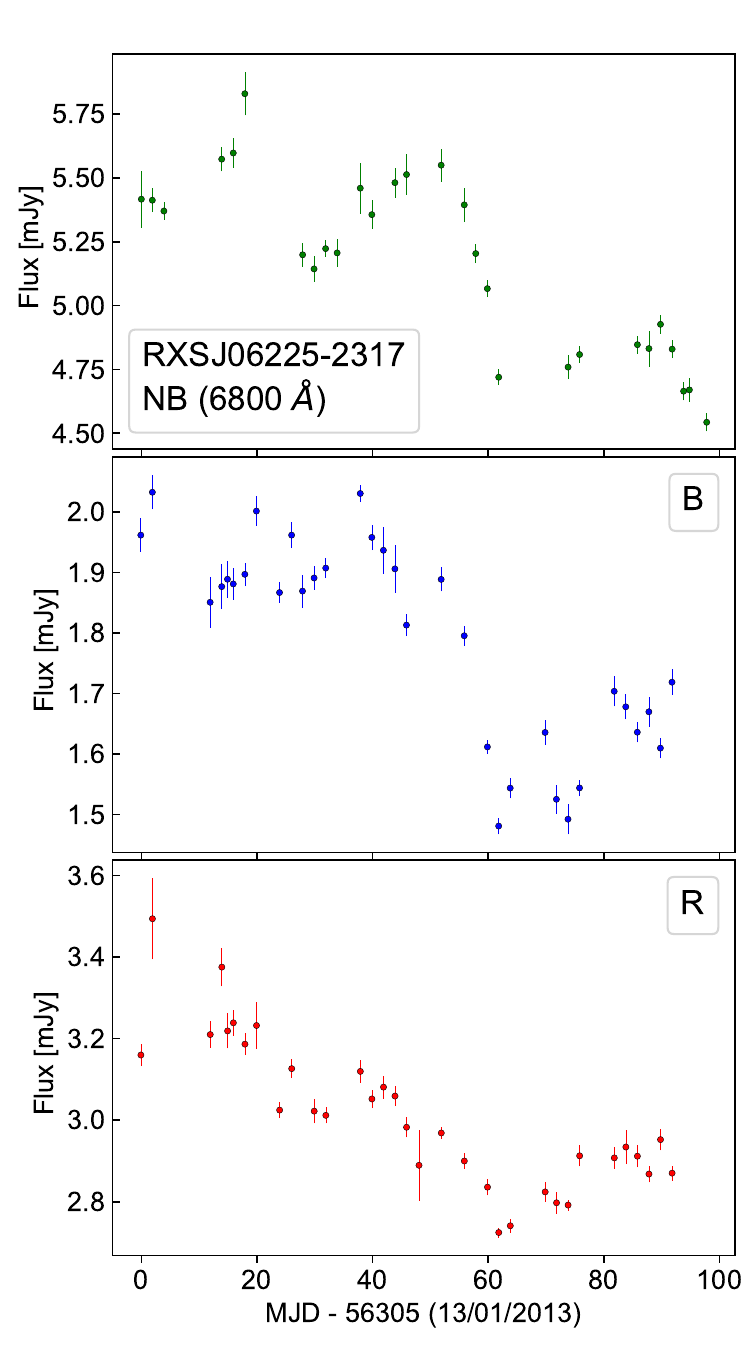}
\includegraphics[width=0.33\columnwidth]{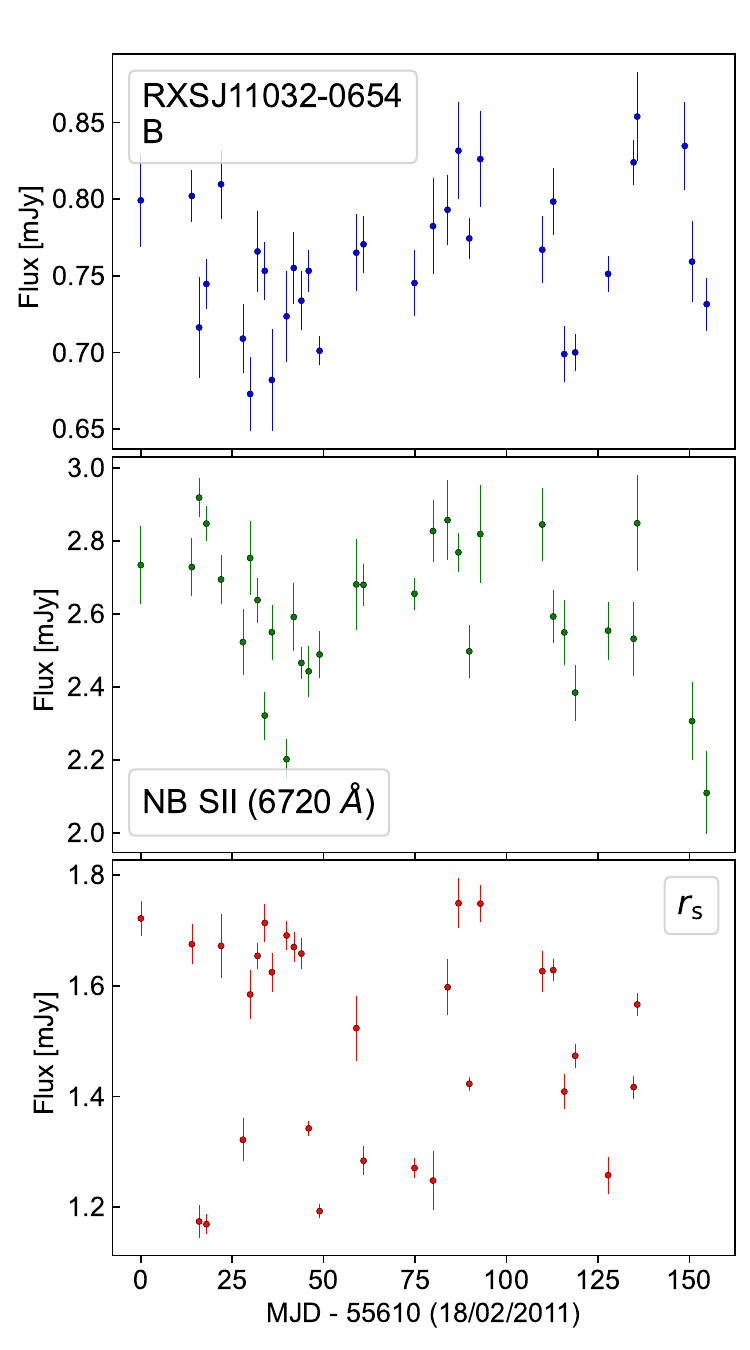}
\includegraphics[width=0.33\columnwidth]{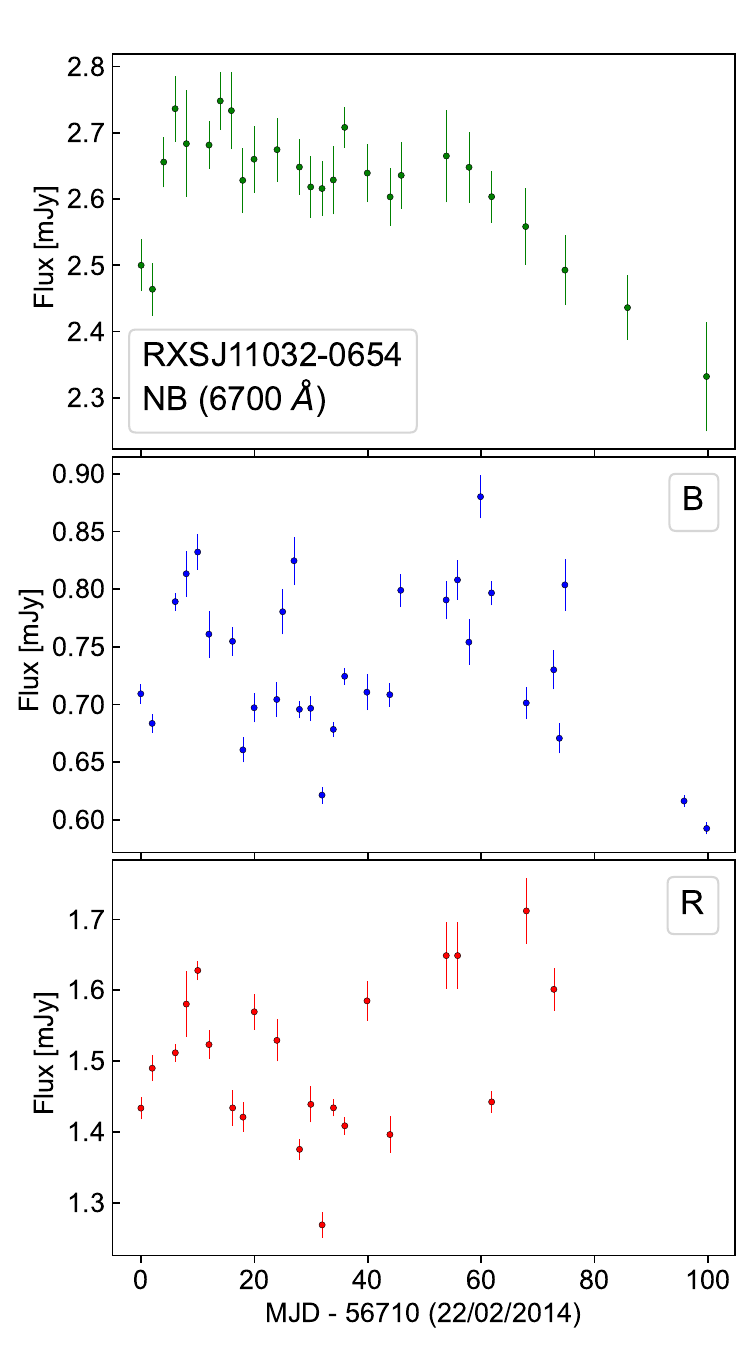}

\includegraphics[width=0.33\columnwidth]{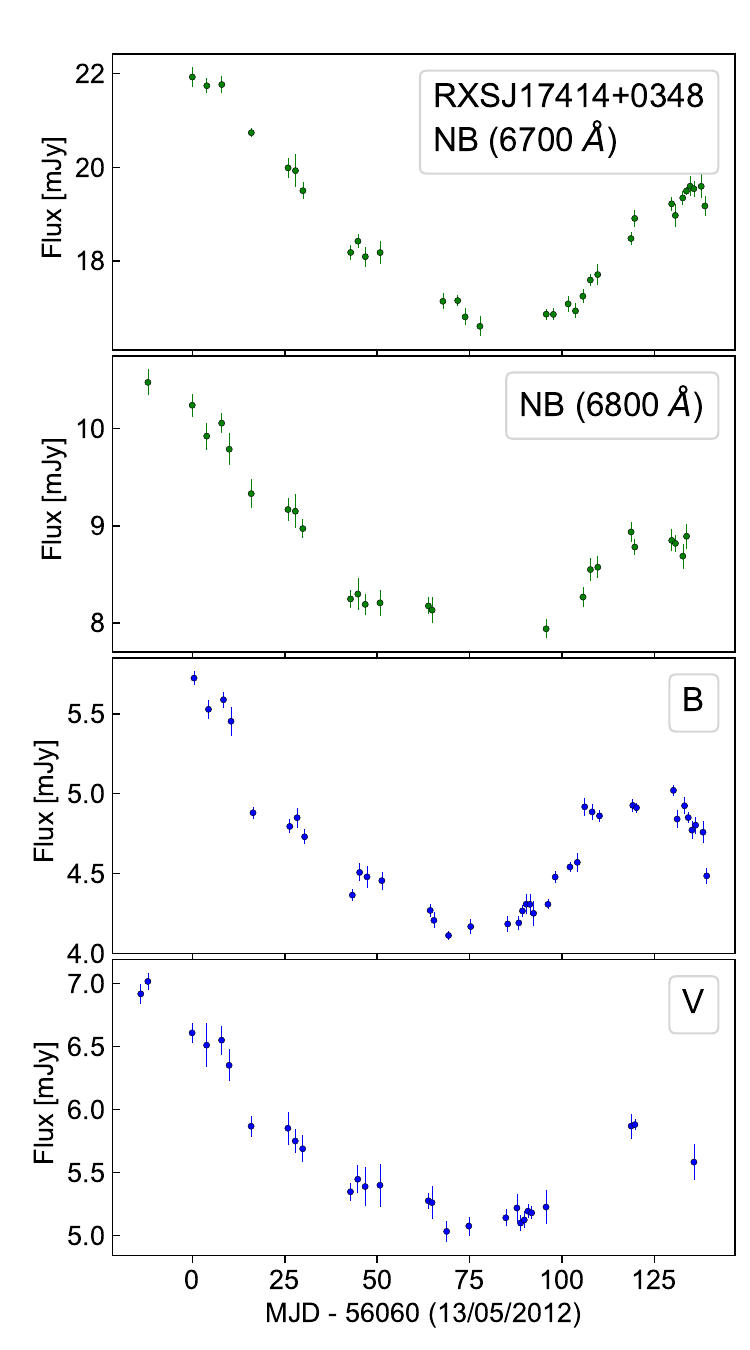}
\includegraphics[width=0.33\columnwidth]{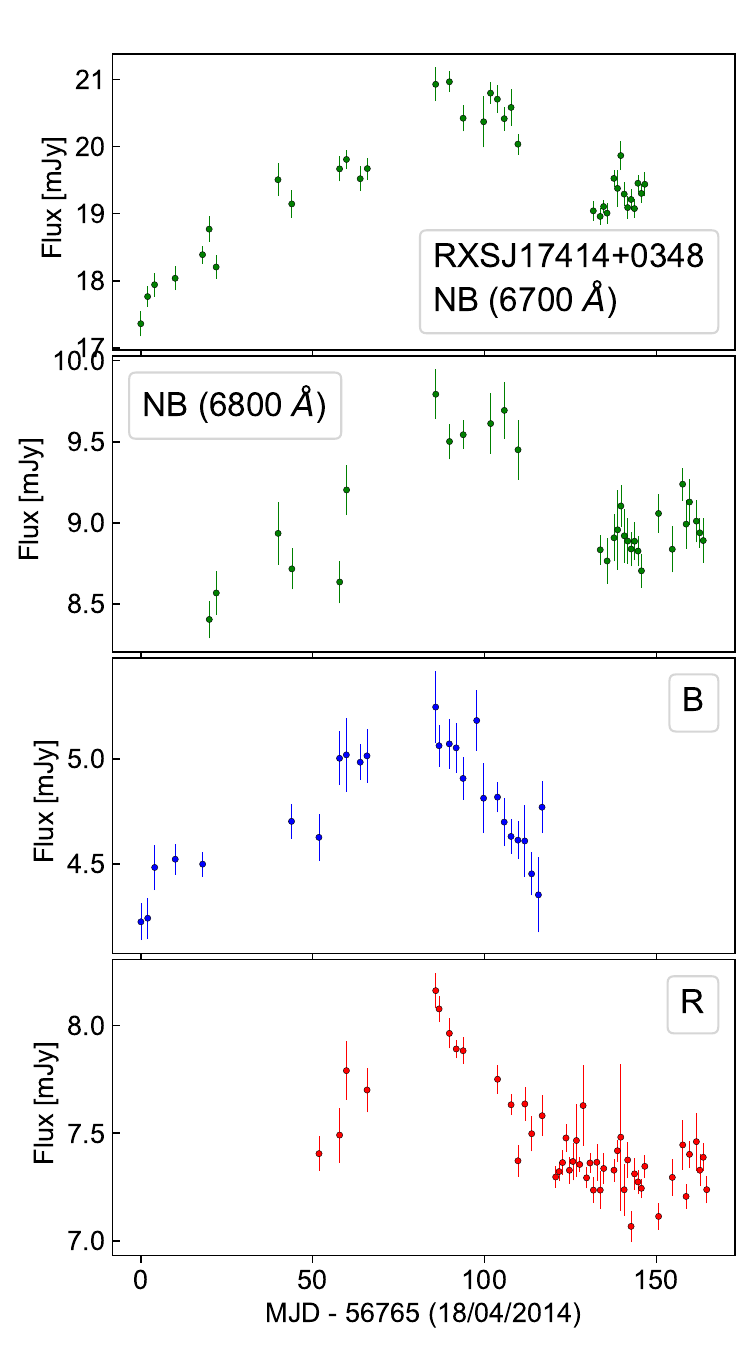}
\includegraphics[width=0.33\columnwidth]{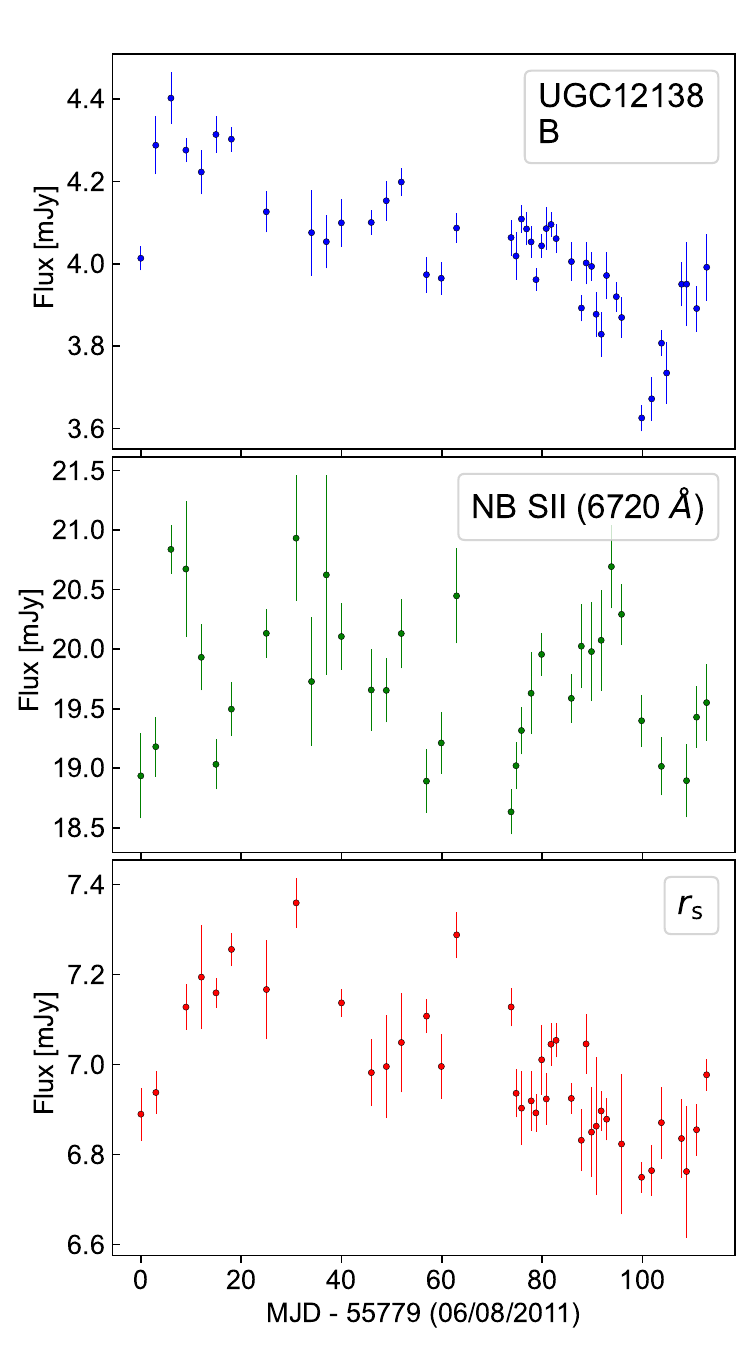}

\includegraphics[width=0.33\columnwidth]{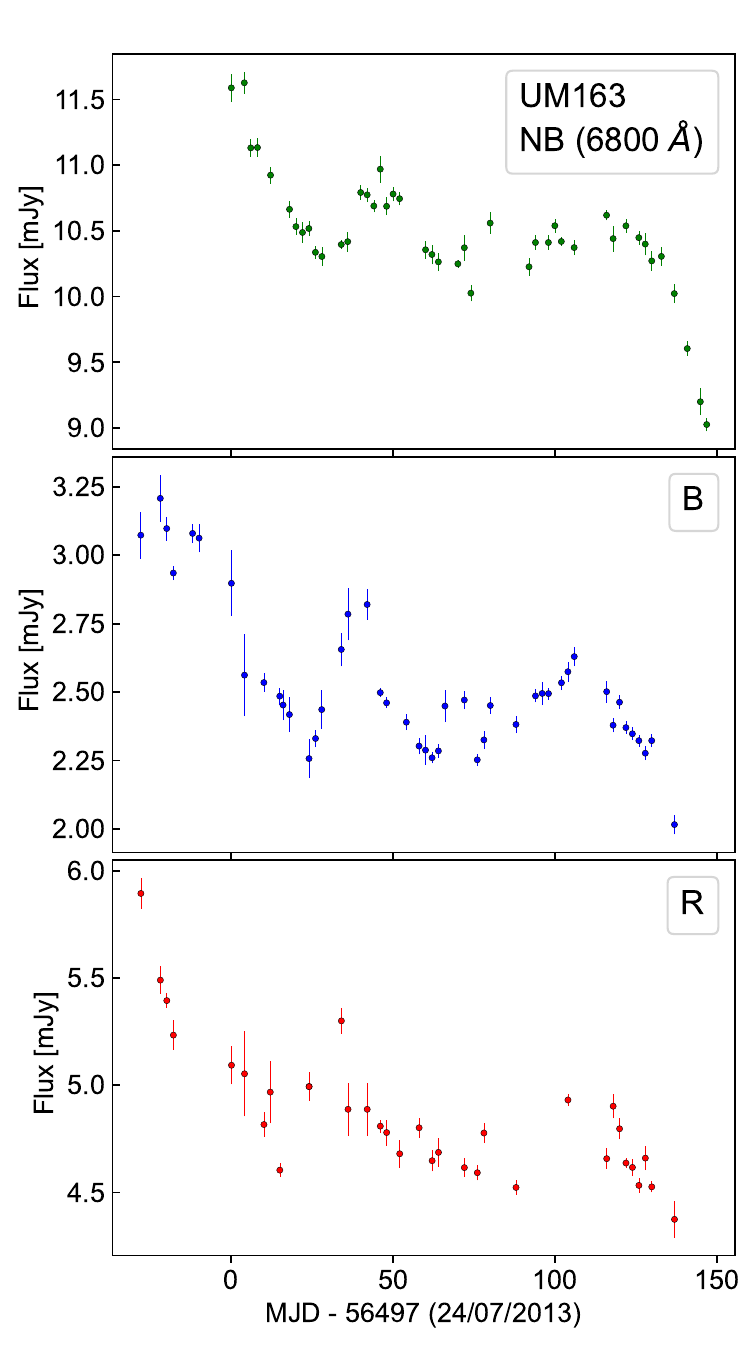}
\includegraphics[width=0.33\columnwidth]{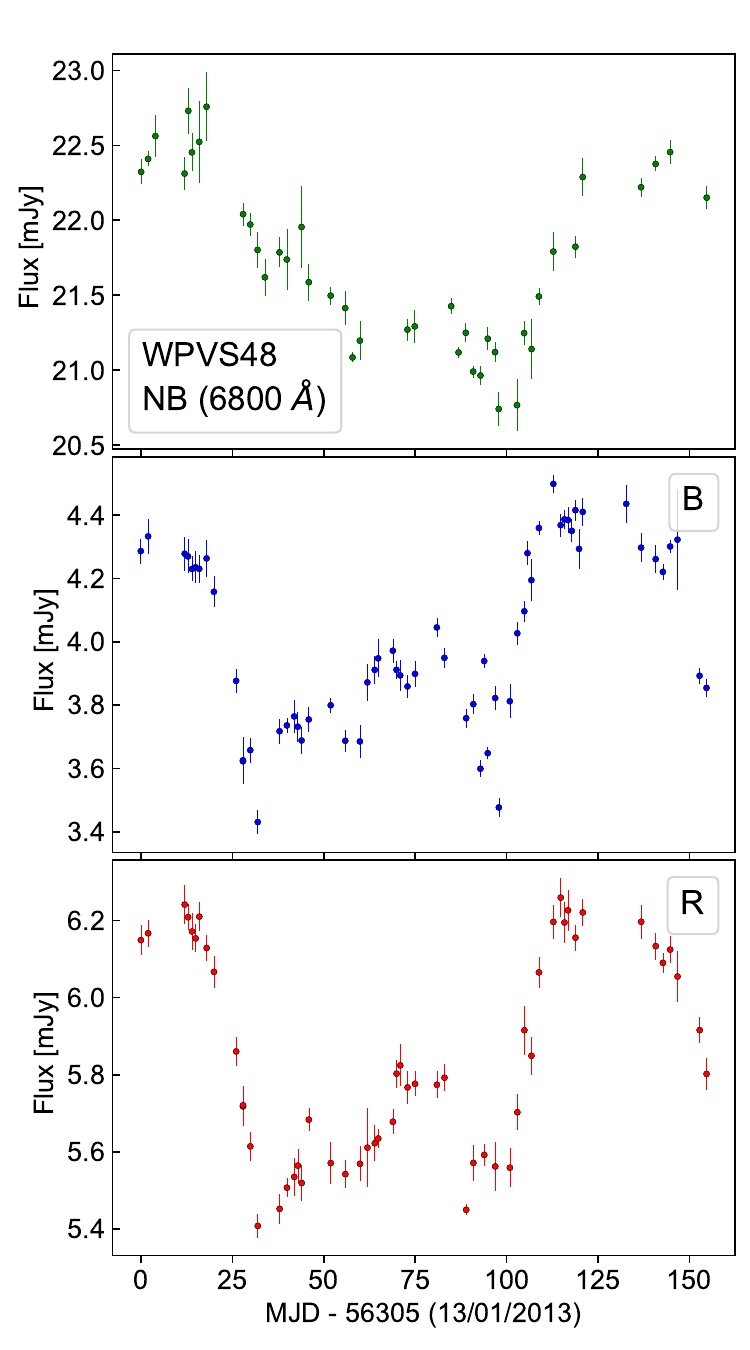}
\includegraphics[width=0.33\columnwidth]{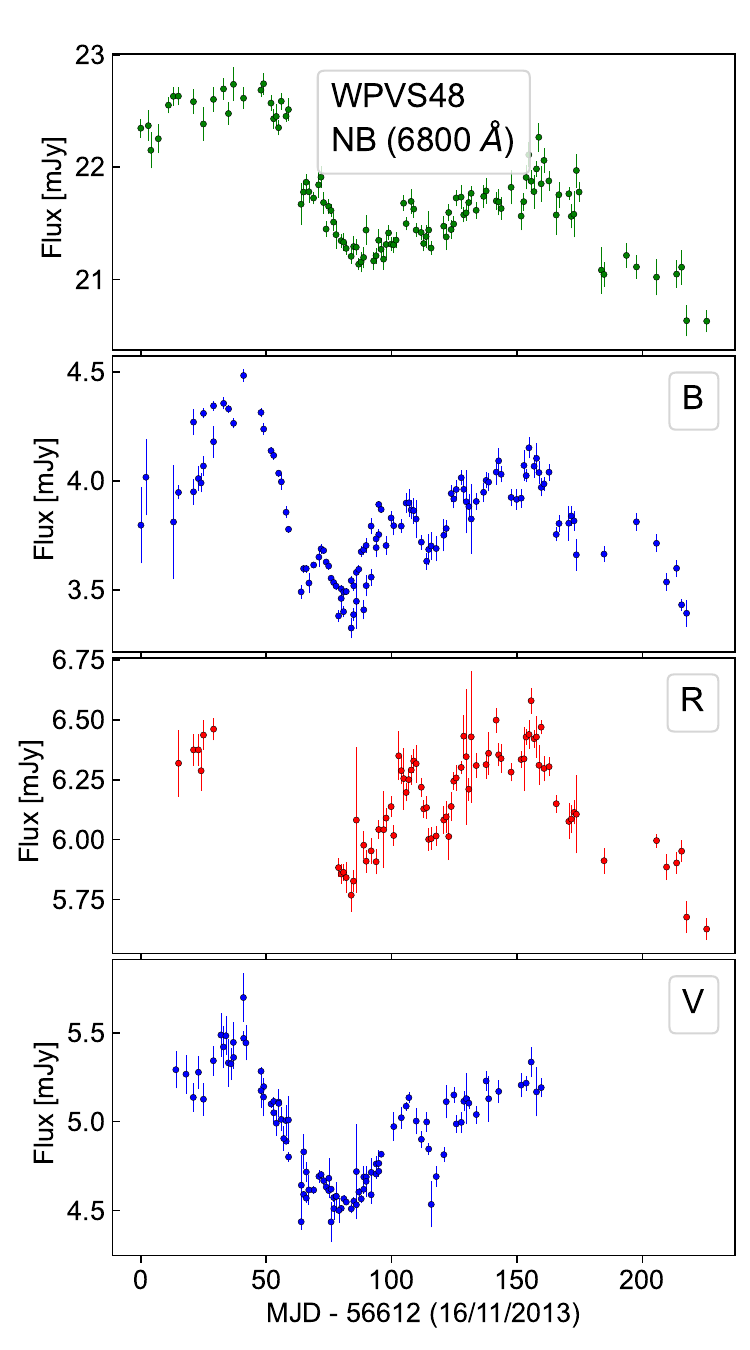}

\includegraphics[width=0.33\columnwidth]{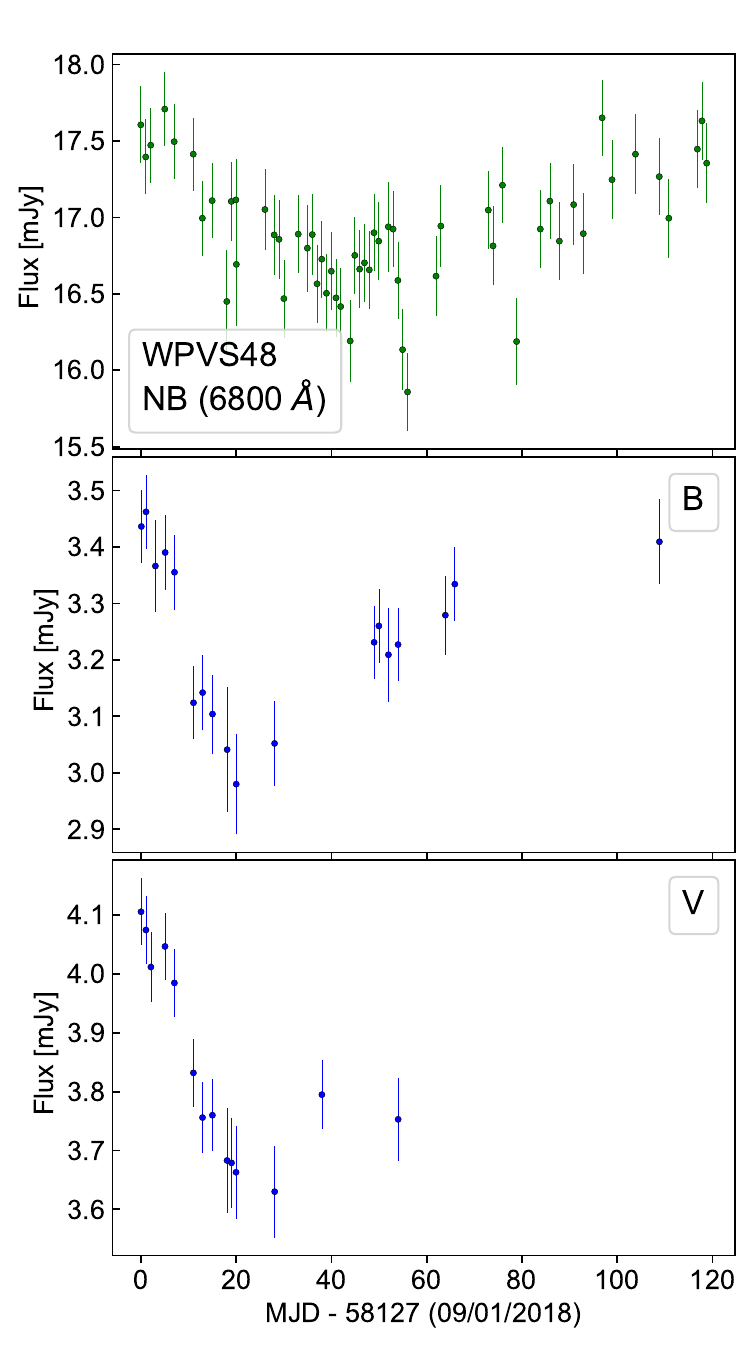}
\includegraphics[width=0.33\columnwidth]{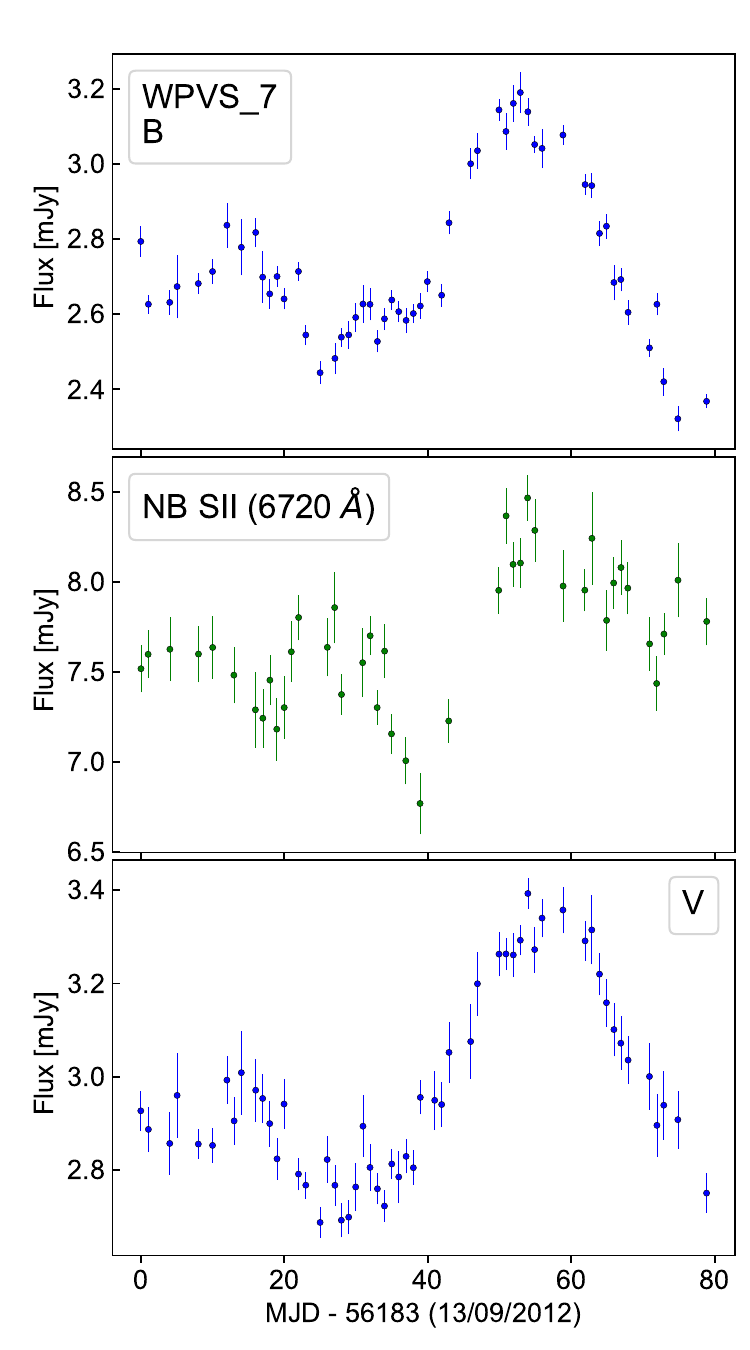}

\newpage

\section{Time-delay measurements formalism}\label{sec:app_time}

In this work we implement the multivariate correlation-function formalism of \citet{1994ApJ...420..806Z} for PRM but do so in a way which is easily implementable with commonly available tools of the trade, namely, cross-correlation functions. As discussed in \citet{2013ApJ...769..124C}, PRM operates on two light curves which differ by the relative contribution of (delayed) BLR emission to them. Following the nomenclature of \citet{2012ApJ...747...62C}, the band for which the BLR contribution is small is denoted by $f_c$ while the relatively BLR-rich light curve is denoted by $f_{lc}$. As in \citet{2013ApJ...769..124C}, we approximate $f_{lc}$ to first order \citep[see][for the inclusion of finite transfer function effects]{2013ApJ...769..124C} by the following model:
\begin{equation}
f_{lc}^{m}(t)=(1-\alpha)f_c(t)+\alpha f_c(t-\tau).
\label{model}
\end{equation}
Here, $\alpha$ is a parameter representing the contribution of the lagging component to the signal ($0 \le \alpha\le 1$), and $\tau>0$ is the time-delay (continuum time-delays between the different wavelength bands are ignored here, which seems to provide a good approximation for quasar data over the optical band; \citealt{2013ApJ...772....9C}). 

The model parameters, $\alpha$ and $\tau$, may be deduced by searching for the best agreement between the model, $f_{lc}^m$, and the data, $f_{lc}$. In RM studies, this is often accomplished by searching for the parameter values that maximize the Pearson correlation coefficient, $r$, which for our model yields:
\begin{equation}
r(\tau,\alpha)= \frac{1}{N\sigma_{lc}\sigma_{lc}^m} \sum_{n=1}^{N} f_{lc}(t_n) f_{lc}^m(t_n-\tau)=\frac{1-\alpha}{N\sigma_{lc}\sigma_c} \sum_{n=1}^{N} f_{lc}(t_n) f_c(t_n)+\frac{\alpha}{N(\tau)\sigma_{lc}(\tau)\sigma_c(\tau)} \sum_{n=1}^{N(\tau)} f_{lc}(t_n) f_c(t_n-\tau),
\end{equation}
where we work with light curve averages set to zero. For {\it stationary} lightcurves, the standard deviations\footnote{Here, $\sigma_{c}=\sqrt{N_c^{-1}\sum_{i=1}^{N_c} f_{c}(t_n)^2}$ and \\ $\sigma_{lc}=\sqrt{N_{lc}^{-1}\sum_{i=1}^{N_{lc}} f_{lc}(t_n)^2}$} $\sigma_c(\tau)=\sigma_c,~\sigma_{lc}(\tau)=\sigma_{lc}$, and the light curve averages do not depend on the part of the light curve sampled (i.e., is independent of the time-shift), where it is understood that light curve extrapolations are ignored. In the limit $N\to \infty$, $N(\tau) \to N$ and the expressions simplify considerably. For realistic quasar light curves, which are often characterized by large-amplitude variations at the lowest frequencies probed, this requires de-trending \citep{1999PASP..111.1347W}. For finite $N$, finite deviations from the true definition of the correlation function will arise whose effect on the time-lag determination depends much on the particular light curve and its sampling. 
In the limit of stationary infinite time-series, 
\begin{equation}
\sigma_{lc}^m=\sqrt{(1-2\alpha+2\alpha^2)\sigma_c^2+\frac{2\alpha(1-\alpha)}{N} \sum_{i=1}^{N}f_c(t_n)f_c(t_n-\tau)}=\sigma_c\sqrt{1-2\alpha+2\alpha^2+2\alpha(1-\alpha){\rm ACF}_c(\tau)}.
\end{equation}
Here, the autocorrelation function, ACF$_c$ is evaluated for $f_c$, and we obtain
\begin{equation}
r(\tau,\alpha)= \frac{(1-\alpha) \sum_{n=1}^N f_{lc}(t_n)f_c(t_n)+ \alpha \sum_{n=1}^{N} f_{lc}(t_n)f_c(t_n-\tau) }{N \sigma_{lc}\sigma_c\sqrt{1-2\alpha+2\alpha^2+2\alpha(1-\alpha){\rm ACF}_c(\tau)}}=\frac{(1-\alpha){\rm CCF}(0)+\alpha {\rm CCF}(\tau)}{\sqrt{1-2\alpha+2\alpha^2+2\alpha(1-\alpha){\rm ACF}_c(\tau)}},
\label{rrr}
\end{equation}
where CCF$(\tau)$ is the cross correlation function between the lightcurves in the bands evaluated at time-delay $\tau$, and specifically,  CCF$(0)$=CCF$(\tau=0)$. 

The above expression for the Pearson correlation coefficient has two noteworthy limits: in case $f_{lc}$ is just the lagging emission signal (e.g., continuum emission has been subtracted from band containing the BLR signal) then $\alpha=1$ and we obtain the standard cross-correlation function term commonly used in spectroscopic RM,
\begin{equation}
r(\tau,\alpha=1)={\rm CCF}(\tau).
\end{equation}
In the opposite limit, $\alpha \ll 1$, and Taylor-expanding Equation \ref{rrr} to first order in $\alpha$ yields 
\begin{equation}
r(\tau,\alpha)\simeq {\rm CCF}(0)+\alpha \left [ {\rm CCF}(\tau)-{\rm CCF}(0){\rm ACF}_c(\tau) \right ].
\end{equation}
Neglecting the normalization factors CCF(0) and $\alpha$, the time-delay may be obtained by searching for the maximum of ${\rm CCF}(\tau)-{\rm CCF}(0){\rm ACF}_c(\tau)$. Interestingly, for $\alpha \ll 1$ the light curves in both bands across the optical range are highly correlated by virtue of continuum emission processes dominating them\footnote{This is true over the optical wavelength range, but may not be true for bands that are significantly removed in wavelength for which continuum time-delays may be relevant.} hence ${\rm CCF}(0)\simeq 1$. In this limit, we therefore seek to maximize the expression ${\rm CCF}(\tau)-{\rm ACF}_c(\tau)$, which is the estimator considered by \citet{2012ApJ...747...62C} for performing RM using broadband data. 

Recalling the general expression (Eq. \ref{rrr}), one may significantly reduce the computational cost of searching for an extremum in 2D space, to searching for a maximum in $r$ along a path by requiring that $\partial r(\tau,\alpha)/\partial \alpha =0$. The latter requirement defines a path in 2D space defined by
\begin{equation}
\alpha(\tau)=\frac{{\rm CCF}(0){\rm ACF}_c(\tau)-{\rm CCF}(\tau)}{[{\rm CCF}(\tau)+{\rm CCF}(0)][{\rm ACF}_c(\tau)-1]}.
\label{alpha_e}
\end{equation}
Estimating the Pearson correlation coefficient over this path results in a one-dimensional calculation for which $r_e(\tau)\equiv r(\tau,\alpha(\tau))$ is extremal (maximal), and is given by
\begin{equation}
R_e(\tau)=\sqrt{\frac{{\rm CCF}^2(0)-2{\rm CCF}(0){\rm CCF}(\tau){\rm ACF}_c(\tau)+{\rm CCF}^2(\tau)}{1-{\rm ACF}_c^2(\tau)}}.
\label{R_e}
\end{equation}
The above expressions are simple to code and efficient to evaluate by standard means, as they include combinations of auto- and cross-correlation functions, which are commonly used in RM studies. 

Implementing the above for non-uniformly sampled time-series with gaps, we use the linearly interpolated scheme \citep[and references therein]{1986ApJ...305..175G,1998PASP..110..660P} often used in RM. Specifically, we verify that the sampled light curve (when calculating auto-/cross-correlations) are characterized by zero mean with their appropriate standard deviations evaluated at each time step \citep[the local correlation function formalism defined by][]{1999PASP..111.1347W}. Further, as noted in \citet{1999PASP..111.1347W} in the context of spectroscopic RM and in \citet[see also \citealt{2013ApJ...769..124C}]{2012ApJ...747...62C} in the context of PRM, detrending of the light curves by a first degree polynomial is recommended as it leads to more stationary behavior for sources with soft power-density spectra, and to improved time-delays. We therefore de-trend the light curves by subtracting a first order polynomial.

\subsection{Light curves simulations}\label{app:simulations}

We test the time lag determination formalism presented in Equations~\ref{alpha_e} and \ref{R_e} via simulations. Specifically, we simulate continuum light curves as a sum of modes drawn from a power spectrum of the form $f (\omega) \propto \omega^{\beta}$ ($\omega$ is the frequency) with $\beta \sim -2$, as is typical for describing AGN variability  \citep{1995A&A...300..707T,2003A&A...412..317C,2009ApJ...698..895K,2018ApJ...857..141S}, and is consistent with damped random-walk models \citep{2013ApJ...765..106Z}. However, more recent studies \citep[e.g.,][using \textit{Kepler} data]{2018ApJ...857..141S}  indicate somewhat steeper slopes ($\beta\sim -3$) over lightcrossing timescales of the BLR. For our simulations we therefore assume $-3 \le \beta \le -1$. We simulate the line-rich band as assumed in equation~\ref{model}, impliying a transfer function of the Dirac's $\delta$-function form. The resulting light curve is determined by specifying the delay, $\tau_{\rm echo}$, and the relative contribution of the delayed component to the line-rich band, $\alpha_\mathrm{echo}$. 

We define the total duration, $N_{\rm tot}$, of the continuum light curves as 100 days and consider two different time-delays, $\tau_{\rm echo}=10$\,days and $\tau_{\rm echo}=30$\,days. To mimic the observed signal, we add Gaussian noise  at different levels corresponding to 1\%, 5\%, and 10\% of the signal, and distinguish between uniform and non-uniform sampling. We analyze the mock light curves with our PRM formalism, by computing $R_{\rm e}(\tau)$ as well as $\alpha(\tau)$ following expressions~\ref{alpha_e} and \ref{R_e}, thereby attempting to recover the input parameters. The correlation functions are implemented using a linear interpolation scheme, as commonly employed in reverberation studies \citep{1987ApJS...65....1G,1993PASP..105..247P,1999PASP..111.1347W}. The time-delay recovered from the mock data is identified with the lag, which maximizes the coefficient $Re(\tau)$, $\tau_{\rm peak}$, and for which $0<\alpha(\tau_{\rm peak}) \le 1$.

\begin{figure}
 \centering
	\includegraphics[width=0.44\columnwidth]{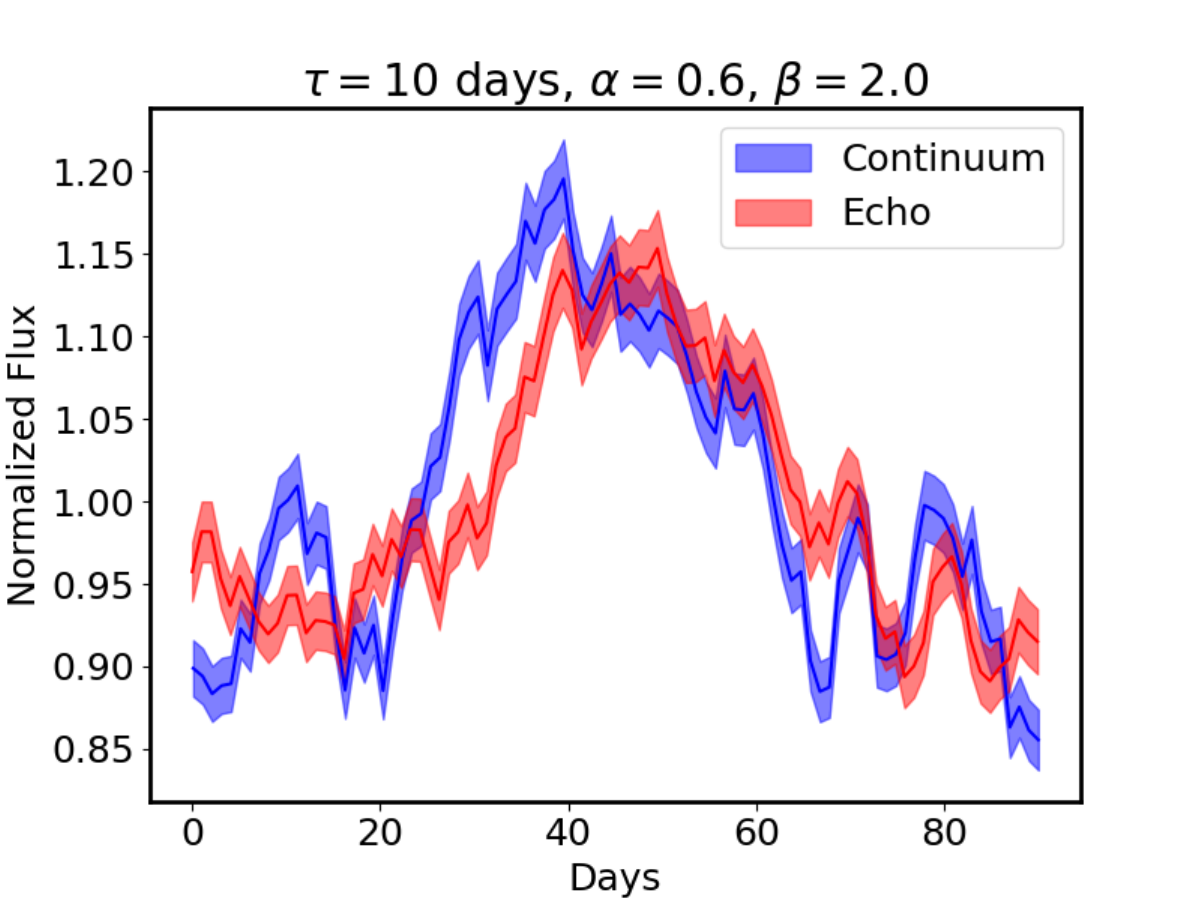}
	\includegraphics[width=0.44\columnwidth]{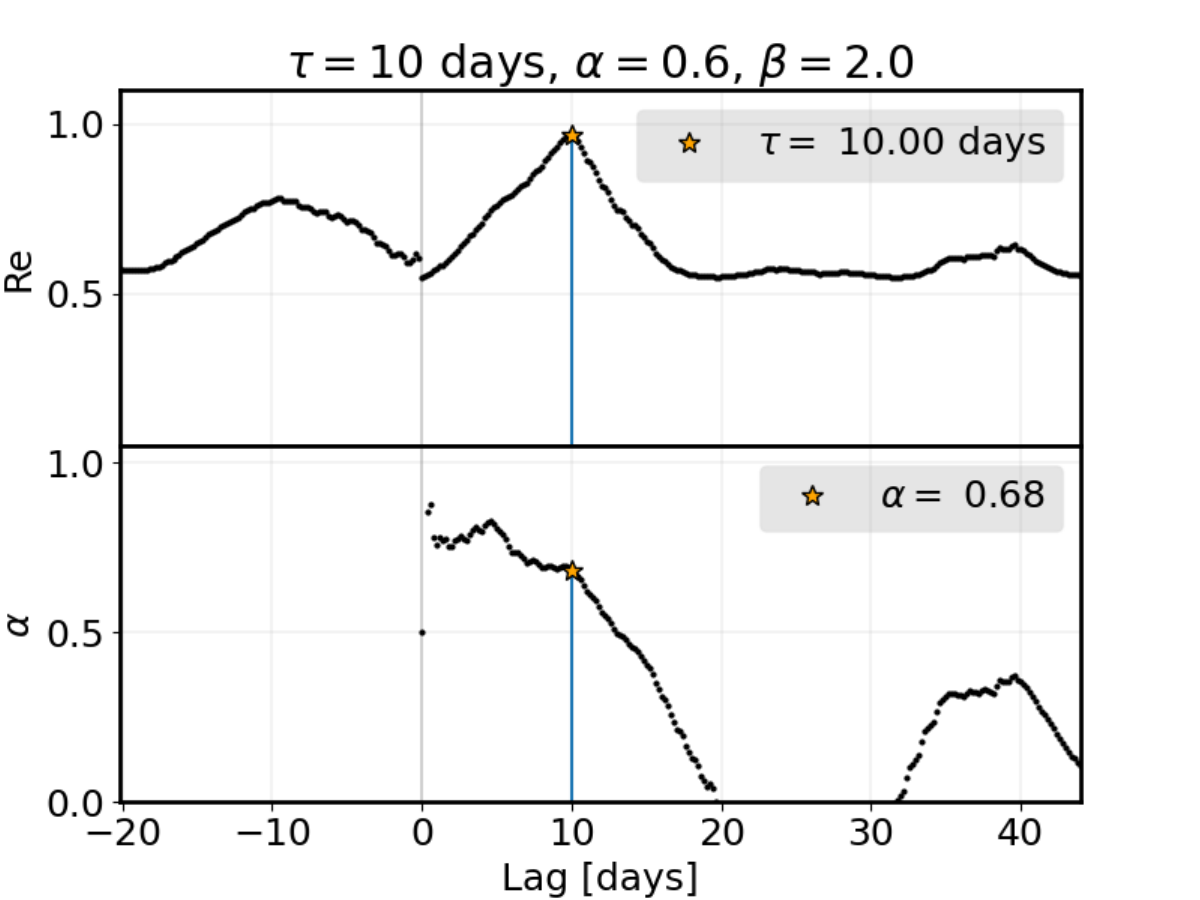}
 \caption{{Left: 100 days long simulated continuum light curve with an exponent $\beta=-2$ in blue and simulated echo light curve with $\alpha=0.6$ and $\tau=10$ days in red. Right: Correlation coefficient $Re$ in the upper pannel with its corresponding $\alpha$ value in the bottom pannel. The peak delay $\tau_{\rm peak}$ and $\alpha_{\rm peak}$ are marked with an orange star and labeled.}}
 \label{fig:sim_example}
\end{figure}

The left panel of Figure~\ref{fig:sim_example} shows an example for simulated continnuum and line-rich light curves with uniform sampling and a total duration $N_{\rm tot}$ of 100 days ($\beta =-2$, $F_\mathrm{var}=0.1$, and $\tau_{\rm echo}=$ 10\,days, $\alpha_{\rm echo} = 0.6$ were assumed). 
Figure~\ref{fig:sim_example} (right panel) shows the resulting $R_e(\tau)$, which peaks at  $\tau_{\rm echo} \simeq 10$\,days, and implies $\alpha_{\rm peak}\simeq 0.68$. This example shows that the time delay is well reproduced with this formalism, while $\alpha_{\rm peak}$ is less accurately constrained.

\begin{figure}
 \centering
	\includegraphics[width=0.49\columnwidth]{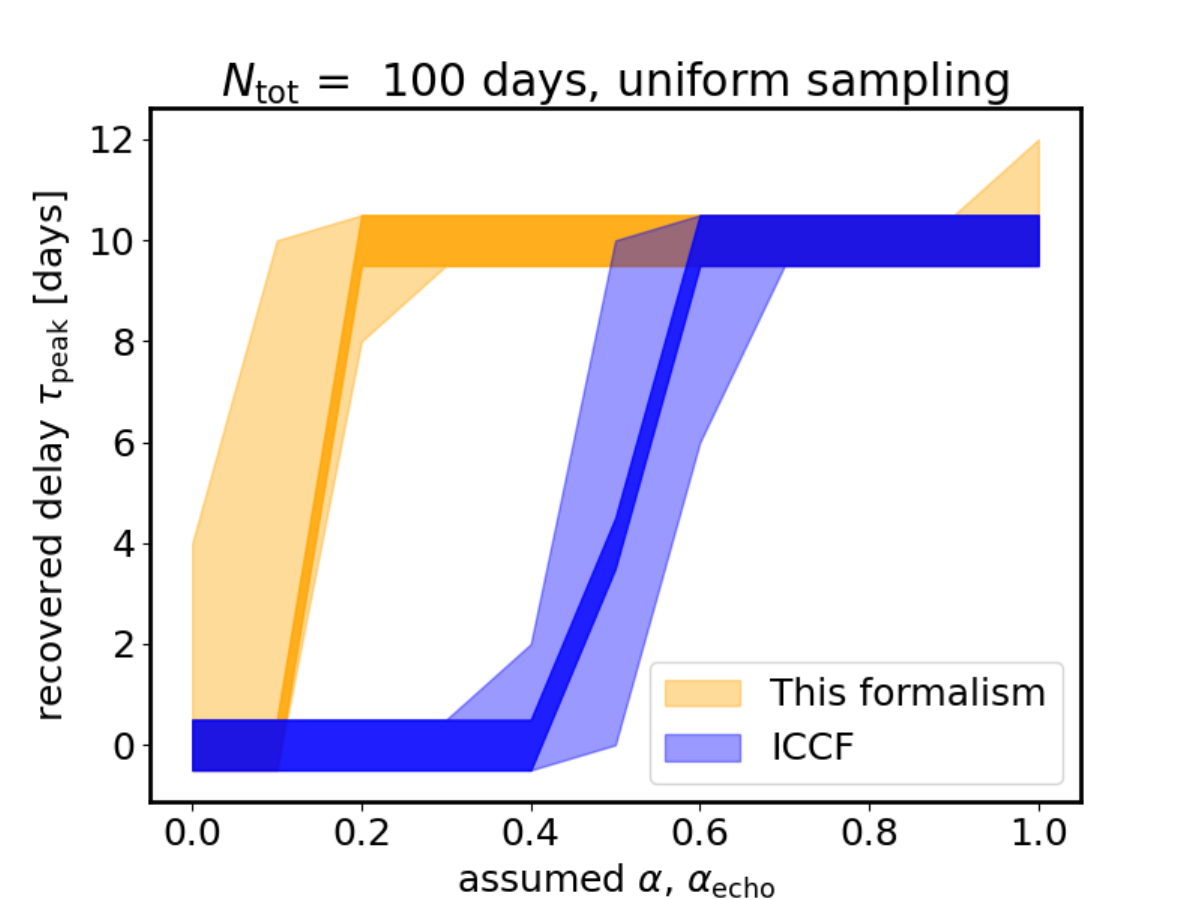}
 	\includegraphics[width=0.49\columnwidth]{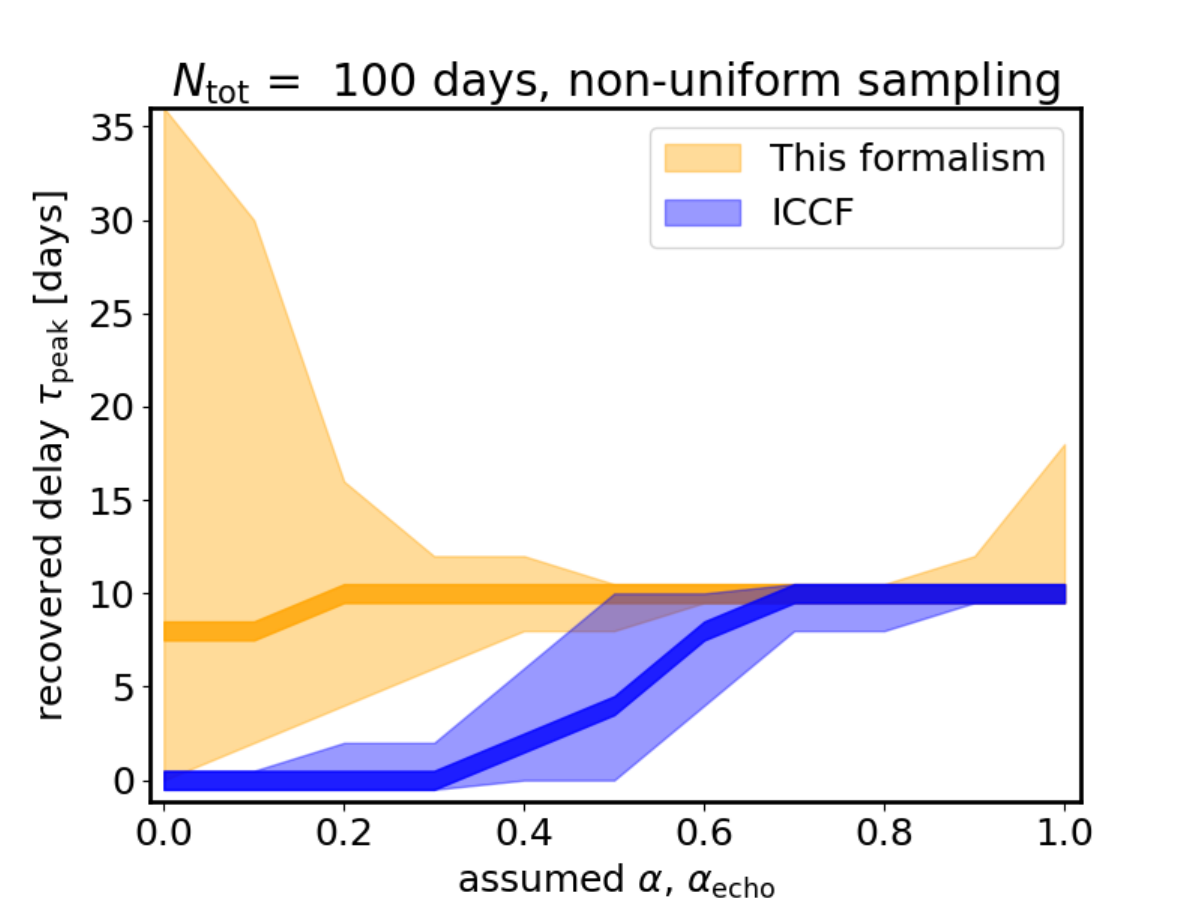}
 \caption{{Recovered time delay $\tau_{\rm peak}$ versus the assumed value for $\alpha_{\rm echo}$. The results obtained with this new formalism are shown in orange, while the results obtained directly with the ICCF method are shown in blue. Left pannel: results for uniform light curves with $N_{rm tot}=100$ days and noise of 1\%. Right pannel: Same for non-uniform light curves.}}
 \label{fig:simlc}
\end{figure}

To identify potential deviations of the measured time-delay from the input lag, we repeat the above calculation for $10^3$ different light curves. Figure~\ref{fig:simlc} shows $\tau_{\rm peak}$ for uniformly sampled light curves with $N_{\rm tot}=100$ days and an assumed noise level of 1\%. The Figure shows the recovered time delay, $\tau_{\rm peak}$, versus the input value for the relative contribution of the delayed component to the band, $\alpha_{\rm echo}$. The results obtained with the PRM formalism introduced here are compared to those obtained via the Interpolated Cross-Correlation Function (ICCF) technique: while the results from both methods converge for high $\alpha_{\rm echo}$ values (above 0.6), for lower values the new method is able to recover the input lag down to much lower values, which often characterize broadband data ($\alpha_{\rm echo}=$ 0.1; \citealt{2012ApJ...747...62C}), whereas the time delay obtained with the ICCF method is biased towards lower values for $\alpha_{\rm echo} < 0.6$, and hence inapplicable for narrowband data\footnote{Note that the comparison takes the two photometric light curves as they are, without explicit subtraction of an estimate of the continuum contribution from the line-carrying light curve, which is done in a properly performed standard PRM analysis.} (at $\alpha_{\rm echo}\lesssim$ 0.4, as appropriate for some narrow-band and broadband data, the ICCF method leads to zero delays for our model). Figure~\ref{fig:simlc} also shows the case for non-uniformly sampled data, for which $\sim 30$\% of the points have been eliminated using a random subset selection process \citep{1998PASP..110..660P}. Evidently, the scatter in the recovered lags by our PRM method is substantial for small values of $\alpha_{\rm echo}$ which are typical of broadband data, although the mean is still comparable to the input lag. For $\alpha_{\rm echo}$ values typical of narrowband data, our PRM formalism correctly recovers the input lag. This contrasts the ICCF results, which lead to erroneous measurements also for broad and narrowband data.   

\begin{figure}
\centering
\includegraphics[width=\columnwidth]{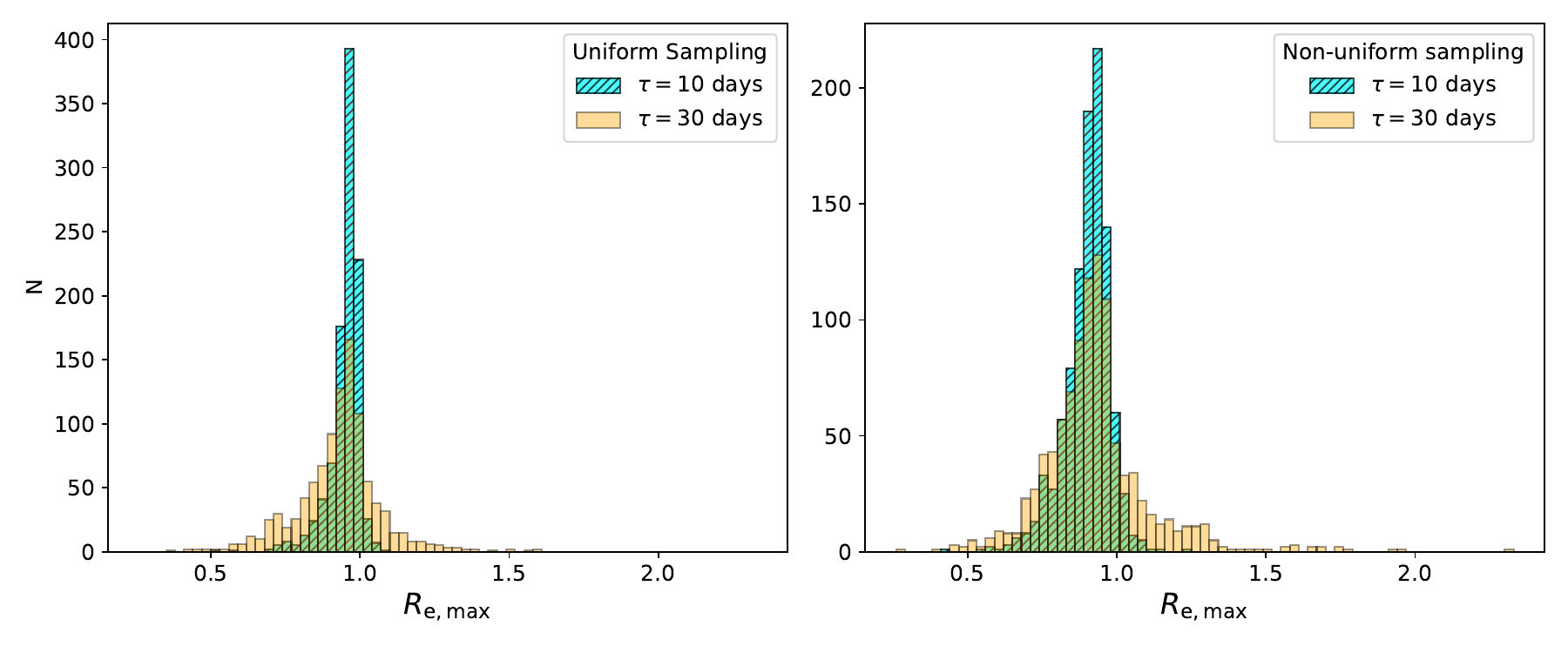}
\caption{Maximal value reached by the correlation coefficient $R_e$ for different kind of simulation settings. {Left panel: Simulated light curves with a total duration $N_{\rm tot} = 100$ days and uniform sampling, $\tau_{\rm echo} = 10$ days (cyan) and $\tau = 30$ days (orange); right panel: for $N_{\rm tot} = 100$ days and non-uniform sampling, $\tau_{\rm echo} = 10$ days (cyan) and $\tau = 30$ days (orange).}}% Simulated light curves with a total duration $N_{\rm tot}= 100$ days, $\tau_{\rm echo}= 10$ days and uniform sampling (top left); for $N_{\rm tot}= 100$ days, $\tau_{\rm echo}= 10$ days and non-uniform sampling (top right); for $N_{\rm tot}= 100$ days, $\tau_{\rm echo}= 30$ days and uniform sampling (bottom left) and for $N_{\rm tot}= 100$ days, $\tau_{\rm echo}= 30$ days and non-uniform sampling (bottom right).}}
 \label{fig:hist_re}
\end{figure}

Formally, $R_e(\tau) \le 1$ and yet in certain cases, our implementation of the formalism can lead to $R_e(\tau)>1$. This may occur when the cadence is non-uniform, and/or the time delay represents a considerable fraction of the total duration of the time series. Figure~\ref{fig:hist_re} shows the statistics of the maximum $R_e(\tau)$ value for simulated data. Clearly, a higher fraction of $R_e(\tau)>1$ occurrences is obtained when the time-delay is a significant fraction of the total duration of the time-series, and when considerable gaps exist, as in the case of non-uniform sampling. Therefore, great care must be taken when interpreting the results of the correlation functions at long time-delays, wherein the period overlap between the shifted time-series reduces by $\sim 50$\%. The reason for such effects is due to the implementation of the formalism with using ICCF 'building blocks' to evaluate equation \ref{R_E}.  Partial remedy would be to define a common grid to all auto- and cross-correlation functions, and yet this may lead to a greater reliance on interpolated data than on real data, leading in some cases to erroneous results, which we opt to avoid. A further advantage of our current scheme is the use of readily available ICCF codes when calculating $R_e$. A more in-depth investigation of the limitations of this approach is beyond the scope of the present work.

\begin{figure}
\centering
\includegraphics[width=0.44\columnwidth]{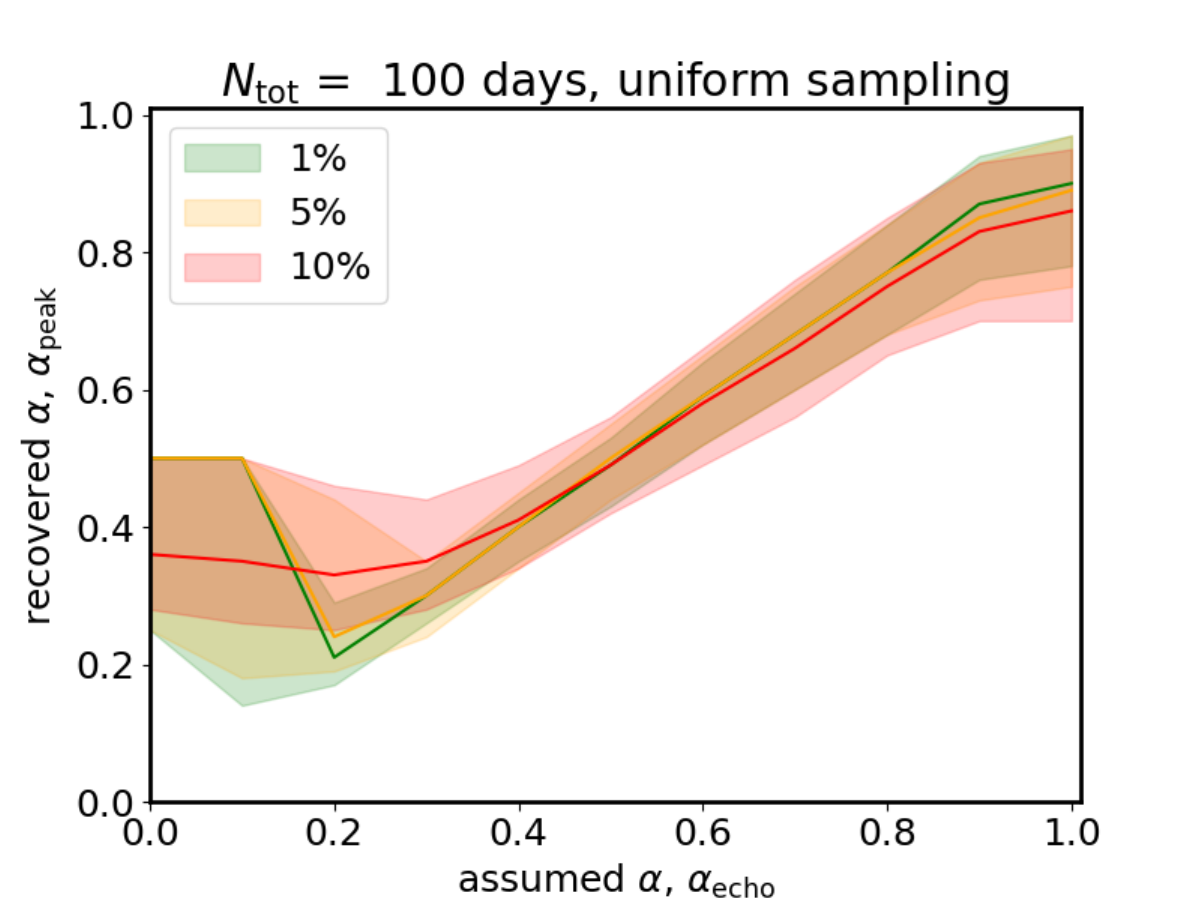}
\includegraphics[width=0.44\columnwidth]{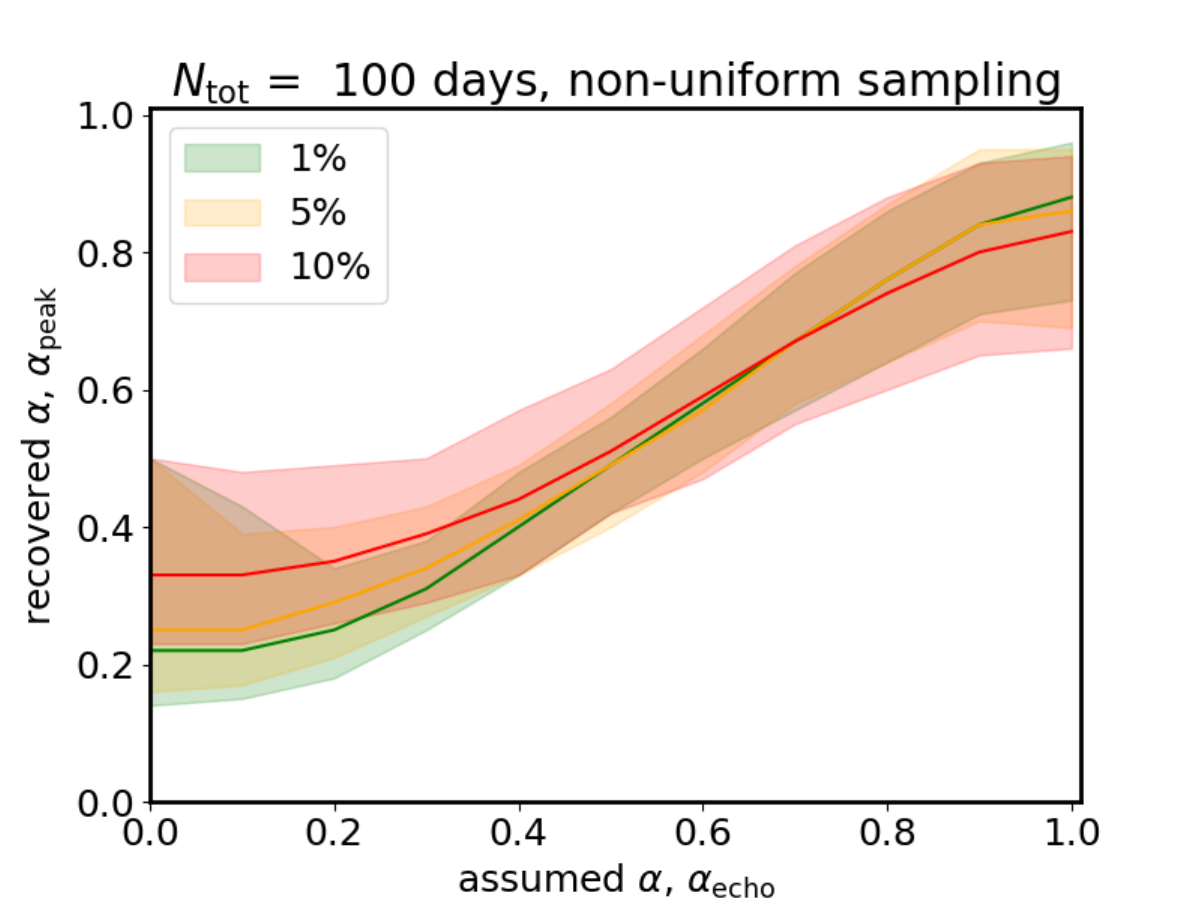}
\caption{{Recovered $\alpha_{\rm peak}$ versus assumed $\alpha_{\rm echo}$ values for different noise values and uniform sampling (left) and non-uniform sampling(right).}}
\label{fig:alpha_sim}
\end{figure}

Lastly, we note that the recovered contribution of the delayed component,  $\alpha_{\rm peak}$, agrees with the simulated values for $\alpha_{\rm echo} \gtrsim 0.4$; see Figure~\ref{fig:alpha_sim}. This conclusion depends little on the noise level, and is relatively insensitive to the sampling. It shows that $\alpha_{\rm echo}$ is well recovered down to $\alpha_\mathrm{echo}\sim 0.2$, where it plateaus for the simulations conducted here. Thus, while time delays can be relatively well reproduced, the corresponding values for $\alpha$ may be biased to higher values for broadband data. 

{We carried out additional simulations to account for the effect of continuum time delays within the modeled echo light curve. To this end, we assume that the light curve, which photometrically traces line emission, is a combination of two different delayed components: one is delayed continuum signal by $\tau_{\rm cont}$ and the other is a delayed emission-line signal by $\tau_{\rm line}$. As before, the simulated ratio between the contributions of the line and continuum components equals $\alpha_\mathrm{echo}/(1-\alpha_\mathrm{echo})$. We set $\tau_{\rm cont}$ to be between 5\% and 15\% of the assumed $\tau_{\rm line}$, as supported by continuum RM studies \citep{2022A&A...659A..13F,2023A&A...672A.132F}. By incorporating these effects, we aim to quantify the biases introduced in the emission-line time-delay, when the underlying PRM model used (i.e., Eq. \ref{model} as well as the current JAVELIN implementation) is incomplete. As previously described, we conducted $10^3$ simulations for various assumed values of $\alpha_{\rm echo}$ and analyzed the recovered delays obtained using our formalism.} 

\begin{figure}
\centering
\includegraphics[width=\columnwidth]{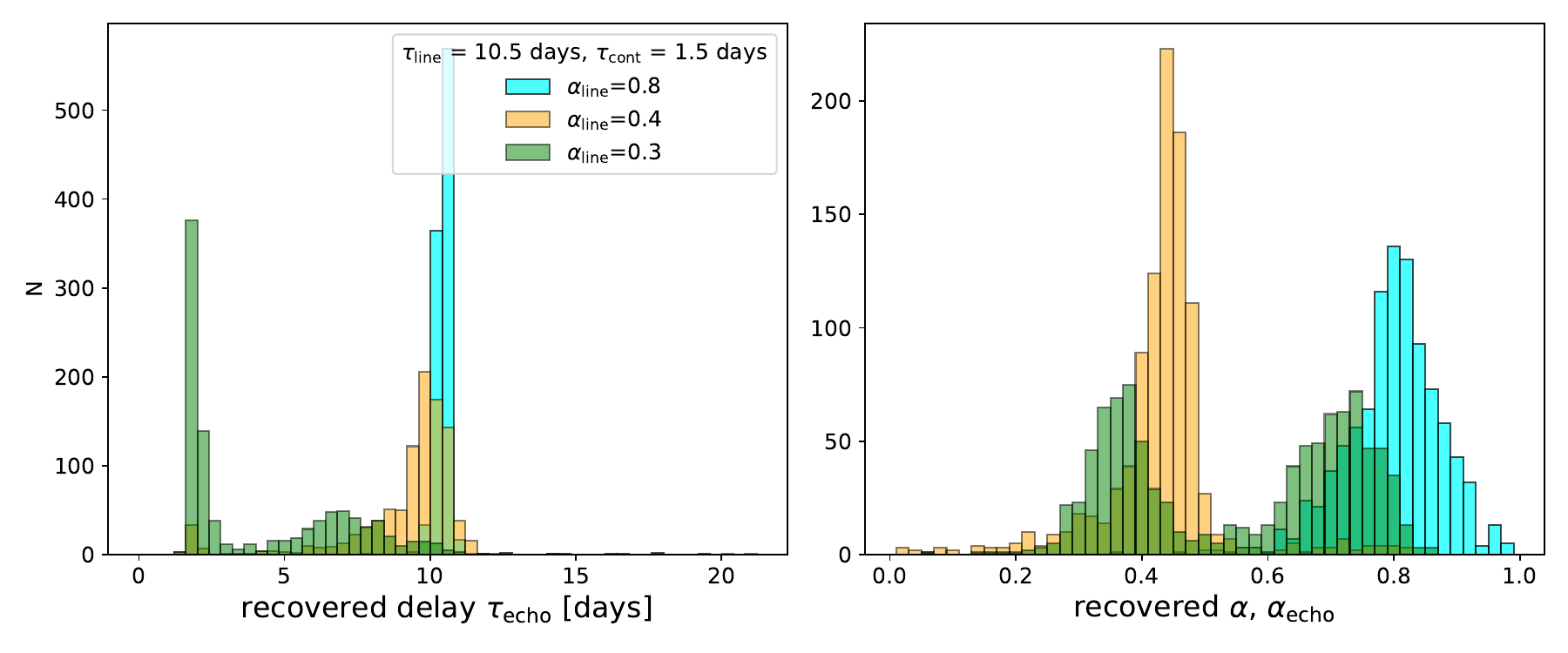}
\caption{{Statistics for the recovered time delay (left) and $\alpha$ (right) with a continuum delay ($\tau_{\rm cont}$ = 1.5 days) set at 15\% of the assumed line delay ($\tau_{\rm line}$ = 10.5 days). Results are presented for three different assumed values of $\alpha$: 0.3 in green, 0.4 in orange, and 0.8 in cyan.}}
\label{fig:hist_tcont}
\end{figure}

{Figure~\ref{fig:hist_tcont} illustrates an example of the statistics for the recovered time delay (left panel) alongside $\alpha$ (right panel), with a continuum delay ($\tau_{\rm cont}=$1.5 days) set at 15\% of the assumed line delay ($\tau_{\rm line}=$10.5 days). The figure presents results for three different assumed values of $\alpha_\mathrm{echo}$: 0.3, 0.4, and 0.8. It demonstrates that for larger values of $\alpha_\mathrm{echo}$, both the recovered echo delay and the recovered $\alpha$ align closely with the assumptions. However, for the smaller value ($\alpha_\mathrm{echo}=$ 0.3), the recovered delay displays two distinct peaks, resulting in an overall shift to lower values. The corresponding recovered $\alpha$ value also shows two components: one associated with the continuum delay and the other with the line delay.}

\begin{figure}
\centering
\includegraphics[width=\columnwidth]{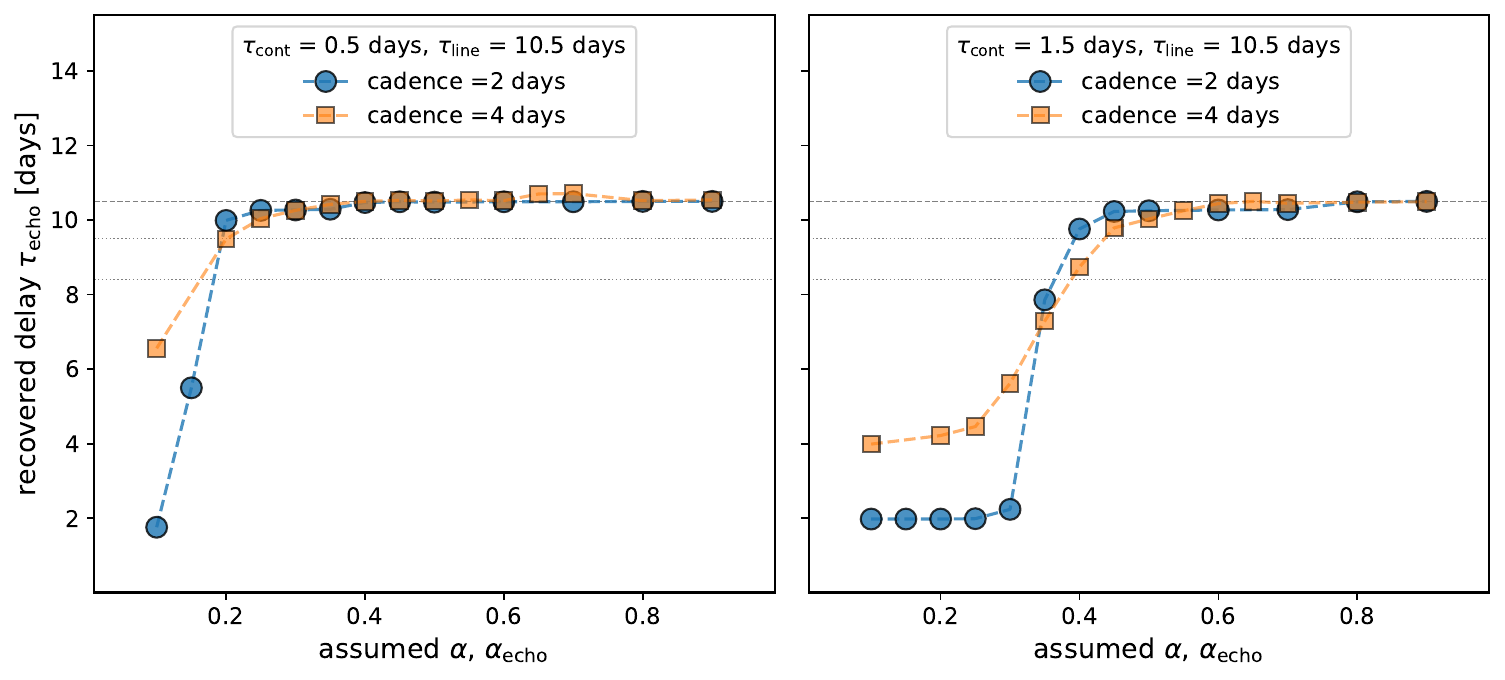}
\caption{{Recovered time delay as a function of the assumed value for $\alpha_{\rm echo}$. The left panel displays results obtained with an additional continuum delay of 10\% of the line delay, while the right panel presents results for a continuum delay of 15\% of the line delay. Blue circles represent results for a cadence of 2 days, and orange squares indicate a cadence of 4 days. The dashed black horizontal line marks the assumed $\tau_{\rm line}$, while the dotted grey line indicates a 10\% deviation towards lower values.}}
\label{fig:stats_tcont}
\end{figure}

{Additionally, Figure~\ref{fig:stats_tcont} illustrates the results for the recovered delay across various $\alpha_\mathrm{echo}$ values and different continuum delay settings. Our analysis reveals that continuum contamination can affect the accuracy of the recovered echo time delay in certain cases. Specifically, when the line fraction covered by the observing filter is below 20\% ($\alpha_\mathrm{echo}<$ 0.2), the continuum contamination leads to an underestimation of the recovered echo delay (see the left panel of Figure~\ref{fig:stats_tcont}). As the continuum time delay increases to 15\% of the emission line (shown in the right panel of Figure~\ref{fig:stats_tcont}), this effect becomes more pronounced, particularly when the line fraction falls below 40\% ($\alpha_\mathrm{echo}<$ 0.4). Additionally, the cadence of the light curves also influences the results, especially when the continuum time delay is sufficiently large. This implies that great care must be taken when attempting to achieve high-fidelity PRM using broadband data.}

\newpage
\section{H\texorpdfstring{$\alpha$}{alpha} correlations and continuum contribution to the emission line}\label{app:corr}
This appendix presents time-lag estimations for all the sources in our sample. Each of the plots contains five panels: the top panel shows the normalized and detrended light curves for the continuum signal (blue) and the NB band (red). $F_{\rm var}$ and $\eta$ values associated with the light curves is also noted (see Appendix~\ref{app:lc}). The left panels in the second and third rows illustrate the individual correlation results. The top section displays the correlation coefficient $R_{\rm e}$, while the bottom section shows the corresponding $\alpha$ value. The range of $\alpha$ values under consideration for detecting delay is highlighted in cyan. Details on how the $\alpha$ value range was selected for each object are provided in Appendix~\ref{sec:comments_objects}. The right panel on the second row of each figure displays the distribution of the time-lag estimations following $10^3$ flux-randomization iterations; the blue histogram represents the peak delays and the red histogram represents the centroid delays. The resulting values for the 50\% percentile and the errors (15\% and 85\% percentiles) are labeled. The right panel on the third row of each depicts the distribution for the deduced $\alpha$ values (see Appendix~\ref{sec:app_time}), with a similar meaning as for the time delays. The bottom panel in each figure presents the distribution for the maximal correlation coefficient, $R_{\rm e,max}$, hence is a measure of the goodness of fit from which time-delay determination follows. The blue histogram shows the distribution obtained after many flux-randomization iterations are performed, and the orange histogram represents the $R_{\rm e,max}$ value for the time-permutation (TP) scheme (see Section~\ref{sec:time_det}). The confidence level of our results is denoted in per-cent in the bottom panel.

\begin{deluxetable}{llccc|llccc}
\renewcommand{\arraystretch}{0.85}
\tablecaption{Values for the peak and center of gravity ($\alpha_{\rm peak}$ and $\alpha_{\rm cent}$) obtain with the formalism and $\alpha_{\rm phot}$ expected from the photometry estimation. }\label{tab:alpha_results}
 \tablehead{Object & Year & $\alpha_{\rm peak}$ & $\alpha_{\rm cent}$& $\alpha_{\rm phot}$&Object & Year &  $\alpha_{\rm peak}$ & $\alpha_{\rm cent}$& $\alpha_{\rm phot}$}
 \startdata
1H2107-097&2012 &$0.83^{+0.02}_{-0.02}$&$0.54^{+0.04}_{-0.12}$& & MRK335&2010 &$0.73^{+0.1}_{-0.08}$ &$0.74^{+0.11}_{-0.09}$& \\ 
3C120 &2014 &$0.55^{+0.05}_{-0.05}$&$0.48^{+0.05}_{-0.05}$&$0.63\pm0.05$& MRK335&2011 &$0.61^{+0.07}_{-0.05}$&$0.61^{+0.06}_{-0.07}$& \\ 
AKN120&2018 &$0.57^{+0.05}_{-0.05}$&$0.62^{+0.04}_{-0.06}$& & MRK335&2014 &$0.66^{+0.13}_{-0.06}$&$0.61^{+0.06}_{-0.05}$&$0.65\pm0.05$\\ 
CTSG03\_04&2013 &$0.5^{+0.02}_{-0.02}$ &$0.44^{+0.03}_{-0.03}$& 0.7 & MRK509&2014 &$0.62^{+0.05}_{-0.04}$&$0.65^{+0.03}_{-0.03}$&$0.76\pm0.05$\\ 
ESO141-G55&2015 &$0.46^{+0.01}_{-0.01}$&$0.46^{+0.01}_{-0.01}$&$0.73\pm0.04$& MRK705&2013 &$0.59^{+0.05}_{-0.02}$&$0.45^{+0.03}_{-0.03}$&$0.52\pm0.07$\\ 
ESO141-G55&2013 &$0.77^{+0.06}_{-0.34}$&$0.49^{+0.03}_{-0.03}$&$0.68\pm0.05$& MRK841&2014 &$0.4^{+0.04}_{-0.04}$ &$0.36^{+0.04}_{-0.02}$& \\ 
ESO323-G77&2015 &$0.74^{+0.07}_{-0.06}$&$0.73^{+0.06}_{-0.07}$&$0.61\pm0.06$& NGC1019&2011 &$0.75^{+0.08}_{-0.08}$&$0.74^{+0.07}_{-0.11}$& \\ 
ESO374-G25&2011 &$0.4^{+0.09}_{-0.08}$ &$0.3^{+0.03}_{-0.03}$ & & NGC4726&2013 &$0.78^{+0.04}_{-0.04}$&$0.79^{+0.03}_{-0.02}$&$0.27\pm0.09$\\ 
ESO399-IG20&2011 &$0.56^{+0.04}_{-0.04}$&$0.58^{+0.04}_{-0.04}$& & NGC5940&2014 &$0.58^{+0.07}_{-0.06}$&$0.55^{+0.05}_{-0.06}$&$0.49\pm0.1$ \\ 
ESO438-G09&2011 &$0.32^{+0.15}_{-0.03}$&$0.31^{+0.03}_{-0.03}$& & NGC6860&2015 &$0.43^{+0.02}_{-0.03}$&$0.38^{+0.04}_{-0.03}$&$0.46\pm0.15$\\ 
ESO438-G09&2015 &$0.35^{+0.02}_{-0.02}$&$0.3^{+0.04}_{-0.03}$ &$0.48\pm0.11$& NGC7214&2011 &$0.61^{+0.05}_{-0.05}$&$0.5^{+0.09}_{-0.1}$& \\ 
ESO490-IG26&2011 &$0.55^{+0.05}_{-0.04}$&$0.53^{+0.05}_{-0.04}$& & NGC7469&2012 &$0.3^{+0.05}_{-0.03}$ &$0.26^{+0.06}_{-0.01}$&$0.0 \pm0.0$ \\ 
ESO511-G030&2013 &$0.62^{+0.02}_{-0.02}$&$0.54^{+0.02}_{-0.01}$&$0.53\pm0.13$& NGC7603&2014 &$0.2^{+0.0}_{-0.01}$&$0.18^{+0.01}_{-0.02}$&$0.43\pm0.14$\\ 
ESO511-G030&2014 &$0.52^{+0.01}_{-0.01}$&$0.51^{+0.01}_{-0.01}$&$0.37\pm0.17$& NGC985&2014 &$0.78^{+0.05}_{-0.05}$&$0.72^{+0.05}_{-0.04}$&$0.77 \pm0.0$ \\ 
ESO549-G49&2012 &$0.49^{+0.05}_{-0.05}$&$0.41^{+0.12}_{-0.14}$& & PG1149-110&2013 &$0.8^{+0.04}_{-0.09}$ &$0.73^{+0.08}_{-0.05}$&$0.41\pm0.06$\\ 
ESO578-G09&2014 &$0.51^{+0.11}_{-0.07}$&$0.48^{+0.03}_{-0.03}$&$0.53\pm0.12$& PGC50427&2014 &$0.81^{+0.03}_{-0.06}$&$0.74^{+0.04}_{-0.03}$&$0.58\pm0.1$ \\ 
F1041 &2013 &$0.59^{+0.02}_{-0.02}$&$0.6^{+0.02}_{-0.02}$ &$0.46\pm0.1$ & PGC50427&2011 &$0.65^{+0.04}_{-0.04}$&$0.65^{+0.03}_{-0.03}$& \\ 
HE0003-5023&2014 &$0.58^{+0.07}_{-0.25}$&$0.36^{+0.05}_{-0.05}$&$0.56\pm0.05$& PGC64989&2013 &$0.63^{+0.03}_{-0.05}$&$0.6^{+0.04}_{-0.07}$ &$0.34\pm0.07$\\ 
HE1136-2304&2016 &$0.47^{+0.04}_{-0.04}$&$0.38^{+0.09}_{-0.07}$& & PGC64989&2014 &$0.42^{+0.01}_{-0.01}$&$0.42^{+0.01}_{-0.01}$&$0.38\pm0.12$\\ 
HE1136-2304&2018 &$0.62^{+0.11}_{-0.09}$&$0.59^{+0.09}_{-0.07}$& & RXSJ06225-2317&2013 &$0.41^{+0.01}_{-0.01}$&$0.41^{+0.01}_{-0.02}$&$0.41\pm0.15$\\ 
HE1136-2304&2015 &$0.65^{+0.05}_{-0.03}$&$0.69^{+0.04}_{-0.04}$& & RXSJ11032-0654& 2011 & $0.53^{+0.06}_{-0.06}$& $0.50^{+0.09}_{-0.19}$&\\ 
HE1143-1810&2016 &$0.43^{+0.02}_{-0.02}$&$0.43^{+0.02}_{-0.03}$& & RXSJ11032-0654&2014 & $0.52^{+0.20}_{-0.21}$& $0.33^{+0.17}_{-0.06}$&\\ 
HE2128-0221&2016 &$0.44^{+0.23}_{-0.06}$&$0.43^{+0.05}_{-0.05}$& & RXSJ17414+0348&2012 &$0.52^{+0.01}_{-0.01}$&$0.44^{+0.02}_{-0.02}$& \\ 
IC4329A&2015 &$0.52^{+0.03}_{-0.12}$&$0.38^{+0.02}_{-0.02}$&$0.7\pm0.07$ & RXSJ17414+0348&2014 &$0.53^{+0.06}_{-0.02}$&$0.57^{+0.03}_{-0.04}$& 0.60\\ 
IRAS01089-4743&2013 &$0.45^{+0.04}_{-0.03}$&$0.36^{+0.13}_{-0.12}$& 0.30& UGC12138&2012 &$0.61^{+0.03}_{-0.03}$&$0.64^{+0.05}_{-0.04}$& \\ 
IRAS09595-0755&2013 &$0.84^{+0.01}_{-0.02}$&$0.81^{+0.03}_{-0.05}$&$0.5\pm0.08$ & UM163 &2013 &$0.84^{+0.01}_{-0.01}$&$0.79^{+0.02}_{-0.02}$&$0.53\pm0.13$\\ 
IRAS23226-3843&2013 &$0.47^{+0.1}_{-0.04}$ &$0.27^{+0.11}_{-0.02}$& 0.25 & WPVS00&2012 &$0.5^{+0.03}_{-0.03}$ &$0.48^{+0.03}_{-0.03}$& \\ 
MCG+03.47-002&2013 &$0.46^{+0.01}_{-0.01}$&$0.45^{+0.01}_{-0.01}$& 0.44 & WPVS48&2014 &$0.48^{+0.02}_{-0.02}$&$0.46^{+0.02}_{-0.02}$&$0.71\pm0.04$\\ 
MCG-02.12.050&2014 &$0.54^{+0.03}_{-0.04}$&$0.53^{+0.04}_{-0.04}$&$0.43\pm0.12$& WPVS48&2013 &$0.45^{+0.01}_{-0.01}$&$0.42^{+0.01}_{-0.01}$&$0.73\pm0.03$\\ 
MRK1239&2015 &$0.79^{+0.05}_{-0.11}$&$0.79^{+0.04}_{-0.13}$&$0.73\pm0.03$& WPVS48&2018 &$0.52^{+0.02}_{-0.02}$&$0.52^{+0.02}_{-0.02}$& \\ 
MRK1347&2014 &$0.47^{+0.18}_{-0.05}$&$0.39^{+0.07}_{-0.07}$&$0.49\pm0.06$& \\
\enddata
\end{deluxetable}

\includegraphics[width=0.49\columnwidth]{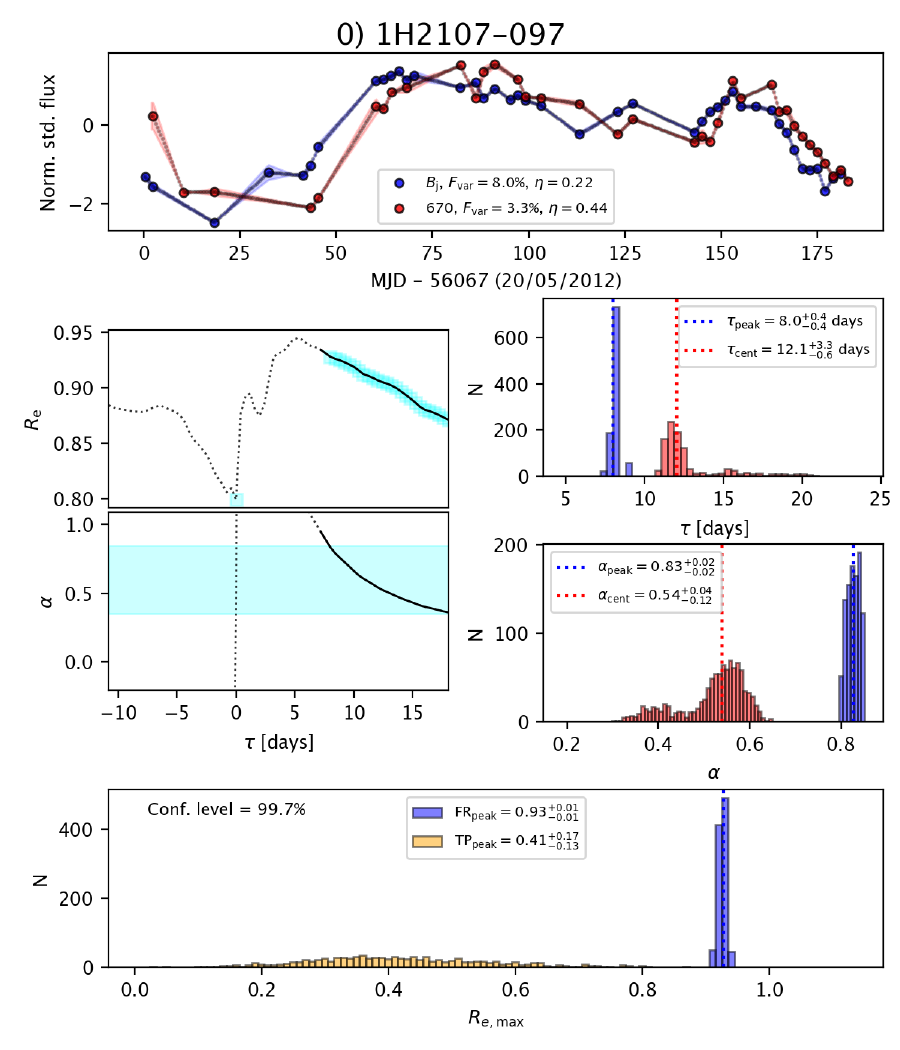}
\includegraphics[width=0.49\columnwidth]{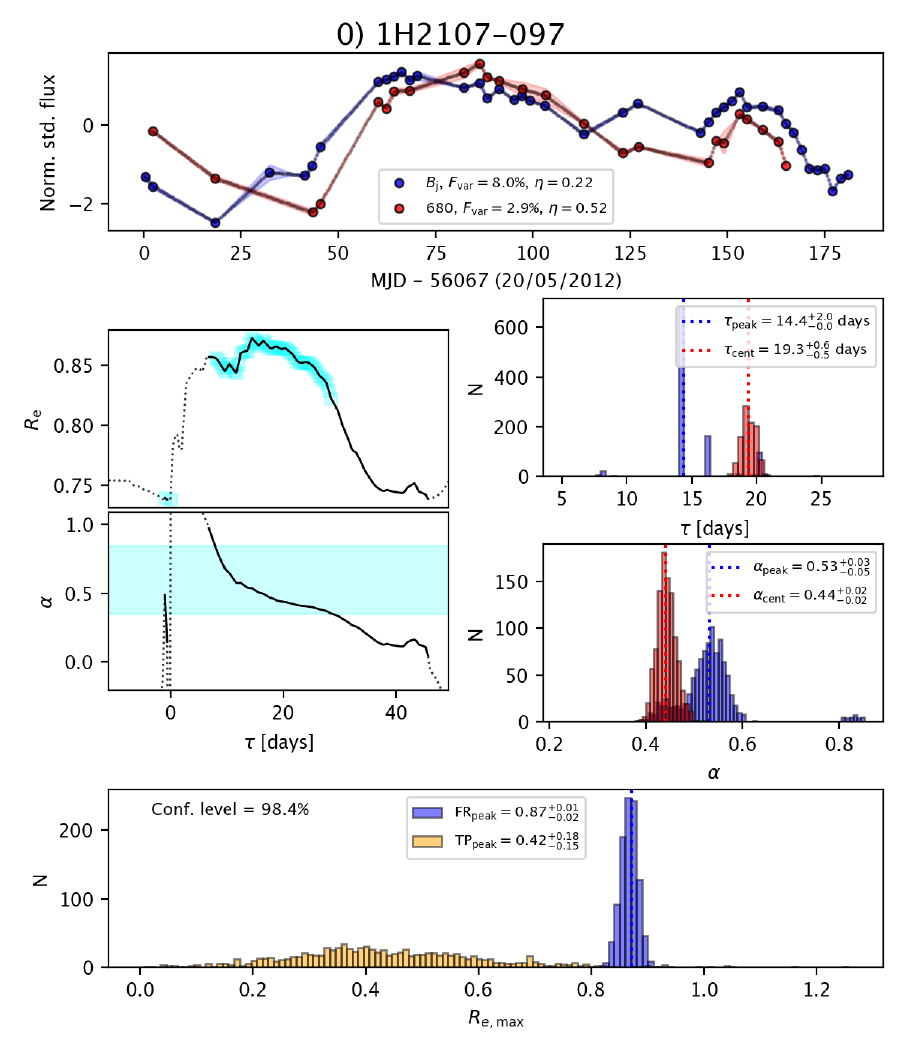}
\includegraphics[width=0.49\columnwidth]{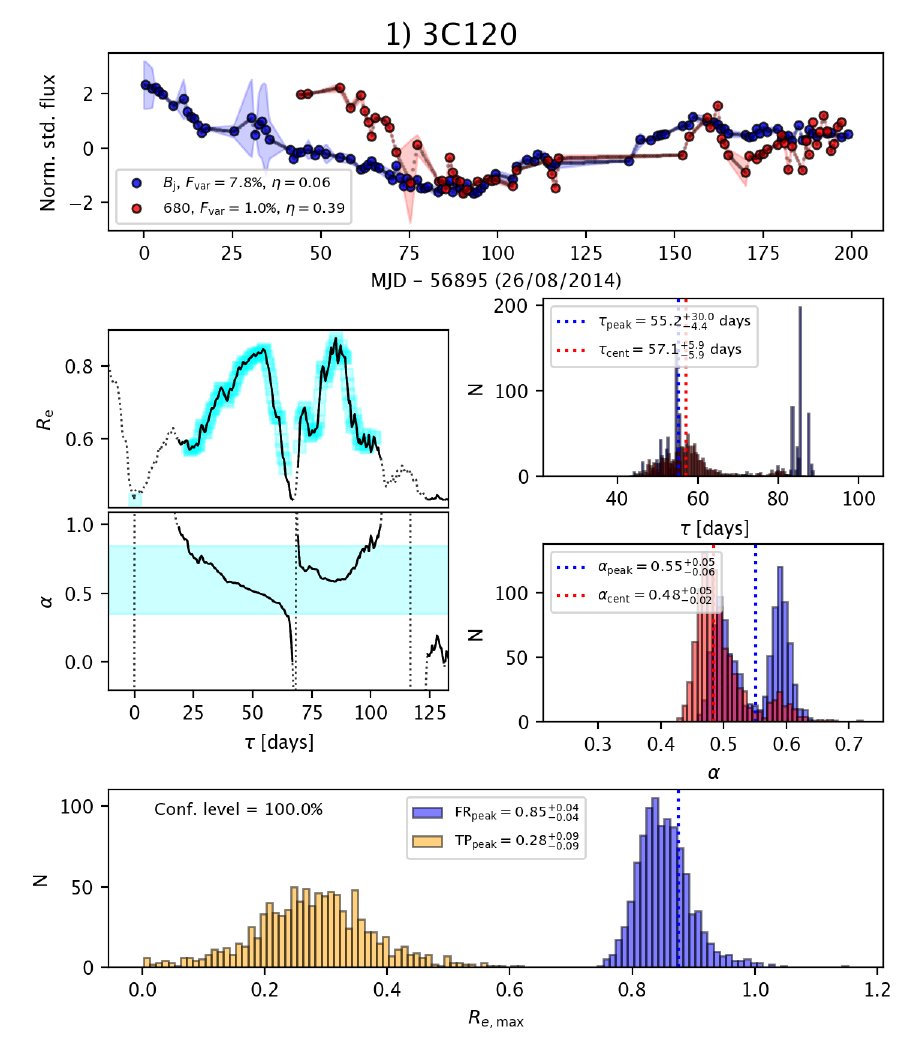}
\includegraphics[width=0.49\columnwidth]{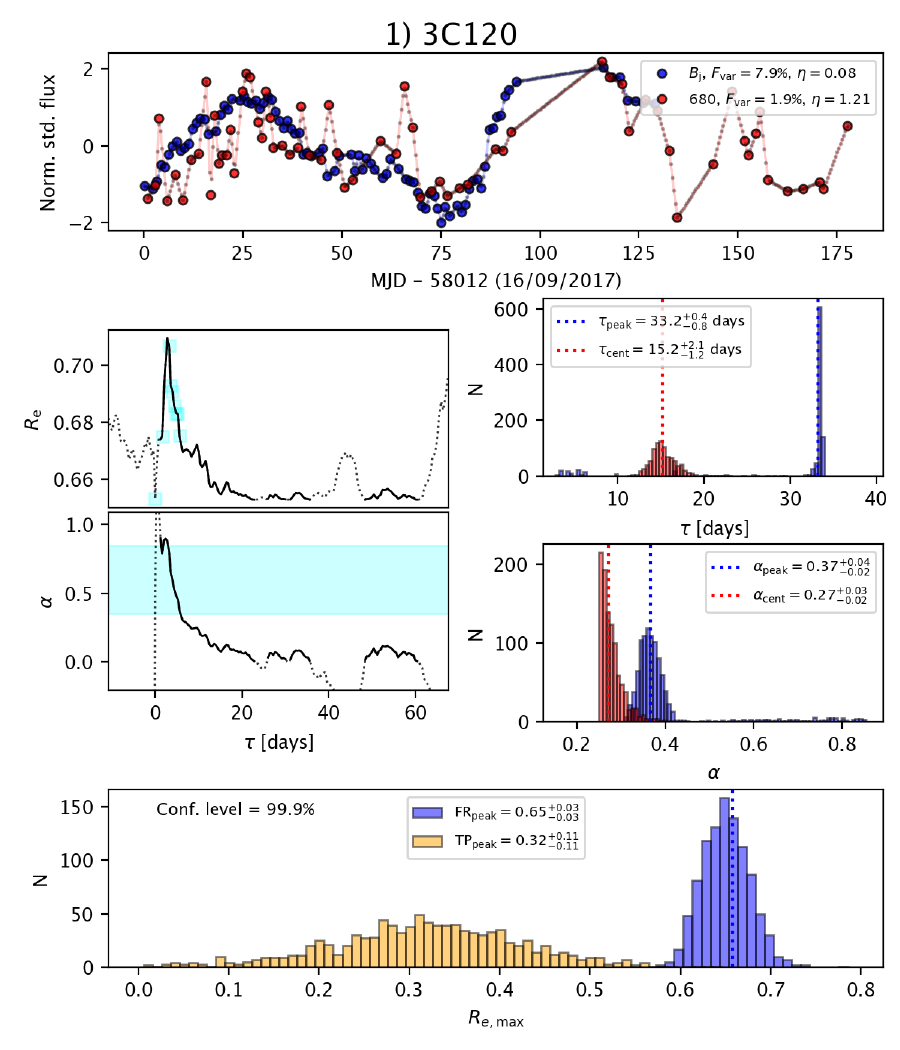}
\includegraphics[width=0.49\columnwidth]{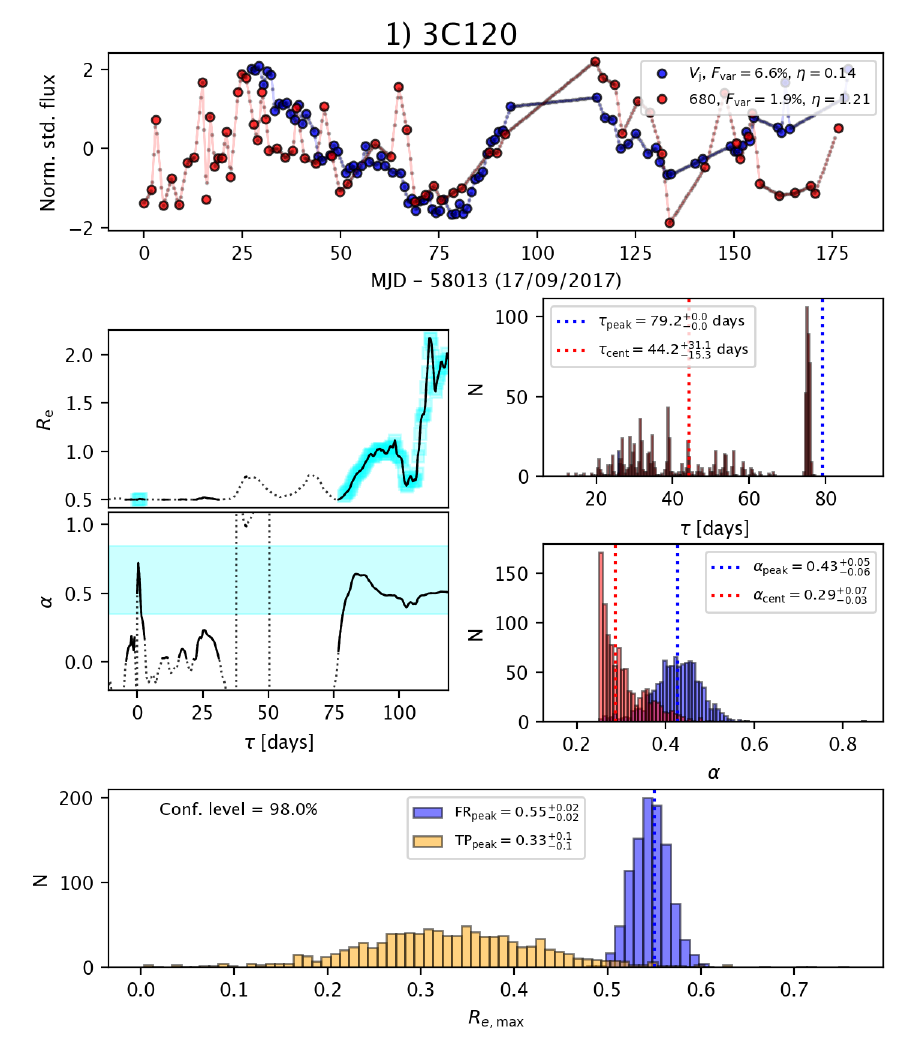}
\includegraphics[width=0.49\columnwidth]{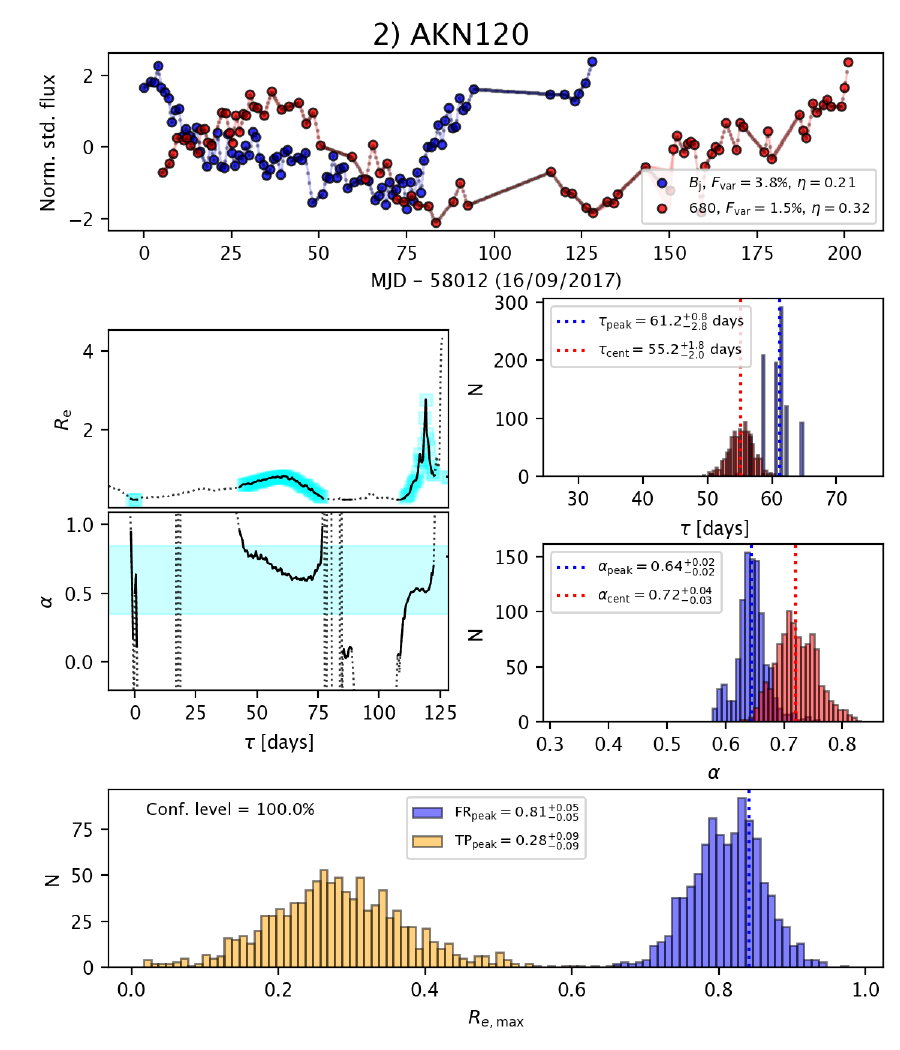}
\includegraphics[width=0.49\columnwidth]{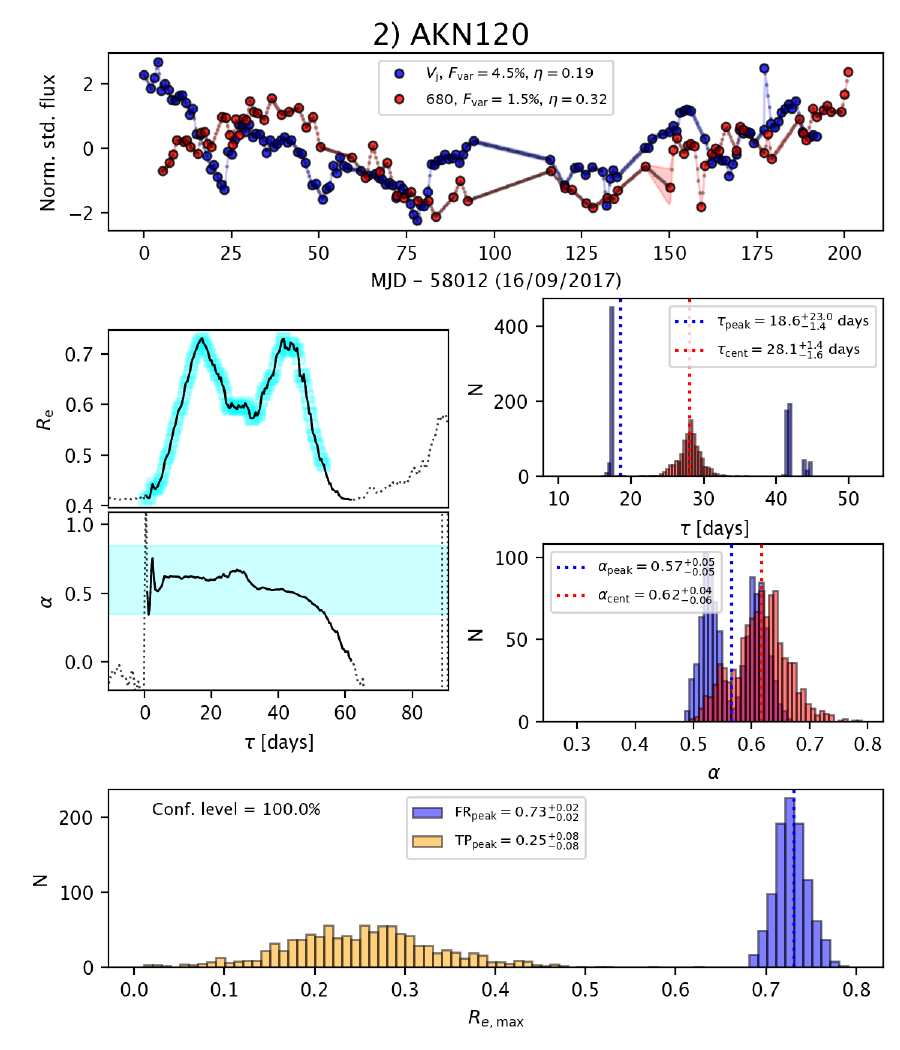}
\includegraphics[width=0.49\columnwidth]{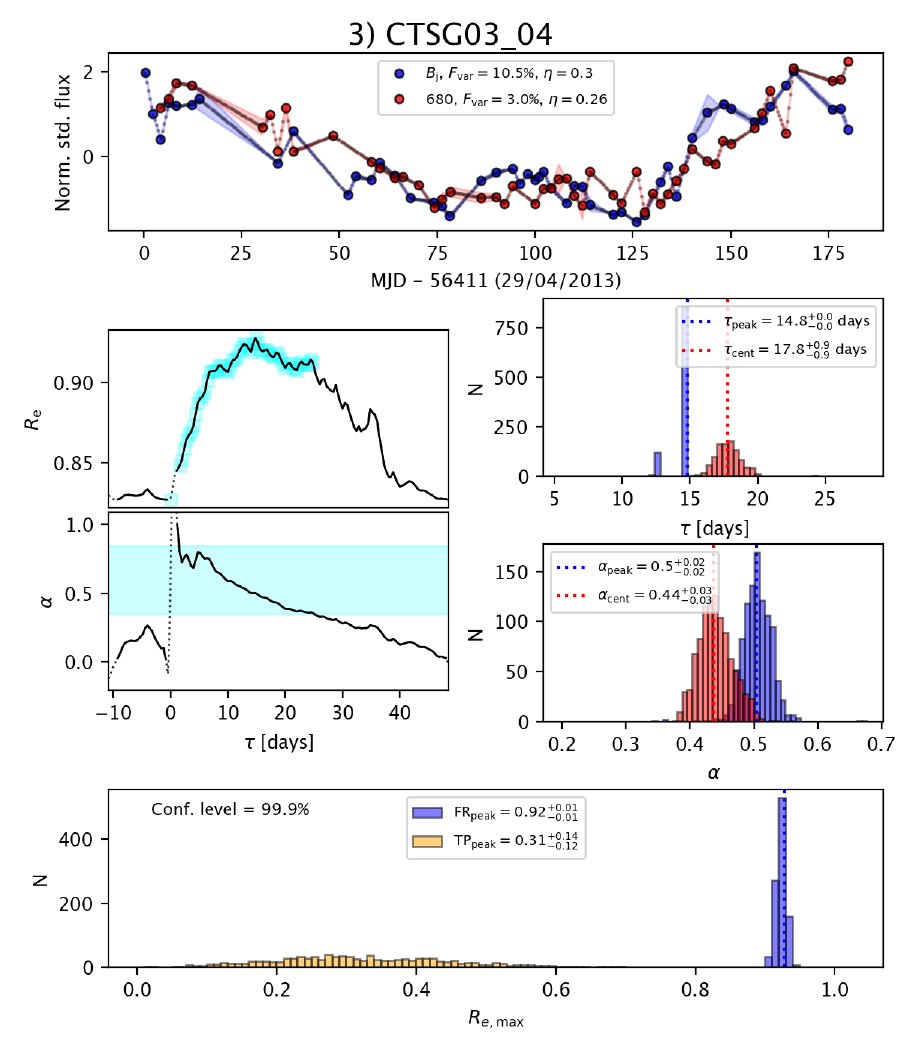}
\includegraphics[width=0.49\columnwidth]{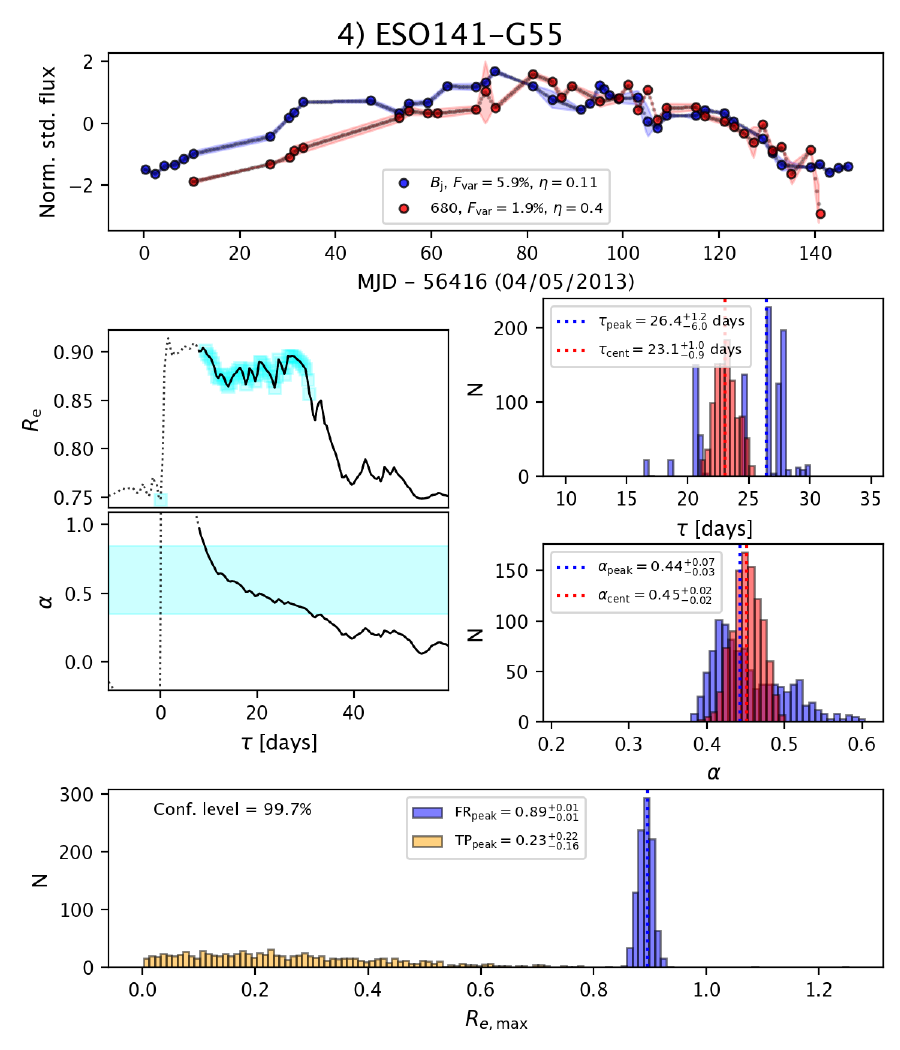}
\includegraphics[width=0.49\columnwidth]{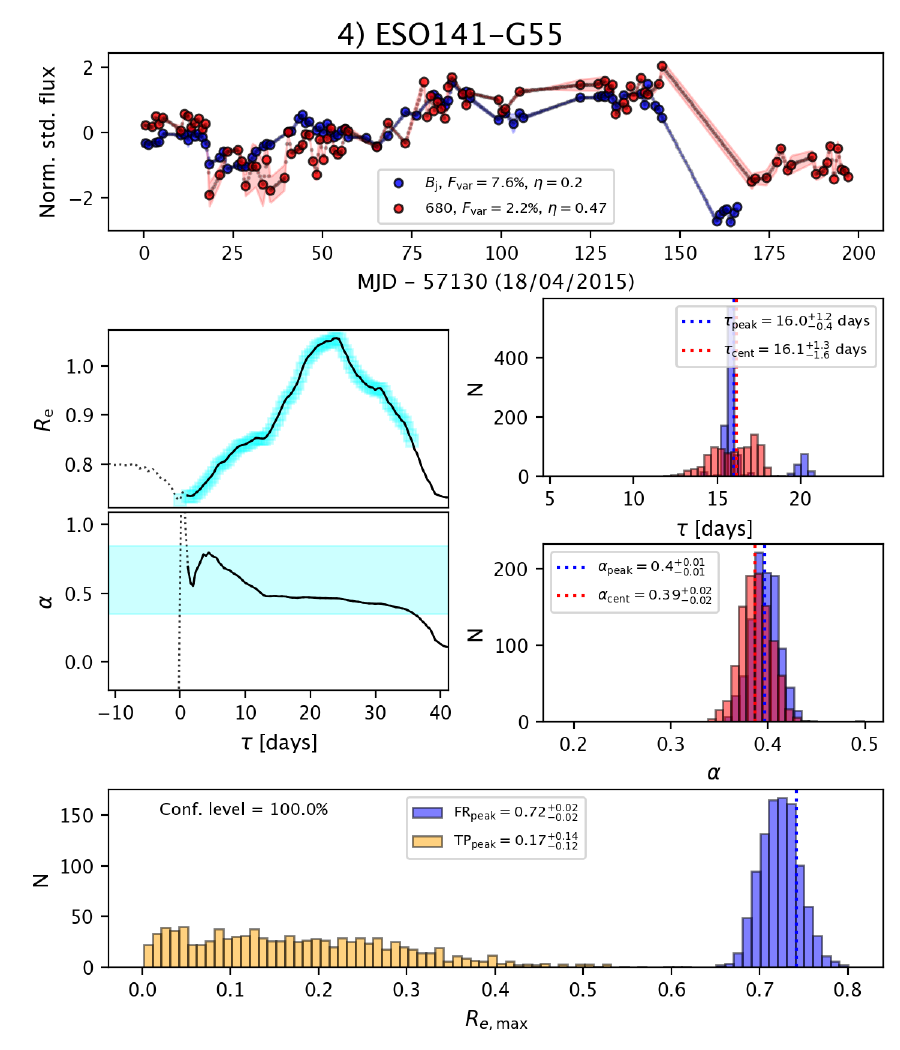}
\includegraphics[width=0.49\columnwidth]{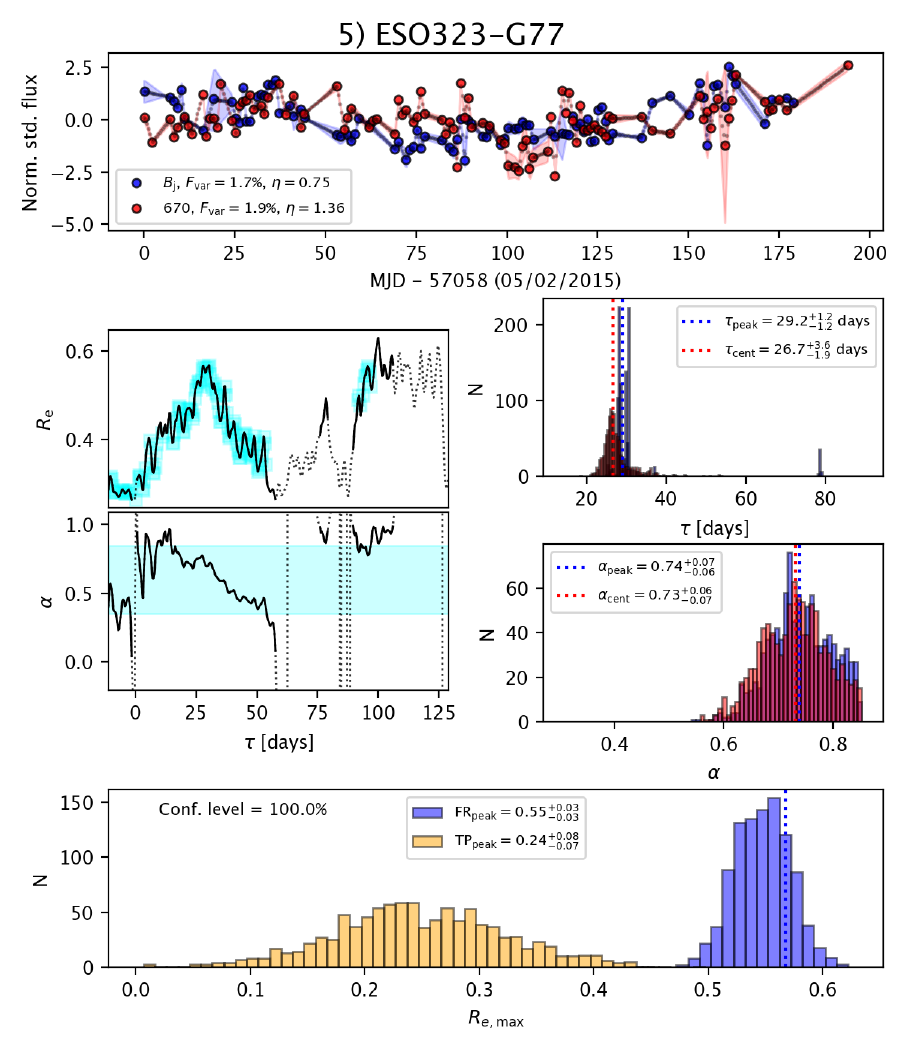}
\includegraphics[width=0.49\columnwidth]{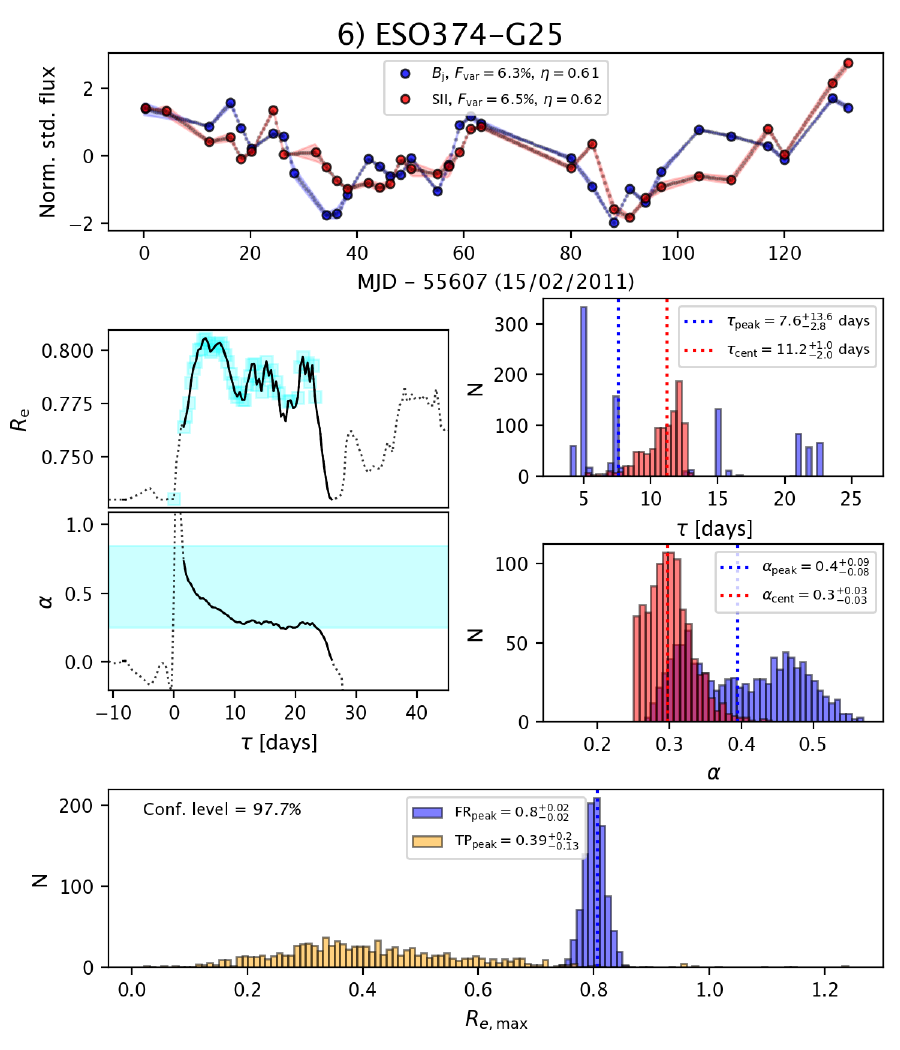}
\includegraphics[width=0.49\columnwidth]{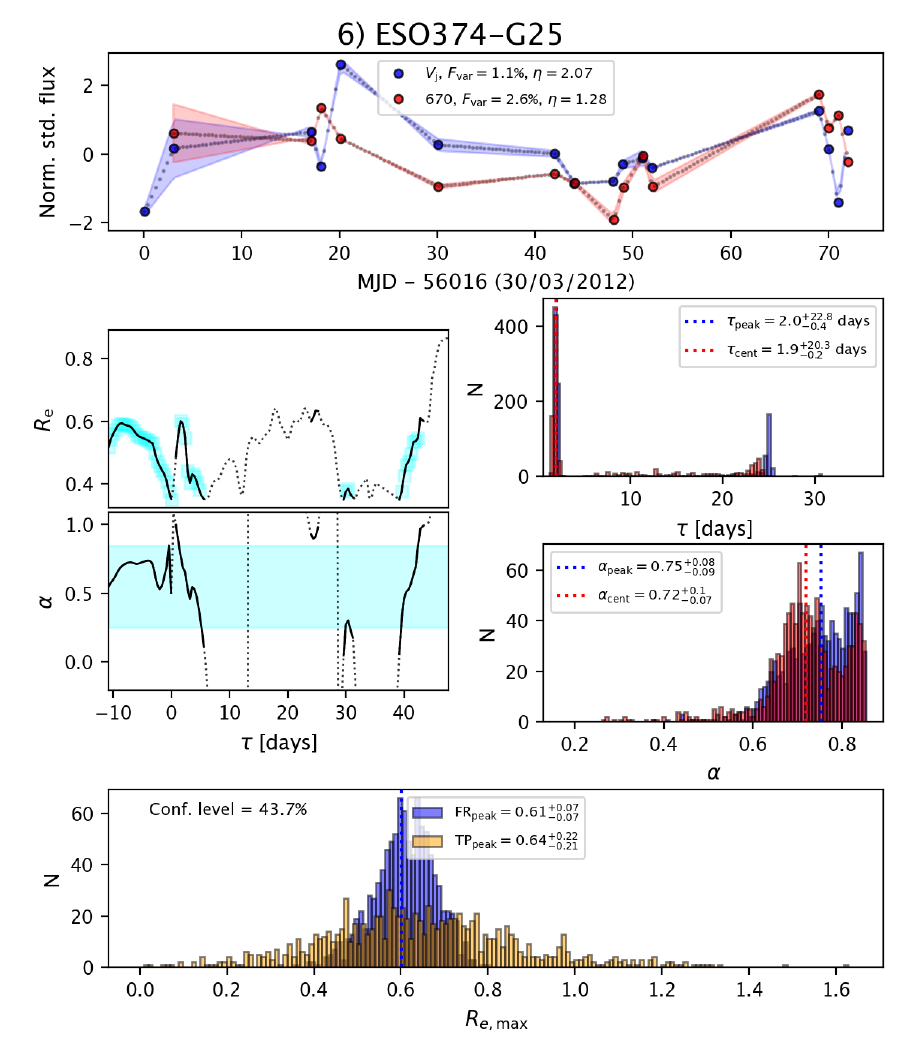}
\includegraphics[width=0.49\columnwidth]{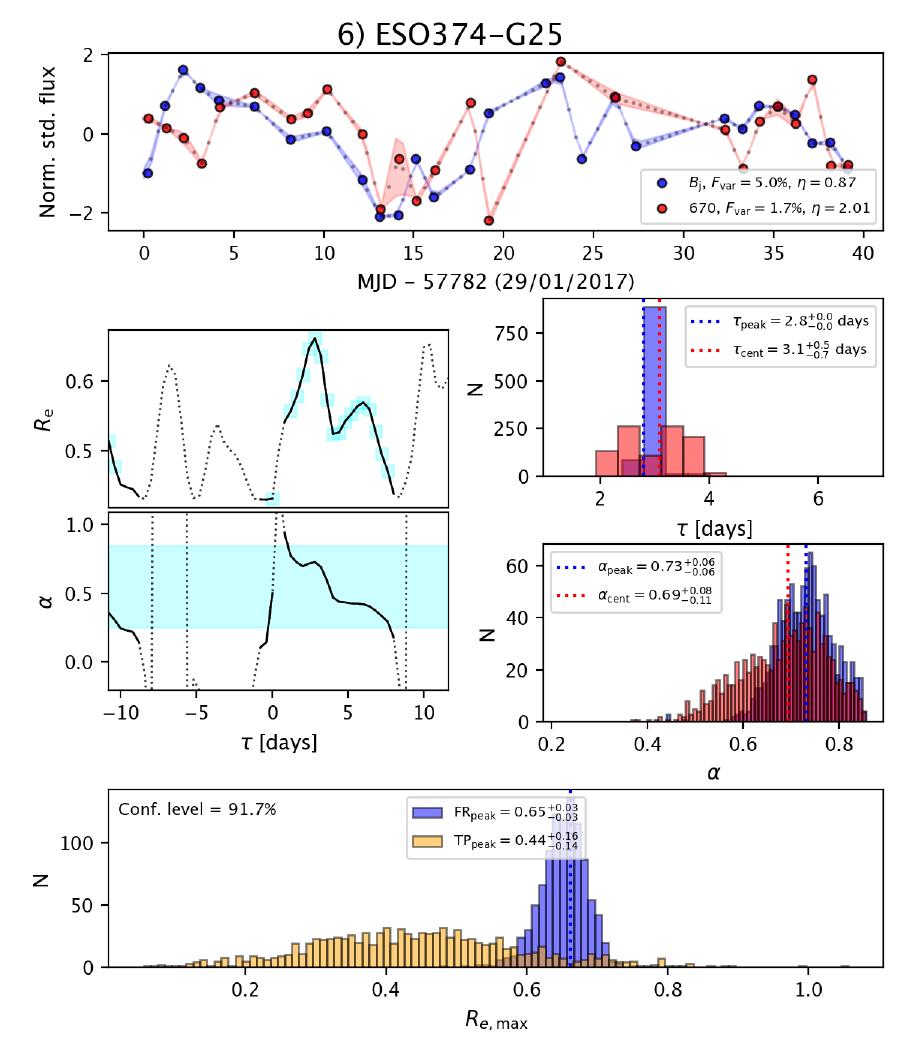}
\includegraphics[width=0.49\columnwidth]{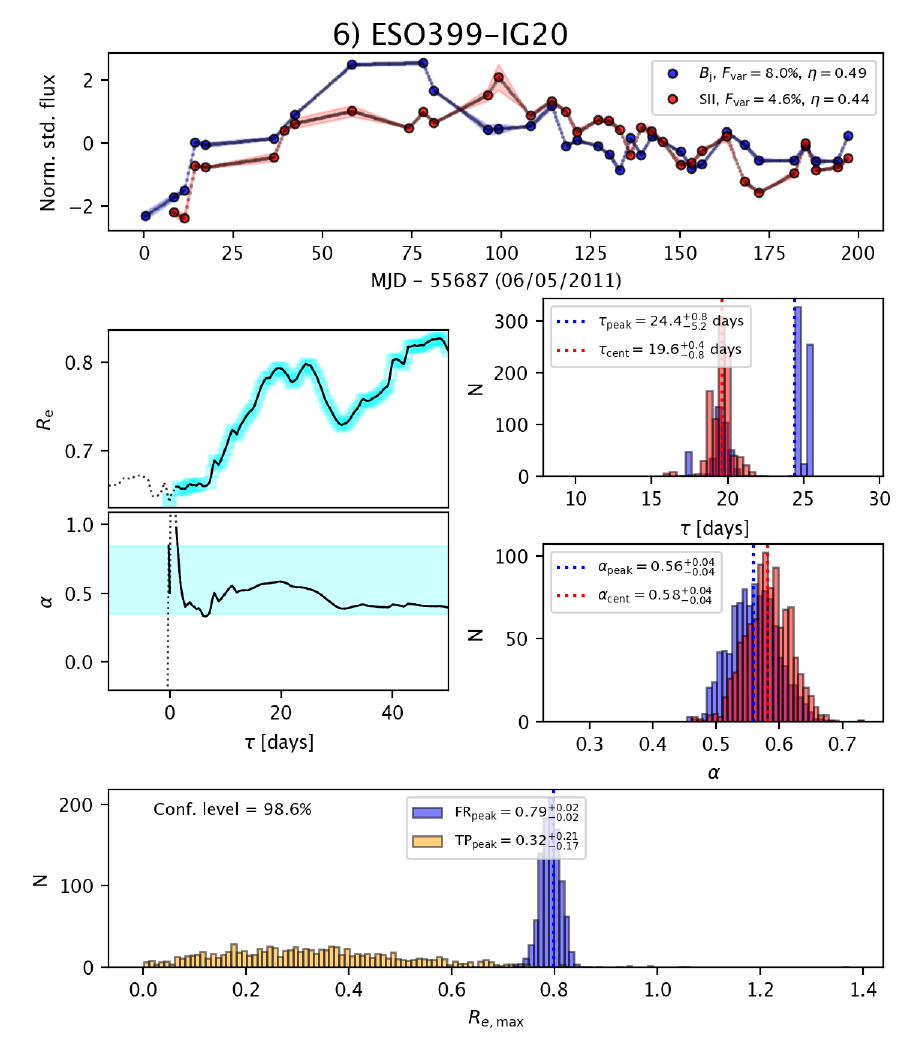}
\includegraphics[width=0.49\columnwidth]{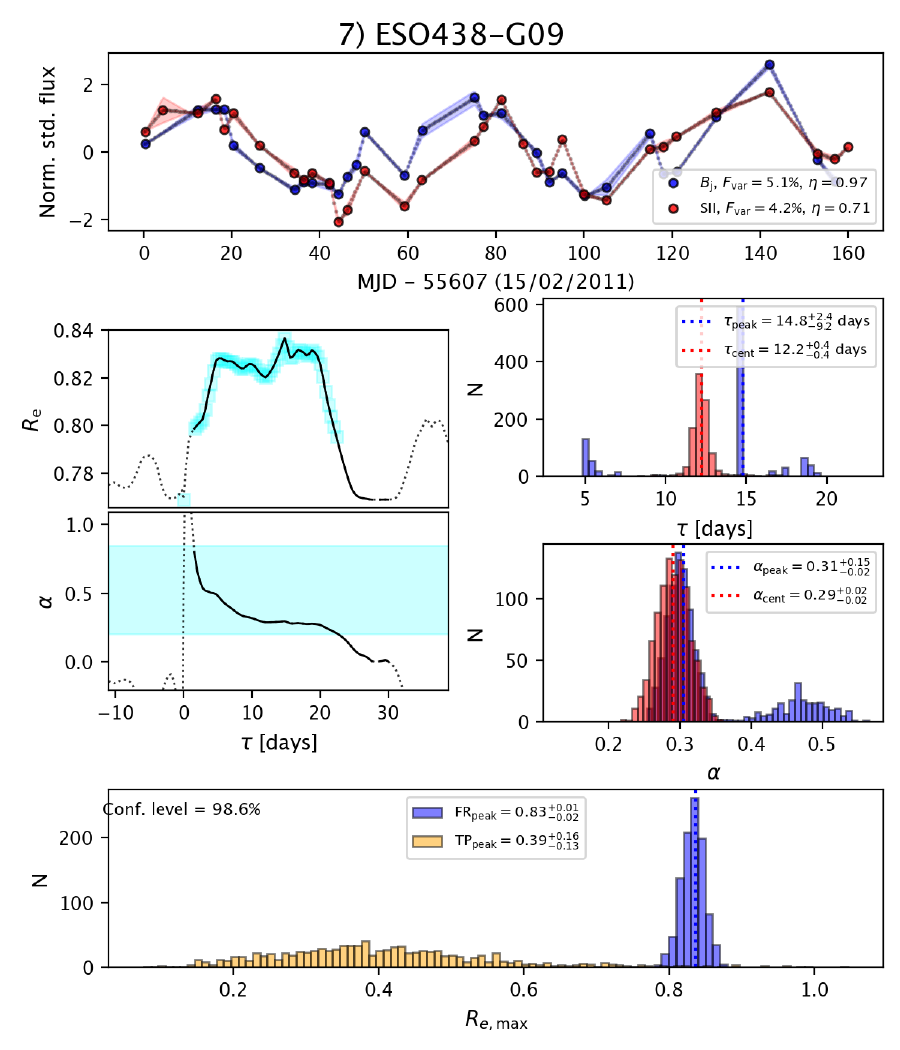}
\includegraphics[width=0.49\columnwidth]{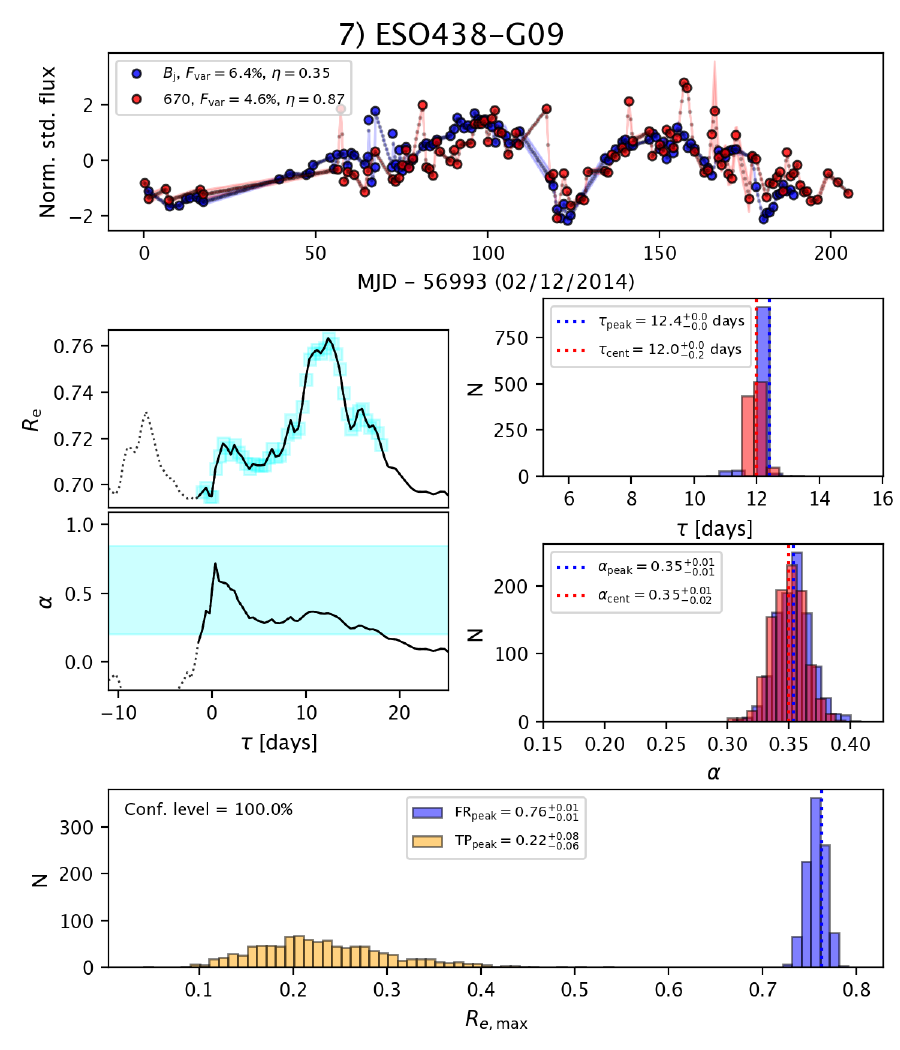}
\includegraphics[width=0.49\columnwidth]{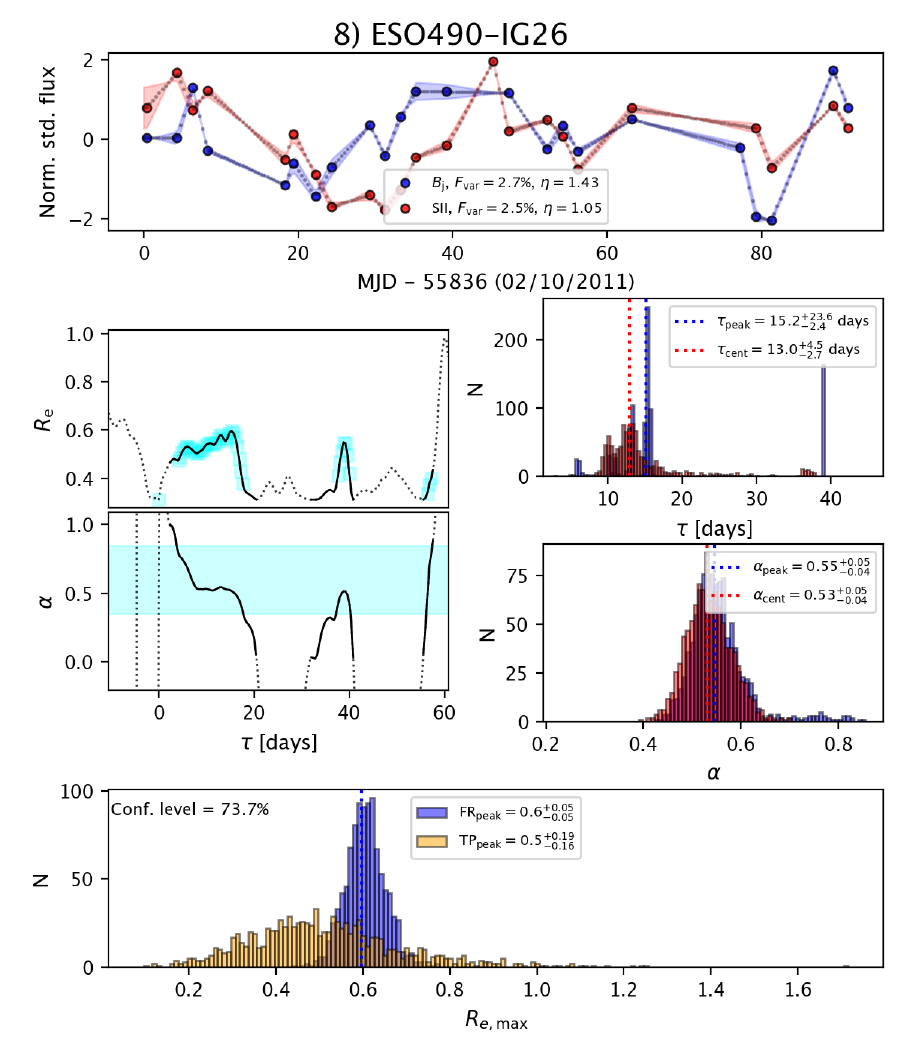}
\includegraphics[width=0.49\columnwidth]{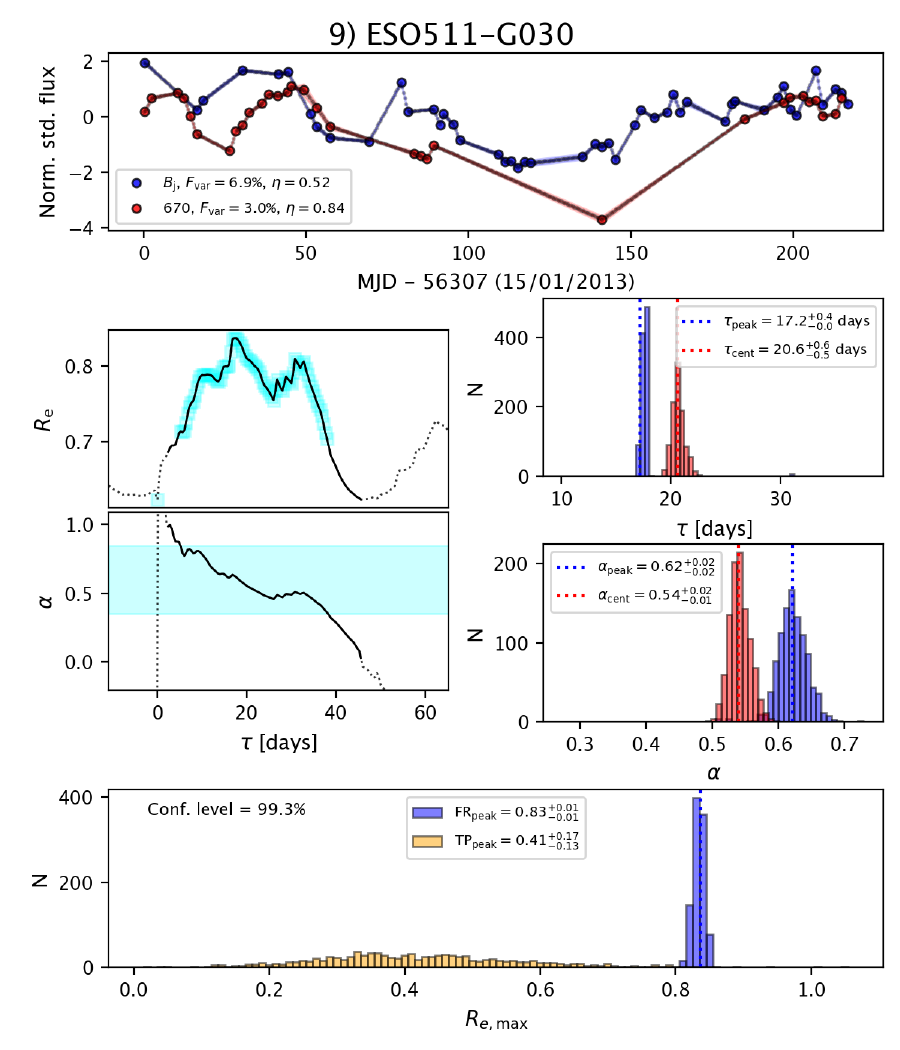}
\includegraphics[width=0.49\columnwidth]{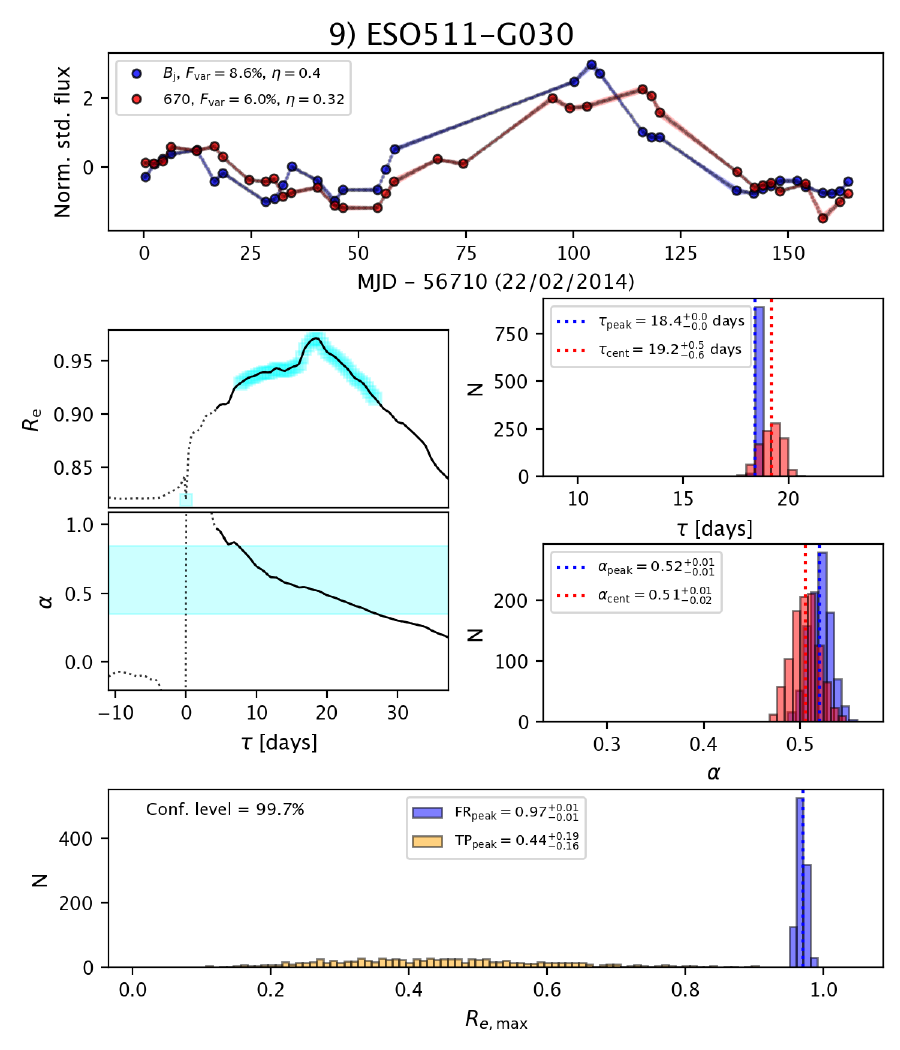}
\includegraphics[width=0.49\columnwidth]{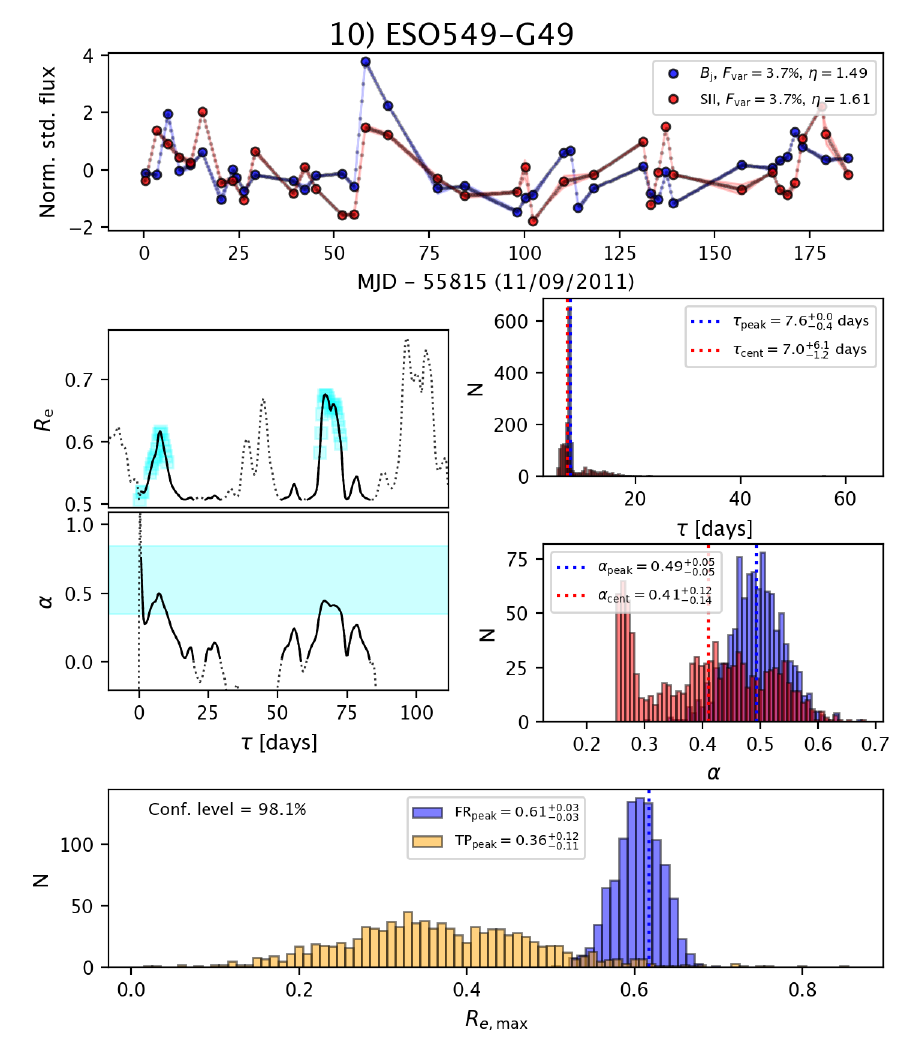}
\includegraphics[width=0.49\columnwidth]{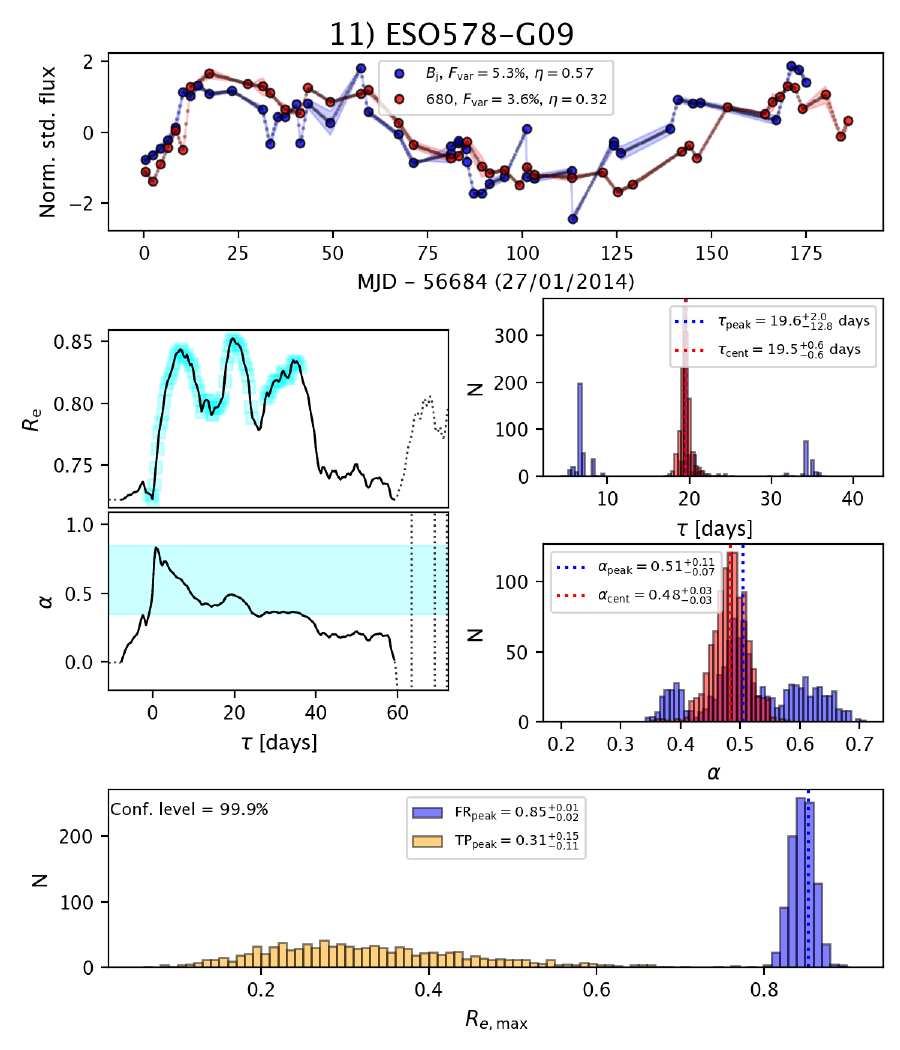}
\includegraphics[width=0.49\columnwidth]{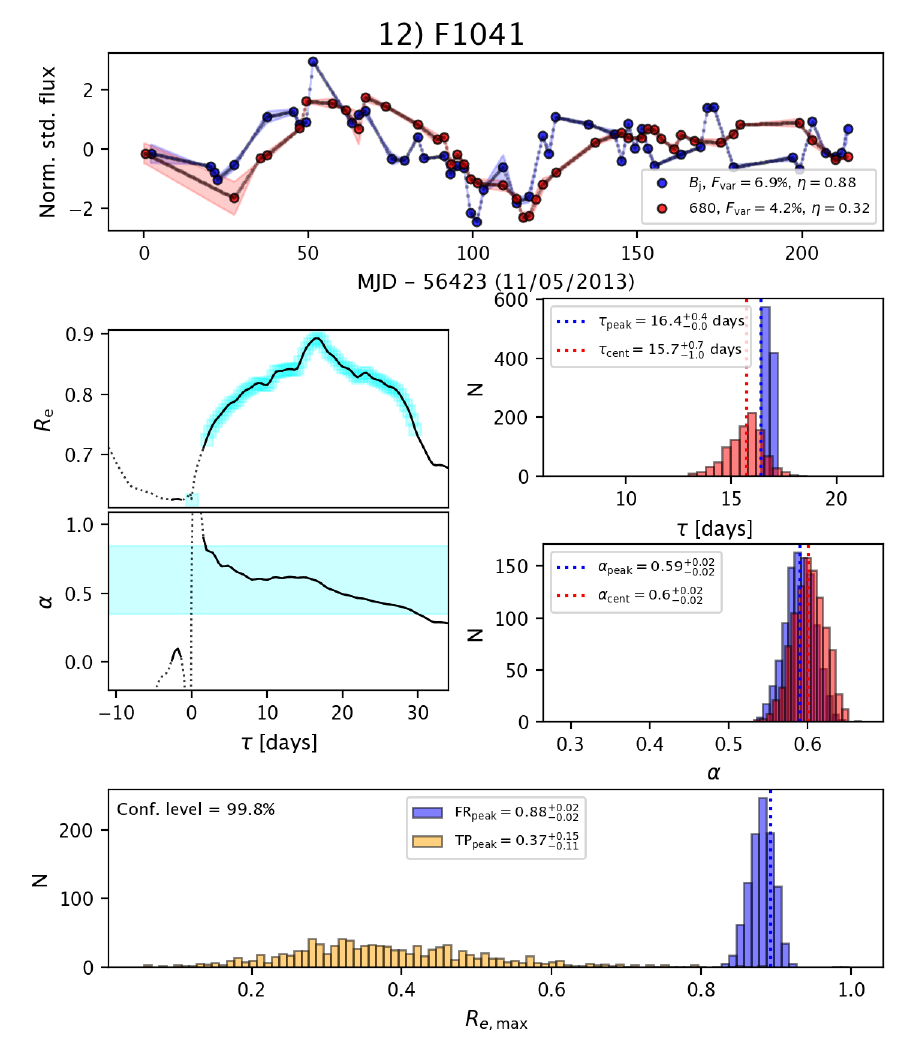}
\includegraphics[width=0.49\columnwidth]{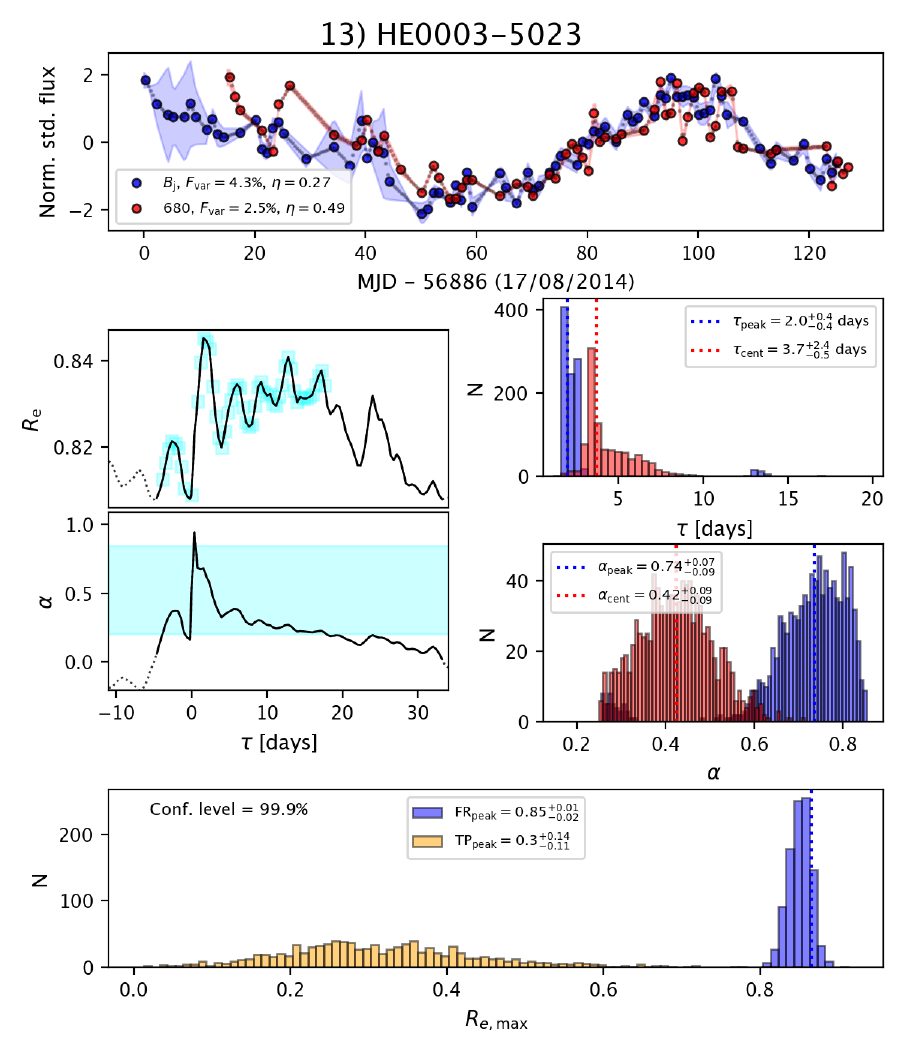}
\includegraphics[width=0.49\columnwidth]{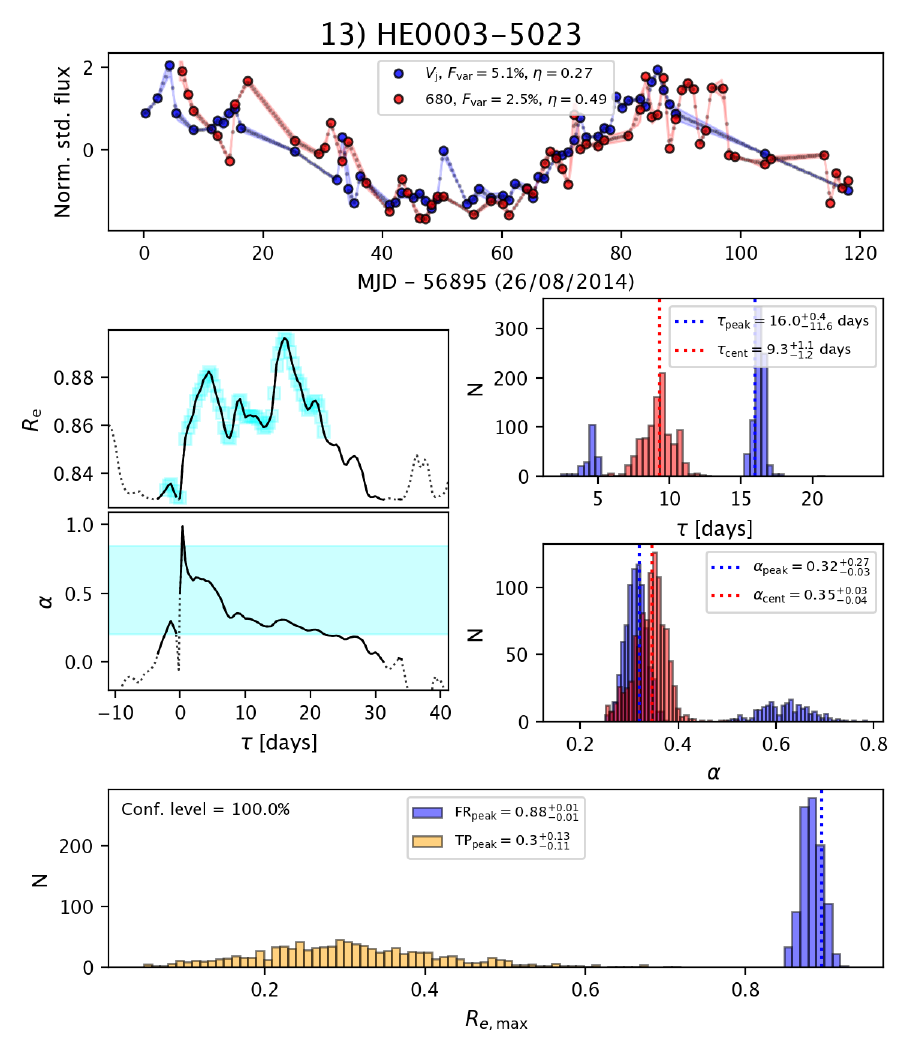}
\includegraphics[width=0.49\columnwidth]{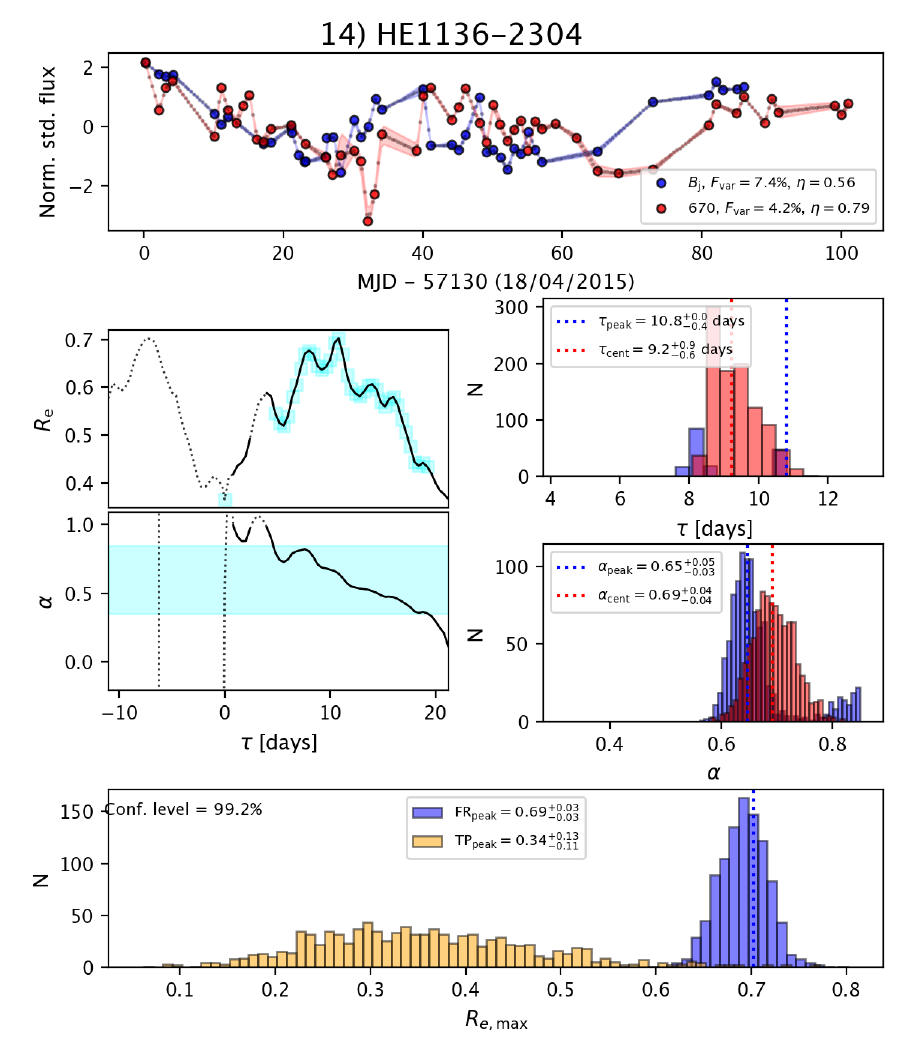}
\includegraphics[width=0.49\columnwidth]{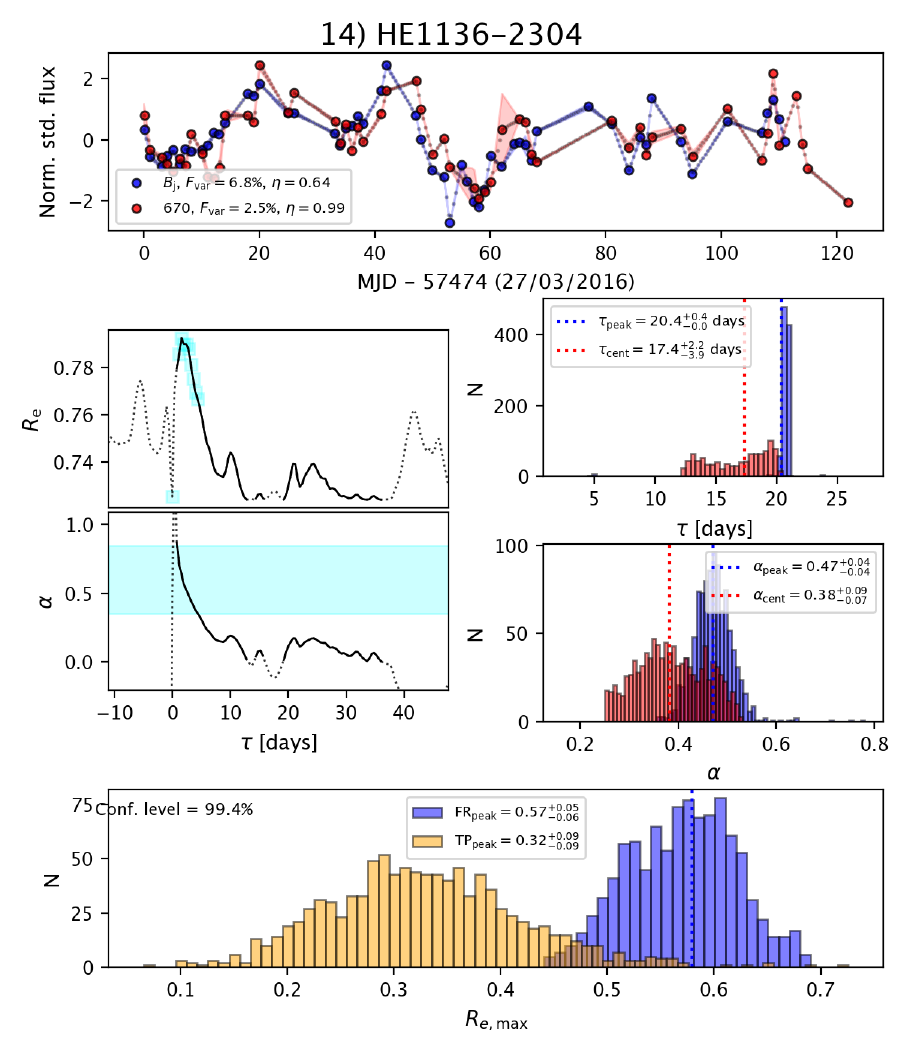}
\includegraphics[width=0.49\columnwidth]{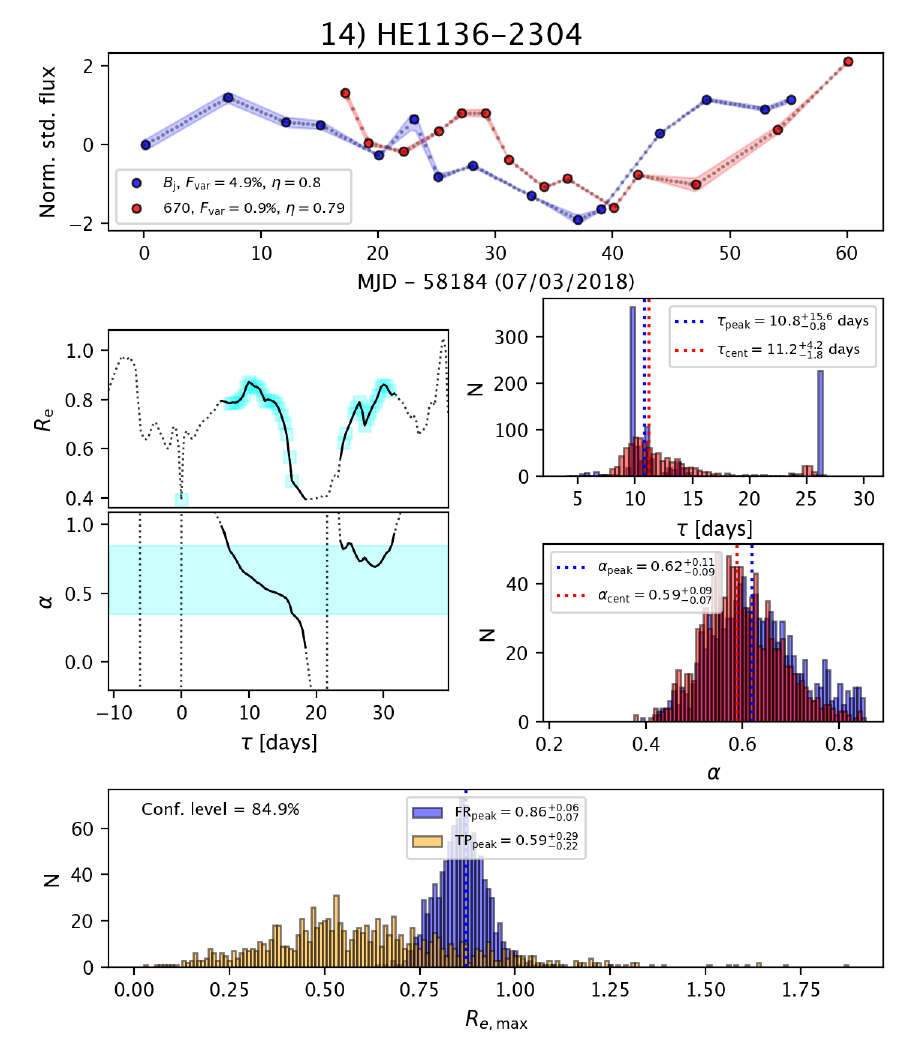}
\includegraphics[width=0.49\columnwidth]{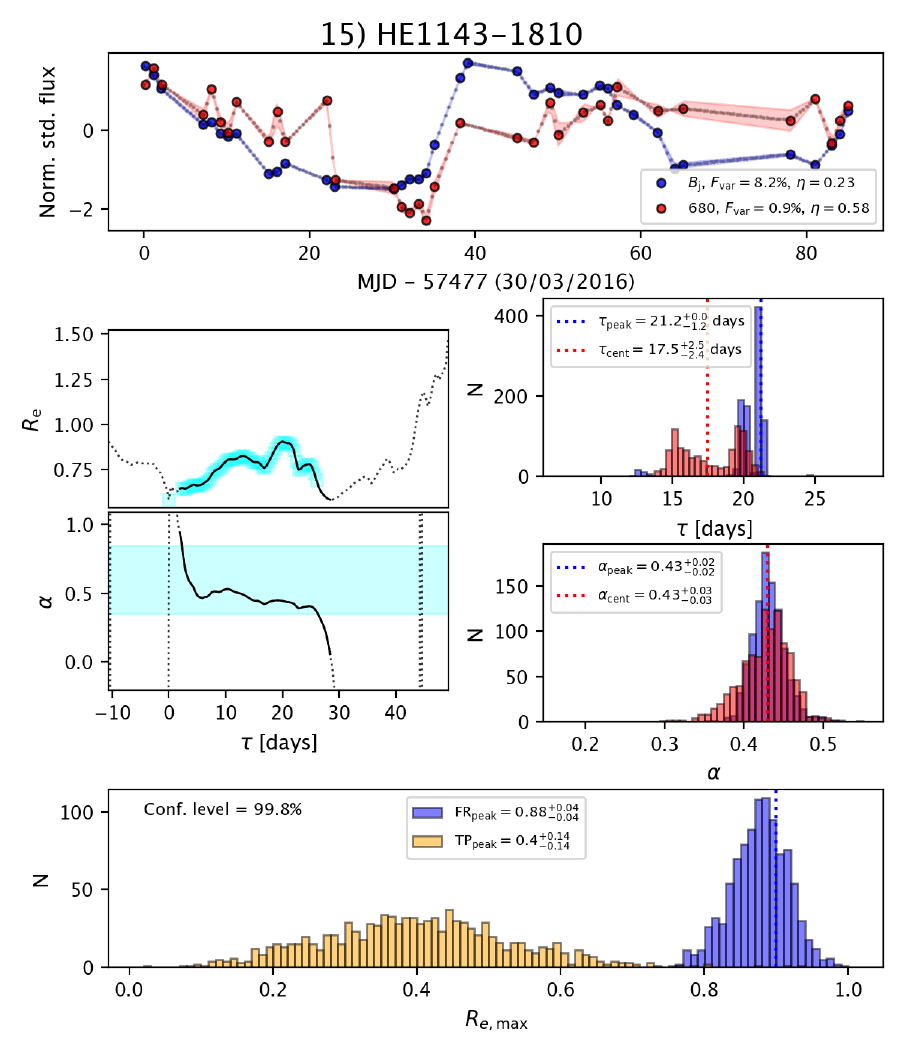}
\includegraphics[width=0.49\columnwidth]{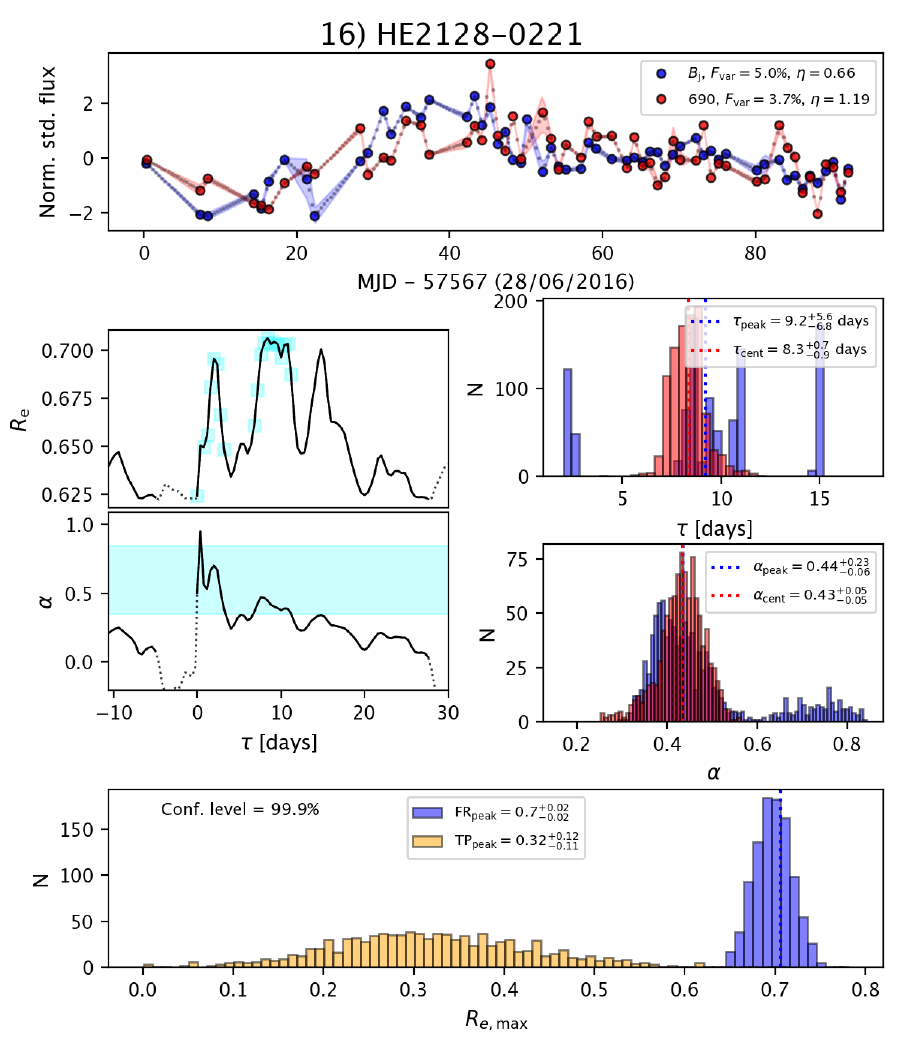}
\includegraphics[width=0.49\columnwidth]{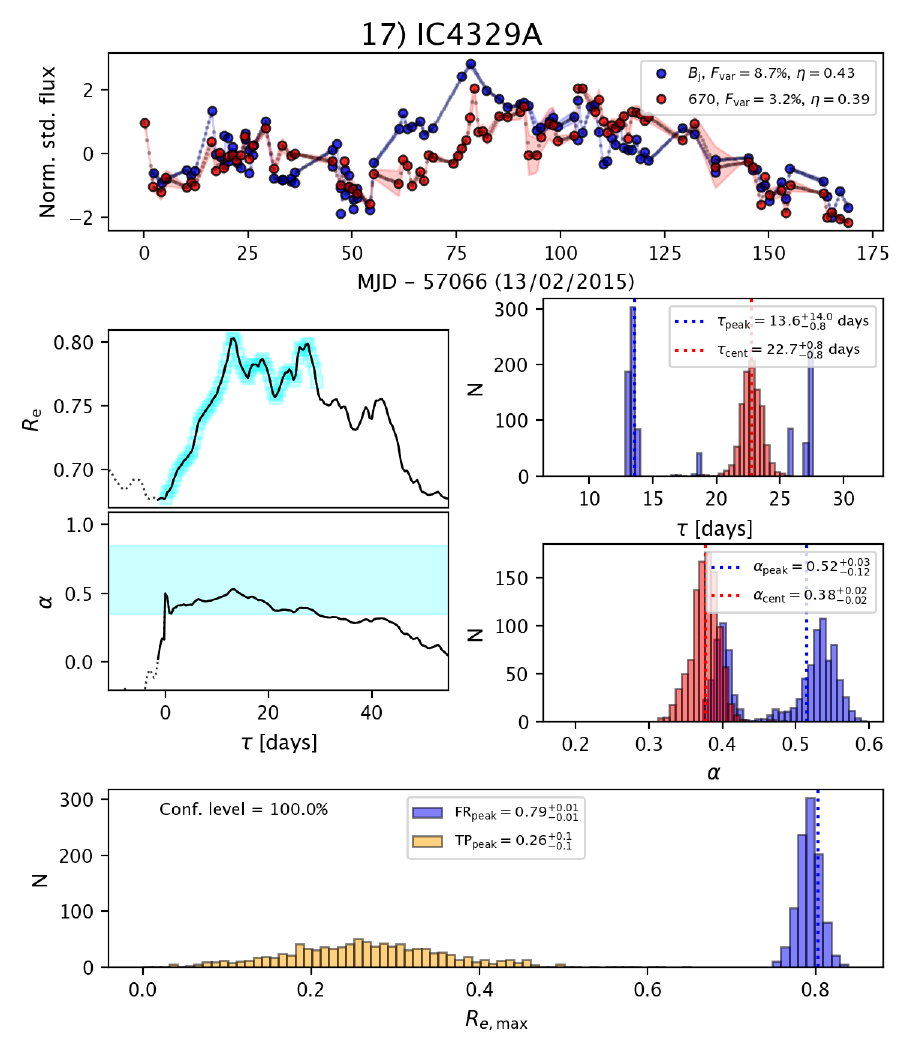}
\includegraphics[width=0.49\columnwidth]{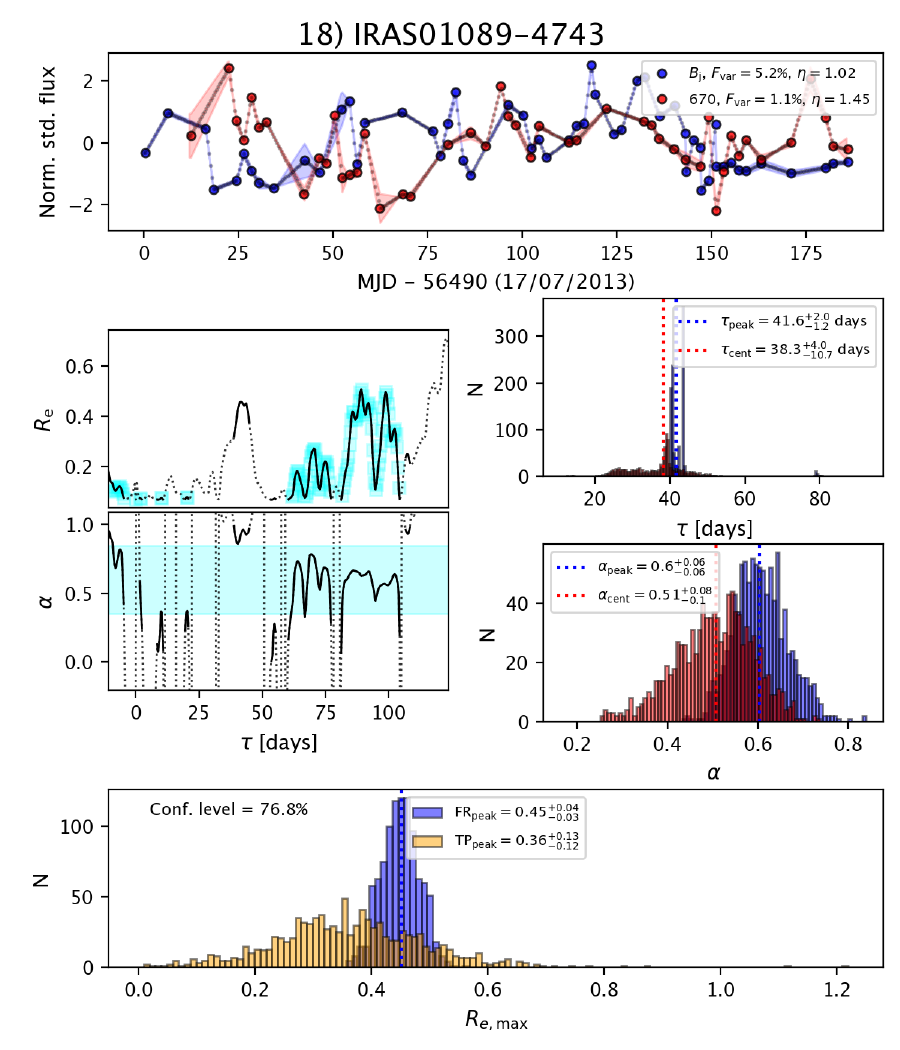}
\includegraphics[width=0.49\columnwidth]{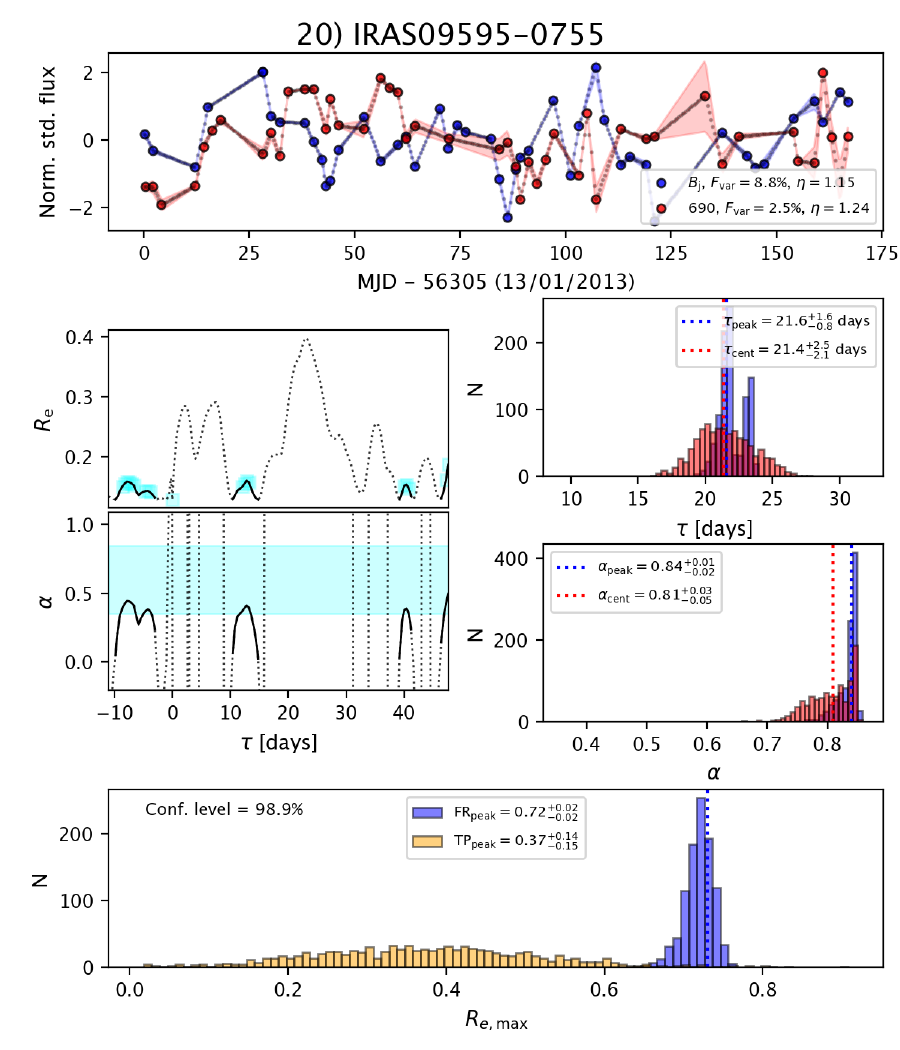}
\includegraphics[width=0.49\columnwidth]{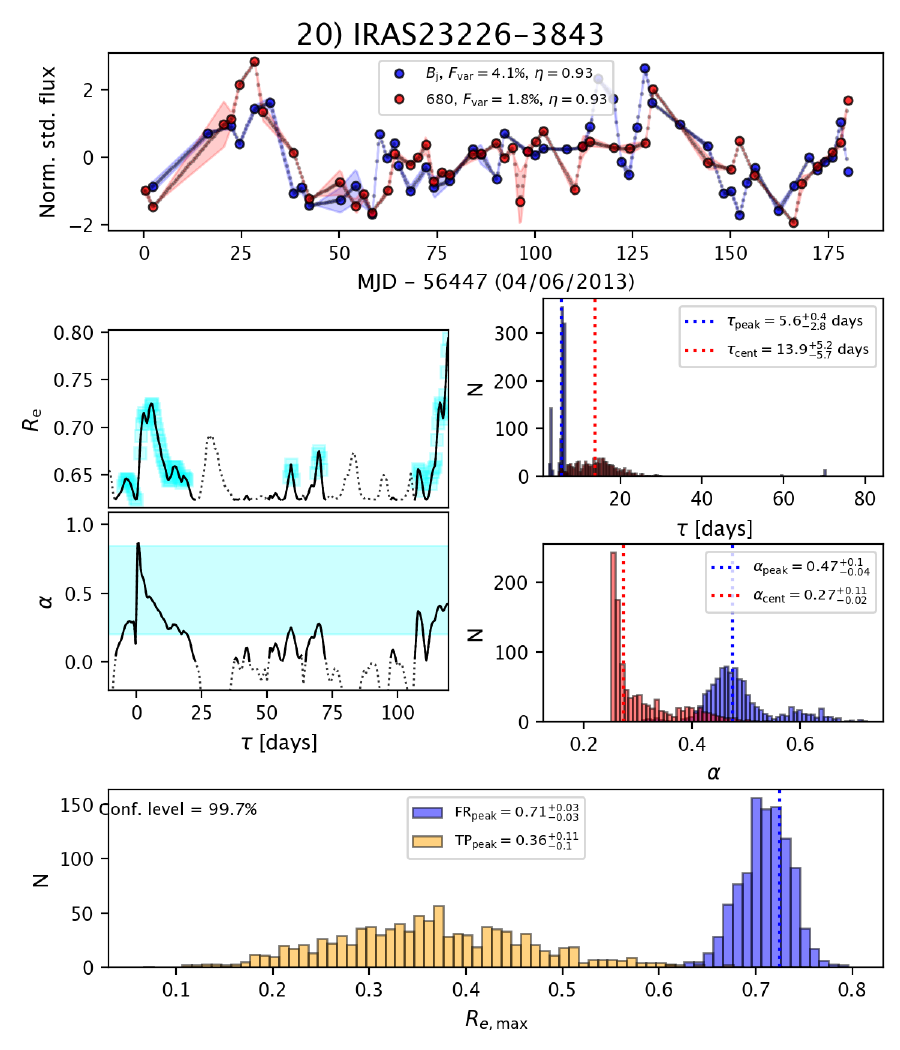}
\includegraphics[width=0.49\columnwidth]{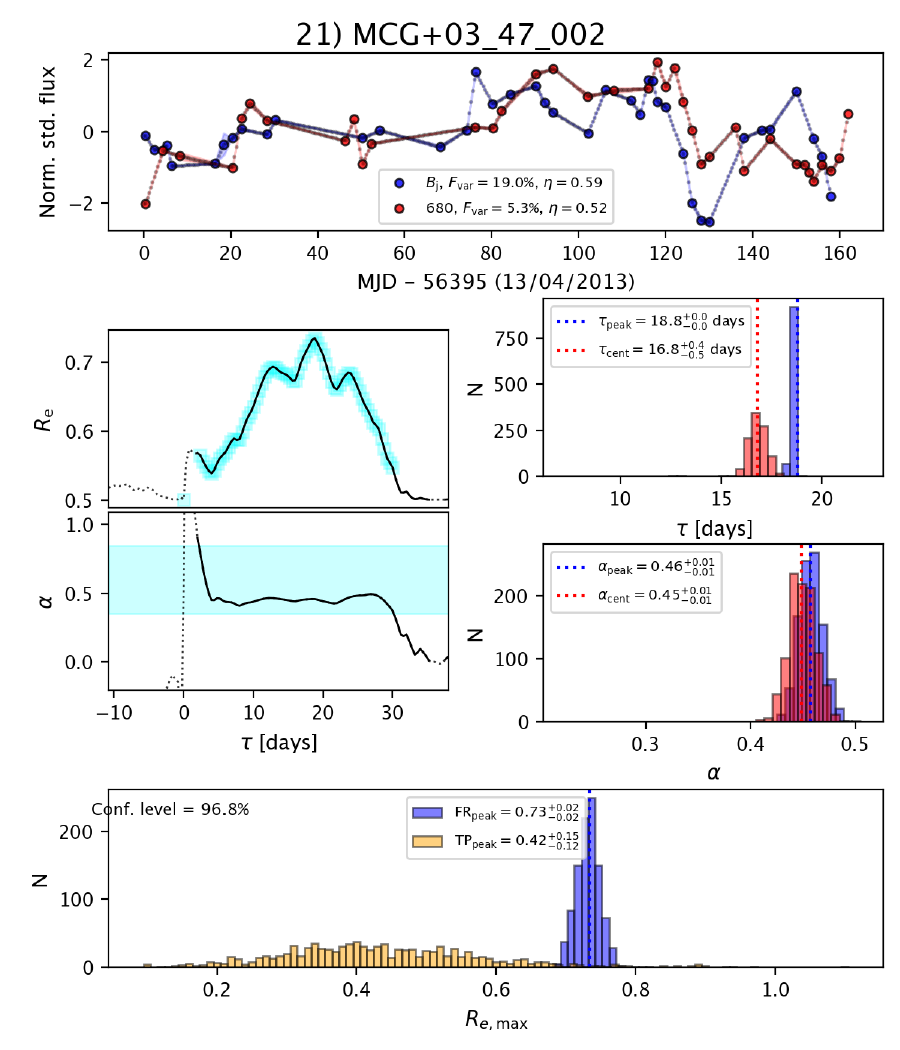}
\includegraphics[width=0.49\columnwidth]{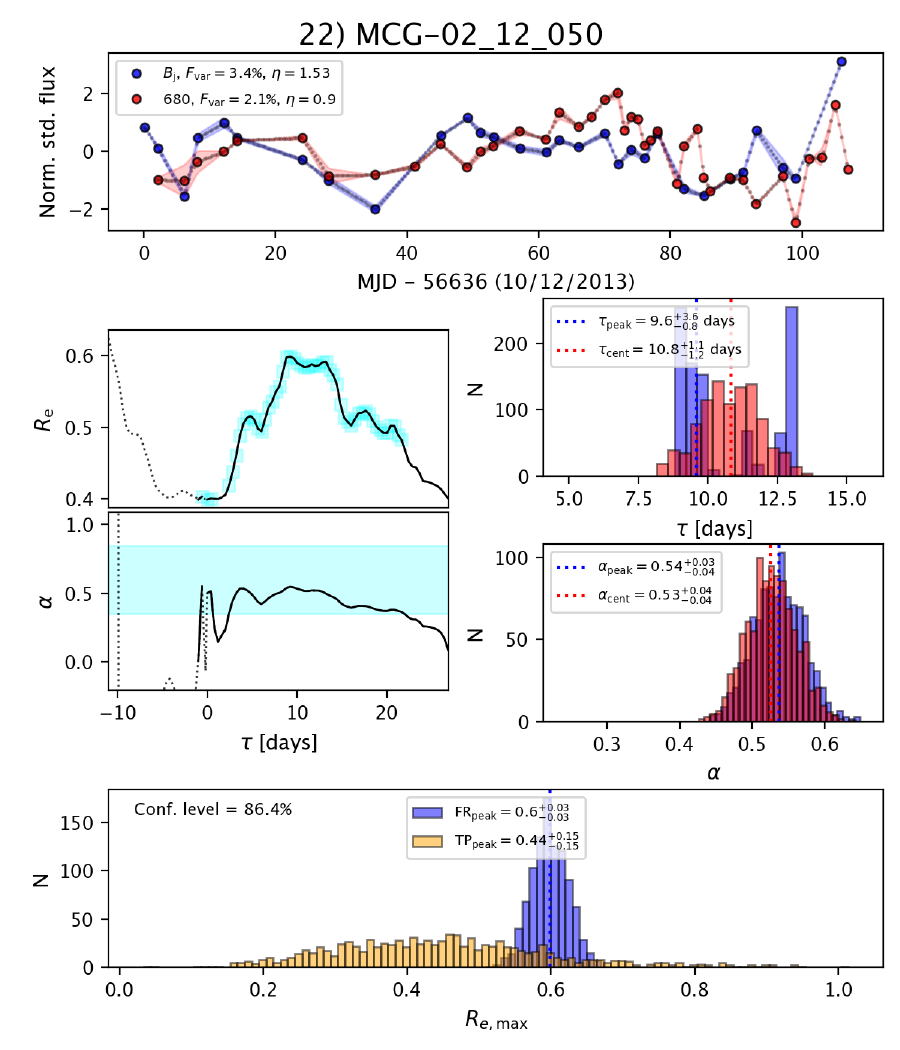}
\includegraphics[width=0.49\columnwidth]{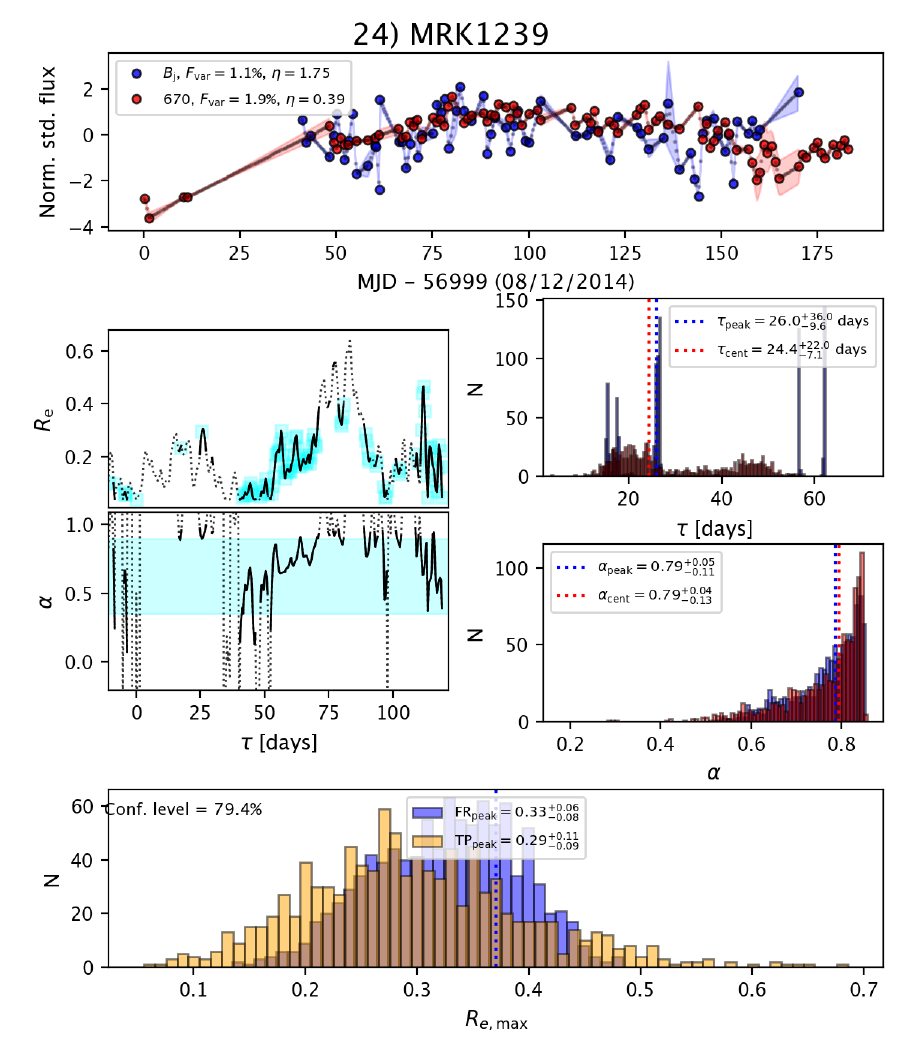}
\includegraphics[width=0.49\columnwidth]{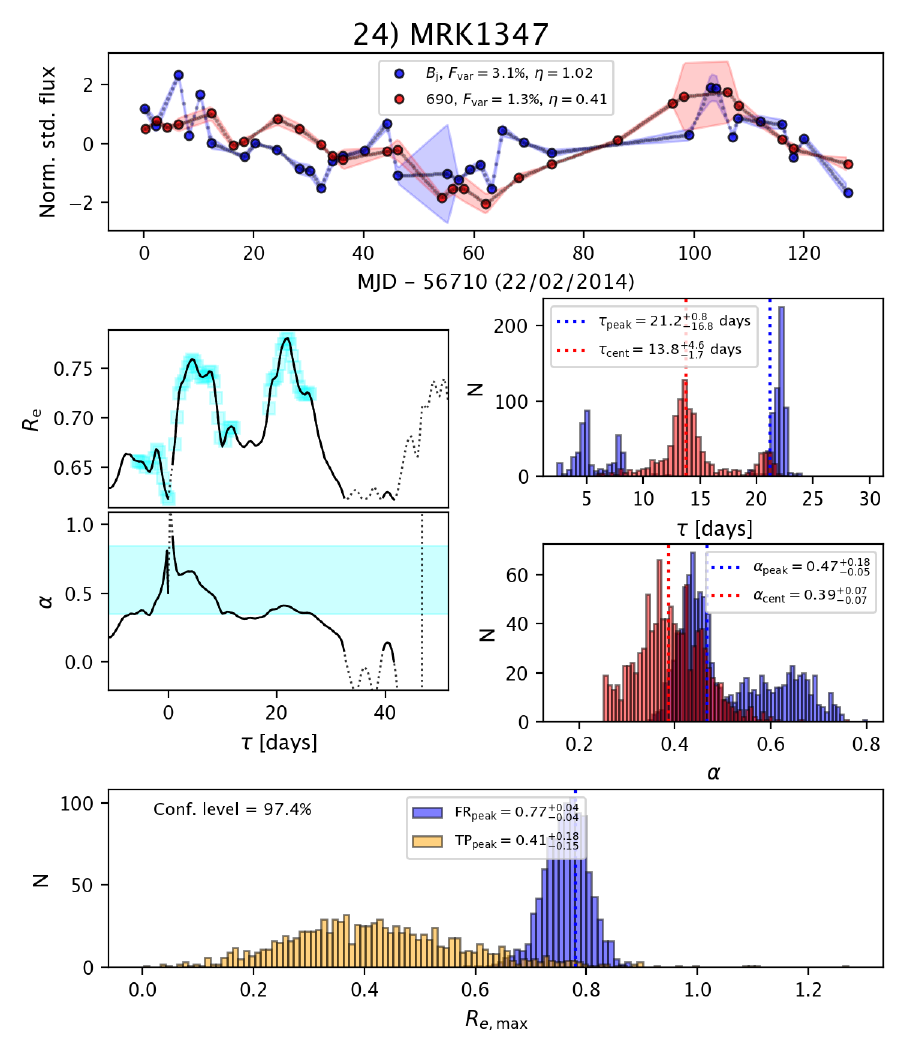}
\includegraphics[width=0.49\columnwidth]{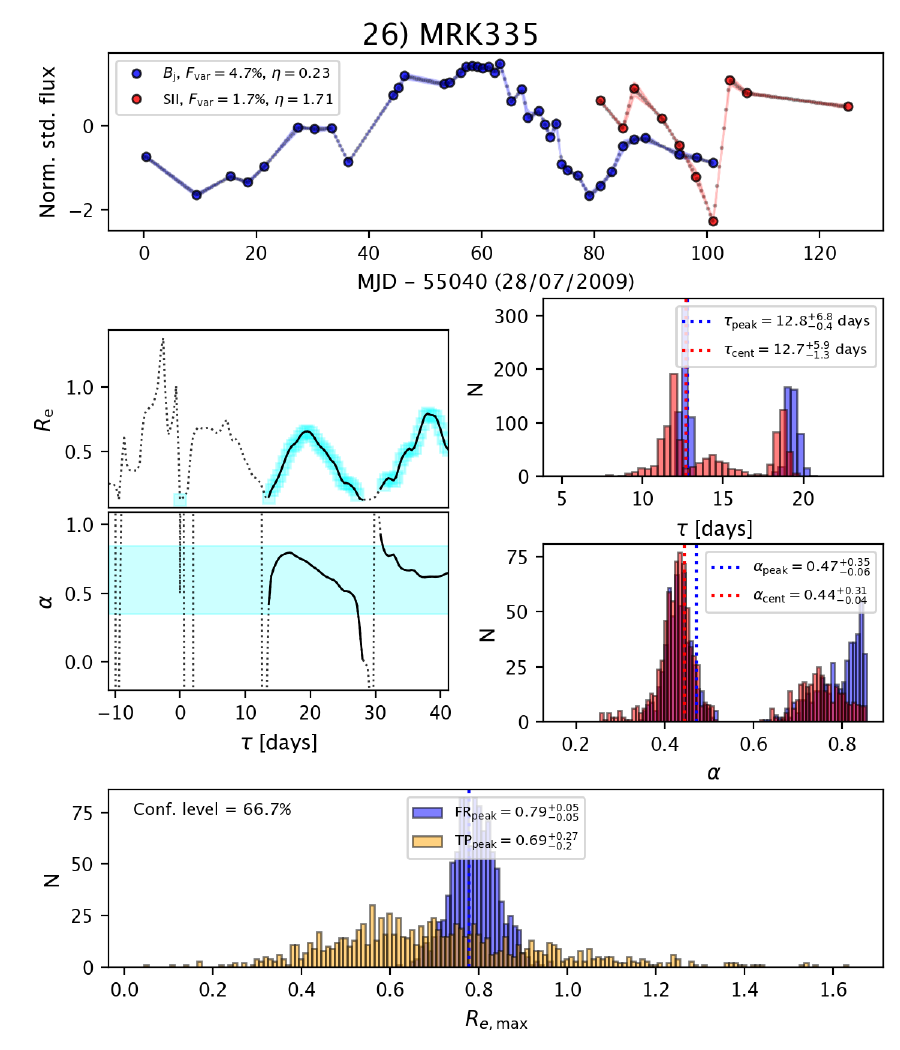}
\includegraphics[width=0.49\columnwidth]{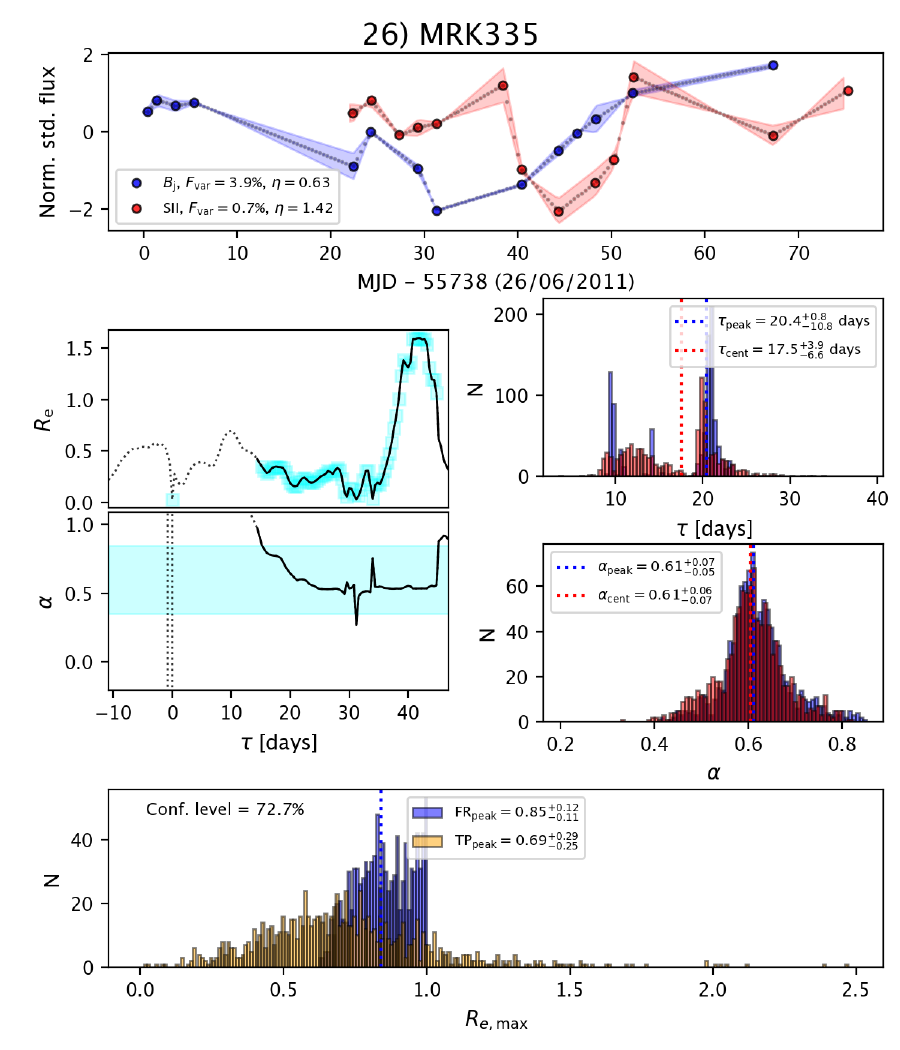}
\includegraphics[width=0.49\columnwidth]{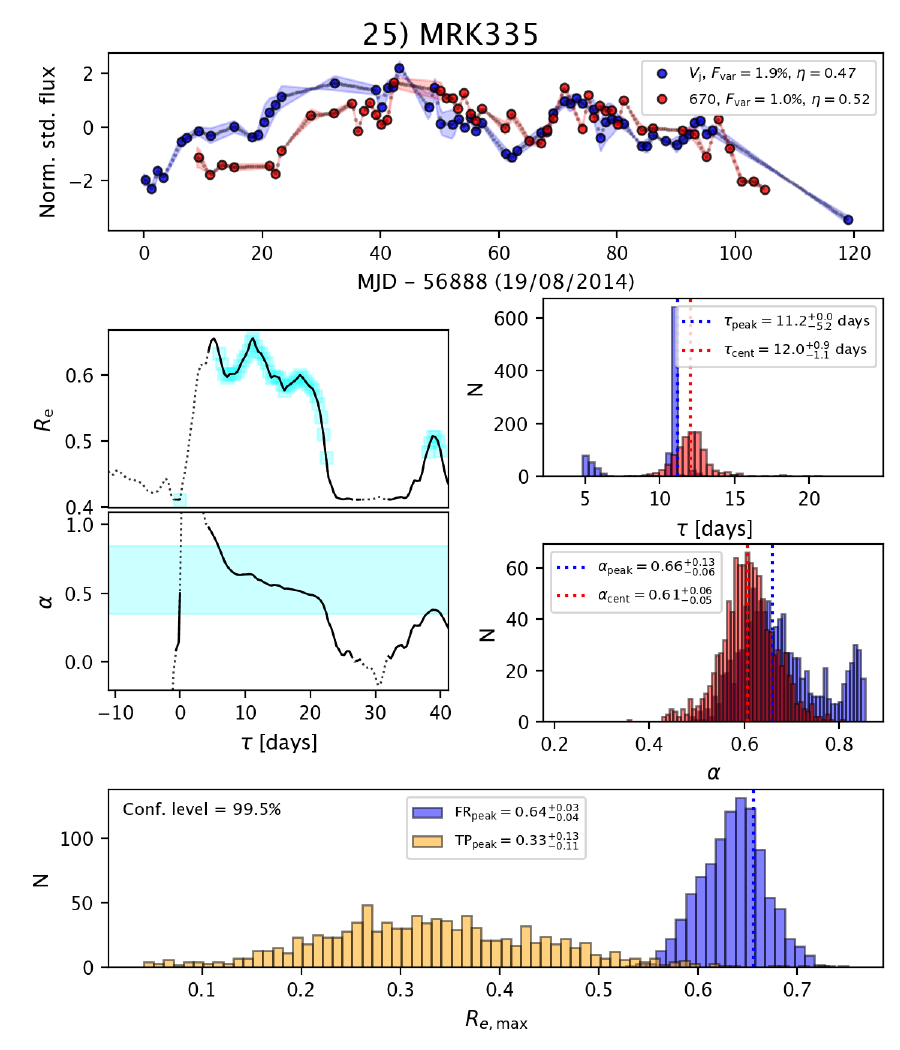}
\includegraphics[width=0.49\columnwidth]{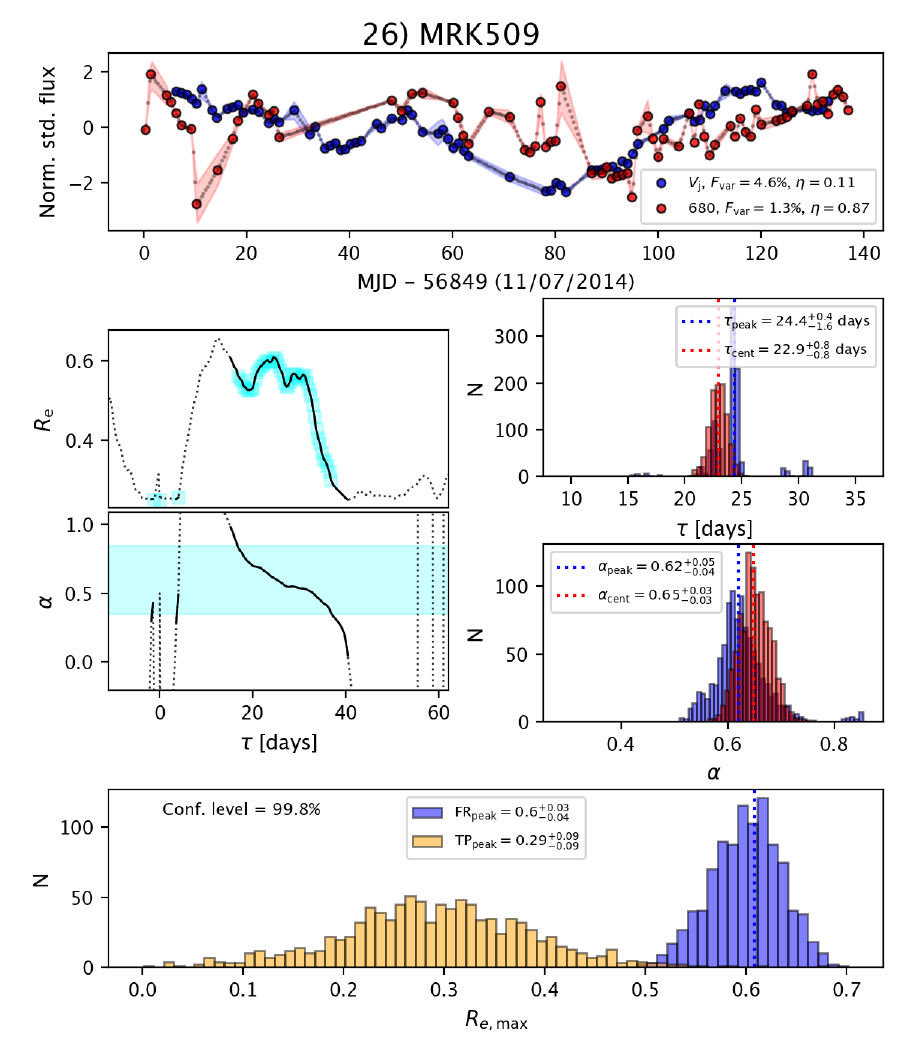}
\includegraphics[width=0.49\columnwidth]{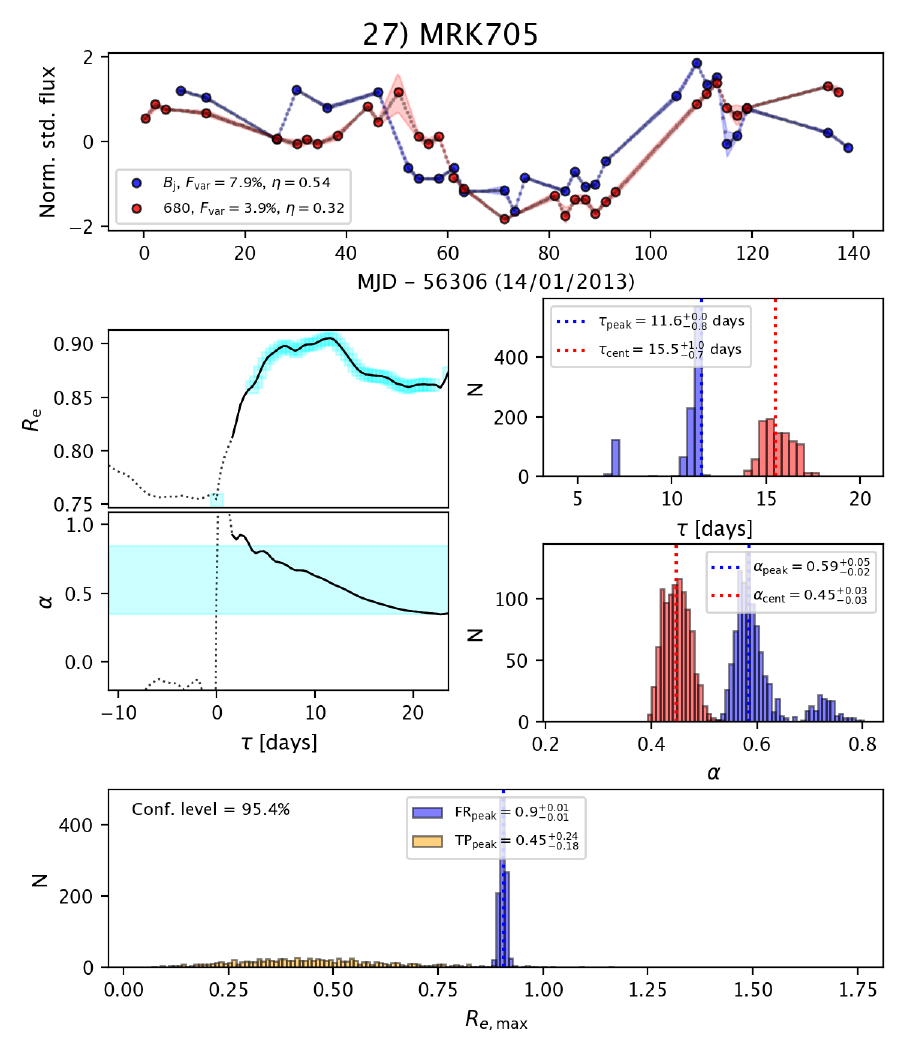}
\includegraphics[width=0.49\columnwidth]{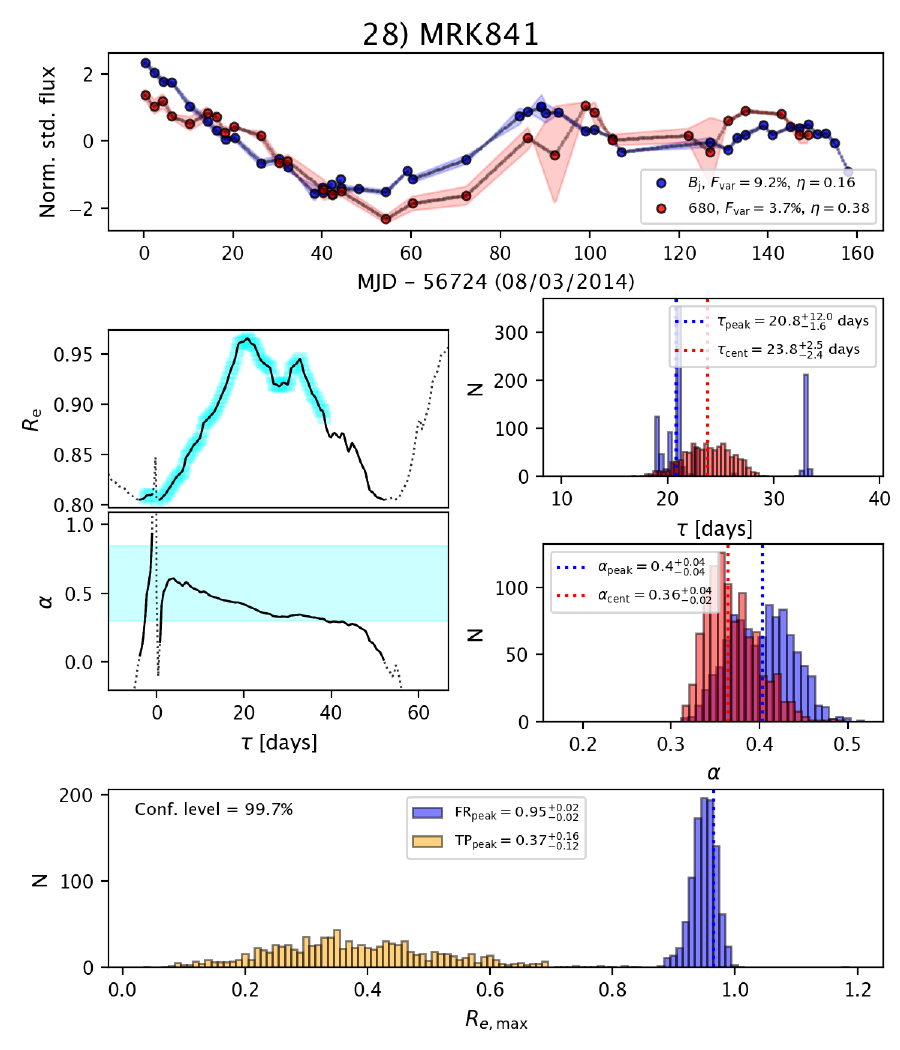}
\includegraphics[width=0.49\columnwidth]{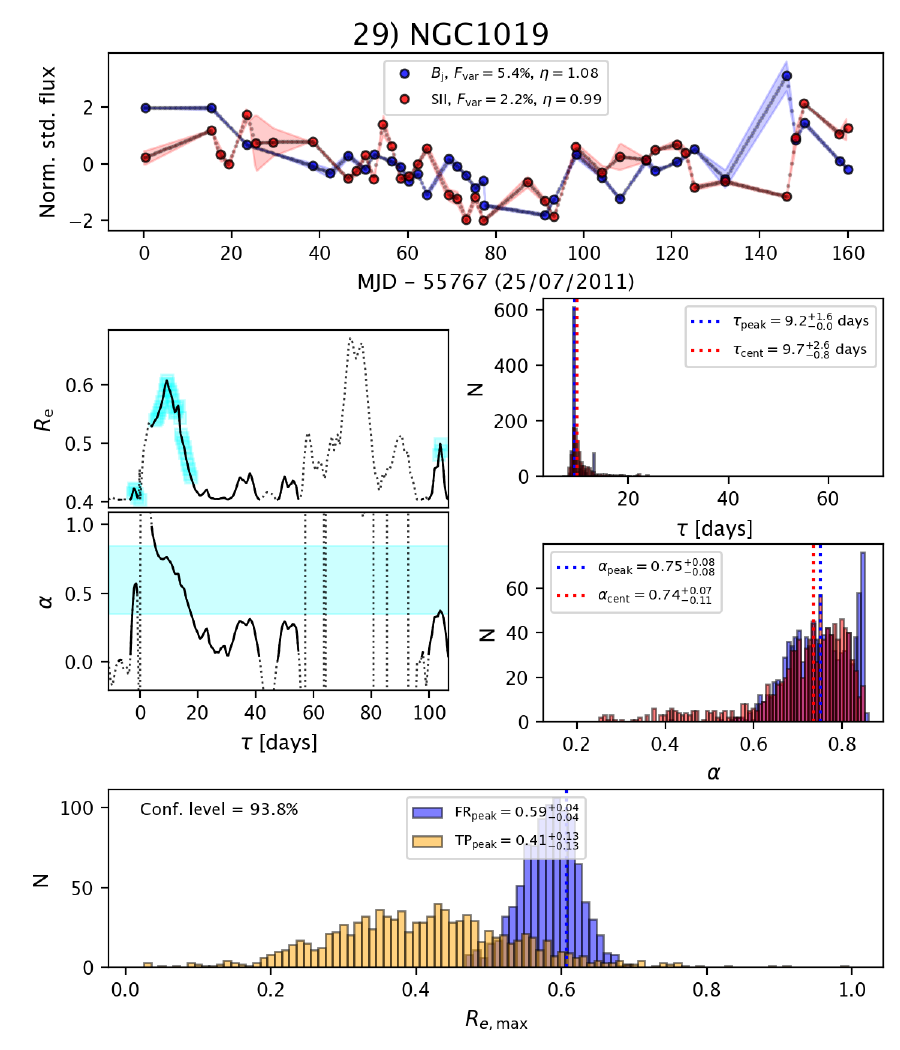}
\includegraphics[width=0.49\columnwidth]{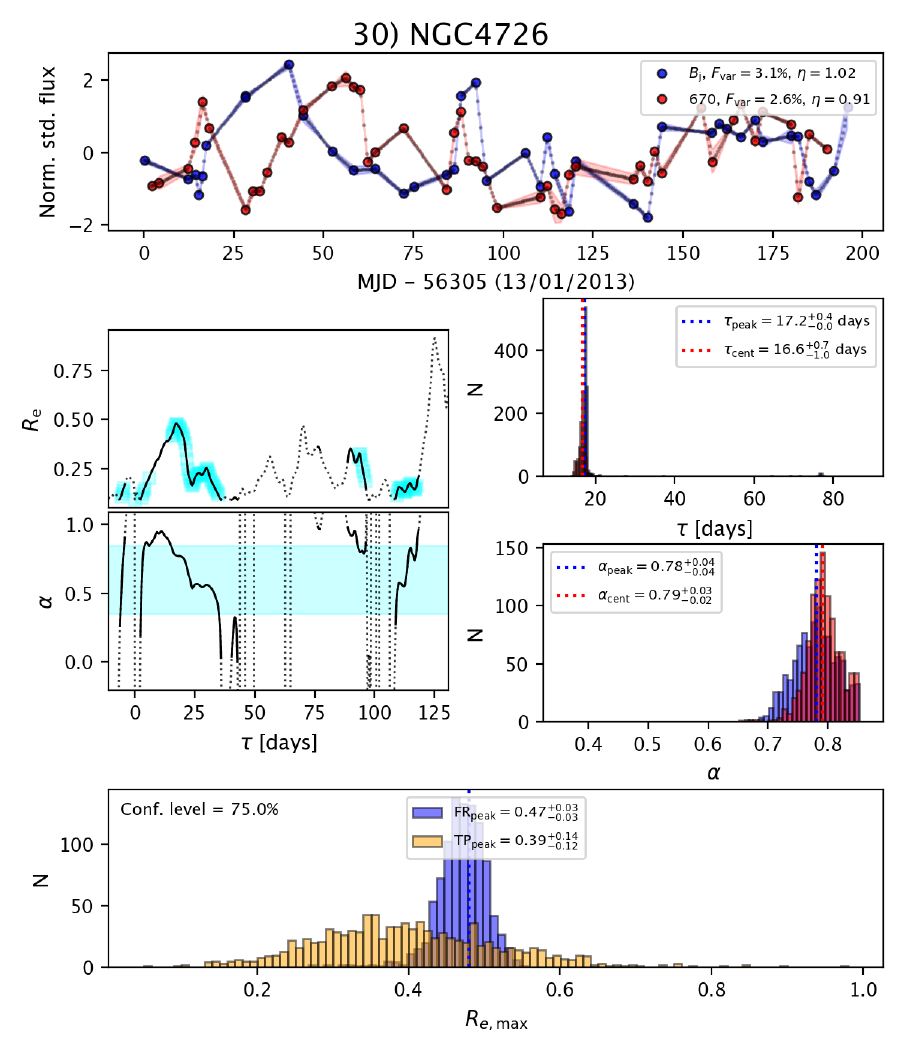}
\includegraphics[width=0.49\columnwidth]{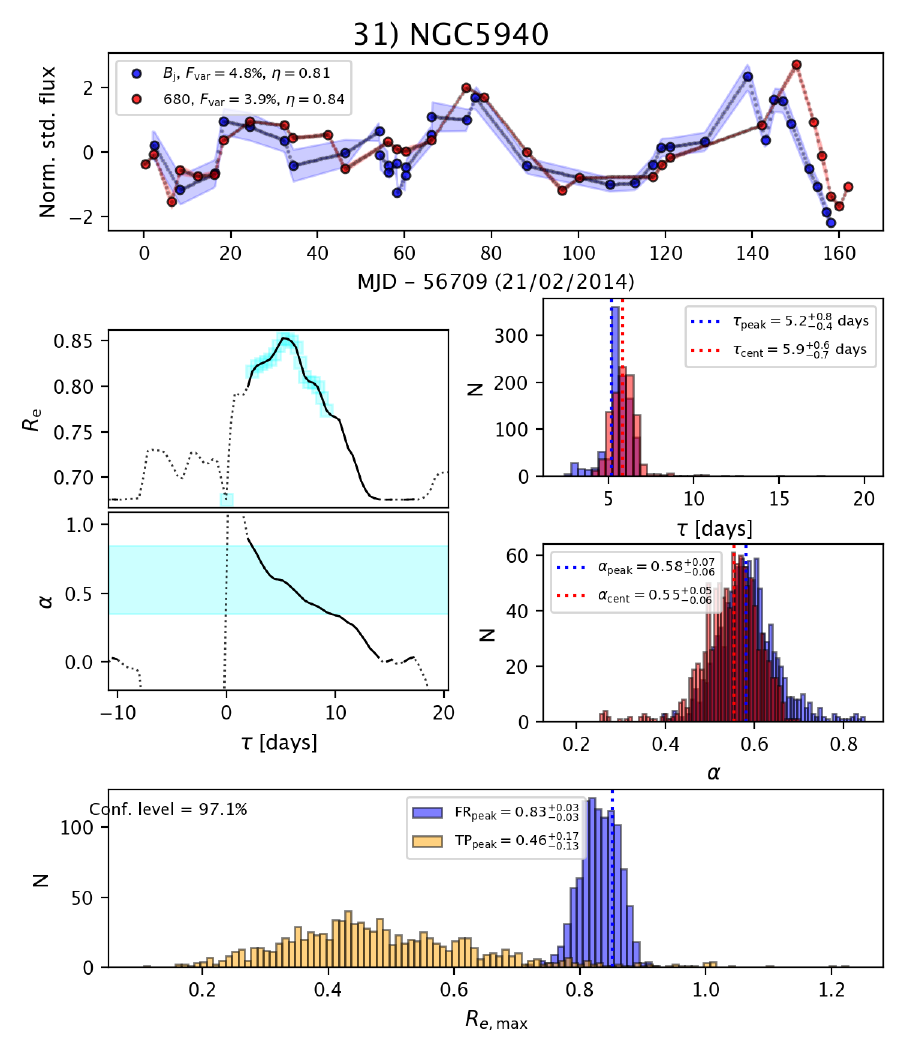}
\includegraphics[width=0.49\columnwidth]{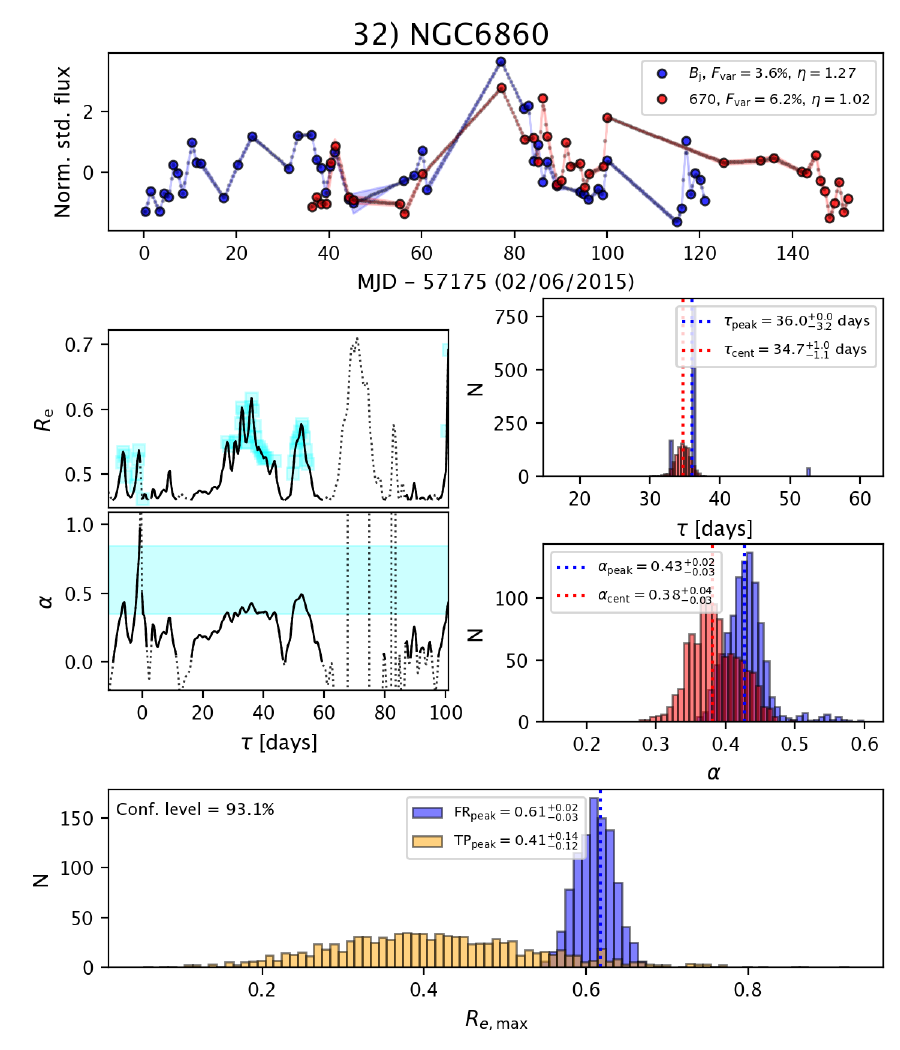}
\includegraphics[width=0.49\columnwidth]{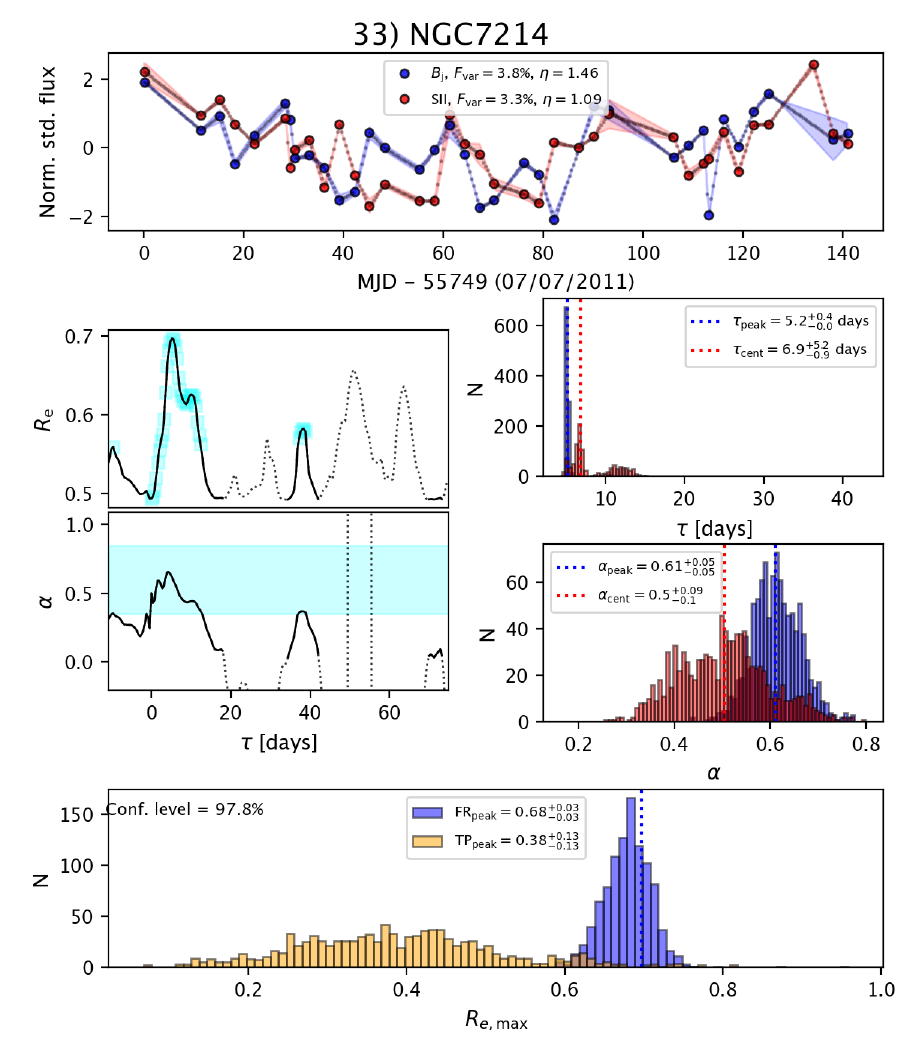}
\includegraphics[width=0.49\columnwidth]{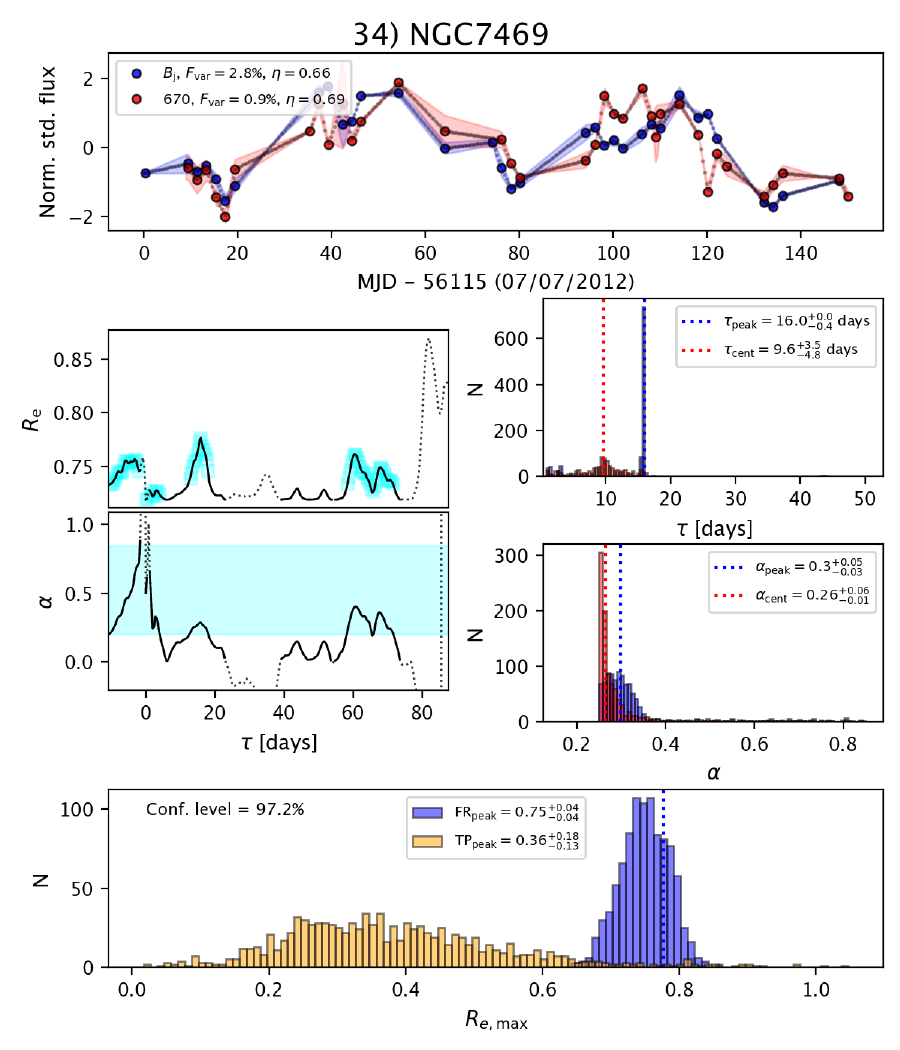}
\includegraphics[width=0.49\columnwidth]{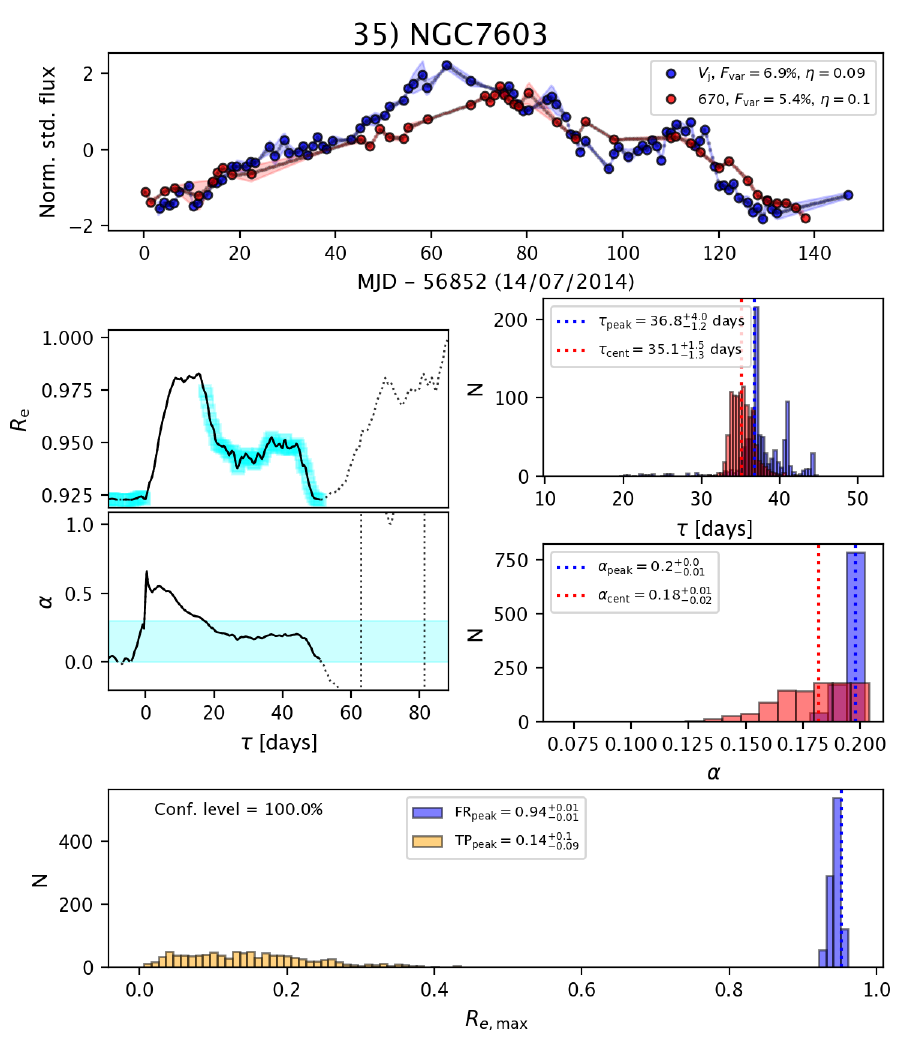}
\includegraphics[width=0.49\columnwidth]{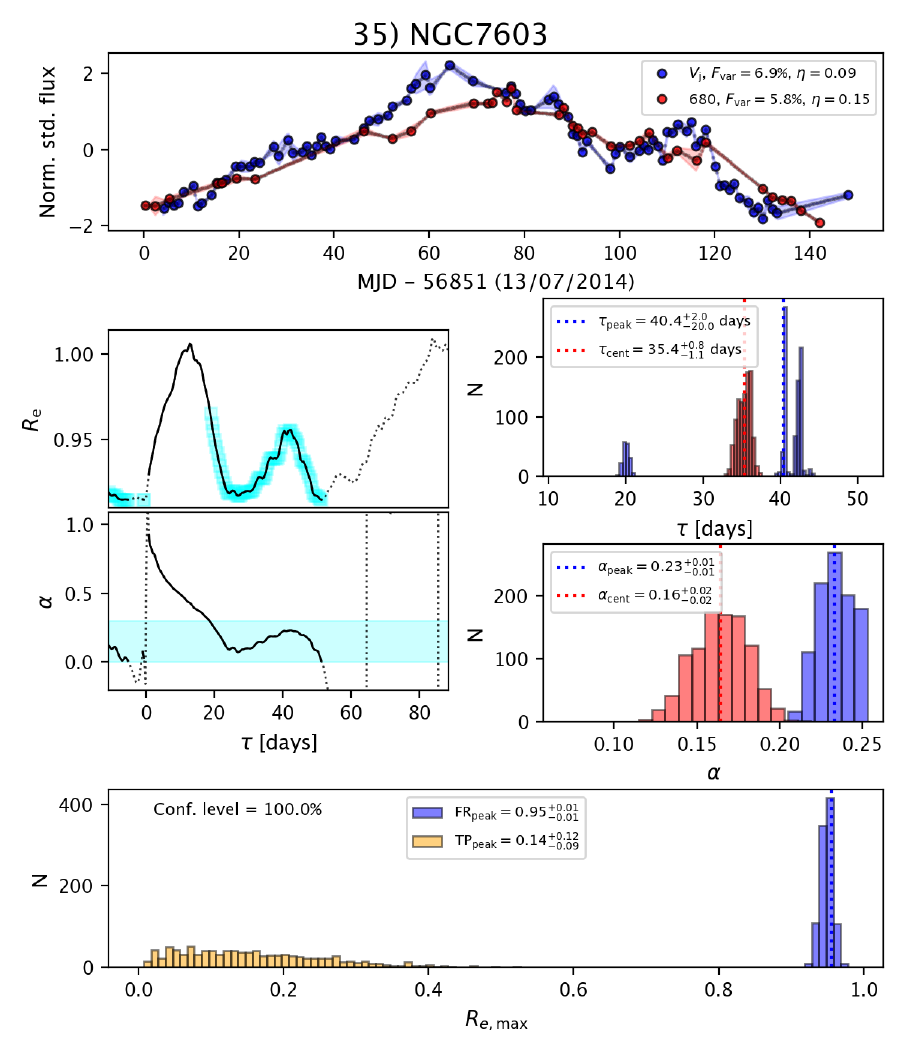}
\includegraphics[width=0.49\columnwidth]{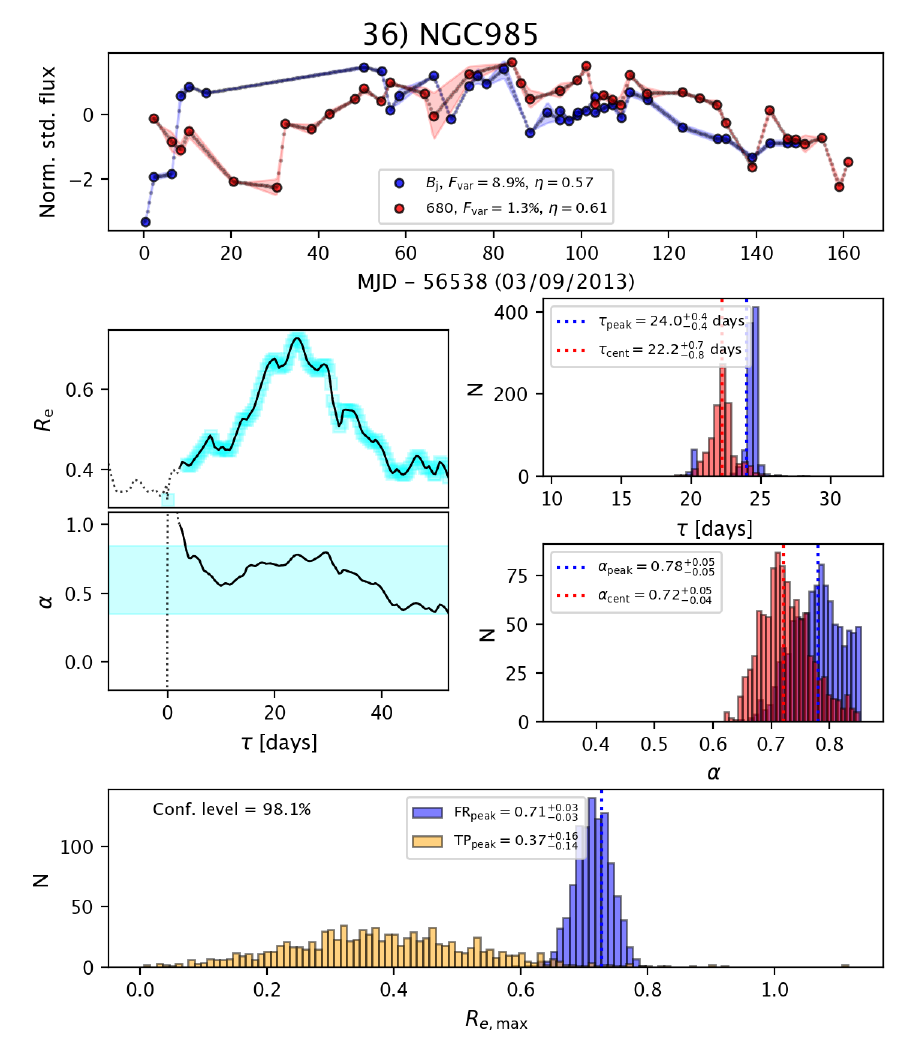}
\includegraphics[width=0.49\columnwidth]{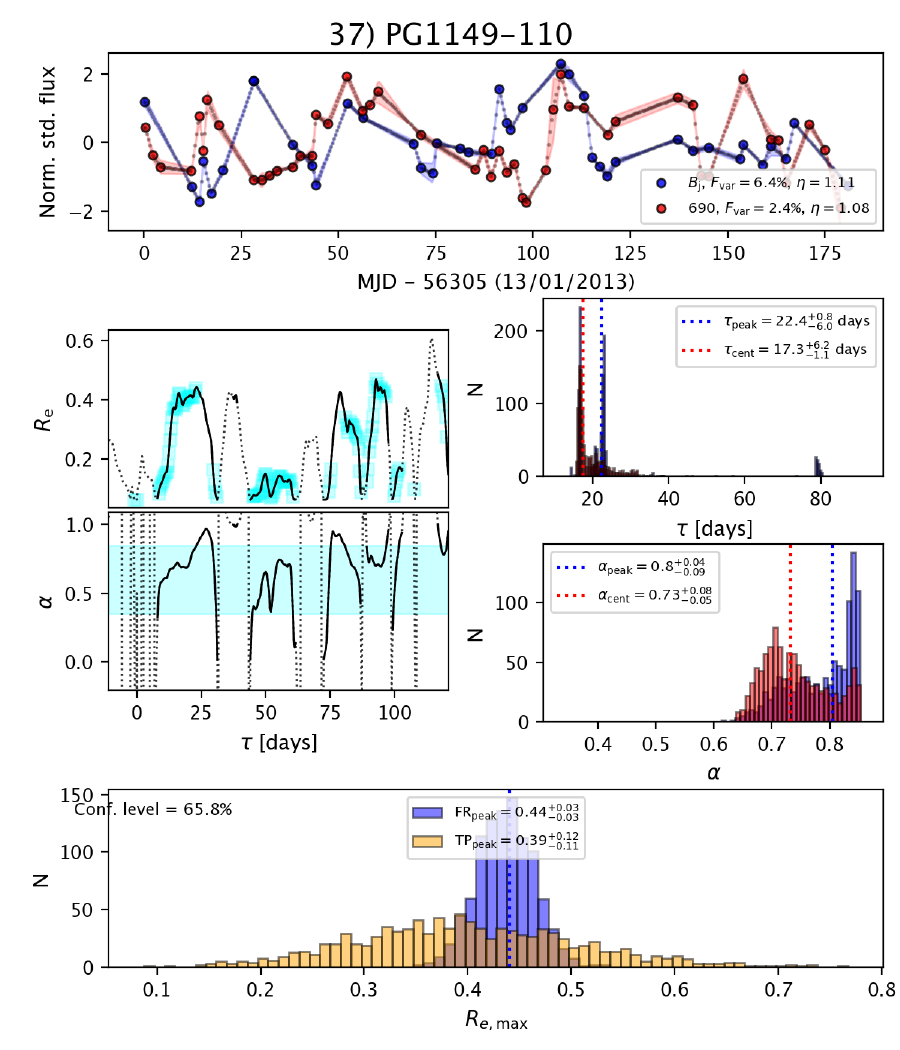}
\includegraphics[width=0.49\columnwidth]{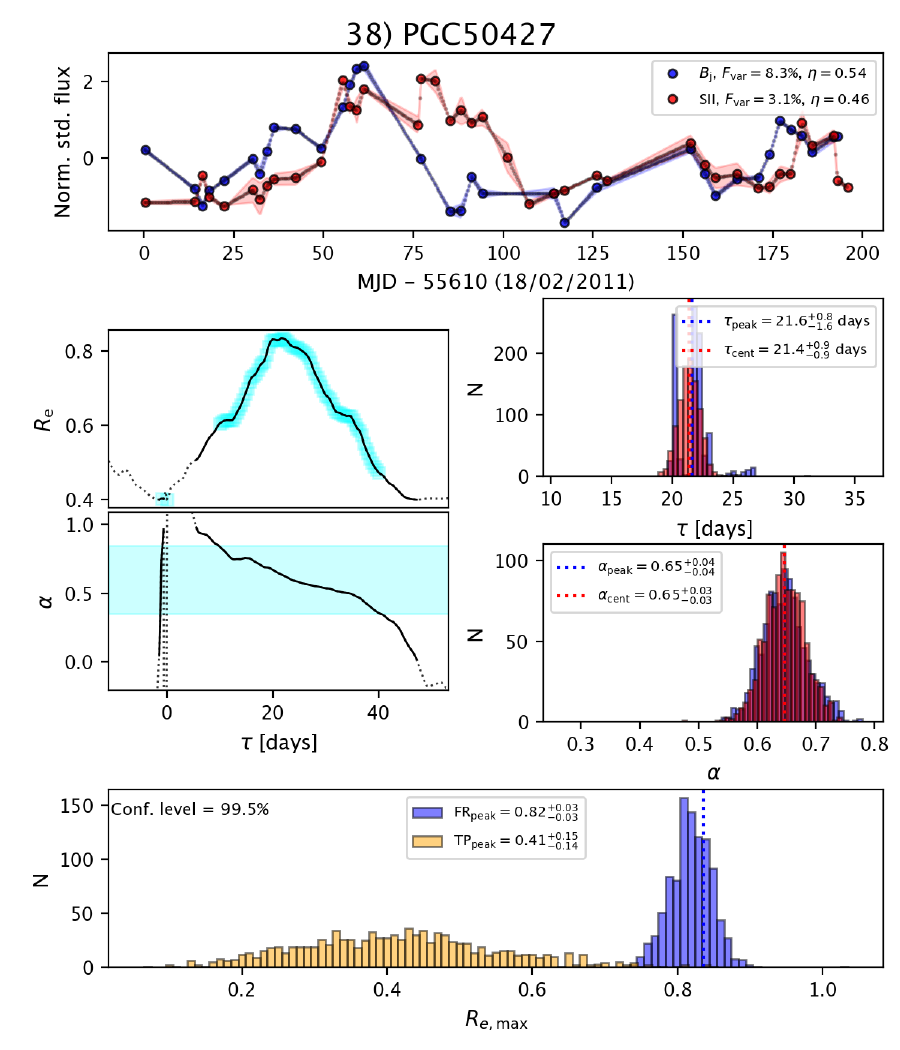}
\includegraphics[width=0.49\columnwidth]{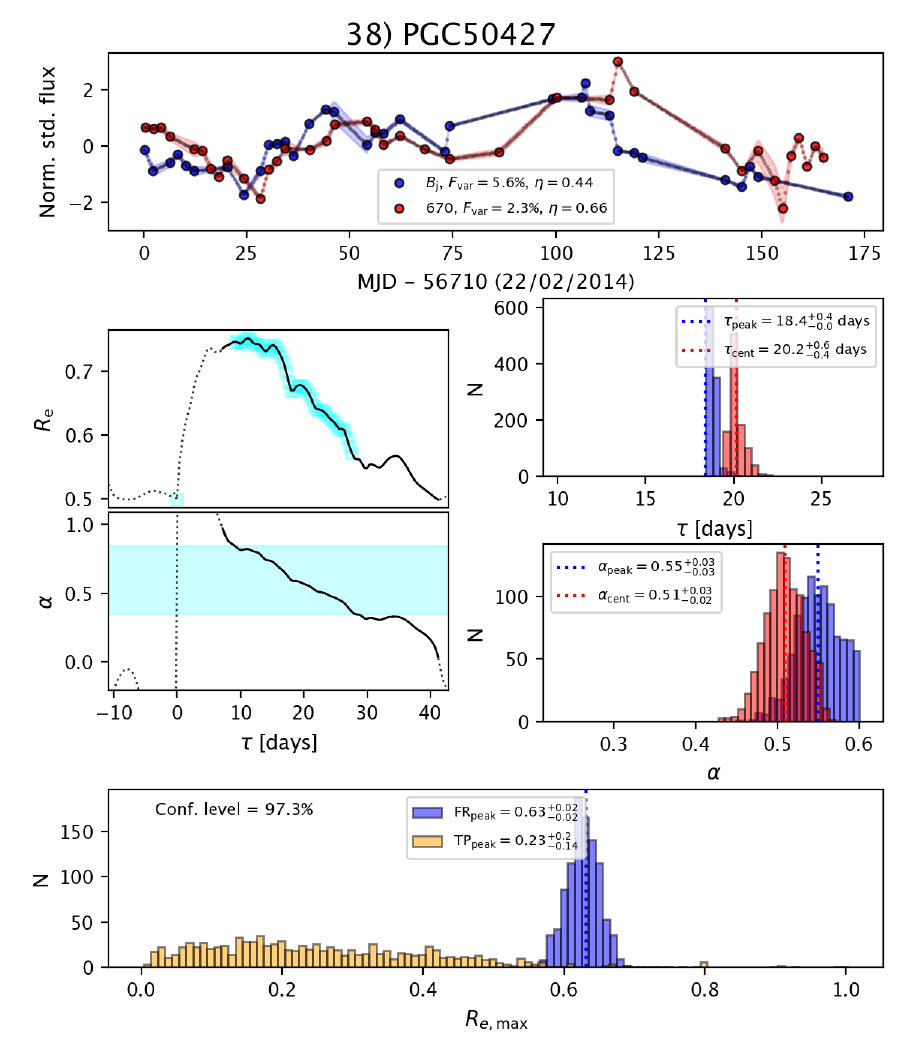}
\includegraphics[width=0.49\columnwidth]{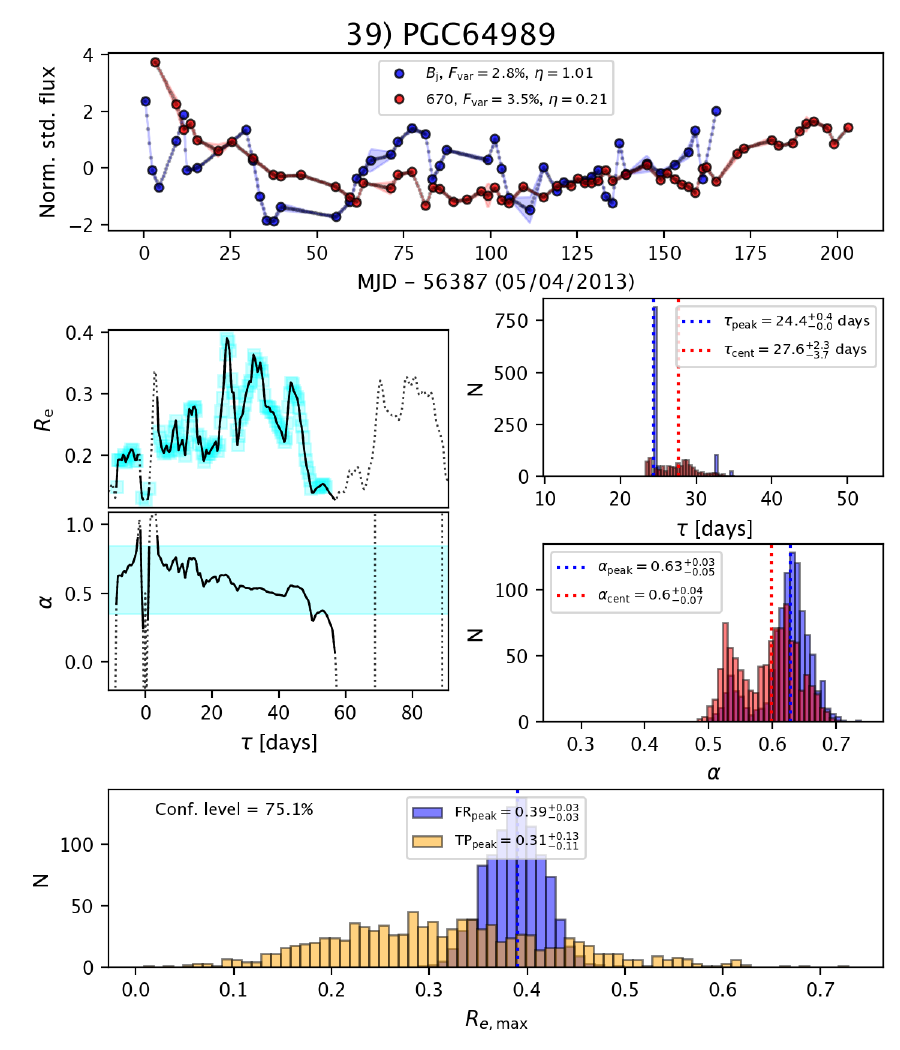}
\includegraphics[width=0.49\columnwidth]{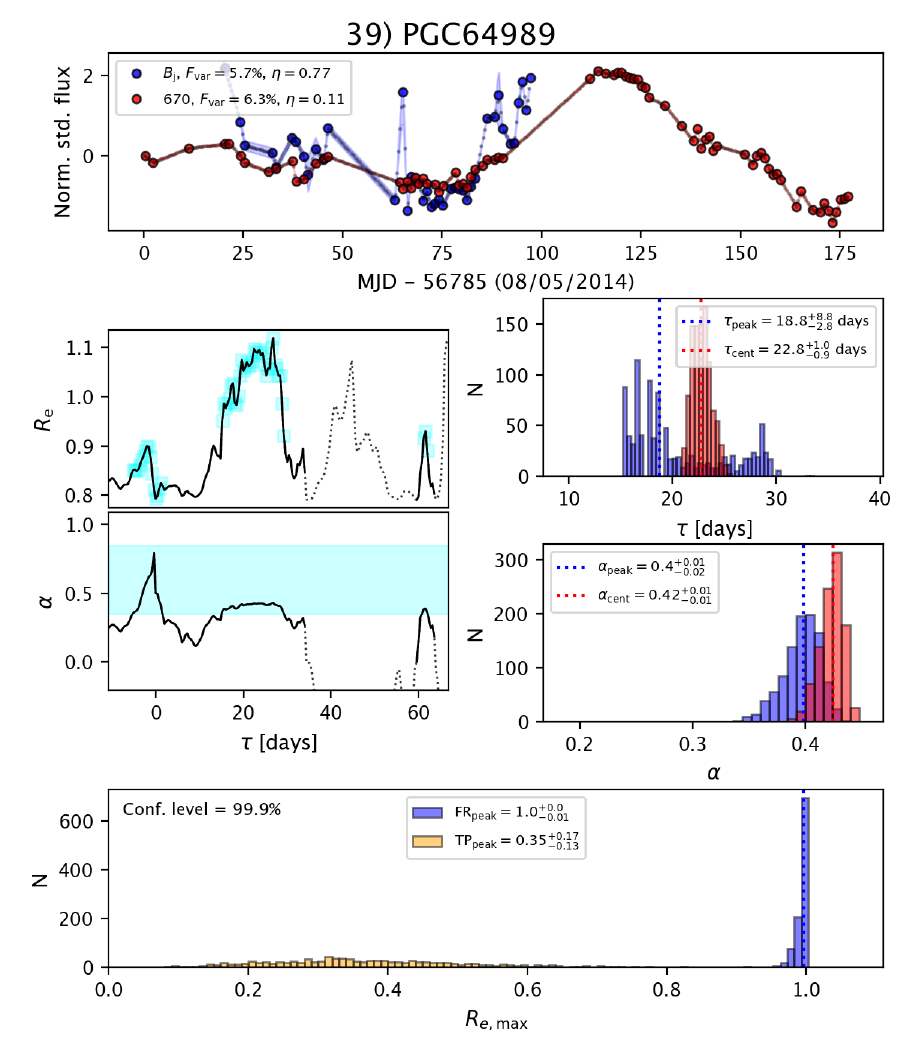}
\includegraphics[width=0.49\columnwidth]{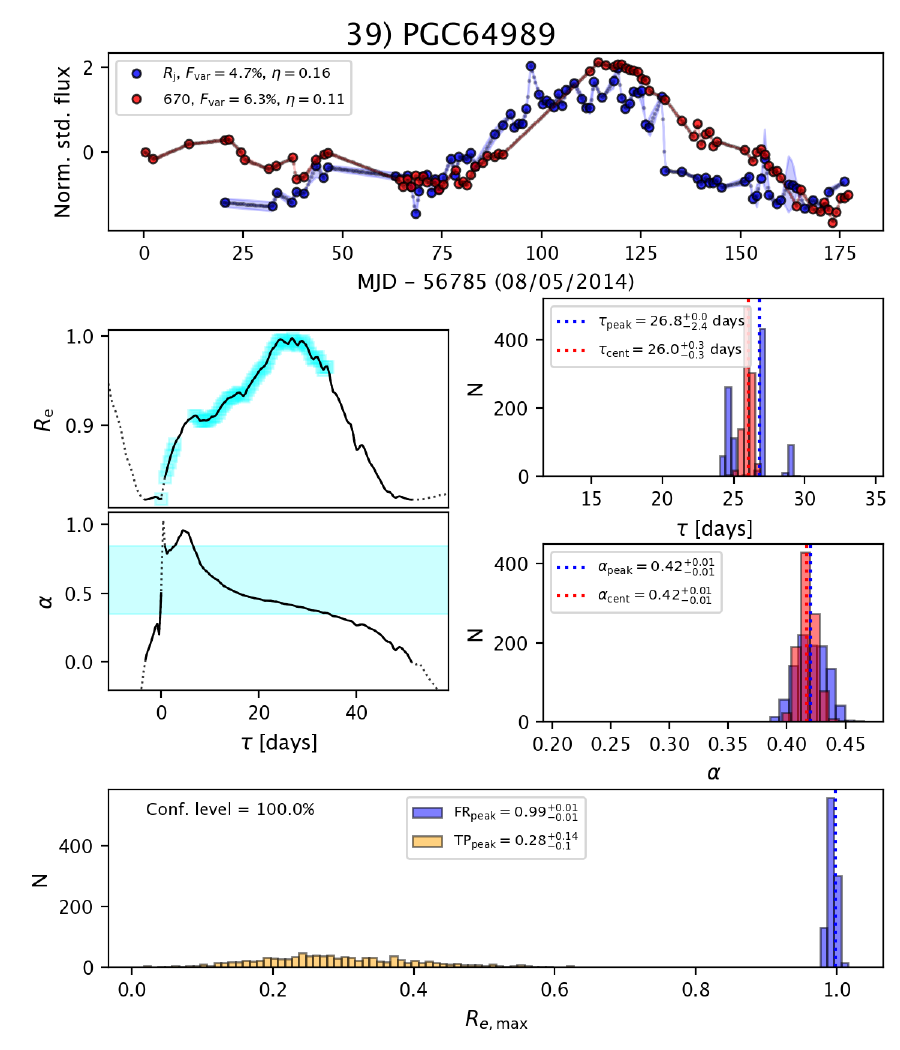}
\includegraphics[width=0.49\columnwidth]{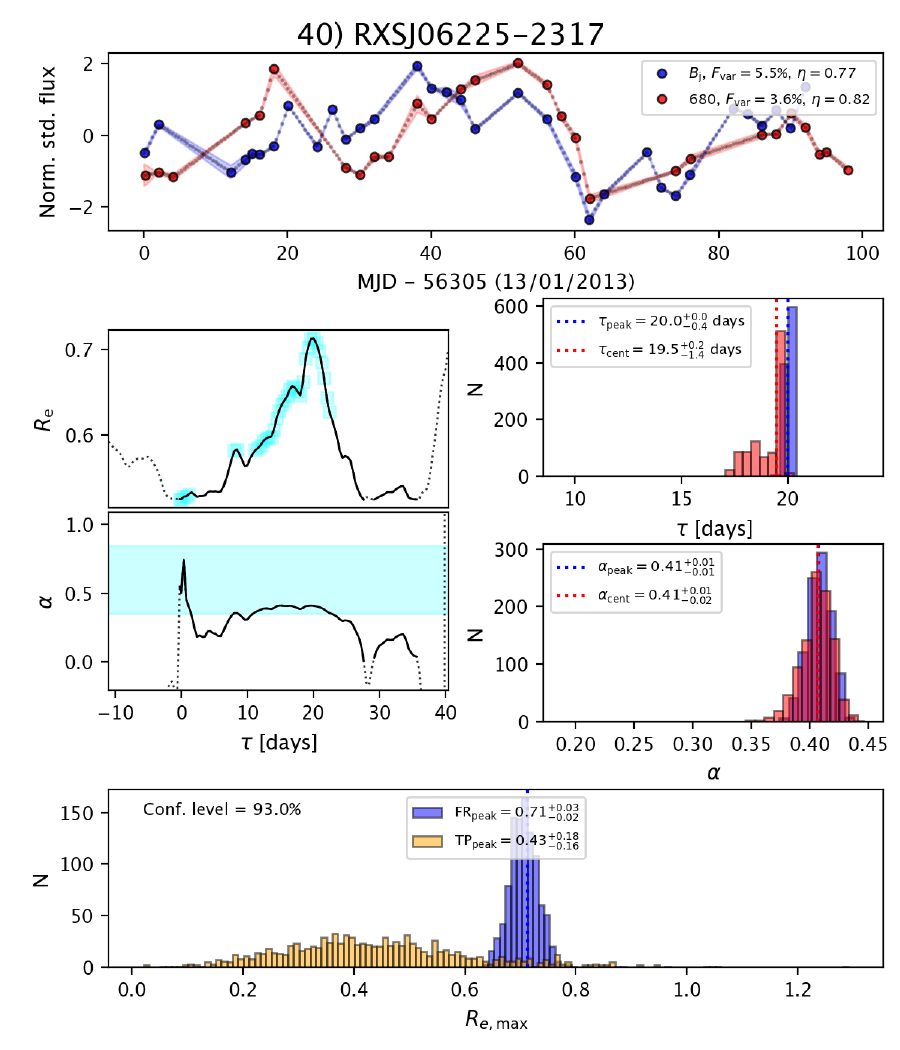}
\includegraphics[width=0.49\columnwidth]{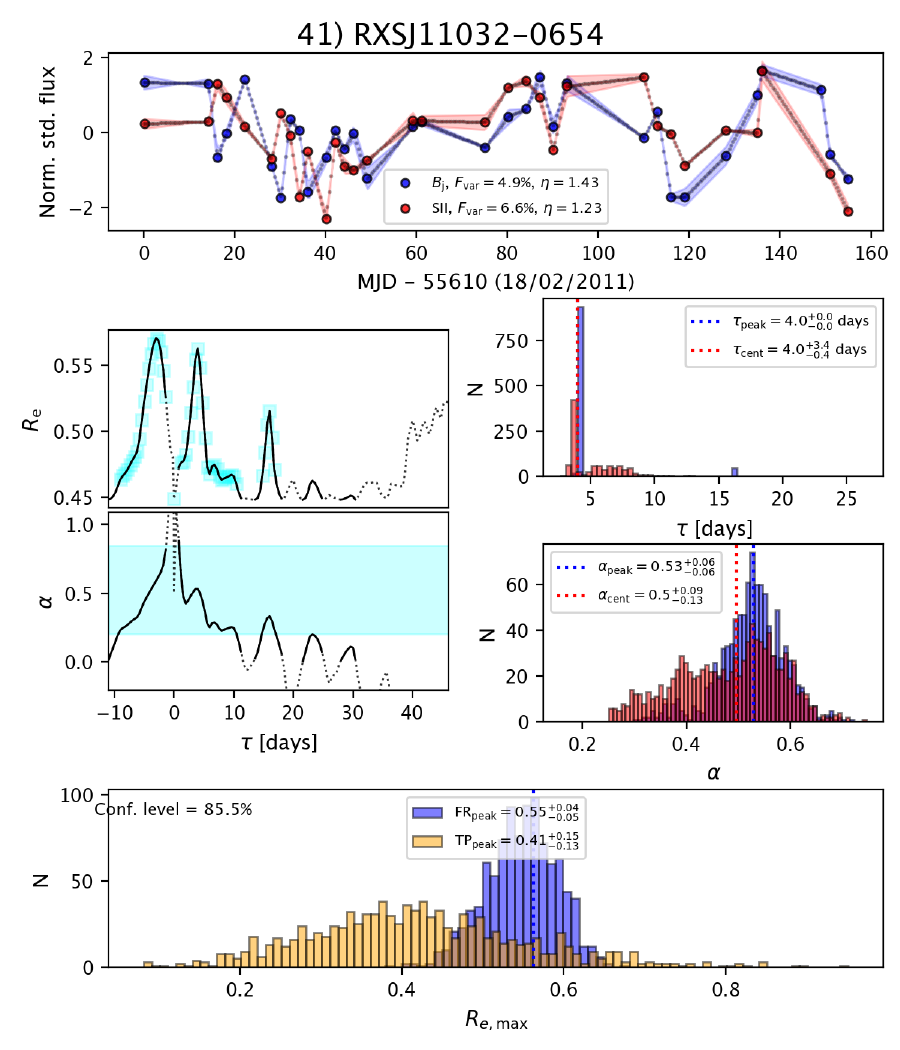}
\includegraphics[width=0.49\columnwidth]{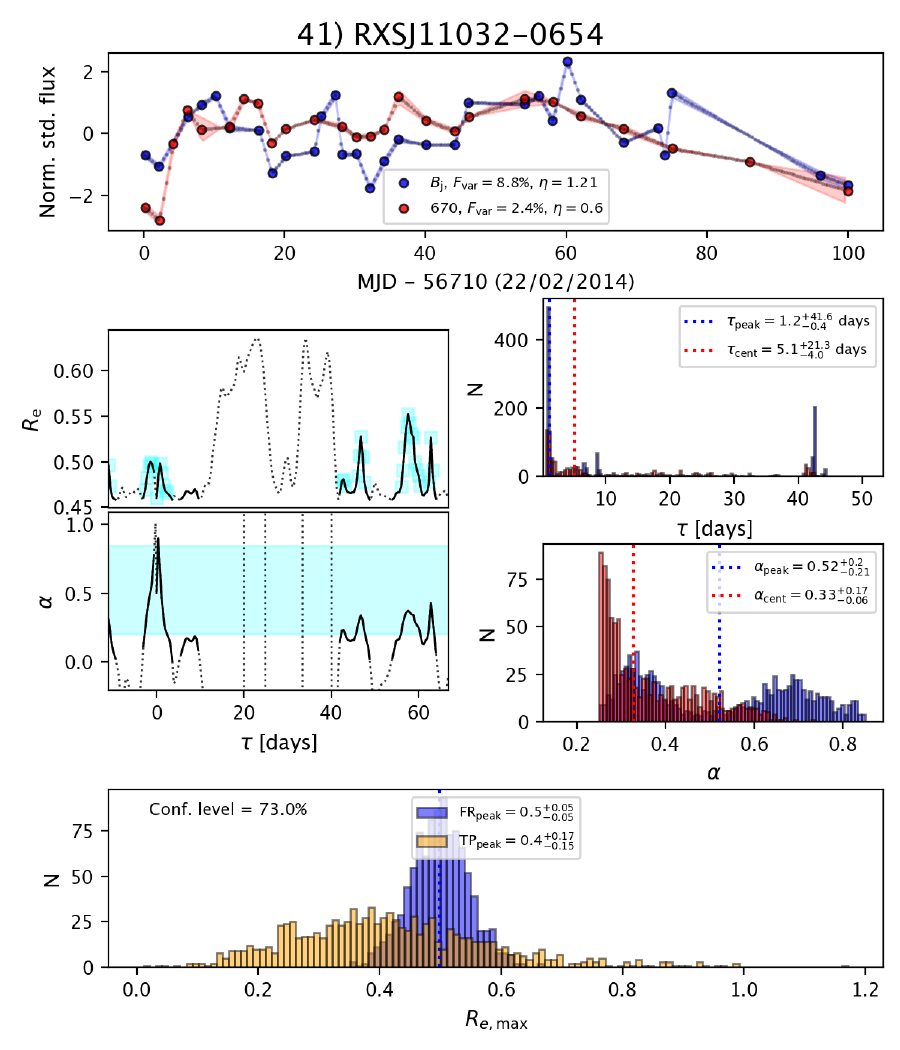}
\includegraphics[width=0.49\columnwidth]{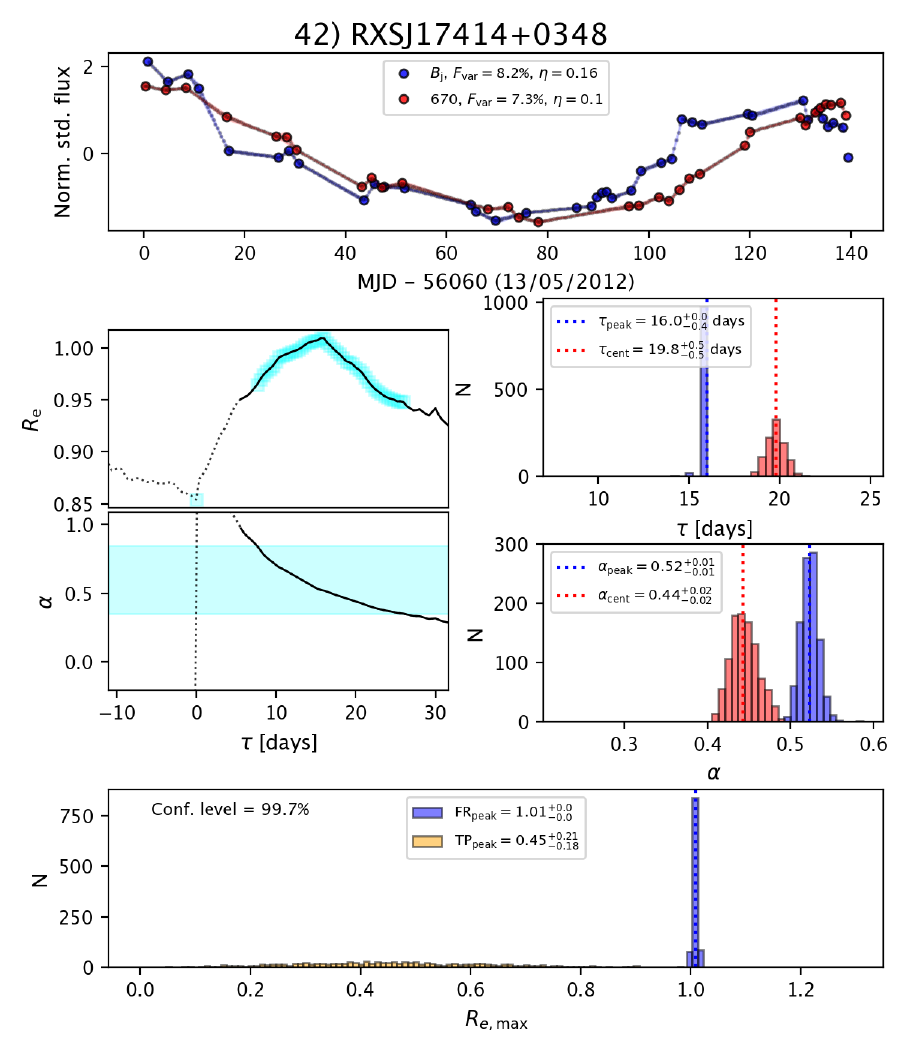}
\includegraphics[width=0.49\columnwidth]{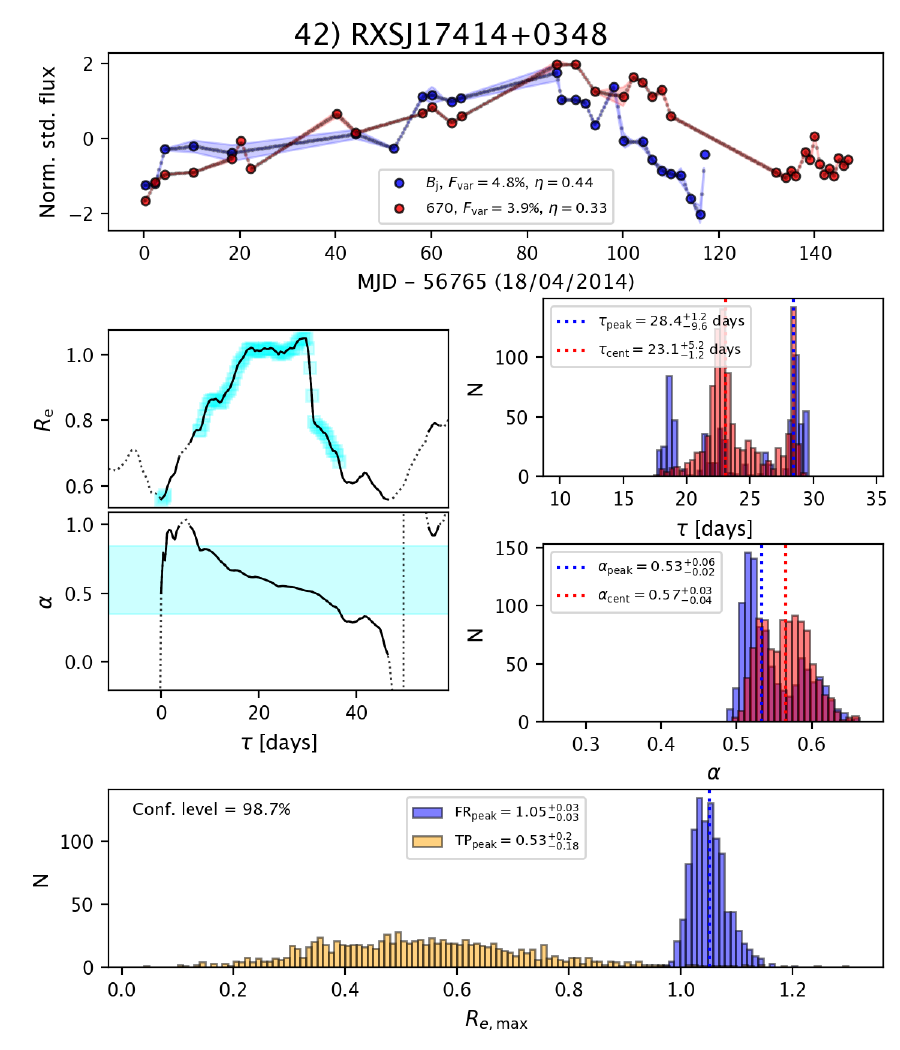}
\includegraphics[width=0.49\columnwidth]{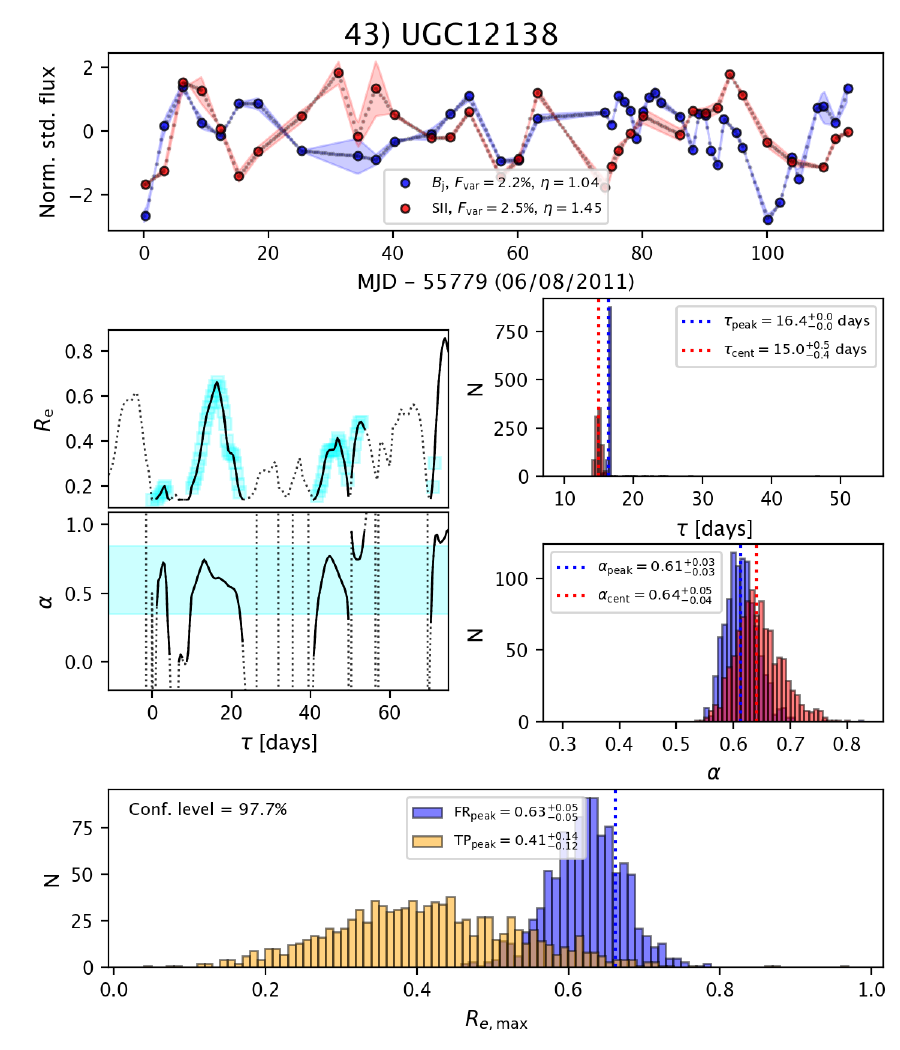}
\includegraphics[width=0.49\columnwidth]{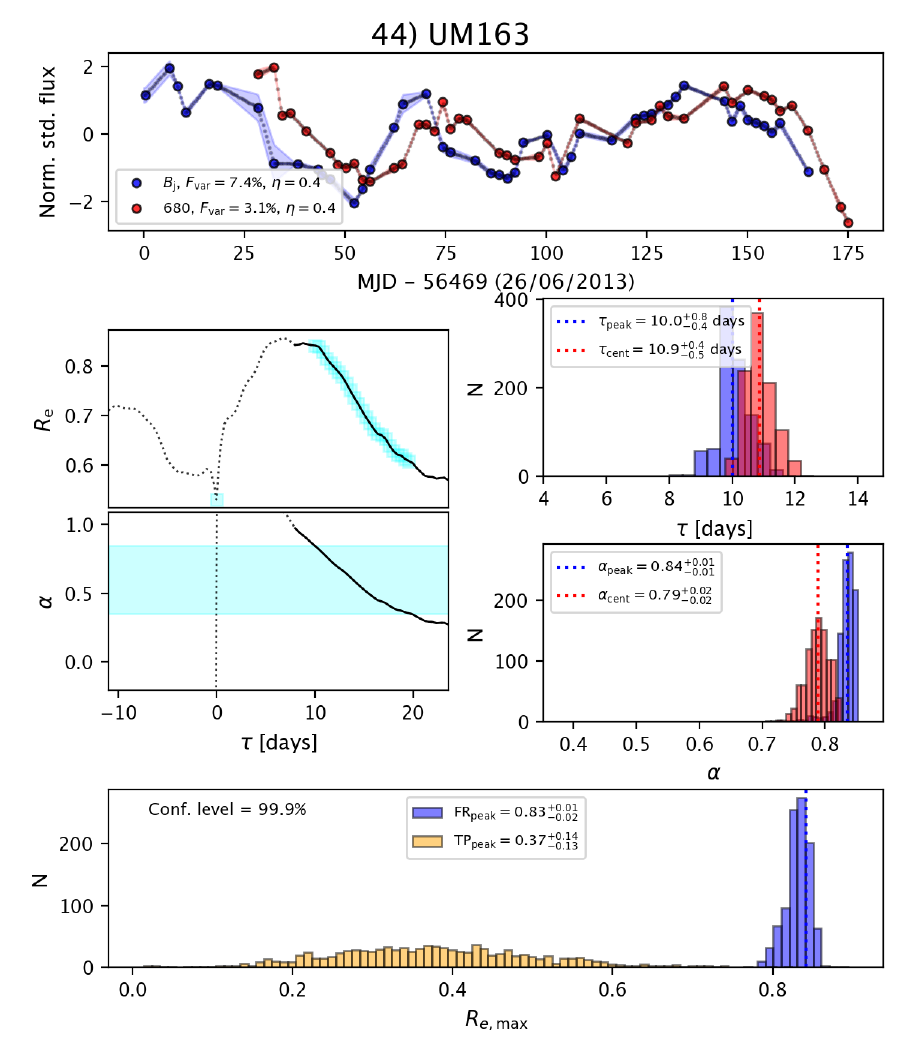}
\includegraphics[width=0.49\columnwidth]{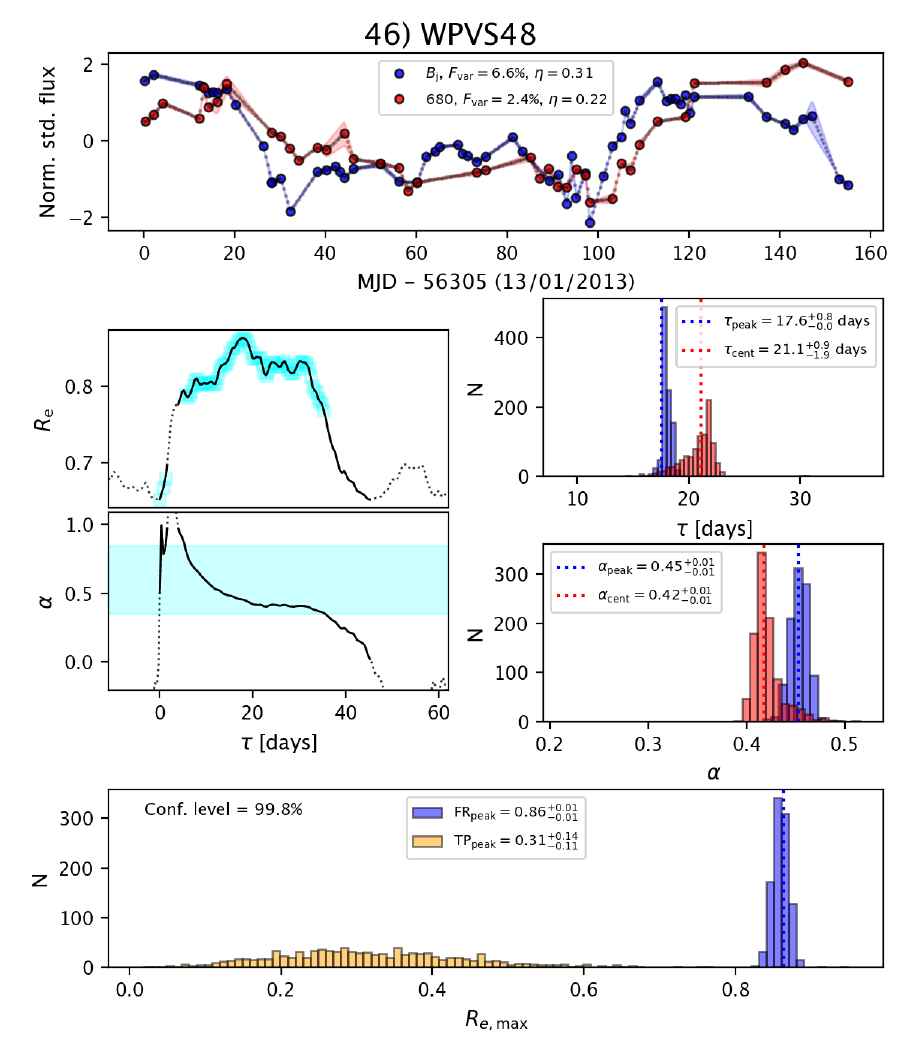}
\includegraphics[width=0.49\columnwidth]{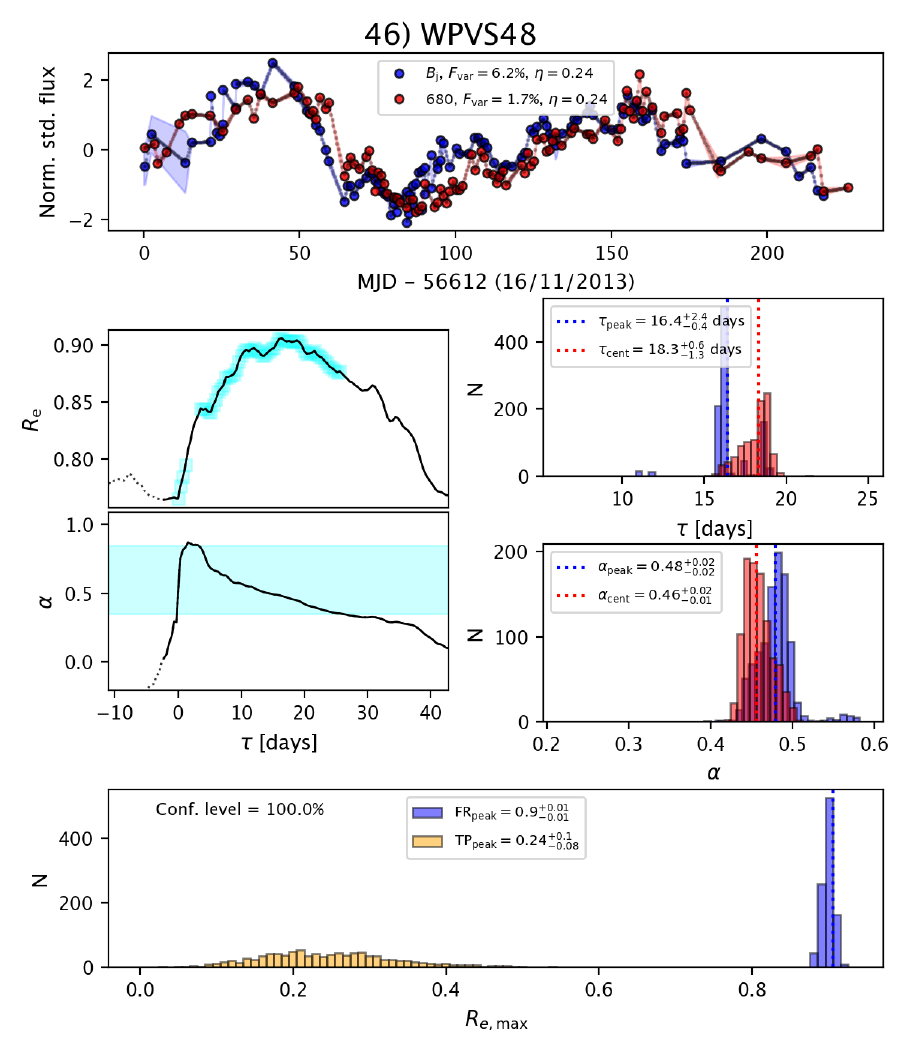}

\includegraphics[width=0.49\columnwidth]{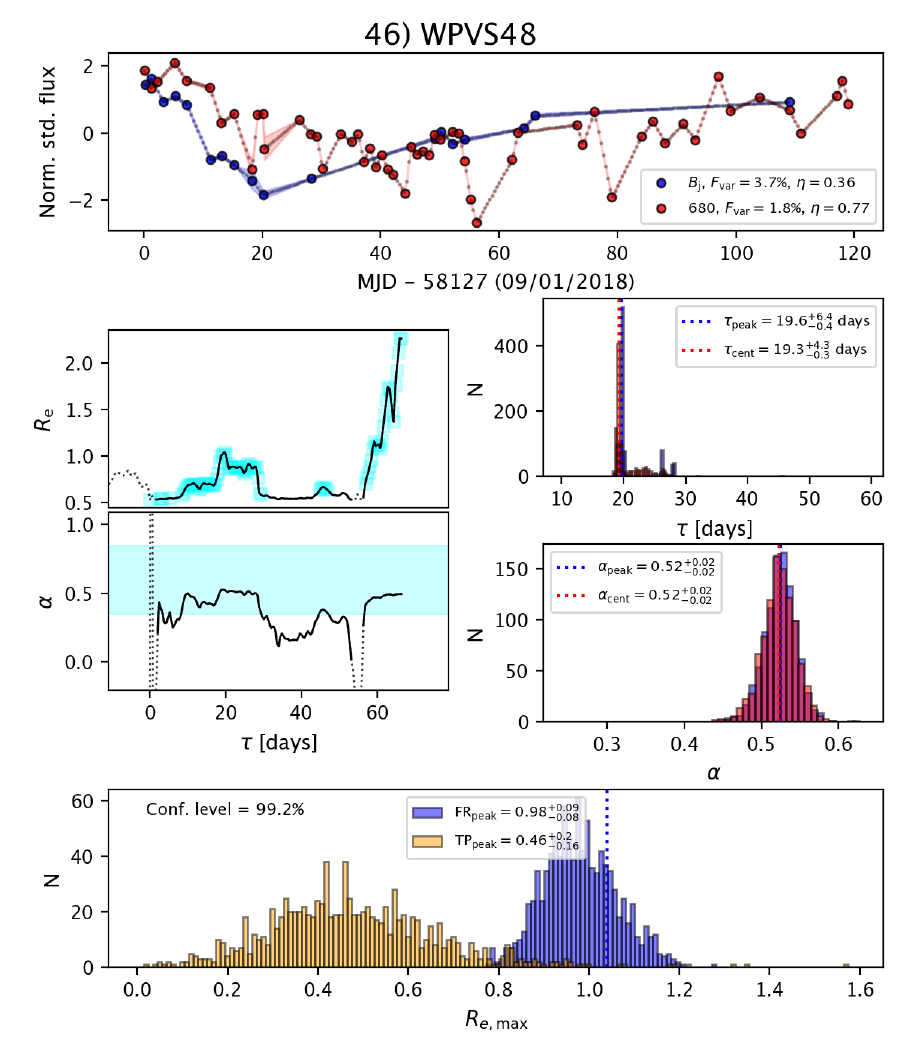}
\includegraphics[width=0.49\columnwidth]{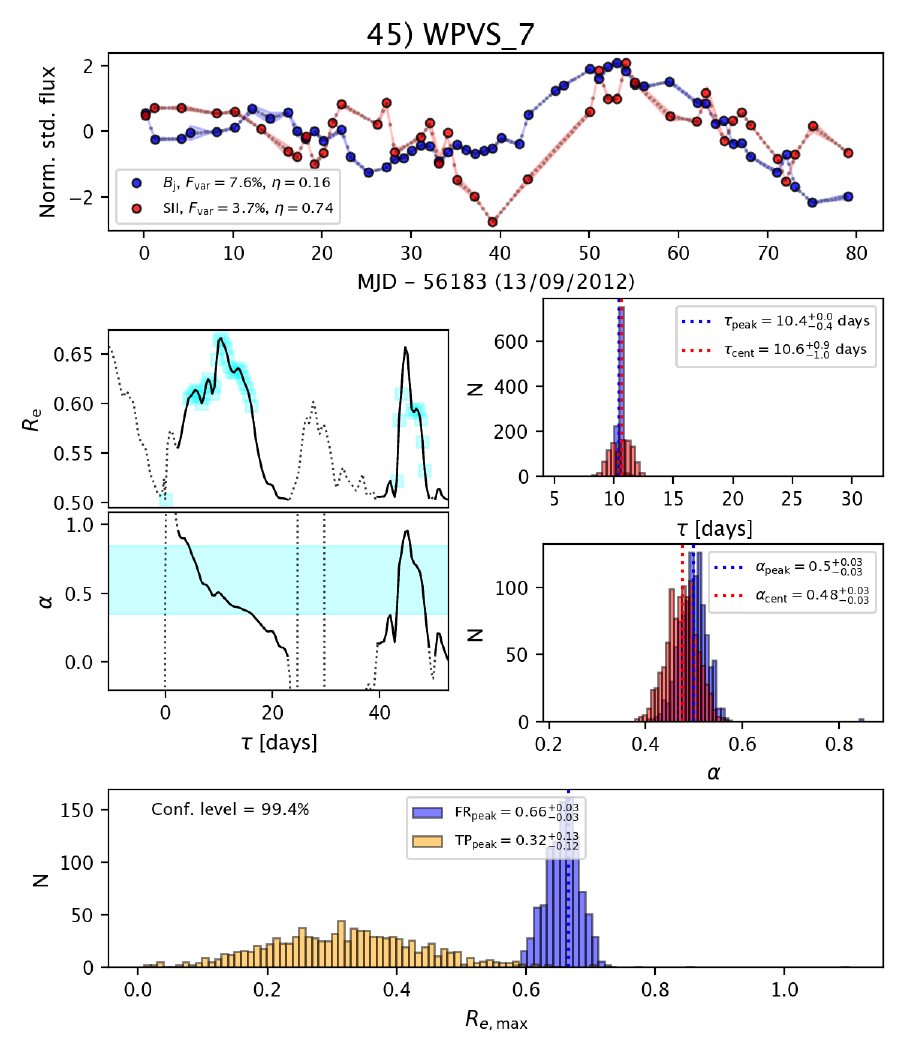}

\newpage
\section{Flux Variation Gradient Diagrams}\label{app:fvg}

Here we present $BV$, $BR$ and $Br_{\rm s}$ flux-flux diagrams for each object and filter combination. For objects with multi-epoch observations, the different colors mean different observation seasons. The bisector fit for the AGN slope is shown as a blue line and its value is denoted in the figure legend. Host-slope ranges were derived from \citet{2010ApJ...711..461S}, and are delineated by grey wedges. The intersection between the host and AGN slopes, provides an estimate for the host contribution to the bands, and is marked by a red star. A summary for all the FVG slopes is provided in Table~\ref{tab:fvg}, and the complete list of flux values (total and host) for each source is summarized in Table~\ref{tab:ad_all}.

\includegraphics[width=0.33\columnwidth]{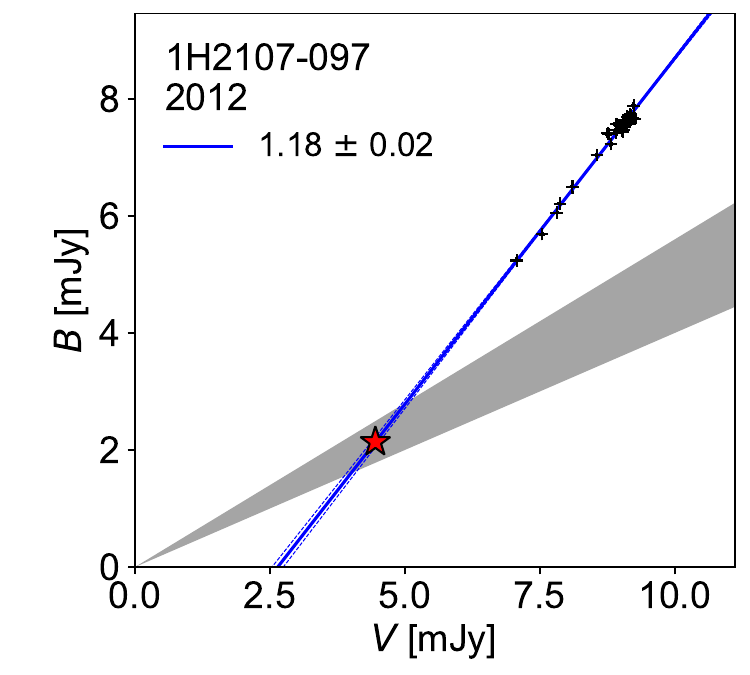}
\includegraphics[width=0.33\columnwidth]{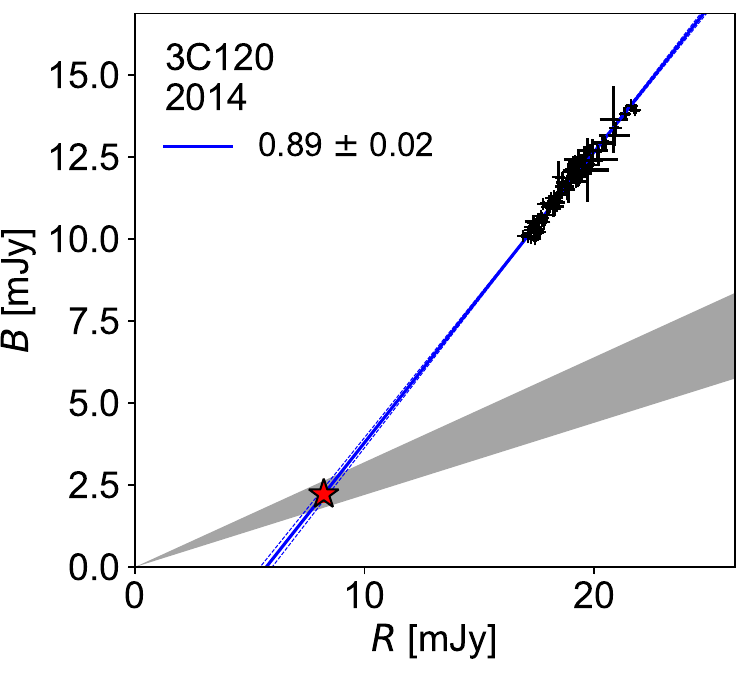}
\includegraphics[width=0.33\columnwidth]{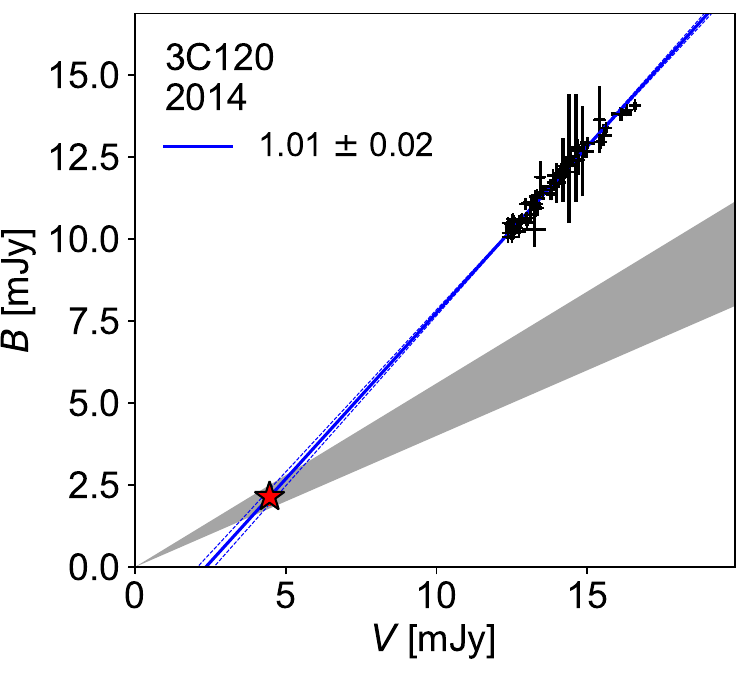}

\includegraphics[width=0.33\columnwidth]{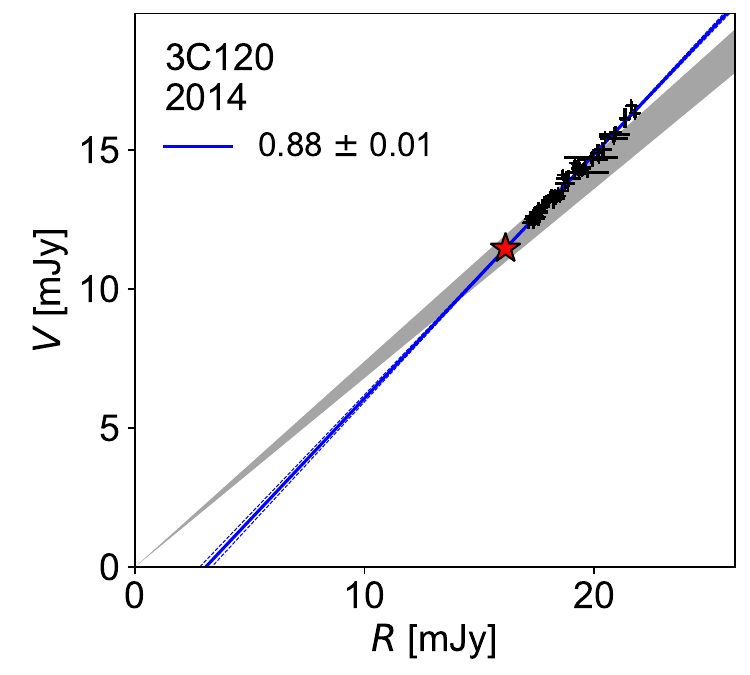}
\includegraphics[width=0.33\columnwidth]{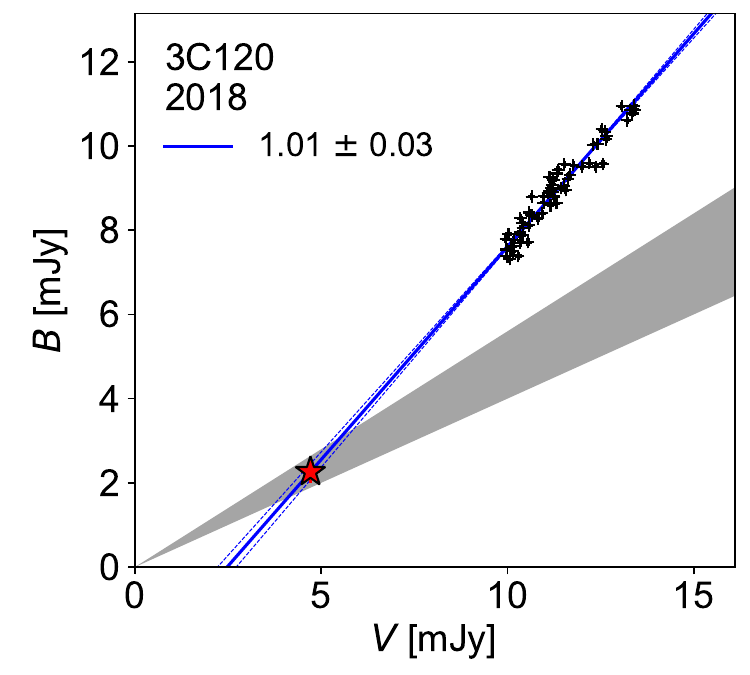}
\includegraphics[width=0.33\columnwidth]{fig_pdf/3C120_mix_FVG_B_j_V_j.pdf}

\includegraphics[width=0.33\columnwidth]{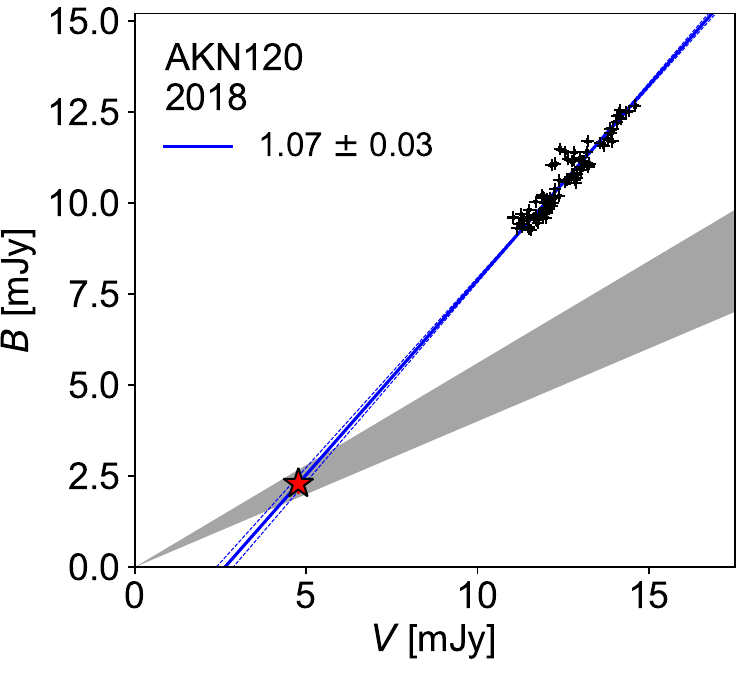}
\includegraphics[width=0.33\columnwidth]{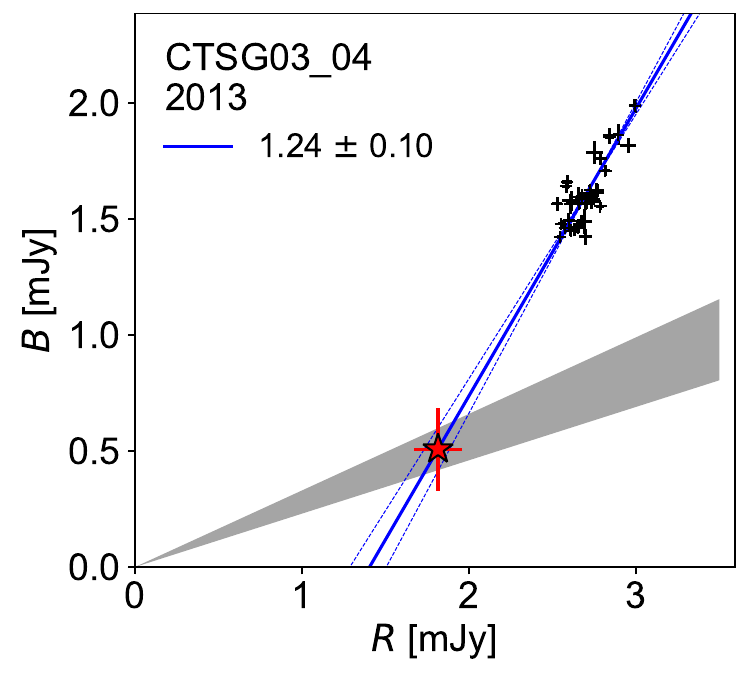}
\includegraphics[width=0.33\columnwidth]{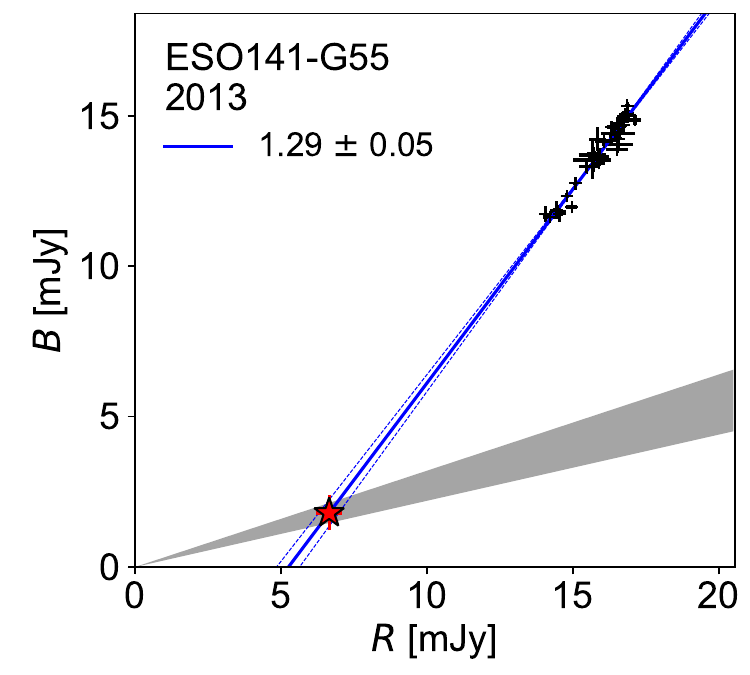}

\includegraphics[width=0.33\columnwidth]{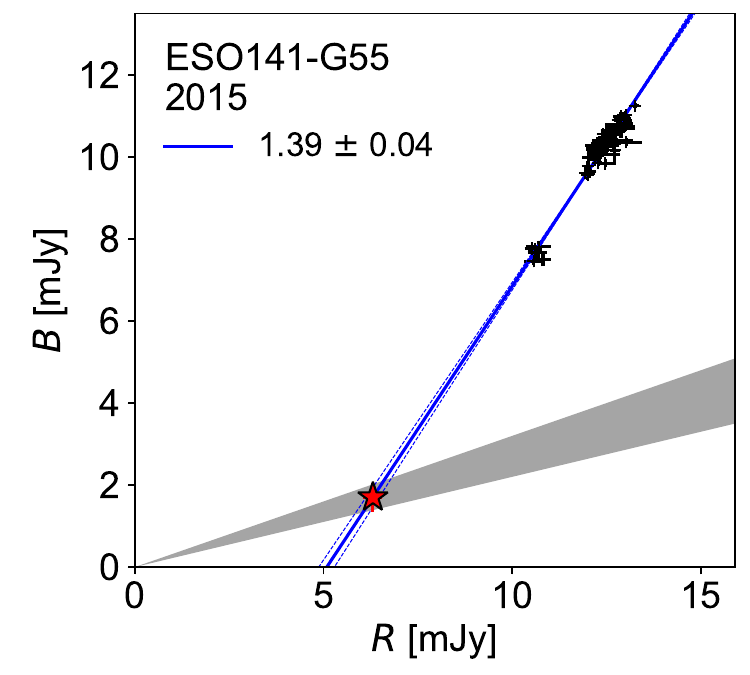}
\includegraphics[width=0.33\columnwidth]{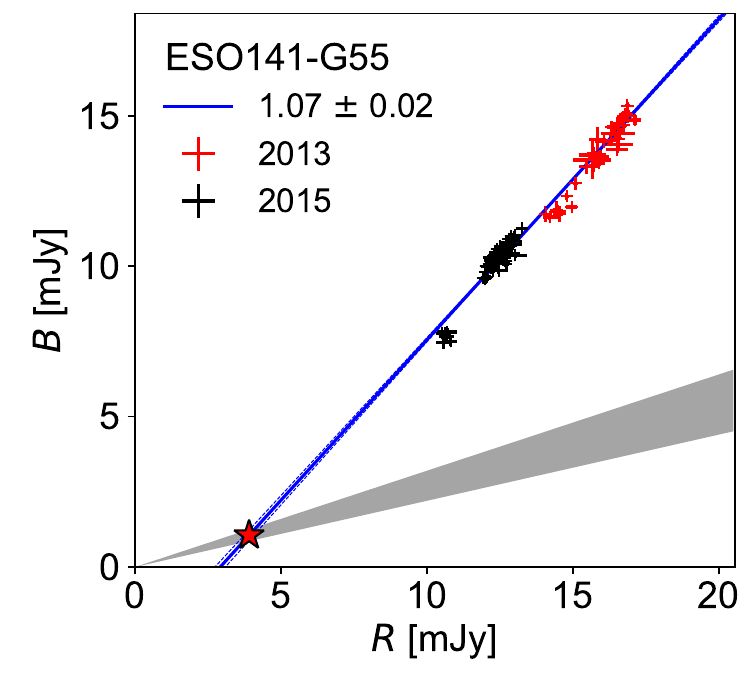}
\includegraphics[width=0.33\columnwidth]{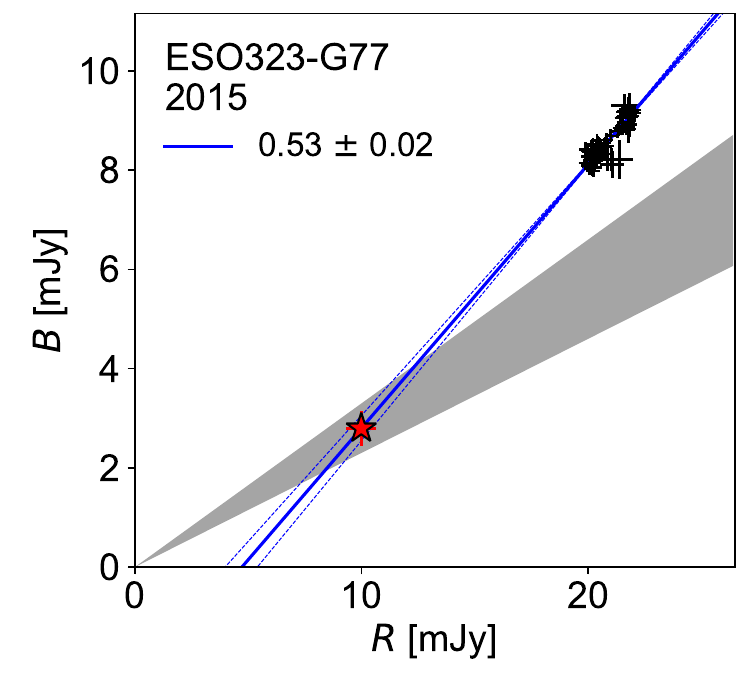}

\includegraphics[width=0.33\columnwidth]{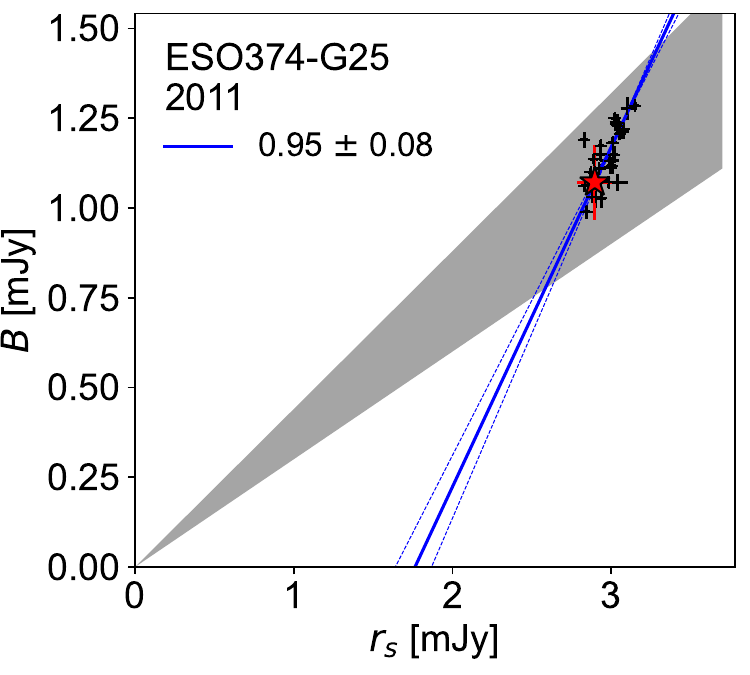}
\includegraphics[width=0.33\columnwidth]{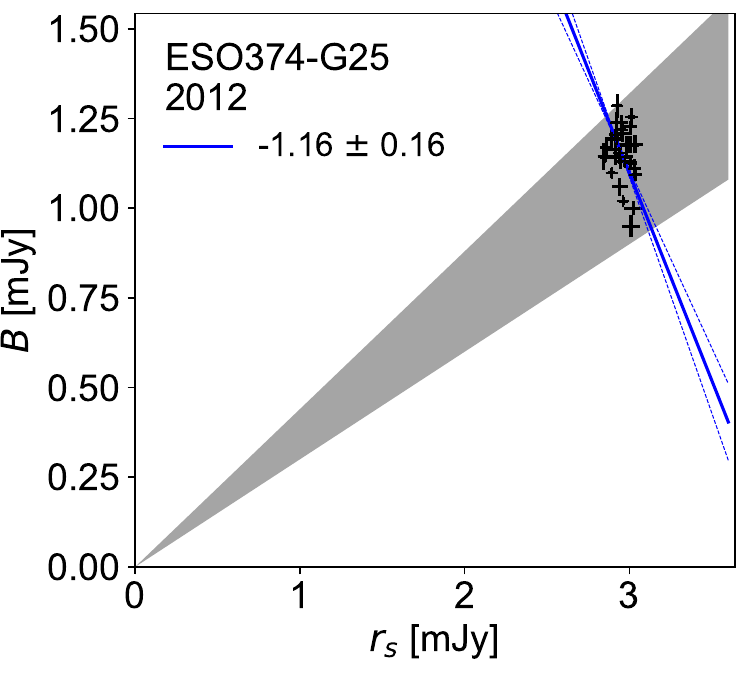}
\includegraphics[width=0.33\columnwidth]{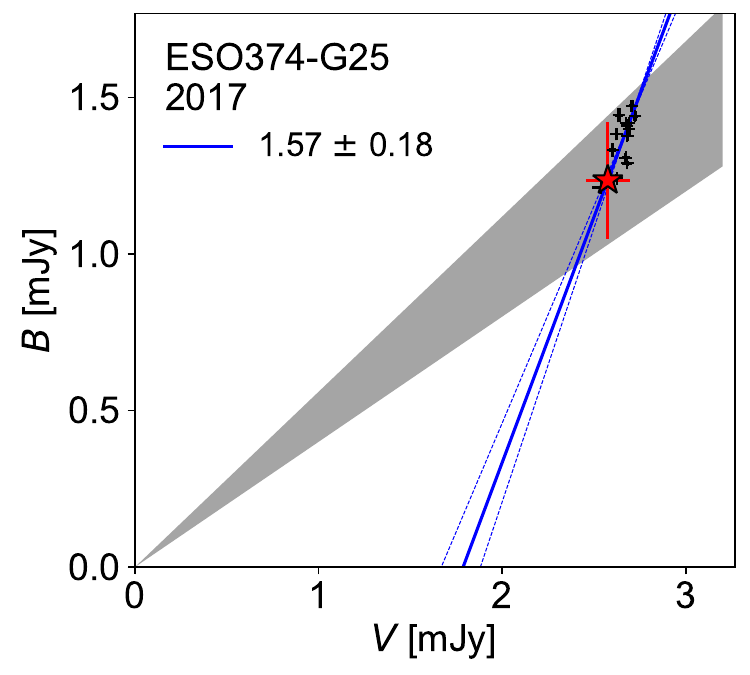}

\includegraphics[width=0.33\columnwidth]{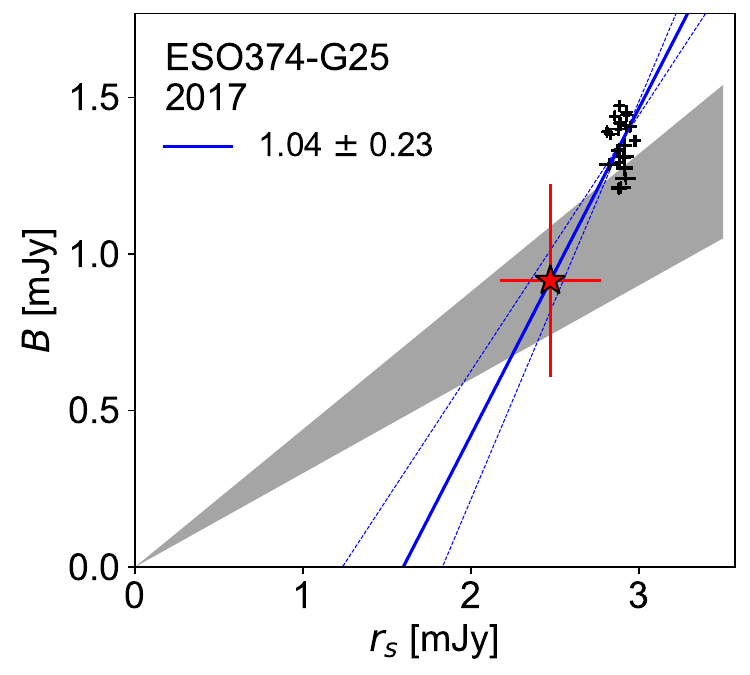}
\includegraphics[width=0.33\columnwidth]{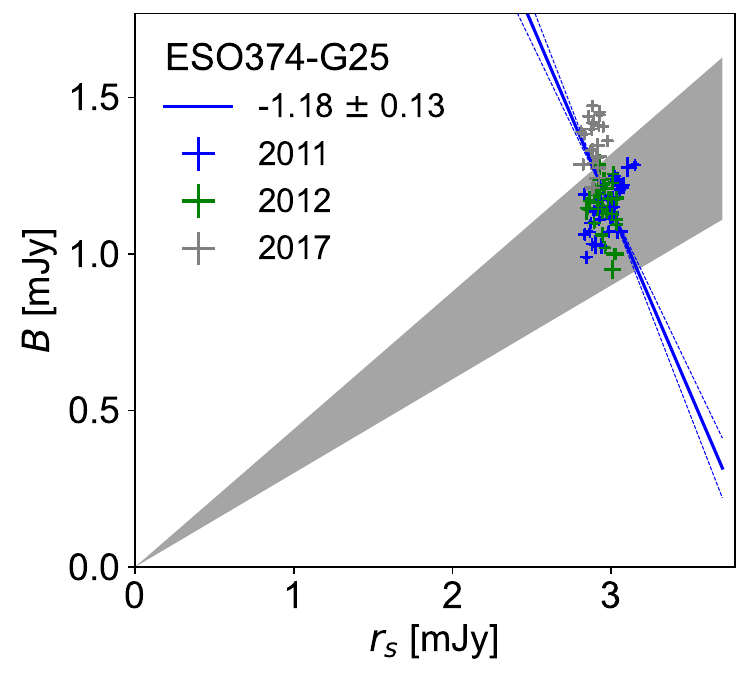}
\includegraphics[width=0.33\columnwidth]{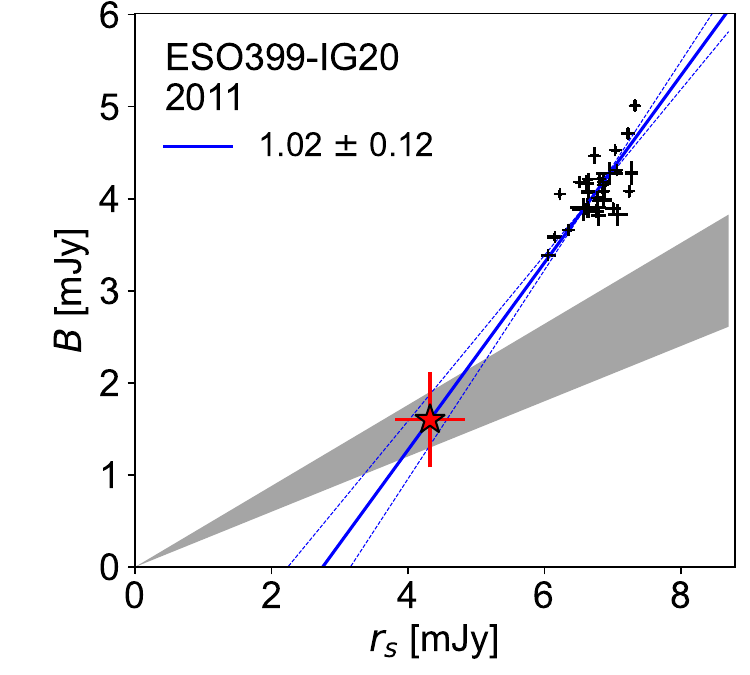}

\includegraphics[width=0.33\columnwidth]{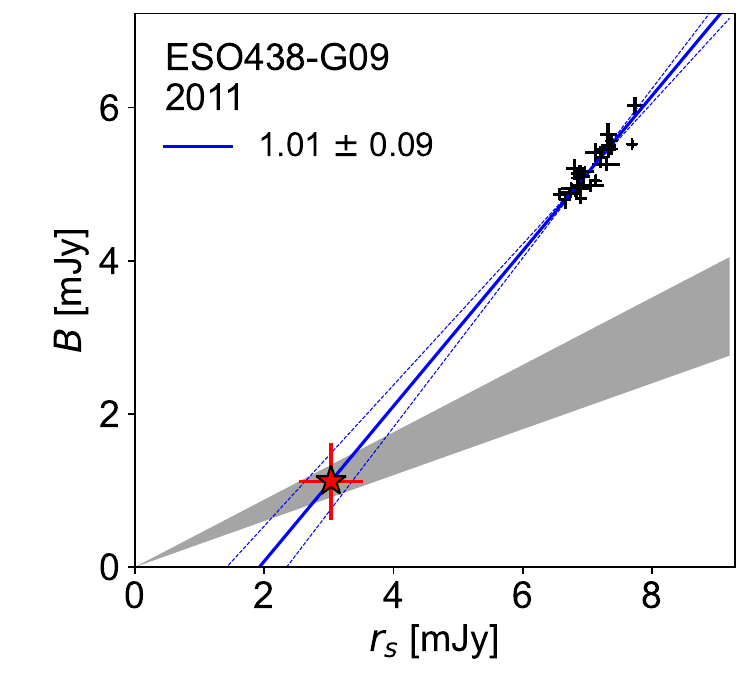}
\includegraphics[width=0.33\columnwidth]{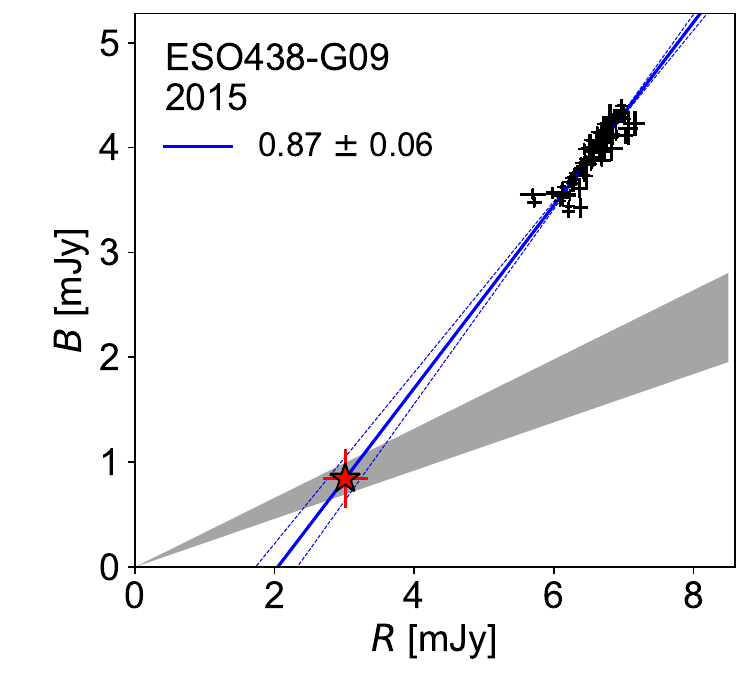}
\includegraphics[width=0.33\columnwidth]{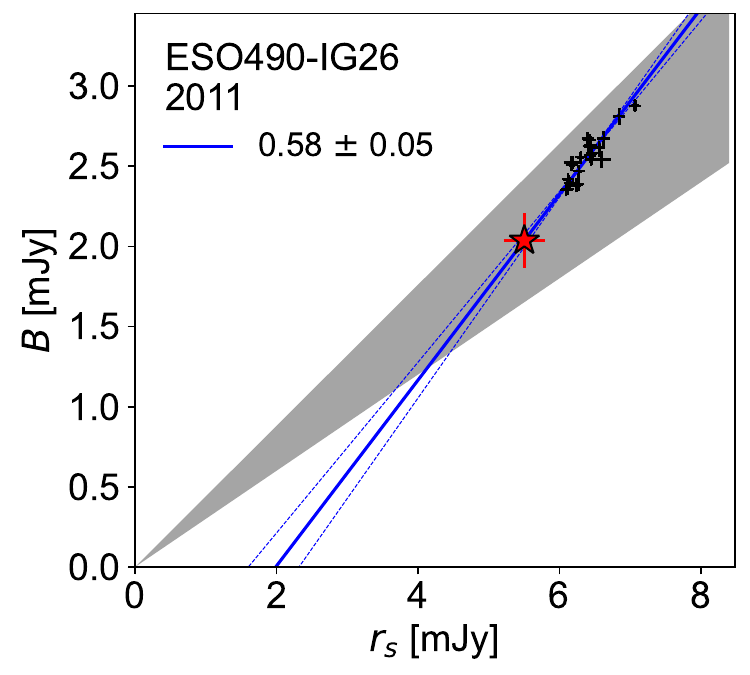}

\includegraphics[width=0.33\columnwidth]{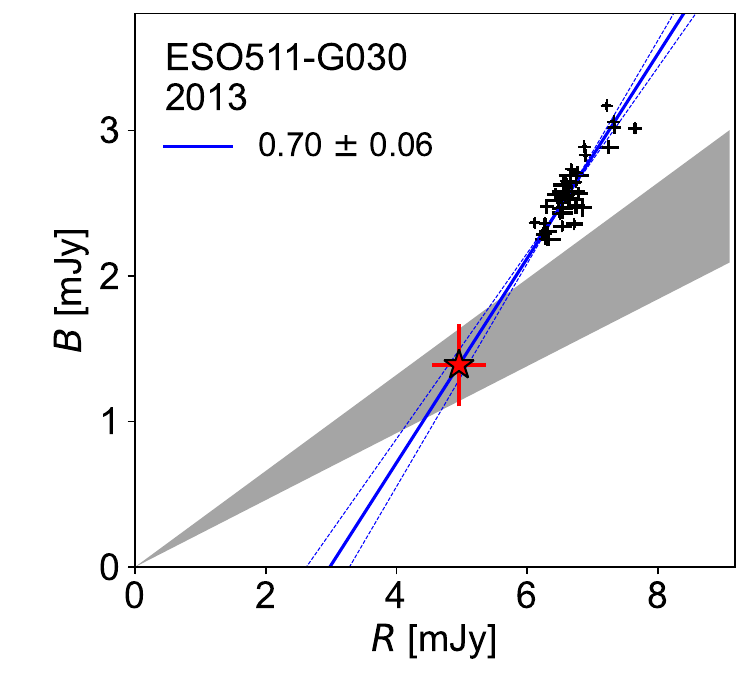}
\includegraphics[width=0.33\columnwidth]{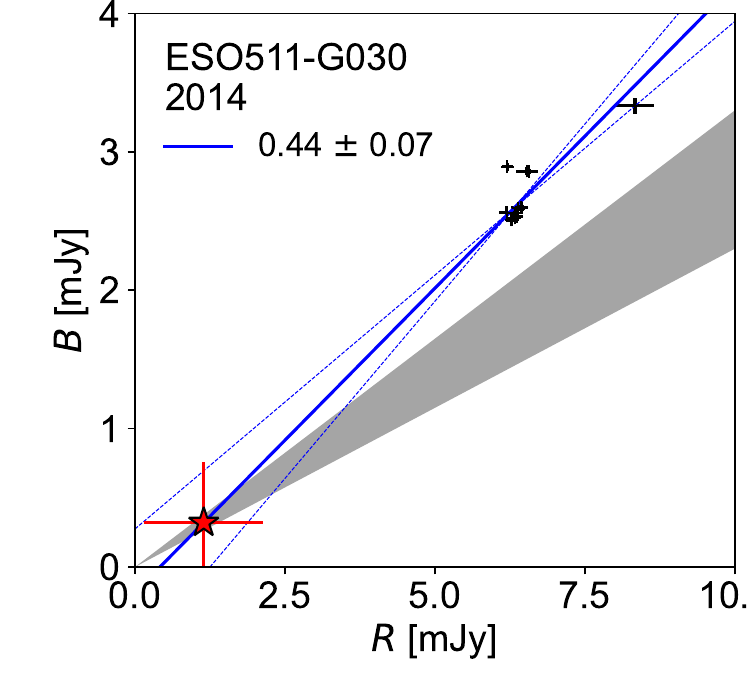}
\includegraphics[width=0.33\columnwidth]{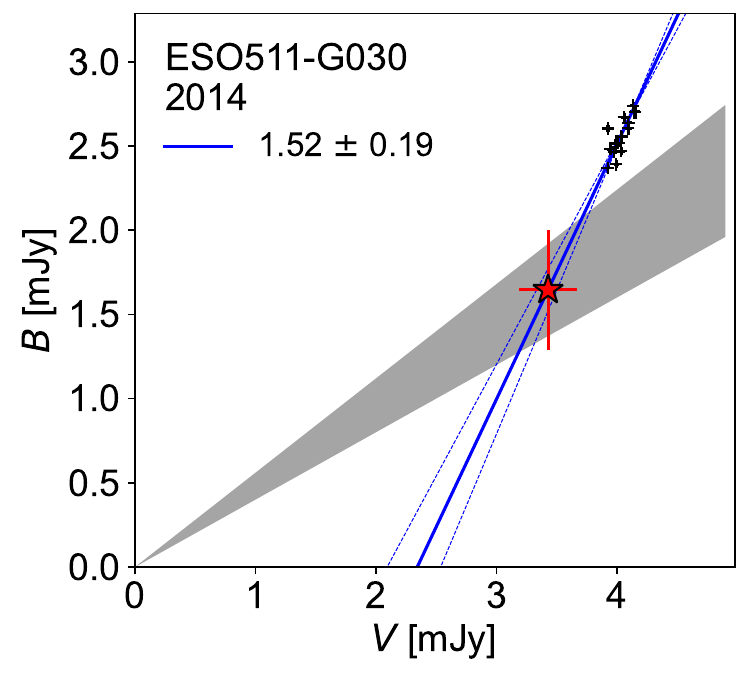}

\includegraphics[width=0.33\columnwidth]{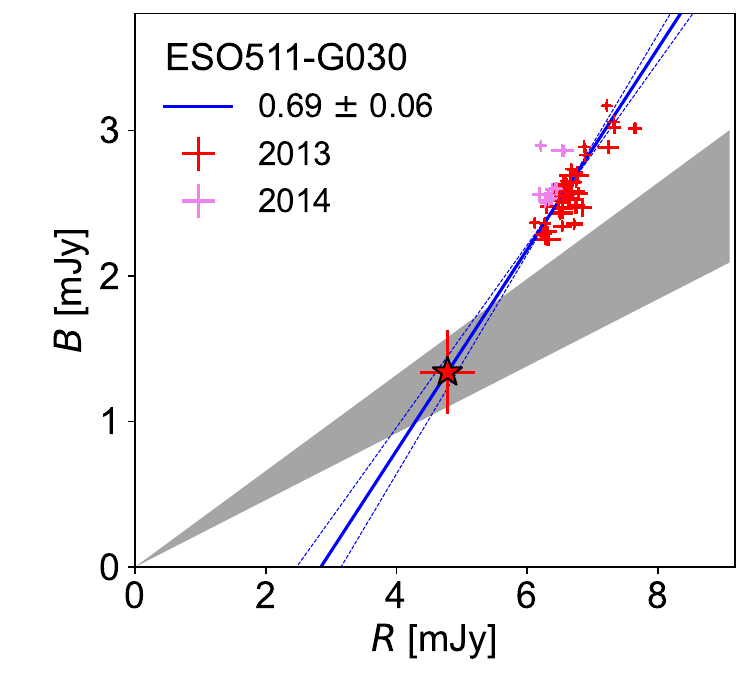}
\includegraphics[width=0.33\columnwidth]{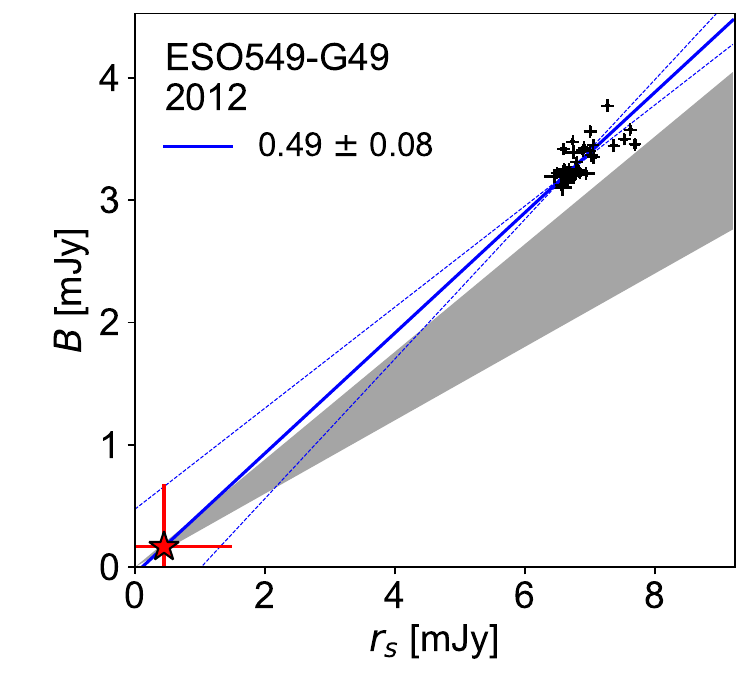}
\includegraphics[width=0.33\columnwidth]{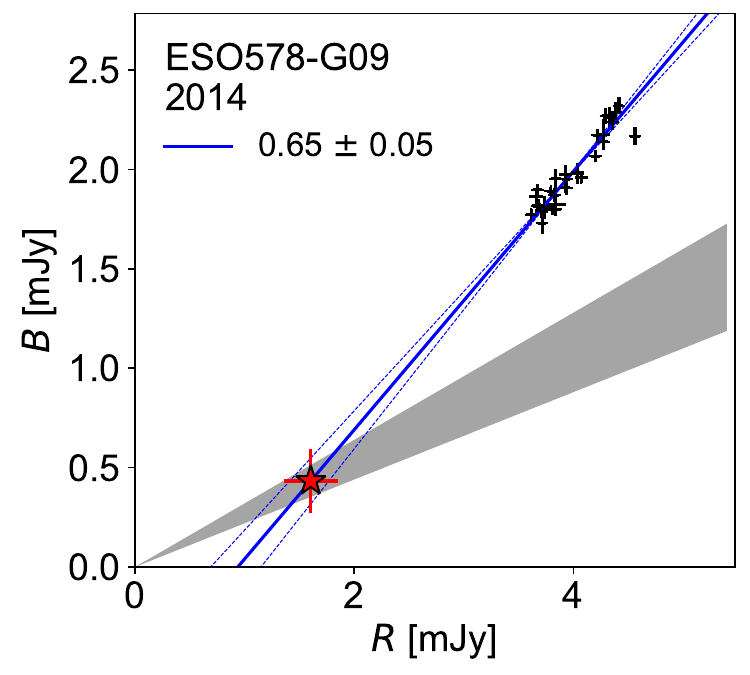}

\includegraphics[width=0.33\columnwidth]{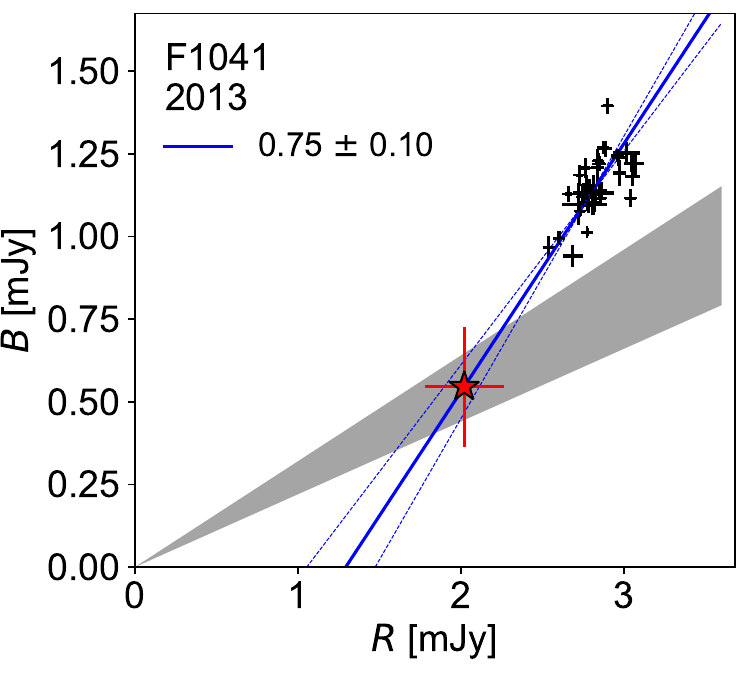}
\includegraphics[width=0.33\columnwidth]{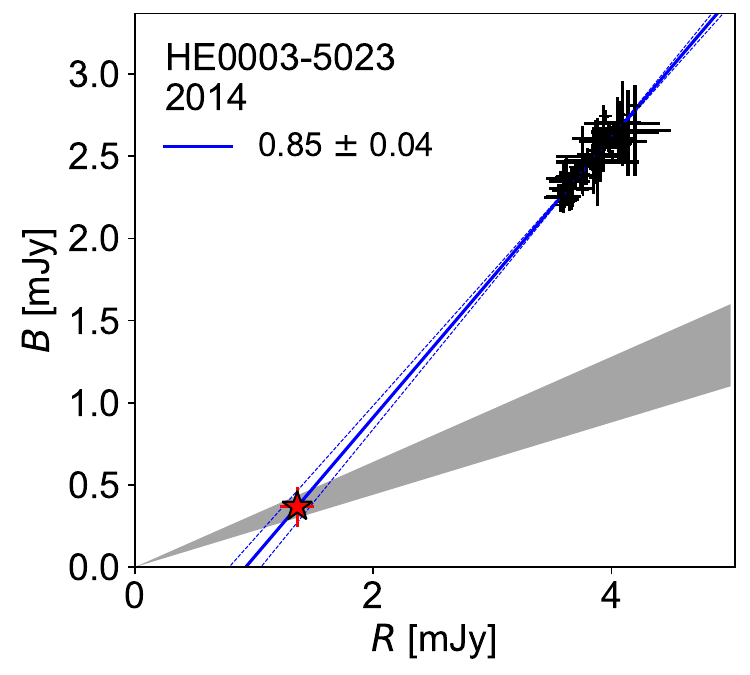}
\includegraphics[width=0.33\columnwidth]{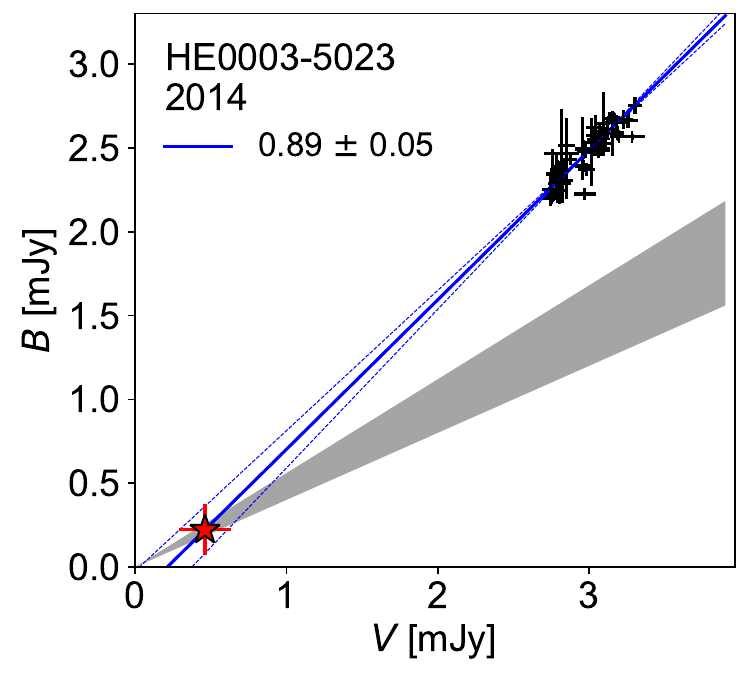}

\includegraphics[width=0.33\columnwidth]{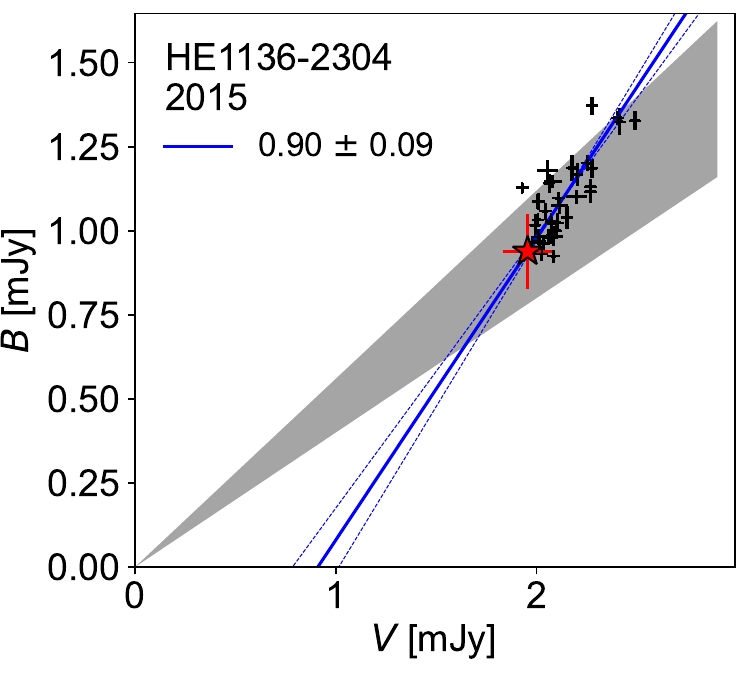}
\includegraphics[width=0.33\columnwidth]{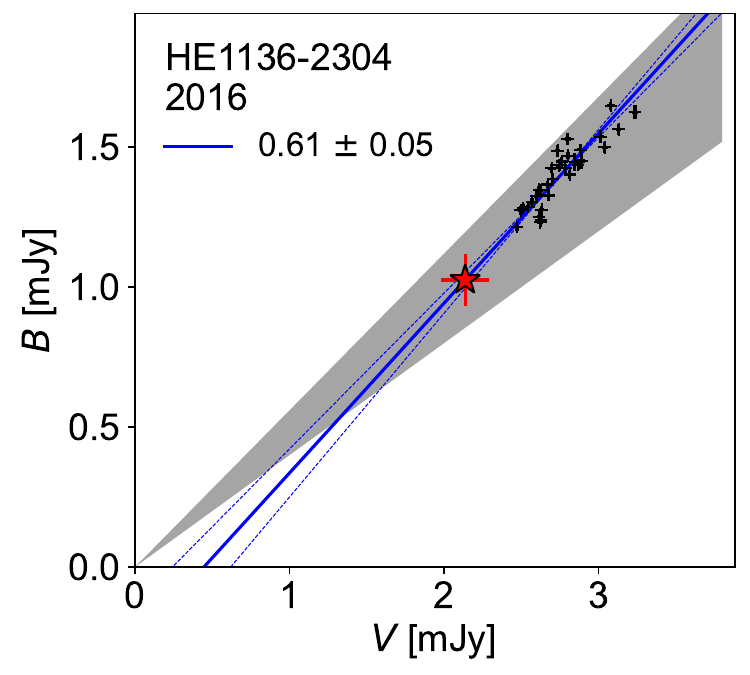}
\includegraphics[width=0.33\columnwidth]{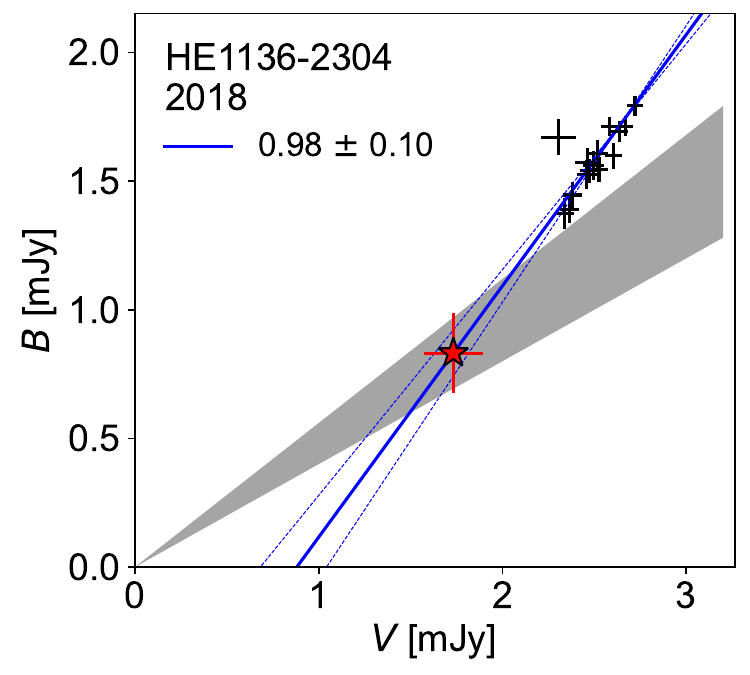}

\includegraphics[width=0.33\columnwidth]{fig_pdf/HE1136-2304_mix_FVG_B_j_V_j.pdf}
\includegraphics[width=0.33\columnwidth]{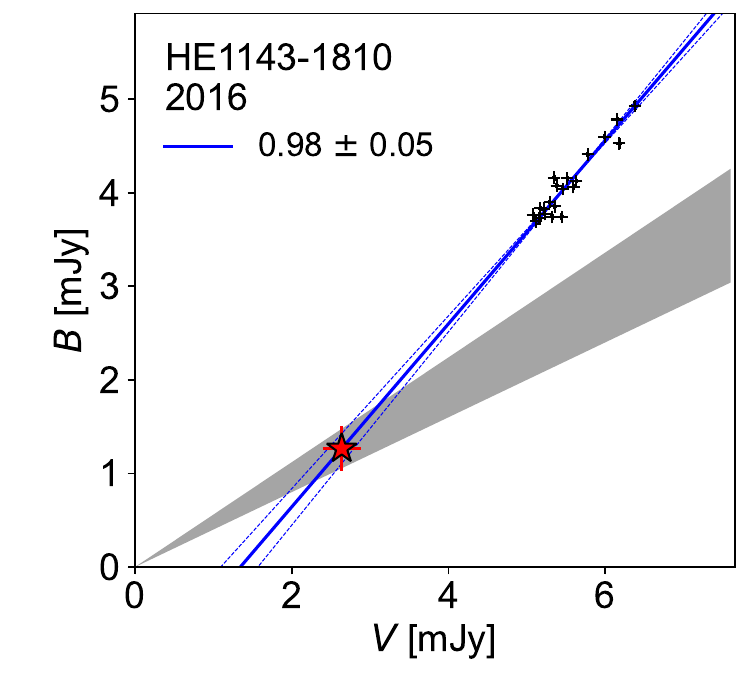}
\includegraphics[width=0.33\columnwidth]{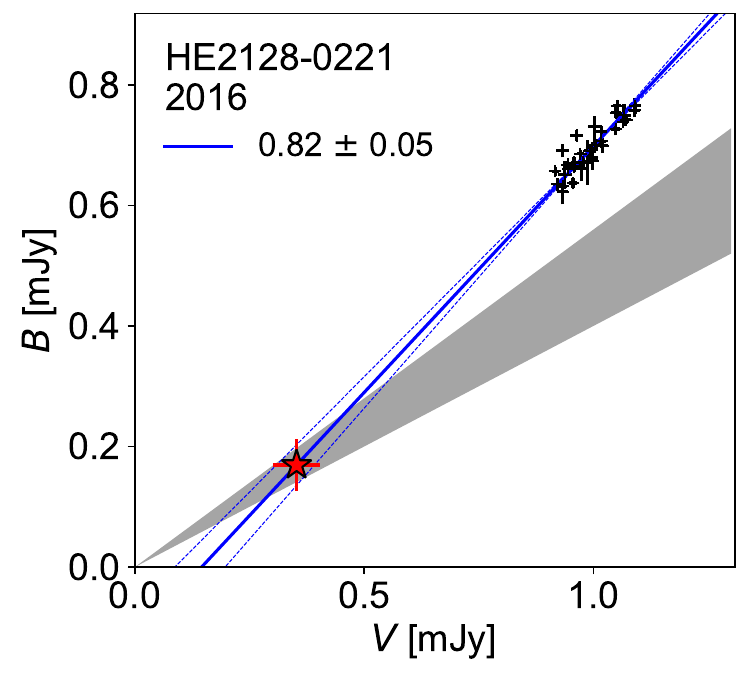}

\includegraphics[width=0.33\columnwidth]{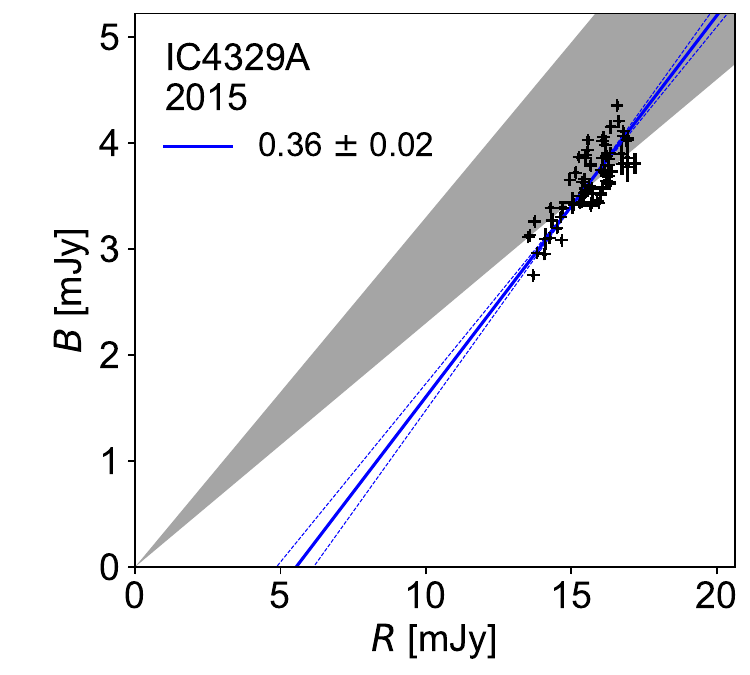}
\includegraphics[width=0.33\columnwidth]{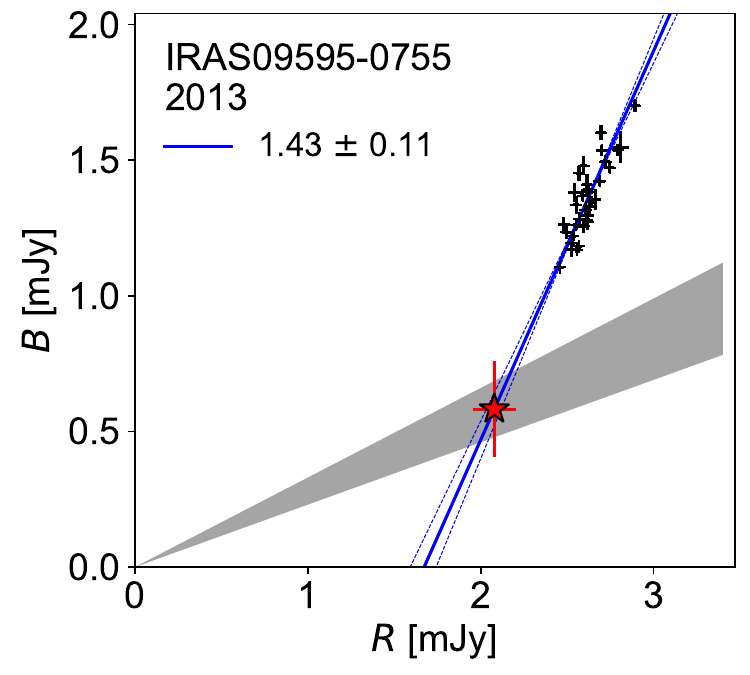}
\includegraphics[width=0.33\columnwidth]{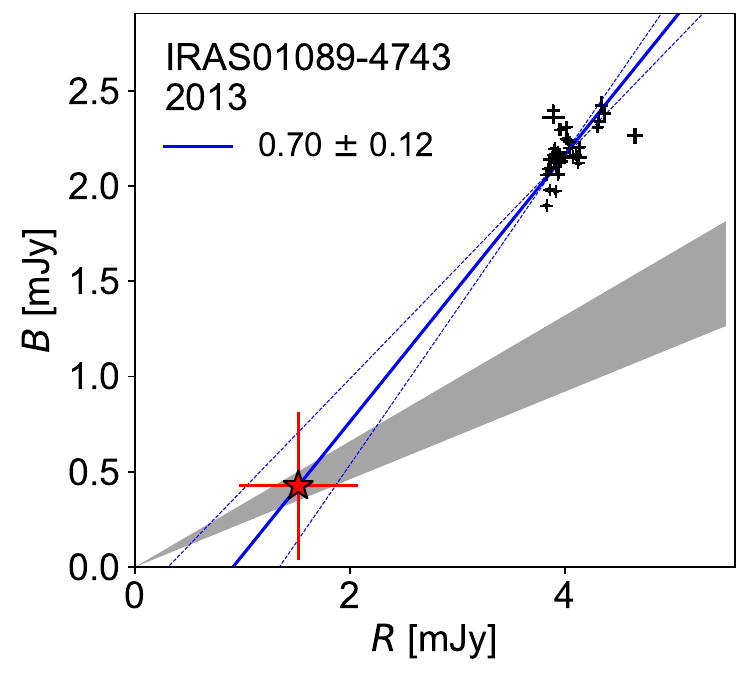}

\includegraphics[width=0.33\columnwidth]{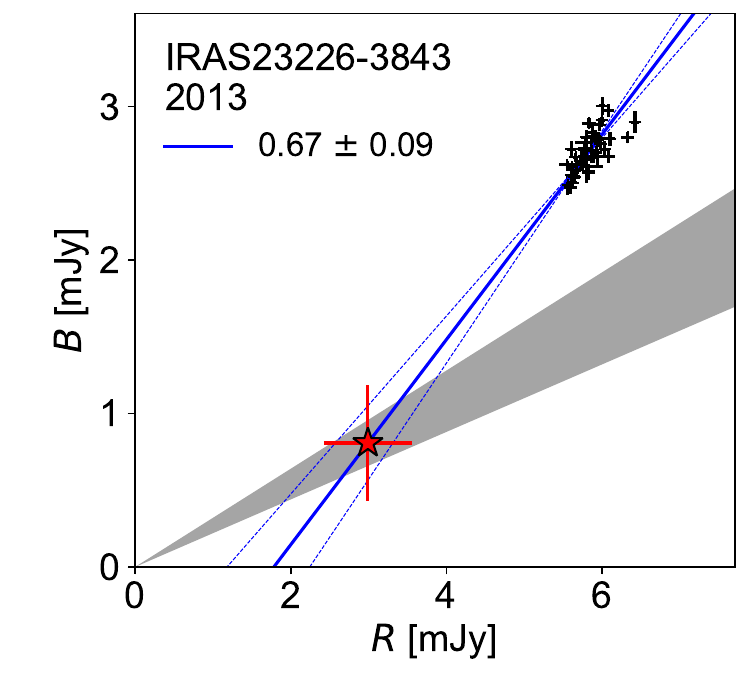}
\includegraphics[width=0.33\columnwidth]{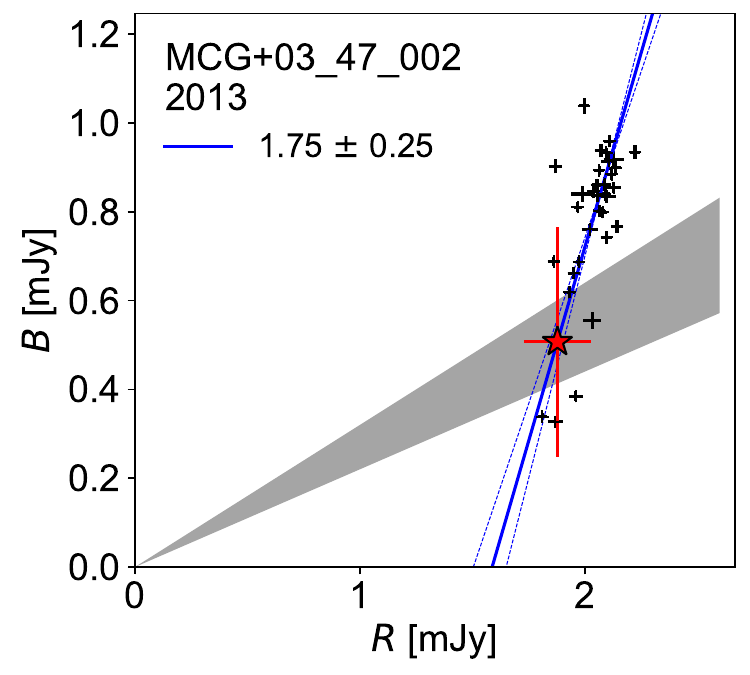}
\includegraphics[width=0.33\columnwidth]{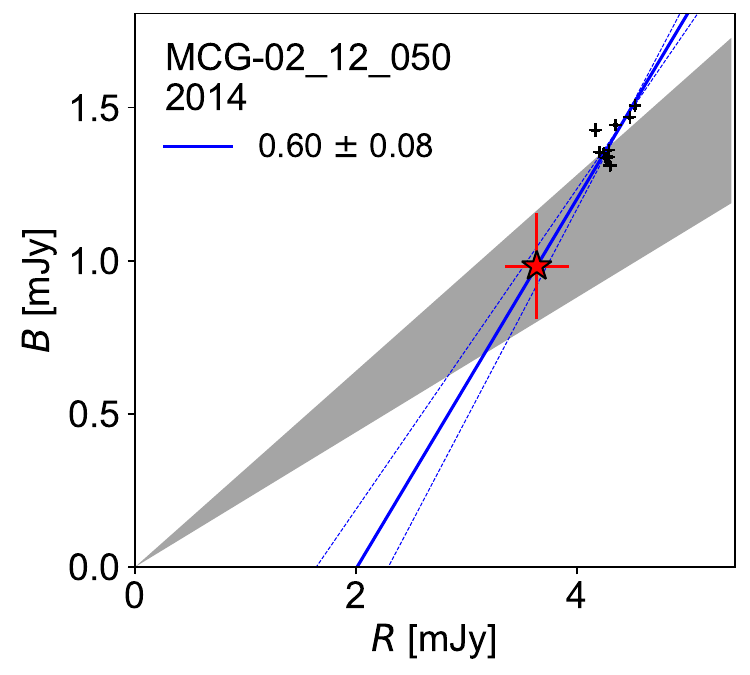}

\includegraphics[width=0.33\columnwidth]{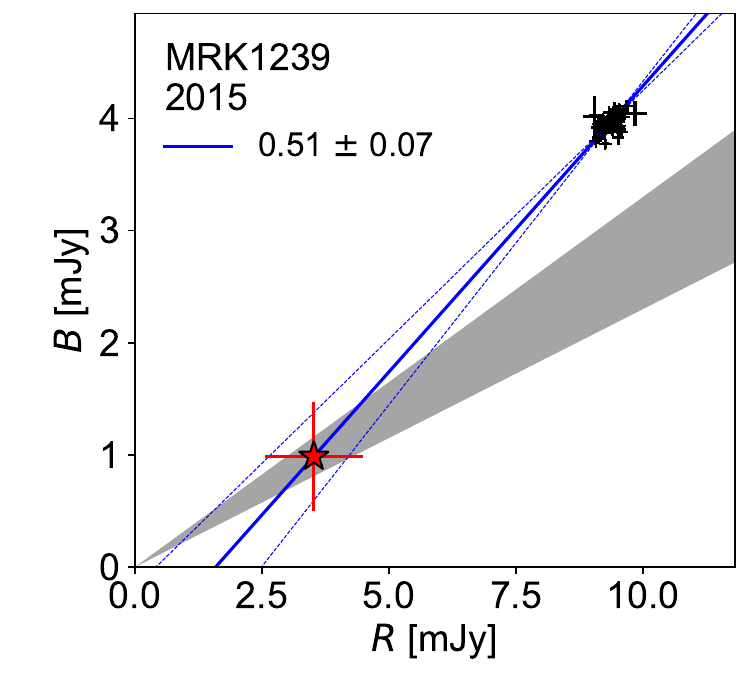}
\includegraphics[width=0.33\columnwidth]{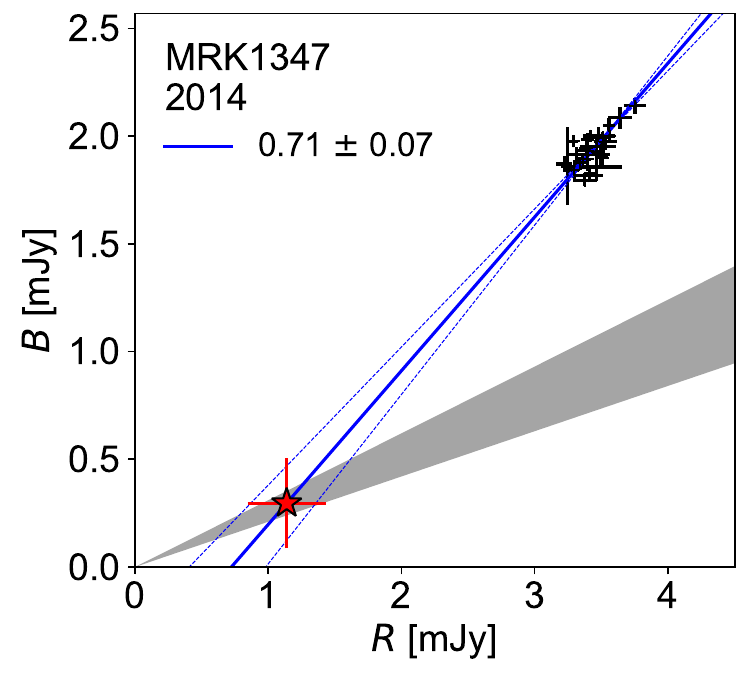}
\includegraphics[width=0.33\columnwidth]{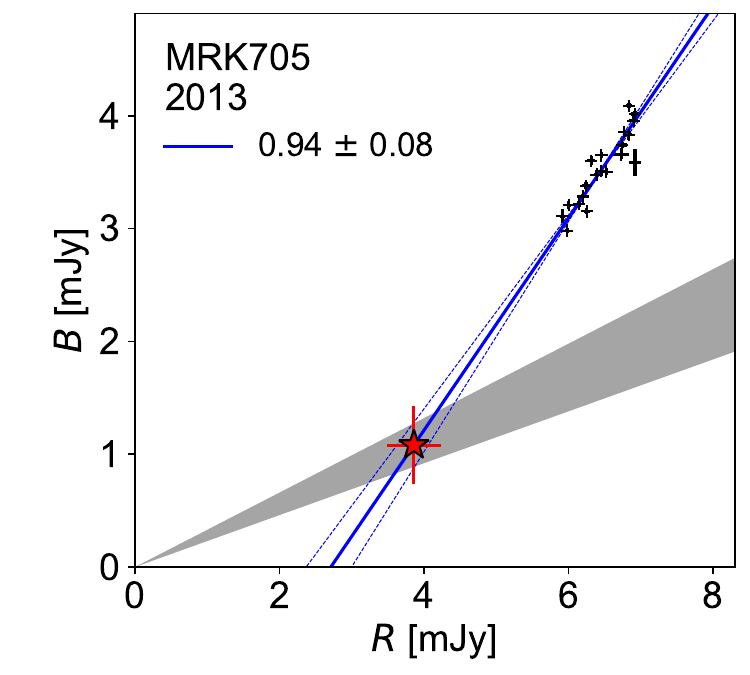}

\includegraphics[width=0.33\columnwidth]{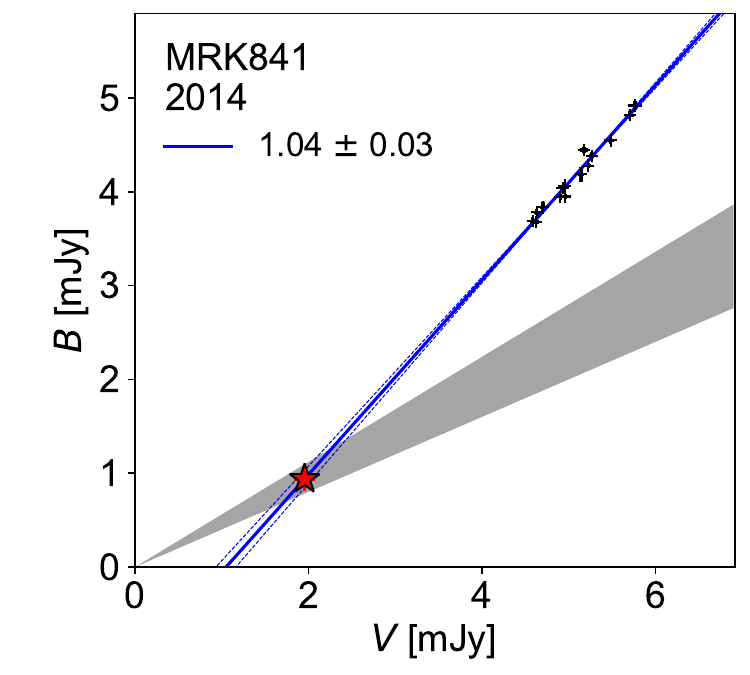}
\includegraphics[width=0.33\columnwidth]{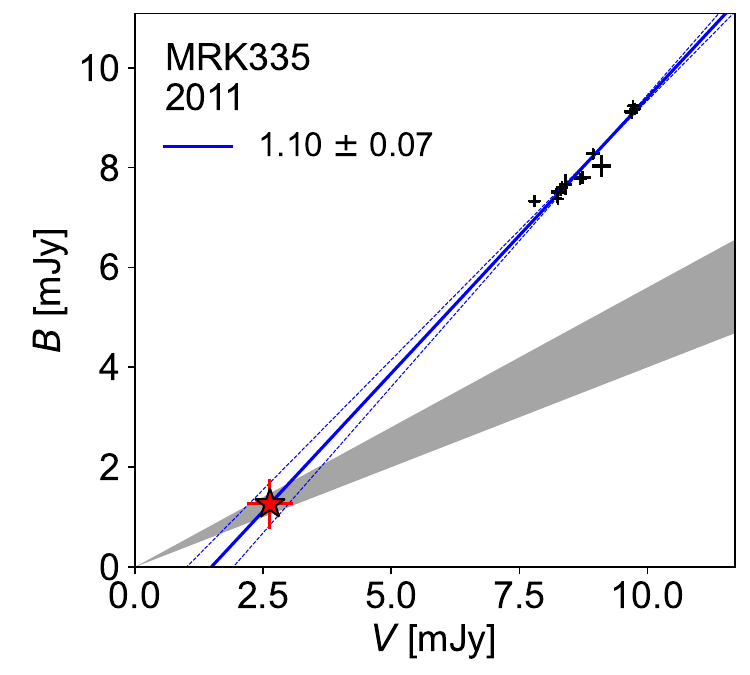}
\includegraphics[width=0.33\columnwidth]{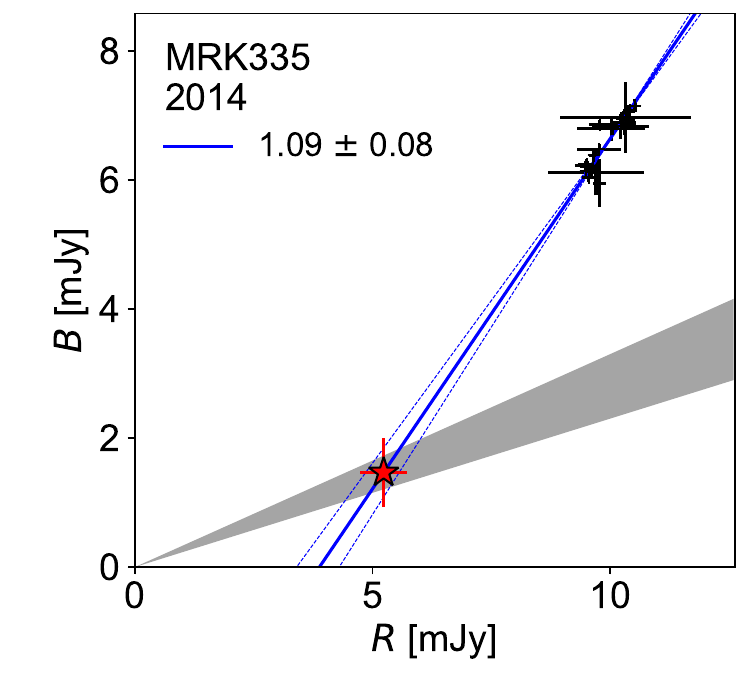}

\includegraphics[width=0.33\columnwidth]{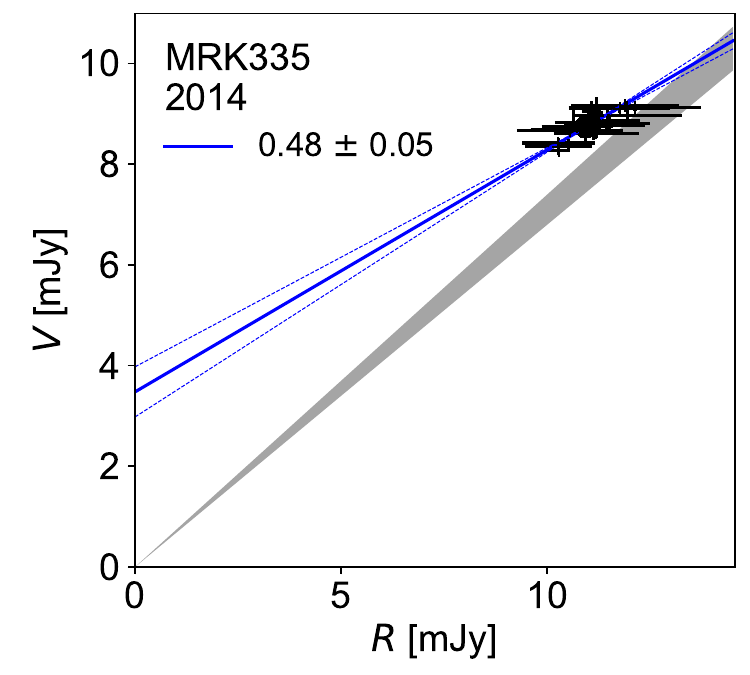}
\includegraphics[width=0.33\columnwidth]{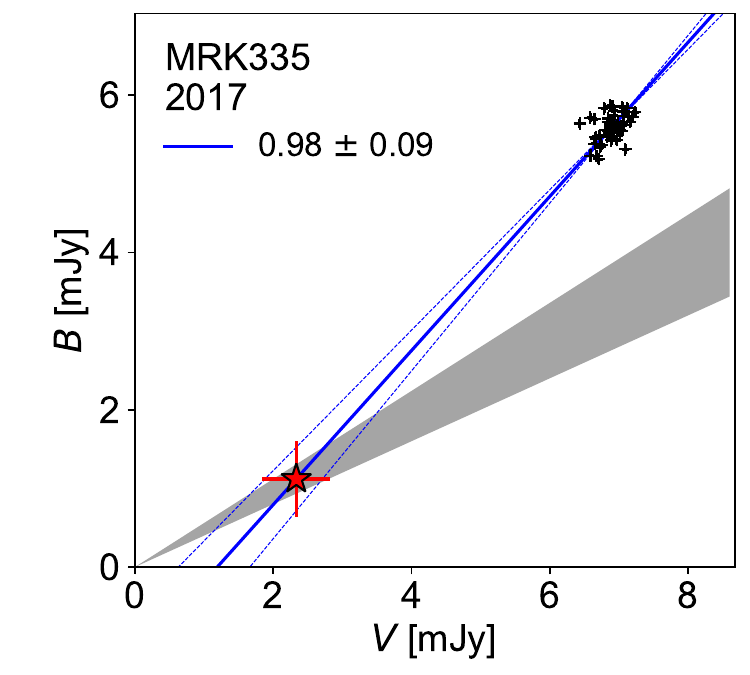}
\includegraphics[width=0.33\columnwidth]{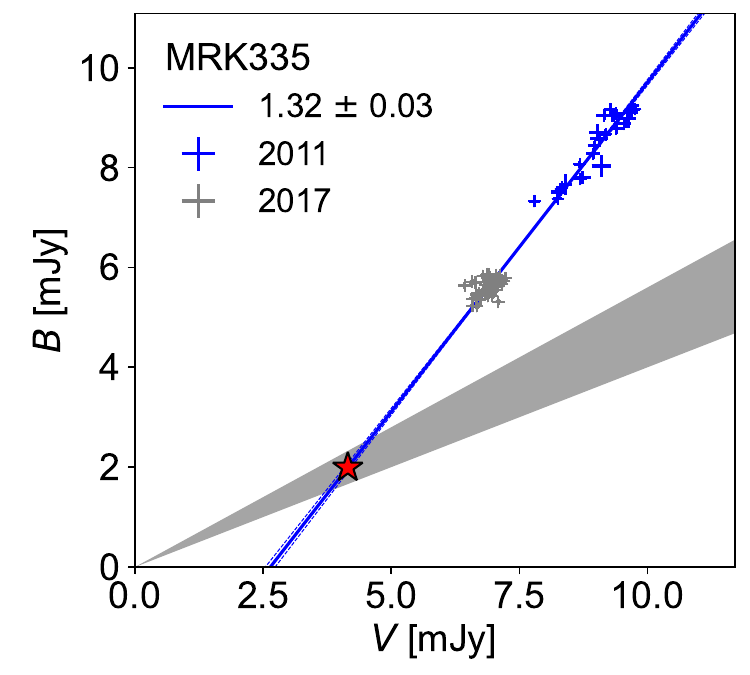}

\includegraphics[width=0.33\columnwidth]{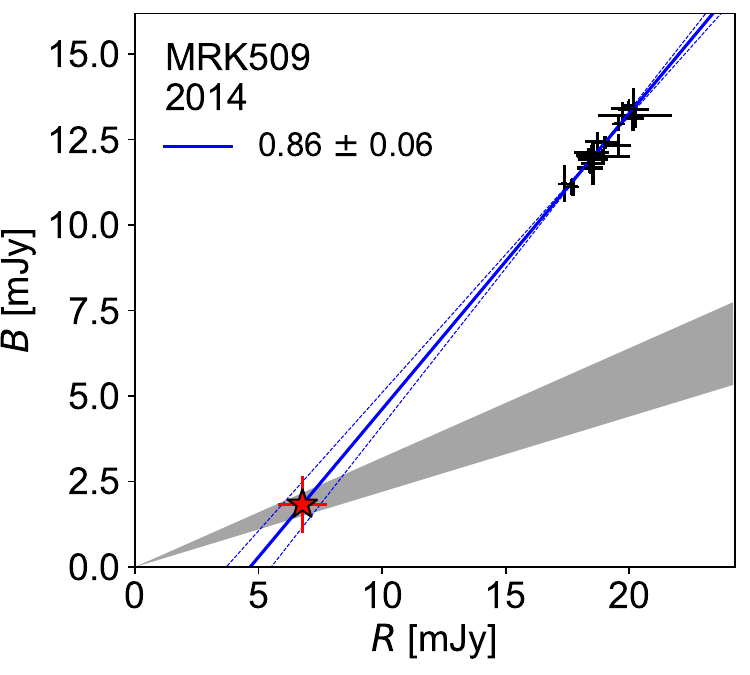}
\includegraphics[width=0.33\columnwidth]{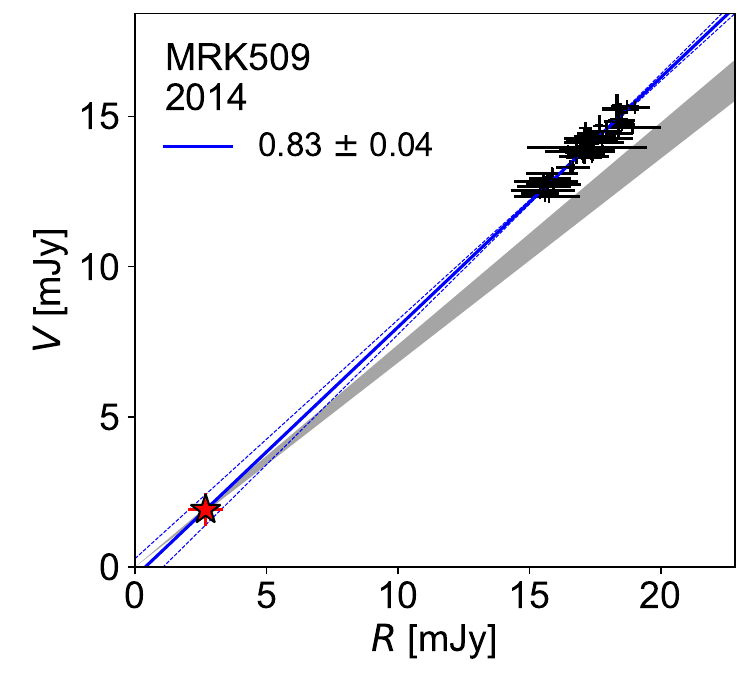}
\includegraphics[width=0.33\columnwidth]{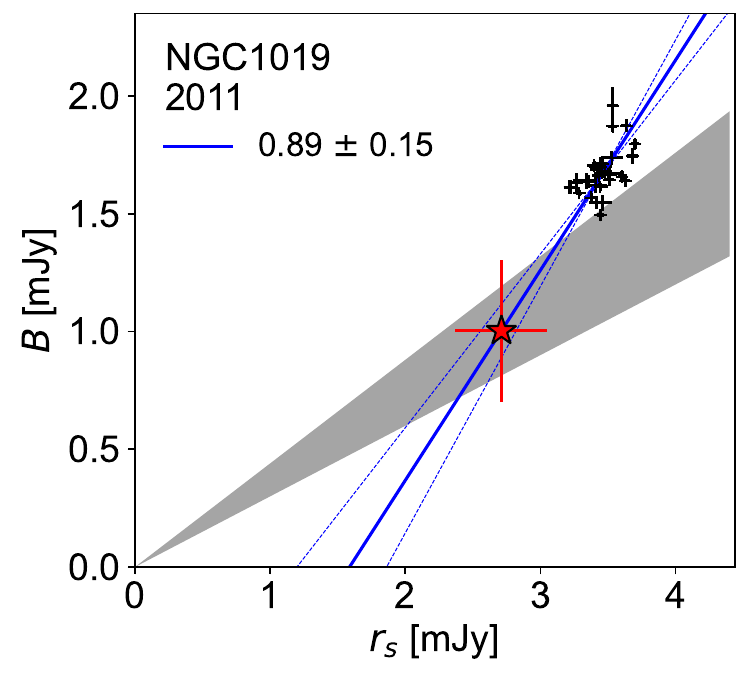}

\includegraphics[width=0.33\columnwidth]{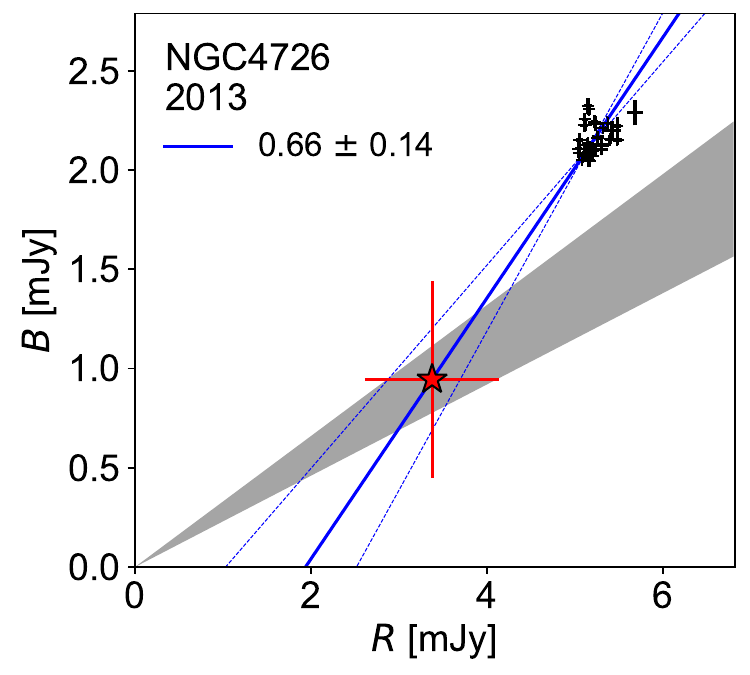}
\includegraphics[width=0.33\columnwidth]{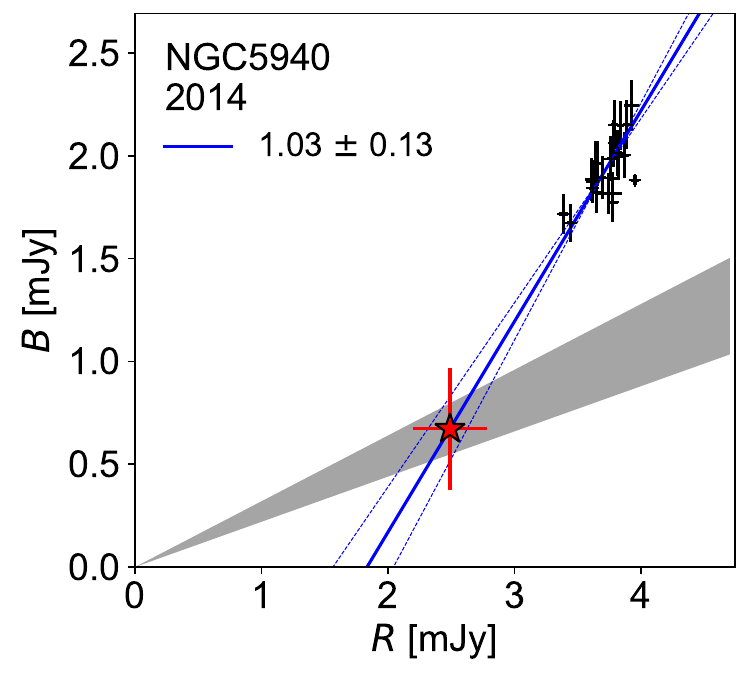}
\includegraphics[width=0.33\columnwidth]{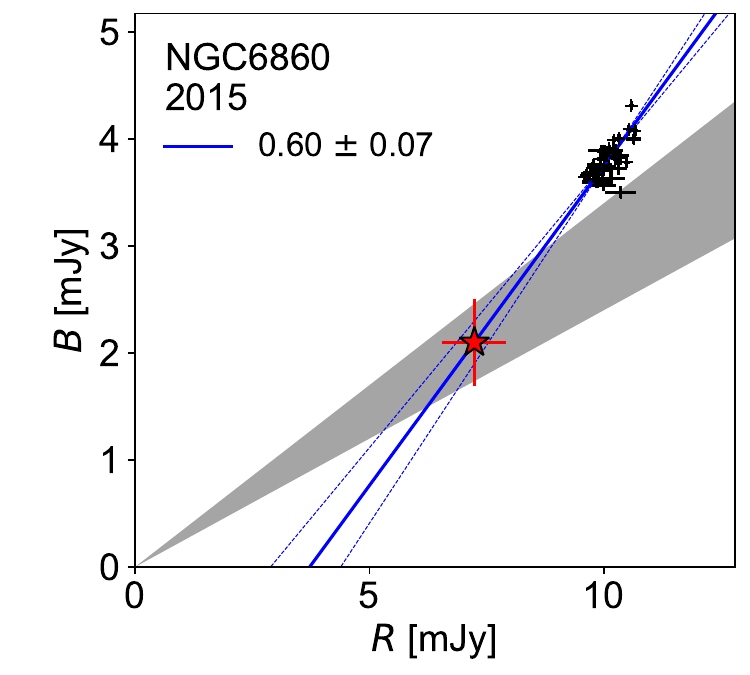}

\includegraphics[width=0.33\columnwidth]{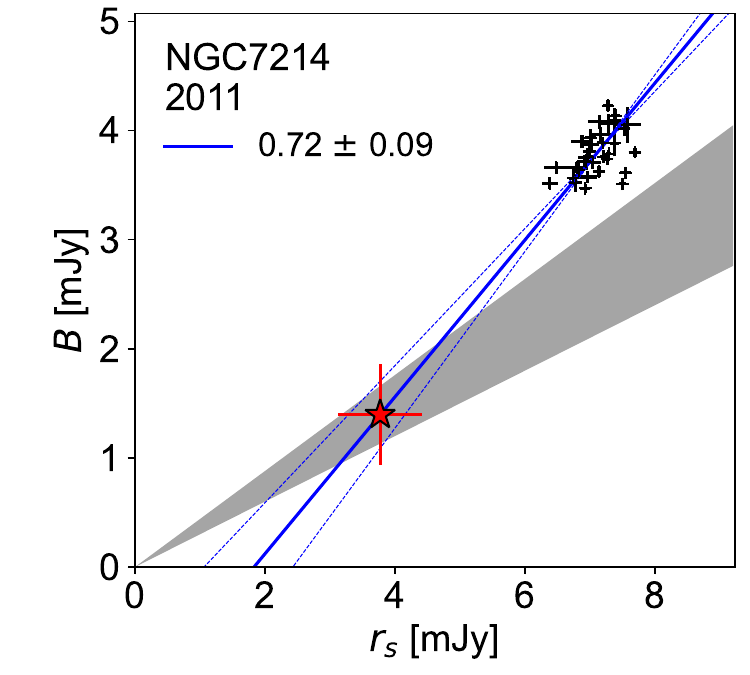}
\includegraphics[width=0.33\columnwidth]{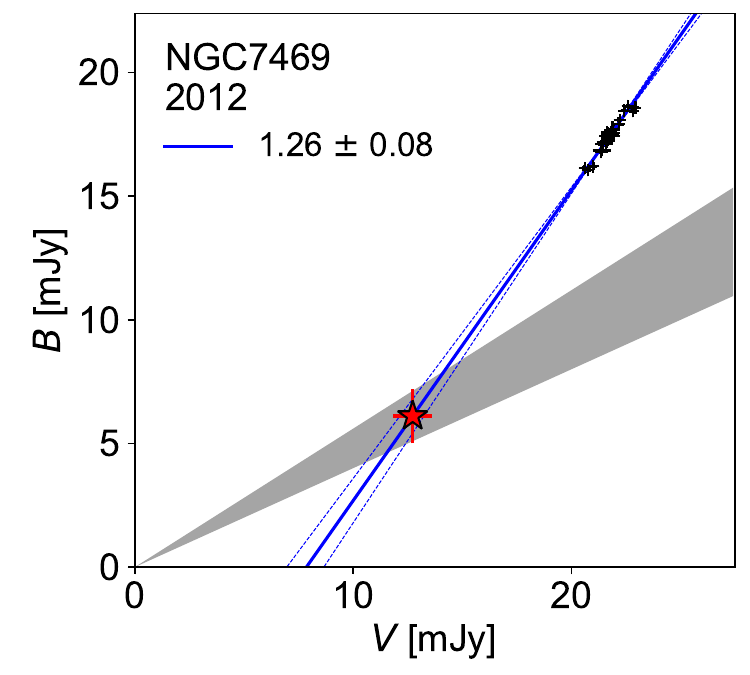}
\includegraphics[width=0.33\columnwidth]{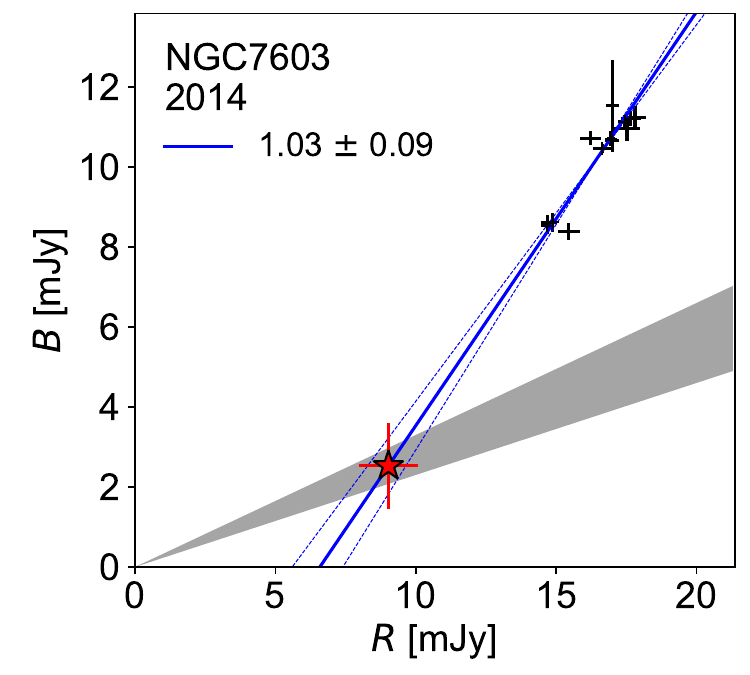}

\includegraphics[width=0.33\columnwidth]{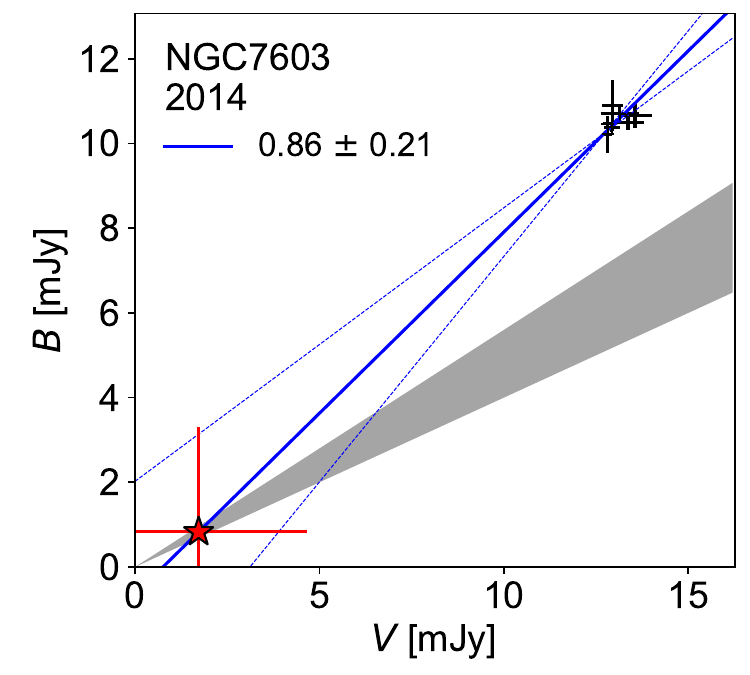}
\includegraphics[width=0.33\columnwidth]{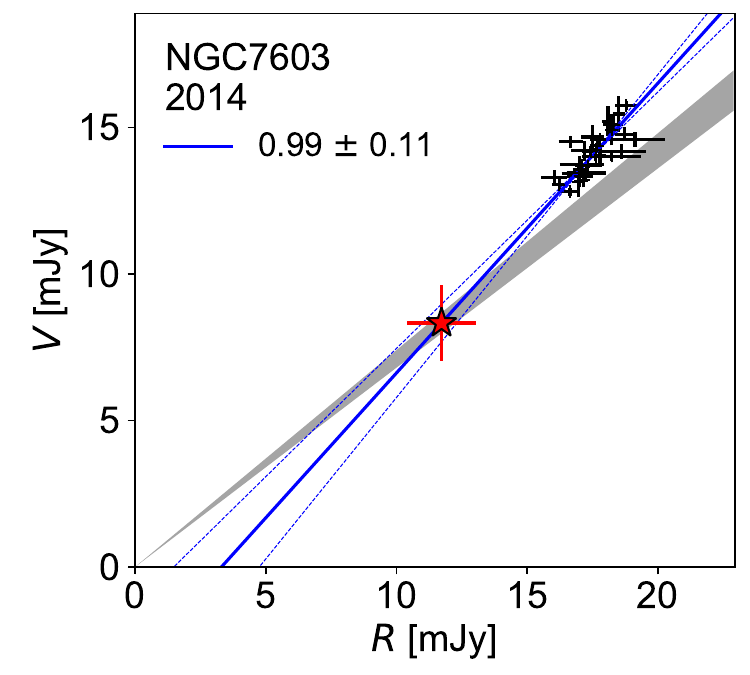}
\includegraphics[width=0.33\columnwidth]{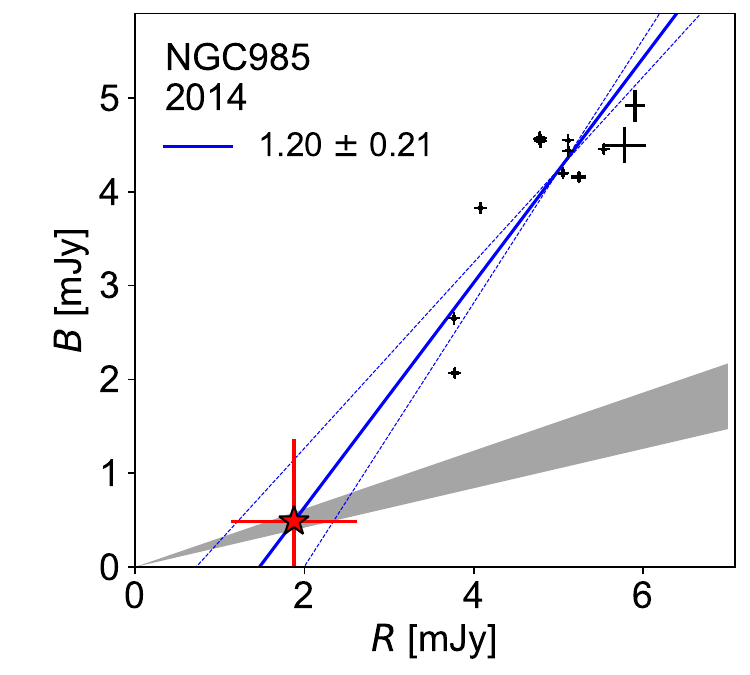}

\includegraphics[width=0.33\columnwidth]{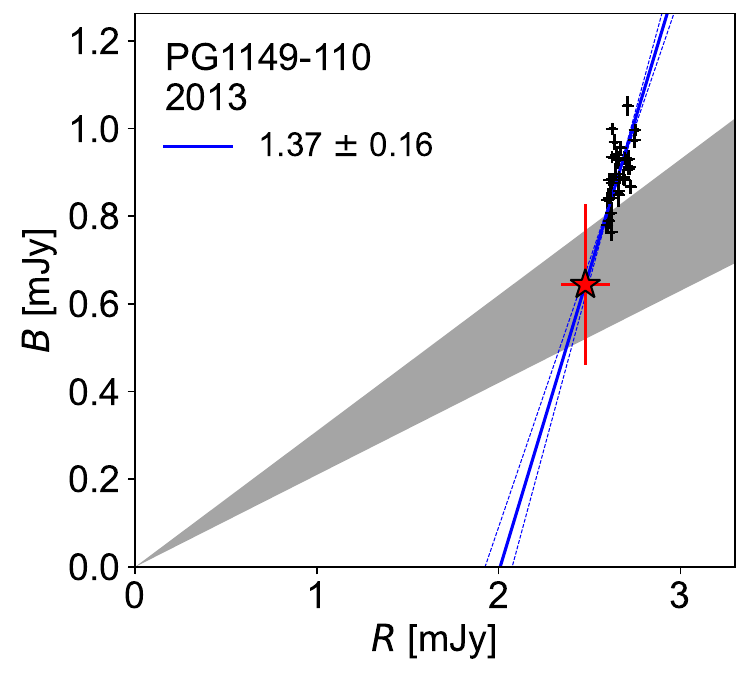}
\includegraphics[width=0.33\columnwidth]{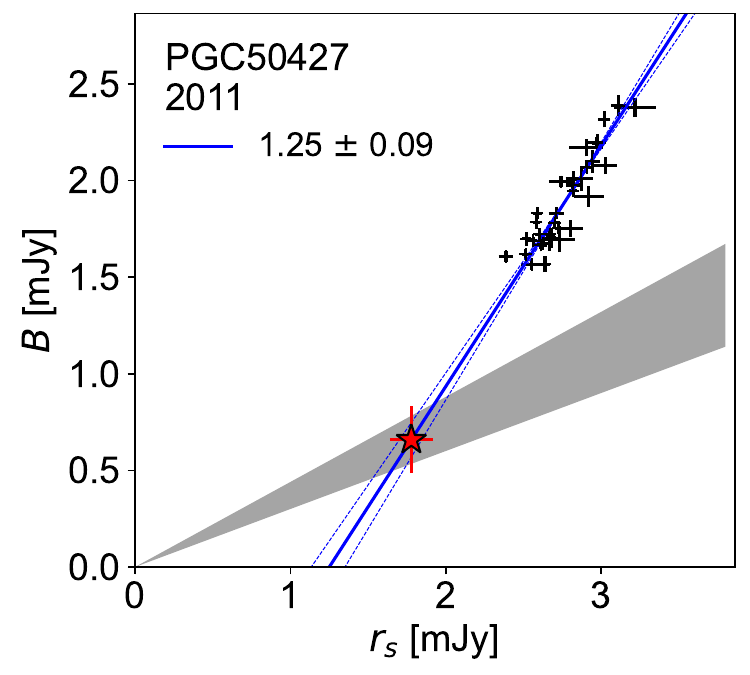}
\includegraphics[width=0.33\columnwidth]{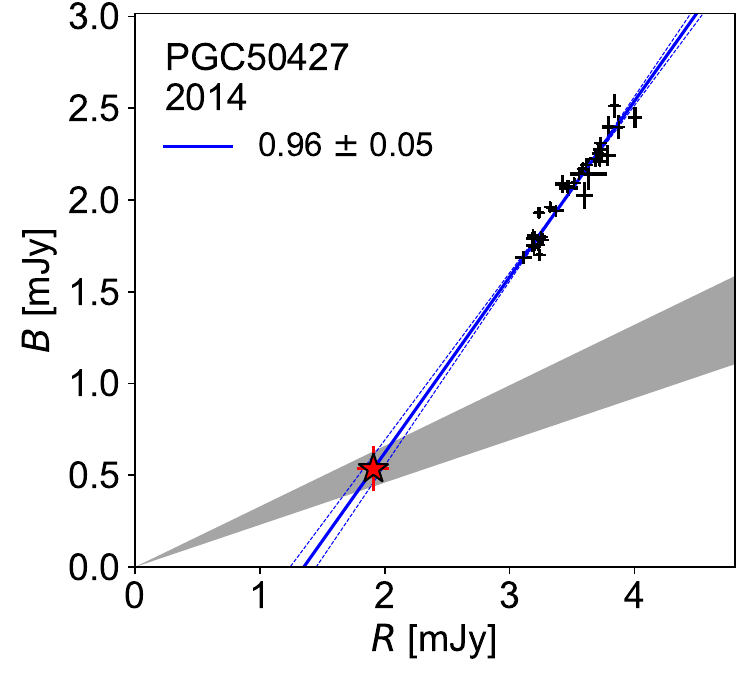}

\includegraphics[width=0.33\columnwidth]{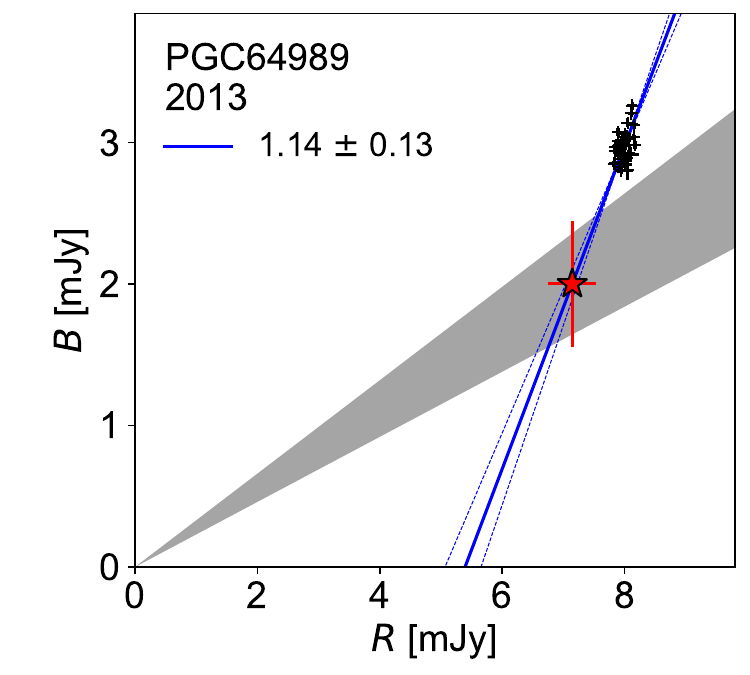}
\includegraphics[width=0.33\columnwidth]{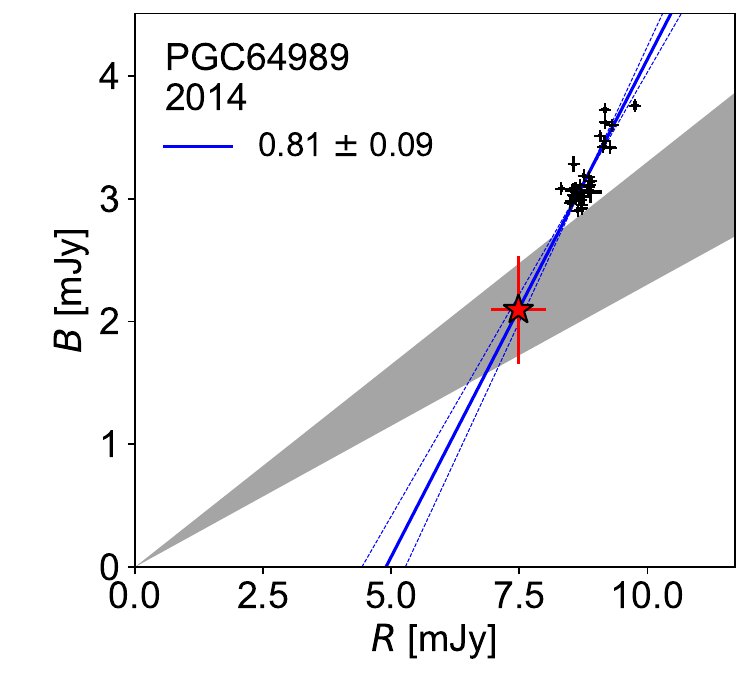}
\includegraphics[width=0.33\columnwidth]{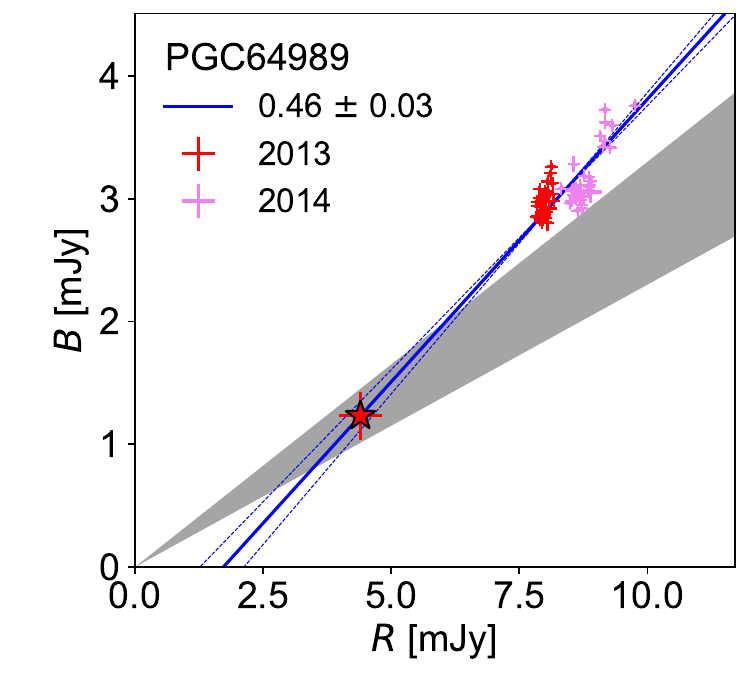}

\includegraphics[width=0.33\columnwidth]{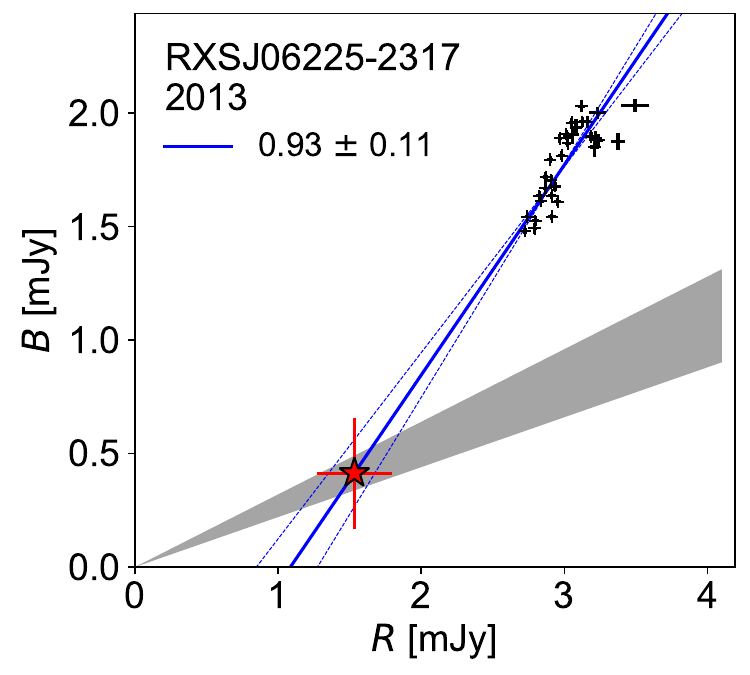}
\includegraphics[width=0.33\columnwidth]{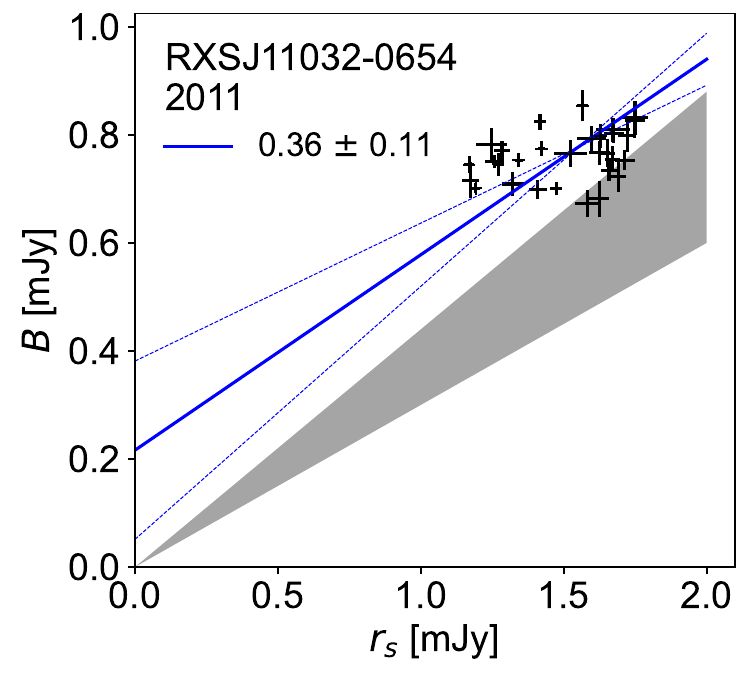}
\includegraphics[width=0.33\columnwidth]{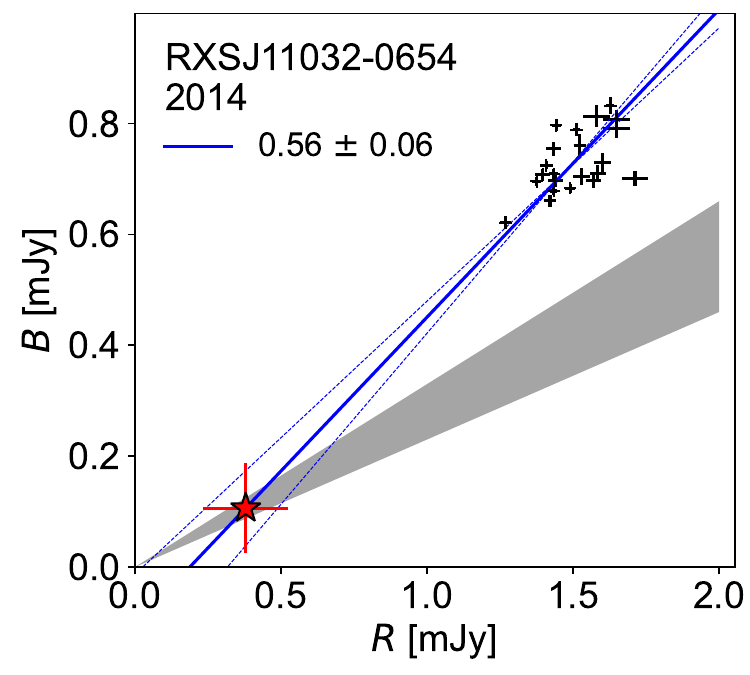}

\includegraphics[width=0.33\columnwidth]{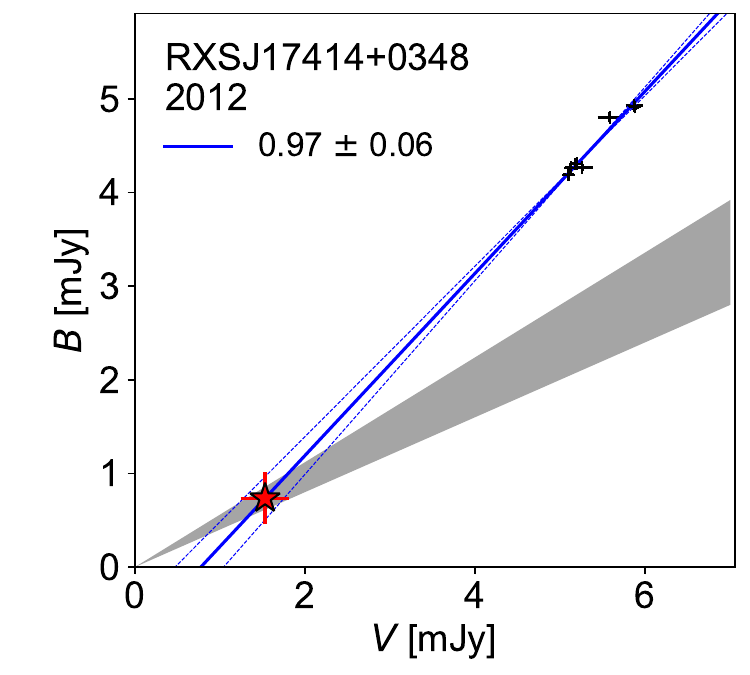}
\includegraphics[width=0.33\columnwidth]{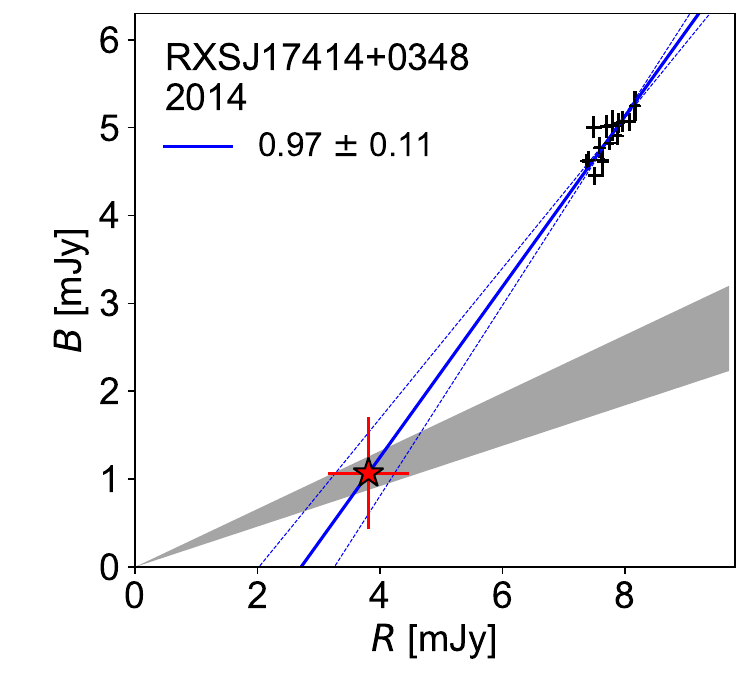}
\includegraphics[width=0.33\columnwidth]{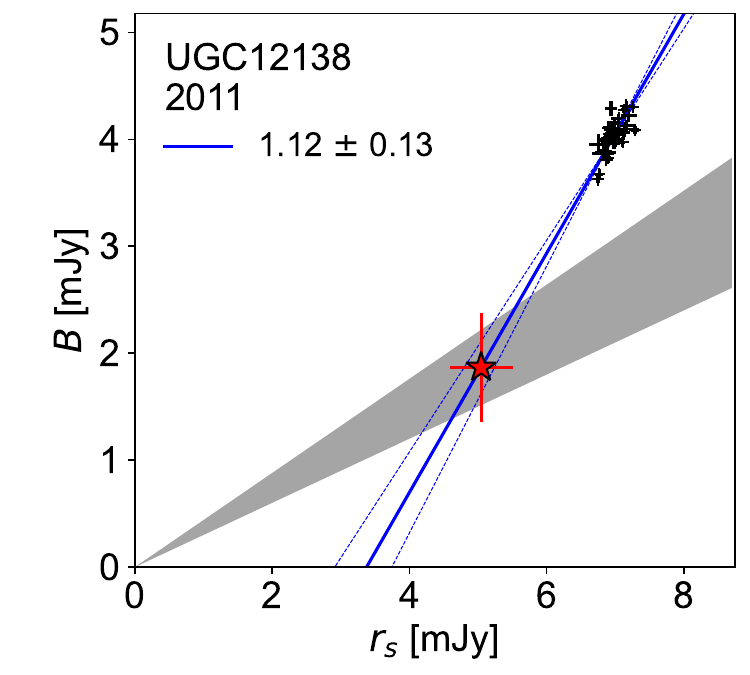}

\includegraphics[width=0.33\columnwidth]{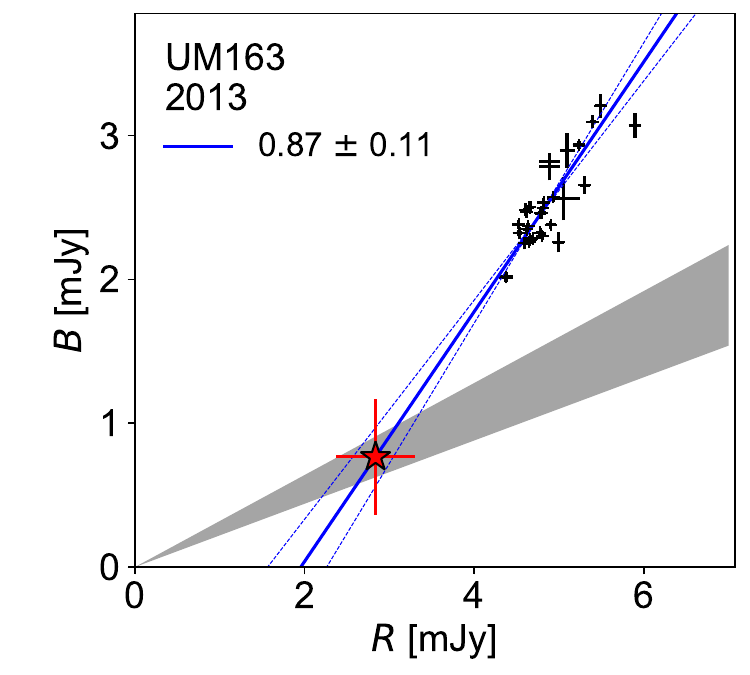}
\includegraphics[width=0.33\columnwidth]{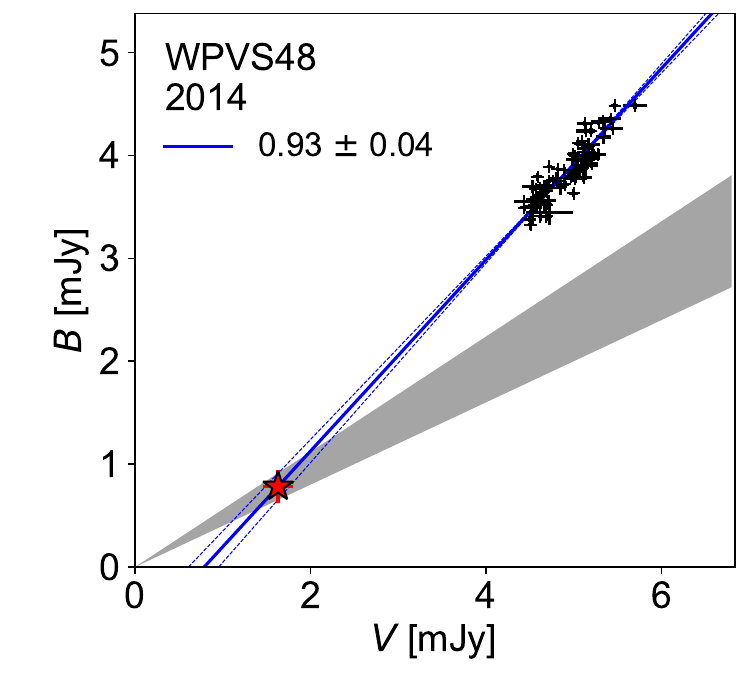}
\includegraphics[width=0.33\columnwidth]{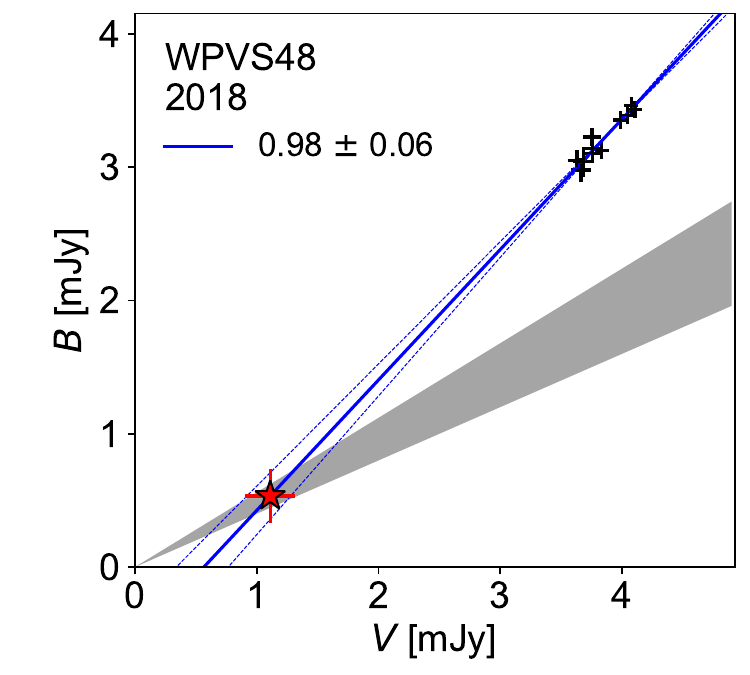}

\includegraphics[width=0.33\columnwidth]{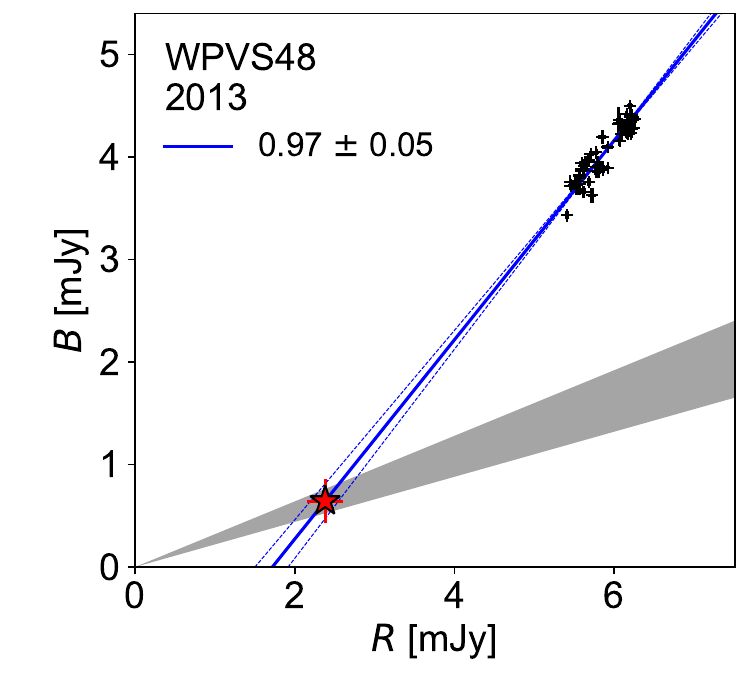}
\includegraphics[width=0.33\columnwidth]{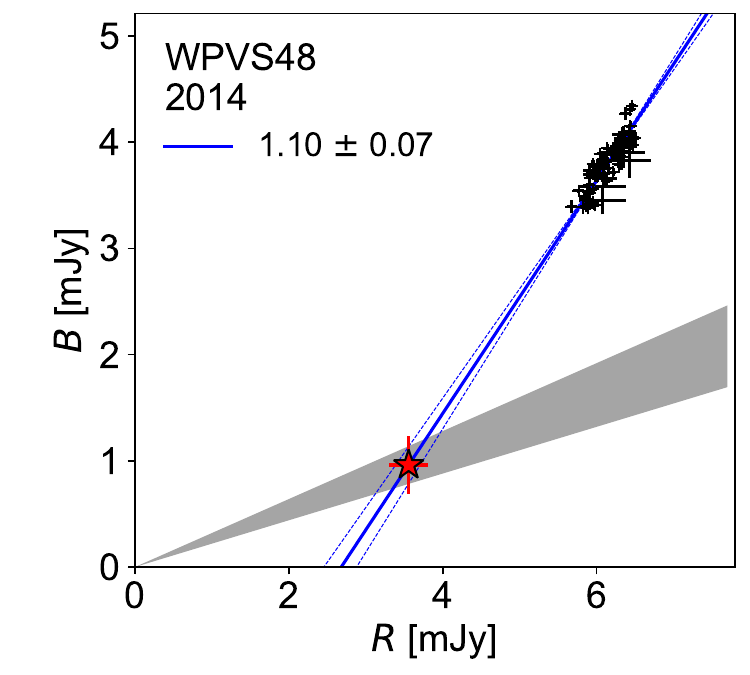}
\includegraphics[width=0.33\columnwidth]{fig_pdf/WPVS48_mix_FVG_B_j_R_j.pdf}

\includegraphics[width=0.33\columnwidth]{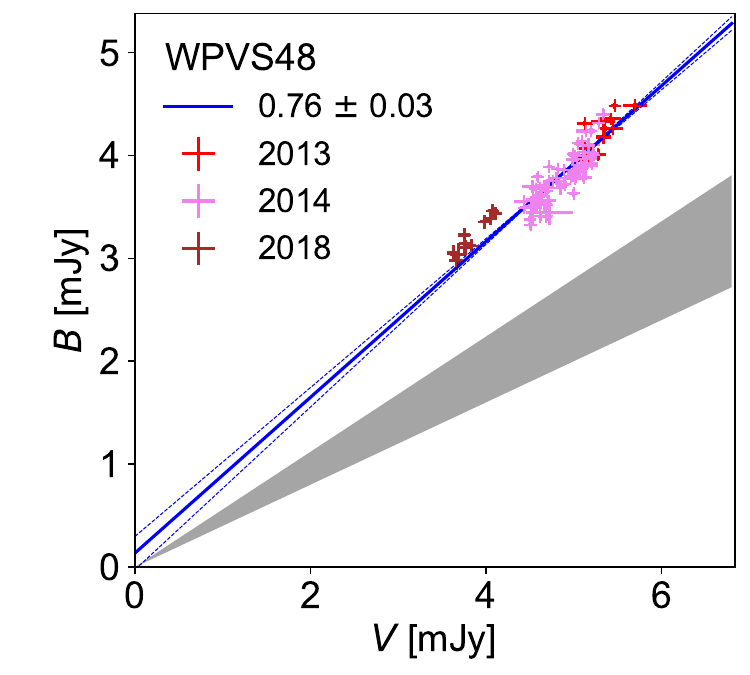}
\includegraphics[width=0.33\columnwidth]{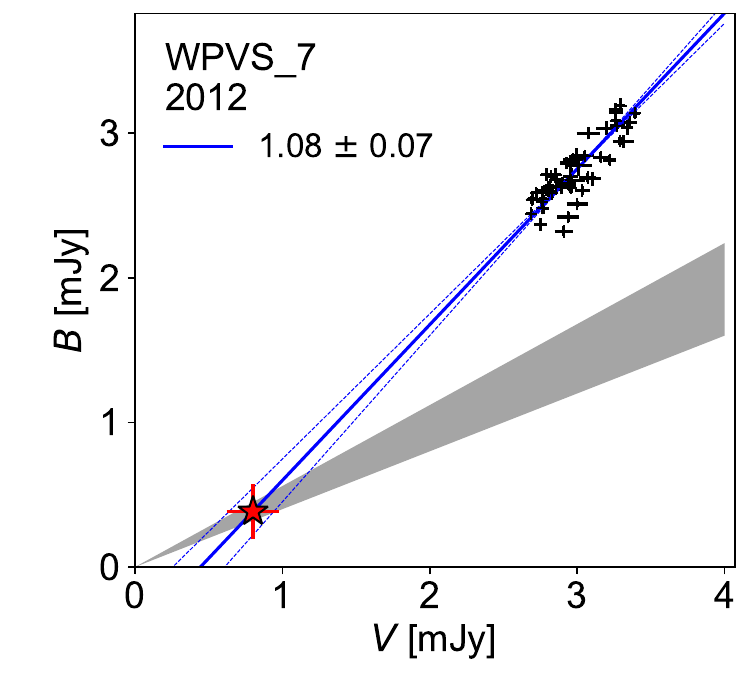}

\begin{deluxetable}{llc|llc|llc}[ht]
\renewcommand{\arraystretch}{0.70}
 \tablecaption{FVG slope results for filter combinations $BR$, $Br$, $BV$ and $VR$ for single observation campaigns and for combined observations}
 \label{tab:fvg}
 \tablehead{Object&Year & $B/R$ & Object & year & $B/V$ & Object & Year & $B/r_s$}
 \startdata
HE0003-5023&2014&0.85$\pm$0.04&HE0003-5023&2014&0.89$\pm$0.05&NGC1019&2011&0.89$\pm$0.15\\
MRK335&2014&1.09$\pm$0.08&MRK335&2011&1.1$\pm$0.07&ESO549-G49&2012&0.49$\pm$0.08\\
IRAS01089-4743&2013&0.7$\pm$0.12&MRK335&2017&0.98$\pm$0.09&ESO490-IG26&2011&0.58$\pm$0.05\\
NGC985&2014&1.2$\pm$0.21&MRK335&comb&1.32$\pm$0.03&ESO374-G25&2011&0.95$\pm$0.08\\
3C120&2014&0.89$\pm$0.02&WPVS\_7&2012&1.08$\pm$0.07&ESO374-G25&2012&-1.16$\pm$0.16\\
MCG-02\_12\_050&2014&0.6$\pm$0.08&3C120&2014&1.01$\pm$0.02&ESO374-G25&2017&1.04$\pm$0.23\\
RXSJ06225-2317&2013&0.93$\pm$0.11&3C120&2018&1.01$\pm$0.03&ESO374-G25&comb&-1.18$\pm$0.13\\
MRK705&2013&0.94$\pm$0.08&3C120&comb&1.05$\pm$0.01&RXSJ11032-0654&2011&0.36$\pm$0.11\\
MRK1239&2015&0.51$\pm$0.07&AKN120&2018&1.07$\pm$0.03&ESO438-G09&2011&1.01$\pm$0.09\\
WPVS48&2013&0.97$\pm$0.05&WPVS48&2014&0.93$\pm$0.04&PGC50427&2011&1.25$\pm$0.09\\
WPVS48&2014&1.1$\pm$0.07&WPVS48&2018&0.98$\pm$0.06&ESO399-IG20&2011&1.02$\pm$0.12\\
WPVS48&comb&0.95$\pm$0.06&WPVS48&comb&0.76$\pm$0.03&NGC7214&2011&0.72$\pm$0.09\\
IRAS09595-0755&2013&1.43$\pm$0.11&ESO374-G25&2017&1.57$\pm$0.18&UGC12138&2011&1.12$\pm$0.13\\
RXSJ11032-0654&2014&0.56$\pm$0.06&HE1136-2304&2015&0.9$\pm$0.09& ---- & ---- &V/R\\
ESO438-G09&2015&0.87$\pm$0.06&HE1136-2304&2016&0.61$\pm$0.05&MRK335&2014&0.48$\pm$0.05\\
PG1149-110&2013&1.37$\pm$0.16&HE1136-2304&2018&0.98$\pm$0.1&3C120&2014&0.88$\pm$0.01\\
NGC4726&2013&0.66$\pm$0.14&HE1136-2304&comb&0.71$\pm$0.05&MRK509&2014&0.83$\pm$0.04\\
ESO323-G77&2015&0.53$\pm$0.02&HE1143-1810&2016&0.98$\pm$0.05&NGC7603&2014&0.99$\pm$0.11\\
MRK1347&2014&0.71$\pm$0.07&ESO511-G030&2014&1.52$\pm$0.19& ---- & ---- &----\\
IC4329A&2015&0.36$\pm$0.02&MRK841&2014&1.04$\pm$0.03\\
ESO578-G09&2014&0.65$\pm$0.05&RXSJ17414+0348&2012&0.97$\pm$0.06\\
PGC50427&2014&0.96$\pm$0.05&1H2107-097&2012&1.18$\pm$0.02\\
ESO511-G030&2014&0.44$\pm$0.07&HE2128-0221&2016&0.82$\pm$0.05\\
NGC5940&2014&1.03$\pm$0.13&NGC7469&2012&1.26$\pm$0.08\\
RXSJ17414+0348&2014&0.97$\pm$0.11&NGC7603&2014&0.86$\pm$0.21\\
MCG+03\_47\_002&2013&1.75$\pm$0.25&---& ----& ----\\
ESO141-G55&2013&1.29$\pm$0.05&\\
ESO141-G55&2015&1.39$\pm$0.04&\\
ESO141-G55&comb&1.07$\pm$0.02&\\
CTSG03\_04&2013&1.24$\pm$0.1&\\
NGC6860&2015&0.6$\pm$0.07&\\
PGC64989&2013&1.14$\pm$0.13&\\
PGC64989&2014&0.81$\pm$0.09&\\
PGC64989&comb&0.46$\pm$0.03&\\
MRK509&2014&0.86$\pm$0.06&\\
F1041&2013&0.75$\pm$0.1&\\
NGC7603&2014&1.03$\pm$0.09&\\
IRAS23226-3843&2013&0.67$\pm$0.09&\\
UM163&2013&0.87$\pm$0.11&\\
\enddata
\end{deluxetable}

\clearpage
\startlongtable
\begin{deluxetable}{llcccccccr}
\renewcommand{\arraystretch}{0.7}
\tablecaption{Total and host fluxes for $B$ $V$ $r_s$ and $R$ obtained via the FVG method (see text). The interpolated nuclear optical flux at 5100\AA, $f_{5100\rm ,obs}$, is noted as are the values for $\beta$ from the fit to the nuclear optical broadband spectrum ($f_\lambda\propto \lambda^{\beta}$).}
\label{tab:ad_all}
\tablehead{Object&Year & $B_{\rm tot}$ & $B_{\rm host}$ & $V_{\rm tot}$ & $V_{\rm host}$ & $R_{\rm tot}$/$r_{\rm s,tot}$ & $R_{\rm host}$/$r_{\rm s,host}$ & $f_{5100,\rm obs}$ & $\beta$ \\
& & [mJy] & [mJy] & [mJy] & [mJy] & [mJy] & [mJy] & [mJy] & }
\startdata
1H2107-097 & 2012 &$ 7.45 \pm 0.03 $&$ 2.14 \pm 0.13 $&$ 9.01 \pm 0.02 $&$ 4.45 \pm 0.11 $&--&--&$ 4.78 \pm 0.12 $& -0.64 \\
3C120 & 2014 &$ 11.79 \pm 0.34 $&$ 2.22 \pm 0.27 $&--&--&$ 18.76 \pm 0.18 $&$ 8.22 \pm 0.31 $&$ 9.89 \pm 0.41 $& 0.2 \\
3C120 & 2014 &$ 11.79 \pm 0.34 $&$ 2.14 \pm 0.29 $&$ 13.83 \pm 0.07 $&$ 4.46 \pm 0.29 $&--&--&$ 9.46 \pm 0.35 $& -0.12 \\
3C120 & 2014 &$ 11.79 \pm 0.34 $&$ 2.47 \pm 0.15 $&$ 13.83 \pm 0.07 $&$ 5.15 \pm 0.14 $&--&--&$ 8.88 \pm 0.23 $& -0.3 \\
3C120 & 2018 &$ 9.34 \pm 0.0 $&$ 2.26 \pm 0.27 $&$ 10.97 \pm 0.0 $&$ 4.71 \pm 0.27 $&--&--&$ 6.51 \pm 0.27 $& -0.51 \\
3C120 & comb &$ 10.75 \pm 0.27 $&$ 2.47 \pm 0.15 $&$ 12.67 \pm 0.06 $&$ 5.15 \pm 0.14 $&--&--&$ 7.75 \pm 0.21 $& -0.4 \\
AKN120 & 2018 &$ 10.29 \pm 0.01 $&$ 2.29 \pm 0.28 $&$ 11.98 \pm 0.04 $&$ 4.77 \pm 0.26 $&--&--&$ 7.45 \pm 0.27 $& -0.43 \\
CTSG03\_04 & 2013 &$ 1.62 \pm 0.02 $&$ 0.51 \pm 0.18 $&--&--&$ 2.7 \pm 0.01 $&$ 1.82 \pm 0.14 $&$ 1.03 \pm 0.17 $& -0.48 \\
ESO141-G55 & 2013 &$ 13.96 \pm 0.11 $&$ 1.8 \pm 0.58 $&--&--&$ 16.0 \pm 0.22 $&$ 6.65 \pm 0.45 $&$ 11.12 \pm 0.56 $& -0.55 \\
ESO141-G55 & 2013 &$ 13.96 \pm 0.11 $&$ 1.75 \pm 0.66 $&--&--&$ 16.0 \pm 0.22 $&$ 6.5 \pm 0.5 $&$ 11.21 \pm 0.63 $& -0.52 \\
ESO141-G55 & 2015 &$ 10.34 \pm 0.05 $&$ 1.7 \pm 0.35 $&--&--&$ 12.43 \pm 0.1 $&$ 6.31 \pm 0.25 $&$ 7.68 \pm 0.33 $& -0.72 \\
ESO141-G55 & 2015 &$ 10.34 \pm 0.05 $&$ 1.75 \pm 0.66 $&--&--&$ 12.43 \pm 0.1 $&$ 6.5 \pm 0.5 $&$ 7.57 \pm 0.62 $& -0.77 \\
ESO141-G55 & comb &$ 10.63 \pm 0.09 $&$ 1.06 \pm 0.22 $&--&--&$ 12.8 \pm 0.16 $&$ 3.91 \pm 0.2 $&$ 9.33 \pm 0.24 $& -0.15 \\
ESO323-G77 & 2015 &$ 8.4 \pm 0.06 $&$ 2.8 \pm 0.35 $&--&--&$ 20.53 \pm 0.13 $&$ 9.99 \pm 0.66 $&$ 6.95 \pm 0.45 $& 1.32 \\
ESO374-G25 & 2011 &$ 1.15 \pm 0.01 $&$ 1.07 \pm 0.1 $&--&--&$ 3.0 \pm 0.01 $&$ 2.9 \pm 0.11 $&$ 0.09 \pm 0.1 $& 0.57 \\
ESO374-G25 & 2011 &$ 1.15 \pm 0.01 $&--&--&--&$ 3.0 \pm 0.01 $&--&$ 1.72 \pm 0.01 $& 2.45 \\
ESO374-G25 & 2017 &$ 1.36 \pm 0.0 $&$ 0.91 \pm 0.31 $&--&--&$ 2.87 \pm 0.01 $&$ 2.47 \pm 0.3 $&$ 0.43 \pm 0.31 $& -0.3 \\
ESO374-G25 & 2017 &$ 1.36 \pm 0.0 $&$ 1.24 \pm 0.19 $&$ 2.68 \pm 0.01 $&$ 2.57 \pm 0.12 $&--&--&$ 0.11 \pm 0.14 $& -0.36 \\
ESO399-IG20 & 2011 &$ 4.08 \pm 0.03 $&$ 1.6 \pm 0.52 $&--&--&$ 6.78 \pm 0.05 $&$ 4.32 \pm 0.51 $&$ 2.47 \pm 0.52 $& -0.02 \\
ESO438-G09 & 2011 &$ 5.1 \pm 0.03 $&$ 1.12 \pm 0.5 $&--&--&$ 6.96 \pm 0.04 $&$ 3.03 \pm 0.5 $&$ 3.96 \pm 0.5 $& -0.03 \\
ESO438-G09 & 2011 &$ 5.1 \pm 0.03 $&$ 1.12 \pm 0.5 $&--&--&$ 6.96 \pm 0.04 $&$ 3.03 \pm 0.5 $&$ 3.96 \pm 0.5 $& -0.03 \\
ESO438-G09 & 2015 &$ 4.0 \pm 0.02 $&$ 0.84 \pm 0.28 $&--&--&$ 6.61 \pm 0.03 $&$ 3.01 \pm 0.32 $&$ 3.3 \pm 0.29 $& 0.27 \\
ESO438-G09 & 2015 &$ 4.0 \pm 0.02 $&$ 1.12 \pm 0.5 $&--&--&$ 6.61 \pm 0.03 $&$ 3.86 \pm 0.31 $&$ 2.84 \pm 0.45 $& -0.1 \\
ESO490-IG26 & 2011 &$ 2.54 \pm 0.01 $&$ 2.04 \pm 0.17 $&--&--&$ 6.41 \pm 0.03 $&$ 5.5 \pm 0.29 $&$ 0.64 \pm 0.22 $& 1.53 \\
ESO490-IG26 & 2011 &$ 2.54 \pm 0.01 $&--&--&--&$ 6.41 \pm 0.03 $&--&$ 3.74 \pm 0.02 $& 2.37 \\
ESO511-G030 & 2013 &$ 2.54 \pm 0.01 $&$ 1.39 \pm 0.28 $&--&--&$ 6.59 \pm 0.05 $&$ 4.96 \pm 0.41 $&$ 1.3 \pm 0.32 $& 0.73 \\
ESO511-G030 & 2013 &$ 2.54 \pm 0.01 $&$ 1.46 \pm 0.66 $&--&--&$ 6.59 \pm 0.05 $&$ 5.03 \pm 0.5 $&$ 1.22 \pm 0.61 $& 0.77 \\
ESO511-G030 & 2014 &$ 2.56 \pm 0.01 $&$ 0.32 \pm 0.44 $&--&--&$ 6.49 \pm 0.06 $&$ 1.15 \pm 0.99 $&$ 3.01 \pm 0.6 $& 1.81 \\
ESO511-G030 & 2014 &$ 2.56 \pm 0.01 $&$ 1.65 \pm 0.36 $&$ 4.03 \pm 0.01 $&$ 3.43 \pm 0.24 $&--&--&$ 0.68 \pm 0.28 $& -1.74 \\
ESO511-G030 & 2014 &$ 2.56 \pm 0.01 $&$ 1.46 \pm 0.5 $&$ 4.03 \pm 0.01 $&$ 3.04 \pm 0.5 $&--&--&$ 1.02 \pm 0.5 $& -0.44 \\
ESO511-G030 & comb &$ 2.56 \pm 0.01 $&$ 1.34 \pm 0.29 $&--&--&$ 6.57 \pm 0.05 $&$ 4.78 \pm 0.42 $&$ 1.39 \pm 0.33 $& 0.8 \\
ESO549-G49 & 2012 &$ 3.24 \pm 0.01 $&$ 0.17 \pm 0.51 $&--&--&$ 6.74 \pm 0.04 $&$ 0.45 \pm 1.05 $&$ 4.15 \pm 0.71 $& 1.84 \\
ESO578-G09 & 2014 &$ 1.97 \pm 0.01 $&$ 0.43 \pm 0.16 $&--&--&$ 3.83 \pm 0.01 $&$ 1.6 \pm 0.25 $&$ 1.75 \pm 0.19 $& 0.77 \\
F1041 & 2013 &$ 1.14 \pm 0.01 $&$ 0.55 \pm 0.18 $&--&--&$ 2.82 \pm 0.02 $&$ 2.02 \pm 0.24 $&$ 0.65 \pm 0.2 $& 0.63 \\
HE0003-5023 & 2014 &$ 2.5 \pm 0.07 $&$ 0.37 \pm 0.12 $&--&--&$ 3.88 \pm 0.08 $&$ 1.36 \pm 0.14 $&$ 2.26 \pm 0.15 $& 0.35 \\
HE0003-5023 & 2014 &$ 2.5 \pm 0.07 $&$ 0.22 \pm 0.15 $&$ 2.98 \pm 0.02 $&$ 0.46 \pm 0.17 $&--&--&$ 2.44 \pm 0.17 $& 0.42 \\
HE1136-2304 & 2015 &$ 1.06 \pm 0.01 $&$ 0.94 \pm 0.11 $&$ 2.08 \pm 0.01 $&$ 1.96 \pm 0.12 $&--&--&$ 0.12 \pm 0.12 $& 0.0 \\
HE1136-2304 & 2015 &$ 1.06 \pm 0.01 $&$ 0.9 \pm 0.11 $&$ 2.08 \pm 0.01 $&$ 1.87 \pm 0.12 $&--&--&$ 0.19 \pm 0.12 $& 1.14 \\
HE1136-2304 & 2016 &$ 1.42 \pm 0.01 $&$ 1.03 \pm 0.09 $&$ 2.73 \pm 0.01 $&$ 2.14 \pm 0.15 $&--&--&$ 0.52 \pm 0.13 $& 1.73 \\
HE1136-2304 & 2016 &$ 1.42 \pm 0.01 $&$ 0.9 \pm 0.09 $&$ 2.73 \pm 0.01 $&$ 1.87 \pm 0.15 $&--&--&$ 0.73 \pm 0.13 $& 2.1 \\
HE1136-2304 & 2018 &$ 1.57 \pm 0.01 $&$ 0.83 \pm 0.16 $&$ 2.52 \pm 0.02 $&$ 1.73 \pm 0.16 $&--&--&$ 0.77 \pm 0.16 $& 0.27 \\
HE1136-2304 & 2018 &$ 1.57 \pm 0.01 $&$ 0.9 \pm 0.16 $&$ 2.52 \pm 0.02 $&$ 1.87 \pm 0.16 $&--&--&$ 0.66 \pm 0.16 $& -0.13 \\
HE1136-2304 & comb &$ 1.35 \pm 0.02 $&$ 0.9 \pm 0.09 $&$ 2.46 \pm 0.02 $&$ 1.87 \pm 0.13 $&--&--&$ 0.54 \pm 0.12 $& 1.13 \\
HE1136-2304 & comb &$ 1.35 \pm 0.02 $&$ 0.9 \pm 0.09 $&$ 2.46 \pm 0.02 $&$ 1.87 \pm 0.13 $&--&--&$ 0.54 \pm 0.12 $& 1.13 \\
HE1143-1810 & 2016 &$ 4.37 \pm 0.01 $&$ 1.27 \pm 0.24 $&$ 5.41 \pm 0.01 $&$ 2.64 \pm 0.25 $&--&--&$ 2.87 \pm 0.25 $& -0.47\\
HE2128-0221 & 2016 &$ 0.67 \pm 0.01 $&$ 0.17 \pm 0.04 $&$ 0.99 \pm 0.0 $&$ 0.35 \pm 0.05 $&--&--&$ 0.59 \pm 0.05 $& 1.03 \\
IC4329A & 2015 &$ 3.58 \pm 0.02 $&--&--&--&$ 15.73 \pm 0.09 $&--&$ 5.93 \pm 0.04 $& 3.08 \\
IRAS01089-4743 & 2013 &$ 2.15 \pm 0.02 $&$ 0.43 \pm 0.39 $&--&--&$ 4.0 \pm 0.02 $&$ 1.52 \pm 0.56 $&$ 1.95 \pm 0.44 $& 0.76 \\
IRAS09595-0755 & 2013 &$ 1.34 \pm 0.01 $&$ 0.58 \pm 0.18 $&--&--&$ 2.61 \pm 0.01 $&$ 2.08 \pm 0.12 $&$ 0.67 \pm 0.16 $& -0.75 \\
IRAS23226-3843 & 2013 &$ 2.68 \pm 0.02 $&$ 0.81 \pm 0.38 $&--&--&$ 5.8 \pm 0.03 $&$ 2.99 \pm 0.57 $&$ 2.15 \pm 0.44 $& 0.85 \\
MCG-02\_12\_050 & 2014 &$ 1.52 \pm 0.01 $&$ 0.98 \pm 0.17 $&--&--&$ 4.3 \pm 0.03 $&$ 3.64 \pm 0.29 $&$ 0.58 \pm 0.21 $& 0.42 \\
MCG-02\_12\_050 & 2014 &$ 1.52 \pm 0.01 $& --&--&--&$ 4.3 \pm 0.03 $&--&$ 2.17 \pm 0.02 $& 2.16 \\
MCG+03\_47\_002 & 2013 &$ 0.84 \pm 0.01 $&$ 0.51 \pm 0.26 $&--&--&$ 2.06 \pm 0.01 $&$ 1.88 \pm 0.15 $&$ 0.27 \pm 0.23 $& -1.26 \\
MRK1239 & 2015 &$ 3.96 \pm 0.02 $&$ 0.99 \pm 0.49 $&--&--&$ 9.36 \pm 0.03 $&$ 3.52 \pm 0.96 $&$ 3.74 \pm 0.63 $& 1.41 \\
MRK1347 & 2014 &$ 1.94 \pm 0.03 $&$ 0.3 \pm 0.21 $&--&--&$ 3.41 \pm 0.03 $&$ 1.14 \pm 0.29 $&$ 1.83 \pm 0.24 $& 0.68 \\
MRK335 & 2011 &$ 7.79 \pm 0.05 $&$ 1.26 \pm 0.5 $&$ 9.08 \pm 0.03 $&$ 2.63 \pm 0.45 $&--&--&$ 6.48 \pm 0.47 $& -0.05 \\
MRK335 & 2011 &$ 7.79 \pm 0.05 $&$ 1.99 \pm 0.5 $&$ 9.08 \pm 0.03 $&$ 4.15 \pm 0.45 $&--&--&$ 5.19 \pm 0.47 $& -0.68 \\
MRK335 & 2014 &$ 6.79 \pm 0.14 $&$ 1.47 \pm 0.54 $&--&--&$ 10.8 \pm 0.41 $&$ 5.23 \pm 0.5 $&$ 5.4 \pm 0.58 $& 0.1 \\
MRK335 & 2014 &$ 6.79 \pm 0.14 $&$ 1.99 \pm 0.51 $&--&--&$ 10.8 \pm 0.41 $&$ 6.86 \pm 0.47 $&$ 4.49 \pm 0.56 $& -0.41 \\
MRK335 & 2017 &$ 5.61 \pm 0.0 $&$ 1.12 \pm 0.48 $&$ 6.9 \pm 0.0 $&$ 2.33 \pm 0.49 $&--&--&$ 4.54 \pm 0.49 $& 0.07 \\
MRK335 & 2017 &$ 5.61 \pm 0.0 $&$ 1.99 \pm 0.48 $&$ 6.9 \pm 0.0 $&$ 4.15 \pm 0.49 $&--&--&$ 3.0 \pm 0.49 $& -1.15 \\
MRK335 & comb &$ 5.73 \pm 0.03 $&$ 1.99 \pm 0.16 $&$ 7.05 \pm 0.04 $&$ 4.15 \pm 0.12 $&--&--&$ 3.14 \pm 0.14 $& -1.06 \\
MRK335 & comb &$ 5.73 \pm 0.03 $&$ 1.99 \pm 0.16 $&$ 7.05 \pm 0.04 $&$ 4.15 \pm 0.12 $&--&--&$ 3.14 \pm 0.14 $& -1.06 \\
MRK509 & 2014 &$ 12.06 \pm 0.14 $&$ 1.83 \pm 0.84 $&--&--&$ 17.19 \pm 0.4 $&$ 6.78 \pm 0.98 $&$ 10.29 \pm 0.91 $& 0.04 \\
MRK705 & 2013 &$ 3.43 \pm 0.02 $&$ 1.08 \pm 0.35 $&--&--&$ 6.45 \pm 0.02 $&$ 3.86 \pm 0.37 $&$ 2.43 \pm 0.36 $& 0.2 \\
MRK841 & 2014 &$ 4.92 \pm 0.06 $&$ 0.94 \pm 0.12 $&$ 4.96 \pm 0.01 $&$ 1.96 \pm 0.12 $&--&--&$ 3.28 \pm 0.13 $& -1.18 \\
NGC1019 & 2011 &$ 1.66 \pm 0.01 $&$ 1.0 \pm 0.3 $&--&--&$ 3.47 \pm 0.02 $&$ 2.71 \pm 0.34 $&$ 0.7 \pm 0.32 $& 0.36 \\
NGC4726 & 2013 &$ 2.15 \pm 0.01 $&$ 0.95 \pm 0.5 $&--&--&$ 5.16 \pm 0.02 $&$ 3.38 \pm 0.77 $&$ 1.37 \pm 0.58 $& 0.82 \\
NGC5940 & 2014 &$ 1.92 \pm 0.04 $&$ 0.67 \pm 0.3 $&--&--&$ 3.76 \pm 0.02 $&$ 2.49 \pm 0.29 $&$ 1.26 \pm 0.3 $& 0.03 \\
NGC6860 & 2015 &$ 3.77 \pm 0.01 $&$ 2.1 \pm 0.41 $&--&--&$ 10.08 \pm 0.08 $&$ 7.24 \pm 0.68 $&$ 2.0 \pm 0.49 $& 1.11 \\
NGC7214 & 2011 &$ 3.76 \pm 0.03 $&$ 1.4 \pm 0.46 $&--&--&$ 7.15 \pm 0.05 $&$ 3.77 \pm 0.64 $&$ 2.74 \pm 0.53 $& 0.92 \\
NGC7469 & 2012 &$ 17.51 \pm 0.1 $&$ 6.11 \pm 1.1 $&$ 21.69 \pm 0.06 $&$ 12.72 \pm 0.88 $&--&--&$ 9.68 \pm 0.96 $& -1.0 \\
NGC7603 & 2014 &$ 10.58 \pm 0.24 $&$ 0.83 \pm 2.47 $&$ 13.66 \pm 0.12 $&$ 1.73 \pm 2.94 $&--&--&$ 11.19 \pm 2.78 $& 0.84 \\
NGC7603 & 2014 &$ 10.58 \pm 0.24 $&$ 2.53 \pm 1.08 $&--&--&$ 17.28 \pm 0.28 $&$ 9.03 \pm 1.05 $&$ 8.12 \pm 1.1 $& 0.05 \\
NGC985 & 2014 &$ 4.6 \pm 0.04 $&$ 0.49 \pm 0.88 $&--&--&$ 4.97 \pm 0.06 $&$ 1.88 \pm 0.74 $&$ 3.73 \pm 0.84 $& -0.59 \\
PG1149-110 & 2013 &$ 0.9 \pm 0.01 $&$ 0.64 \pm 0.18 $&--&--&$ 2.65 \pm 0.01 $&$ 2.48 \pm 0.13 $&$ 0.22 \pm 0.17 $& -0.88 \\
PG1149-110 & 2013 &$ 0.9 \pm 0.01 $&--&--&--&$ 2.65 \pm 0.01 $&--&$ 1.3 \pm 0.01 $& 2.25 \\
PGC50427 & 2011 &$ 1.83 \pm 0.01 $&$ 0.66 \pm 0.17 $&--&--&$ 2.72 \pm 0.03 $&$ 1.78 \pm 0.14 $&$ 1.07 \pm 0.16 $& -0.56 \\
PGC50427 & 2011 &$ 1.83 \pm 0.01 $&$ 0.66 \pm 0.17 $&--&--&$ 2.72 \pm 0.03 $&$ 1.88 \pm 0.14 $&$ 1.02 \pm 0.16 $& -0.85 \\
PGC50427 & 2014 &$ 2.09 \pm 0.02 $&$ 0.53 \pm 0.12 $&--&--&$ 3.57 \pm 0.03 $&$ 1.91 \pm 0.13 $&$ 1.59 \pm 0.13 $& 0.13 \\
PGC50427 & 2014 &$ 2.09 \pm 0.02 $&$ 0.66 \pm 0.12 $&--&--&$ 3.57 \pm 0.03 $&$ 2.27 \pm 0.12 $&$ 1.38 \pm 0.12 $& -0.2 \\
PGC64989 & 2013 &$ 2.94 \pm 0.01 $&$ 2.0 \pm 0.45 $&--&--&$ 7.98 \pm 0.02 $&$ 7.15 \pm 0.39 $&$ 0.9 \pm 0.43 $& -0.26 \\
PGC64989 & 2014 &$ 3.08 \pm 0.04 $&$ 2.1 \pm 0.44 $&--&--&$ 8.72 \pm 0.07 $&$ 7.48 \pm 0.54 $&$ 1.06 \pm 0.47 $& 0.49 \\
PGC64989 & comb &$ 3.02 \pm 0.03 $&$ 1.23 \pm 0.2 $&--&--&$ 8.49 \pm 0.06 $&$ 4.4 \pm 0.42 $&$ 2.37 \pm 0.27 $& 1.72 \\
RXSJ06225-2317 & 2013 &$ 1.87 \pm 0.01 $&$ 0.41 \pm 0.24 $&--&--&$ 2.98 \pm 0.02 $&$ 1.53 \pm 0.26 $&$ 1.46 \pm 0.25 $& -0.01 \\
RXSJ11032-0654 & 2011 &$ 0.76 \pm 0.01 $&--&--&--&$ 1.58 \pm 0.01 $&--&$ 1.03 \pm 0.01 $& 1.87 \\
RXSJ11032-0654 & 2014 &$ 0.72 \pm 0.01 $&--&--&--&$ 1.5 \pm 0.01 $&--&$ 0.92 \pm 0.01 $& 1.53 \\
RXSJ17414+0348 & 2012 &$ 4.73 \pm 0.01 $&$ 0.73 \pm 0.28 $&$ 5.42 \pm 0.04 $&$ 1.53 \pm 0.29 $&--&--&$ 3.92 \pm 0.29 $& -0.12 \\
RXSJ17414+0348 & 2012 &$ 4.73 \pm 0.01 $&$ 1.03 \pm 0.28 $&$ 5.42 \pm 0.04 $&$ 2.14 \pm 0.29 $&--&--&$ 3.41 \pm 0.29 $& -0.5 \\
RXSJ17414+0348 & 2014 &$ 4.74 \pm 0.03 $&$ 1.07 \pm 0.64 $&--&--&$ 7.37 \pm 0.05 $&$ 3.81 \pm 0.66 $&$ 3.63 \pm 0.65 $& -0.06 \\
RXSJ17414+0348 & 2014 &$ 4.74 \pm 0.03 $&$ 1.07 \pm 0.61 $&--&--&$ 7.37 \pm 0.05 $&$ 3.81 \pm 0.63 $&$ 3.63 \pm 0.62 $& -0.06 \\
UGC12138 & 2011 &$ 4.03 \pm 0.02 $&$ 1.87 \pm 0.51 $&--&--&$ 6.94 \pm 0.03 $&$ 5.04 \pm 0.46 $&$ 2.05 \pm 0.49 $& -0.33 \\
UM163 & 2013 &$ 2.46 \pm 0.03 $&$ 0.77 \pm 0.4 $&$ 4.8 \pm 0.04 $&$ 2.84 \pm 0.46 $&--&--&$ 1.78 \pm 0.42 $& 0.31 \\
WPVS\_7 & 2012 &$ 2.68 \pm 0.01 $&$ 0.38 \pm 0.19 $&$ 2.94 \pm 0.01 $&$ 0.8 \pm 0.18 $&--&--&$ 2.19 \pm 0.18 $& -0.3 \\
WPVS48 & 2013 &$ 3.95 \pm 0.02 $&$ 0.64 \pm 0.22 $&--&--&$ 5.8 \pm 0.01 $&$ 2.38 \pm 0.22 $&$ 3.35 \pm 0.22 $& 0.07 \\
WPVS48 & 2014 &$ 3.81 \pm 0.04 $&$ 0.78 \pm 0.16 $&$ 4.98 \pm 0.04 $&$ 1.63 \pm 0.18 $&--&--&$ 3.25 \pm 0.18 $& 0.42 \\
WPVS48 & 2014 &$ 3.81 \pm 0.04 $&$ 0.96 \pm 0.27 $&--&--&$ 6.21 \pm 0.05 $&$ 3.56 \pm 0.25 $&$ 2.78 \pm 0.27 $& -0.15 \\
WPVS48 & 2018 &$ 3.25 \pm 0.01 $&$ 0.53 \pm 0.2 $&$ 3.78 \pm 0.01 $&$ 1.11 \pm 0.21 $&--&--&$ 2.69 \pm 0.21 $& -0.08 \\
WPVS48 & comb &$ 3.82 \pm 0.03 $&$ 0.73 \pm 0.27 $&--&--&$ 6.09 \pm 0.04 $&$ 2.7 \pm 0.28 $&$ 3.19 \pm 0.27 $& 0.19 \\
\enddata
\tablecomments{a = $r_s$ filter; b = no host subtraction}
\end{deluxetable}

\newpage
\section{Single-epoch spectra}\label{app:spec}
{Below we present the set of the spectra used in this work to constrain the FWHM of the broad H$\alpha$ emission line in different sources (Table. \ref{tab:spectroscopy_log}). The black lines show the continuum-subtracted H$\alpha$ emission line in the rest frame of the source. Individual best-fit narrow components for the [NII] and [SII] lines, as well as the broad H$\alpha$, are shown in green. In cases where the H$\alpha$ line is fitted with three components (narrow, medium, broad), the sum of the middle and broad components is represented by a blue line. The sum of all Gaussian emission components is traced by a red line. Positions of the NIIa, NIIb, SIIa, and SIIb lines are marked. The orange colored surfaces trace the throughput curves of the NB filters used (at 670nm, 680nm and 690nm)}, while that of the narrower SII-band (at 672nm) is shown in gray.

{We also calculated the ratio of the broad H$\alpha$ line flux to the total flux. First, we determined the fraction of the total H$\alpha$ flux captured by the narrowband observation filter. Next, we estimated the ratio of the broad component—obtained from our fitting process—to the total H$\alpha$ emission line flux, which involved integrating the broad component fit across the entire H$\alpha$ line. The product of the fraction of H$\alpha$ captured by the filter and the ratio of the broad component provides an estimate of the broad component captured within the narrowband filter, denoted as H$\alpha_{\rm NB}$. The fraction H$\alpha_{\rm NB}$ is listed in Table~\ref{tab:sample} for all the objects.}

\begin{figure*}
\includegraphics[width=0.98\textwidth]{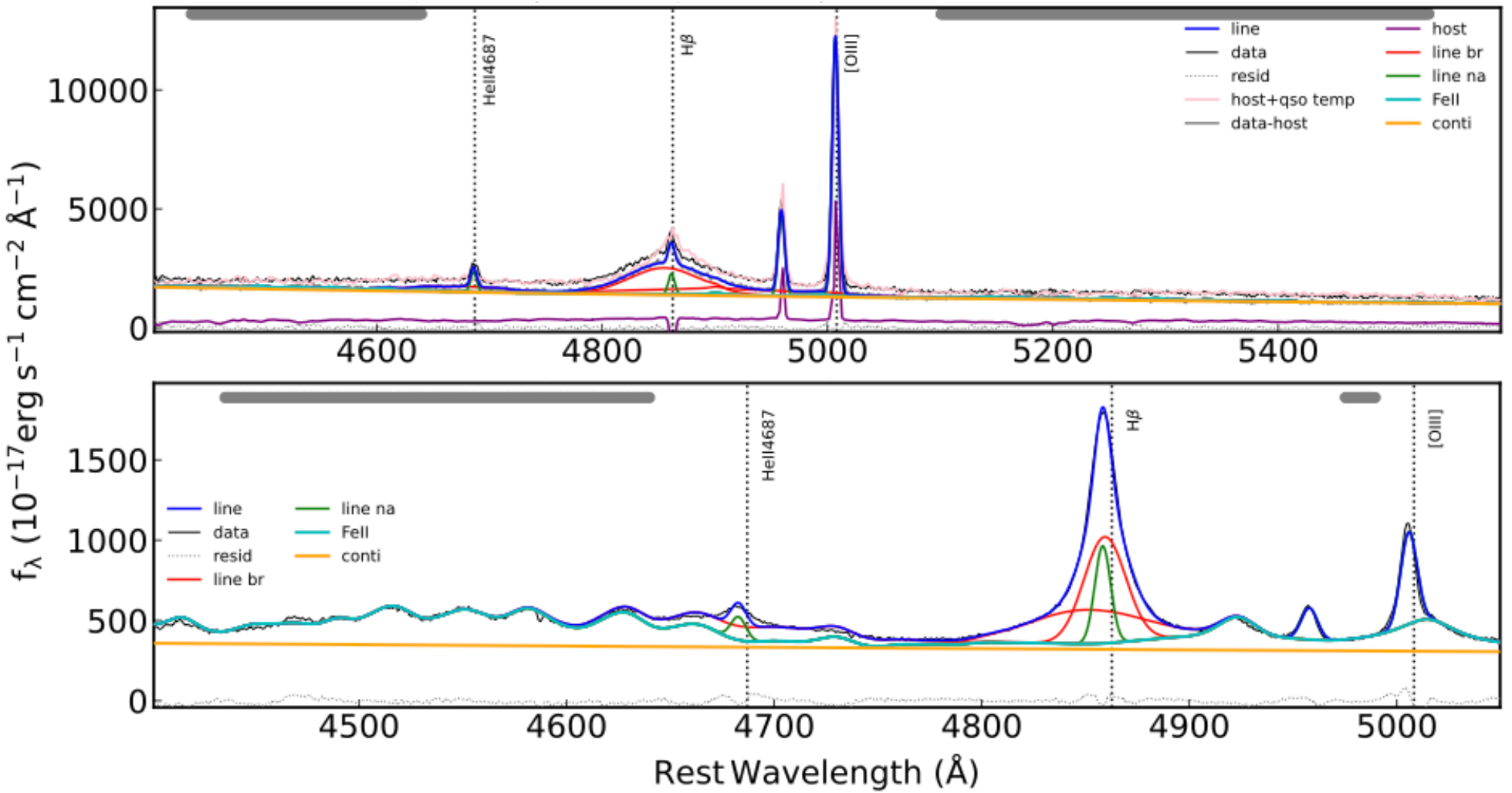}
\caption{{Iron-template and multi-component H$\beta$ model fit to the optical spectrum of Mrk841 (top) and RXSJ17414+2304 (bottom). The fit includes a model for the host, where applicable. The derived fluxes for the iron blend and H$\beta$ are then used to calculate their ratio, $R_\mathrm{Fe}$, which may be used to estimate the mass-accretion rate (\S\ref{sec:rblr}).}}
\label{fig:rfe}
\end{figure*}

{Additionally, two examples illustrating the fit for extracting the ratio of the iron-blend flux to the broad-H$\beta$ flux, $R_{\mathrm{Fe{}}}$, are shown in Figure~\ref{fig:rfe}. The top panel shows Mrk841 having a low $R_{\mathrm{Fe{}}}$ value with iron-blend emission being hardly discernible, while the bottom panel depicts the spectrum of RXSJ17414+2304 with a significant iron-blend contribution, thereby implying high $R_{\mathrm{Fe{}}}$ values. The $R_{\mathrm{Fe{}}}$ values for all objects are presented in Table~\ref{tab:results_rfe}.}

\begin{table}[]
\renewcommand{\arraystretch}{0.70}
\caption{Log of optical spectroscopic observations with SALT, FAST  and HET.}
\centering
 \begin{tabular}{lccrr}
  \hline \hline
  \noalign{\smallskip}
  Object  &  Julian Day & UT Date & {Exp. Time} & {Tel.} \\ 
    &   $+ 2\,400\,000$   &     & {[s]}  &  \\
  \noalign{\smallskip}
  \hline 
  \noalign{\smallskip}
3C\,120 &(1)         & -    & -  &   HET\\
3C\,120 &  56983.31    &  2014-11-22  & x  &   FAST\\
AKN\,120  &   58821.74 &   2019-12-04  &  848 &   HET\\
{ESO\,141-G55} &   56417.54 &   2013-05-05  &  900 &   SALT\\
ESO\,374-G25 &   55972.57 &   2012-02-15  &  2 $\times$ 460 &   SALT\\
ESO\,399-IG20   &   56095.41 &   2012-06-16  &   700 &   SALT\\
ESO\,438-G09 &   55981.61 &   2012-02-24  &   2 $\times$ 480 &   SALT\\
ESO578-G09& 57044.53& 2015-01-22 &x& FAST\\
HE\,1136-2304 & (2)    &   -   &   - &   SALT\\
IRAS09595-0755&  57005.54  &   2014-12-14  &   x &   FAST\\
MCG-02.12.050 &    56983.33  &  2014-11-22  & x  &   FAST\\
MRK1347& 57010.53 & 2014-12-19 & x & FAST\\
MRK\,335  & 56983.24 & 2014-11-22 & x& FAST\\
MRK\,509  &   56064.55 &   2012-05-17  &   700 &   SALT\\
MRK\,509  &   56984.11 & 2014-11-23 & &   FAST\\
MRK841& 57044.54 &2015-01-22& x& FAST\\
NGC1019  &   56240.41 &   2012-11-08  &   x   &   SALT\\
NGC1019  &   56983.25 &  2014-11-22   &   x   &   FAST\\
NGC5940&57044.54&2015-01-22&x&FAST\\
NGC\,7603    &   56984.13 &  2014-11-23  & &   FAST\\
NGC\,7603    &   56235.31 &   2012-11-03  &   2 $\times$ 480 &   SALT\\
NGC\,985  &   56235.31 &  2014-11-21   & &   FAST\\
PGC\,50427   &   56071.47 &   2012-05-23  &   700 &   SALT\\
PGC64989 & 56985.06 & 2014-11-24& x& FAST \\
RXS\,J06225-2317&   56983.43 &   2014-11-22  &   x &   FAST\\
RXS\,J06225-2317&   58122.54 &   2018-01-04  &   864 &   SALT\\
RXS\,J11032-0654&   55976.41 &   2012-02-18  &   2 $\times$ 480 &   SALT\\
RXS\,J11032-0654&   57012.54 &  2014-12-21   &   x &   FAST\\
RXS\,J17414+034 &   56064.53 &   2012-05-17  &   700 &   SALT\\
  WPVS\,48 &   56423.34 &   2013-05-10  &   600 &   SALT\\
  WPVS\,48&  56985.53  &   2014-11-24  &   x &   FAST\\
  \hline 
 \end{tabular}
 \tablecomments{(1) We used the average spectrum from a spectroscopic campaign on 3C\,120 \citep{kollatschny14}. (2) We used the average spectrum from a spectroscopic  campaign on HE\,1136-2304 \citep{kollatschny18}.}
\label{tab:spectroscopy_log}
\end{table}

\includegraphics[width=0.33\columnwidth]{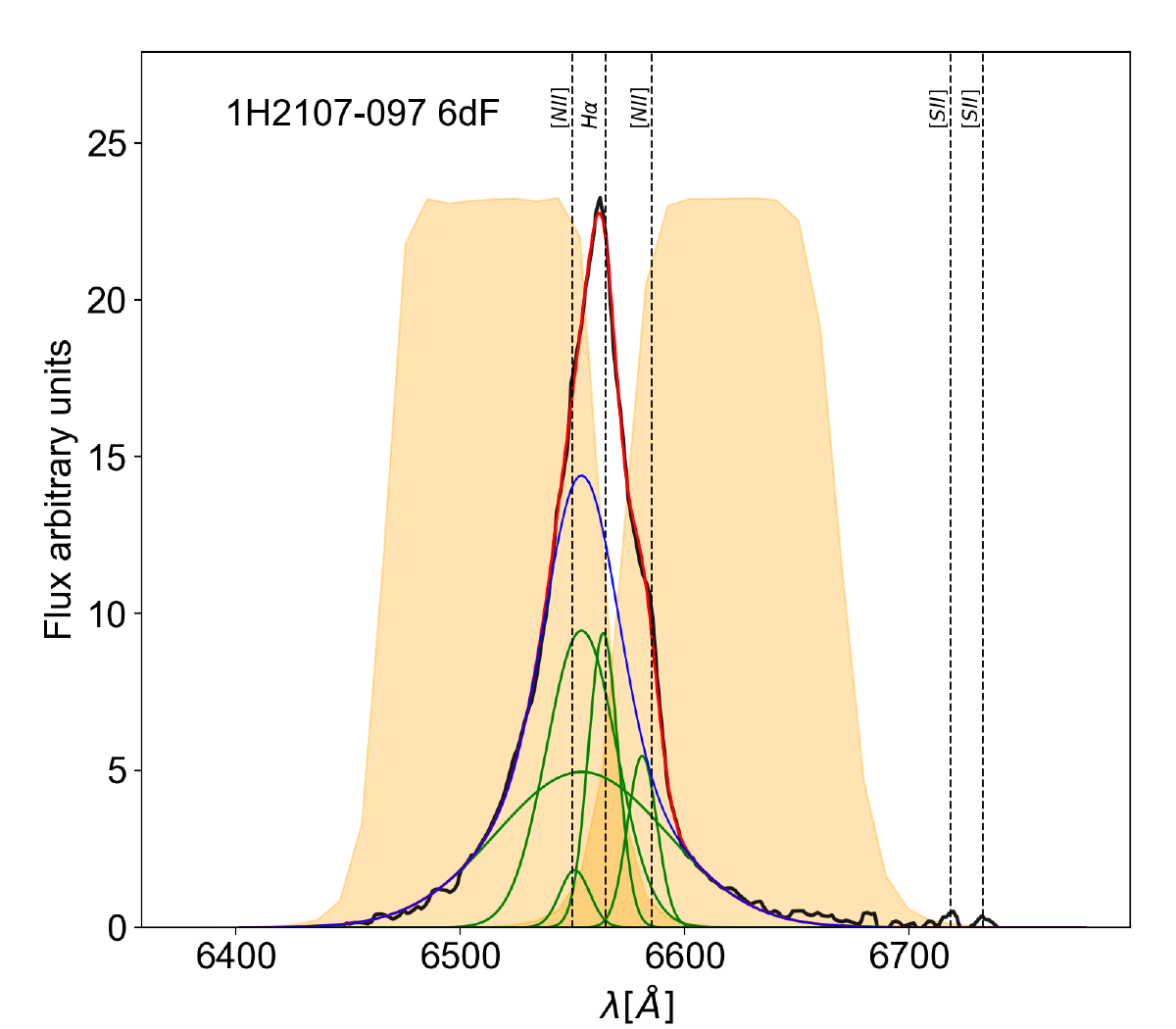}
\includegraphics[width=0.33\columnwidth]{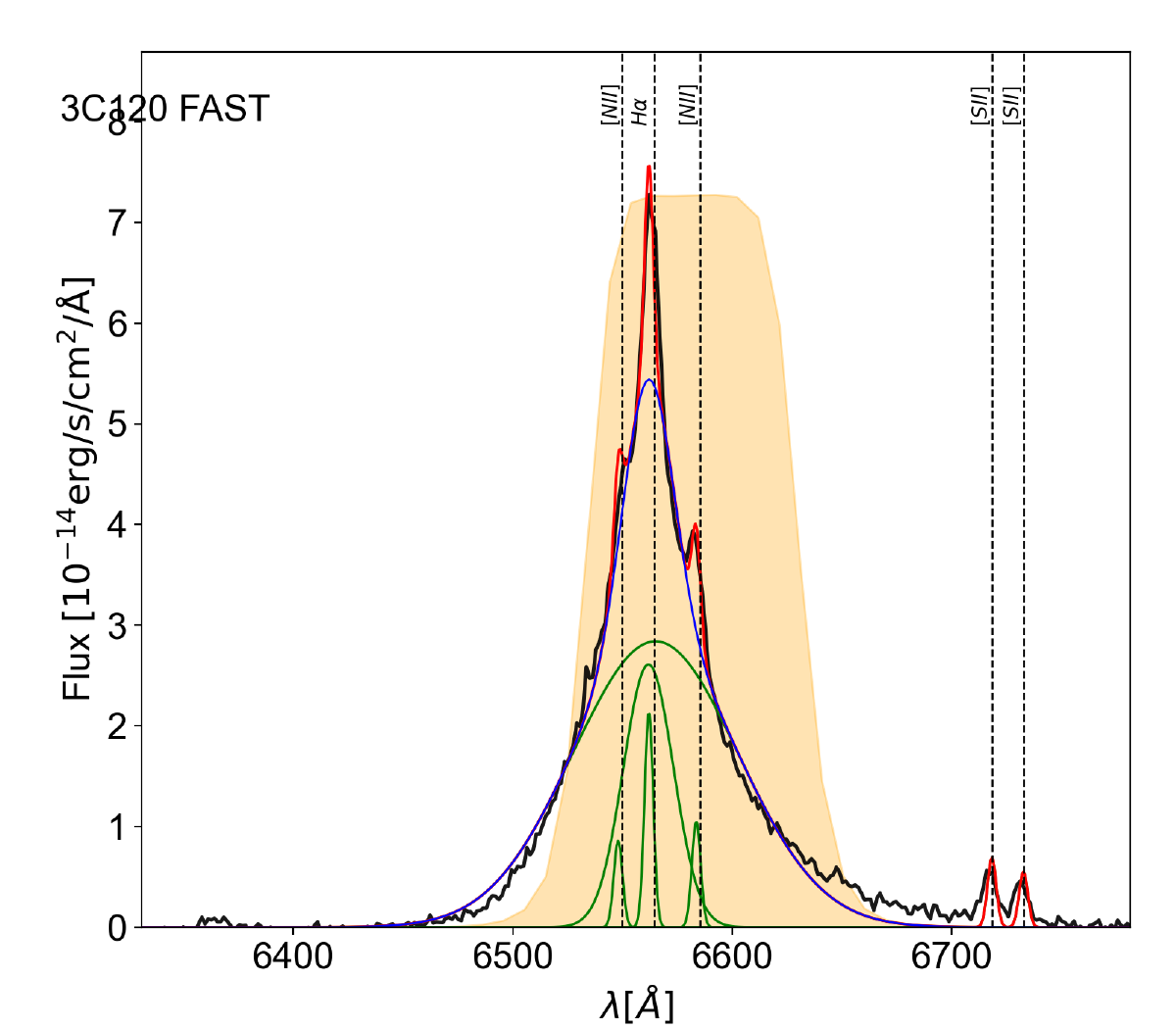}
\includegraphics[width=0.33\columnwidth]{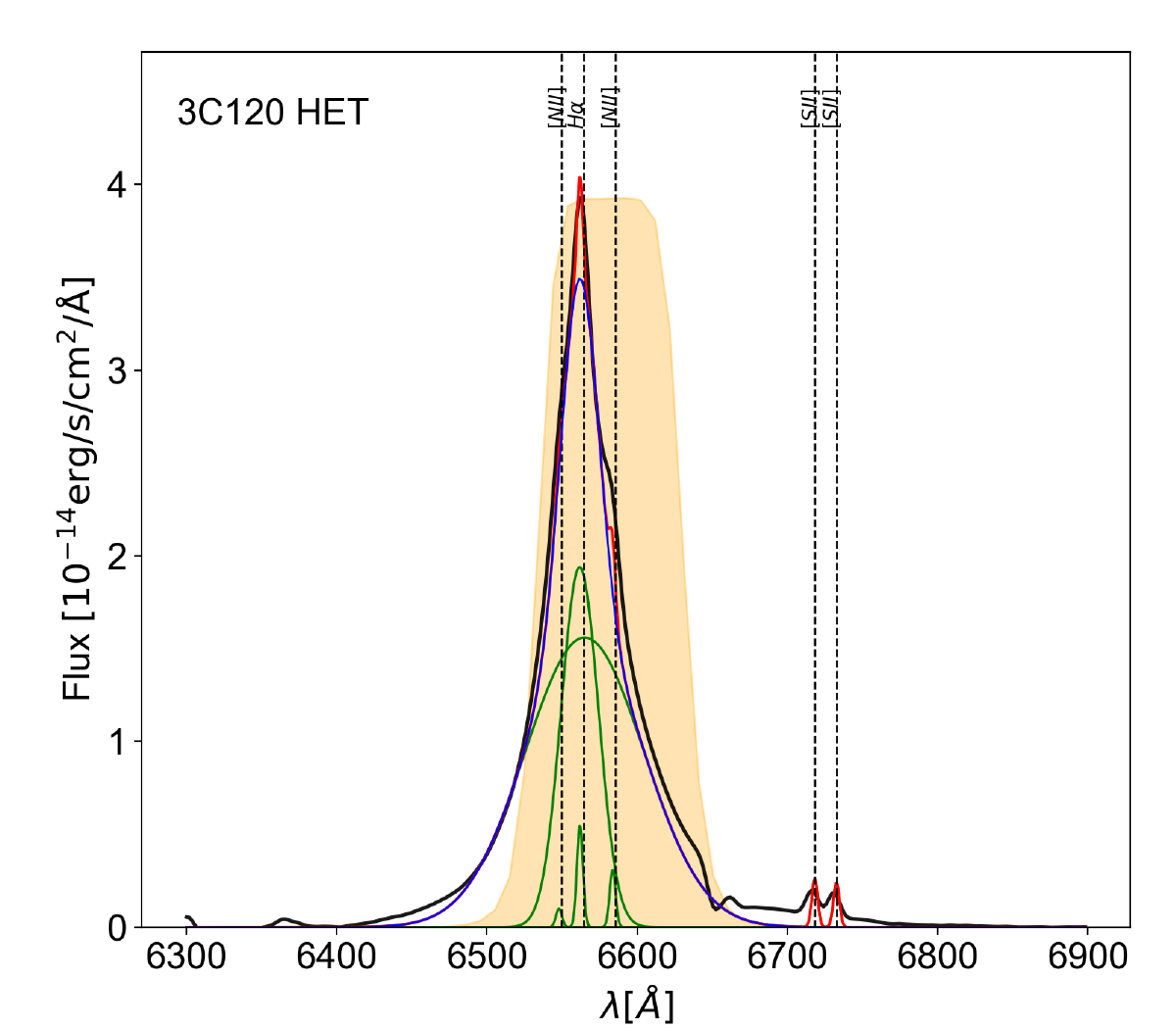}

\includegraphics[width=0.33\columnwidth]{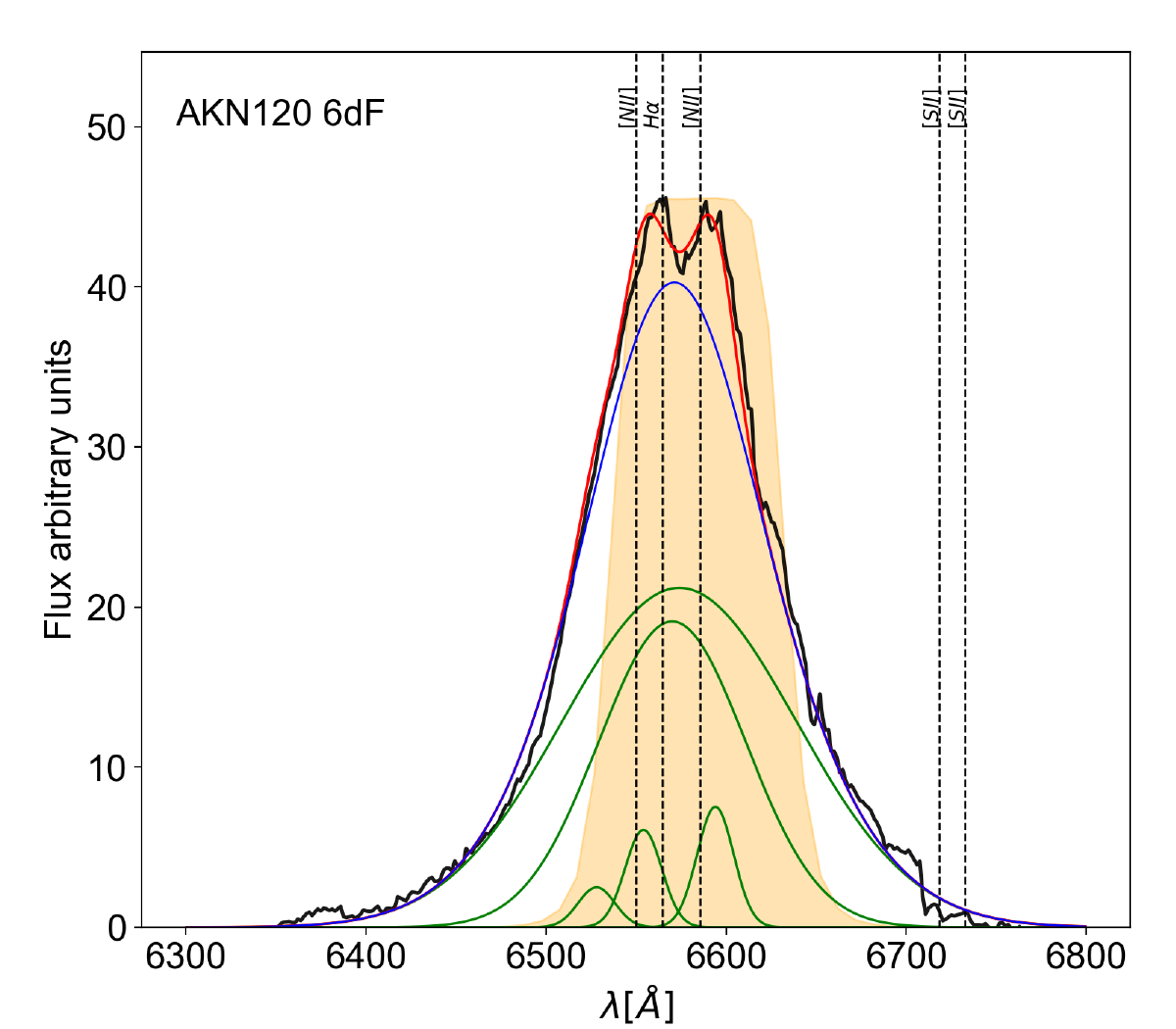}
\includegraphics[width=0.33\columnwidth]{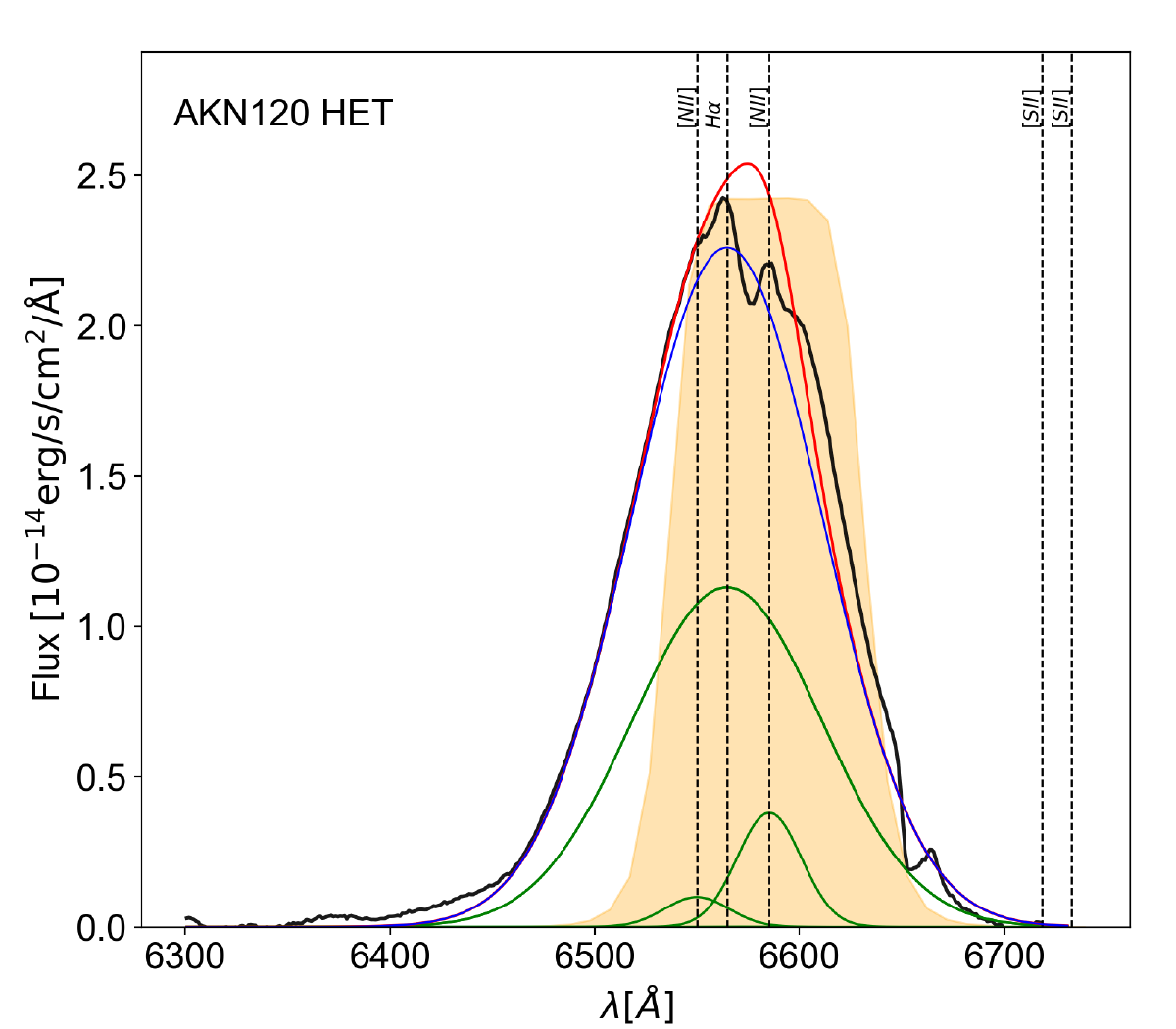}
\includegraphics[width=0.33\columnwidth]{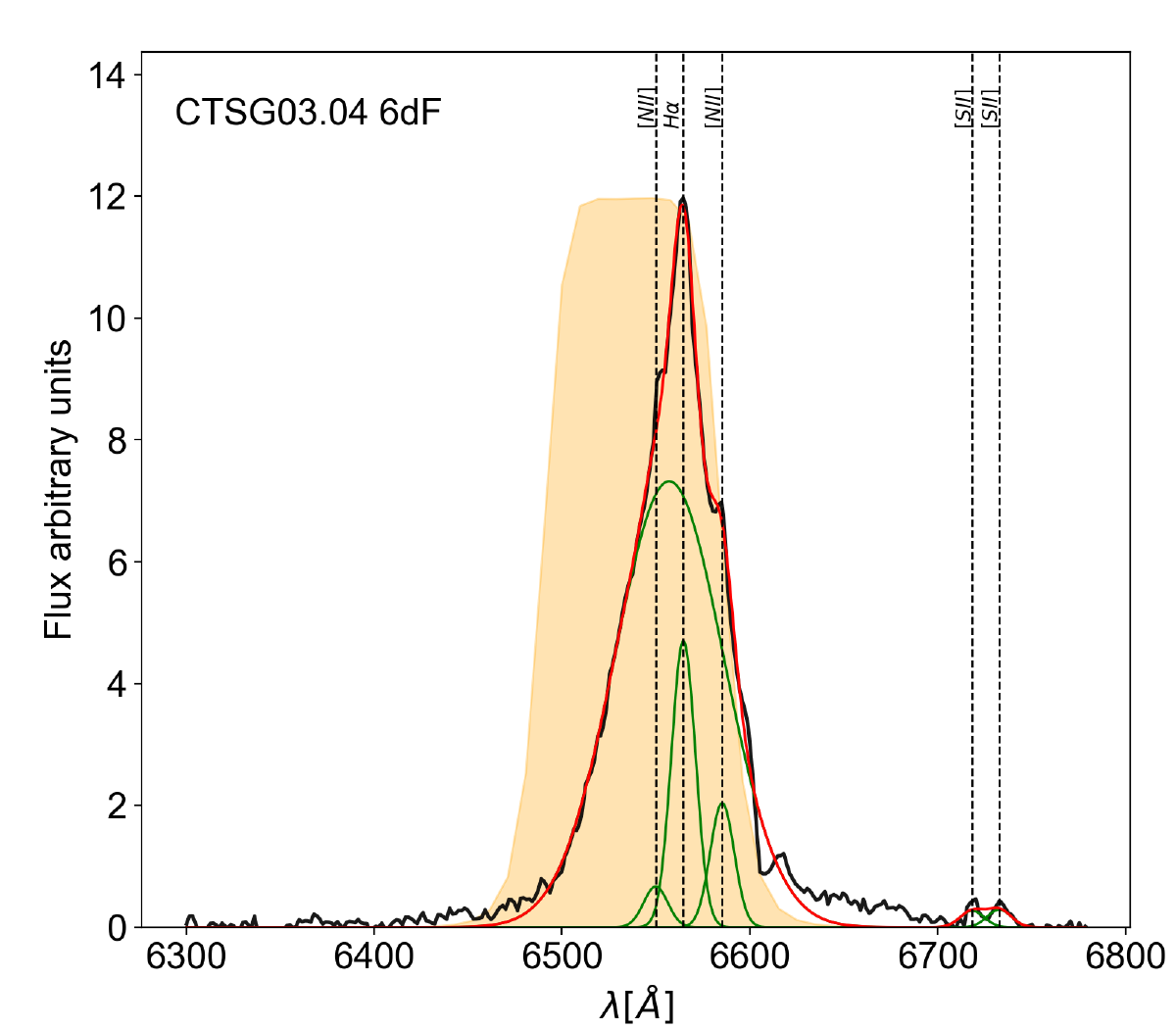}

\includegraphics[width=0.33\columnwidth]{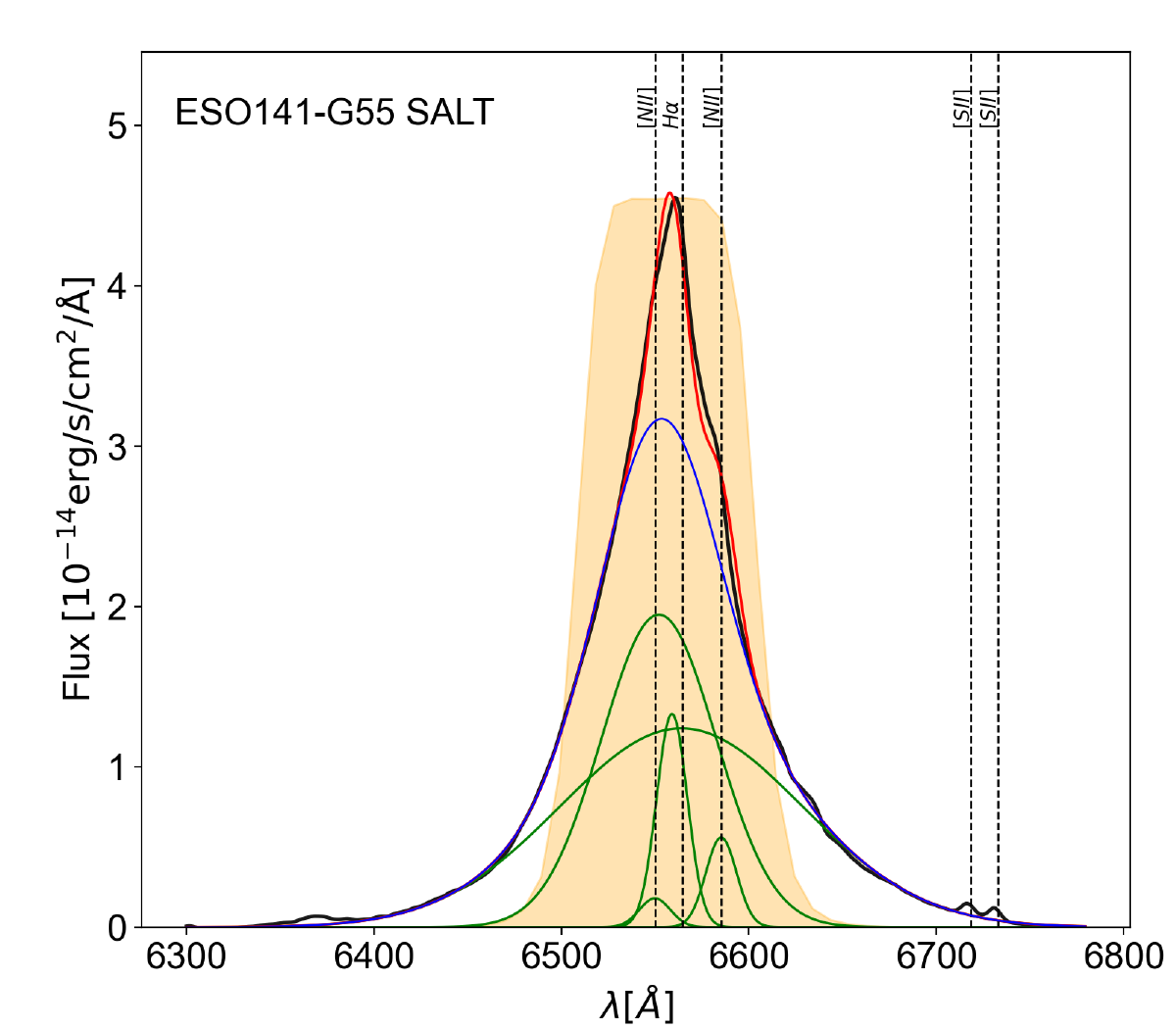}
\includegraphics[width=0.33\columnwidth]{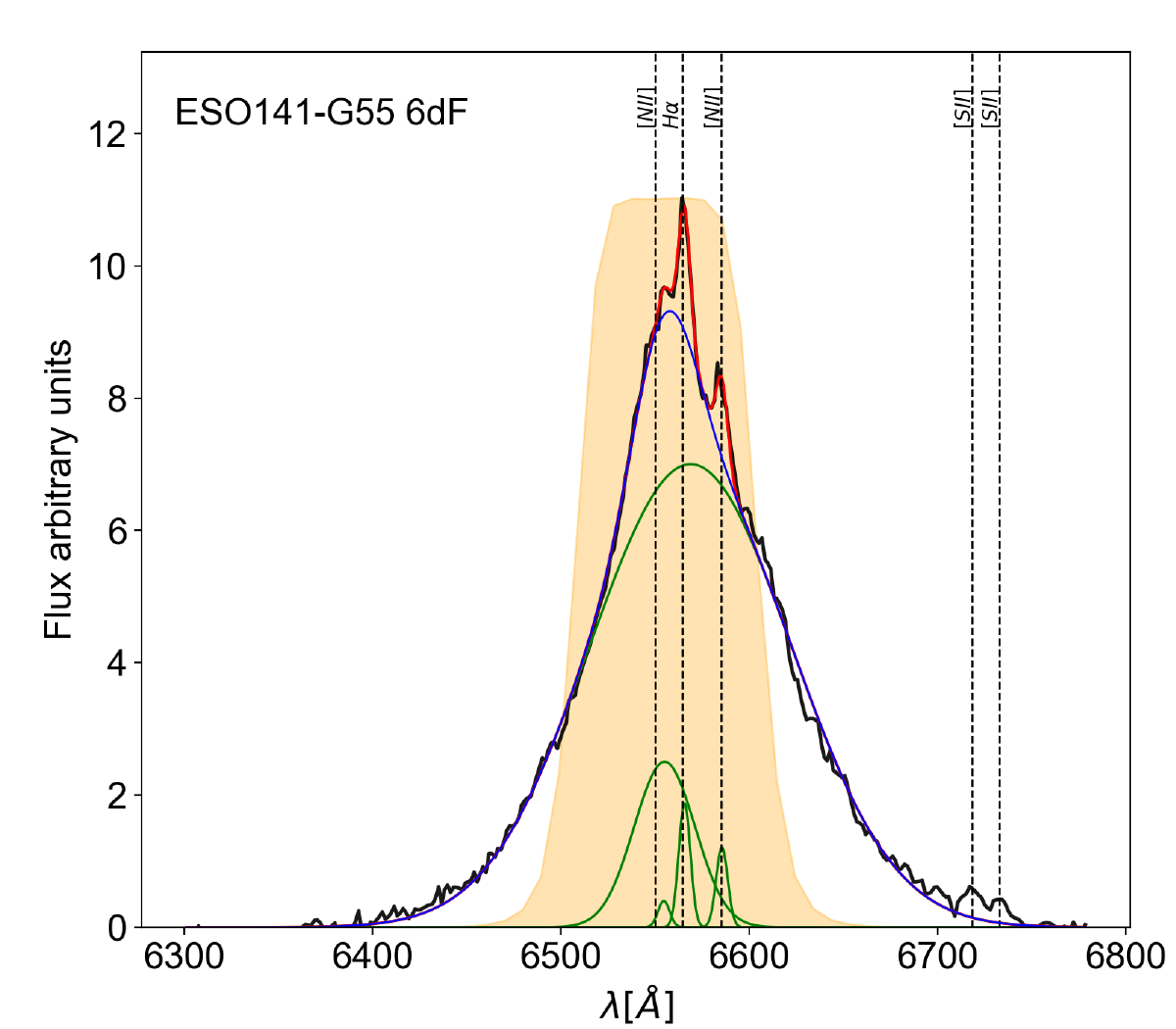}
\includegraphics[width=0.33\columnwidth]{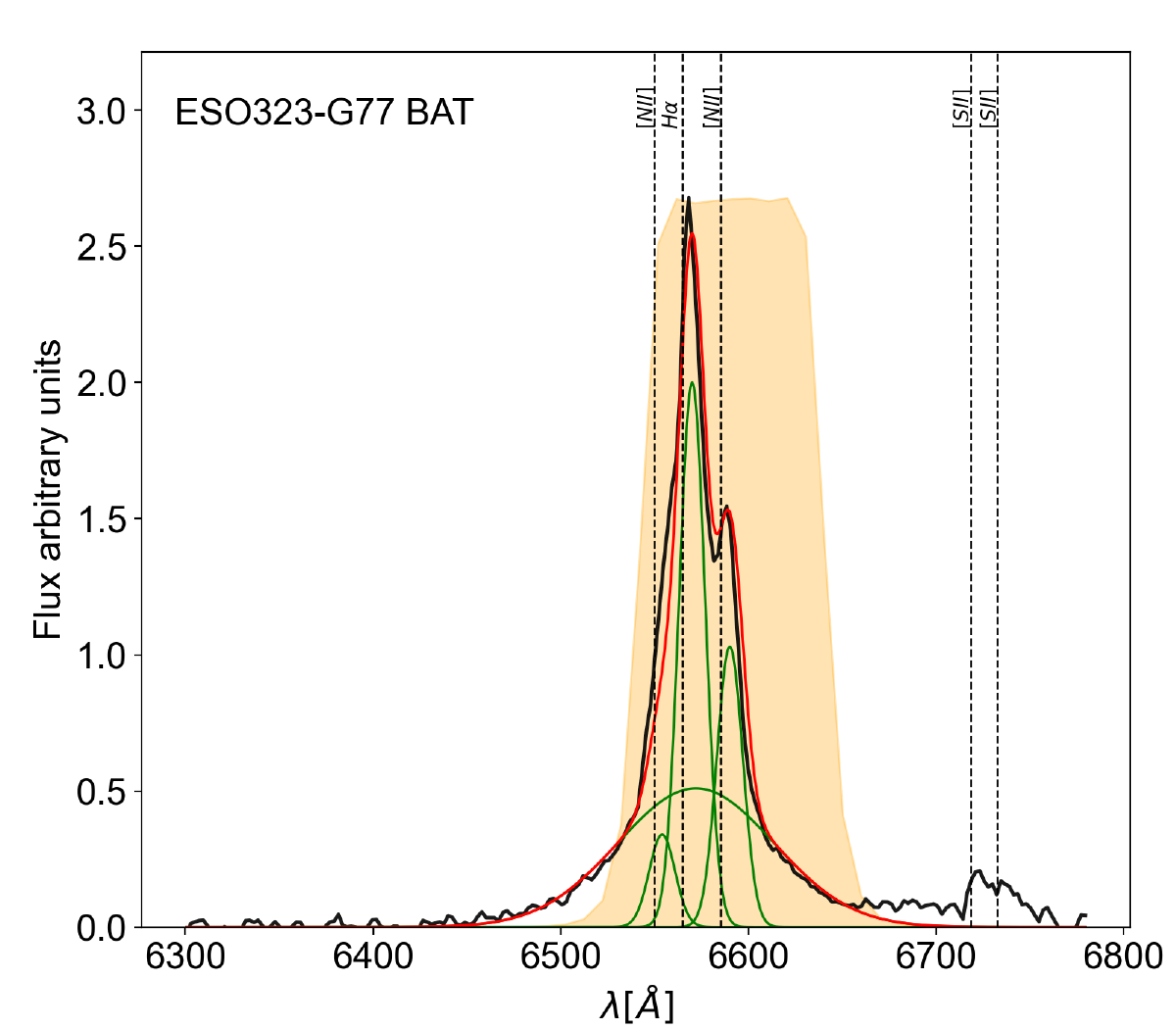}

\includegraphics[width=0.33\columnwidth]{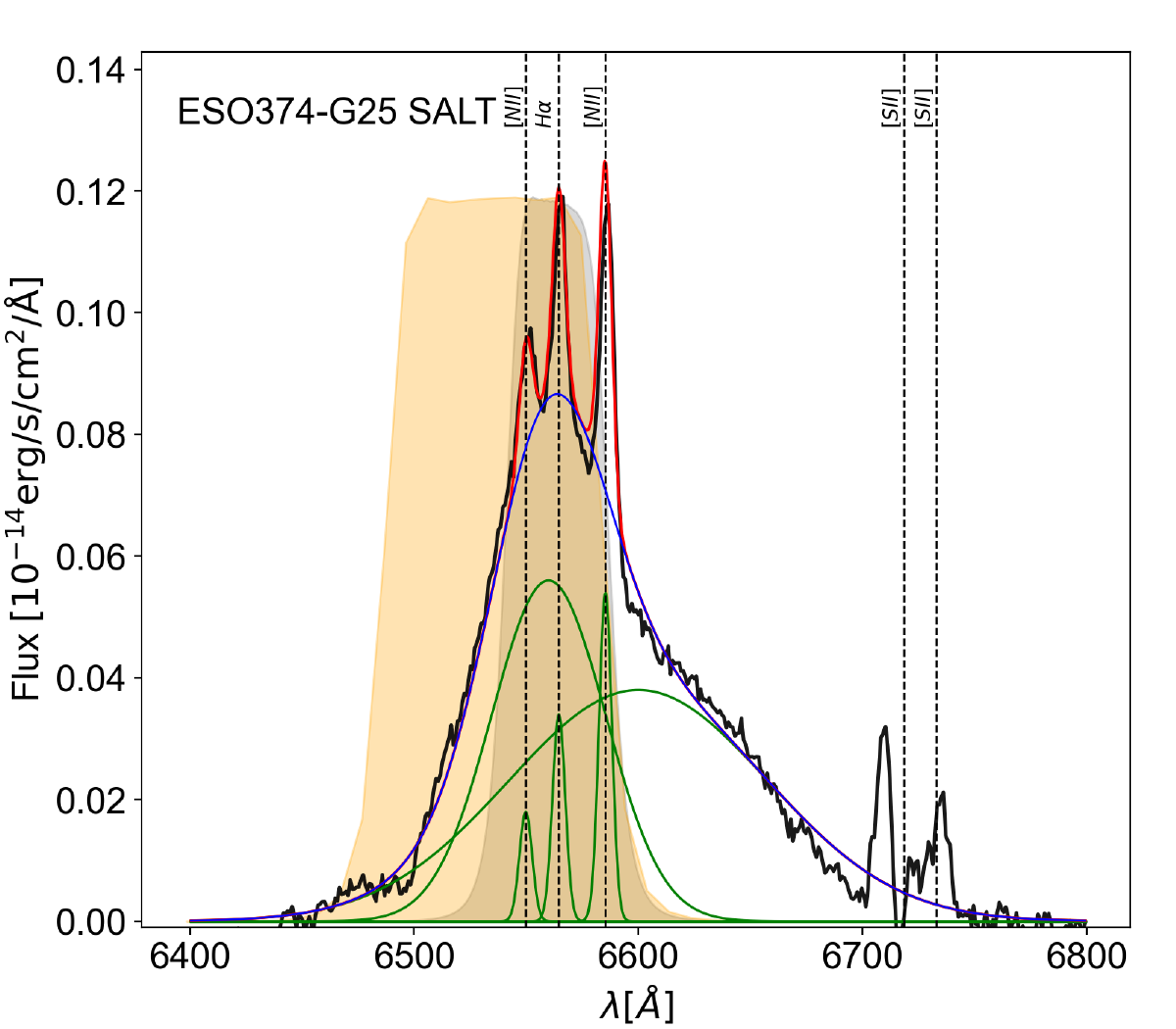}
\includegraphics[width=0.33\columnwidth]{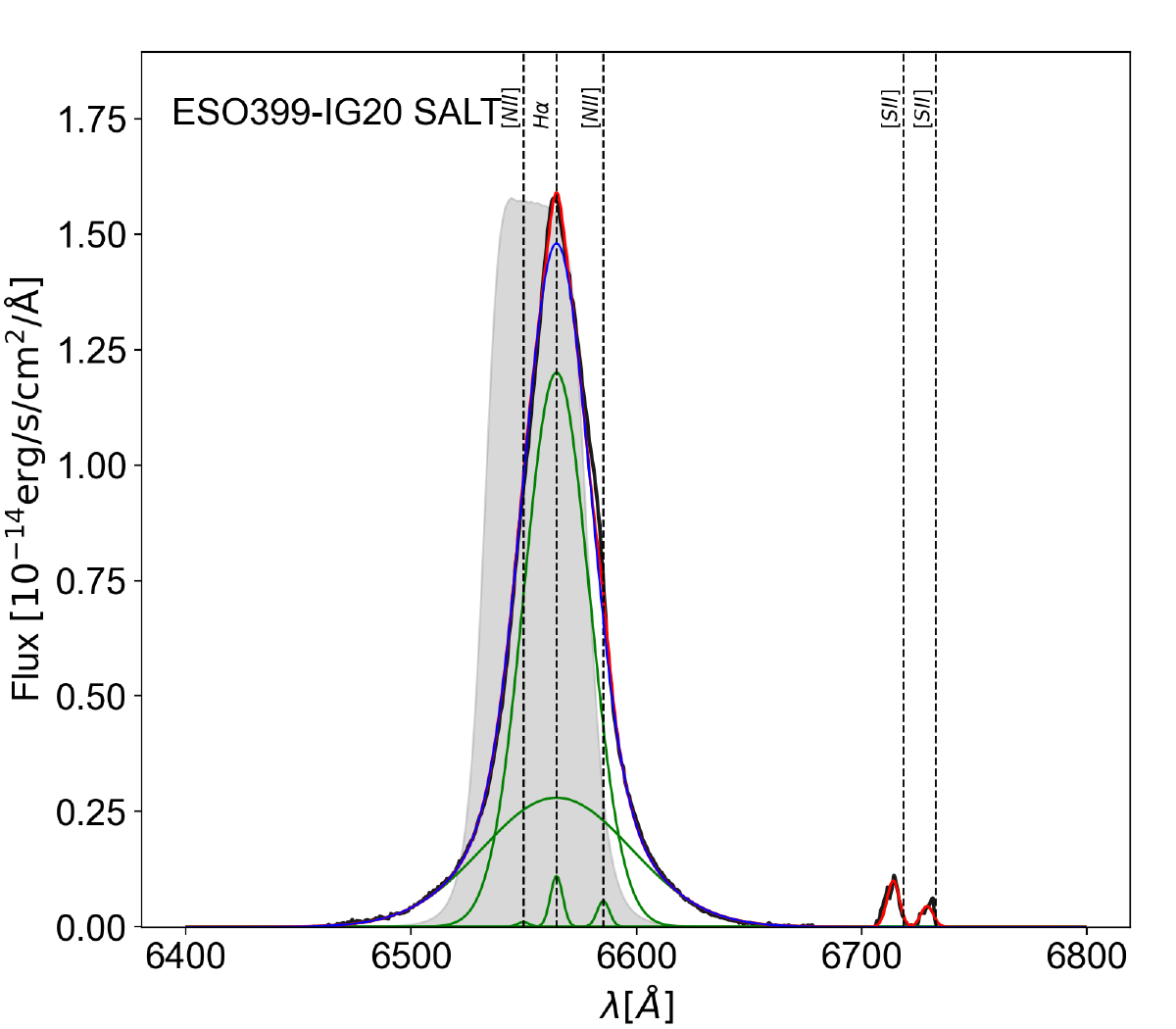}
\includegraphics[width=0.33\columnwidth]{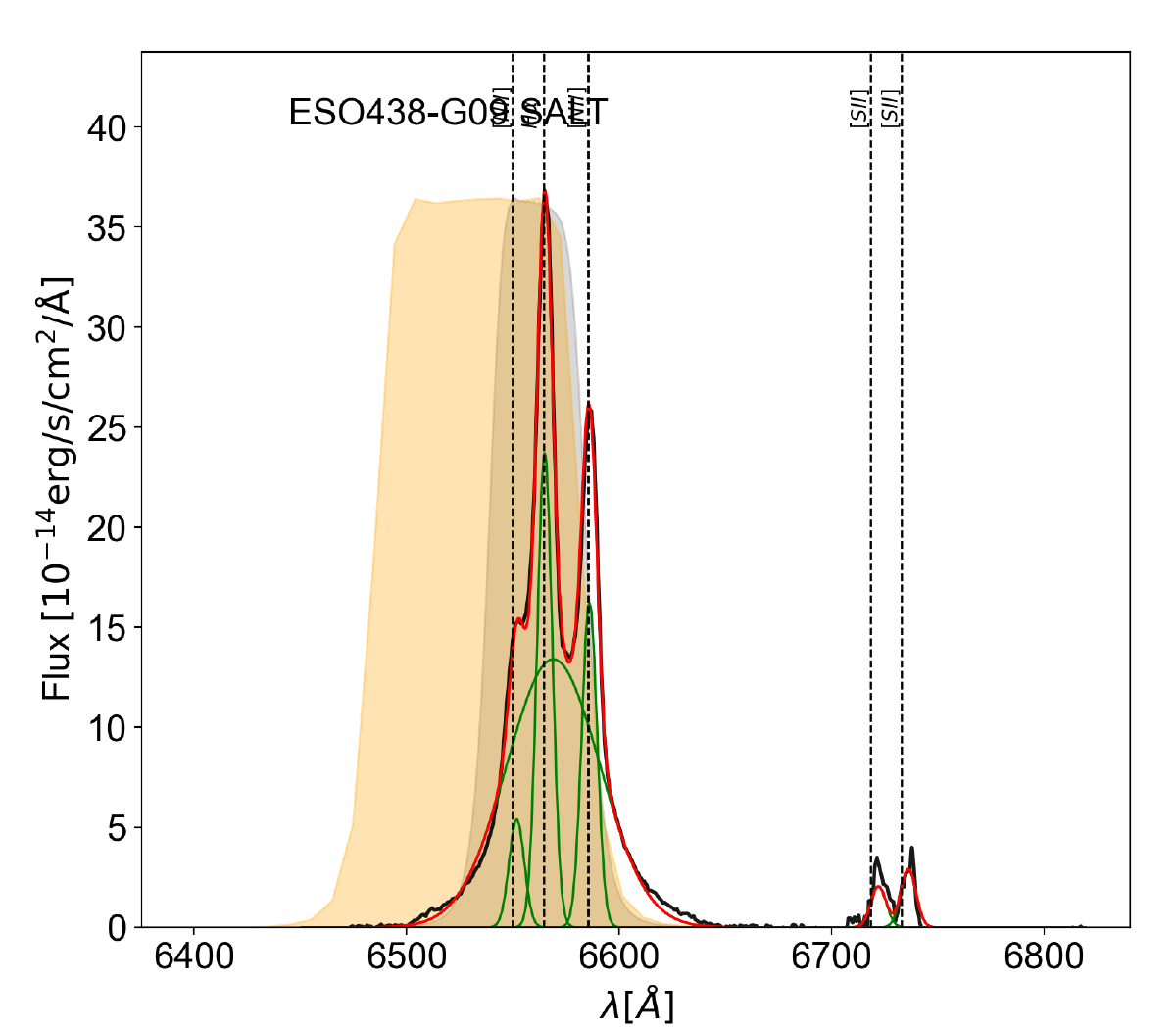}

\includegraphics[width=0.33\columnwidth]{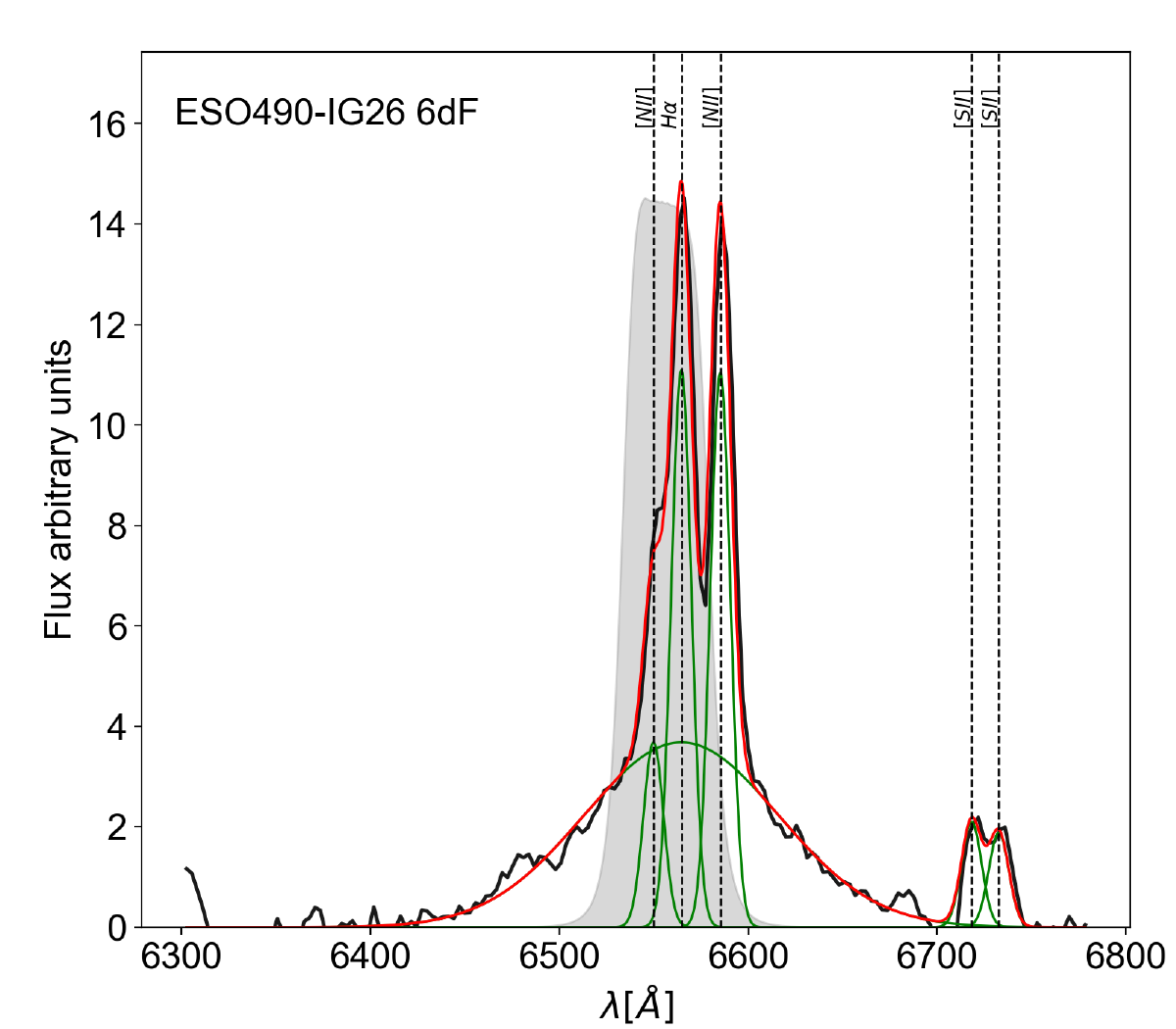}
\includegraphics[width=0.33\columnwidth]{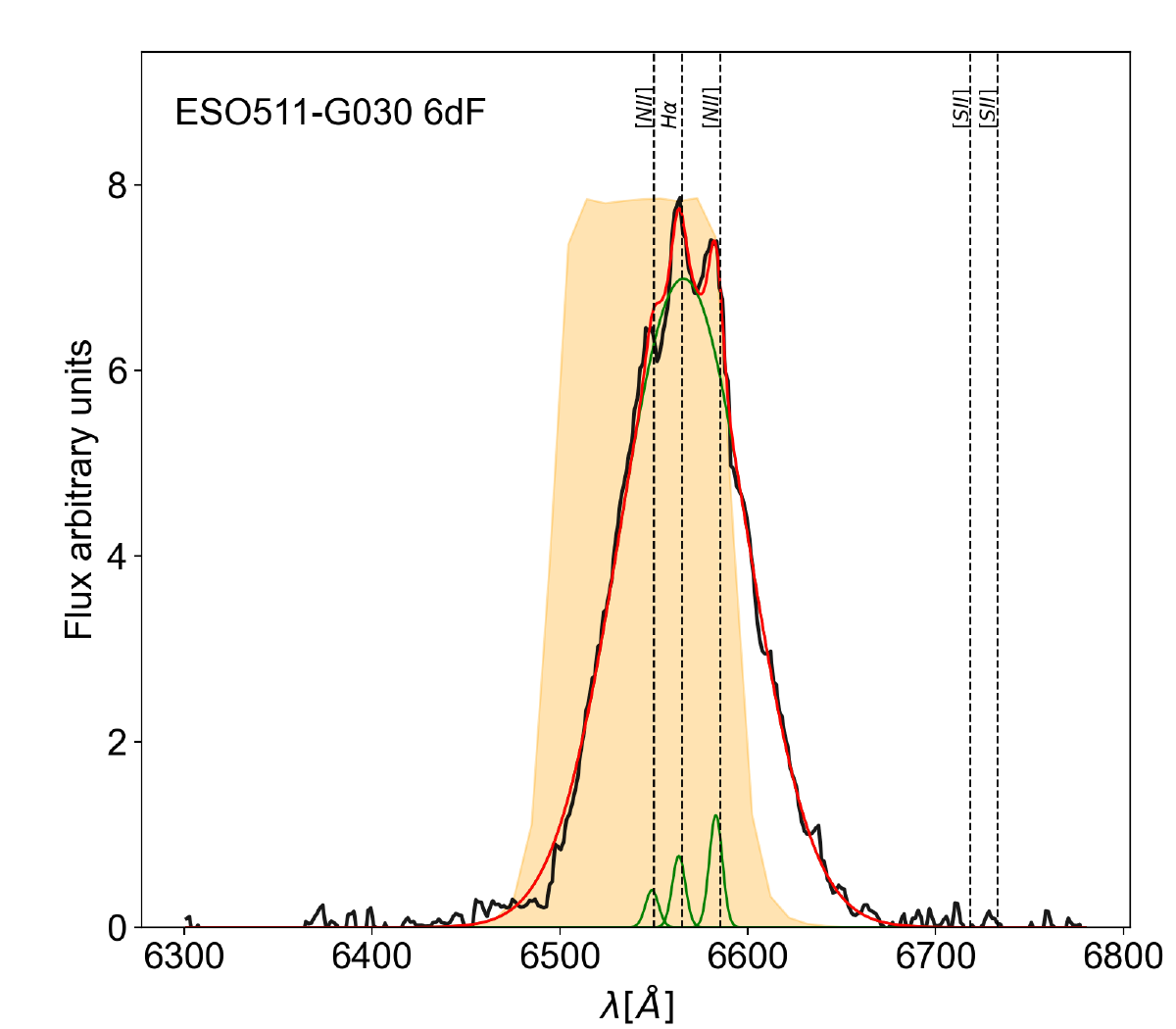}
\includegraphics[width=0.33\columnwidth]{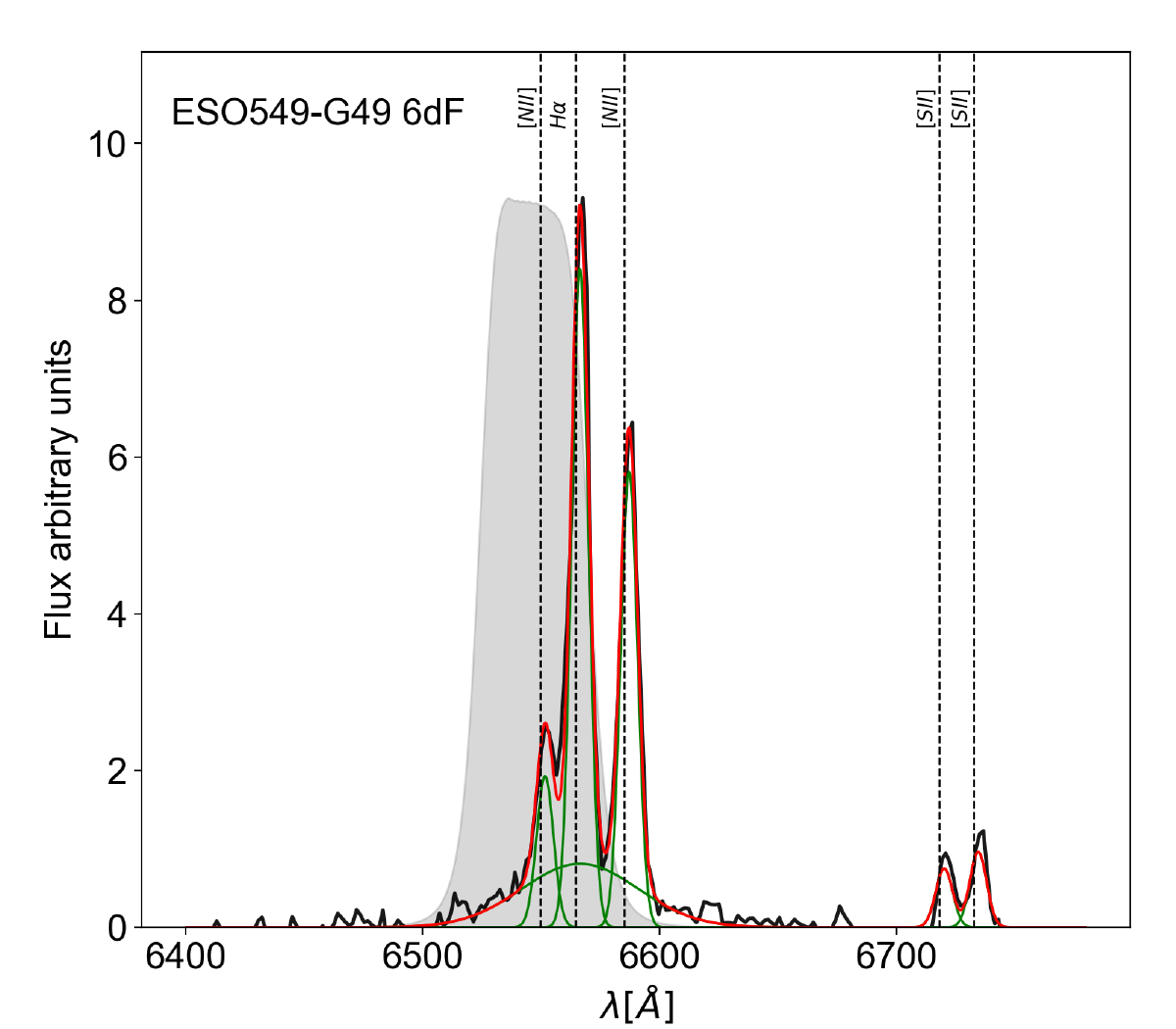}

\includegraphics[width=0.33\columnwidth]{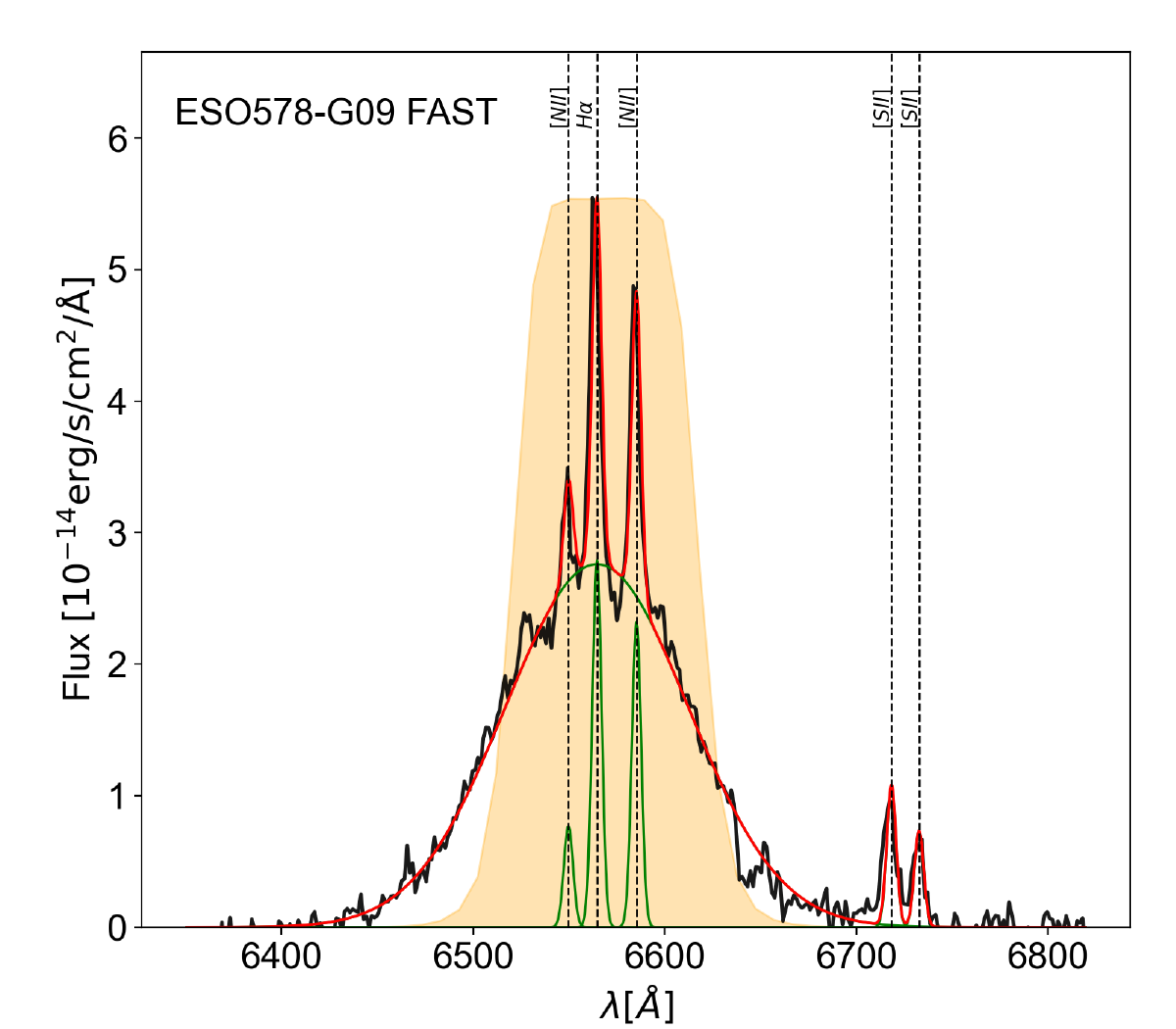}
\includegraphics[width=0.33\columnwidth]{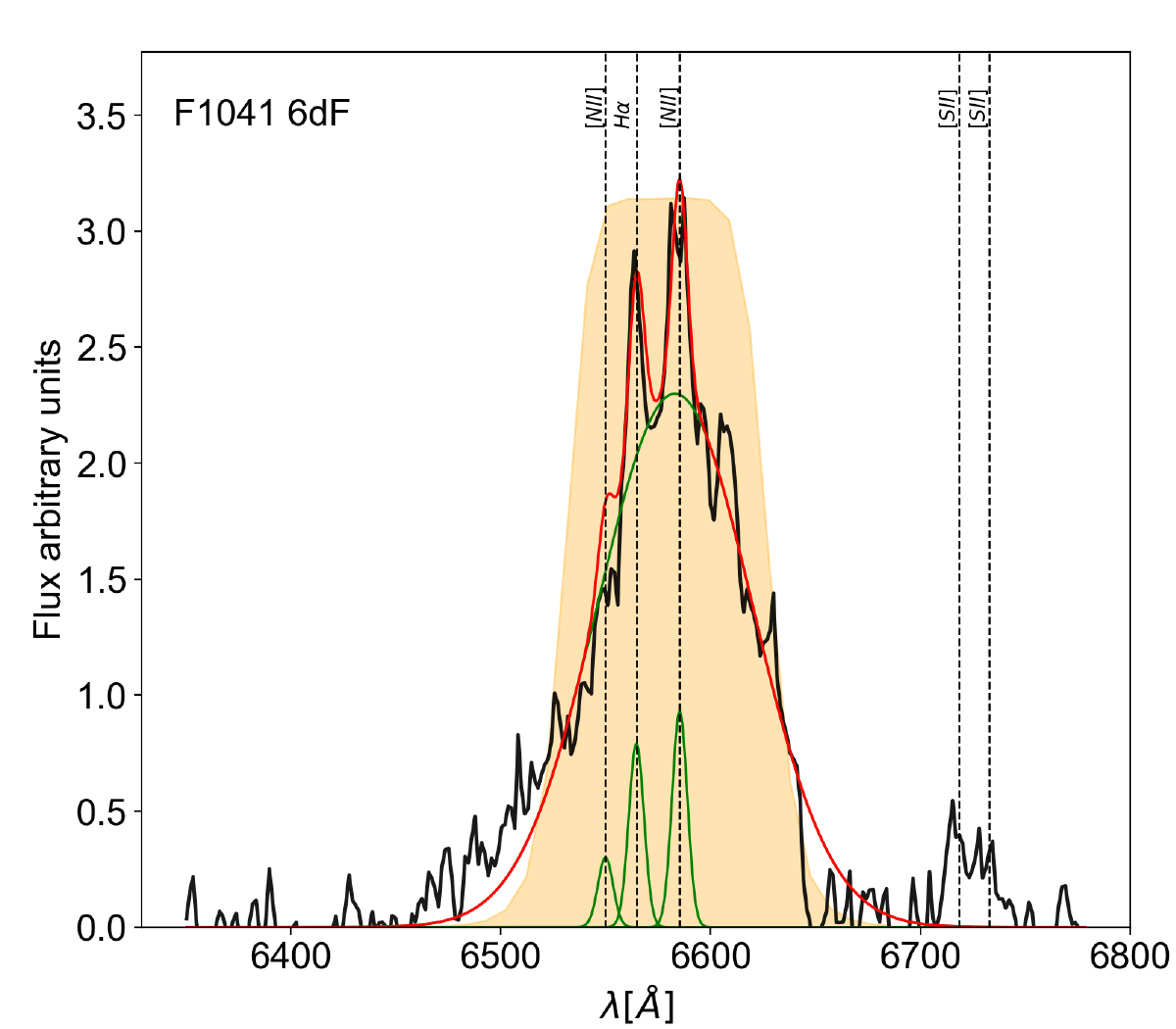}
\includegraphics[width=0.33\columnwidth]{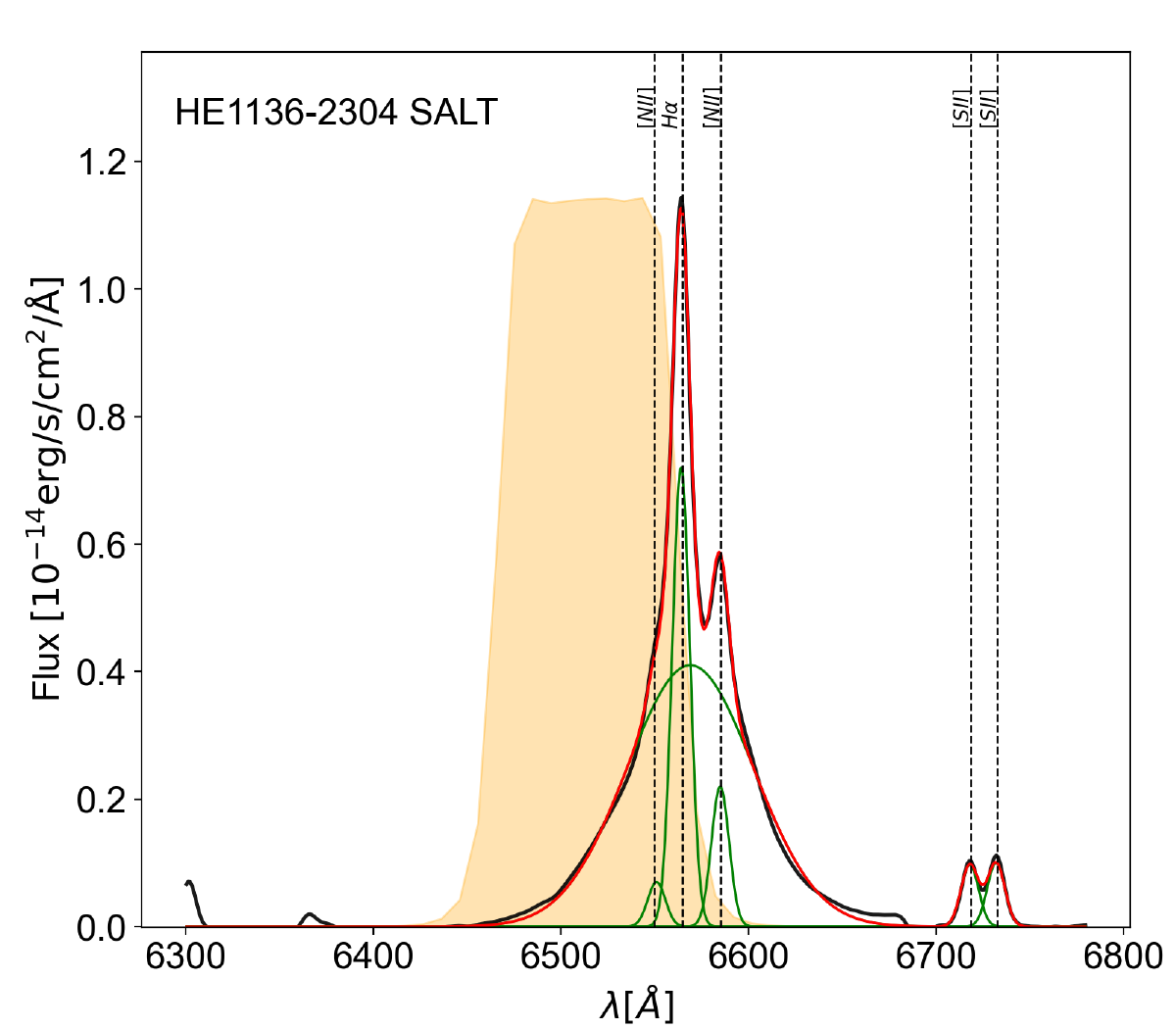}

\includegraphics[width=0.33\columnwidth]{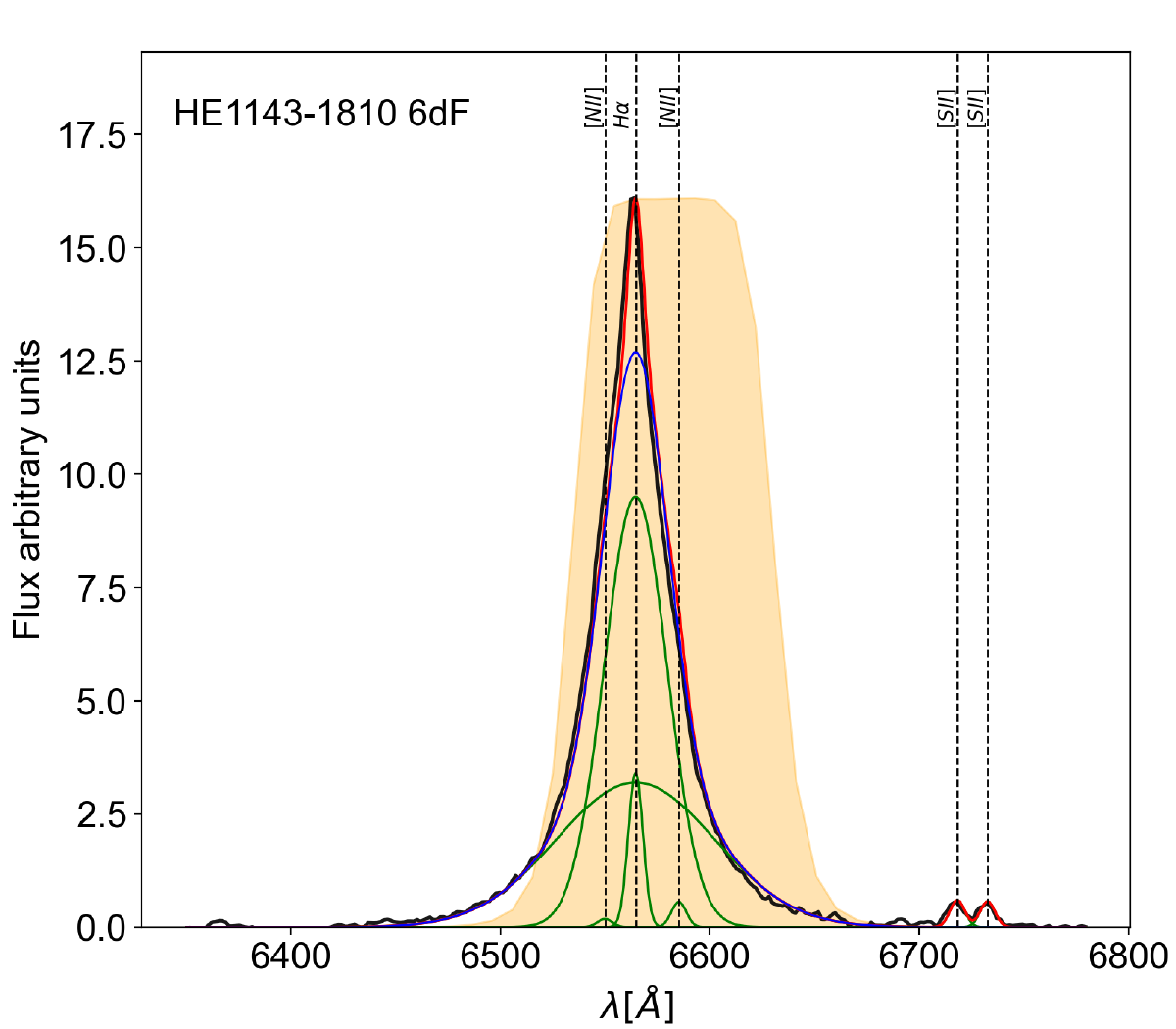}
\includegraphics[width=0.33\columnwidth]{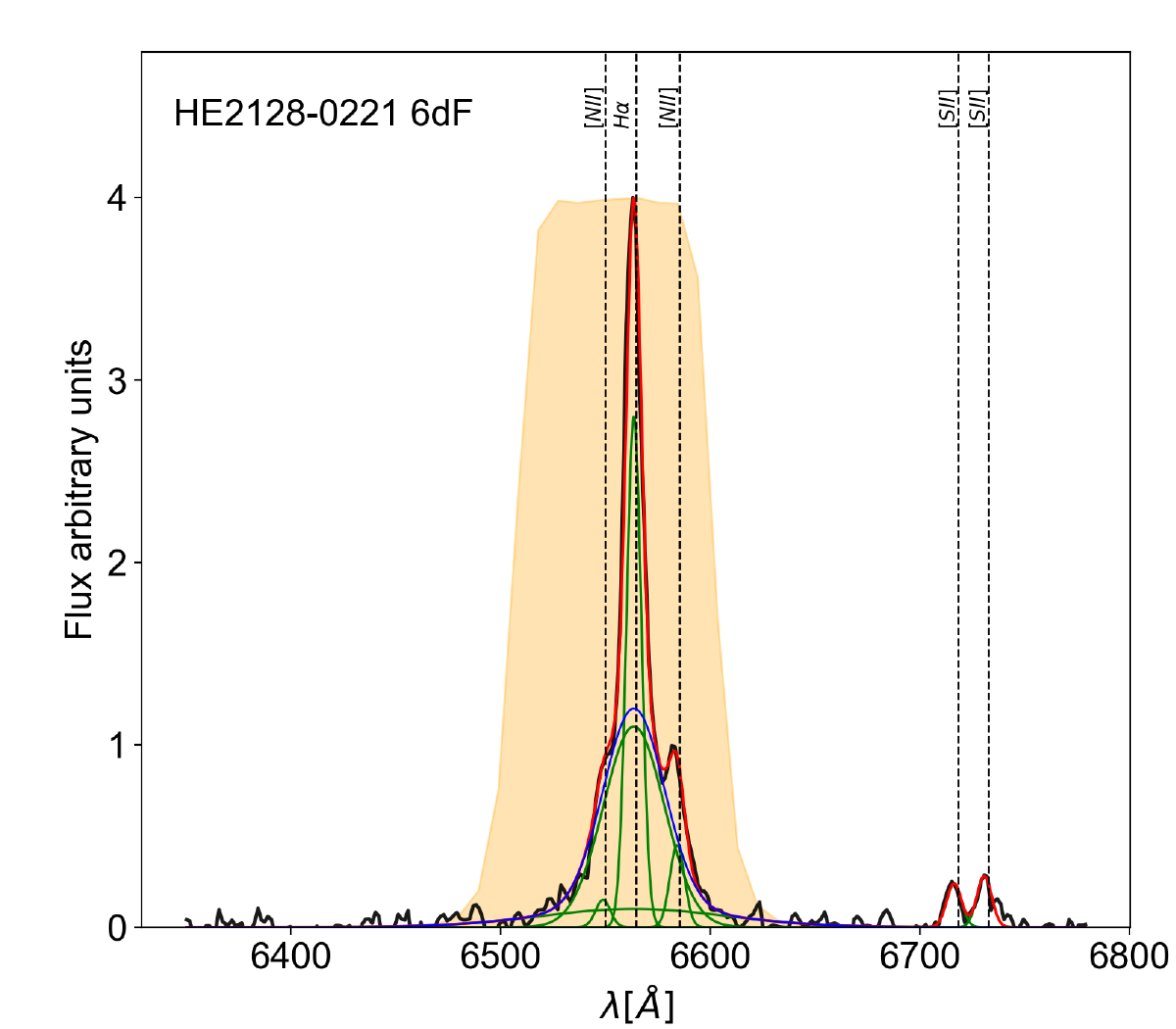}
\includegraphics[width=0.33\columnwidth]{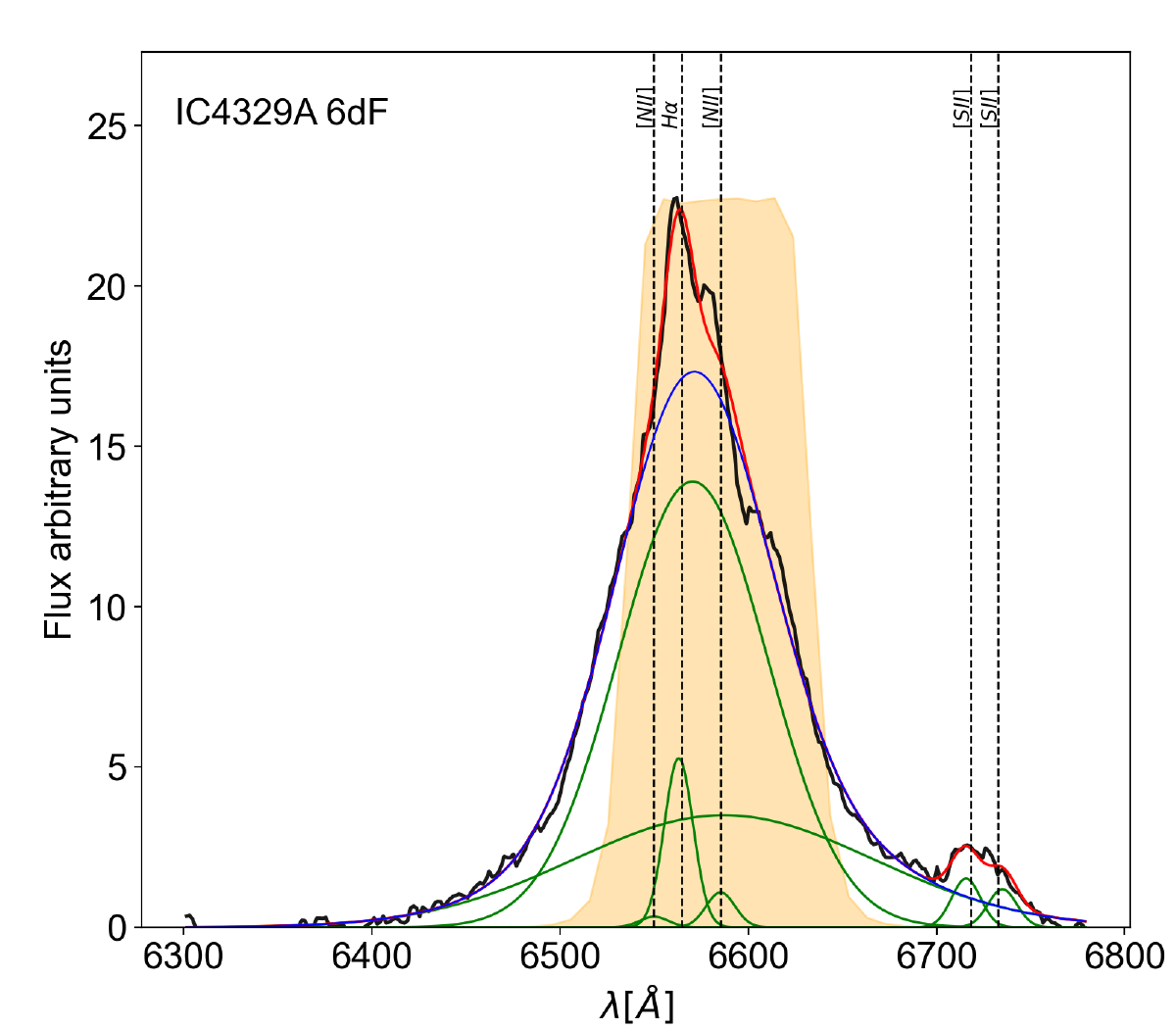}

\includegraphics[width=0.33\columnwidth]{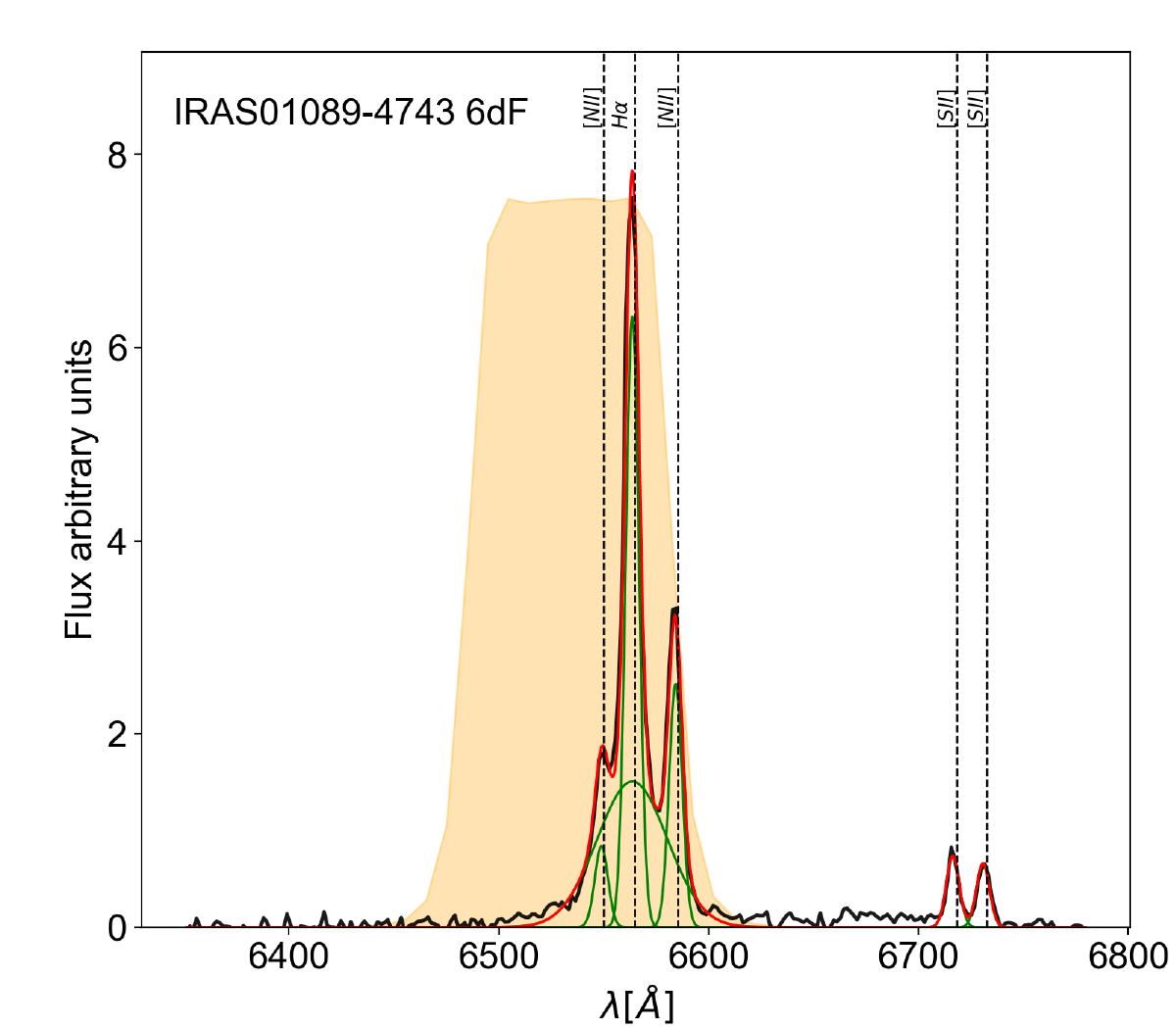}
\includegraphics[width=0.33\columnwidth]{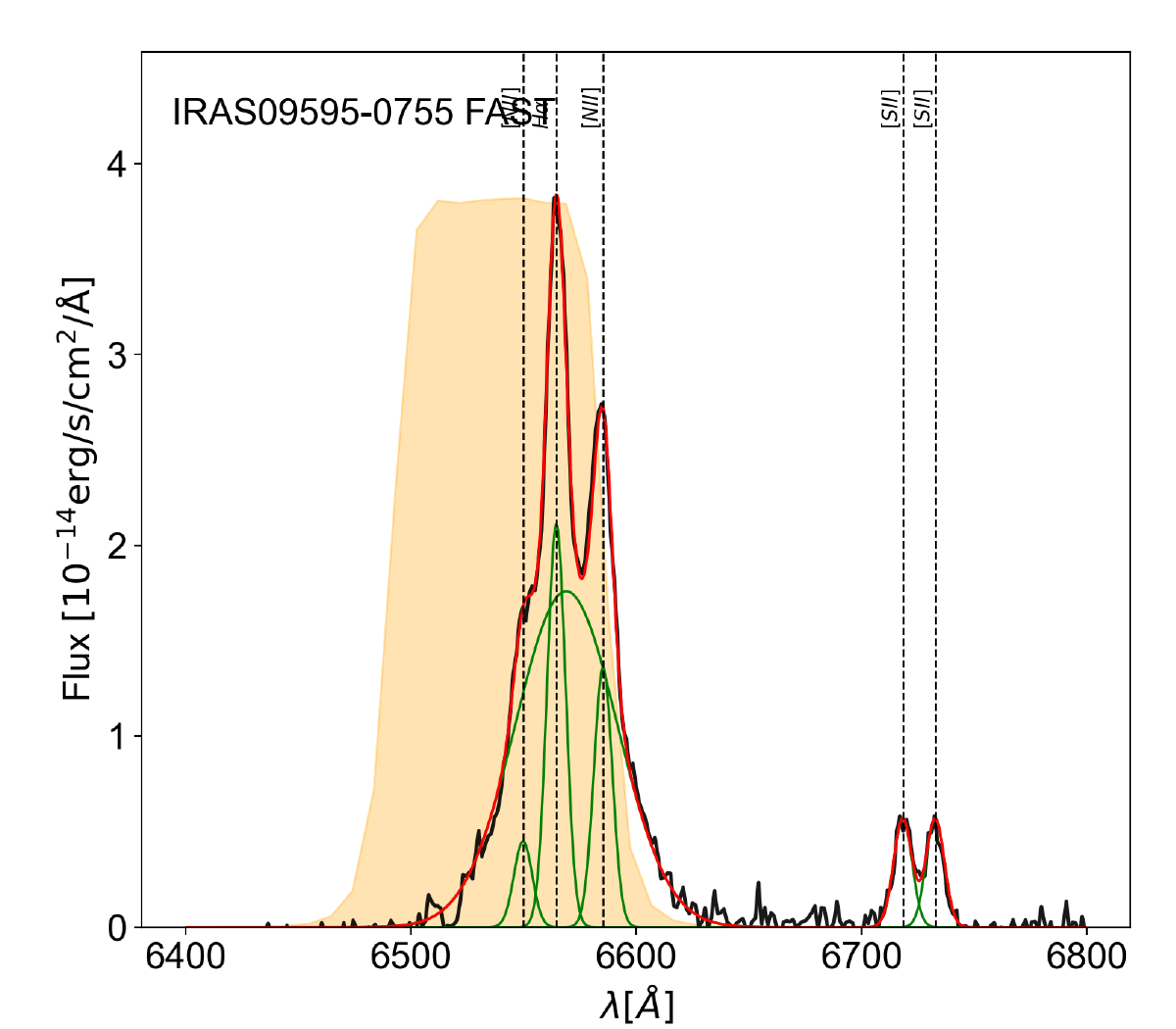}
\includegraphics[width=0.33\columnwidth]{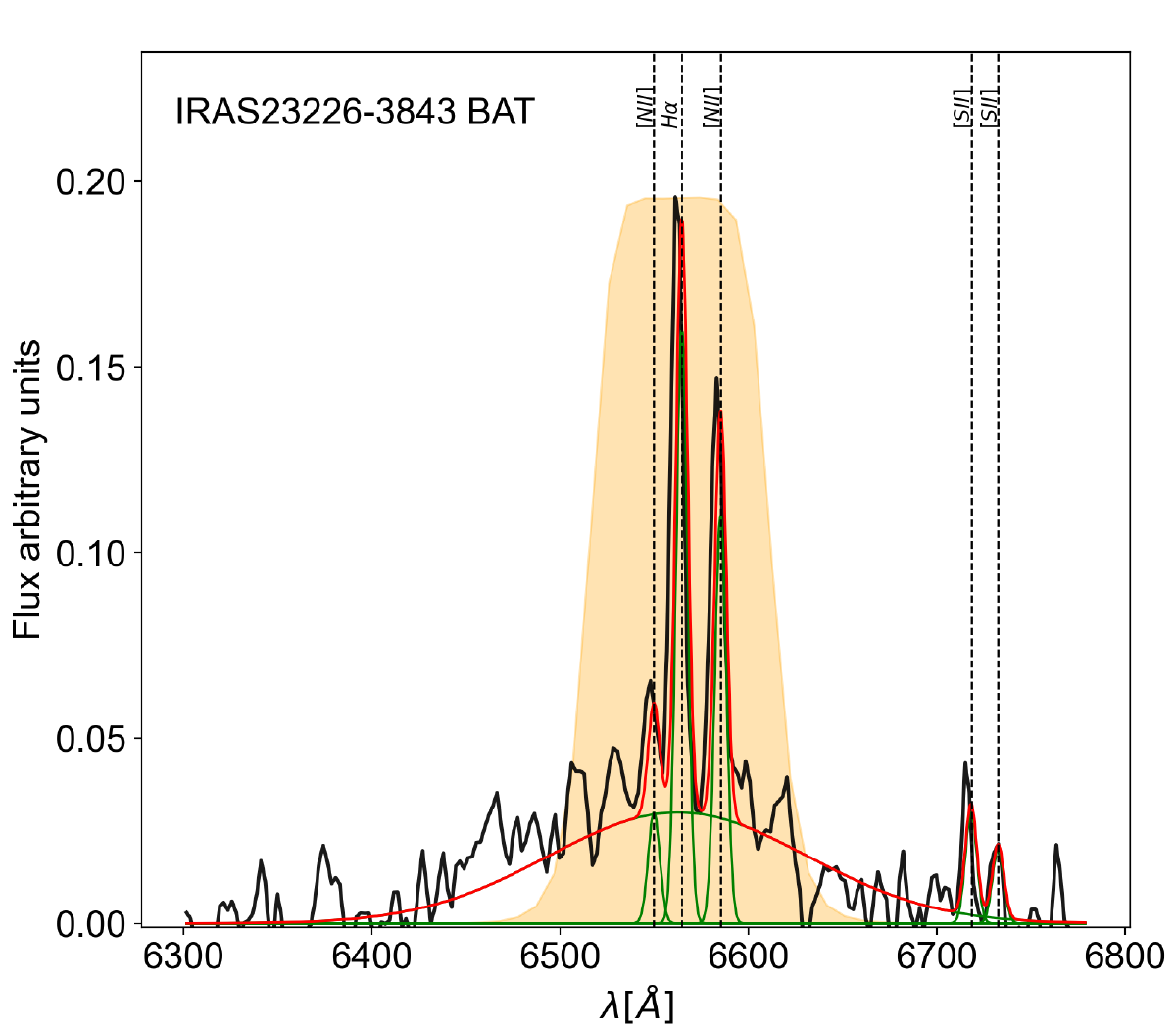}

\includegraphics[width=0.33\columnwidth]{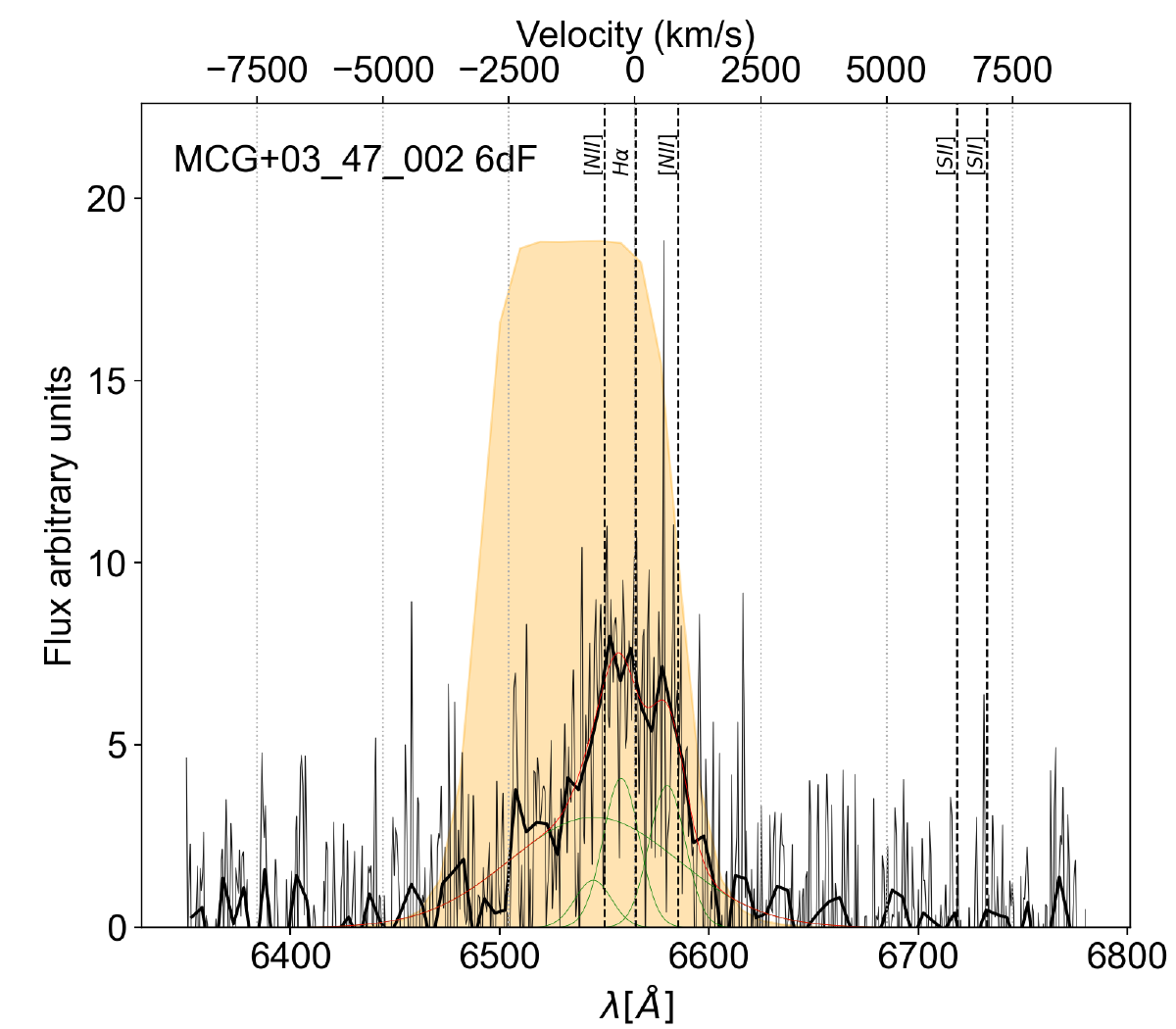}
\includegraphics[width=0.33\columnwidth]{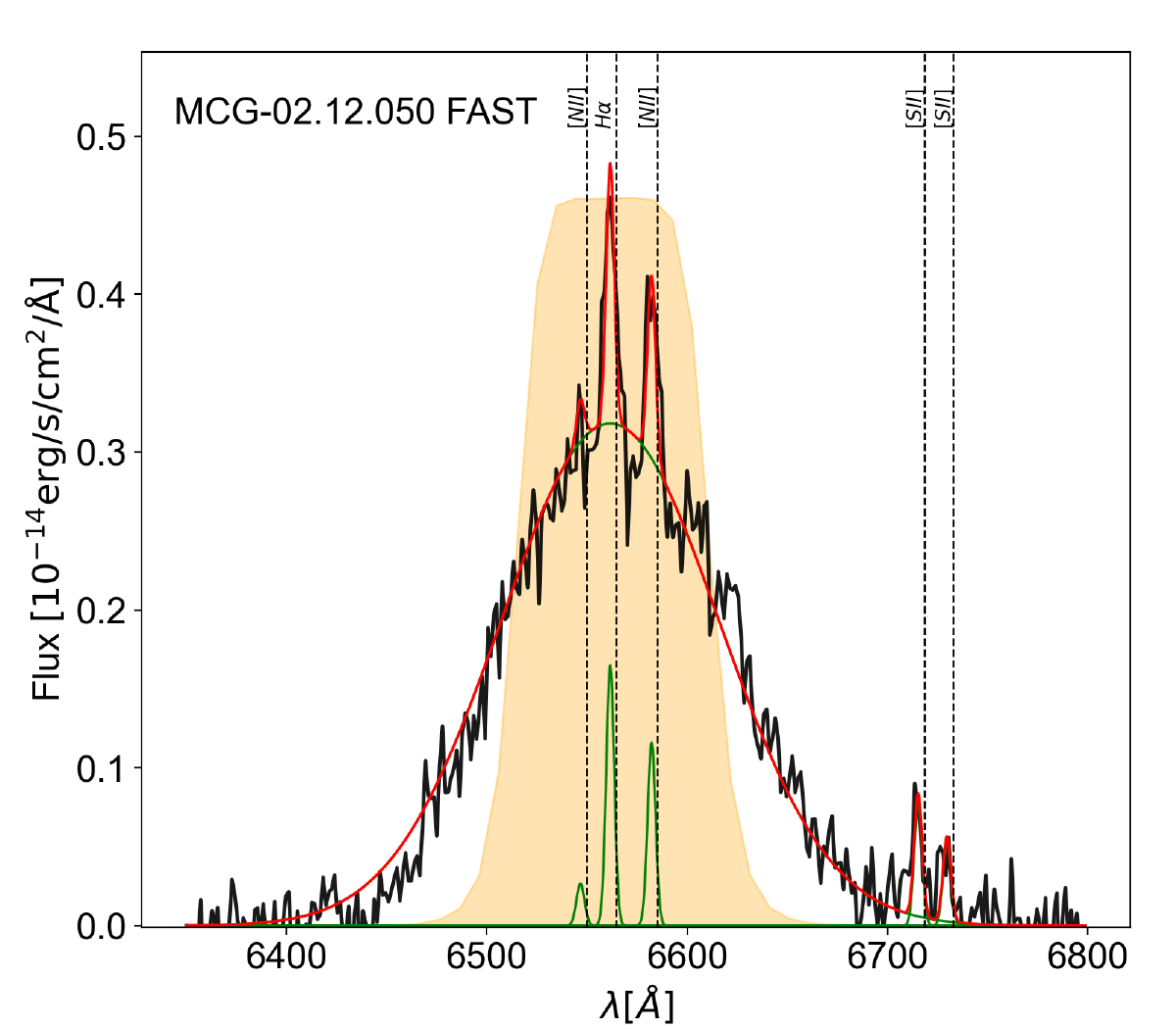}
\includegraphics[width=0.33\columnwidth]{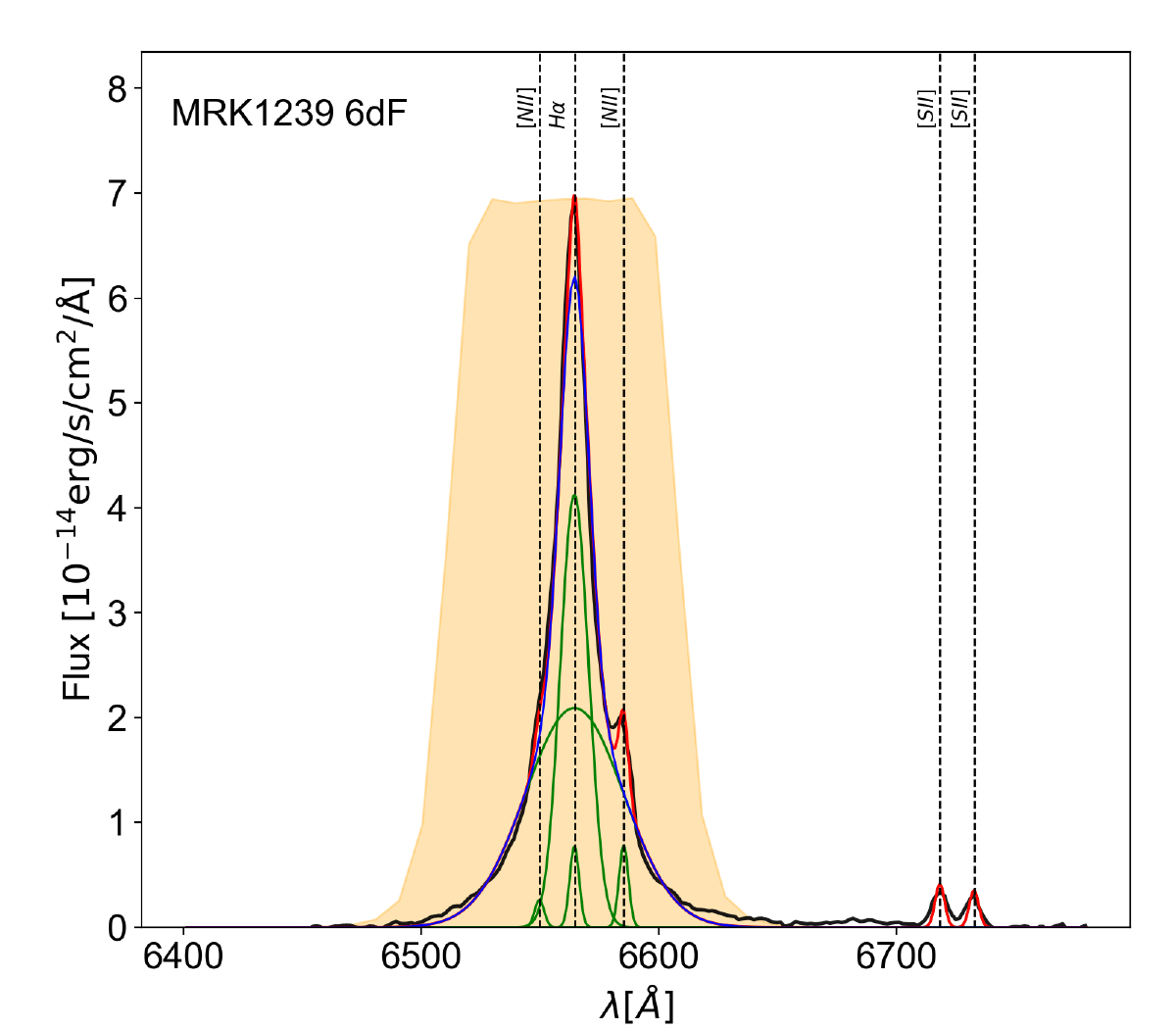}

\includegraphics[width=0.33\columnwidth]{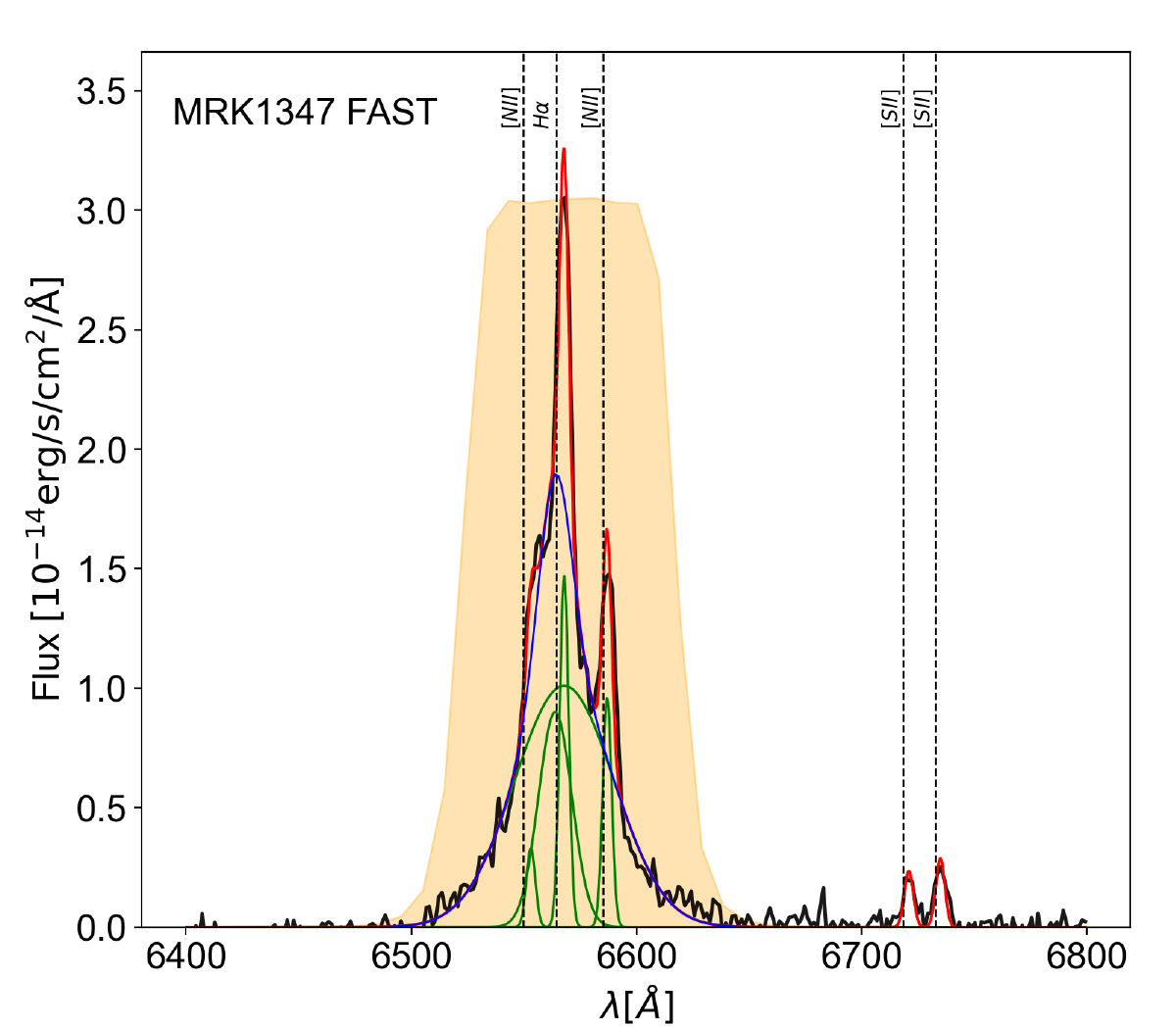}
\includegraphics[width=0.33\columnwidth]{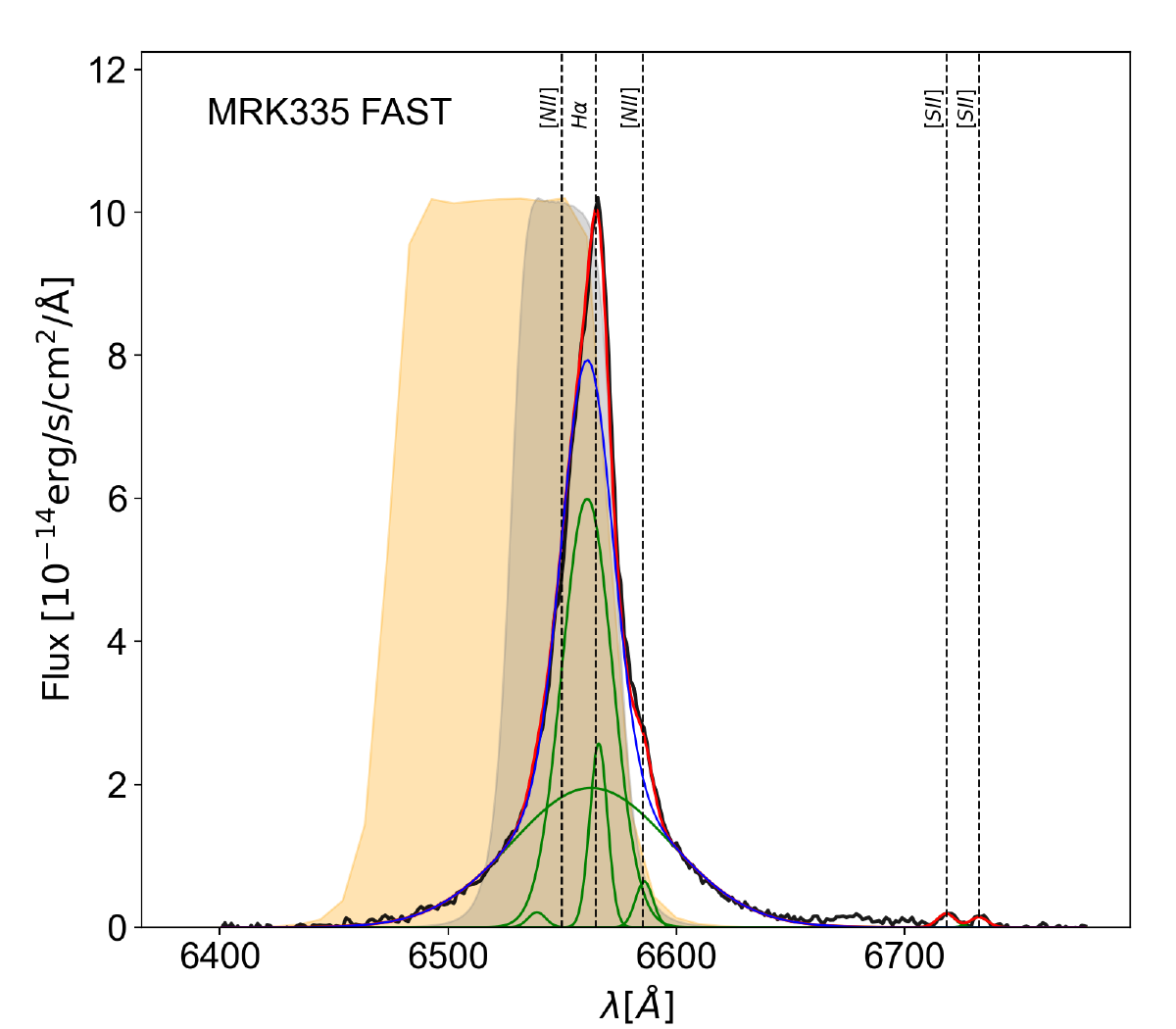}
\includegraphics[width=0.33\columnwidth]{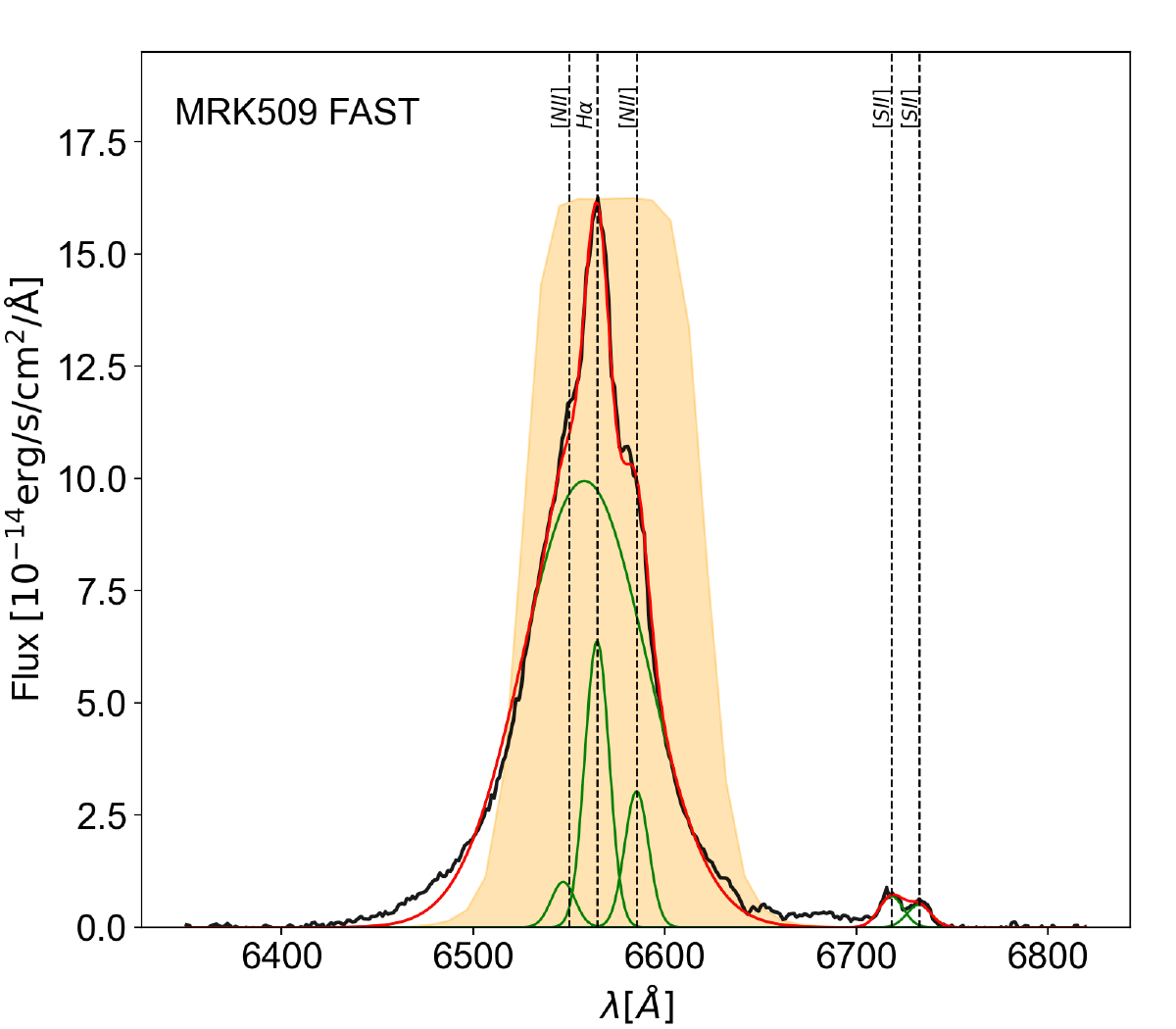}

\includegraphics[width=0.33\columnwidth]{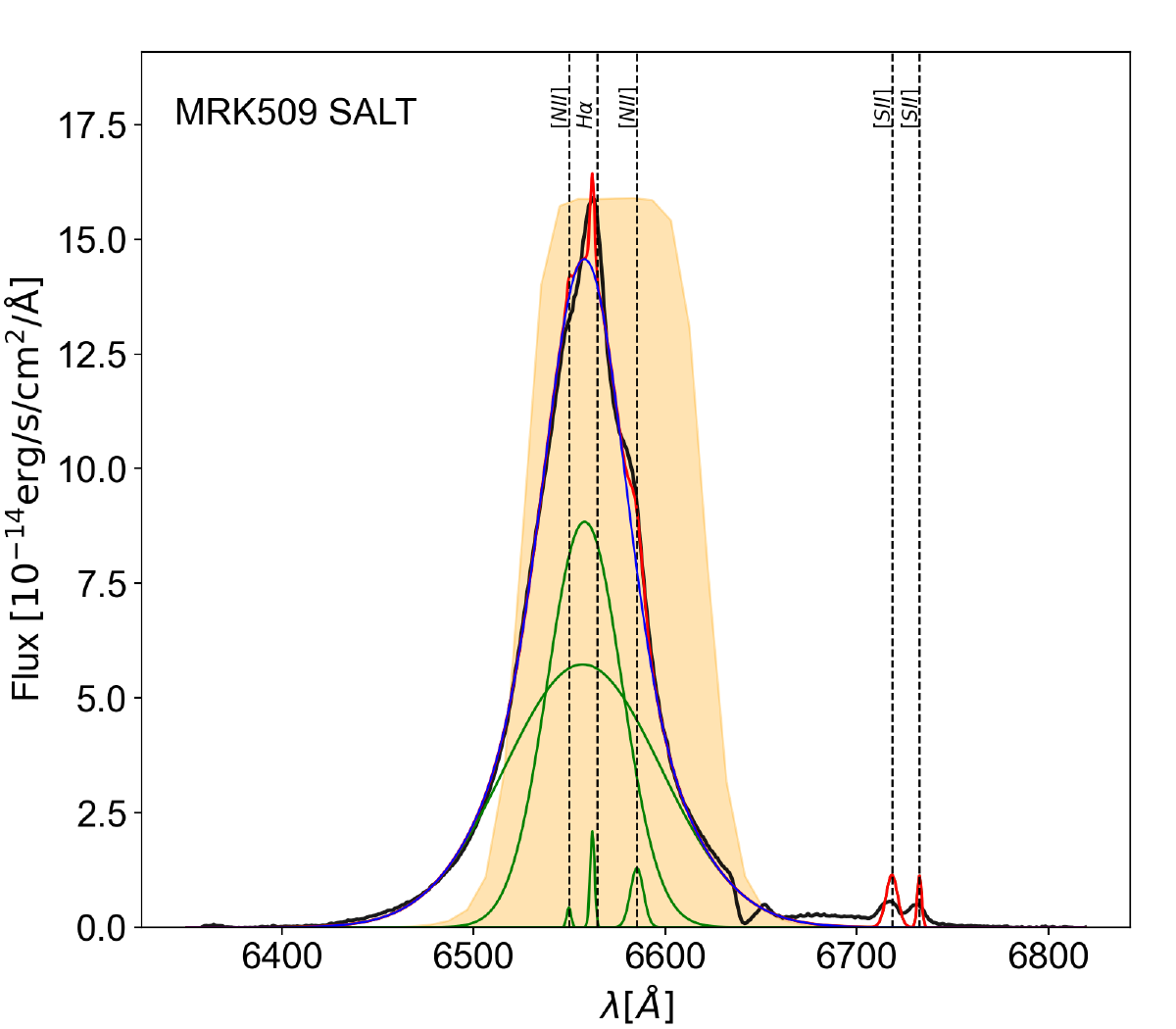}
\includegraphics[width=0.33\columnwidth]{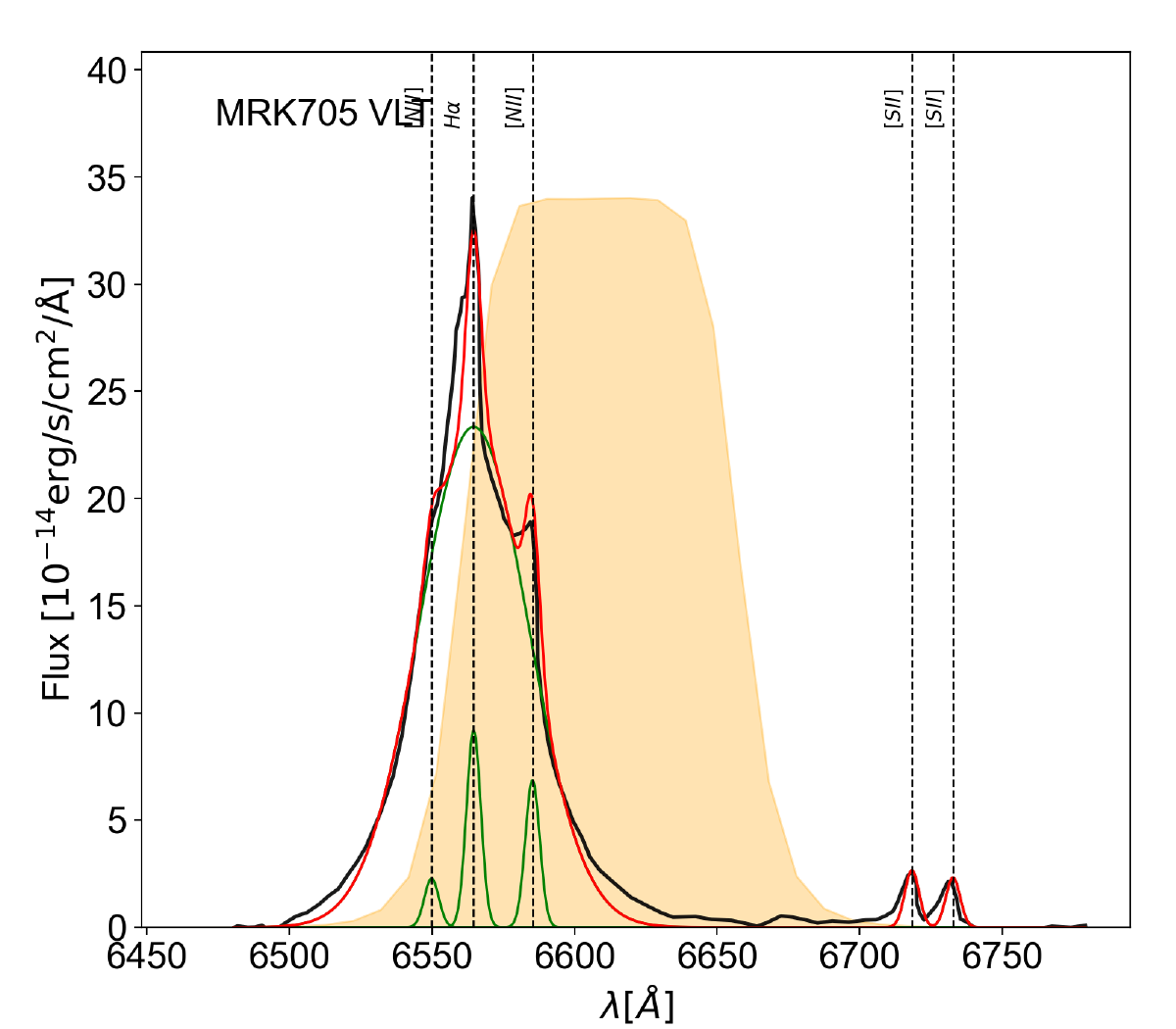}
\includegraphics[width=0.33\columnwidth]{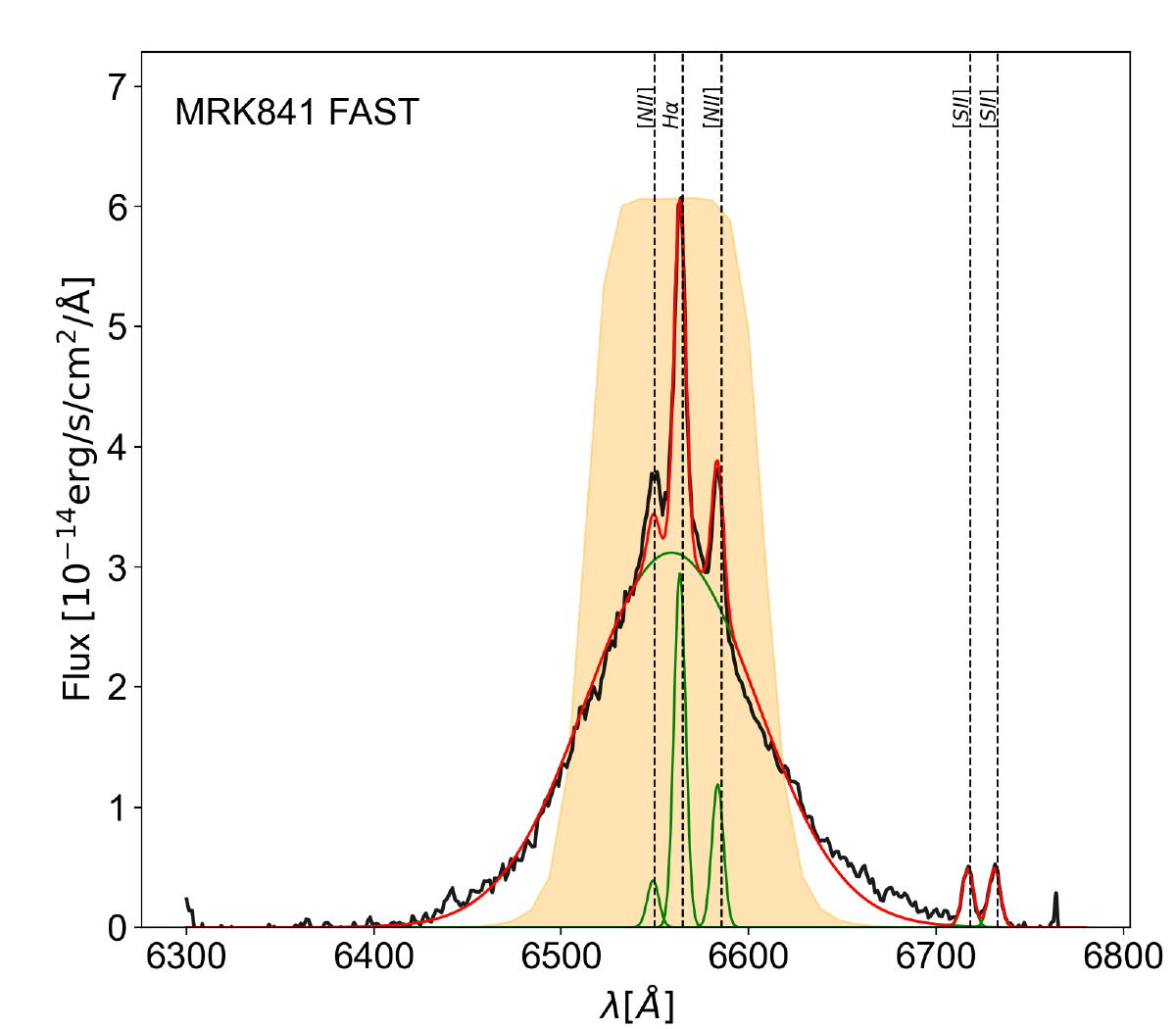}

\includegraphics[width=0.33\columnwidth]{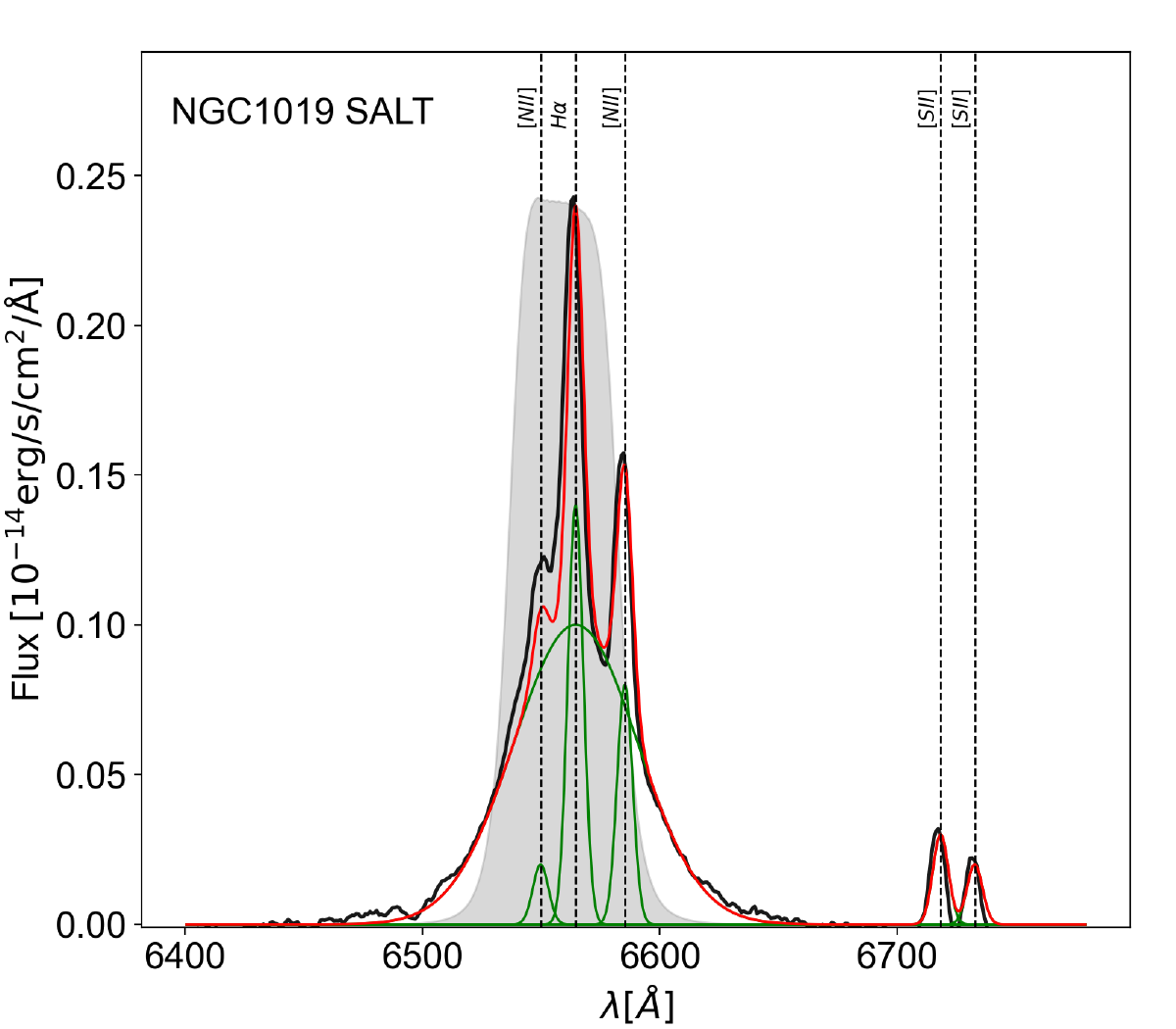}
\includegraphics[width=0.33\columnwidth]{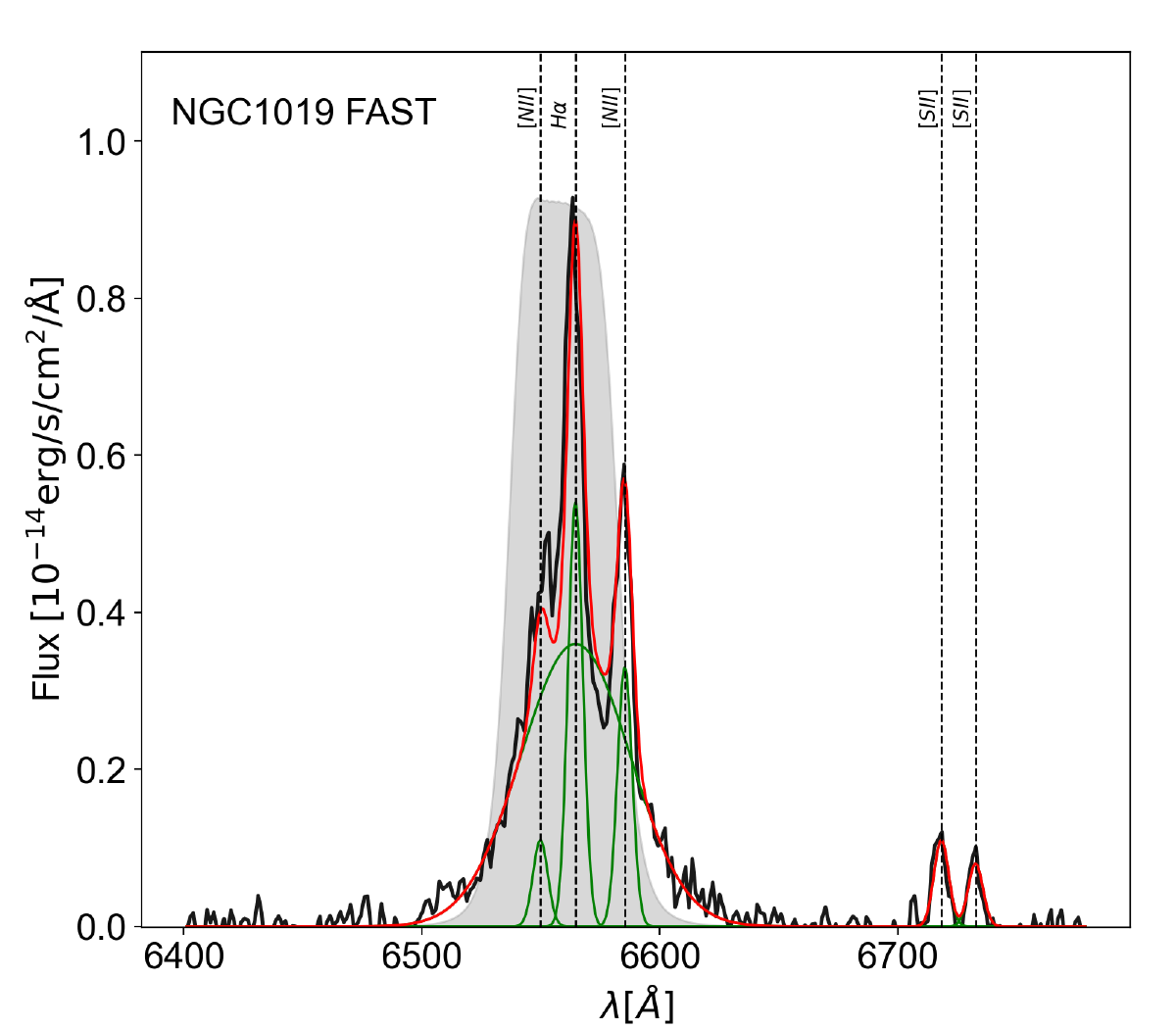}
\includegraphics[width=0.33\columnwidth]{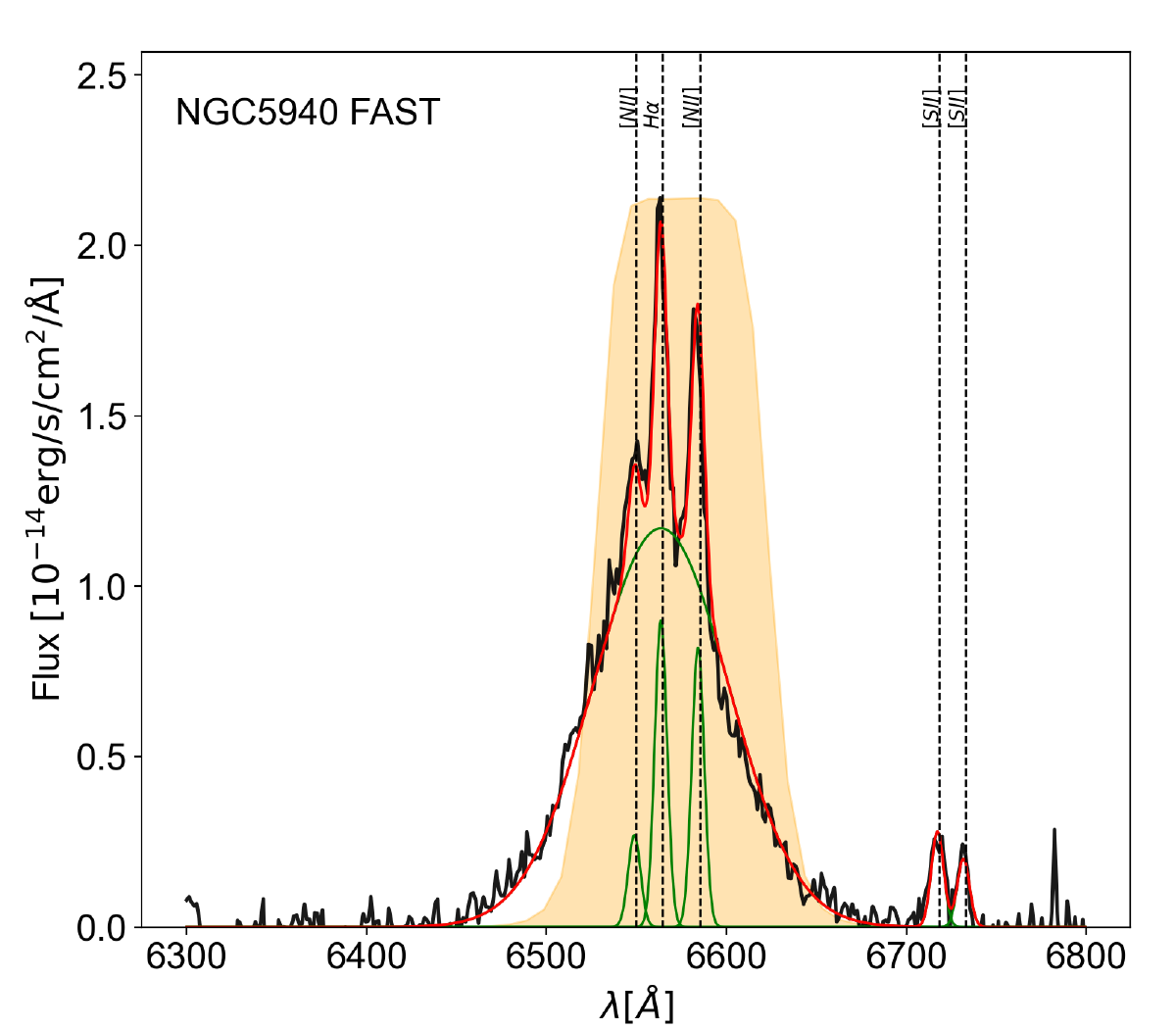}

\includegraphics[width=0.33\columnwidth]{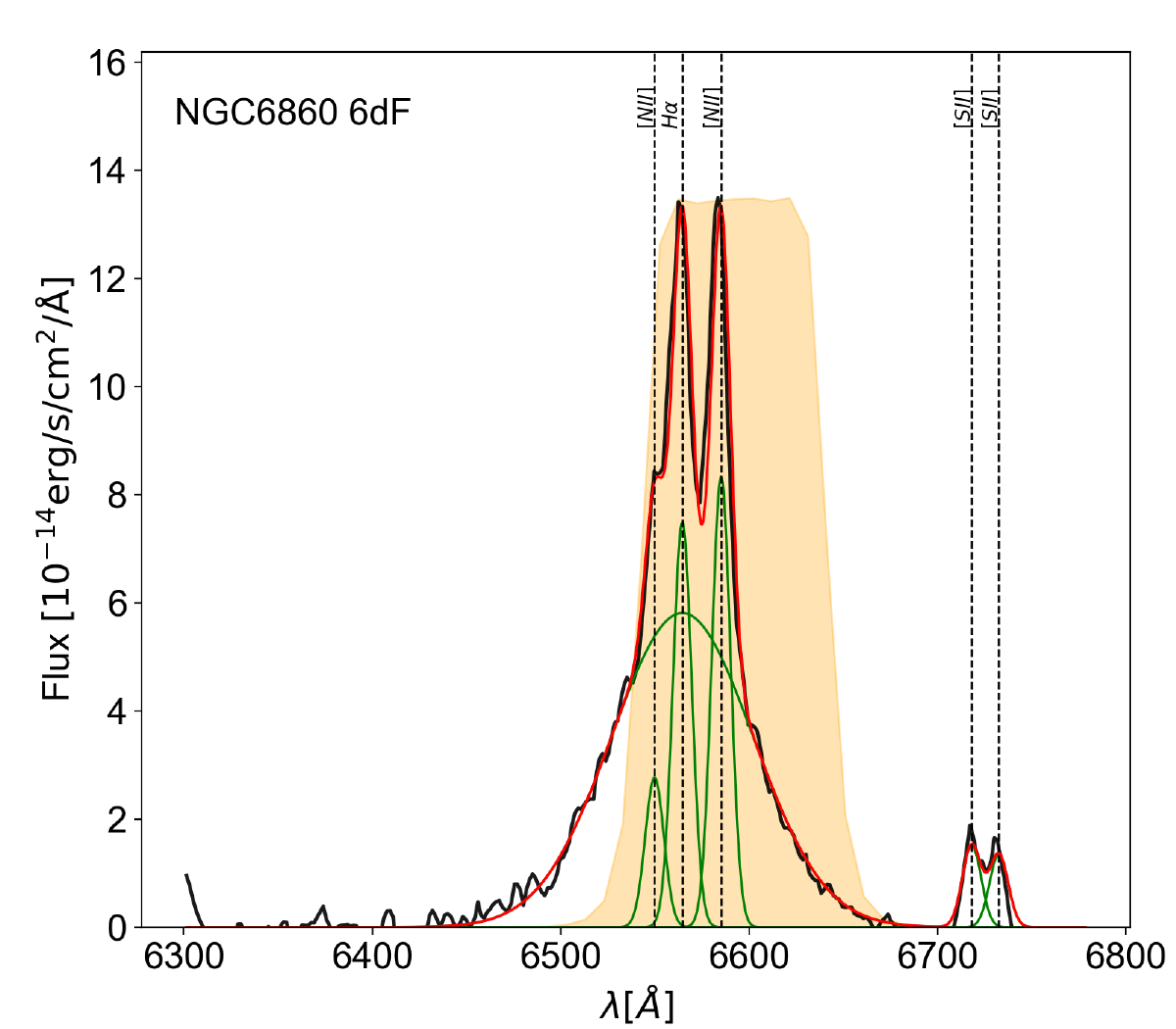}
\includegraphics[width=0.33\columnwidth]{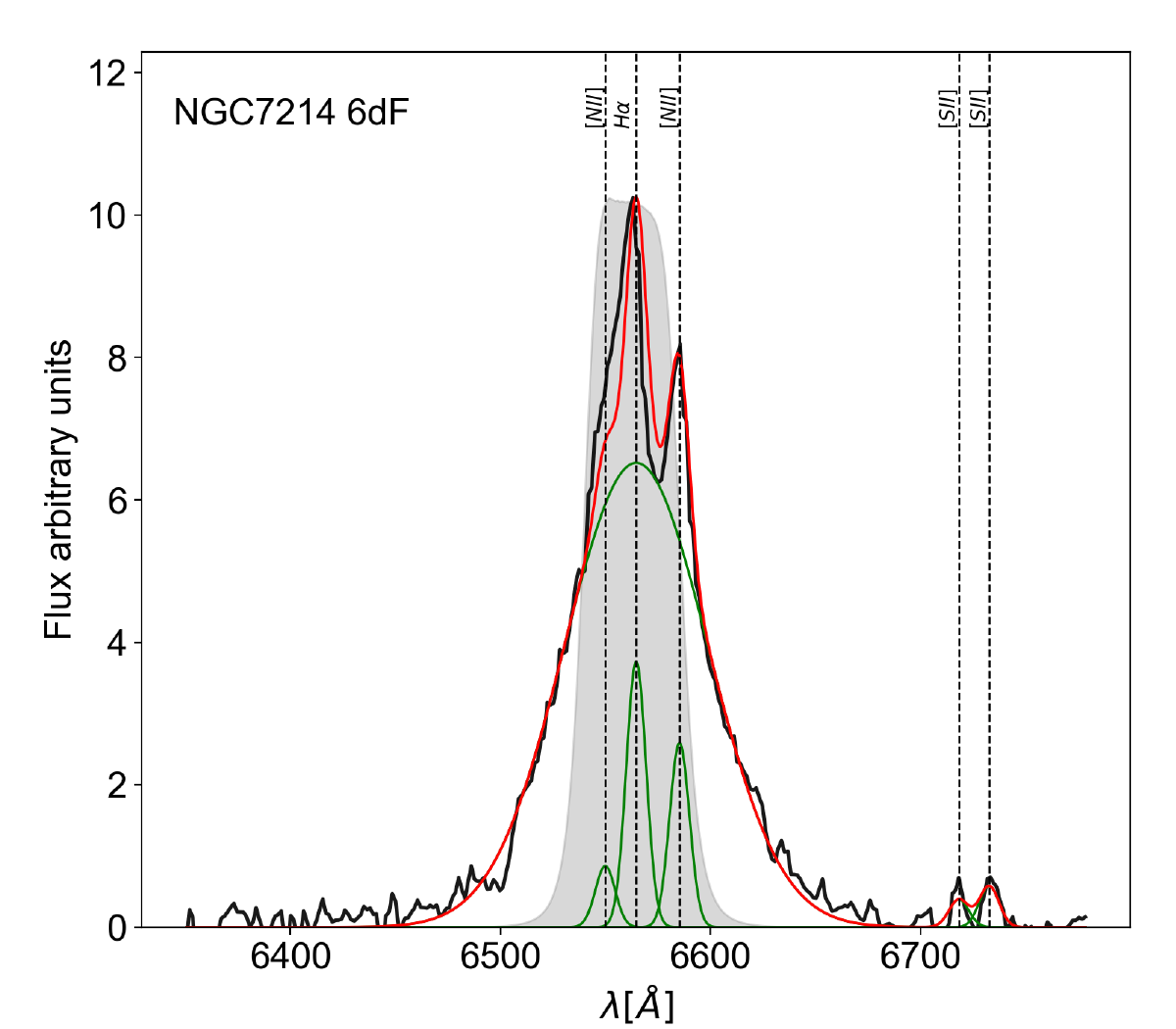}
\includegraphics[width=0.33\columnwidth]{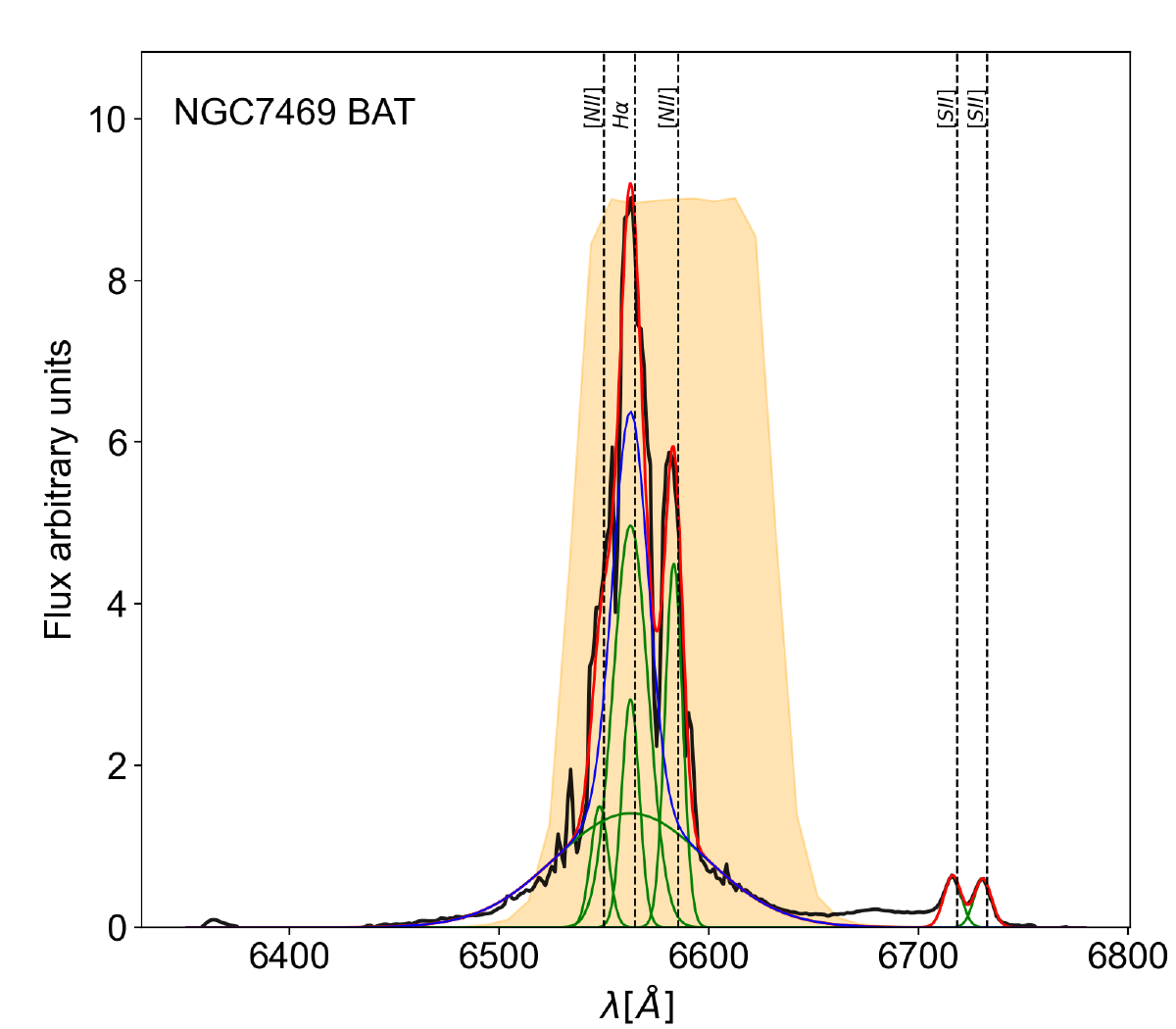} 

\includegraphics[width=0.33\columnwidth]{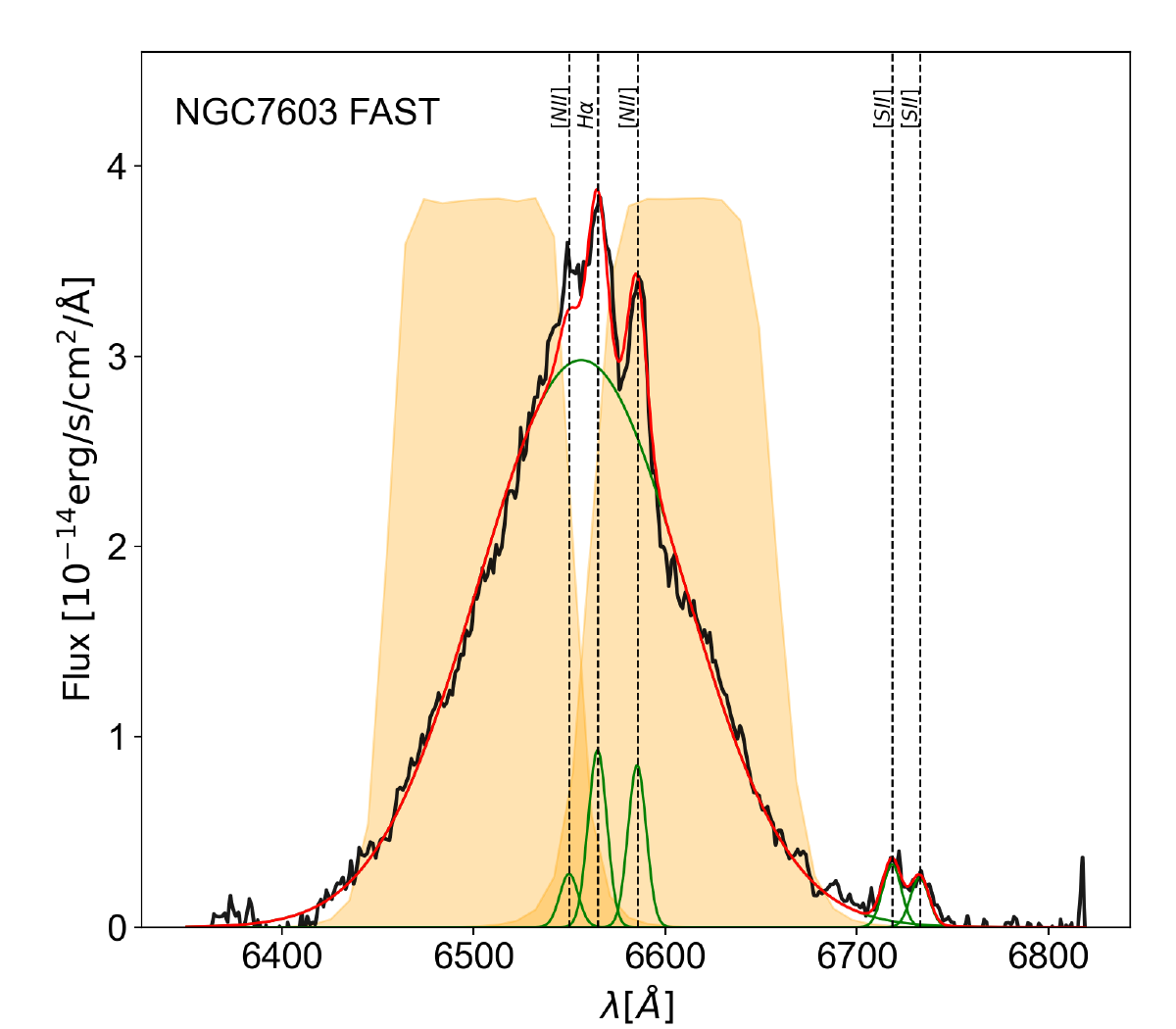}
\includegraphics[width=0.33\columnwidth]{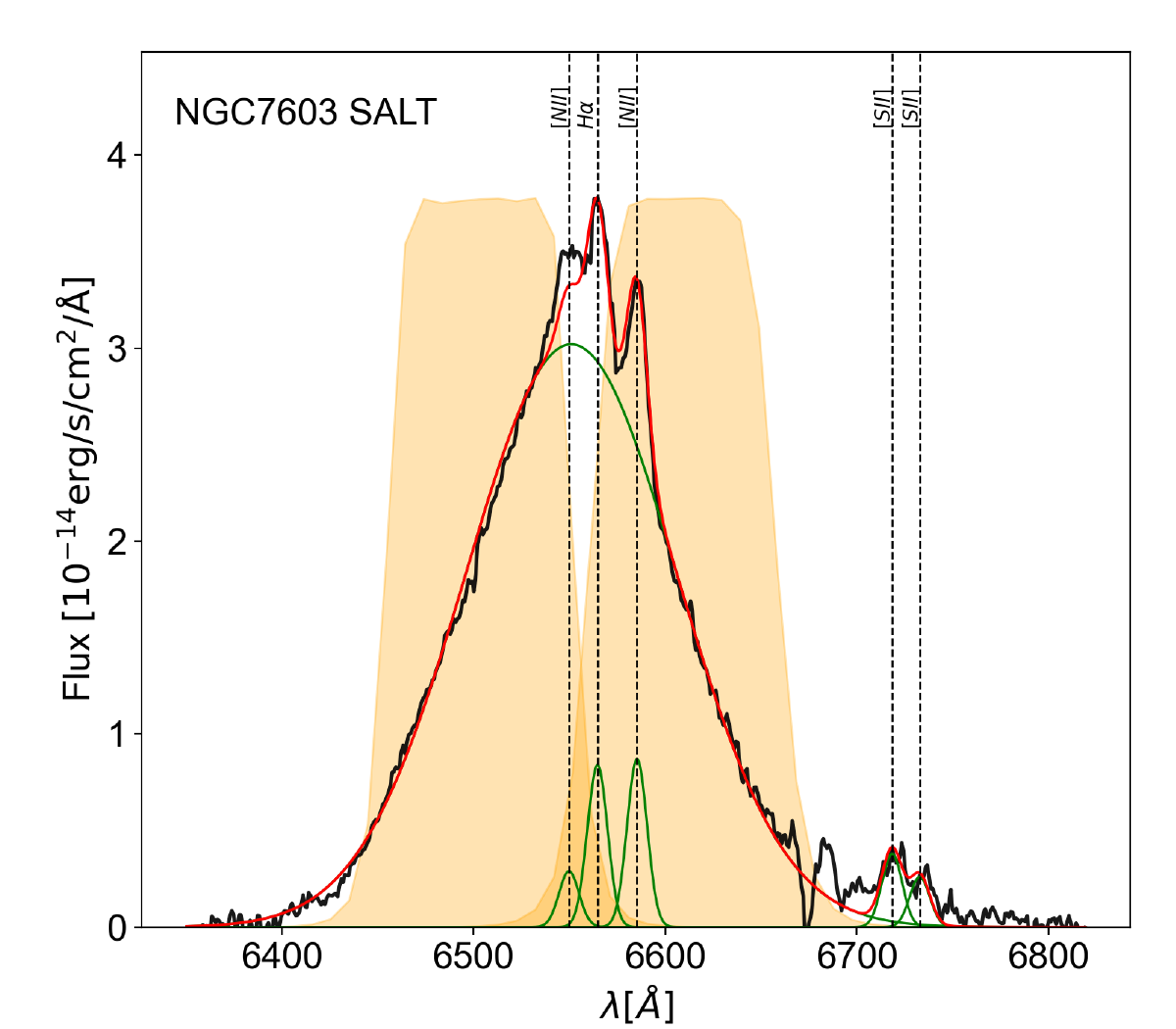}
\includegraphics[width=0.33\columnwidth]{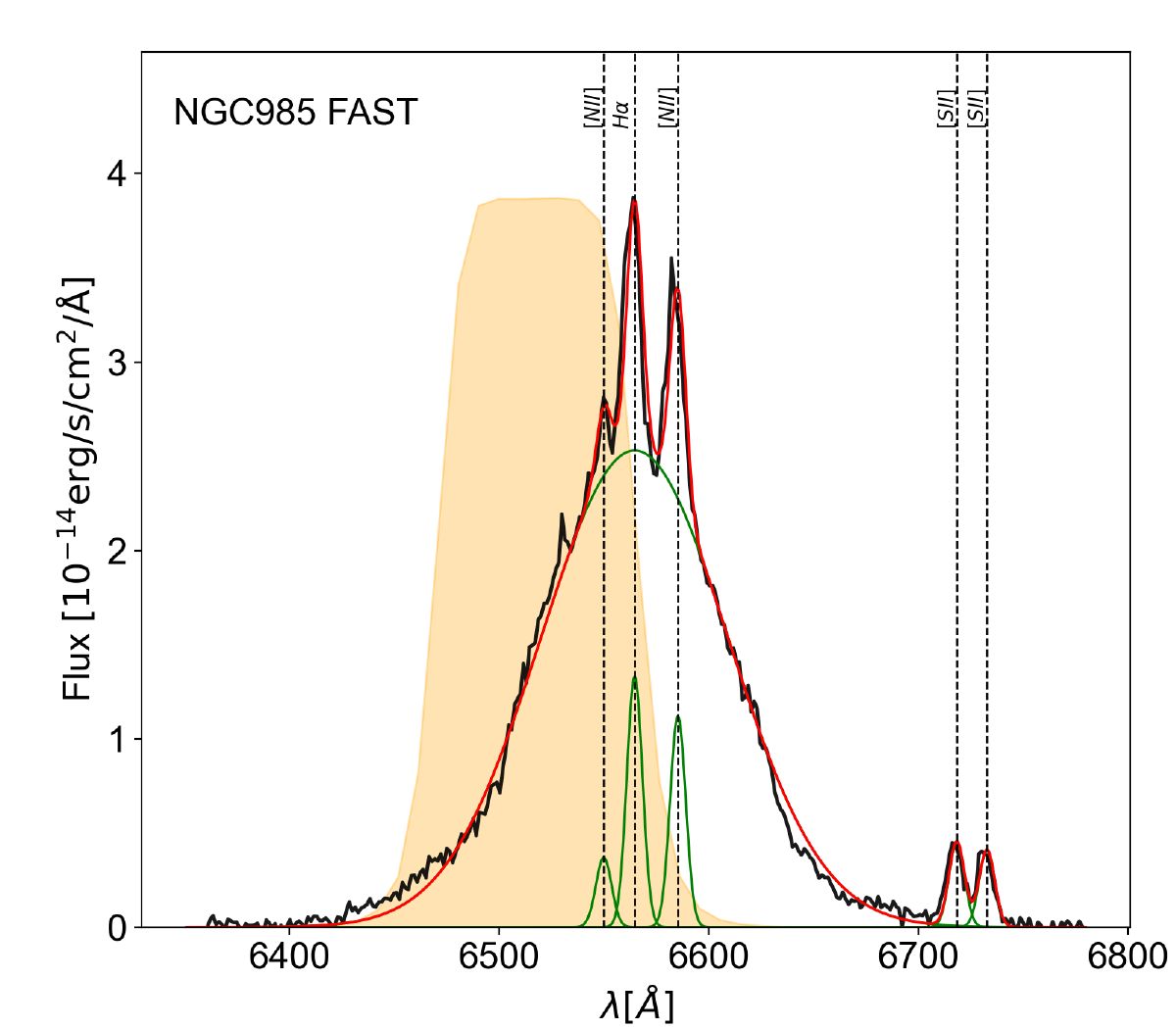}

\includegraphics[width=0.33\columnwidth]{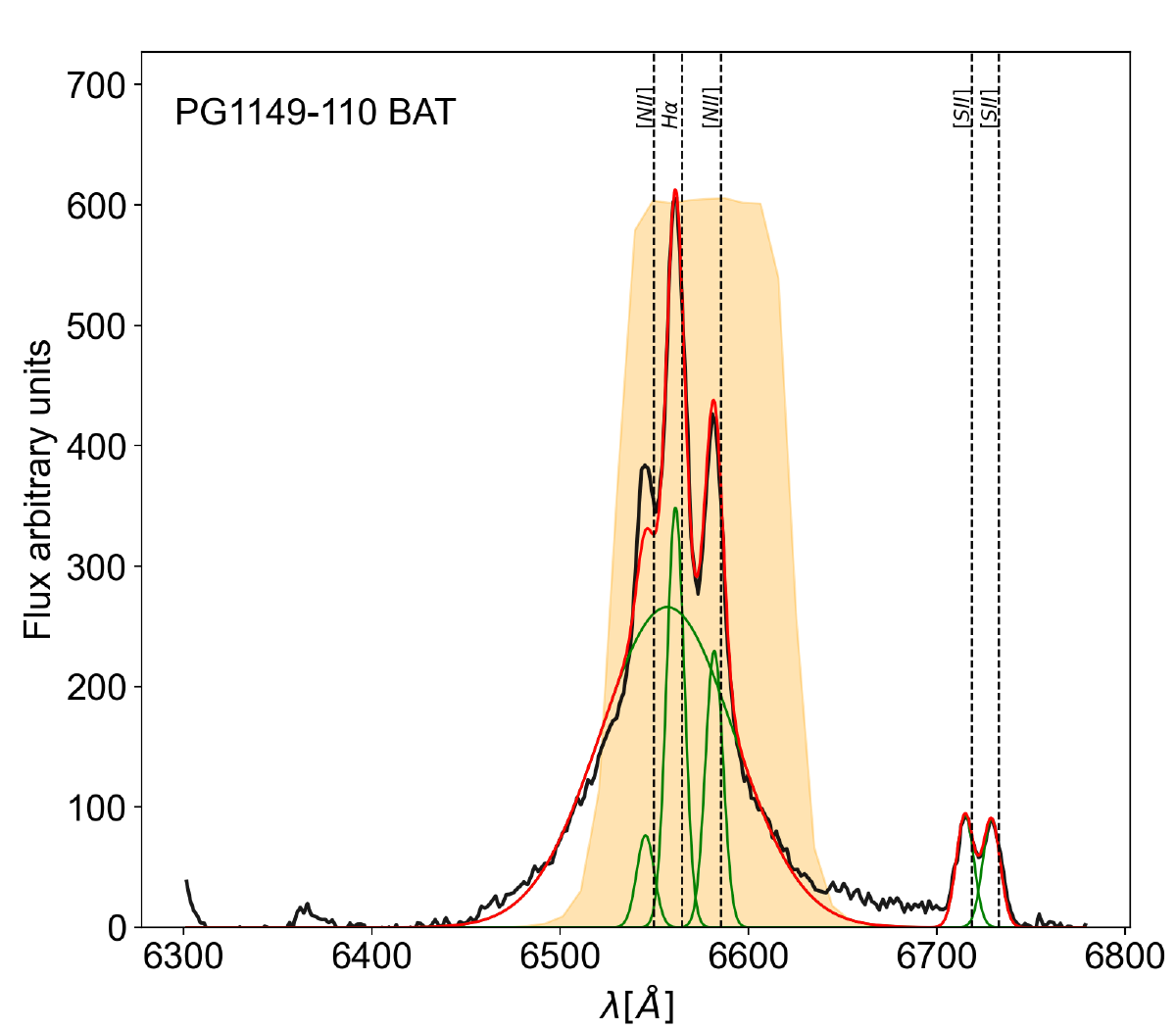}
\includegraphics[width=0.33\columnwidth]{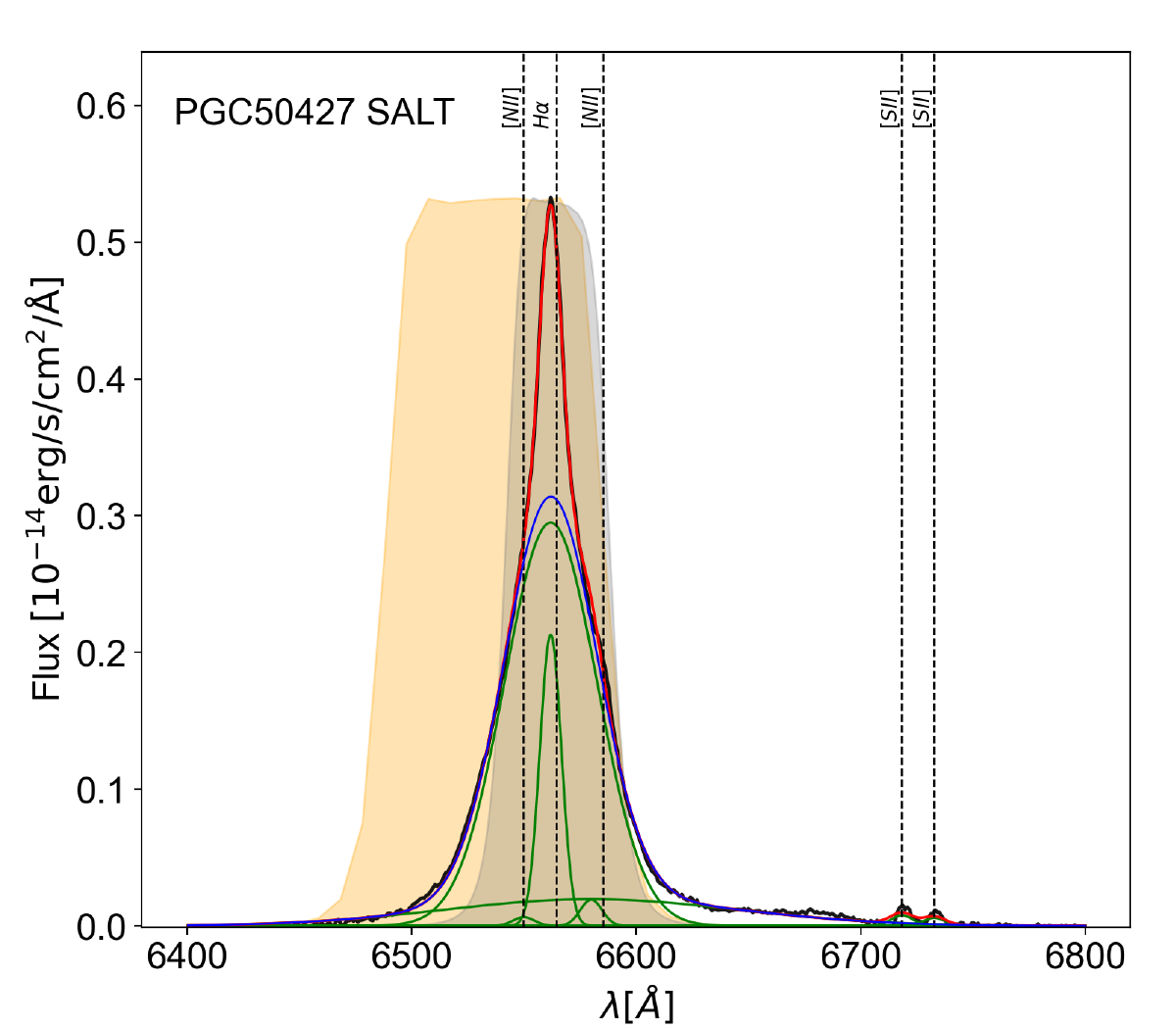}
\includegraphics[width=0.33\columnwidth]{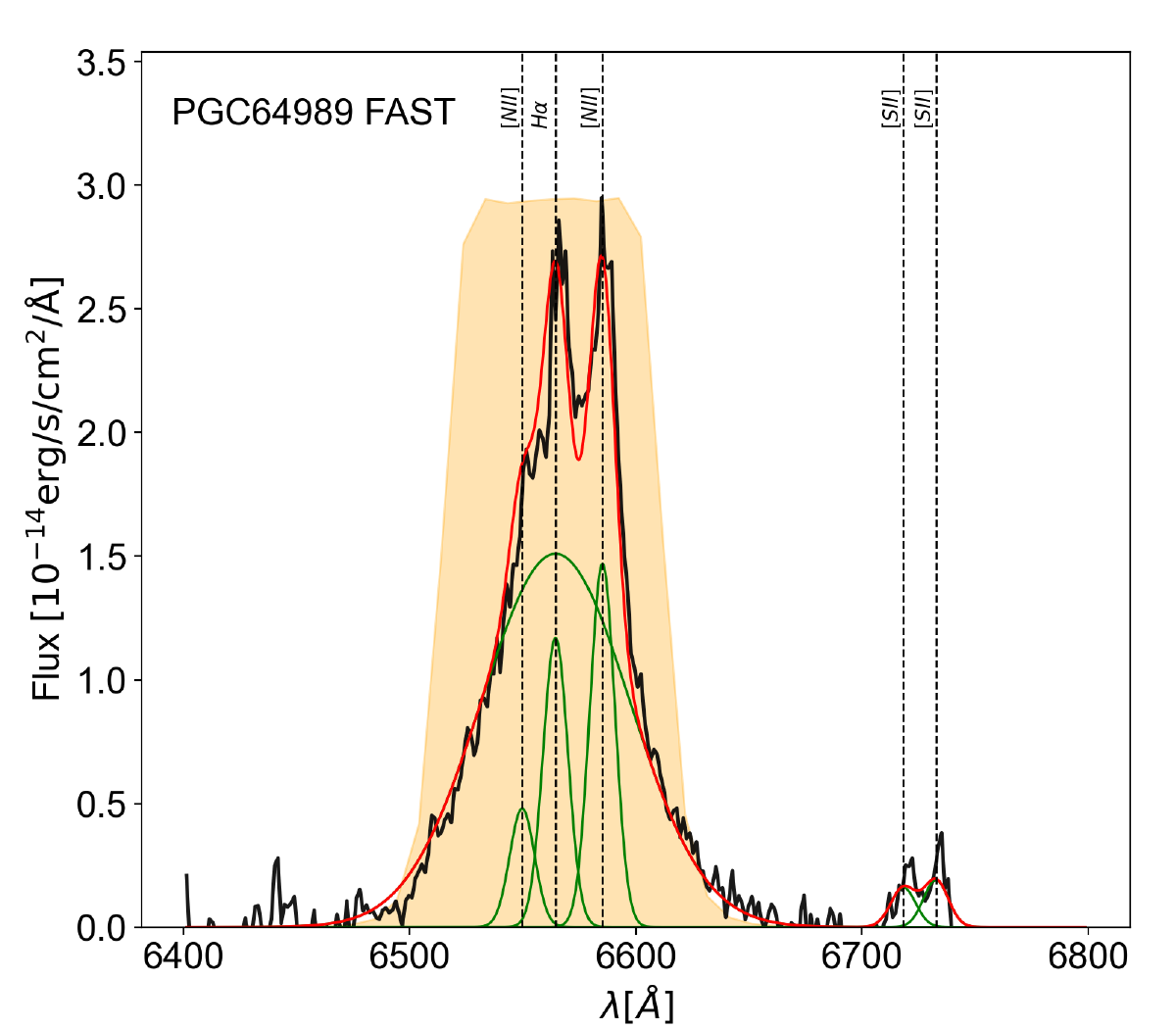}

\includegraphics[width=0.33\columnwidth]{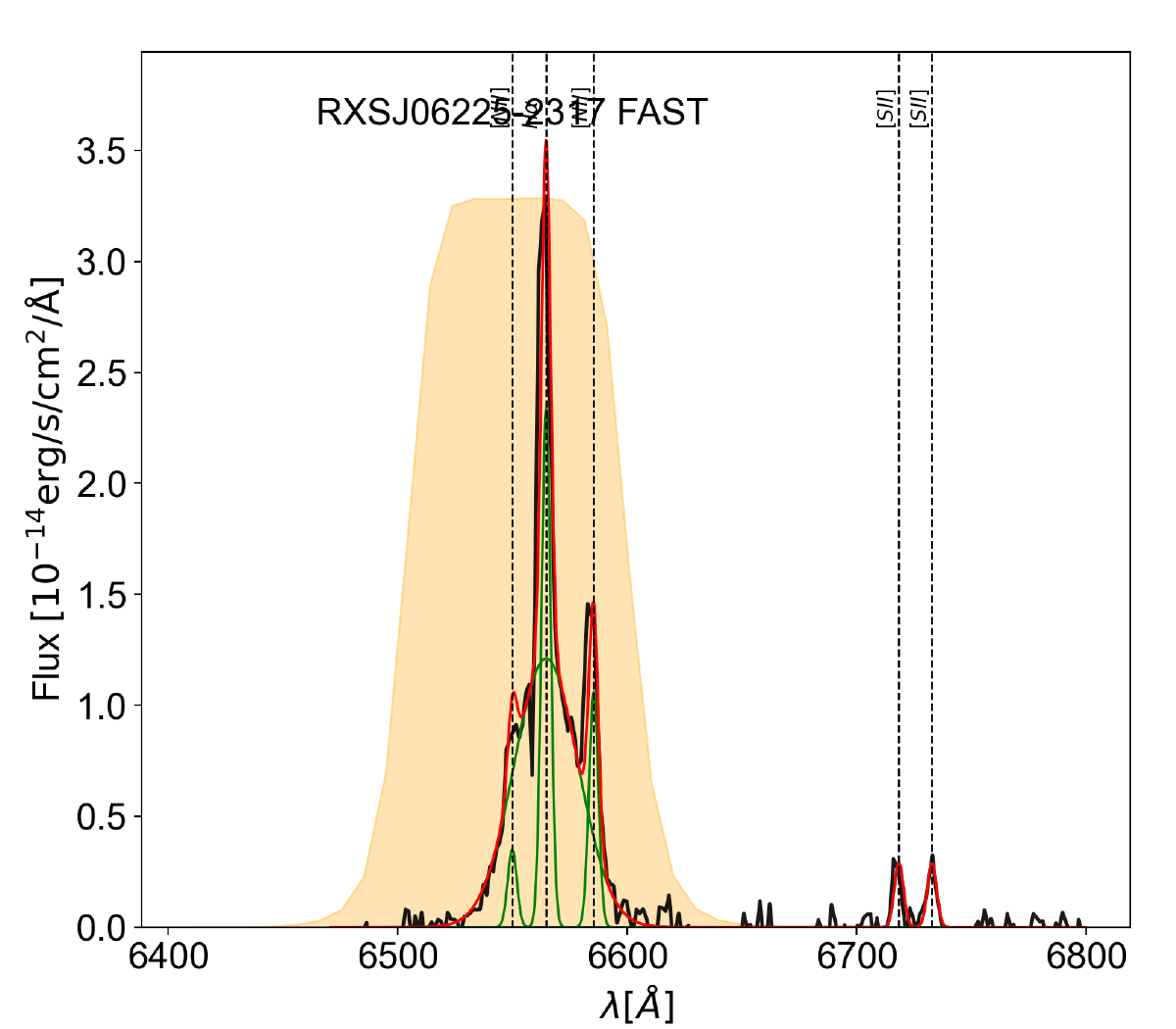}
\includegraphics[width=0.33\columnwidth]{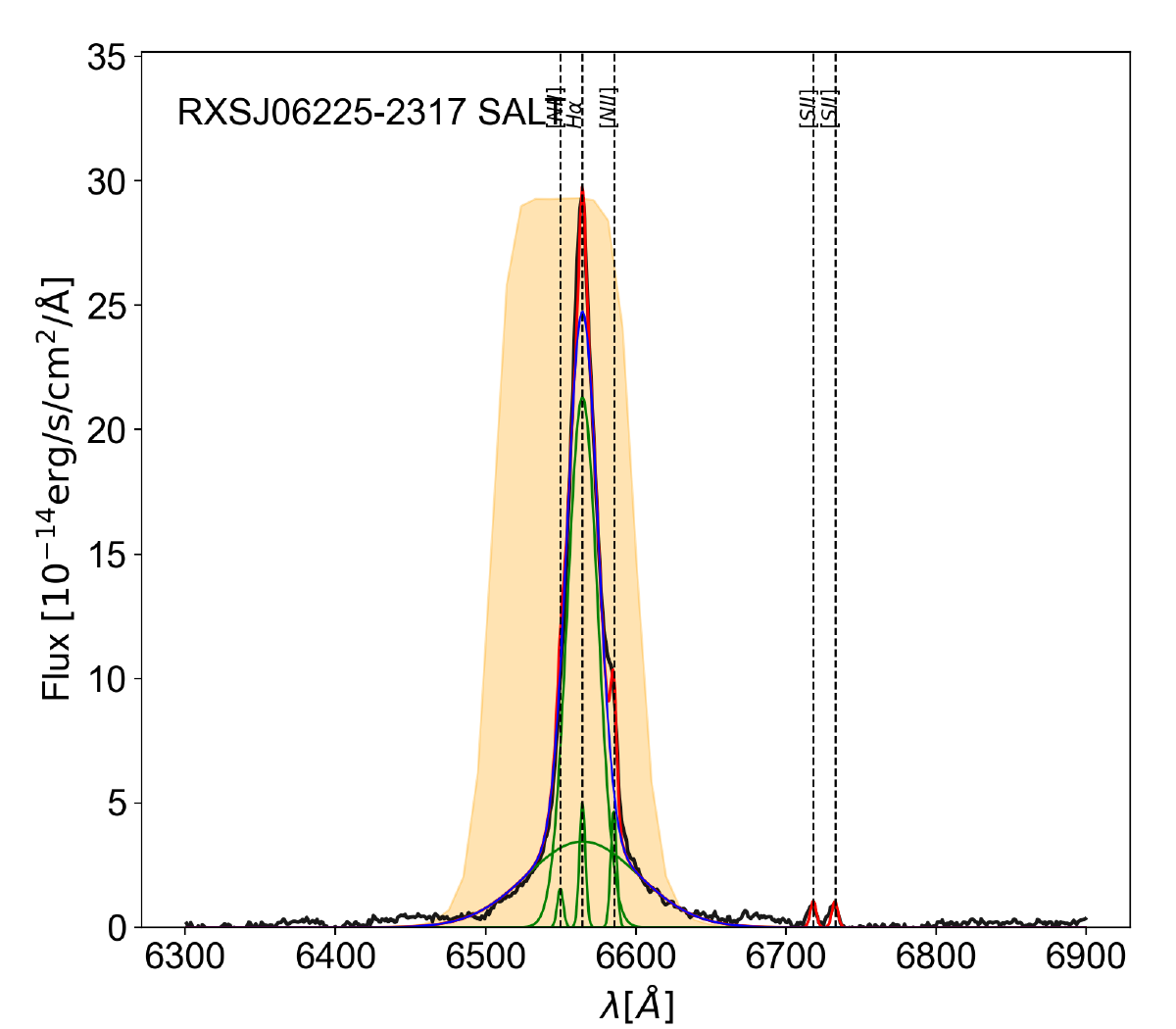}
\includegraphics[width=0.33\columnwidth]{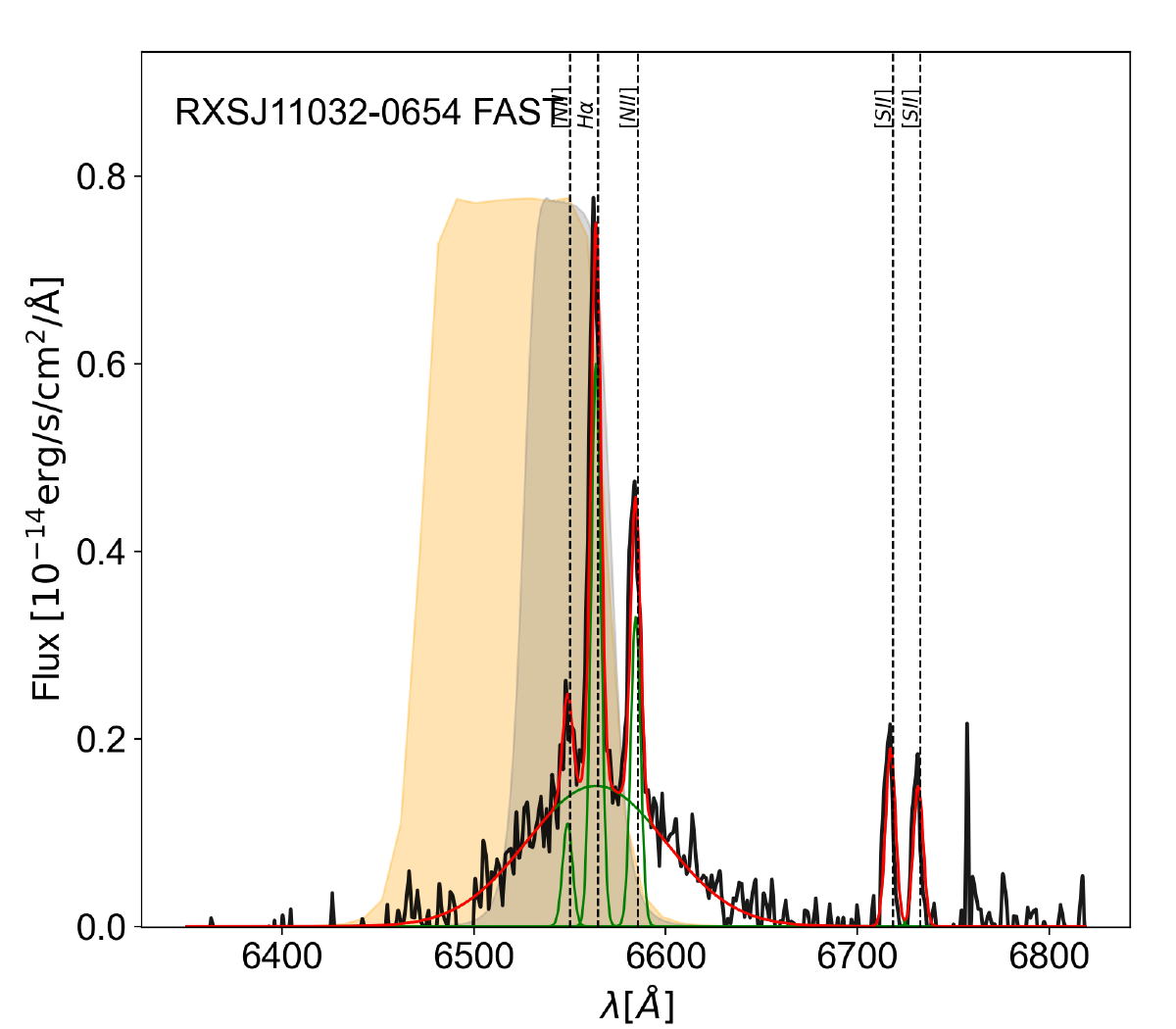}

\includegraphics[width=0.33\columnwidth]{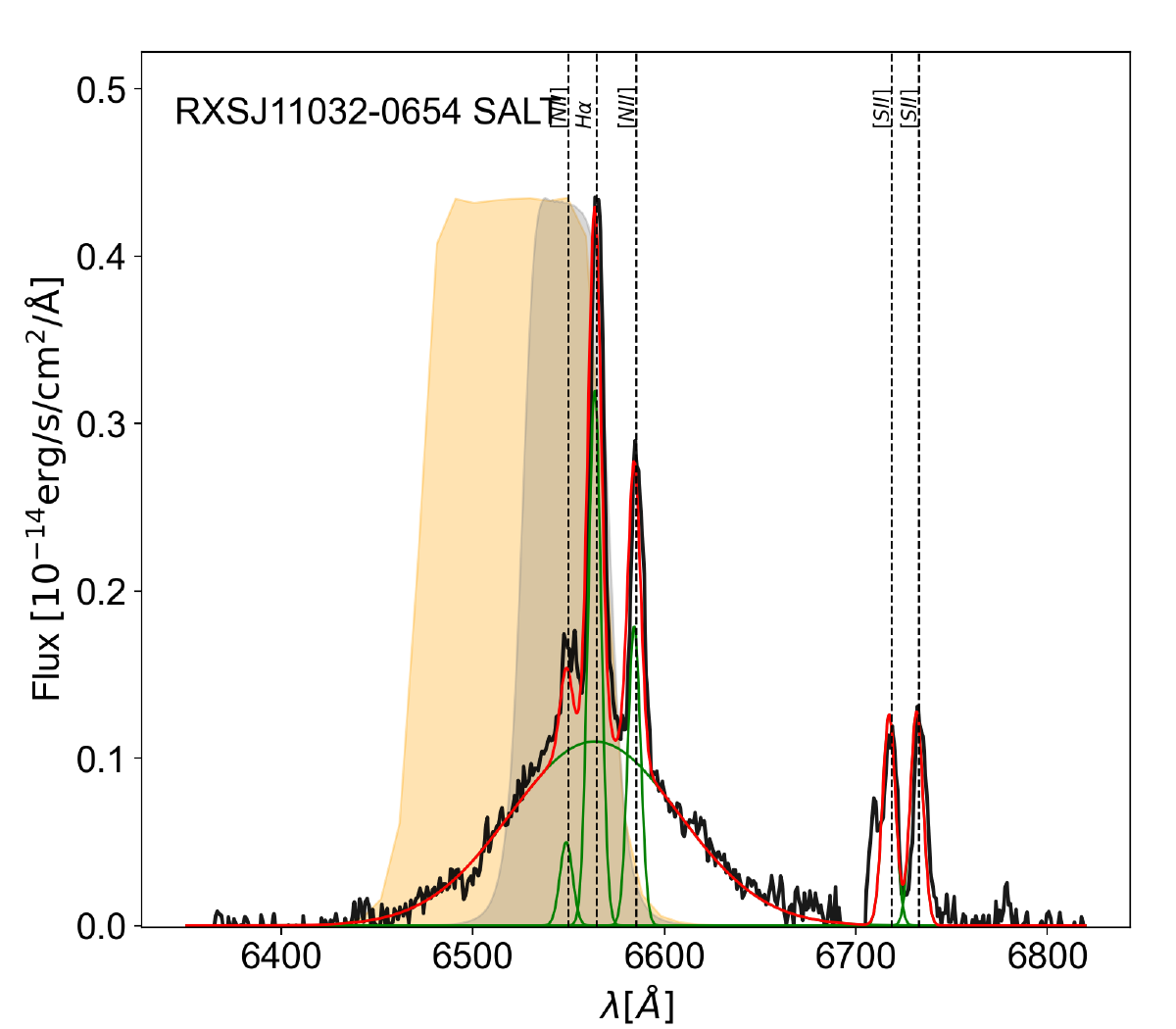}
\includegraphics[width=0.33\columnwidth]{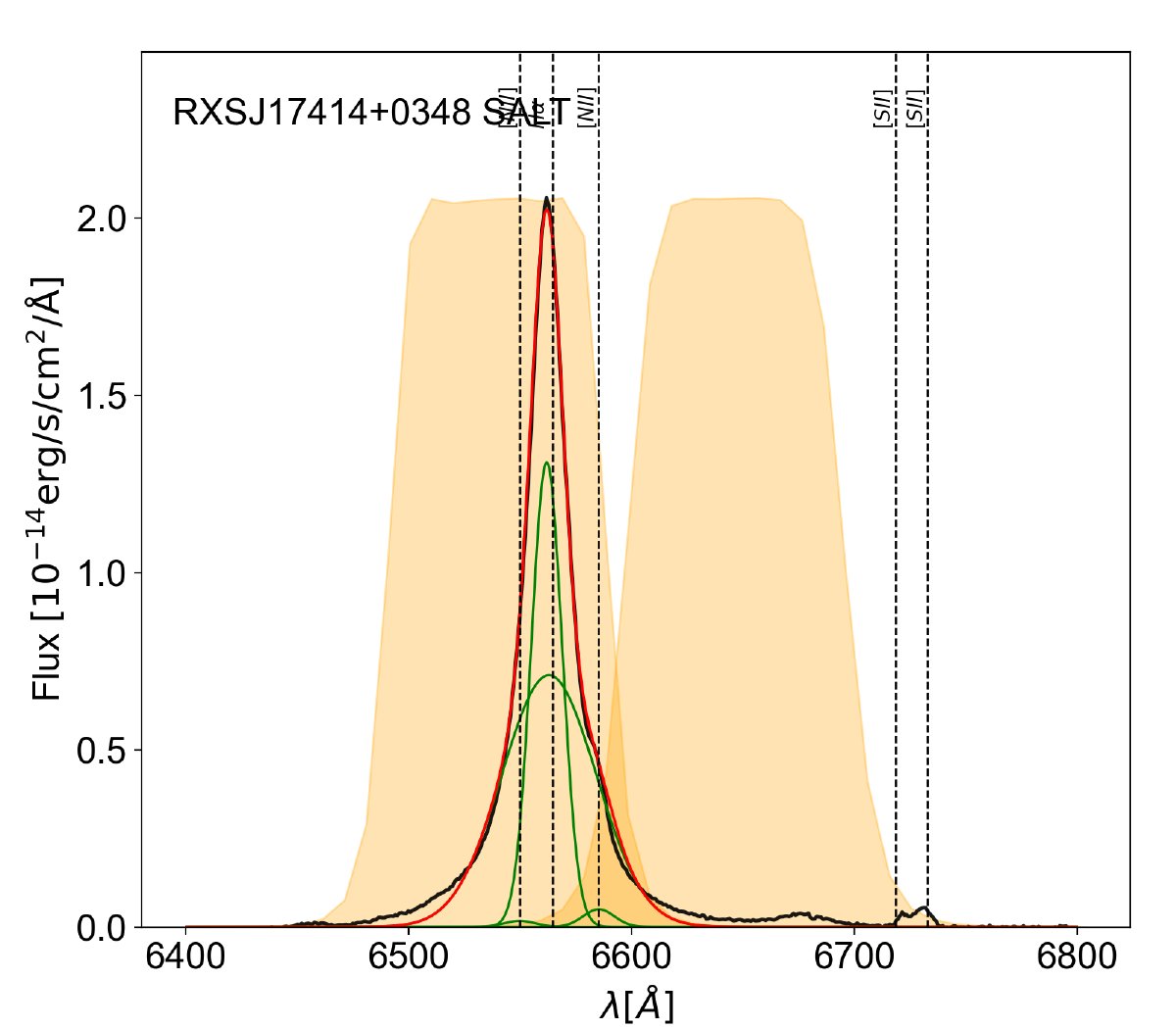}
\includegraphics[width=0.33\columnwidth]{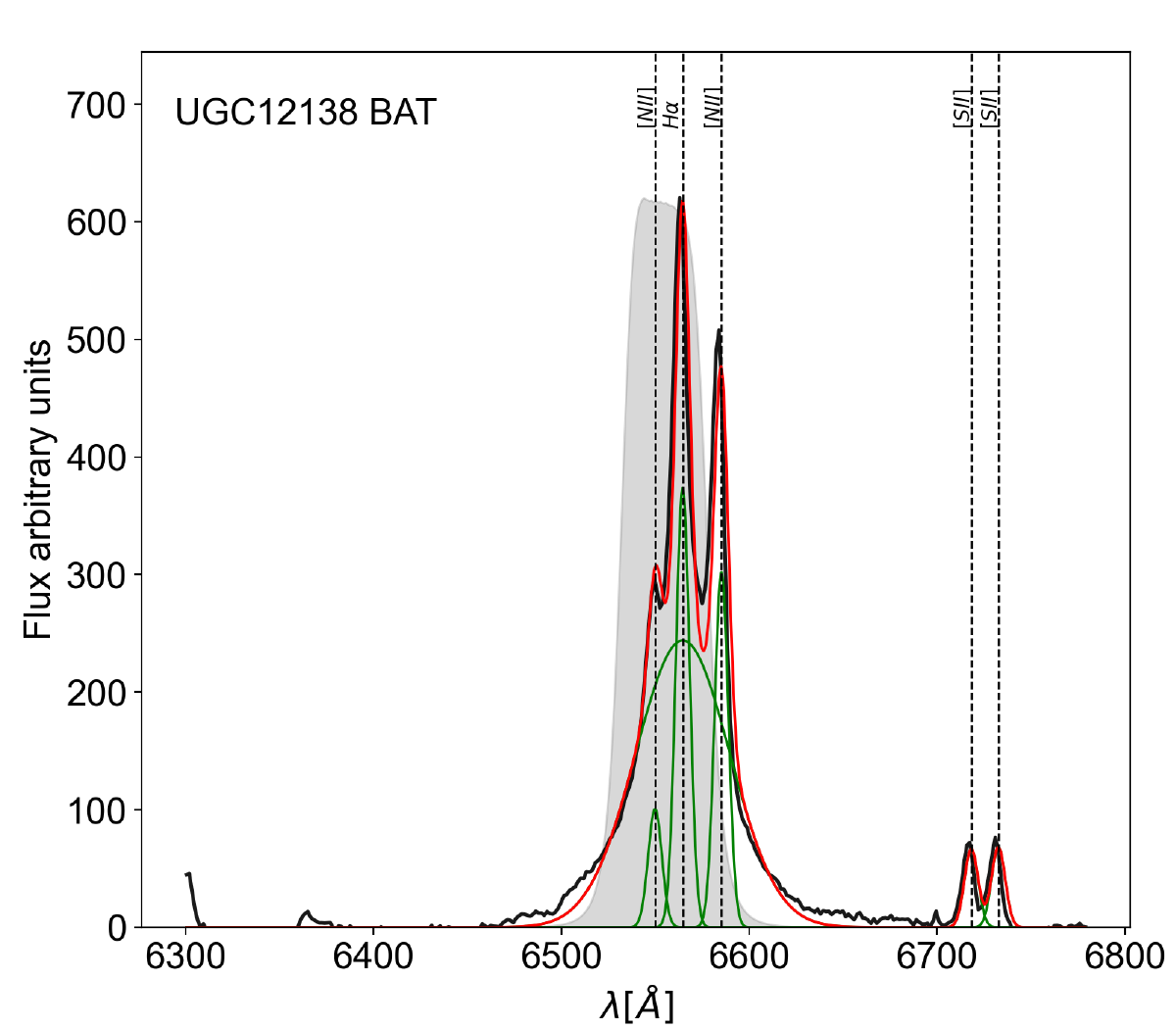}

\includegraphics[width=0.33\columnwidth]{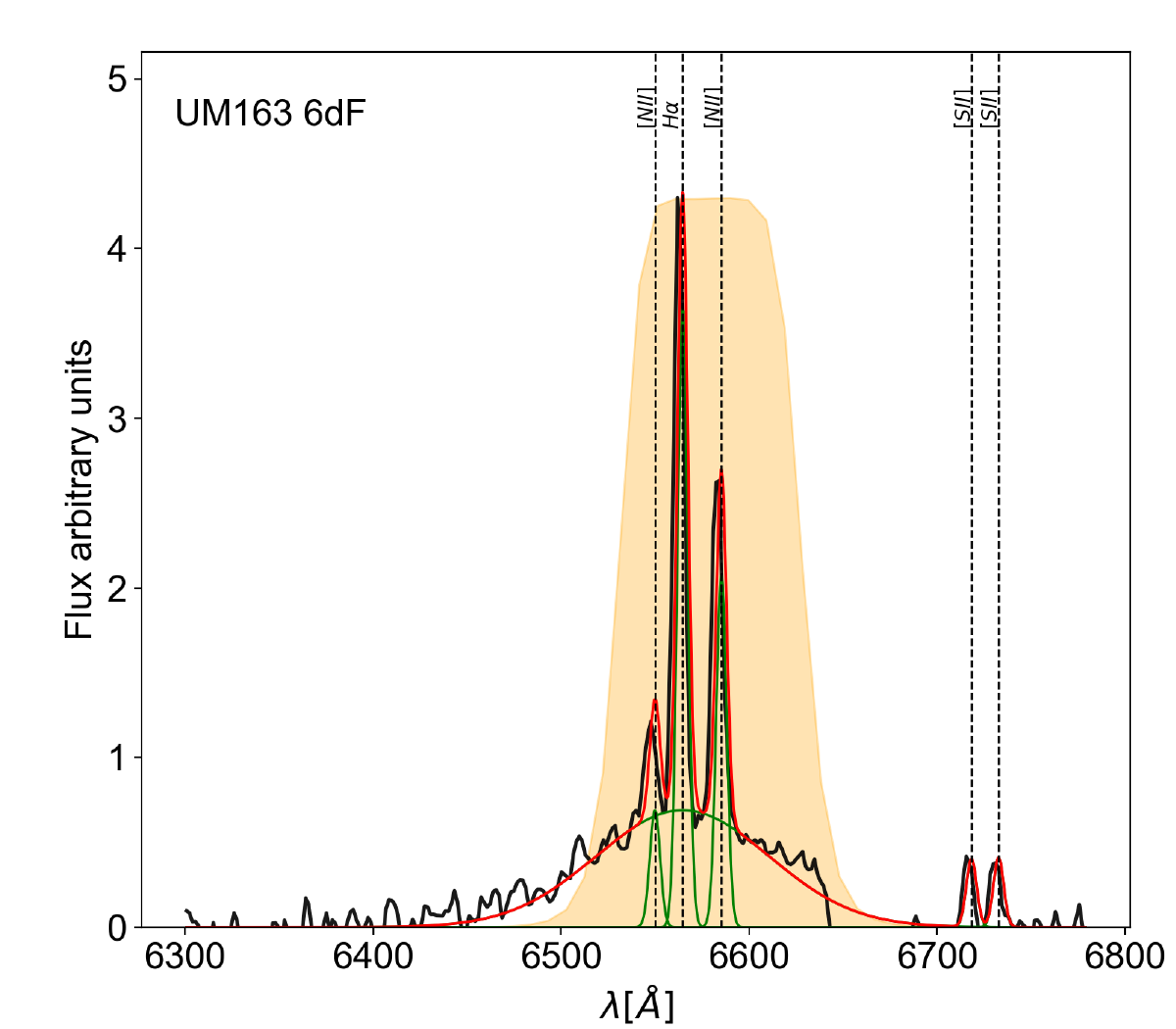}
\includegraphics[width=0.33\columnwidth]{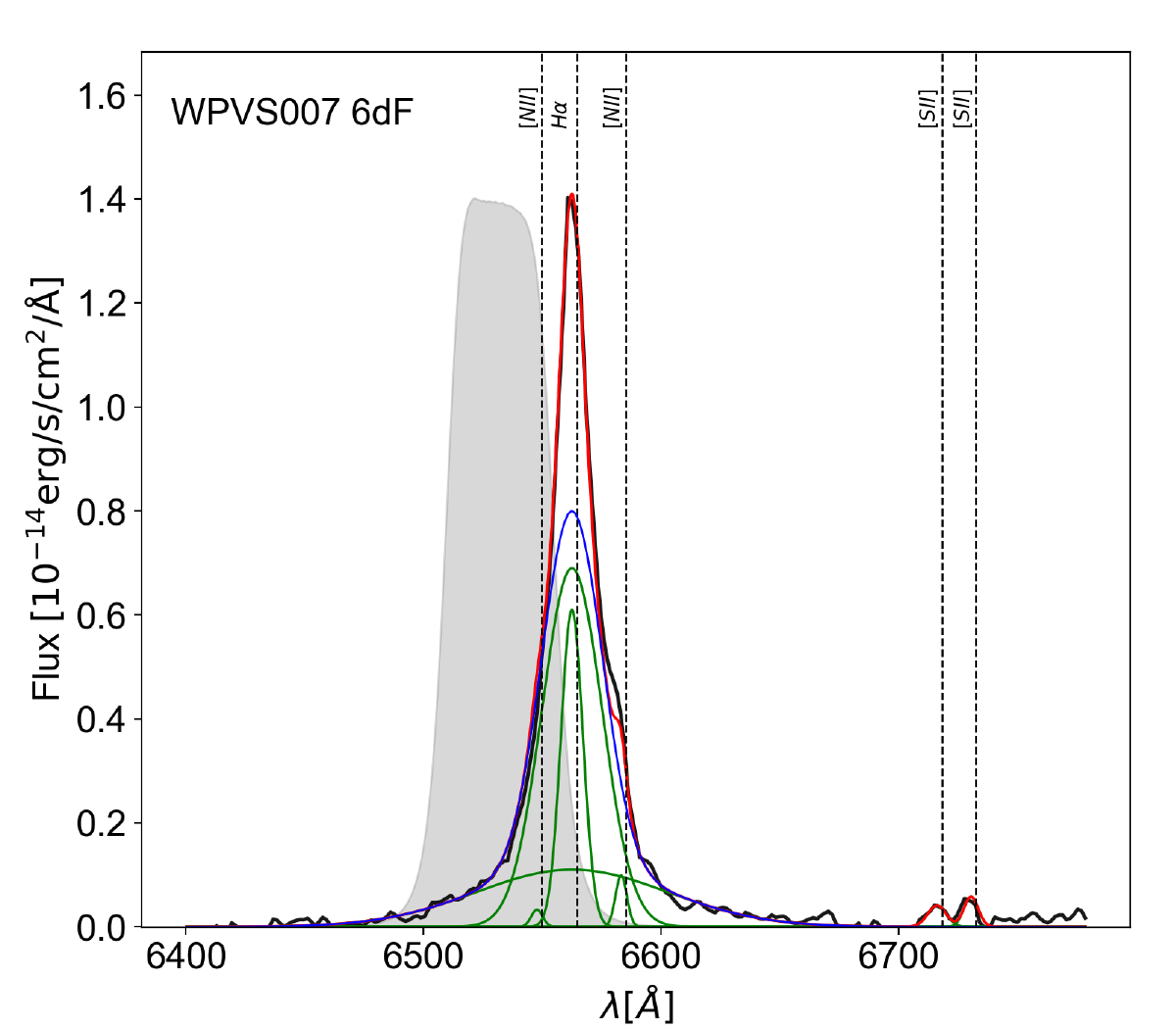}
\includegraphics[width=0.33\columnwidth]{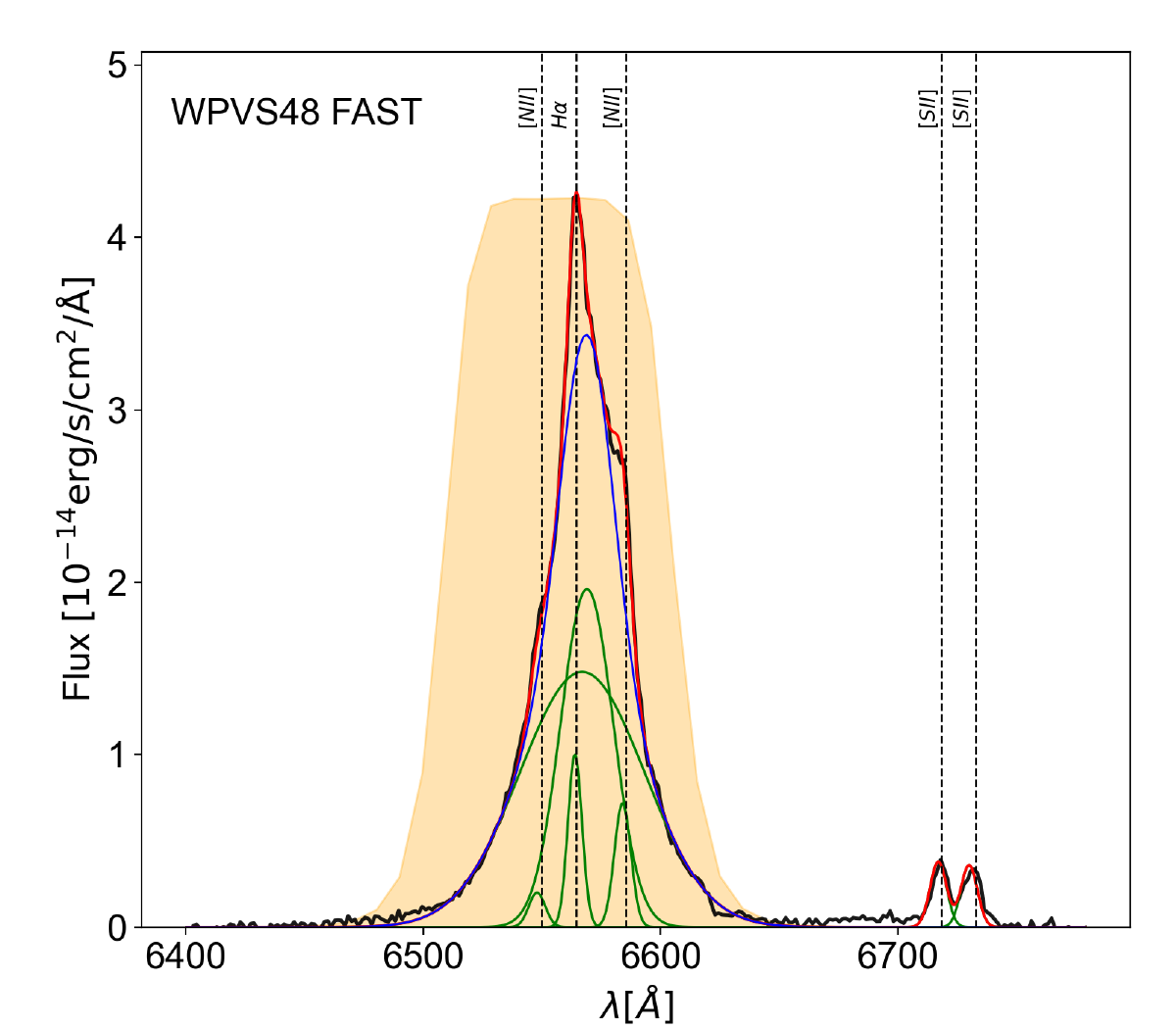}

\includegraphics[width=0.33\columnwidth]{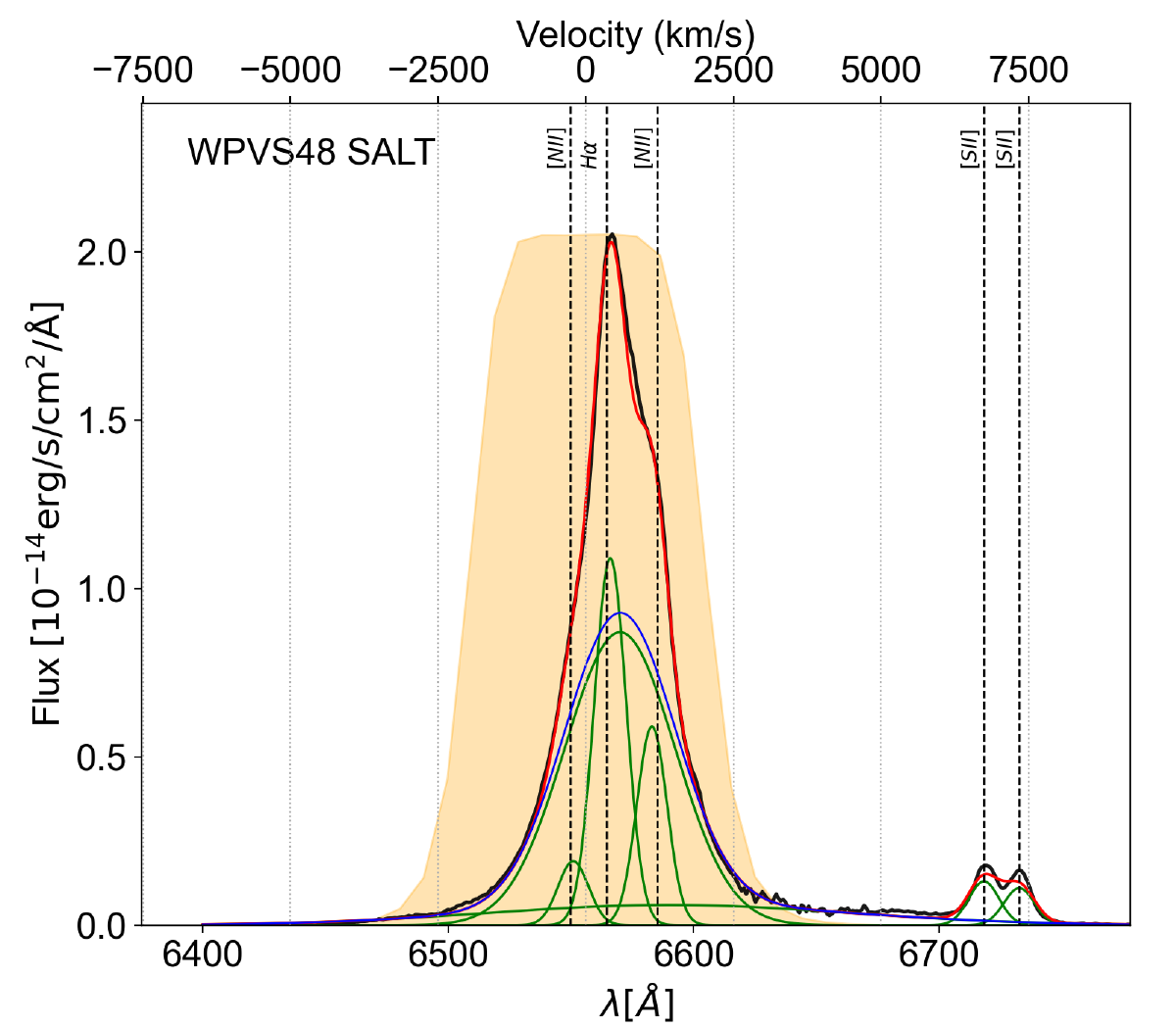}

\begin{deluxetable}{lll|lll}
\renewcommand{\arraystretch}{0.8}
\tablecaption{Values for $R_{\rm Fe}$} 
\label{tab:results_rfe}
\tablehead{{Object} & {Spectra$^1$} & $R_{\rm Fe}$ &  {Object} & {Spectra$^1$} & $R_{\rm Fe}$  }
\startdata
1H2107-097	&	 	&	0.494	$\pm$	0.029	&	MRK509	&	 SALT	&	0.126	$\pm$	0.062	\\
3C120	&	 HET	&	0.394	$\pm$	0.007	&	MRK509	&	 FAST	&	0.184	$\pm$	0.01	\\
3C120	&	 FAST	&	0.095	$\pm$	0.006	&	MRK705	&	 VLT	&	--	\\
AKN120	&	 	&	0.829	$\pm$	0.013	&	MRK705	&	 KPNO	&	0.527	$\pm$	0.016	\\
CTSG03$\_$04	&	 	&	0.002	$\pm$	0.001	&	MRK841	&	 	&	0.121	$\pm$	0.009	\\
ESO141-G55	&	 SALT	&	0.474	$\pm$	0.001	&	NGC1019	&	 FAST	&	0.509	$\pm$	0.08	\\
ESO141-G55	&	6dF	&	0.534	$\pm$	0.077	&	NGC1019	&	 SALT	&	0.482	$\pm$	0.036	\\
ESO323-G77	&	 	&	1.605	$\pm$	0.049	&	NGC4726	&	 	&	--	\\
ESO374-G25	&	 	&	1.101	$\pm$	0.355	&	NGC5940	&	 	&	0.755	$\pm$	0.061	\\
ESO399-IG20	&	 	&	0.570	$\pm$	0.023	&	NGC6860	&	 	&	0.459	$\pm$	0.156	\\
ESO438-G09	&	 	&	2.217	$\pm$	0.932	&	NGC7214	&	 	&	1.609	$\pm$	0.283	\\
ESO490-IG26	&	 --	&	--	&	NGC7469	&	 	&	0.326	$\pm$	0.077\\
ESO511-G030	&	 	&	0.326	$\pm$	0.059	&	NGC7603	&	 SALT	&	0.961	$\pm$	0.147	\\
ESO549-G49	&	 	&	--	&	NGC7603	&	 FAST	&	0.753	$\pm$	0.025	\\
ESO578-G09	&	 	&	0.567	$\pm$	0.058	&	NGC985	&	 	&	0.03	$\pm$	0.002	\\
F1041	&	 	&	--	&	PG1149-110	&	 	&	0.053	$\pm$	0.005	\\
HE1136-2304	&	 	&	0.235	$\pm$	0.001	&	PGC50427	&	 	&	0.188	$\pm$	0.032	\\
HE1143-1810	&	 	&	0.111	$\pm$	0.029	&	PGC64989	&	 	&	0.184	$\pm$	0.025	\\
HE2128-0221	&	 	&	0.695	$\pm$	0.131	&	RXSJ06225-2317	&	 	&	0.066	$\pm$	0.006	\\
IC4329A	&	 	&	0.195	$\pm$	0.059	&	RXSJ11032-0654	&	 FAST	&	0.383	$\pm$	0.052	\\
IRAS01089-4743	&	 	&	0.703	$\pm$	0.468	&	RXSJ11032-0654	&	 SALT	&	0.347	$\pm$	0.326	\\
IRAS09595-0755	&	 	&	0.358	$\pm$	0.062	&	RXSJ17414+034	&	 	&	0.953	$\pm$	0.003	\\
IRAS23226-3843	&	 	&	1.204	$\pm$	0.05	&	UGC12138	&	 	&	 --	\\
MCG+03-47-002	&	 	&			&	UM163	&	 	&	0.418	$\pm$	0.14	\\
MCG-02-12-050	&	 	&	0.405	$\pm$	0.018	&	WPVS007	&	 	&	1.933	$\pm$	1.22	\\
MRK1239	&	 	&	0.768	$\pm$	0.116	&	WPVS48	&	 FAST	&	0.436	$\pm$	0.021	\\
MRK1347	&	 	&	0.888	$\pm$	0.068	&	WPVS48	&	 SALT	&	0.162	$\pm$	0.009	\\
MRK335	&	 	&	0.343	$\pm$	0.017	&							\\
\enddata
\tablecomments{(1) Spectra noted if two spectra are available (see Table~\ref{tab:sample} for all the available spectra). }
\end{deluxetable}

\clearpage
\section{Time lag, luminosities for all objects}

\renewcommand{\arraystretch}{0.70}
\startlongtable
\begin{deluxetable}{llcccccccrr}
\tablecaption{\parbox{0.9\textwidth}{All time lags in observer frame. $\mathscr{R}_{Re}$ denotes the ratio $R_{\rm e FR}/ R_{\rm e, TP}$ and $\mathscr{C}$ the confidence level for the time lag determination. Delays showing  $\mathscr{R}_{Re}> 1.7$ and $\mathscr{C}>90\%$ are considered as good results (see section~\ref{sec:rblr} for general details on the confidence of the results and Appendix~\ref{sec:comments_objects} for individual cases). }}
\label{tab:results}
\setlength{\tabcolsep}{3pt} 
\tablehead{Object & year & $\tau_{\rm peak}$ & $\tau_{\rm cent}$ & $\mathscr{R}_{Re}/\mathscr{C}$ & $f_{5100\rm,obs}$ & $L_{5100}$  & FWHM & $M_{\rm BH}$ &$log\dot{\mathscr{M}}$ & $log\dot{\mathscr{M}}_{R_\mathrm{Fe{}}}$   \\
& & [days] & [days] & -- / \% & [mJy] & [$10^{43}$erg s$^{-1}$] & [km s$^{-1}$] & [$10^{6}M_{\sun}$] & &
}
\startdata
1H2107-097 & 2012 & $8.0^{+0.4}_{-0.4}$ & $12.1^{+3.3}_{-0.8}$ & 2.3/99.7 & $4.82 \pm 0.12$&$4.77 \pm 0.12$ & $2333\pm 447$ & $1.26^{+0.35}_{-0.09}$ & $0.809^{+0.244}_{-0.06}$ & $-0.553\pm 0.012$\\
1H2107-097 & 2012 & $14.4^{+2.0}_{-0.0}$ & $19.3^{+0.6}_{-0.5}$ & 2.1/98.4 & $4.82 \pm 0.12$&$4.77 \pm 0.12$ & $2333\pm 447$ & $2.01^{+0.06}_{-0.05}$ & $0.404^{+0.03}_{-0.026}$ & $-0.553\pm 0.012$\\
1H2107-097 & avg & $11.0^{+2.0}_{-0.4}$ & $15.7^{+3.4}_{-0.9}$ & -- & $4.82 \pm 0.12$&$4.77 \pm 0.12$ & $2333\pm 447$ & $1.63^{+0.36}_{-0.1}$ & $0.583^{+0.194}_{-0.053}$ & $-0.553\pm 0.012$\\ \hline
3C120 & 2014 & $55.2^{+30.0}_{-4.4}$ & $57.1^{+5.9}_{-5.9}$ & 3.0/100 & $8.88 \pm 0.23$&$14.34 \pm 0.37$ & $2924\pm 66$ & $9.27^{+0.99}_{-0.99}$ & $-0.209^{+0.094}_{-0.094}$ & $-1.883\pm 0.219$\\
AKN120 & 2018 & $18.6^{+23.0}_{-1.4}$ & $28.1^{+1.4}_{-1.6}$ & 2.6/100 & $7.48 \pm 0.27$&$11.91 \pm 0.43$ & $5759\pm 12$ & $17.7^{+0.91}_{-1.04}$ & $-0.891^{+0.048}_{-0.054}$ & $0.563\pm 0.001$\\
CTSG03\_04 & 2013 & $14.8^{+0.4}_{-1.4}$ & $17.8^{+0.9}_{-0.9}$ & 3.0/99.9 & $1.03 \pm 0.17$&$2.47 \pm 0.41$ & $3042\pm 242$ & $3.11^{+0.16}_{-0.16}$ & $-0.405^{+0.093}_{-0.093}$ & $-2.193\pm 0.431$\\ \hline
ESO141-G55 & 2013 & $26.4^{+1.2}_{-6.0}$ & $23.1^{+1.0}_{-0.9}$ & 3.9/100 & $11.21 \pm 0.63$&$23.1 \pm 1.3$ & $4981\pm 578$ & $10.84^{+0.49}_{-0.44}$ & $-0.034^{+0.048}_{-0.045}$ & $-0.62\pm 0.012$\\
ESO141-G55 & 2015 & $16.0^{+1.2}_{-0.4}$ & $16.1^{+1.3}_{-1.6}$ & 4.2/100 & $7.57 \pm 0.62$&$15.6 \pm 1.28$ & $4981\pm 578$ & $7.55^{+0.63}_{-0.78}$ & $0.024^{+0.083}_{-0.098}$ & $-0.62\pm 0.012$\\ 
ESO141-G55 & avg & $21.2^{+1.7}_{-6.0}$ & $19.6^{+1.6}_{-1.8}$ & -- & $8.76 \pm 0.88$&$18.05 \pm 1.81$ & $4981\pm 578$ & $9.2^{+0.78}_{-0.88}$ & $-0.052^{+0.088}_{-0.096}$ & $-0.62\pm 0.012$\\ \hline
ESO323-G77 & 2015 & $29.2^{+1.2}_{-1.2}$ & $26.7^{+3.6}_{-1.9}$ & 2.3/100 & $6.95 \pm 0.45$&$2.52 \pm 0.16$ & $4246\pm 460$ & $9.3^{+1.27}_{-0.67}$ & $-1.343^{+0.123}_{-0.07}$ & $3.15\pm 0$\\
ESO374-G25 & 2011 & $7.6^{+14.2}_{-2.5}$ & $11.2^{+1.0}_{-2.0}$ & 2.1/97.7 & $1.72 \pm 0.01$&$<1.58$ & $4481\pm 969$ & $4.31^{+0.39}_{-0.79}$ & $-1.002^{+0.079}_{-0.159}$ & $1.47\pm 0.001$\\
ESO399-IG20 & 2011 & $24.4^{+0.8}_{-5.2}$ & $19.6^{+0.4}_{-0.8}$ & 2.2/98.6 & $2.47 \pm 0.52$&$2.16 \pm 0.45$ & $1843\pm 81$ & $1.27^{+0.03}_{-0.05}$ & $0.281^{+0.104}_{-0.109}$ & $-0.3\pm 0.007$\\ \hline
ESO438-G09 & 2011 & $14.8^{+2.4}_{-9.2}$ & $12.2^{+0.4}_{-0.4}$ & 2.1/98.6 & $3.96 \pm 0.5$&$3.65 \pm 0.46$ & $2300\pm 167$ & $1.24^{+0.04}_{-0.04}$ & $0.649^{+0.068}_{-0.068}$ & $5.19\pm 0$\\
ESO438-G09 & 2015 & $12.4^{+0.4}_{-0.4}$ & $12.0^{+0.2}_{-0.2}$ & 3.5/100 & $2.84 \pm 0.28$&$2.61 \pm 0.26$ & $2300\pm 167$ & $1.22^{+0.02}_{-0.02}$ & $0.447^{+0.05}_{-0.05}$ & $5.19\pm 0$\\
ESO438-G09 & avg & $13.6^{+2.4}_{-9.2}$ & $12.1^{+0.4}_{-0.4}$ & -- & $3.44 \pm 0.57$&$3.17 \pm 0.52$ & $2300\pm 167$ & $1.23^{+0.04}_{-0.04}$ & $0.565^{+0.086}_{-0.086}$ & $5.19\pm 0$\\ \hline
ESO490-IG26 & 2011 & $15.2^{+2.2}_{-3.0}$ & $13.0^{+4.5}_{-2.7}$ & 1.2/73.7 & $3.74 \pm 0.04$&$<3.64$ & $5588\pm 412$ & $7.77^{+2.76}_{-1.65}$ & $-0.973^{+0.308}_{-0.185}$ & --\\
ESO549-G49 & 2012 & -- & -- & 1.7/98.1 & $4.73 \pm 0.05$&$<4.55$ & $2766\pm 270$ & -- & -- & --\\ \hline
ESO511-G030 & 2013 & $17.2^{+0.2}_{-0.2}$ & $20.9^{+0.7}_{-0.5}$ & 2.0/99.3 & $1.3 \pm 0.32$&$1.01 \pm 0.25$ & $3656\pm 12$ & $5.36^{+0.18}_{-0.13}$ & $-1.46^{+0.124}_{-0.122}$ & $-1.113\pm 0.056$\\ 
ESO511-G030 & 2014 & $18.7^{+0.2}_{-0.2}$ & $19.2^{+0.5}_{-0.6}$ & 2.2/99.7 & $1.02 \pm 0.5$&$0.79 \pm 0.39$ & $3656\pm 12$ & $4.92^{+0.13}_{-0.16}$ & $-1.544^{+0.241}_{-0.241}$ & $-1.113\pm 0.056$\\ 
ESO578-G09 & 2014 & $19.6^{+2.0}_{-12.8}$ & $19.5^{+0.6}_{-0.6}$ & 2.7/99.9 & $1.75 \pm 0.19$&$3.39 \pm 0.37$ & $5125\pm 14$ & $9.7^{+0.31}_{-0.31}$ & $-1.188^{+0.06}_{-0.06}$ & $-0.31\pm 0.009$\\
F1041 & 2013 & $16.4^{+0.4}_{-0.0}$ & $15.7^{+0.7}_{-1.0}$ & 2.4/99.8 & $0.65 \pm 0.2$&$1.04 \pm 0.32$ & $3676\pm 886$ & $4.03^{+0.19}_{-0.27}$ & $-1.196^{+0.156}_{-0.161}$ & --\\
HE0003-5023 & 2014 & $2.0^{+8.5}_{-0.4}$ & $6.6^{+1.4}_{-1.7}$ & 3.1/99.9 & $2.24 \pm 0.61$&$3.62 \pm 0.99$ & $3396\pm 0$ & $1.44^{+0.32}_{-0.38}$ & $0.509^{+0.232}_{-0.267}$ & --\\ \hline
HE1136-2304 & 2015 & $10.6^{+1.0}_{-1.0}$ & $9.1^{+0.5}_{-0.2}$ & 2.1/99.2 & $0.19 \pm 0.12$&$0.22 \pm 0.14$ & $3544\pm 221$ & $2.18^{+0.12}_{-0.05}$ & $-1.669^{+0.312}_{-0.309}$ & $-1.417\pm 0.072$\\
HE1136-2304 & 2016 & $20.4^{+1.0}_{-1.0}$ & $17.4^{+2.2}_{-3.9}$ & 1.8/99.4 & $0.74 \pm 0.12$&$0.86 \pm 0.14$ & $3544\pm 221$ & $4.17^{+0.54}_{-0.96}$ & $-1.346^{+0.138}_{-0.215}$ & $-1.417\pm 0.072$\\
HE1136-2304 & 2018 & $11.0^{+15.6}_{-1.0}$ & $11.2^{+4.2}_{-1.8}$ & 1.5/84.9 & $0.66 \pm 0.12$&$0.77 \pm 0.14$ & $3544\pm 221$ & $2.69^{+1.03}_{-0.44}$ & $-1.038^{+0.346}_{-0.169}$ & $-1.417\pm 0.072$\\ \hline
HE1143-1810 & 2016 & $21.2^{+0.2}_{-1.2}$ & $17.5^{+2.5}_{-2.4}$ & 2.2/99.8 & $2.87 \pm 0.25$&$5.01 \pm 0.44$ & $2143\pm 15$ & $1.53^{+0.23}_{-0.22}$ & $0.674^{+0.135}_{-0.13}$ & $-1.83\pm 0.231$\\
HE2128-0221 & 2016 & $9.2^{+5.6}_{-5.8}$ & $8.3^{+0.7}_{-0.9}$ & 2.2/99.9 & $0.59 \pm 0.05$&$2.43 \pm 0.21$ & $1660\pm 124$ & $0.43^{+0.04}_{-0.05}$ & $1.311^{+0.088}_{-0.107}$ & $0.117\pm 0.006$\\
IC4329A & 2015 & $13.4^{+14.0}_{-0.6}$ & $22.7^{+0.8}_{-0.8}$ & 3.0/100 & $7.09 \pm 0.06$&$<2.87$ & $4940\pm 274$ & $10.69^{+0.38}_{-0.38}$ & $-1.369^{+0.031}_{-0.031}$ & $-1.55\pm 0.153$\\
IRAS01089-4743 & 2013 & -- &-- & 1.2/76.8 & $1.95 \pm 0.44$&$1.55 \pm 0.35$ & $1731\pm 124$ & -- & -- & $0.143\pm 0.072$\\
IRAS09595-0755 & 2013 & $61.6^{+12.0}_{-0.4}$ & $57.4^{+1.6}_{-1.8}$ & 1.2/68.5 & $0.67 \pm 0.16$&$3.03 \pm 0.72$ & $2402\pm 18$ & $6.16^{+0.18}_{-0.2}$ & $-0.865^{+0.119}_{-0.12}$ & $-1.007\pm 0.045$\\
IRAS09595-0755 & 2013 &-- &-- & 1.2/68.5 & $0.67 \pm 0.16$&$3.03 \pm 0.72$ & $2402\pm 18$ & -- & -- & --\\
IRAS23226-3843 & 2013 & -- & -- & 2.0/99.7 & $2.15 \pm 0.44$&$3.96 \pm 0.81$ & -- & -- & -- & $1.813\pm 0$\\
MCG+03-47-002 & 2013 & $18.8^{+0.0}_{-0.2}$ & $16.8^{+0.4}_{-0.5}$ & 1.7/96.8 & $0.28 \pm 0.22$&$0.66 \pm 0.52$ & -- & -- & -- & --\\
MCG-02.12.050 & 2014 & $9.6^{+3.6}_{-0.8}$ & $10.8^{+1.3}_{-1.2}$ & 1.4/86.4 & $2.31 \pm 0.02$&$<4.37$ & $5585\pm 782$ & $6.38^{+0.8}_{-0.73}$ & $-0.634^{+0.108}_{-0.1}$ & $-0.85\pm 0.022$\\
MRK1239 & 2015 &-- & -- & 1.1/79.4 & $5.51 \pm 0.04$&$<3.47$ & $1043\pm 358$ & -- & -- & $0.36\pm 0.003$\\
MRK1347 & 2014 & $21.2^{+1.0}_{-16.8}$ & $13.8^{+4.6}_{-1.7}$ & 1.9/97.4 & $1.83 \pm 0.24$&$7.41 \pm 0.97$ & $1576\pm 540$ & $0.64^{+0.22}_{-0.08}$ & $1.682^{+0.311}_{-0.129}$ & $0.76\pm 0.001$\\ \hline
MRK335 & 2010 & $19.6^{+0.0}_{-0.4}$ & $19.0^{+0.4}_{-0.3}$ & 1.0/57.0 & --&-- & $1611\pm 259$ & $0.94^{+0.02}_{-0.02}$ & -- & $-1.057\pm 0.036$\\
MRK335 & 2011 & $20.4^{+0.8}_{-10.8}$ & $17.5^{+3.9}_{-6.6}$ & 1.2/72.7 & $5.19 \pm 0.2$&$4.62 \pm 0.18$ & $1611\pm 259$ & $0.87^{+0.2}_{-0.34}$ & $1.11^{+0.199}_{-0.337}$ & $-1.057\pm 0.036$\\
MRK335 & 2014 & $11.2^{+0.2}_{-5.2}$ & $12.0^{+0.9}_{-1.1}$ & 2.0/99.5 & $4.49 \pm 0.25$&$3.99 \pm 0.22$ & $1611\pm 259$ & $0.6^{+0.05}_{-0.06}$ & $1.343^{+0.072}_{-0.086}$ & $-1.057\pm 0.036$\\ \hline
MRK509 & 2014 & $24.4^{+0.4}_{-1.6}$ & $22.9^{+0.8}_{-0.8}$ & 2.1/99.8 & $10.29 \pm 0.91$&$17.31 \pm 1.53$ & $3451\pm 32$ & $5.17^{+0.19}_{-0.19}$ & $0.421^{+0.053}_{-0.053}$ & $-1.587\pm 0.114$\\
MRK705 & 2013 & $11.6^{+0.0}_{-0.8}$ & $15.5^{+1.0}_{-0.7}$ & 2.0/95.4 & $2.43 \pm 0.36$&$3.21 \pm 0.48$ & $1919\pm 332$ & $1.09^{+0.07}_{-0.05}$ & $0.677^{+0.093}_{-0.083}$ & $-0.443\pm 0.009$\\
MRK841 & 2014 & $20.8^{+12.0}_{-1.6}$ & $23.8^{+2.5}_{-2.4}$ & 2.6/99.7 & $3.28 \pm 0.13$&$6.75 \pm 0.27$ & $4645\pm 734$ & $9.72^{+1.06}_{-1.02}$ & $-0.74^{+0.097}_{-0.093}$ & $-1.797\pm 0.184$\\
NGC1019 & 2011 & $9.2^{+1.6}_{-0.0}$ & $9.7^{+2.0}_{-0.8}$ & 1.5/93.8 & $0.7 \pm 0.32$&$0.57 \pm 0.26$ & $2755\pm 80$ & $1.41^{+0.3}_{-0.12}$ & $-0.677^{+0.289}_{-0.235}$ & $-0.593\pm 0.014$\\
NGC4726 & 2013 & -- & -- & 1.2/75.0 & $3.01 \pm 0.03$&$<3.05$ & $3119\pm 0$ & -- & -- & $-1.037\pm 0.041$\\
NGC5940 & 2014 & $5.2^{+0.8}_{-0.4}$ & $5.9^{+0.8}_{-0.7}$ & 1.8/97.1 & $1.26 \pm 0.3$&$2.26 \pm 0.54$ & $4033\pm 22$ & $1.82^{+0.26}_{-0.22}$ & $0.002^{+0.168}_{-0.158}$ & $0.317\pm 0.002$\\
NGC6860 & 2015 & $36.0^{+0.0}_{-3.2}$ & $34.7^{+1.0}_{-1.1}$ & 1.5/93.1 & $2.0 \pm 0.49$&$0.61 \pm 0.15$ & $3668\pm 1016$ & $9.02^{+0.26}_{-0.29}$ & $-2.243^{+0.122}_{-0.123}$ & $-0.67\pm 0.042$\\
NGC7214 & 2011 & $5.2^{+0.4}_{-0.4}$ & $6.9^{+5.2}_{-0.9}$ & 1.8/97.8 & $2.74 \pm 0.53$&$2.1 \pm 0.41$ & $3662\pm 100$ & $1.77^{+1.37}_{-0.24}$ & $-0.025^{+0.677}_{-0.15}$ & $3.163\pm 0$\\
NGC7469 & 2012 & $16.0^{+0.0}_{-0.4}$ & $9.6^{+3.5}_{-4.8}$ & 2.1/97.2 & $9.68 \pm 0.96$&$3.13 \pm 0.31$ & $1615\pm 119$ & $0.48^{+0.18}_{-0.25}$ & $1.365^{+0.325}_{-0.444}$ & --\\
NGC7603 & 2014 & $36.8^{+4.0}_{-1.2}$ & $35.1^{+1.5}_{-1.3}$ & 6.8/100 & $8.12 \pm 1.1$&$9.04 \pm 1.22$ & $5778\pm 10$ & $22.34^{+0.98}_{-0.85}$ & $-1.273^{+0.076}_{-0.074}$ & $0.31\pm 0.002$\\
NGC985 & 2014 & $24.0^{+0.4}_{-0.4}$ & $22.2^{+0.7}_{-0.8}$ & 1.9/98.1 & $3.73 \pm 0.84$&$10.2 \pm 2.3$ & $4675\pm 347$ & $9.12^{+0.3}_{-0.34}$ & $-0.416^{+0.114}_{-0.115}$ & $-2.1\pm 0.35$\\
PG1149-110 & 2013 & $22.4^{+0.8}_{-6.0}$ & $17.3^{+6.2}_{-1.1}$ & 1.1/65.8 & $1.4 \pm 0.02$&$<5.21$ & $3579\pm 700$ & $4.14^{+1.56}_{-0.28}$ & $-0.137^{+0.327}_{-0.058}$ & $-2.023\pm 0.3$\\ \hline
PGC50247 & 2011 & $21.8^{+1.0}_{-1.8}$ & $21.6^{+0.9}_{-0.9}$ & 2.0/99.5 & $1.02 \pm 0.16$&$0.87 \pm 0.14$ & $2377\pm 11$ & $2.34^{+0.1}_{-0.1}$ & $-0.836^{+0.085}_{-0.085}$ & $-1.573\pm 0.131$\\
PGC50247 & 2014 & $18.4^{+0.4}_{-0.0}$ & $20.2^{+0.6}_{-0.4}$ & 1.7/97.4 & $1.38 \pm 0.12$&$1.18 \pm 0.1$ & $2377\pm 11$ & $2.19^{+0.07}_{-0.04}$ & $-0.581^{+0.05}_{-0.046}$ & $-1.573\pm 0.131$\\ \hline
PGC64989 & 2013 & $24.4^{+0.4}_{-0.0}$ & $27.6^{+2.3}_{-3.7}$ & 1.3/75.1 & $0.9 \pm 0.04$&$0.45 \pm 0.02$ & $3275\pm 770$ & $5.7^{+0.48}_{-0.78}$ & $-2.04^{+0.077}_{-0.121}$ & $-1.587\pm 0.128$\\
PGC64989 & 2014 & $26.8^{+0.2}_{-2.4}$ & $26.0^{+0.3}_{-0.3}$ & 3.5/100 & $1.06 \pm 0.04$&$0.53 \pm 0.02$ & $3275\pm 770$ & $5.37^{+0.06}_{-0.06}$ & $-1.882^{+0.021}_{-0.021}$ & $-1.587\pm 0.128$\\ \hline
RXSJ06225-2317 & 2013 & $20.0^{+0.0}_{-0.4}$ & $19.5^{+0.2}_{-1.4}$ & 1.7/93.0 & $1.46 \pm 0.25$&$3.23 \pm 0.55$ & $1506\pm 30$ & $0.84^{+0.01}_{-0.06}$ & $0.91^{+0.084}_{-0.106}$ & $-1.98\pm 0.274$\\ \hline
RXJ1103.2-0654 & 2011 &-- & -- & 1.3/85.5 & --&--& $3849\pm 46$ & -- & -- & $-1.043\pm 0.37$\\
RXJ1103.2-0654 & 2014 & -- & -- & 1.2/73.0 & --&-- & $3849\pm 46$ &-- & -- & --\\ \hline 
RXSJ17414+0348 & 2012 & $16.0^{+0.0}_{-0.4}$ & $15.4^{+0.6}_{-0.8}$ & 2.2/99.7 & $3.63 \pm 0.62$&$2.77 \pm 0.47$ & $2222\pm 129$ & $1.46^{+0.06}_{-0.08}$ & $0.328^{+0.09}_{-0.095}$ & $0.977\pm 0$\\
RXSJ17414+0348 & 2014 & $28.4^{+1.2}_{-9.6}$ & $20.5^{+1.2}_{-1.2}$ & 1.9/98.7 & $3.41 \pm 0.29$&$2.61 \pm 0.22$ & $2222\pm 129$ & $1.94^{+0.12}_{-0.12}$ & $0.039^{+0.067}_{-0.067}$ & $0.977\pm 0$\\
UGC12138 & 2012 & $16.4^{+0.2}_{-0.2}$ & $15.0^{+0.5}_{-0.4}$ & 1.6/97.7 & $2.05 \pm 0.49$&$1.69 \pm 0.4$ & $2693\pm 269$ & $2.08^{+0.07}_{-0.06}$ & $-0.303^{+0.12}_{-0.119}$ & --\\
UM163 & 2013 & $10.0^{+0.8}_{-0.4}$ & $10.9^{+0.4}_{-0.5}$ & 2.3/99.9 & $1.37 \pm 0.2$&$2.12 \pm 0.31$ & $4901\pm 77$ & $4.97^{+0.19}_{-0.24}$ & $-0.911^{+0.079}_{-0.082}$ & $-0.807\pm 0.051$\\
WPVS007 & 2012 & $10.4^{+0.0}_{-0.4}$ & $10.6^{+0.9}_{-1.0}$ & 2.1/99.4 & $2.19 \pm 0.18$&$2.56 \pm 0.21$ & $1557\pm 163$ & $0.49^{+0.04}_{-0.05}$ & $1.222^{+0.086}_{-0.093}$ & $4.243\pm 0.002$\\ \hline
WPVS48 & 2013 & $17.6^{+0.8}_{-0.0}$ & $21.1^{+0.9}_{-1.9}$ & 2.8/99.8 & $2.69 \pm 0.2$&$5.88 \pm 0.44$ & $1917\pm 24$ & $1.47^{+0.06}_{-0.14}$ & $0.812^{+0.053}_{-0.089}$ & $-1.66\pm 0.134$\\
WPVS48 & 2014 & $16.4^{+2.4}_{-0.4}$ & $18.3^{+0.6}_{-1.3}$ & 3.8/100 & $2.69 \pm 0.54$&$5.88 \pm 1.18$ & $1917\pm 24$ & $1.27^{+0.04}_{-0.09}$ & $0.936^{+0.102}_{-0.117}$ & $-0.747\pm 0.018$\\
WPVS48 & 2018 & $19.6^{+6.4}_{-0.8}$ & $19.3^{+4.3}_{-0.3}$ & 2.1/99.2 & $2.29 \pm 0.4$&$5.0 \pm 0.87$ & $1917\pm 24$ & $1.34^{+0.31}_{-0.02}$ & $0.784^{+0.218}_{-0.086}$ & $-0.747\pm 0.018$\\
WPVS48 & avg & $18.9^{+6.4}_{-0.8}$ & $18.9^{+2.7}_{-2.7}$ & -- & $2.56 \pm 0.7$&$5.59 \pm 1.53$ & $1917\pm 24$ & $1.31^{+0.19}_{-0.19}$ & $0.875^{+0.185}_{-0.185}$ & $-0.747\pm 0.018$\\
\hline
\enddata
\end{deluxetable}

\begin{deluxetable}{lllll}
\renewcommand{\arraystretch}{0.45}
\tablecaption{Literature values for H$\alpha$ Reverberation Mapping }
\tablehead{{Object} & {z} & {$t_{\rm H\alpha}$} & {Log$L_{5100}$} & {Ref.} \\ 
 & & days & } 
\startdata
Mrk142   & 0.04494 & $2.78^{+1.17}_{-0.88}$ & $43.54 \pm 0.02$ & B       \\
SBS1116+583A   & 0.02787 & $4.01^{+1.37}_{-0.95}$ & $42.07 \pm 0.28$ & B       \\
Arp151   & 0.02109 & $7.84^{+1.03}_{-0.98}$ & $42.48 \pm 0.11$ & B       \\
Mrk1310  & 0.01956 & $4.51^{+0.66}_{-0.61}$ & $42.23 \pm 0.17$ & B       \\
NGC4253  & 0.01293 & $25.17^{+0.65}_{-0.85}$   & $42.51 \pm 0.13$ & B       \\
NGC4748  & 0.01463 & $7.50^{+2.97}_{-4.57}$ & $42.49 \pm 0.13$ & B       \\
NGC6814  & 0.00521 & $9.46^{+1.90}_{-1.56}$ & $42.05 \pm 0.29$ & B       \\ \hline
Mrk1501  & 0.0893  & $67^{+24}_{-38}$    & $44.14 \pm 0.02$ & C       \\
J0101+422   & 0.1900  & $118^{+17}_{-17}$  & $44.89 \pm 0.01$ & C       \\
PG0947+396  & 0.2059  & $71^{+16}_{-35}$   & $44.71 \pm 0.01$ & C       \\
VIIIZw218   & 0.1274  & $140^{+26}_{-26}$  & $44.53 \pm 0.01$ & C       \\
PG1440+356  & 0.0791  & $80^{+63}_{-30}$   & $44.63 \pm 0.01$ & C       \\
NGC4395  & 0.00106 & $0.058^{+0.010}_{-0.010}$ & $39.76 \pm 0.01$ & (C)    \\ \hline
17    & 0.457   & $119.2^{+5.2}_{-7.9}$ & $43.99 \pm 0$    & S       \\
85    & 0.238   & $61.3^{+13.6}_{-10.3}$   & $43.37 \pm 0$    & S       \\
101   & 0.458   & $75.7^{+7.1}_{-8.3}$  & $44.4 \pm 0$  & S       \\
126   & 0.192   & $136.1^{+6.9}_{-6.4}$ & $43.29 \pm 0$    & S       \\
160   & 0.36    & $90.2^{+4.1}_{-3.9}$  & $43.82 \pm 0$    & S       \\
184   & 0.193   & $145.4^{+58.5}_{-42.2}$  & $43.73 \pm 0$    & S       \\
291   & 0.532   & $95.3^{+20.1}_{-19.6}$   & $43.9 \pm 0$  & S       \\
305   & 0.527   & $54.6^{+21.3}_{-16.5}$   & $44.22 \pm 0$    & S       \\
320   & 0.265   & $30.7^{+7.1}_{-11.7}$ & $43.44 \pm 0$    & S       \\
371   & 0.473   & $32.4^{+2.7}_{-2.9}$  & $44.13 \pm 0$    & S       \\
645   & 0.474   & $25.2^{+2.3}_{-2.5}$  & $44.1 \pm 0$  & S       \\
733   & 0.455   & $27.4^{+13.5}_{-28.7}$   & $43.91 \pm 0$    & S       \\
766   & 0.165   & $13.9^{+3.9}_{-3.9}$  & $43.75 \pm 0$    & S       \\
767   & 0.527   & $33.4^{+14.5}_{-13.0}$   & $43.91 \pm 0$    & S       \\
768   & 0.259   & $27.7^{+2.8}_{-2.9}$  & $43.4 \pm 0$  & S       \\
769   & 0.187   & $14.0^{+3.4}_{-3.3}$  & $42.96 \pm 0$    & S       \\
772   & 0.249   & $22.9^{+2.1}_{-2.3}$  & $43.46 \pm 0$    & S       \\
781   & 0.264   & $16.0^{+4.6}_{-5.5}$  & $43.63 \pm 0$    & S       \\
789   & 0.425   & $47.4^{+4.9}_{-5.3}$  & $43.74 \pm 0$    & S       \\
790   & 0.238   & $10.4^{+3.6}_{-3.6}$  & $43.33 \pm 0$    & S       \\
798   & 0.423   & $17.5^{+6.5}_{-9.7}$  & $44.07 \pm 0$    & S       \\
840   & 0.244   & $9.6^{+1.3}_{-1.4}$   & $43.31 \pm 0$    & S       \\
845   & 0.273   & $9.4^{+4.6}_{-4.4}$   & $42.78 \pm 0$    & S       \\ \hline
PG0026   & 0.142   & $116^{+25}_{-27}$  & $44.91 \pm 0.02$ & K       \\
PG0052   & 0.155   & $183^{+57}_{-38}$  & $44.75 \pm 0.03$ & K       \\
PG0804   & 0.1  & $175^{+18}_{-15}$  & $44.85 \pm 0.02$ & K       \\
PG0844   & 0.064   & $37^{+15}_{-15}$   & $44.23 \pm 0.06$ & K       \\
PG1211   & 0.085   & $107^{+35}_{-42}$  & $44.69 \pm 0.06$ & K       \\
PG1226   & 0.158   & $444^{+56}_{-55}$  & $45.90 \pm 0.02$ & K       \\
PG1229   & 0.064   & $67^{+37}_{-43}$   & $43.64 \pm 0.06$ & K       \\
PG1307   & 0.155   & $155^{+81}_{-13}$  & $44.79 \pm 0.02$ & K       \\
PG1351   & 0.087   & $227^{+149}_{-72}$    & $44.64 \pm 0.06$ & K       \\
PG1411   & 0.089   & $95^{+37}_{-34}$   & $44.50 \pm 0.02$ & K       \\
PG1426   & 0.086   & $83^{+42}_{-48}$   & $44.57 \pm 0.02$ & K       \\
PG1613   & 0.129   & $38^{+35}_{-19}$   & $44.71 \pm 0.03$ & K       \\
PG1617   & 0.114   & $100^{+28}_{-33}$  & $44.33 \pm 0.02$ & K       \\
PG2130   & 0.061   & $223^{+50}_{-26}$  & $44.14 \pm 0.03$ & K       \\
\hline
\enddata
\tablecomments{Delays are in restframe. Reference columns - B: H$\alpha$ time delay from \cite{2010ApJ...716..993B} and luminosity from \cite{2013ApJ...767..149B}. S: Values from \cite{2024ApJS..272...26S}, object corresponds to the RMID from SDSS and luminosity is the median luminosity from whole campaign and host-fraction subtracted. K: H$\alpha$ lag values and luminosity from \cite{2000ApJ...533..631K} and if available luminosity from \cite{2013ApJ...767..149B}. C: \cite{2023ApJ...953..142C} and (C): \cite{2021ApJ...921...98C}.}
\label{tab:lit_ha}
\end{deluxetable}

\section{Comments on individual objects}\label{sec:comments_objects}

Here we elaborate on the idiosyncrasies of the sources in our sample with particular emphasis on the time-lag measurements and host-galaxy subtraction. We compare our results to previously reported measurements, when available.

\begin{enumerate}[leftmargin=0.55cm]

\item {1H2107-097:} 
 Due to its redshift (z=0.02698), the H$\alpha$ emission line is not fully covered by a single narrowband (NB) filter. Our observations use two NB filters: NB670 and NB680, each covering distinct parts of the H$\alpha$ line profile. Specifically, the NB670 filter covers the blue wing and a substantial portion of the H$\alpha$ core and the NB680 band mainly covers the red wing of the line. With no restrictions to the time delay formalism we derived a time delay of $\tau = 12.1^{+3.3}_{-0.6}$ days and $\alpha = 0.54^{+0.04}_{-0.12}$ for the NB670 band and $\tau = 19.3^{+0.6}_{-0.5}$ days, with $\alpha = 0.44^{+0.02}_{-0.02}$ for the NB680 filter.
While these results hint at a rotating disk model for the BLR, a more thorough study of the geometry of the H$\alpha$ line is beyond of the scope of this paper. Therefore, we present an average delay $\tau = 15.7^{+4.2}_{-4.2}$ days for this source. 
The $BV$-band FVG reveals a blue trend with a slope of $1.18 \pm 0.02$, intersecting precisely within the host galaxy's color range. Consequently, we derive host-subtracted flux values, leading to a host-subtracted luminosity of $4.77 \pm 0.57 \times 10^{43} , \text{erg s}^{-1}$.

\item {3C120:} The correlation exhibits two peaks, at 55 days and at 80 days, both with $\alpha \sim 0.5$, aligning with the $\alpha$ value derived from $\alpha_{\rm phot}$. For a comprehensive analysis of the same photometric observations conducted by our group, refer to \cite{2018AaA620A137R}. Regarding the H$\alpha$ line, they reported a delay of $71.2^{+12.4}_{-13.3}$ days, which is consistent with our findings within the measurement uncertainties. 
3C 120 has been the subject of numerous reverberation mapping (RM) campaigns. Notably, delays for H$\beta$ from observations conducted between 1989 and 1996 were reported as $38.1^{+21.3}_{-15.3}$ days in \cite{1998PASP..110..660P}. For the period 2008-2009, \cite{kollatschny14} reported delays for various emission lines: $23.9^{+4.6}_{-3.9}$ days for H$\gamma$, $12.0^{+7.5}_{-7.0}$ days for He II, $26.8^{+6.7}_{-7.3}$ days for He I, $27.9^{+7.1}_{-5.9}$ days for H$\beta$, and $28.5^{+9.0}_{-8.5}$ days for H$\alpha$. Additionally, \cite{2012ApJ...755...60G} provided a delay of $25.9^{+2.3}_{-2.3}$ days for H$\beta$ based on data from 2010-2011. 
Given the stability of the $BV$ FVG across all observations, we derived host values within our aperture as $B = 2.47 \pm 0.14$ and $V = 5.15 \pm 0.13$ mJy, providing host-subtracted luminosities. For a detailed analysis of the FVG in 3C120, refer to \cite{2015A&A...581A..93R}.

\item {AKN120:} The correlation coefficient for $V$ and NB680 reveals two peaks: one at 20 days and another at 45 days, both with $\alpha \sim 0.6$. The final centroid delay, derived using our formalism, is $28.1^{+1.4}_{-1.6}$ days, also with $\alpha \sim 0.6$. The peak delay is slightly shorter at $18.6^{+23.0}_{-1.4}$ days, with larger uncertainties reflecting the secondary peak in the correlation. The low value of $R_{\rm e, FR} \sim 0.75$ is primarily due to a 20-day gap in observations and weak features in the light curves. The NB680 light curve shows only a more or less monotonic decline between a high and low constant plateaus, i.e., without clear structures and flux minima/maxima. The onset of the decline clearly lags the decline seen in the $B$ and $V$ bands by about 30 days, indicating that it is reasonable to adopt a lower limit of about 30 days for the H$\alpha$ lag (consistent with the centroid delay determined). 
 Previous spectroscopic $H\beta$ campaigns reported delays of $50^{+9.0}_{-12}$ days in 1990 and $36^{+16}_{-9.0}$ days in 1995 \cite{1998PASP..110..660P}.
\cite{2011AaA535A73H} conducted a photometric RM campaign in 2010, focusing on H$\beta$ and using the $NB$ filter $OIII$. They initially reported an H$\beta$ delay of $48.0 \pm 3.0$ days. However, after interpolating missing data points, the delay was revised to $30.2^{+1.0}_{-1.2}$ days.
The $H\alpha$ delay from this work appears significantly shorter than the previously reported $H\beta$ delays, but within the uncertainties, the peak delay could be consistent with $\tau_{H\beta} < \tau_{H\alpha}$.
The 2018 $BV$ FVG exhibits a tight slope and a blue color, with a slope of $1.07 \pm 0.03$, consistent with the slopes reported by \cite{1997MNRAS.292..273W}.

\item {CTSG03\_04:} The correlation coefficient is broad, spanning from 5 to 25 days, with a peak around 15 days and an average $\alpha \sim 0.5$, which is smaller than the $\alpha_{\rm phot}$ of approximately 0.7. The final reported delay is $t_{\rm cent} = 17.8 \pm 0.9$ days and $t_{\rm peak} = 14.8^{+0.4}_{-1.4}$ days, with a high confidence level. The $BR$ FVG shows a slope of 1.24, indicating a very blue color, consistent with the findings in the color analysis by \cite{2002A&A...384..780B}.

 \item {ESO141-G55:} 
 This source was observed in two campaigns. For the 2013 campaign, the PRM formalism revealed a broad correlation ridge spanning from 2 to 30 days and two regions can be identified: the first region centered around ten days with $\alpha>0.6$, and a secondary region extending from almost 15 to 30 days with $\alpha\sim~0.4$. Applying the formalism without restrictions, we determined a delay of $19.5^{+0.7}_{-0.5}$ days and $\alpha\sim0.5$, but showing large discrepancies between $\alpha_{\rm cent}$ and $\alpha_{\rm peak}$, due to the broad correlation and the fast increase of $\alpha$ at low delays. Therefore we additionally adjusted the search to focus on delays where $\alpha<0.6$ and found a delay of $23^{+1.0}_{-0.9}$ days with $\alpha=0.45^{+0.02}_{-0.02}$.  The spectrum shows a prominent H$\alpha$ line with an estimated FWHM of 4980$\pm$578 km s$^{-1}$, therefore $\alpha_{\rm phot} \sim$ 0.70 might be taken as an upper limit due to the broad nature of the line. 
For the 2015 campaign, when analyzing the complete light curve (with a 15-day gap), we report a delay of approximately 24 days. Due to the gap in the light curve, the value for $R_{\rm e,max}$ exceeds unity and therefore we restrict the delay focusing only in the initial part of the light curve, which reduces the delay slightly with a value of $16.1^{+1.3}_{-1.6}$ days with an $\alpha$ value of 0.39. The shorter delay in 2015 is qualitatively consistent with the smaller AGN luminosity in 2015. The averaged delay for both campaigns is $19.5^{+1.6}_{-1.8}$ days. 
The $BR$ FVGs appear stable; in 2013, the $BR$ FVG slope was approximately 1.3, and in 2015 it was around 1.4. While combining both campaigns, the slope decreases slightly to 1.07.  \cite{1992MNRAS.257..659W} report a $BR$ slope of about 1.16, consistent with our findings.
When considering the host color assumption and using the single-epoch FVGs, we derived a host $B$ value of $1.75\pm0.58$~mJy and a host $R$ value of $6.5\pm 0.45$~mJy. On average, this resulted in an exponent for the host-subtracted SED of $\beta \sim -0.64$, showing the blue character of this AGN.
After subtracting the host contribution, the interpolated flux at 5100\AA~is 11.21 $\pm$ 1.11 mJy for 2013 and 7.74 $\pm$ 0.83 mJy for 2015, indicating a decrease in luminosity of nearly 30\%. The average luminosity is $19.52\times10^{43}$erg s$^{-1}$. 
 \cite{1996ApJ...469..648W} conducted a spectroscopic RM study but encountered challenges in accurately determining the time delay for this AGN.

\item {ESO323-G77:} The correlation shows a main peak at around $30$ days. The time delay determination exhibits a 100\% confidence level, but with a low $R_{\rm e,max}$ value ($\sim 0.5$), probably due to the noisy lightcurves, therefore the delay is discarded from further analysis. The $BR$ slope is $0.53$ which indicates reddened AGN, and the slope is consistent with previous AGN $BR$ color found in \cite{1992MNRAS.257..659W, 1997MNRAS.292..273W} 0.45, 0.53 indicating that the slope remained stable over the years.

 \item {ESO374-G25:} The 2011 and 2012 lightcurves were anayzed in \cite{2012PhDT343R} with the classical RM method. For the 2011 campaign they found a time delay of $12.5_{-2.6}^{+2.0}$ days, similar to that deduced for the 2012 campaign with $11.4\pm2.7$ days by a combination of the SII and NB670 filter, and in agreement with our results. 

 \item {ESO399-IG20:} There are two possible solutions depending on the value of $\alpha$. If we restrict $\alpha$ to be greater than 0.5, the main peak occurs at around 20 days. If we allow for smaller values of $\alpha$, the main peak delay is around 40 days. We argue for the delay at 20 days with a larger $\alpha$ value due to the narrow SII filter used, which is narrower and suggests a stronger contribution of the  H$\alpha$ line. An analysis of the photometric data previously presented in \cite{2013A&A...552A...1P} deduces a delay of $18.7^{+2.5}_{-2.2}$ days, which is in agreement with this study.

\item {ESO438-G09:} This source was observed in two campaigns, both showing well structured light curves with pronounced minima and maxima, (allowing already by eye to identify a lag between the continuum filters and NB670). For 2011, we observe a broad correlation coefficient ranging from 5 days to 25 days, with a peak at 15 days. We might distinguish between a shorter delay with a higher $\alpha$ and a longer delay with $\alpha$ of 0.3. Since $\alpha_{\rm phot}$ is approximately 0.50, we discard the shorter delays but note that $\alpha$ might be higher for the 2011 epoch since the narrower SII filter was used. The delay found for 2011 is $\tau = 12.2\pm0.4$ with $\alpha=0.29^{+0.02}_{-0.02}$. The 2015 campaign exhibits a distinct peak at 12 days with an alpha of 0.35. The averaged delay for the two epochs is $12.1^{+0.4}_{-0.5}$ days. 
The $B$ flux shows a decrease of around 20\% from 2011 to 2015. 
For 2015, we found a $BR$ slope of 0.87, which is consistent with the slope found in \cite{1992MNRAS.257..659W}. 
For 2011, the $Br_s$ slope is $\sim$ 1.0, which is also consistent with a stable FVG, as the $r_s$ filter covers shorter wavelengths, resulting in an expected higher slope as $BR$.
Observing the single epoch FVGs, we found a $B$-host value of 1.12 $\pm$ 0.5 mJy for 2011 and $0.84 \pm 0.28$ mJy for 2015. Therefore, we assume a $B$-host value of $1.09 \pm 0.53$, which is the average between the minimal and maximal values for both epochs.
With $B_{host}$ we extrapolate and find a $r_{s,host}=  2.94 \pm 1.47$ mJy and $R_{host} = 3.89 \pm 1.94$ mJy.
Finally the $f_{5100}$ host-subtracted flux is  $4.1\pm 0.94$ mJy for 2011 and $2.81 \pm 0.99$ for 2015, which translates to a luminosity of  $3.77 \pm 0.87$ and $2.59 \pm 0.91\times10^{43}$erg s$^{-1}$, showing a decrease of almost 30\% between the two epochs. The average luminosity for both epochs is $3.17\pm1.23\times10^{43}$erg s$^{-1}$. 
Although we observed a decrease in the luminosity that is not obvious translated in a shorter delay. 
We note that this object is the highest accreting source in our sample according to the $R_\mathrm{Fe}$ method for estimating $\mathscr{\dot{M}}$ (see text).

\item {ESO490-IG26:} The H$\alpha$ line is dominated by narrow components but also shows a broad component. The correlation coefficient shows two possible lag solutions, with peaks at around 15 days and 40 days, both with $\alpha \sim 0.5$. Given that the duration of the light curve is approximately 80 days, the second delay is at the limit of the search range. Therefore, the delay is likely around 15 days.
Nevertheless, the delay is excluded from further analysis since the $R_{\rm e,max}$ value is around 0.5, and the confidence level is below 85\%.
The $Br_{\rm s}$ FVG is red with a slope of 0.60. We are unable to disentangle the host galaxy's contribution from the total flux, so we report an upper limit for the luminosity.

\item {ESO511-G030:} This source was observed in two campaigns. The light curves show a gap of $\sim 20$ days in both campaigns, with a mean cadence $\sim$4 days for the continuum and $\sim$6 days for the NB. In 2013, the correlation coefficient exhibits two main peaks, one at around 17 days with $\alpha\sim0.6$ and another adjacent one at 32 days with $\alpha\sim0.5$. 
Carrying out the analysis without $\alpha$ restrictions, we find $\tau_{\rm cent} = 20.6^{+0.6}_{-0.5}$ days and $\alpha = 0.54^{+0.01}_{-0.01}$, which agrees with $\alpha_{\rm phot}\sim 0.5$. 
 For 2014, without any restrictions on $\alpha$, we find $\tau = 16.4\pm0.4$ days and $\alpha=0.54^{+0.01}_{-0.01}$, in agreement with the previous campaign. The value for $\alpha_{\rm phot}$ for this campaign is reduced to $\sim0.4$ since the NB670 reduces by a factor of 20\% but the $B$ and $R$ fluxes remained stable. Nevertheless, this fact is not reflected in the recovered $\alpha$ in the correlation and we are not able to restrict further $\alpha$ value due to the broad nature of the correlation. We report an average delay of $\tau=18.5^{+0.7}_{-0.64}$ days. 
The $BR$ FVG slope for 2013 is well defined with $0.70\pm0.06$ showing a slightly red AGN color with $B_{\rm host}=1.39\pm0.28$mJy and $R_{host}=4.96\pm0.41$mJy. On the other hand, the 2014 FVG is not well covered but the 2013-2014 combination yields a slope of $0.61\pm0.07$ with a host value in $B$ of $1.22\pm0.35$ mJy and $R_{\rm host} = 4.35\pm0.58$mJy. For 2014 $BV$ yield $B_{\rm host}=1.65\pm0.36$mJy and $V_{host}=3.43\pm0.24$. Since the intersection between the different epochs shows consistent results we average those results and interpolate for the filters $V$ and $R$ to obtain the average host values. We report $B_{\rm host} = 1.42\pm0.33$~mJy, $V_{\rm host} = 2.96\pm0.70$~mJy and $R_{\rm host} = 4.05\pm1.00$~mJy, yielding an interpolated flux $f_{5100}$ of $1.56\pm0.58$mJy and $1.61\pm0.60$ mJy for 2013 and 2014 respectively, which translates to an average luminosity of $1.23\pm0.65\times10^{43}$erg s$^{-1}$. 
An X-ray study by \cite{2021ApJ...908..198G} reports an accretion rate of 0.004-0.008 Eddington units. Similarly, our analysis of $\dot{\mathscr{M}}_{R\mathrm{Fe}}$ and $\dot{\mathscr{M}}$ also indicates a low level of accretion for this source.

\item {ESO549-G49:} There is a possible solution indicating a delay of around 7 days with $\alpha \sim 0.5$, but the $R_{\rm e,max}$ value is quite low, so this delay is not included in the final results. Additionally, the H$\alpha$ line is dominated by narrow components, with the broad component being extremely weak, making the $\alpha \sim 0.5$ value unrealistic for this object. Further, the FVG is poorly constrained.

\item {ESO578-G09:} The H$\alpha$ line is well covered by the NB filter and exhibits a very broad component with an FWHM of approximately 5000 km s$^{-1}$. The correlation shows three peaks at around 10, 20, and 30 days, with $\alpha$ values of 0.7, 0.5, and 0.4. Given that $\alpha_{\rm phot} \sim 0.5$, we restrict the $\alpha$ value to around 0.5 and report a centroid delay of $t_{\rm cent} = 19.5 \pm 0.6$ days. 
The $BR$ FVG is slightly red with a slope of 0.65 but is tight and shows a clear intersection with the host galaxy's color, allowing us to report a robust host-subtracted luminosity.% We note that the host galaxy in this source is viewed relatively edge-on.

\item {F1041:} The correlation analysis yield a lag between 5 and 25 days with $\alpha \sim 0.5$, in agreement with $\alpha_{\rm phot} \sim 0.5$. The H$\alpha$ emission line is well covered by the NB filter and exhibits a strong broad component with FWHM $\sim~3600$ km s$^{-1}$. The FVG is slightly red with a slope of 0.75 for the $BR$ combination. The poor quality of the 6dF spectra around H$\beta$ prevents us from estimating the accretion rate based on the $R_{\rm Fe}$ method.

\item {HE0003-5023:} The confidence level in the lag determination is high; however, substantial differences in delays arise depending on the filter used for the continuum. The continuum light curve is particularly noisy at the beginning of the campaign, which can lead to spurious correlations, such as those observed at very short time lags. Even after removing the noisy data, discrepancies in the lags persist, leading us to exclude this object from further analysis. Additionally, $\alpha$ values vary between 0.7 and 0.3 depending on the filter used. We also lack spectrosopic data to properly assess and corroborate the emission-line contribution to the band. The FVG is well constrained for this source, and we report a host-subtracted luminosity.

\item {HE1136-2304:} The object is classified as a CL type and has been observed in three different campaigns: 2015, 2016, and 2018. The time delay reported from the 2015 campaign has a high confidence level, with a delay of approximately 10 days, consistent with the value reported by \cite{kollatschny18}. However, the other campaigns exhibit lower confidence levels due to their shorter duration and are not included in the results for multi-epoch objects.
We constructed a multi-epoch FVG and observed a change in the AGN's slope (the source became bluer in 2018; see text).
The spectroscopic reverberation mapping study by \cite{2018A&A...618A..83Z} reported a delay of $15.0^{+4.2}_{-3.8}$ days for the integrated H$\alpha$ line and $7.5^{+4.6}_{-5.7}$ days for H$\beta$. Our narrowband filter only covers the bluer part of the H$\alpha$ line, so the delays reported are consistent with these findings. For further details on this object, refer to \cite{kollatschny18} and \cite{2018A&A...618A..83Z}.

\item {HE1143-1810:} The H$\alpha$ emission line is well covered by the narrowband filter, with a FWHM of approximately 2000 km~s$^{-1}$. The correlation analysis reveals a  delay between 15 and 20 days with $\alpha$ around 0.5. The $BV$ FVG is well constrained and exhibits a blue characteristic with a slope close to unity. An X-ray study by \cite{2020A&A...634A..92U} reports an accretion rate of 0.7-0.9 Eddington units, suggesting high accretion activity. However, our analysis of $\dot{\mathscr{M}}_{R\mathrm{Fe}}$ indicates low accretion, while $\dot{\mathscr{M}}$ shows moderate accretion. 

\item {HE2128-0221:} The broad component of the H$\alpha$ line is weak, with an FWHM less than 2000 km s$^{-1}$, showing a predominance of narrow lines. The correlation is modest, with $R_{\rm e,max} \sim 0.7$, showing a primary peak around 10 days with $\alpha \sim 0.5$. This value might be overestimated given the nature of the broad component in the spectra. Additionally, there is a secondary peak at 15 days with a lower $\alpha \sim 0.3$, and a potential further delay at 25 days with an even lower $\alpha$. Due to the low $R_{\rm e,max}$ values at these delays, we report the main peak at approximately 10 days. The FVG is well-constrained and slightly red. This object belongs to the class of moderate accretors based on the $R_{\rm Fe}$.

\item {IC4329A:} The correlation coefficient presents two main peaks: one at 15 days with $\alpha\sim 0.5$ and another at 25 days with $\alpha\sim 0.4$. The  centroid delay that we report is 23 days, with a peak delay at 14 days. \cite{1996ApJ...469..648W} analyzed spectra from 1992, reporting an upper limit delay of 25 days for H$\alpha$, which aligns with our findings. Recently, \cite{2023ApJ...944...29B} conducted a spectroscopic RM campaign in 2022, finding an H$\beta$ delay ($\tau_{H\beta}$) of $16.3^{+2.6}_{-2.3}$ days, slightly shorter than the centroid H$\alpha$ delay reported in this work.
The $BR$ FVG shows a very red AGN slope of $0.35 \pm 0.02$, with no possible intersection with the adopted host color. The FVG highlights the extremely red nucleus of this source, which is also identified as an example of an extremely reddened galaxy in \cite{1992MNRAS.257..659W}, exhibiting a high range of extinction values ($A_V$ of 2-6 mag). Therefore, the luminosity presented in Table~\ref{tab:results_good}, albeit not being host subtracted, could be significantly under-estimated due to poorly constrained but substantial extinction.

\item {IRAS01089-4743:} We do not report a delay for this source due to a low $R_{\rm e,max}$ value and low confidence level. The $BR$ FVG is not well constrained, nevertheless we report a host-subtracted luminosity. The spectrum is strongly dominated by narrow lines, with a very weak broad component, as inferred also by $\alpha_{\rm phot} \sim 0.3$.

\item {IRAS09595-0755:} The main correlation-coefficient peak does not yield physical values for $\alpha$, thus no delay is reported. The H$\alpha$ line, well covered by the NB filter, displays prominent narrow lines alongside a substantial broad component with an FWHM of approximately 2400 km s$^{-1}$. The FVG is notably blue, with a slope of around 1.4 and a tight intersection with the host color. Consequently, we report a host-subtracted luminosity for ths source.

\item {IRAS23226-3843:} From the spectrum, the H$\alpha$ line is dominated by narrow-line emission, with the broad component being very weak; hence, we do not report a FWHM value for the H$\alpha$ line. The weakness of the broad component is manifested by $\alpha_{\rm phot}$, which is around 0.25. The correlation coefficient shows a peak at approximately 5 days with $\alpha$ around 0.5. There is another peak at 15 days with a smaller $\alpha$, but the correlation coefficient $R_{\rm e}$ is low, almost 0.7, so we do not report this delay. The FVG is well constrained and shows a red slope of approximately 0.7. Although IRAS23226-3843 is classified as Sy1 object, it is known to change its spectral type and is classified as CL. For details on this source, refer to \cite{2020A&A...638A..91K,2023A&A...670A.103K}.

\item {MCG+03\_47-002:} The correlation coefficient indicates a delay at $\sim 20$ days with $\alpha\sim 0.5$, which aligns with $\alpha_{\rm phot}$. The spectra do not clearly display the line, so we do not report a FWHM measurement. The $BR$ FVG slope shows a blue trend, but the intersection with the host color is unclear. Consequently, the host-fraction for this source is likely overestimated, and the actual AGN luminosity is higher than what is reported in Table~\ref{tab:results_good} (note also the large uncertainties in the luminosity).

\item {MGC-02.12.050:} The peak delay is at $\sim 10$ days with $\alpha \sim 0.5$, which is consistent with $\alpha_{\rm phot} \sim 0.43$. The value for $R_{\rm e,max}$ is less than 0.7, and the ratio $R_{\rm e, FR}$/$R_{\rm e, FP}$ is low. Therefore, we exclude this object from further analysis. The AGN slope appears reddened, but we have limited matching points in the light curves. The total flux falls within the adopted range for the host-color, so we do not report a host-subtracted luminosity for this source. Instead, we provide an interpolation between the observed fluxes.

\item {MRK1239:} We could not find any significant delay/meaningful value for $\alpha$. Due to the low variability, the FVG is also poorly constrained.

\item {MRK1347:} The H$\alpha$ line emission is well covered by the NB filter and exhibits prominent narrow lines, with the FWHM for the broad component approximately 1500 km s$^{-1}$. While \citet{2006AaA...455..773V} classifies it as a S1 galaxy, \cite{2016MNRAS.462.1256C} define it as a NLS1, noting a FWHM of the broad H$\beta$ component of 1614 km s$^{-1}$, similar to our findings. 
The correlation coefficient shows two peaks: one between 5 and 10 days with $\alpha > 0.6$ and another at 20 days with $\alpha \sim 0.45$. The second peak agrees better with $\alpha_{\rm phot}$, which is around 0.5. The FVG slope of 0.7 indicates a slightly red nucleus.

 \item {MRK335:} For the 2010 campaign, we find 2 peaks, one at 5 days with no physically meaningful $\alpha$ value and one at 20 days with $\alpha \sim 0.7-0.8$, with the result agreeing with $20.5_{-2.8}^{+2.0}$ days reported in \cite{2011AaA535A73H}. Due to the short duration of the NB light curve, the $R_{e,\max}$ value is low, and therefore we do not obtain a high confidence level for this dataset. As a result, we do not include it in the final analysis, but for details on this observing campaign, refer to \cite{2011AaA535A73H}. For the 2011 campaign, the same pattern occurs: we find a peak at 20 days with $\alpha \sim$ 0.7, consistent with the previous campaign. Due to the duration of the campaign, as before, the confidence level is not high. On the other hand, for 2014 the $R_{e}$ correlation obtained is broad and 3 peaks can be distinguished. One at 5 days with an unphysical value for $\alpha$, another at around 15 days with $\alpha \sim$0.6 and an additional peak at around 20 days with $\alpha\sim0.5$.

\item {MRK509:} The correlation coefficient shows a peak at 15 days, but the associated $\alpha$ value exceeds unity, and hence is unphysical, and therefore discarded. Physical peaks are observed at 25 days with $\alpha \sim 0.6$, and at 35 days with $\alpha$ between 0.4 and 0.5. Since $\alpha_{\rm phot}$ is approximately 0.75, we favor the 25-day delay.
In comparison, the $H\alpha$ delay reported here is $22.9^{+0.8}_{-0.8}$ days, which is notably shorter than the H$\beta$ lag of $76.0 \pm 7.0$ days reported by \cite{1998PASP..110..660P} and \cite{2004ApJ...613..682P}, who used much sparser light curves with an average sampling interval of 7 days. If we assume that the H$\alpha$ lag is on average a factor 1.3 longer than H$\beta$, then the predicted H$\alpha$ lag is 103 days. The time span of our light curves in 2014 is about 150 days, which is only marginally longer than that predicted H$\alpha$ lag. 
The $BR$ FVG slope is approximately 0.90, slightly lower than the 1.2 reported by \cite{1992MNRAS.257..659W,1997MNRAS.292..273W}. Additionally, \cite{2002A&A...384..780B} found that in MRK509, the continuum emission from the nucleus contributes nearly 60\% to the total flux, which is consistent with the results from the $BR$ FVG.

\item {MRK705:} The NB filter does not fully cover the broad H$\alpha$ line, with the blue wing missing, though the $\alpha_{\rm phot}$ is approximately 0.52. The correlation-coefficient ridge is broad spanning the range 5-25 days, with a peak at 11 days. A secondary peak appears around 50 days, which is associated with an unphysical value for $\alpha$ and hence discarded. 

\item {MRK841:} The H$\alpha$ line in Mrk 841 exhibits a very broad component with a FWHM of approximately 4500 km s$^{-1}$. Its core is well covered by the NB filter, although the signal in the extended line wings is  not fully captured. The correlation coefficient $R_{\rm e}$  reveals two peaks: one around 20 days and another at 30 days, with corresponding $\alpha$ values of 0.5 and 0.4, respectively (see Figure~\ref{fig:corr_example}). These values are somewhat lower but consistent with $\alpha_{\rm phot}\sim$ 0.6 for the broad line. The $BV$ FVG analysis shows a well-constrained slope near unity, indicating a typical blue AGN. The $R_{\rm Fe}$ analysis suggests a low accretion rate for this object.
Mrk841 is part of the Lick AGN Monitoring Project \citep{2015ApJS..217...26B}, and a previous study reported an H$\beta$ delay of $11.2^{+4.8}_{-1.9}$ days \citep{2022ApJ...925...52U},  indicating a (non-contemporaneous) H$\alpha$ to H$\beta$ delay ratio of approximately 2.

\item {NGC1019:} The $R_{\rm e}$ correlation coefficient exhibits a peak around 10 days, with $\alpha\simeq 0.75$. This value for $\alpha$ reasonable given that  line emission in this source was traced using the SII filter, which is narrower than NB6700 and captures a higher emission line fraction (i.e.,, relatively less continuum). However, since the maximum $R_{\rm e, max}$ value is below 0.7, the source is excluded from the final analysis.

\item {NGC4726:} The correlation coefficient shows a peak around 15 days, with an $\alpha$ value of approximately 0.7. The $\alpha_{\rm phot}$ value is below 0.3, reflecting the weak nature of the H$\alpha$ line in the mean spectrum, as also evident in the 6dF spectrum where no line could be reliably fitted. This delay is discarded due to the low confidence level and the low $R_{\rm e, FR}/R_{\rm e, TP}$ ratio. Additionally, the FVG is not well constrained.

\item {NGC5940:} The correlation coefficient shows a primary peak at 5 days with an $\alpha$ of approximately 0.6, while $\alpha_{\rm phot}$ is 0.5. The H$\alpha$ line is well covered by the NB filter and exhibits a broad profile with a FWHM of around 4000 km s$^{-1}$.
A previous spectroscopic RM campaign reported an $H\beta$ delay of $5.70^{+0.90}_{-0.82}$ days in \cite{2013ApJ...769..128B}, consistent with the delay found in this study.
The FVG for the $BR$ filters is around unity and well constrained. A moderate accretion rate, $\mathscr{\dot{M}}_\mathrm{R_Fe}$, is implied.

\item {NGC6860:} The correlation coefficient shows a peak at around 35 days, but since the corresponding $R_{\rm e, FR}$ value is approximately 0.6, we exclude this object from further analysis.

\item {NGC7214:} The correlation coefficient shows a peak around 5 days with an $\alpha$ value of approximately 0.6. The core of the H$\alpha$ line is well covered by the SII filter, making this value plausible. While $R_{\rm e, max}<0.7$, the ratio of $R_{\rm e, FR}/R_{\rm e, TP}$ is high ($>1.6$), and we include it in our final analysis. The FVG is slightly red, with the $Br$ slope around 0.70.

\item {NGC7469:} Previous spectroscopic reverberation mapping studies  reported an H$\beta$ delay of $4.5^{+0.7}_{-0.8}$ days and an H$\alpha$ delay of $4.7^{+1.6}_{-1.3}$ days \citep{1998ApJ...500..162C}. A later campaign \citep{2014ApJ...795..149P} observed an H$\beta$ delay of $10.8^{+3.4}_{-1.3}$ days.
Here, the $R_{\rm e}$ correlation exhibits a peak around 15 days, with $R_{\rm e} \approx 0.8$ and a corresponding $\alpha$ value of approximately 0.3. This $\alpha$ value is consistent with the weak broad component observed in the spectra, where the H$\alpha$ line is primarily dominated by narrow lines. The FVG indicates that the AGN is blue, with a $BV$ slope of 1.27.

\item {NGC7603:} The H$\alpha$ line is not covered in full by a single NB filter, and was therefore monitored using two NB filters. In the spectrum, the broad component of the line is prominent, with a FWHM of approximately 6000 km s$^{-1}$, with $\alpha_{\rm phot} \simeq 0.4$. The $V$-band was used as a tracer of the continuum emission since the $B$ filter does not cover the entire observing season. The correlation coefficient reveals a peak at 13 days with $0.4< \alpha <0.5$. A secondary peak exists around 40 days with $\alpha\simeq 0.25$. Both peaks exhibit high correlation coefficient values, with $R_{\rm e, max} \sim 0.95$. The delay between the $V$ band and NB670/NB680 is $12.6^{+1.0}_{-4.0}$\,days/ $12.8^{+0.4}_{-0.2}$\,days, with $\alpha=0.40^{+0.10}_{-0.03}/0.44^{+0.01}_{-0.01}$. When restricting the search to the lower values of $\alpha$, the centroid delays are $35.1^{+1.5}_{-1.3}$\,days and $35.4^{+0.8}_{-1.1}$\,days for the NB670 and the NB680 bands, respectively. Because the center of the  H$\alpha$ line is not completely covered, i.e., likely lost at the edges of the two NB filters, one expects values for $\alpha$ which are smaller than $\alpha_{\rm phot}=0.4$. This indeed argues in favor of the longer delay. This delay is also consistent with the delay of $41.7^{+3.7}_{-9.0}$ days obtained by \cite{blex} who had analyzed the NGC7603 light curves with the recipes in \cite{2012PhDT343R}, see ESO374-G25 above. 
In a previous 20-year monitoring, \cite{2000A&A...361..901K} had discovered line profile variations and that the lines became more asymmetric and exhibited a stronger Balmer decrement when the galaxy
was in a lower state. Our recent spectra show quite symmetric line profiles, indicating a higher activity state. The $BR$ FVG exhibits a blue color with a slope of approximately 1.0 and the accretion rate estimated from $R_{\text{Fe}}$ is moderate.

\item {NGC985}: At redshift z$=$0.04314 the NB680 filter covers almost exactly the blue half of the strong broad H$\alpha$-line, with $\alpha_{\rm phot} = 0.77$. The $R_{\rm e}$ correlation shows a main peak at 24 days with $\alpha\sim 0.7$, which is consistent with $\alpha_{\rm phot}$. The $R_{e}$ correlation shows two additional correlation peaks at 50 and 60 days but with $\alpha\sim0.4$, and with $R_{\rm e,max}<0.6$, and are therefore discarded. 
For comparison, monitored as part of the LAMP project \citep{2015ApJS..217...26B}, \cite{2022ApJ...925...52U} reported a H$\beta$ delay of $7.8^{+10.1}_{-9.8}$ days, based on a 220 days long monitoring with a median sampling of 8 days.

\item {PG1149-110:} The H$\alpha$ line is well covered by the NB, with $\alpha_{\rm phot}\simeq 0.4$. However, the correlation scheme does not lead to a clear lag, as  $R_{\rm e}<0.5$ and  $\alpha \simeq 1$. Consequently, this delay is excluded from the final analysis. The $BV$ FVG slope is notably blue, with a value of 1.37~$\pm$~0.16.

\item {PGC50427:} This source was observed in two campaigns, both showing well structured light curves with pronounced minima and maxima. For the 2011 campaign we find a robust lag of $21.4^{+0.9}_{-0.9}$ days with $\alpha\sim0.65$. 
For the 2014 campaign we obtain a broad correlation ridge extending from 3 days to 18 days with $\alpha>0.6$, and a second ridge extending from 18 until 30 days with $\alpha<0.6$. Without restricting $\alpha$, we obtain a delay of $14.2^{+1.6}_{-0.9}$ and $\alpha=0.74^{+0.04}_{-0.04}$. The $\alpha_{\rm phot}$ of 0.58 for the 2014 campaign (observed with NB670) sets a limit to the possible $\alpha$ values. Since the SII filter (2011 campaign) is narrower the H$\alpha$ variation carried within it should be intuitive larger than in the NB670 filter, therefore for the 2014 dataset we can set the restriction of $\alpha<0.6$ and find $\tau = 20.2^{+0.6}_{-0.4}$ and $\alpha=0.51^{+0.03}_{-0.02}$, being similar to the result in the first campaign.  
The average delay for both campaigns delay is $17.8^{+1.8}_{-1.3}$ days without any restriction for $\alpha$. Requiring that $\alpha<0.6$ for 2014 results on average $\tau = 20.8^{+1.1}_{1.0}$ days. 
The photometric datasets from the 2011 and 2014 campaigns were also analyzed in \cite{2015AaA576A73P}, where they report $\tau=20.4^{+0.4}_{-1.0}$ and $\tau=18.7^{0.6}_{-1.6}$ days, respectively, and in agreement with the results reported here using a different PRM scheme. 
Taking into account the AGN and host slope intersection for $Br_s$ and $BR$ we derived an average $B_{\rm host} = 0.62\pm0.20$~mJy, extrapolating we find $r_{s,\rm host} = 1.67\pm0.53$ mJy and $R_{\rm host} = 2.21\pm0.70$~mJy. Interpolating the host subtracted fluxes we report $f_{5100} = 1.2\pm0.41$ for 2011 and $f_{5100} = 1.42\pm0.44$mJy for 2014 with an average luminosity of $1.12\pm0.40\times10^{43}$erg s$^{-1}$, in agreement with \cite{2015AaA576A73P} with $\sim1.20\times10^{43}$erg s$^{-1}$.

\item {PGC64989:} This source was observed in two campaigns (2013 and 2014). 
The first campaign does not yield a reliable delay due to a low confidence level. For the second campaign we obtained a robust delay, based on the correlation between $R$ and the NB filter; we used $R$, because the $B$ light curve covers only the first half of the campaign. 
Regarding the AGN luminosity, we performed the FVG analysis for each campaigns yielding consistent $BR$ slopes close to 1 and a host component of about 2~mJy in the $B$ band. 
In the $R$ band, however, the host components of 2013 and 2014 differ, being  7.25 and 7.5 mJy, respectively.
The difference may be explained by a strong contribution of the H$\alpha$ line to the $R$ band in 2014 (a similar shift was observed for 3C 120 by \citet{2018AaA620A137R}, their Fig. 6).
Indeed, for PGC64989 the NB670 light curves show an increase by about 2.5 mJy which corresponds to 0.25 mJy in R, taking into account that the bandpass of NB670 is a factor 10 smaller than in $R$. 
Given that the $B$ data points are missing during the brightest phase, the listed AGN luminosity of PGC64989 may be underestimated by about 10\%. We show also the FVG analysis of the combined 2013 and 2014 data, 
which however leads to a questionable red slope.

\item {RXSJ06225-2317:} The broad component of the H$\alpha$ line is weak, with a FWHM of $\sim 1500$ km s$^{-1}$. The  correlation shows a clear peak at 20 days with $\alpha\simeq \alpha_{\rm phot} \sim 0.4$. The FVG is well constrained.

\item {RXJ1103.2-0654:} Due to the low source variability and the unreliable FVG analysis, we discard this source from further analysis.

\item {RXSJ17414+0348:} This source was observed in two campaigns. The H$\alpha$ line is well covered by the NB670 filter, with $\alpha_{\rm phot}\sim 0.6$ for the 2014 campaign. During the 2012 campaign, a peak is observed at 16 days with $\alpha \sim 0.5$. For the 2014 campaign, the correlation coefficient is broader, spanning between 15 and 25 days. The $R_{\rm e}$ value for this campaign exceeds unity due to the long observational gap and lack of overlapping data towards the end; however, the delays are still well-recovered.
The FVG is well constrained, and although the matched data for the 2012 campaign is less consistent, the host values are consistent with those from 2014, and a host-subtracted luminosity is reported.

\item {UGC12138:} The SII filter covers well the H$\alpha$ line, which is dominated by narrow components but also features a strong broad component with a FWHM$\simeq 2600$ km s$^{-1}$. The correlation coefficient shows a clear peak at 15 days with $\alpha \sim 0.6$, but with $R_{e,FR}<0.7$. We therefore do not include the source in the final analysis (Table~\ref{tab:results_good}). The FVG is well-constrained and indicates a blue nucleus.

\item {UM163:} The 6dF spectrum is dominated by narrow lines and the broad component appears weak, consistent with $\alpha_{\rm phot} \sim 0.5$. There is a significant increase in $\alpha$ for the at short lags, ranging from 0.9 to 0.45. A secondary peak appears at 40 days, but the value of $\alpha$ is unphysical. We report a delay despite the high $\alpha$ value. The FVG is well-constrained.

\item {WPVS48:} We present data for three  campaigns. The photometric data for the first two campaigns were previously analyzed in the context of dust RM in \cite{2014A&A...561L...8P}, and a combined dust and BLR RM in \cite{2018rnls.confE..57S}. \cite{2014A&A...561L...8P} reports NIR delays of $64\pm4$ and $71\pm5$ days for $J$ and $K$, respectively.
The delays and luminosities remain stable across all campaigns, with an average $\tau_{\rm cent}$ of $20.3\pm4.6$ days, considering the maximum and minimum values for the three campaigns. The last campaign shows a significant gap in the data, causing the correlation coefficient $R_{\rm e}$ to exceed unity and the delay to have larger errors.
The FVGs appear stable; however, the combined $BV$ FVG does not provide a host value. We obtained the host values by combining the individual FVGs, which are all in agreement.
The FWHM is less than 2000 km s$^{-1}$. Nevertheless, the accretion rate based on $R_{\rm Fe}$ implies low accretion rates.

\item {WPVS007:} The SII band does not adequately cover the H$\alpha$-line core, but does include the blue wing. The correlation coefficient is broad, with $0.6<R_{\rm e,max}<0.7$, peaking at 12 days. There is a secondary peak at 25 days, but with an unphysical value for  $\alpha$. The high $R_{\rm Fe}$ indicates a high-accreting object. The $BV$ FVG does not suggest a red nucleus, since the AGN slope is steep, with a value of $1.08 \pm 0.08$.

\end{enumerate}

\section{Validation of Time Delay Results}\label{app:peak_rl}

\begin{figure*}
    \centering
    \includegraphics[width=1.0\textwidth]{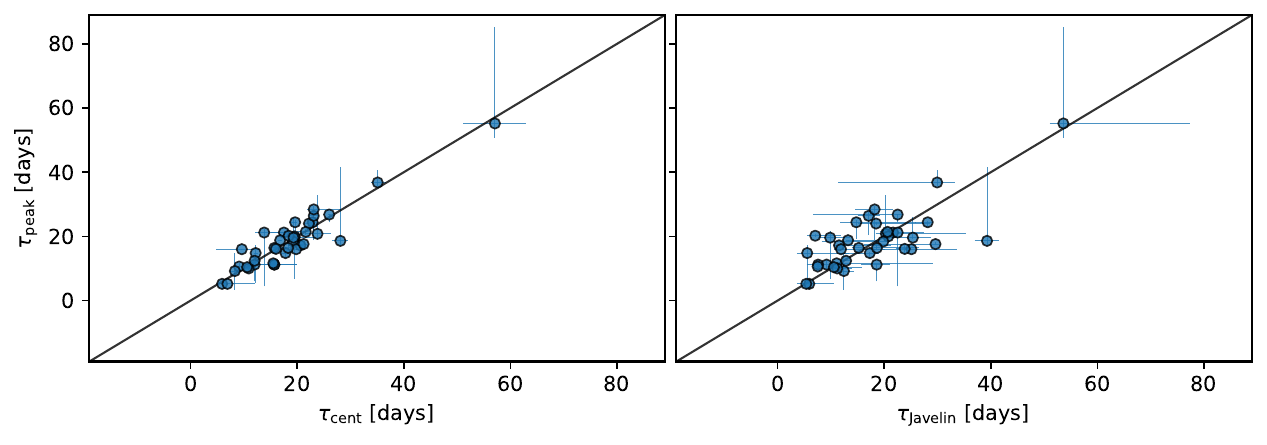}
    \caption{{Left: Peak values versus centroid values obtained from our formalism. Right: Peak lag values versus lags derived from Javelin's software. A one-to-one ratio is depicted in both plots.}}
\label{fig:delay_comparison}
\end{figure*}
We replicated the analysis conducted in the main paper, which used centroid-lag values, by applying the same methodology to peak-lag values.  {Additionally, time-delays were obtained using the JAVELIN code in photometric mode \citep{2016ApJ...819..122Z}}, which is an independent approach to the one introduced here. {Figure~\ref{fig:delay_comparison} compares the lag values obtained by the various approaches, which are all in good agreement.}

{To further benchmark the different methods, we study the size luminosity relation obtained by each. Specifically, the scatter in the $r-L$ relation, as obtained using the peak-lag values} is slightly larger than that deduced for the centroid values with $\sigma=0.34$\,dex (see section~\ref{sec:rl}). Restricting the fit to sources in our sample yields $\gamma = 0.16 \pm 0.06$, while further restricting the fit to the 29 most luminous sources in our sample with $\lambda L_\lambda(5100\,\text{\AA})>1.5\times 10^{43}\,\mathrm{erg~s^{-1}}$ yields a slope $\gamma = 0.49 \pm 0.13$. 
{Regarding JAVELIN's results, we found a scatter of $\sigma=0.33dex$ and $\gamma = 0.25\pm0.07$ for all sources, and $\gamma = 0.52\pm0.14$ for the most luminous sources.} These results are consistent with those obtained from the centroid analysis. 

{In light of the simulations carried out in Appendix \ref{app:simulations}, which show a potential time-delay bias due to finite interband continuuum time-delays, Figure~\ref{fig:alpha_ratio_lag} shows the relative deviations between the deduced delays using the centroid scheme ($\tau_{\rm PRM}$) and those expected from the size-luminosity relation ($\tau_{\rm Fit}$) as a function of the fraction of the broad-H$\alpha$ contribution to the flux in the NB. Evidently, there is no clear dependence between the quantities, as might be expected by versions of the model considered here.}

\begin{figure*}
\centering
\includegraphics[width=1.0\textwidth]{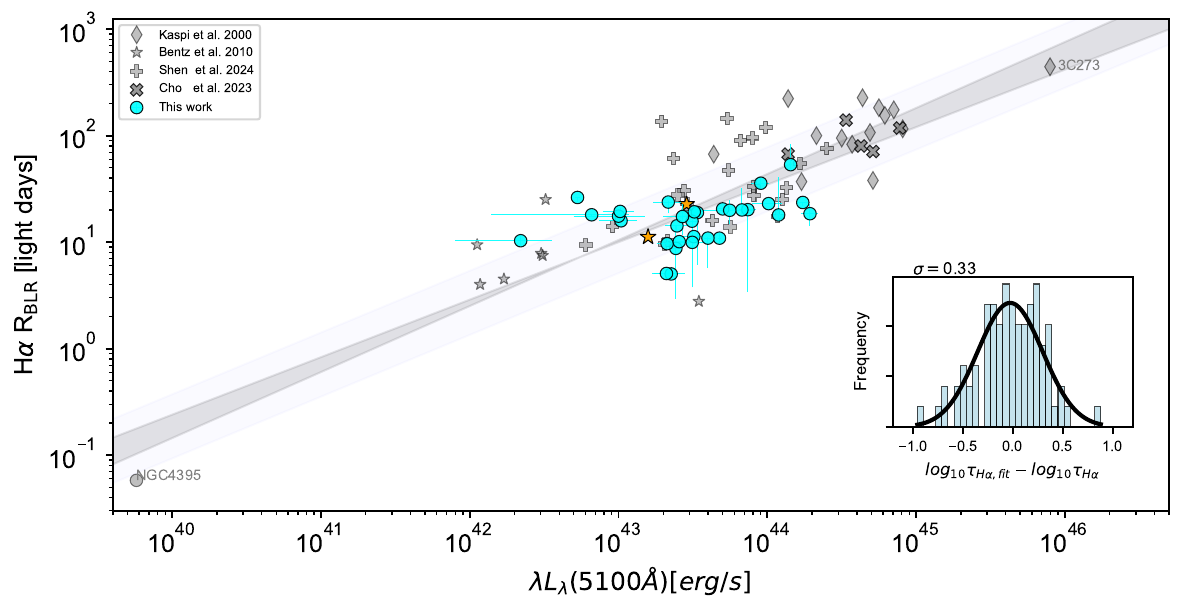}
\caption{Same as Figure~\ref{fig:r_l_ha} but for peak values.}
\label{fig:r_l_ha_peak}
\end{figure*}

\begin{figure*}
\centering
\includegraphics[width=1.0\textwidth]{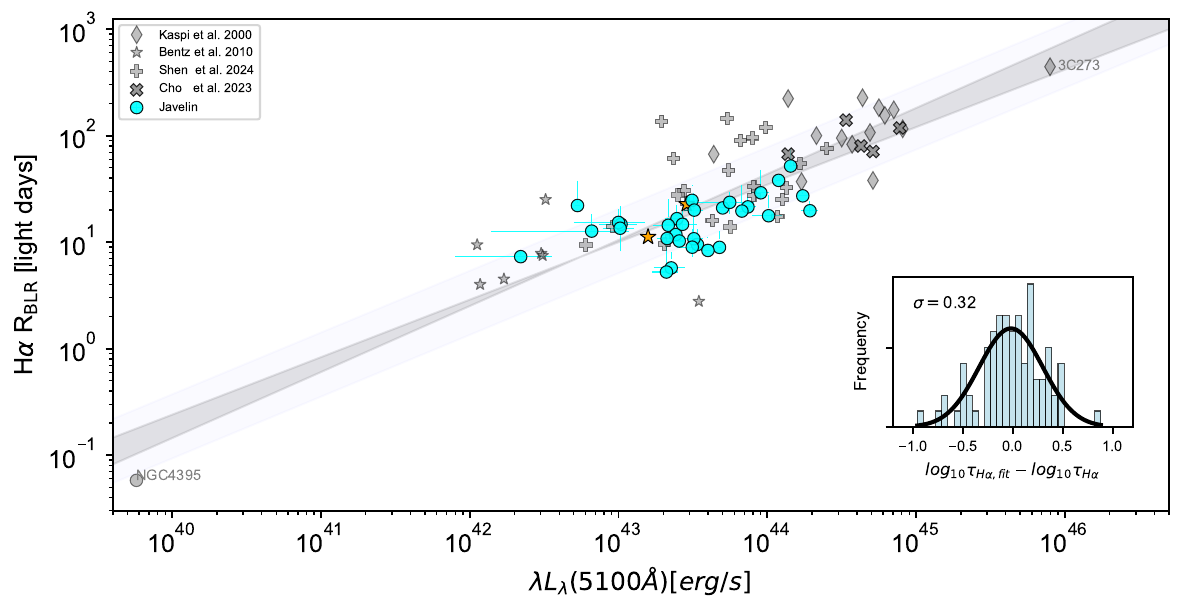}
\caption{{Same as Figure~\ref{fig:r_l_ha} but for delay values obtained with JAVELIN.}}
\label{fig:r_l_ha_javelin}
\end{figure*}
\begin{figure*}
\centering
\includegraphics[width=0.6\textwidth]{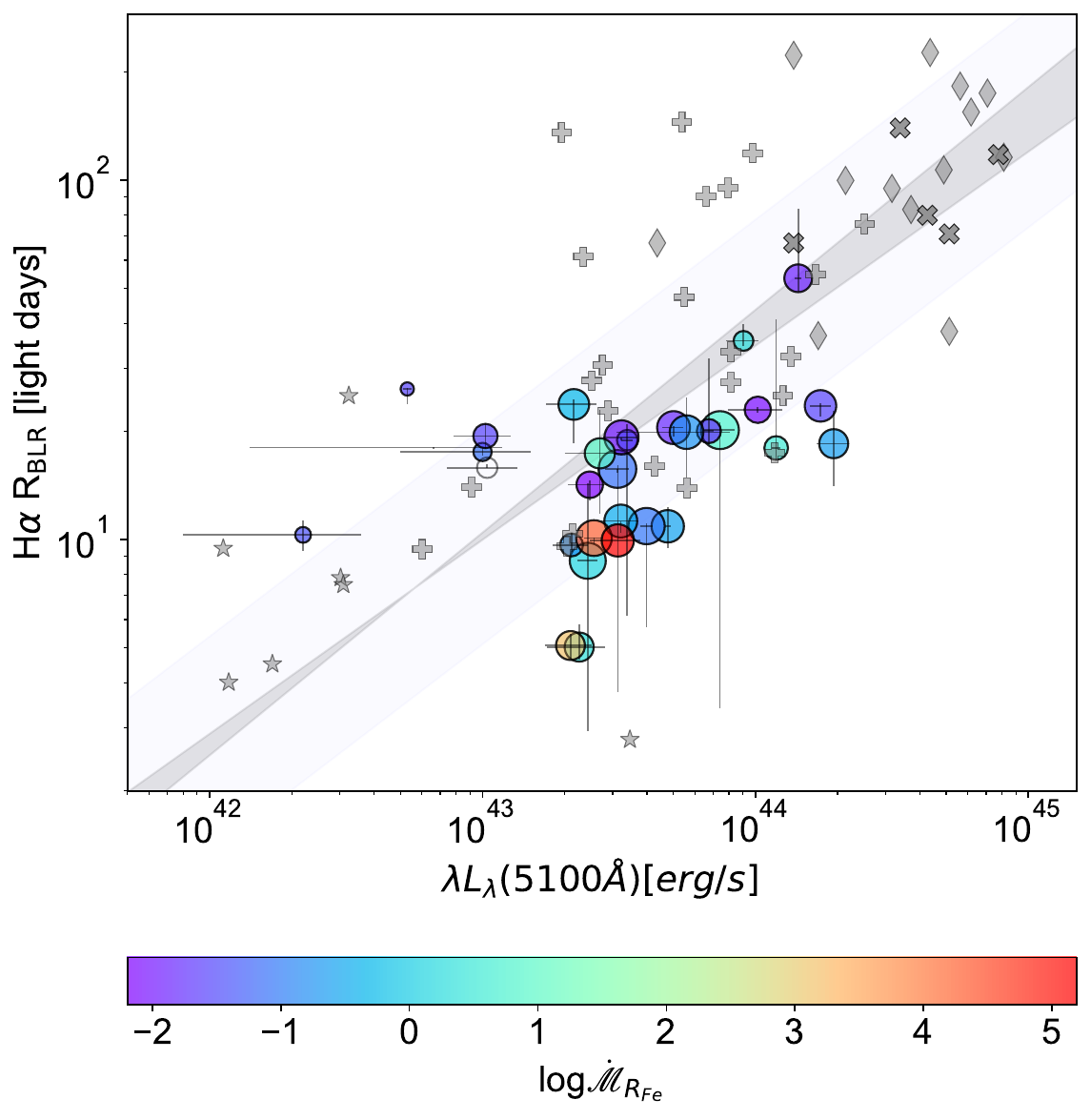}
\caption{Same as Figure~\ref{fig:r_l_acc} but for peak values.}
\label{fig:r_l_acc_peak}
\end{figure*}

\begin{figure*}
    \centering
    \includegraphics[width=0.7\columnwidth]{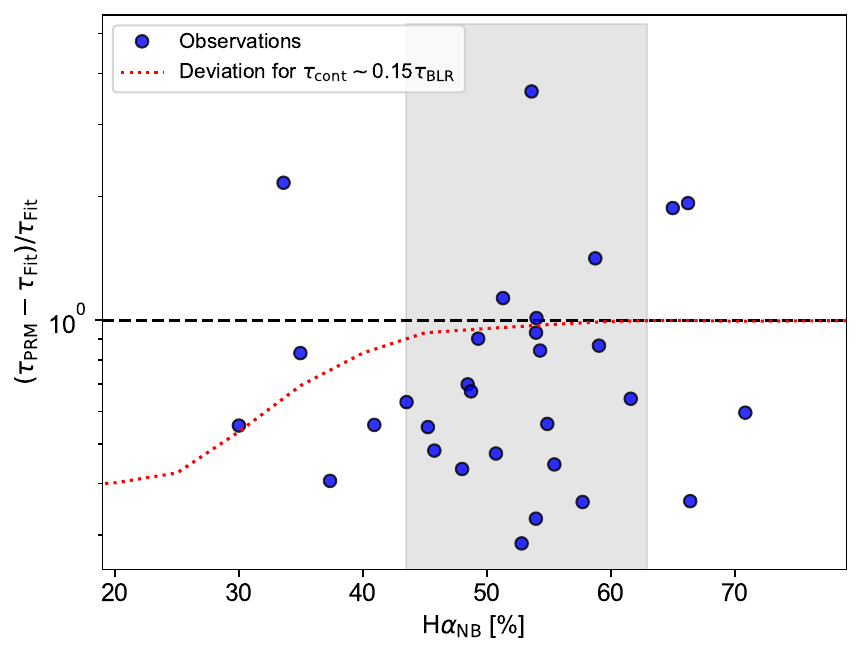}
    \caption{{Difference between the observed delays ($\tau_{\rm PRM}$) and expected delays from the $r_{\rm BLR}$-L best fit ($\tau_{\rm Fit}$) versus the fraction of the H$\alpha$ line within the NB filter in per-cent (H$\alpha_{\rm NB}$). The data derived from the RM formalism is represented by blue circles, while the estimation of the lag deviation due to continuum contamination (where the delay of continuum is about 15\% of the line delay) is marked with a red dotted line. The mean value for the H$\alpha$ contribution, along with the standard deviation, is highlighted as a grey area. }}
    \label{fig:alpha_ratio_lag}
\end{figure*}





\bibliographystyle{aasjournal}
\bibliography{aas_prm_revised} % if your bibtex file is called example.bib

%\begin{thebibliography}{99}
%\end{thebibliography}

\end{document}